# Physics Briefing Book

*Input for the 2026 update of the European Strategy for Particle Physics*

**Electroweak Physics:** Jorge de Blas[1], Monica Dunford[2] *(Conveners)*, Emanuele Bagnaschi[3] *(Scientific Secretary)*, Ayres Freitas[4], Pier Paolo Giardino[5], Christian Grefe[6], Michele Selvaggi[7], Angela Taliercio[8], Falk Bartels[2] *(Contributors)*

**Strong Interaction Physics:** Andrea Dainese[9], Cristinel Diaconu[10] *(Conveners)*, Chiara Signorile-Signorile[11] *(Scientific Secretary)*, Néstor Armesto[12], Roberta Arnaldi[13], Andy Buckley[14], David d'Enterria[7], Antoine Gérardin[15], Valentina Mantovani Sarti[16], Sven-Olaf Moch[17], Marco Pappagallo[18], Raimond Snellings[19,83], Urs Achim Wiedemann[7] *(Contributors)*

**Flavour Physics:** Gino Isidori[20], Marie-Hélène Schune[21] *(Conveners)*, Maria Laura Piscopo[83] *(Scientific Secretary)*, Marta Calvi[84], Yuval Grossman[23], Thibaud Humair[24], Andreas Jüttner[7,25], Jernej F. Kamenik[26,65], Matthew Kenzie[27], Patrick Koppenburg[83], Radoslav Marchevski[28], Angela Papa[29], Guillaume Pignol[30], Justine Serrano[10] *(Contributors)*

**Neutrino Physics & Cosmic Messengers:** Pilar Hernandez[39], Sara Bolognesi[40] *(Conveners)*, Ivan Esteban[41] *(Scientific Secretary)*, Stephen Dolan[7], Valerie Domcke[7], Joseph Formaggio[42], M. C. Gonzalez-Garcia[80,81,82], Aart Heijboer[19], Aldo Ianni[44], Joachim Kopp[7,79], Elisa Resconi[45], Mark Scott[33], Viola Sordini[87] *(Contributors)*

**Beyond the Standard Model Physics:** Fabio Maltoni[8,31], Rebeca Gonzalez Suarez[32] *(Conveners)*, Benedikt Maier[33] *(Scientific Secretary)*, Timothy Cohen[7,28,78,*], Annapaola de Cosa[34,*], Nathaniel Craig[35], Roberto Franceschini[36], Loukas Gouskos[37], Aurelio Juste[38], Sophie Renner[14], Lesya Shchutska[28] *(Contributors)*

**Dark Matter and Dark Sector:** Jocelyn Monroe[54,46], Matthew McCullough[7] *(Conveners)*, Yohei Ema[7,†] *(Scientific Secretary)*, Paolo Agnes[47], Francesca Calore[48], Emanuele Castorina[22], Aaron Chou[49], Monica D'Onofrio[50], Maksym Ovchynnikov[7,†], Tina Pollmann[19], Josef Pradler[59,86], Yotam Soreq[52], Julia Katharina Vogel[53] *(Contributors)*

**Accelerator Science and Technology:** Gianluigi Arduini[7], Philip Burrows[51] *(Conveners)*, Jacqueline Keintzel[7] *(Scientific Secretary)*, Deepa Angal-Kalinin[55], Bernhard Auchmann[76,7], Massimo Ferrario[3], Angeles Faus Golfe[21], Roberto Losito[7], Anke-Susanne Mueller[56], Tor Raubenheimer[57], Marlene Turner[7], Pierre Vedrine[40], Hans Weise[24], Walter Wuensch[7], Chenghui Yu[58] *(Contributors)*

**Detector Instrumentation:** Thomas Bergauer[59], Ulrich Husemann[56] *(Conveners)*, Dorothea vom Bruch[10] *(Scientific Secretary)*, Thea Aarrestad[34], Daniela Bortoletto[54], Shikma Bressler[60], Marcel Demarteau[61], Michael Doser[7], Gabriella Gaudio[62], Inés Gil-Botella[63], Andrea Giuliani[21], Fabrizio Palla[64], Rok Pestotnik[65], Felix Sefkow[24], Frank Simon[56], Maksym Titov[40] *(Contributors)*

**Computing:** Tommaso Boccali[64], Borut Kersevan[26,65] *(Conveners)*, Daniel Murnane[66] *(Scientific Secretary)*, Gonzalo Merino Arevalo[63], John Derek Chapman[27], Frank-Dieter Gaede[24], Stefano Giagu[67], Maria Girone[7], Heather M. Gray[66], Giovanni Iadarola[7], Stephane Jezequel[68], Gregor Kasieczka[15], David Lange[69], Sinéad M. Ryan[70], Nicole Skidmore[71], Sofia Vallecorsa[7] *(Contributors)*

**Theoretical Overview:** Eric Laenen[19,83,85]
**Reviewers:** Anadi Canepa[49], Xinchou Lou[58], Rogerio Rosenfeld[72], Yuji Yamazaki[73]
**Editors:** Roger Forty[7], Karl Jakobs[74], Hugh Montgomery[75], Mike Seidel[28,76], Paris Sphicas[7,77]

[1] Universidad de Granada, Spain
[2] Universität Heidelberg, Germany
[3] INFN Laboratori Nazionali di Frascati, Italy
[4] University of Pittsburgh, US
[5] Universidad Autónoma de Madrid, Spain
[6] Universität Bonn, Germany
[7] CERN, Geneva, Switzerland
[8] Northwestern University, US
[9] INFN Padova, Italy
[10] Aix Marseille University, CNRS/IN2P3, CPPM, Marseille, France
[11] Max Planck Institute for Physics, Germany
[12] IGFAE, Universidade de Santiago de Compostela, Spain
[13] INFN Torino, Italy
[14] Glasgow University, UK
[15] CPT Marseille, France
[16] Technical University Munich, Germany
[17] Hamburg University, Germany
[18] INFN and University of Bari, Italy
[19] Utrecht University, The Netherlands
[20] Zurich University, Switzerland
[21] IJCLab, Paris-Saclay University and IN2P3/CNRS, France
[22] Milan University, Italy
[23] Cornell University, US
[24] DESY Hamburg, Germany
[25] University of Southampton, UK
[26] University of Ljubljana, Slovenia
[27] University of Cambridge, UK
[28] EPFL Lausanne, Switzerland
[29] INFN Cosenza, Italy
[30] LPSC Grenoble, France
[31] Università di Bologna, Italy
[32] Uppsala University, Sweden
[33] Imperial College London, UK
[34] ETH Zurich, Switzerland
[35] UC Santa Barbara, US
[36] Rome III University, Italy
[37] Brown University, US
[38] IFAE Barcelona, Spain
[39] IFIC, University of Valencia and CSIC, Spain
[40] CEA Saclay, Irfu, France
[41] University of the Basque Country UPV/EHU and EHU Quantum Center, Spain
[42] MIT Cambridge, US
[43] Valencia University, Spain
[44] INFN LNGS, Italy
[45] TUM Munich, Germany
[46] Rutherford Appleton Laboratory, UK
[47] GSSI Aquila, Italy
[48] LAPTH Annecy, France
[49] FNAL Batavia, US
[50] Liverpool University, UK
[51] John Adams Institute, Oxford University, UK
[52] Technion Haifa, Israel
[53] Technical University Dortmund, Germany
[54] Oxford University, UK
[55] STFC, Daresbury Laboratory, UK
[56] KIT Karlsruhe, Germany
[57] SLAC, US
[58] IHEP Beijing, China
[59] OEAW-MBI, Vienna, Austria
[60] Weizmann Institute, Israel
[61] Oak Ridge National Laboratory, US
[62] INFN Pavia, Italy
[63] CIEMAT Madrid, Spain
[64] INFN Pisa, Italy
[65] Jožef Stefan Institute, Ljubljana, Slovenia
[66] UC Berkeley and LBNL, US
[67] Rome University La Sapienza, Italy
[68] LAPP Annecy, France
[69] Princeton University
[70] Trinity College Dublin, Ireland
[71] University of Warwick, UK
[72] São Paulo State University, Brazil
[73] Kobe University, Japan
[74] Freiburg University, Germany
[75] Thomas Jefferson National Accelerator Facility, US
[76] PSI, Switzerland
[77] NKUA Athens, Greece
[78] University of Oregon, US
[79] Johannes Gutenberg University Mainz, Germany
[80] University of Barcelona, Spain
[81] ICREA, Spain
[82] C.N. Yang Institute, Stony Brook University, US
[83] Nikhef, Amsterdam, The Netherlands
[84] University of Milano-Bicocca and INFN Milano-Bicocca, Italy
[85] University of Amsterdam, The Netherlands
[86] University of Vienna, Austria
[87] Institut de Physique des Deux Infinis de Lyon, IN2P3, CNRS, France
[*] Also contributed to the Dark Matter and Dark Sector chapter.
[†] Also contributed to the Beyond the Standard Model Physics chapter.

# Contents

# Chapter 1

# Introduction

The European Strategy for Particle Physics (ESPP) is a cornerstone of Europe's decision-making process for the long-term future of the field. Mandated by the CERN Council, it is formed through a broad consultation of the particle physics community and has guided the direction of the field for the past two decades. The ESPP was initiated by the CERN Council in 2005 and updated in 2013 and 2020. It is designed to convey to the CERN Council the views of the community on strategic questions that are key to the future of particle physics. The process involves all CERN Member and Associate Member States as well as international partners.

The most recent update, completed in June 2020, prioritised the full exploitation of the LHC and detector upgrade projects for its high-luminosity phase. It identified an electron-positron Higgs factory as the highest-priority next collider. For the longer term, it presented the ambition of the European particle physics community to operate a proton-proton collider at the highest achievable energy. In addition, it recommended the continued development of scientific programmes beyond colliders at CERN as well as in other laboratories in Europe and worldwide. The strategy also called for roadmaps to be drawn up setting the course of research and development (R&D) in the fields of particle accelerator and detector technology for the coming years.

Since the 2020 Strategy update, excellent progress has been made at CERN and other research centres in preparation for future colliders. A Feasibility Study for the realisation of the Future Circular Collider (FCC) was launched by the CERN Council following the adoption of the strategy update and the final report was released in March 2025. The results have been reviewed by panels which scrutinised both the scientific aspects of the project as well as the cost estimates. The study was reviewed by the CERN Scientific Policy and Finance Committees in September 2025 and will be presented to the CERN Council in November 2025. In addition, the international landscape for future colliders has become clearer: in December 2023 the US P5 prioritisation process was concluded, endorsing an off-shore Higgs factory, located in either Europe or Japan, to advance studies of the Higgs boson following the HL-LHC, while maintaining a healthy on-shore particle physics programme. In addition, the panel recommended targeted collider R&D to establish the feasibility of a 10 TeV muon collider. Shortly afterwards, the Technical Design Report for the proposed Circular Electron Positron Collider (CEPC) in China was released, with the ambition for project approval in the official 5-year funding cycles of the Chinese government. The International Linear Collider (ILC) project, foreseen for realisation in Japan, has meanwhile established an International Technology Network in a bid to increase global support.

Given these developments, as well as the long timescales involved in building new large-scale facilities and the importance of long-term community engagement, the CERN Council called for a third update of the European Strategy for Particle Physics in March 2024. According to the remit, the strategy should develop a visionary and concrete plan that greatly advances knowledge in fundamental physics through the realisation of the next flagship project at CERN. This plan should attract and value international collaboration and allow Europe to continue to play a leading role in the field. Regarding a future collider project, the strategy update should include the preferred option for the next collider at CERN and prioritised alternative options to be pursued if the preferred plan turns out not to be feasible or competitive. In addition, the strategy update should indicate areas of priority for exploration complementary to colliders and other experiments to be considered at CERN and at other European laboratories, as well as participation in projects outside Europe.

### *Organisation of the 2026 ESPP process*

Council established the European Strategy Group (ESG), which is responsible for submitting final recommendations to the Council for approval in early 2026. The Strategy Secretariat is in charge of organising the full process. In deriving the recommendations, the ESPP update should take into account the input of the particle physics community, progress in implementation of the 2020 Strategy update, and the accomplishments in recent years. These should include results from the LHC and other experiments or facilities worldwide, progress in the construction of the High-Luminosity LHC and physics expectations from its operation, the FCC Feasibility Study, and recent technological developments in accelerators, detectors, and computing. The Physics Preparatory Group (PPG) is to distill the community's scientific input and the scientific discussion at the Open Symposium into the current "Physics Briefing Book". This book is provided to the ESG as the foundation for their consideration during their five-day-long session for drafting the recommendations. This is scheduled to take place from 1 to 5 December 2025 at Monte Verità, (Ascona, Switzerland). The Physics Briefing Book is also shared with the particle physics community for their final input into the process.

To organise the work and to cover the full breadth of physics as well as the technology areas of accelerators, detectors and computing, the following nine PPG working groups were formed:

- Electroweak physics
- Strong interactions
- Flavour physics
- BSM physics
- Neutrinos and cosmic messengers
- Dark matter and dark sector
- Accelerator science and technology
- Detector instrumentation
- Computing

The work of each of these groups has been co-organised by two conveners, with one of them being a member of PPG. In addition, Early Career Researchers (ECR), have been appointed to act as Scientific Secretaries. For the evaluation of the physics performance of the proposed colliders and experiments in other areas, benchmark processes and measurements have been defined by the PPG working groups. The focus in this book is on the presentation and discussion of the performance against these benchmarks. The book does not aim to give a complete and comprehensive overview on the different physics and technology areas.

***Involvement of the particle physics community***

To inform the process, the Strategy Secretariat called upon the particle-physics community across universities, laboratories, and national institutes to provide input in various forms. The timeline for submission of input was set to 31 March 2025. To allow the national HEP communities to consider the submissions collected by March 2025, a further opportunity for furnishing input ahead of the Open Symposium, with a deadline of 26 May 2025, was offered. There has been a strong engagement by the community and in total 266 scientific contributions were submitted to the ESPP process. The submissions cover the full spectrum of particle physics and express the priorities of the community at a level ranging from individuals to national communities. An overview on the submissions classified according to their self-attributed themes is displayed in Fig. 1.1. The full list of submissions is given in Appendix C of this book.

A major focus was the preferred project to succeed the HL-LHC as a large-scale accelera-tor facility at CERN. There are several proposals for such a flagship collider: the FCC feasibility study was submitted as input to the Strategy, describing a 91 km circumference infrastructure and the technical realisation of a precision electron-positron “Higgs and electroweak” factory (FCC-ee) followed by an energy-frontier hadron collider (FCC-hh) in a second stage. The possibility of proceeding directly to a FCC-hh as a standalone realisation at an earlier time was also presented. Linear colliders were presented as alternative options, with a Linear Collider Facility (LCF) based on the design of the International Linear Collider (ILC) and CLIC both considered. In addition, smaller collider options were presented, based on re-using the LHC/LEP tunnel. A first proposal, LEP3, suggests accelerating electrons and positrons up to energies of 230 GeV, while a second proposal, LHeC, proposes the realisation of electron-proton collisions at one interaction point of the LHC. LHeC would require the construction of an additional new energy-recovery linac for the acceleration of electrons. Moving from the precision frontier to the energy frontier, several ways to reach the 10 TeV “parton scale” were presented. If the FCC-ee is realised, a natural path is to proceed with proton-proton collisions with a centre-of-mass energies in the range of 85 to 120 TeV, depending on the available high-field magnet technology. A muon collider or wakefield accelerators could perhaps provide a path towards high-energy lepton collisions on a longer timescale.

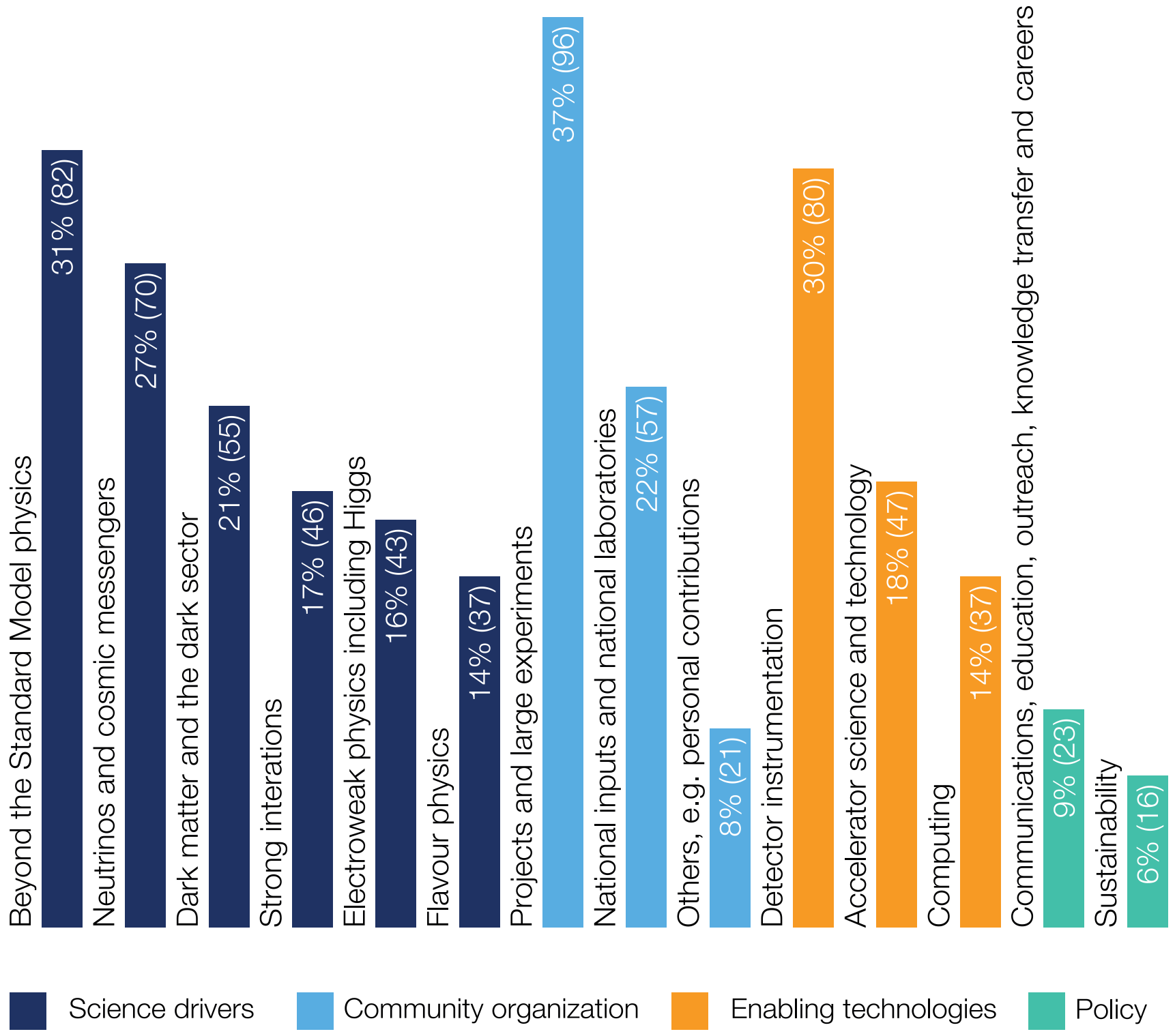

Fig. 1.1: Community Inputs: self-attributed themes of the 266 community inputs to the ESPP (multiple themes can be attributed to a single submission). The themes are ordered by umbrella category and percentage of total submissions in parentheses.

The PPG working groups broadly cover the physics and technology areas discussed in the submissions. Related to non-collider physics, the areas of neutrinos, BSM physics, dark matter, strong interactions and flavour physics attracted a lot of input, both presenting dedicated projects as well as the physics potential. Numerous submissions in the categories of detector instrumentation, accelerator science and technology, and computing complement the community’s physics aspirations. Progress in these technologies is vital for the realisation of a post-LHC collider and non-collider experiments. The full spectrum of the input by the community was discussed at the Open Symposium, which brought together more than 600 physicists from almost 40 countries in Venice, Italy, from 23 to 27 June 2025. The six physics topics as well as the three technology areas on accelerators, detectors and computing, were summarised in rapporteur talks, followed by 45-minute discussions with strong engagement from those present in Venice. An important part of the symposium was devoted to presentations of possible future large-scale accelerator projects at CERN. In addition, the international perspectives from China, Japan and the US were presented.

### *Content of the Physics Briefing Book*

The submissions by the particle physics community as well as the input received during the discussions at the Open Symposium provide the basis for the findings that are summarised in this book. In Chapter 2 an overview on the theory perspectives is given. Addressing the compelling questions and outstanding issues in particle physics requires a broad and diverse experimental programme. This would include the HL-LHC, a new flagship collider, and a wide variety of other experiments including those in neighbouring fields. In Chapters 3 to 9 the outstanding questions in the various physics areas are summarised, together with a discussion of the potential of the different colliders and other experiments to address them. In addition, the potential of the different colliders to search for deviations from the Standard Model is evaluated. Input from several physics areas such as electroweak, flavour and BSM physics are combined. The results are presented in Chapter 3 (Sect. 3.5). The evaluation of the physics potential of the different colliders is made assuming the luminosity and energy scenarios presented in the respective submissions. For the combination of different machines, i.e. carrying out $e^+e^-$ precision physics in a first stage, followed by a higher energy machine at the energy frontier in a second stage, the scenarios presented in Table 1.1 have been considered. It should be noted that the evaluations presented in this book are limited to those accelerator projects proposed for realisation at CERN. The ILC considered as a global project in Japan and the CEPC considered for realisation in China are not included in the evaluation. However, their potential is very close to that of the LCF and FCC-ee colliders, respectively, modulo differences in luminosity.

Throughout the full book natural units $c = \hbar = 1$ are used, leading to masses and momenta being expressed in units of eV.

Table 1.1: Overview on the collider types and energies considered in the evaluation of the physics performance. In the lower part of the table combinations of lower-energy precision machines with colliders reaching higher energies are given.

| | Energies (GeV) | | | | | | | | FCC-hh | CLIC Muon Collider | Muon Collider Plasma Collider |
|---|---|---|---|---|---|---|---|---|---|---|---|
| $e^+e^-$ projects | 91 | 160 | 240 | 365 | 380 | 550 | 1000 | 1500 | 85 TeV | 3 TeV | 10 TeV |
| FCC-ee | X | X | X | X | | | | | | | |
| LCF-250 | X | | X | | | | | | | | |
| LCF-250+550 | X | | X | X | | X | | | | | |
| CLIC-380 | | | | | X | | | | | | |
| CLIC-380+1500 | | | | | X | | | X | | | |
| LEP 3 | X | X | X (230) | | | | | | | | |
| ep collider | | | | | | | | | | | |
| LHeC | 50 GeV (e-) on 7 TeV (p) | | | | | | | | | | |
| High-energy options | | | | | | | | | | | |
| FCC-ee + FCC-hh (85 TeV) | X | X | X | X | | | | | X | | |
| FCC-hh (85 TeV) | | | | | | | | | X | | |
| LCF + 1 TeV option | X | | X | | | X | X | | | | |
| LCF + High-energy lepton collider | X | | X | | | X | | | | | X |
| CLIC + 3 TeV | | | | | X | | | X | | X | |
| Muon Collider | | | | | | | | | | X | X |
| LEP3 + FCC-hh (85 TeV) | X | X | X(230) | | | | | | X | | |
| LEP3 + High-energy lepton collider | X | X | X(230) | | | | | | | | X |
| LHeC + FCC-hh (85 TeV) | 50 GeV (e-) on 7 TeV (p) | | | | | | | | X | | |

The status and developments in the areas of accelerator science and technology, detector instrumentation and computing are presented in Chapters 10 to 12. Chapter 10 contains a brief description of the large-scale accelerator projects proposed for realisation at CERN (Sect. 10.2). The differences in the physics potential between the various collider options, as documented here, along with the technical readiness, risks, timescales and costs will be reviewed by the ESG. Alongside the final national inputs, that are expected to be submitted by 14 November, these assessments will provide the foundation for the final recommendations that will be formulated in the ESPP drafting session in December 2025.

# Chapter 2

# Particle Physics: Open Questions and Opportunities

Is the Higgs boson truly a solitary elementary scalar particle, or the first glimpse of a deeper layer of physics? Why is our Universe made only of matter? What is the nature of dark matter that outweighs visible matter five to one? These fundamental questions are at the heart of our understanding of nature and none can be answered with our present experimental tools.

The Higgs boson plays a particularly special role here, and the needed precise study can only be done at a new flagship collider. This should be complemented by neutrino facilities, dark matter searches, precision experiments and innovative theory. This update of the European Strategy for Particle Physics might be the most consequential since the launch of the LHC, as it must chart the path to this next discovery machine.

We continue to reap rich scientific harvests from the present accelerators and their upgrades, but the open problems in our field are such that we must now prepare for the next generation.

– The Higgs sector: Is there only one Higgs boson? Is it elementary or composite? Do its couplings follow the precise values predicted by the Standard Model? A Higgs factory, recommended already in the previous strategy, is the essential next step to stress-test the Higgs mechanism. In the longer term, determining the Higgs self-coupling with percent-level precision will reveal whether its peculiar potential truly drives electroweak symmetry breaking. The nature of this potential also governs the character of the electroweak phase transition in the early Universe: within the Standard Model it is a smooth crossover, but with new physics it could become first-order, providing the conditions for baryogenesis and connecting the Higgs sector directly to the origin of matter.

– Matter–antimatter asymmetry: In high-energy collisions, matter and antimatter are produced almost equally, yet our Universe is made of matter alone. Precision flavour physics and neutrino experiments, together with collider data, hold the promise of elucidating this profound imbalance.

– Dark matter: We readily observe its gravitational imprint on galaxies and clusters, yet we do not know what it is. Collider searches, direct-detection experiments, and astrophysical observations together may be needed to identify this elusive form of matter.

These and other questions revolve around the outstanding ideas that motivate our field. The chapters that follow in this Briefing Book will explore them in detail. In this introduction, the theoretical ideas and their impact are outlined.

### *Where are we now and where are we going?*

We have now observed all the particles and fields of the Standard Model, but we have not measured all their interactions and parameters: the Higgs self-coupling, many Yukawa couplings, neutrino masses and mixings still await precise determination. Continued, intense scrutiny of the Standard Model, and the search for deviations from it, must remain a central goal of particle physics.

To achieve this, we need a broad programme: insights from the HL-LHC and its successors, in particular a new flagship collider; but also from neutrino experiments, dark matter detectors, and exquisitely sensitive smaller-scale searches. Equally essential are theoretical advances: more precise predictions, new conceptual frameworks, and innovative approaches that connect the different strands of our field. Above all, we will need the many and diverse talents of our community to pursue these goals together.

### *Electroweak and Higgs physics*

Let us start with the ideas behind our main discovery in recent times: the Higgs boson. It is a by-product and thus tell-tale of the Brout-Englert-Higgs mechanism[1]. The Standard Model by definition includes one complex scalar $\mathrm{SU}(2)_{\mathrm{weak}}$ doublet. Three of its components were "seen" already in the 1980s, as part of the massive $W$ and $Z$ bosons, whereas the fourth was finally observed in 2012, lending credence to the Higgs mechanism for mass generation of vector bosons. The non-zero groundstate expectation value $v$ of the doublet following from the assumed form of the Higgs field potential also provides fermion mass terms through Yukawa interactions. It is worth reminding ourselves that standard mass terms are forbidden by SM symmetries. Thus the Higgs field and its potential play a key role in the Standard Model and its physical manifestations.

The potential and its characteristic shape are indeed an assumption. A potential with a non-invariant ground state is not uncommon in physics; condensed matter colleagues deal with such situations in the cases of ferromagnets and Bose-Einstein condensation of Cooper pairs in superconductors. Our case is the Universe. The Higgs boson, a fluctuation around the groundstate manifold of minima, can thus be viewed as a tangible piece of the vacuum itself!

The upshot of the Higgs mechanism, according to the SM, is then that all particle masses are proportional to the groundstate value of the Higgs field, by an amount equal to the coupling strength of the particle to the Higgs field. This includes the Higgs boson itself. Results from the ATLAS and CMS experiments indeed suggest such a linear relation, but there is ample room for deviations. However it may turn out, we, as a field, have established a new concept of mass no less revolutionary than that of Newton and Einstein!

A key part of the future particle physics programme will be to stress-test the SM, and the Higgs sector in particular. To that end vast numbers of Higgs bosons should be created in a new Higgs factory, and their properties and interactions scrutinized. A classic example is the "closure test" of the SM, in which the calculated value of the $W$ mass (a function of the top quark mass, the Higgs boson mass, the Fermi constant etc.) is compared with the measured value. Present uncertainties on the $W$ and top masses will be dramatically reduced, putting the Standard Model under great duress and raising the possibility of detecting interesting deviations.

[1] For brevity I will refer to it below as the Higgs mechanism.

Other highly important parameters to determine are the Higgs boson mass, width and couplings. The best way to do so is in $e^+e^-$ collisions producing an off-shell $Z$-boson radiating a Higgs boson. In such a two-to-two process we can precisely determine the 4-vectors of the incoming leptons and that of the outgoing $Z$-boson, and therefore also that of the Higgs boson. The Higgs boson mass could thus be measured to a few MeV precision, an order of magnitude improvement over the current 110 MeV uncertainty. The same process also yields the total $ZH$ cross-section, and thereby, using again $Z$-boson recoil, the Higgs boson width, which is only 4 MeV in the Standard Model. Thus, with a Higgs factory we would improve our knowledge about the Higgs boson and most of its couplings tremendously, by about an order of magnitude with respect to HL-LHC outcomes. Similar improvements will be possible for other electroweak parameters, providing additional dimensions for stress-testing. In particular, for circular colliders it could be interesting, though very difficult, to measure the electron Yukawa coupling, and test if the electron mass is indeed due to the Higgs mechanism.

The long-term goal of the Higgs programme is the determination of the tri-Higgs self coupling, as a confirmation of the peculiar shape of the Higgs field potential. Recent studies have shown that HL-LHC data on di-Higgs production can do better here than previously thought. ATLAS and CMS expect a $7\sigma$ observation of this process, leading to determination of the coupling to about 30% accuracy. At low-energy Higgs factories the self-coupling is probed only via loop effects, yielding about 20% constraints when combined with HL-LHC data, while higher-energy facilities can directly probe double-Higgs boson production, substantially improving the accuracy. A future collider at the energy frontier will change the game completely: we can then anticipate a few percent-level accuracy on the tri-Higgs coupling, and thus on the potential shape. Yet even with such precision, the striking smallness of the Higgs boson mass itself remains unexplained, leading us to the long-standing naturalness problem.

### *Naturalness*

Some quantities in particle physics appear unnaturally small. Most striking is the Higgs boson mass, lying 17 orders of magnitude below the Planck scale. The Standard Model is understood as a low energy version of a more complete theory. Quantum corrections to scalar masses are quadratically sensitive to the mass of any new particle from such a theory that interacts with the Higgs boson. Unlike fermion masses, which are symmetry-protected, the Higgs boson mass can thus receive large corrections.

This puzzle—why the Higgs boson remains light—lies at the heart of the naturalness problem. It has been a powerful guide before, e.g. divergences in kaon physics led to the prediction of the charm quark. By analogy, the Higgs hierarchy may hint at new degrees of freedom at the TeV scale, such as superpartners, heavy vectorlike quarks, or at a composite Higgs boson as a pseudo–Goldstone boson.

Future colliders should provide decisive tests. A Higgs factory will measure Higgs couplings and electroweak parameters with unprecedented precision, probing for the indirect effects of such new states. At higher energies, direct production of new partners could reveal whether the Higgs boson is indeed elementary or part of a deeper structure.

Thus the naturalness problem, centred on the Higgs boson mass, is a theoretical puzzle but a concrete experimental programme as well, one that future colliders are uniquely positioned to pursue. These considerations naturally motivate theories beyond the Standard Model, and raise the question of how best to describe, and search for, possible new physics.

### *New physics and Effective Field Theory*

Over the past decades, many models have been formulated to address specific shortcomings of the Standard Model. Supersymmetry, for example, offers a natural solution to the hierarchy problem and provides dark matter candidates as well; composite Higgs models reinterpret the Higgs boson as a bound state of more fundamental constituents; extra-dimensional scenarios give new perspectives on unification and gravity. While no single framework has yet been confirmed, each has sharpened our questions and broadened our view of what new physics might look like.

In parallel, more agnostic Effective Field Theory (EFT) approaches have matured into powerful tools, now routinely embedded in event generators and experimental analyses. EFT provides a systematic way to parameterize new physics effects from higher scales, without committing to a specific model, and thus serves as a bridge between theory and experiment. It is central to many performance studies in the chapters that follow, and is treated in depth in Appendix A of this Briefing Book.

The interplay between model building and EFT is especially important: concrete models can guide the choice of EFT operators and benchmark scenarios, while EFT results can be mapped back to constrain or motivate classes of models. Strengthening these connections, and deepening collaboration between theorists and experimentalists, will increase the value of both approaches. In this way, even in the absence of immediate discoveries, we can steadily narrow the space of possibilities and prepare for the surprises that future colliders and experiments may bring.

However, while new physics may emerge at higher energies, the strong interaction itself still poses key puzzles.

### *Understanding the Strong Interaction*

QCD sets the precision benchmark for much of particle physics. Precise determinations of the strong coupling $\alpha_s$, still the least known gauge coupling, and of the quark masses are essential for many observables. Continued advances in high-order perturbation theory, together with lattice-QCD determinations of non-perturbative matrix elements, are key to these determinations and to quantitatively controlled predictions. They are indispensable for interpreting data in terms of Standard Model parameters, for example in extractions of CKM elements from leptonic and semileptonic decays, and in the hadronic contributions entering the calculation of the anomalous magnetic moment of the muon.

A striking open issue in the Standard Model is the Strong CP problem. The QCD Lagrangian may contain a term proportional to $g_s^2 G\widetilde{G}$ which violates P, T and thus CP symmetry. If even one quark were massless, this phase could be rotated away, but lattice QCD essentially rules this out. The term induces an electric dipole moment of the neutron, yet experiments constrain such an EDM to be vanishingly small, implying $|\theta| < 10^{-10}$. Why this parameter is so tiny is the essence of the problem.

Several explanations have been proposed. The most compelling introduces a new global symmetry, the Peccei–Quinn (PQ) symmetry, under which $\theta$ becomes part of a dynamical field. The associated pseudo–Goldstone boson is the axion, whose ground state relaxes $\theta$ to zero, naturally solving the problem.

Axions are light, weakly coupled pseudoscalar particles with interactions with photons (via $\vec{E}\cdot\vec{B}$) and fermions. They could also form dark matter through the misalignment mechani<in the early Universe, leading to a coherently oscillating axion field. Experiments now exploit these couplings in stellar observations, haloscopes, helioscopes, and "light-shining-through-walls" setups to search across a wide mass range. Thus the strong CP problem links nuclear physics, cosmology and astrophysics, and motivates one of the most active experimental programmes in the field today.

The physics of heavy-ion collisions offers different challenges to understanding the strong interaction, and here only a small selection is listed. The Quark Gluon Plasma created in such collisions has the characteristics of a liquid, but when does it change to a collection of (quasi)-particles, with which hard collisions are possible? In similarity to Rutherford scattering, one can look for large angle scattering, which can be tested via jet substructure. A clear observation of such large-angle scattering would provide evidence for quasi-particle behaviour.

At low temperature, chiral symmetry is broken, but can we see it being restored above the critical temperature? Probing this is, in a sense, an exploration of how the QCD vacuum itself changes under extreme conditions. A tell-tale of this restoration could be the "melting" of the $\rho$ mass peak, and increased mixing with the $a_1$ meson, to be tested through careful measurement of the dilepton spectrum.

In heavy-ion collisions, the nuclei contain a high density of partons, especially gluons, at small momentum fraction $x$. Can we see signs of parton saturation, in which partons in addition to splitting have merged before scattering? This is a non-linear effect, and one may look for such deviations from DGLAP scaling at small $x$ values.

These and other studies address both the QCD phase diagram, from the liquid-like quark-gluon plasma to the restoration of chiral symmetry, and the novel dynamics of parton saturation at small $x$, making them a unique laboratory for strong-interaction physics.

### *Flavour physics*

Flavour physics highlights the virtue of virtuality. Through its sensitivity to effects from quantum corrections, it allows us to look, in detail, far beyond collider energy limits, and enables stress-testing the Standard Model in many ways. A particular focus of stress-tests has been the unitarity triangle, where many experiments have pinned the apex already very precisely, and consistently. But with future experiments, including a possible $Z$-pole run at a Higgs factory, much more severe stress tests of the CKM parameters are foreseen, some by as much as a factor 10. Major progress is also envisioned through (new) CP violating observables. For instance, new sensitivity is possible in the charm sector, which is CKM suppressed, but should become ever more accessible. In general, the focus of this research will increasingly shift towards loop-dominated processes, in which penguin diagrams play a key role. The very rarity of such processes, e.g. in the rare decays of the $B^0_{d,s}$ meson to muons, or of $K^+$ to $\pi^+\nu\bar{\nu}$, makes them very sensitive detectors of possible new heavy particles in loops. Further major benefits for precision estimates of flavour observables are expected from lattice calculations of form factors and hadronic matrix elements, towards percent-level accuracy.

A key condition for a successful flavour physics programme is the completion of the HL-LHC programme. To go beyond this at the $e^+e^-$ collider, one would run at the $Z$-pole to collect a few times $10^{12}$ $Z$ bosons. While the number of $b$ hadrons is smaller than in LHC $pp$ collisions, the cleanliness of events can be very valuable in decays with missing energy (e.g. to $\tau$ leptons).

### *Neutrinos*

Finding out the properties and structure of the neutrino sector of the Standard Model will be a major focus of our field in the coming decades. Their tiny interaction strengths can be overcome with intense neutrino beams and very large detectors, as demonstrated already by the discovery of neutrino oscillations; proof that neutrinos have mass, however small. This breakthrough has had a profound impact on particle physics.

The puzzles of their flavour structure and mass ordering are likely to be solved by a variety of worldwide experiments. A particularly compelling question is whether there is CP violation in the lepton sector. Next-generation long-baseline experiments are very likely to provide an answer.

The stakes are high: if CP violation among neutrinos is established, it would strengthen the case for leptogenesis, a mechanism in which an asymmetry in the lepton sector in the early Universe is converted into the observed matter–antimatter asymmetry through electroweak processes. Thus, neutrino physics may hold the key to explaining why anything exists at all.

Numerous other experiments will contribute, often on smaller scales but with profound implications. Chief among them is the search for neutrinoless double-beta decay, which would establish that neutrinos are Majorana fermions, i.e. their own antiparticles, and indicate that lepton number is not a fundamental symmetry of nature. Such a discovery would demonstrate the existence of a new physics scale, that could explain the origin of the cosmic matter excess. At the same time, another cosmic mystery remains entirely open: the identity of the dark matter that dominates the mass of the Universe.

### *Dark Matter*

A major unanswered question pertains to the nature of Dark Matter (DM). We have ample indirect evidence of its existence (e.g. through galaxy rotation curves), but we do not know its nature, only that it is not part of the Standard Model. What could DM be? Some options are the Weakly Interacting Massive Particles (WIMPS), in the GeV–TeV mass range, for which there are many candidates. In the very small (neV–MeV) mass range an option might be dark photons, axions, or axion-like particles (ALPs). These can be found in many Standard Model extensions, and in string theories. Another option is keV sterile neutrinos. An exciting new phase in direct detection of DM will be offered by the next generation of large scale direct detection experiments. Collider sensitivity for WIMPs is complementary to direct/indirect detection and must therefore be part of the experimental Dark Matter search.

### *Theoretical Precision*

Precision is key to future explorations in particle physics. Data from LHC and HL-LHC are moving towards percent-level accuracy, while designs of possible future colliders and their detectors aim for per-mille level accuracy. Such experimental precision must be matched by theory, and indeed theorists have stepped up and achieved great progress, inventing new methods and leveraging increased computing power, in both numerical methods and computer algebra (Mathematica, FORM, etc.). Mathematics in particular has been a source of outstanding ideas in this area, yielding powerful methods such as integration-by-parts identities, finite field methods, differential equations for amplitudes, and much more.

Precision has also found its way into higher order calculations in the Standard Model gauge couplings. Predictions are generally available at least at next-to-leading order (NLO), but the standard is now NNLO calculations and for some cases even NNNLO accuracy. This is often supplemented by all-order resummed contribution of large logarithmic terms, for maximally precise predictions. A major recent innovation is going beyond the Leading Logarithmic (LL) accuracy of parton showers, essential components of event generators, enabled by better treatment of recoil. The emerging new standard is NLL, but NNLL accuracy is likely needed to exploit full physics potential. One clear benefit is that with NNLL showers no anomalously large value of $\alpha_s$ is needed to match, for instance, LEP thrust data. Further significant progress on higher-order corrections is vital to match the experimental precision that will be achieved in future colliders. Moreover, in many cases non-perturbative aspects are limiting factors, hence progress is much needed here as well.

***AI***

A particularly promising part of our future path is Artificial Intelligence, the subject of the 2024 Physics Nobel Prize. Classification, pattern recognition, which are strong points of AI, are already now central to modern particle physics. This has led to marked improvements in several physics studies, such as di-Higgs production signals at the LHC, and plays a key role in determining parton distribution functions. Generative AI and Large Language Models already assist with coding and other tasks, and as such are a major productivity boost. New analysis strategies, calculational methods, and even future detector design may well emerge from this revolutionary new tool, and the further impact of AI will no doubt be immense.

***Final remarks***

The Standard Model has carried us far, but its deepest questions remain unanswered. Answering these demands tools beyond today's accelerators. As recommended in the previous update of the European Strategy for Particle Physics, a future path with a Higgs factory, followed by a collider at the energy frontier, together with advances in neutrino and dark matter experiments, will allow us to stress-test the Standard Model to a breaking point, and to explore entirely new territory. By committing to this path, we can ensure that the coming decades of particle physics are as revolutionary as the last.

# Chapter 3

# Electroweak Physics

The study of EW interactions has been instrumental in shaping the SM, leading to the prediction and in some cases the discovery of its bosonic and fermionic components. As the LHC continues to probe EW interactions, deeper questions about the fundamental nature of the Higgs boson arise. These questions include its role in EWSB, its connection to the origin of fermion masses and flavour dynamics and its potential role in the evolution of the cosmos and the stability of the Universe. Precision measurements in EW physics are powerful tools for probing these questions and new physics beyond the SM.

This chapter discusses the expected status from the HL-LHC and the prospects from future colliders in the areas of Higgs physics (Sect. 3.1), top-quark physics (Sect. 3.2), and precision electroweak observables (Sect. 3.3). The role of and requirements on theoretical aspects are covered in Sect. 3.4. Finally, comparisons across various future colliders using an EFT interpretation are shown in Sect. 3.5. For the $e^+e^-$ machines, the luminosities quoted in Table 10.1 are used, except for LCF where 0.1 ab$^{-1}$ for the $Z$-pole run and 3 ab$^{-1}$ for the 250 GeV run are used. For the LCF 550 GeV, the $e^-/e^+$ polarisation of $\pm 80\%/\mp 30\%$ is used. In addition, a luminosity of 8 ab$^{-1}$ for LCF 1 TeV, 5 ab$^{-1}$ for CLIC 3 TeV, 30 ab$^{-1}$ for FCC-hh, 1 ab$^{-1}$ for LHeC and 10 ab$^{-1}$ for the 10 TeV muon collider (MuC) are assumed here.

## 3.1 Higgs Physics

The SM is an extraordinarily successful model of nature, with its structure guided by principles of symmetry. Despite this, 19 parameters (excluding neutrino masses and mixing) are unpredicted, the majority of which are linked in some way to the properties of the Higgs boson. Uncovering the fundamental principles behind the origin of the masses of all fermions, i.e., including neutrinos, and the structure of the CKM and PMNS matrices, requires a deeper knowledge of the Higgs sector.

One of the many open questions surrounding the Higgs boson is whether it is a fundamental particle. If so, it is surprising that the electroweak scale is so much smaller than other fundamental scales of nature, such as the Planck scale or a possible energy scale of grand unification. This is the essence of the “naturalness” or “hierarchy” problem. Studying the Higgs boson may reveal whether a deeper symmetry, such as supersymmetry —which also includes an extended Higgs sector— explains the fact that the Higgs mass and the electroweak scale are much smaller than the Planck scale. Alternatively, the Higgs boson could also be a composite particle, arising from some new strong dynamics that drive EWSB.

Within the SM, the Higgs boson is responsible for EWSB, which is determined by the structure of the Higgs potential. If new BSM particles couple to the Higgs boson, the form of its potential can change. This has implications not only for particle physics but also for our understanding of the early Universe and its long-term evolution. A strong first-order phase transition in the early Universe could provide one of the conditions needed for electroweak baryogenesis. Additionally, measurements of the top-quark and Higgs-boson masses suggest that the shape of the potential renders our current vacuum metastable, raising fundamental questions about the fate of the Universe. Any dynamics related to the symmetry breaking mechanism would imprint itself on the properties of the Higgs boson.

### 3.1.1 High-Luminosity LHC

Since the Higgs boson discovery in 2012, the LHC has observed its couplings to gauge bosons and third-generation fermions (via measurements of cross section times branching ratios) and achieved evidence of Higgs boson decays to two muons. [ID170]. In $pp$ collisions, all the Higgs boson production modes (gluon-gluon fusion, vector-boson fusion VBF, associated production with vector bosons or pairs of top quarks) have been observed and so far, all measurements are consistent with the SM expectations [1,2].

For the coming two decades, the LHC and the HL-LHC will continue to map the Higgs sector with unprecedented precision. The gauge and 3$^{\text{rd}}$-generation Yukawa coupling parameters are expected to be measured to an experimental precision of 1.5–4% [ID170]. Measurements of rare decays, such as $H \to \mu^+\mu^-$ and $H \to Z\gamma$ will reach a precision of 3% to 7% on the corresponding effective couplings, respectively. Through recent improvements in flavour tagging, the $H \to c\bar{c}$ rate can be constrained to within 1.5 times the SM or better. The uncertainty on the mass of the Higgs boson, $m_H$, is expected to improve to 21 MeV [ID170]. Based on current projections, the dominant uncertainties on the Higgs boson couplings at the HL-LHC are often the theoretical uncertainties on the SM prediction itself, followed by the experimental systematic uncertainties. Apart from the rare decays, the statistical uncertainties are sub-dominant. Improvements in theoretical calculations and reduction of PDF uncertainties would therefore yield major benefit to Higgs physics at the HL-LHC.

At the HL-LHC, the Higgs potential can be probed via measurements of Higgs boson pair production ($HH$). Recent developments in analysis methodologies, detector calibration and jet flavour tagging have resulted in significant improvements in sensitivity for $HH$ final states; the current projection for the measurement of the self-coupling has a precision of 27% which is almost a factor of two improvement compared to projections shown for the last strategy [3]. These recent improvements have also opened the possibility to constrain the Higgs boson quartic self-coupling at the HL-LHC via $HHH$ final states. The expected limit is projected to be 86 times the SM value using the 6 b-quark final state [4]. The quartic couplings with vector bosons ($VVHH$) can be probed via $HH$ production in the VBF channel and are already constrained to be different from zero at 95% CL by current LHC analyses. At the HL-LHC, a precision of 13% on the corresponding coupling can be obtained.

### 3.1.2 Electron-proton colliders

The LHeC, as an electron-positron collider, has sensitivity to Higgs boson couplings via deep-inelastic scattering measurements. It can achieve a first measurement of H $\to c\bar{c}$ and provides a precise determination of EW charged-current Higgs boson production. In addition, LHeC can improve PDF and $\alpha_S$ uncertainties in hadron-collider measurements of Higgs properties, which amounts to a reduction of these uncertainties by a factor of 3 at the HL-LHC [ID214]. These improvements are not included in the HL-LHC projections shown here.

### 3.1.3 Electron-positron colliders

While future $e^+e^-$colliders produce far fewer Higgs bosons than the HL-LHC, they do offer significant advantages due to their precisely known initial state and much cleaner experimental environment. In addition, $e^+e^-$machines can make model-independent measurements of the Higgs boson couplings, unlike interpretations of the HL-LHC measurements, which often assume no exotic decays. At energies around 240 GeV, Higgs bosons are produced mainly via $ZH$ associated production. The VBF production cross section increases significantly with the centre-of-mass energy and becomes the dominant process above roughly 450 GeV. For second-stage linear colliders, running at energies of 550 GeV or above, the $t\bar{t}H$ production modes and di-Higgs boson production modes of $ZHH$ and $\nu\bar{\nu}HH$ become accessible and benefit from the increase in luminosity with the centre-of-mass energy.

Longitudinally polarised beams, which would be available at linear colliders, bring further advantages: s-channel production via $Z/\gamma$ exchange, such as in $ZH$ production, only occurs for Left-Right and Right-Left ($e^-$, $e^+$) beam polarisations. Additionally, Higgs boson production from $WW$ fusion is enhanced for Left-Right and suppressed for Right-Left polarisation, allowing for increased signal cross-sections and reduced background contributions. Depending on the exact beam polarisation scheme used, the effective signal cross-sections changes by a factor ranging from 1.0 to 1.4 for $ZH$ production and 0.2 to 2.4 for $WW$ fusion.

Conversely, circular $e^+e^-$colliders can offset the absence of longitudinal polarisation and the high-energy limitations imposed by synchrotron-radiation losses with far larger instantaneous luminosity and the option to operate several interaction points simultaneously. The FCC-ee (LEP3) yields roughly nine (two) times more Higgs bosons for all interaction points together per unit time compared to the 250 GeV LC.

One of the major advantages of $e^+e^-$colliders is the measurement of the total $ZH$ cross-section. Since the initial state energy is known, the total rate of this process can be measured, independently of the Higgs boson decay mode, by tagging only the $Z$ boson decay in the final state. For the FCC-ee, LEP3 and the LCF-CLIC at 380 GeV, this yields a precision on the total $ZH$ cross-section of 0.31%, 0.66%, and 0.56%, respectively [ID217, ID188, ID140]. The Higgs boson coupling to $Z$ bosons is most precisely measured at $e^+e^-$colliders, and constitutes a powerful test of a potentially composite nature of the Higgs boson [5]. The total $ZH$ cross-section then enables the indirect determination of the total Higgs boson width, by combining measurements of multiple production modes and final states, i.e., $ZH$ production with $H \to ZZ^*$ and $H \to WW^*$ decays or $WW$-fusion production with $H \to WW^*$ decays. Using $H \to ZZ^*$ decays is statistically limited, and $WW$-fusion production at higher energies or alternative decay channels are important.

Precision measurements of Higgs boson couplings, including those to 2$^{\text{nd}}$-generation charged fermions, are also possible [ID217, ID188, ID140]. Couplings to $H \to b\bar{b}$ can be improved compared to the HL-LHC to reach sub-percent-level precision. In addition, couplings to $H \to c\bar{c}$ can be measured at the percent level. Sensitivity to the strange-quark Yukawa coupling, which would give full access to the interactions of the Higgs boson with the 2$^{\text{nd}}$-generation of charged fermions, is within reach of future $e^+e^-$colliders [6], with potential evidence achievable at the FCC-ee [ID217] (see also [ID141]). As with the HL-LHC results, recent advances in flavour tagging have led to significant improvements in these projections. Indirect sensitivity to the top-quark Yukawa coupling can be obtained at the $t\bar{t}$ production threshold through loop corrections. However, measurements of the Higgs boson coupling to top quarks via $t\bar{t}H$ production require a centre-of-mass energy above 480 GeV, with the cross-section doubling by 550 GeV. At linear colliders with an energy of 550 GeV and 1 TeV, the $t\bar{t}H$ cross-section can be measured at a precision of 5.6-8.7% and 3.9-5.7%, respectively [ID140] and depending on the beam polarisation used. The FCC-ee, LEP3 and LCF at 250 GeV will be able to exclude additional invisible Higgs boson decays with a confidence level of 95% for branching fractions above 0.055%, 0.2% and 0.12%, respectively using recoil or missing-mass techniques [ID217, ID188, ID140]. Linear colliders running at higher energies of 3 TeV can improve the precision on the couplings of the Higgs boson with gauge bosons and fermions by a factor of two compared to linear colliders at lower energies. Finally, FCC-ee has the potential to probe the electron Yukawa coupling, however this would require beam monochromatisation and a dedicated run at $\sqrt{s} = m_H$ with five years of extra running time [7].

In addition to the couplings, the Higgs boson mass can be determined to a precision of 4 MeV, 10 MeV, 12 MeV for the FCC-ee, LEP3 and LCF at 250 GeV, respectively. These measurements benefit from the well-defined initial state in $e^+e^-$colliders, enabling reconstruction of the Higgs boson through the recoil technique.

For all the measurements determined via $e^+e^-$colliders discussed above, the dominant uncertainty is assumed to be the statistical uncertainty, implying that all experimental systematic uncertainties as well as theoretical uncertainties are sub-dominant. This assumption places requirements on both the detector design and future improvements in theoretical calculations.

The Higgs potential can be probed at $e^+e^-$colliders either via precision measurements of single Higgs boson production or via $HH$ production for collider options with a centre-of-mass energy at 550 GeV or above. In both cases, the measurements must be connected to the parameter associated with the Higgs boson self-coupling, $\lambda_3$. The production of single Higgs bosons is sensitive to $\lambda_3$ via loop effects and thus the precision of the single-Higgs-boson measurements must be better than 1% to provide sensitivity close to that from $HH$ production [8]. Measurements at multiple energies help break degeneracies in the interpretation of $\lambda_3$. For linear colliders running at higher energies, the precision on the self-coupling can be improved by combining $ZHH$ and $\nu\nu HH$ processes and by exploiting the large cross-section of $\nu\nu HH$ to make differential cross section measurements [9]. To compare the two approaches in a consistent manner requires an EFT analysis, as will be discussed in Sect. 3.5.

### 3.1.4 High-energy colliders

High-energy lepton or hadron machines operating at the multi-TeV scale offer unprecedented energy reach. The advantage of proton-proton colliders is that they produce large samples of Higgs bosons at high centre-of-mass energies, providing access to rare production and decay modes, and also allowing for differential measurements which are particularly sensitive to BSM effects. High-energy muon colliders offer a comparable energy reach, while also providing unique and complementary sensitivity to the EW sector.

In combination with $e^+e^-$collider data, the FCC-hh can achieve sub-percent-level precision on all Higgs-boson couplings, including those that involve rare decay processes such as $H \to 4\ell, \ell\ell\gamma, \mu\mu$. For the studies presented in Refs. [ID227], the absolute determination of the couplings of the Higgs boson is obtained from ratio measurements at FCC-hh that rely on absolute measurements from the FCC-ee (i.e., $H \to 4\ell$). In these ratios, many of the experimental and theoretical systematics uncertainties related to the production cross-section, such as those from PDFs and missing higher orders are assumed to cancel. For the scenarios where FCC-hh follows the HL-LHC directly, the numbers from the HL-LHC in [ID170] have been scaled to a centre-of-mass energy of 85 TeV. For this scenario, the statistical uncertainties are scaled accounting for the higher FCC-hh energy and luminosity, the experimental systematic uncertainties are left unchanged and the assumptions of the theory uncertainties are outlined in Sect. 3.4.

At the FCC-hh, the Higgs boson self-coupling can be measured using $HH$ final states. For the comparisons shown here only $b\bar{b}\gamma\gamma$ and $b\bar{b}\tau^+\tau^-$ [ID227] final states are considered. The FCC-hh is also expected to probe the quartic Higgs self-coupling ($\lambda_4$) via $HHH$ production at the level of 10 times the SM [4]. Furthermore, the FCC-hh will be able to measure the $VVHH$ coupling with a precision at the level of 1% [10].

The FCC-hh offers superior sensitivity to the top-quark Yukawa coupling, achieving a precision of 1% [ID227]. The FCC-hh also provides stringent limits on the branching fraction for Higgs boson decays to invisible particles, reaching down to 0.02%. Finally, the FCC-hh will deliver sensitivity to a large number of EFT operators thanks to the abundant rate of $t\bar{t}H$, $VH$ and VBF processes in the multi-TeV regime [ID227]. For the EFT comparison shown in Sect. 3.5, differential distributions for $WH$ and $t\bar{t}H$ production processes are also considered.

High energy muon colliders can probe Higgs physics via VBF production such as $\mu^+\mu^- \to \nu_\mu \bar{\nu}_\mu H$. The expected precisions on the Higgs boson couplings, including invisible decay modes, are at the percent to sub-percent level and are competitive with those obtained at the FCC-ee. With forward muon tagging, the normalisation of all couplings can also be obtained via neutral-current ($\mu^+\mu^- \to \mu^+\mu^- H$) interactions. The precision of the Higgs boson self-coupling via $HH$, $VVHH$ and $HHH$ final states is also similar to that achieved at the FCC-hh. As with the FCC-hh, the muon collider also has sensitivity to a wide range of EFT operators at high energies.

### 3.1.5 Comparison of $\kappa$ parameters

Modifications of the Higgs boson couplings with respect to their SM values are often parametrised using the so-called $\kappa$-framework [11], with parameters $\kappa_i$ acting as relative modifiers of the respective effective coupling constant. While this parameterisation provides a straight-forward comparison of possible observed SM deviations, it is not a consistent BSM model and, for instance, it cannot readily parameterise BSM effects that modify kinematic distributions. The sensitivity to these kinds of effects are better described in an EFT approach like the one of the Standard Model Effective Field Theory (SMEFT), which is discussed in Sect. 3.5.

The results of the $\kappa$ fits are shown in Fig. 3.1 for the different $e^+e^-$colliders and high energy options, as well as the option of FCC-hh directly after the HL-LHC. These results are obtained for a $\kappa$ scenario including the possibility of Higgs boson decays into additional light degrees of freedom, which could be either invisible ($\text{BR}_{\text{inv}}$) decays or an exotic signal that is not tagged by the experimental analyses ($\text{BR}_{\text{exo}}$). Here, the precision on the Higgs boson width is independent of that of the different SM decays, and allows a comparison of the indirect determination of $\Gamma_H$ across colliders. A determination of the Higgs boson width via this $\kappa$ fit is not possible at $pp$ or $ep$ colliders without additional assumptions. Motivated by several BSM scenarios, for $pp$ and $ep$ colliders $\kappa_{W,Z} \leq 1$ is assumed (see Ref. [12] Sect. 10). For lepton colliders, there is no need for this additional assumption, since the normalisation for all couplings can be obtained via the total $ZH$ cross-section for $e^+e^-$ colliders or via neutral-current ($\mu^+\mu^- \to \mu^+\mu^- H$) interactions with forward muon tagging at a muon collider.

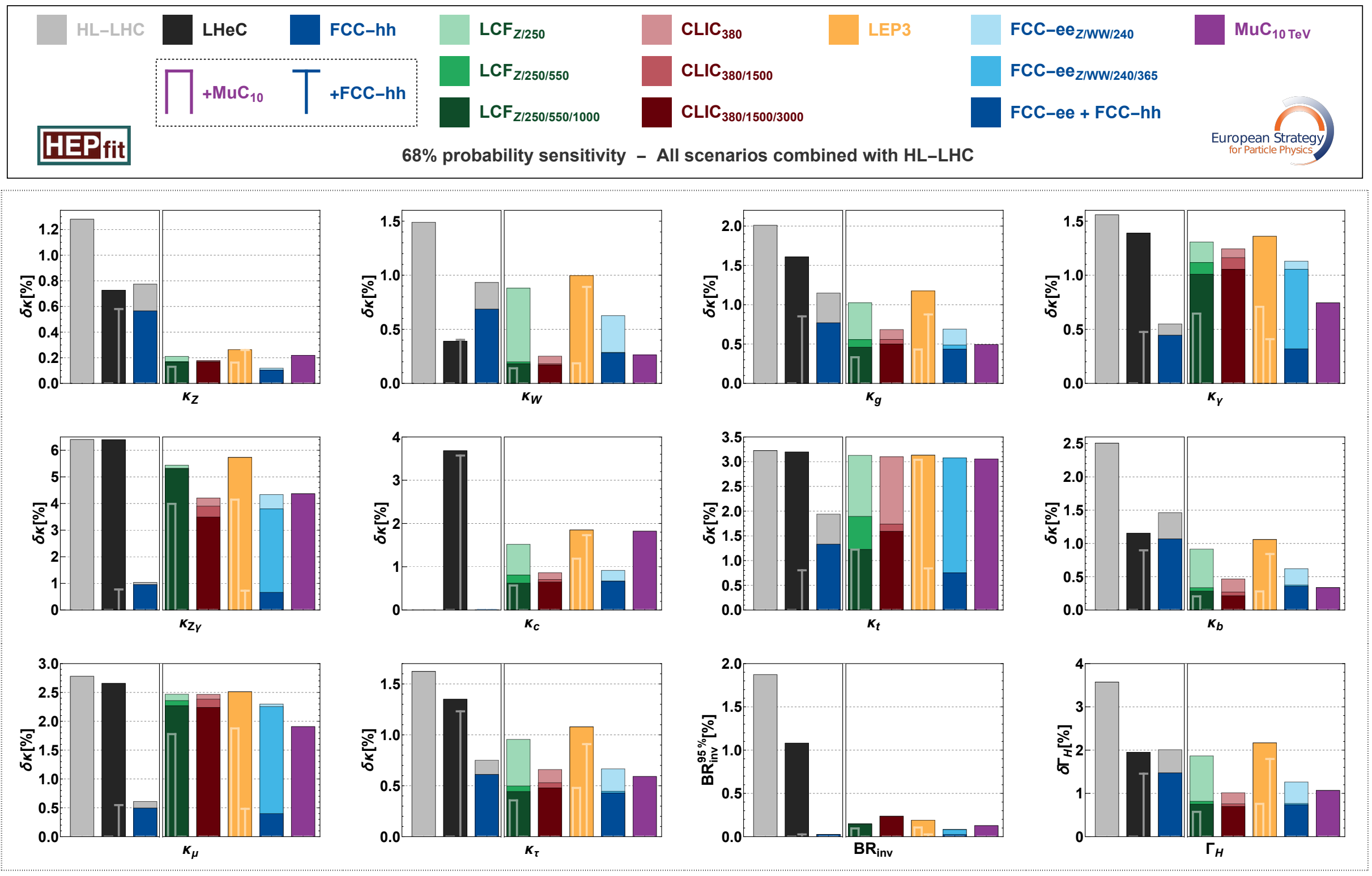

Fig. 3.1: 68% probability sensitivity to deviations from the SM on the $\kappa$ parameters associated to the different Higgs boson couplings compared across different collider types and energies. Also shown are the 95% probability upper limits on the branching fractions of the Higgs boson into invisible (BSM) states and the precision on the indirect determination of the Higgs boson width. The $pp$ and $ep$ scenarios are separated from the ones of lepton colliders to indicate the difference in assumptions that go into the interpretation at the two types of colliders. In particular, the constraint $\kappa_{W,Z} \leq 1$ is assumed for the $pp$ and $ep$ interpretations. In these cases, the reported value for $\kappa_{W,Z}$ refers to the lower bound of the 68% probability interval (i.e. $1 - \kappa_{W,Z} < \delta\kappa$ with 68% probability). The combinations of the LHeC and LEP3 with the FCC-hh are indicated with the "T" bars, whereas combinations of LCF and LEP3 with the 10 TeV muon collider are show with the empty bars. For the FCC-hh only scenario, the results are shown assuming the theory uncertainties stay as in the HL-LHC (in grey) or improve by a factor of two (dark blue).

All $e^+e^-$ and high-energy colliders yield significant improvements in precision of the coupling modifiers compared to the HL-LHC. For $e^+e^-$ energies around 250 GeV, the FCC-ee provides the highest precision compared to LEP3 and the LCF due to its large luminosity. For some rare decays such as $H \to \mu^+\mu^-$ and $H \to Z\gamma$, strong improvement in precision is only obtained by the FCC-hh. All proposed collider projects shown here can perform precise measurements of 2$^{\text{nd}}$ generation couplings such as that to charm quarks, which are difficult to measure at the HL-LHC[1]. The strange quark Yukawa coupling, not shown in the figure, can also be constrained at $e^+e^-$colliders, yielding from the fits at 68% probability, $|\kappa_s| \in [0.4, 1.3]$ for the FCC-ee, and $|\kappa_s| < 1.2, 1.3$ at LCF 250 GeV and LEP3, respectively. Higher centre-of-mass energies, such as $e^+e^-$colliders above 550 GeV, muon colliders or hadron colliders, bring strong improvements to many parameters such as $\kappa_W$ and $\kappa_t$. The LCF with energies up to 1 TeV combined with a 10 TeV muon collider yields comparative results to the FCC-ee +FCC-hh, apart from rare decays and assuming no theory uncertainties. The option of the FCC-hh directly following the HL-LHC bring a strong improvement in many couplings such as $\kappa_g$, $\kappa_t$ and $\kappa_\gamma$ compared to the HL-LHC. The combination of the LHeC and FCC-hh offers complementarity, with the LHeC providing strong constraints on $\kappa_W$ and $\kappa_Z$, while the FCC-hh improves the determination of several other Higgs couplings. Additional benefits from LHeC measurements of PDFs and $\alpha_s$ are not shown here; such improvements in theoretical inputs would further reduce the overall uncertainties, as illustrated by the two different theory assumptions considered for the FCC-hh.

[1]The FCC-hh only projection is not shown because the input data was not available.

The width of the Higgs boson is measured best by FCC-ee when considering lower energy $e^+e^-$options. However, the best precision is obtained when a low energy $e^+e^-$machine is combined with a high-energy option. For the invisible width of the Higgs boson, the highest precision is achieved by a hadron collider - either directly following the HL-LHC or following an $e^+e^-$or $ep$ collider.

For the lepton colliders, theory uncertainties are not shown in the figure. Including these will affect largely the precision of couplings with dependencies on uncertainties from $\alpha_s$ or on the quark masses, such as $\kappa_g$, $\kappa_c$ and $\kappa_b$. At high-energy lepton colliders, the missing higher-order corrections in VBF processes can also affect the precision of $\kappa_W$ and $\kappa_Z$. These uncertainties are more pronounced the higher the collider energy.

## 3.2 Top-quark physics

The top quark, as the heaviest known particle in the SM, plays a pivotal role in the study of electroweak symmetry breaking and the Higgs sector. Its large mass implies a strong coupling to the Higgs boson, and suggests a unique sensitivity to new physics effects. New physics could manifest itself by modifying top-gauge boson interactions and 4-fermion interactions, which can be tested via EFT operators. In addition, the top-quark mass is a critical input to SM consistency tests: it controls electroweak loop corrections that determine the predicted W mass, and enters flavour observables such as meson mixing and rare decays through virtual corrections.

For measurements of the top-quark mass, the HL-LHC could reach a precision of around 200 MeV [ID170] and this result will remain highly competitive for the next two decades. Lepton colliders that can reach the top-quark pair production threshold (the FCC-ee and linear colliders) provide a unique opportunity to measure the mass through threshold scans: the FCC-ee aims for a precision of about 7 MeV, while linear colliders are projected to reach 20 MeV (see Sect. 4.1.2 for more details). However, the experimental uncertainties are smaller compared to the theoretical ones, which are currently around 35 MeV.

To be sensitive to top-gauge boson couplings and new physics via EFT operators, the HL-LHC will probe top-quark interactions through channels such as four-top-quark ($t\bar{t}t\bar{t}$) production and associated $t\bar{t}\gamma$ and $t\bar{t}Z$ production. Differential measurements are important to provide information related to energy-dependent quark operators and will in many cases only be improved beyond the HL-LHC by a future hadron collider [ID227].

At lepton colliders, running at energies around the $t\bar{t}$ threshold enables a precise deter-mination of the top-electroweak couplings. For linear colliders, longitudinal beam polarisation offers an advantage as polarised beams provide direct access to the chiral structure of top-quark couplings. The FCC-ee can also probe the chiral structure of these couplings despite the absence of initial-state polarisation, as the information carried by the final-state polarisation of the top quark is accessible through its decay products. Beyond the threshold region, EFT operators with energy-growing effects become increasingly accessible. Having access to a wide energy range, e.g., linear colliders above 550 GeV or a muon collider, opens up sensitivity to energy-dependent operators. Measurements of cross-sections and forward–backward asymmetries at higher energies, and polarisation asymmetries at linear colliders, constrain four-fermion contact interactions involving the incoming leptons and top quarks ($\ell^+\ell^- t\bar{t}$) and further improve the precision on top-electroweak couplings, as discussed in Sect. 3.5.

In contrast, hadron colliders exhibit complementary sensitivity to different classes of operators, including those for four-quark interactions. The FCC-hh can provide a wide range of differential measurements, such as $t\bar{t}$, single top, $t\bar{t}Z$, which are sensitive to these interactions. In addition, processes such as $t\bar{t}t\bar{t}$, $t\bar{t}\gamma$ and $t\bar{t}Z$ production can be measured to a level of 1% [ID227]. The LHeC also offers unique top-quark probes, particularly of the $Wtb$ vertex. These measurements can complement hadron and lepton collider programmes by isolating specific couplings with reduced QCD backgrounds and different kinematics. A comparison of the physics potential of the different colliders for top-quark physics is discussed in Sect. 3.5.

Studies of FCNCs in the top-quark sector are excellent probes of new physics as they are strongly suppressed in the SM. As these results are often statistically limited, high precision data can yield significant improvements in terms of the reach to the BSM energy scale. Figure 3.2 shows the projected sensitivity to FCNCs to $t \to c$ and $t \to u$ transitions, parameterised via SMEFT operators (see Appendix A). Running at an energy above the top-quark pair threshold is needed to obtain significant improvement compared to the HL-LHC. In general, high energy colliders, such as a muon collider at 10 TeV or a hadron collider provide the best sensitivity.

## 3.3 Electroweak precision observables

Precision measurements of the electroweak sector provide a powerful test of the consistency of the SM itself, as they are sensitive to BSM via off-shell and loop contributions. In addition, if BSM signatures are observed, a broad range of measurements within the EW sector will be crucial for exploring the exact nature of any new physics. As EWPOs are needed to provide meaningful interpretation of other results, adequate precision is also critical in order to maximize the interpretative power within the SM as a whole. For example, quantities such as the effective weak-mixing angle, $\sin^2\theta_W$, and the $W$-boson mass, $m_W$, depend parametrically on other parameters such as the QED and QCD couplings and the masses of the $Z$ boson and the top quark. EWPOs also impact Higgs precision physics, as will be discussed in Sect. 3.5.

The HL-LHC will produce very large samples of $Z$ and $W$ bosons and can improve some EWPOs, for example $\sin^2\theta_W$ and $m_W$. In recent years, there has been significant improvement of the measured precision of these observables via new analysis methodologies [13, 14]. Nonetheless, the precision continues to be limited by the uncertainties in the PDFs, and in the QCD and EW higher-order corrections.

In contrast, lepton colliders can produce large samples of vector bosons, in a cleaner environment. Circular colliders, the FCC-ee and LEP3, benefit from very high luminosities, yielding $6 \times 10^{12}$ and $2 \times 10^{12}$ $Z$ bosons for all interaction points, respectively. The linear col-lider operates at lower luminosities, yielding $5 \times 10^{9}$ $Z$ bosons but has polarised beams, which partially compensate the lower luminosity via measurements that are sensitive to chiral observables. This section discusses traditional EWPOs determined by $e^{+}e^{-}$machines running at the $Z$-pole and the $WW$ threshold. High-energy lepton colliders ($e^{+}e^{-}$ or $\mu^{+}\mu^{-}$) also have enhanced sensitivity to some of the interactions that affect EWPOs via vector-boson scattering and production measurements and will be compared in Sect. 3.5 using the SMEFT framework. Table 3.1 highlights the projected uncertainties on a selection of EWPOs for the FCC-ee, LCF and LEP3 [ID217, ID140, ID188]. The large luminosity of the FCC-ee at the $Z$ pole leads to improvements of roughly a factor 30 in observables that depend primarily on the total rate, compared to the LCF. For chiral observables, where beam polarisation provides additional sensitivity at LCs, the relative gain of the FCC-ee is smaller, at the level of a factor 5, as illustrated in the table.

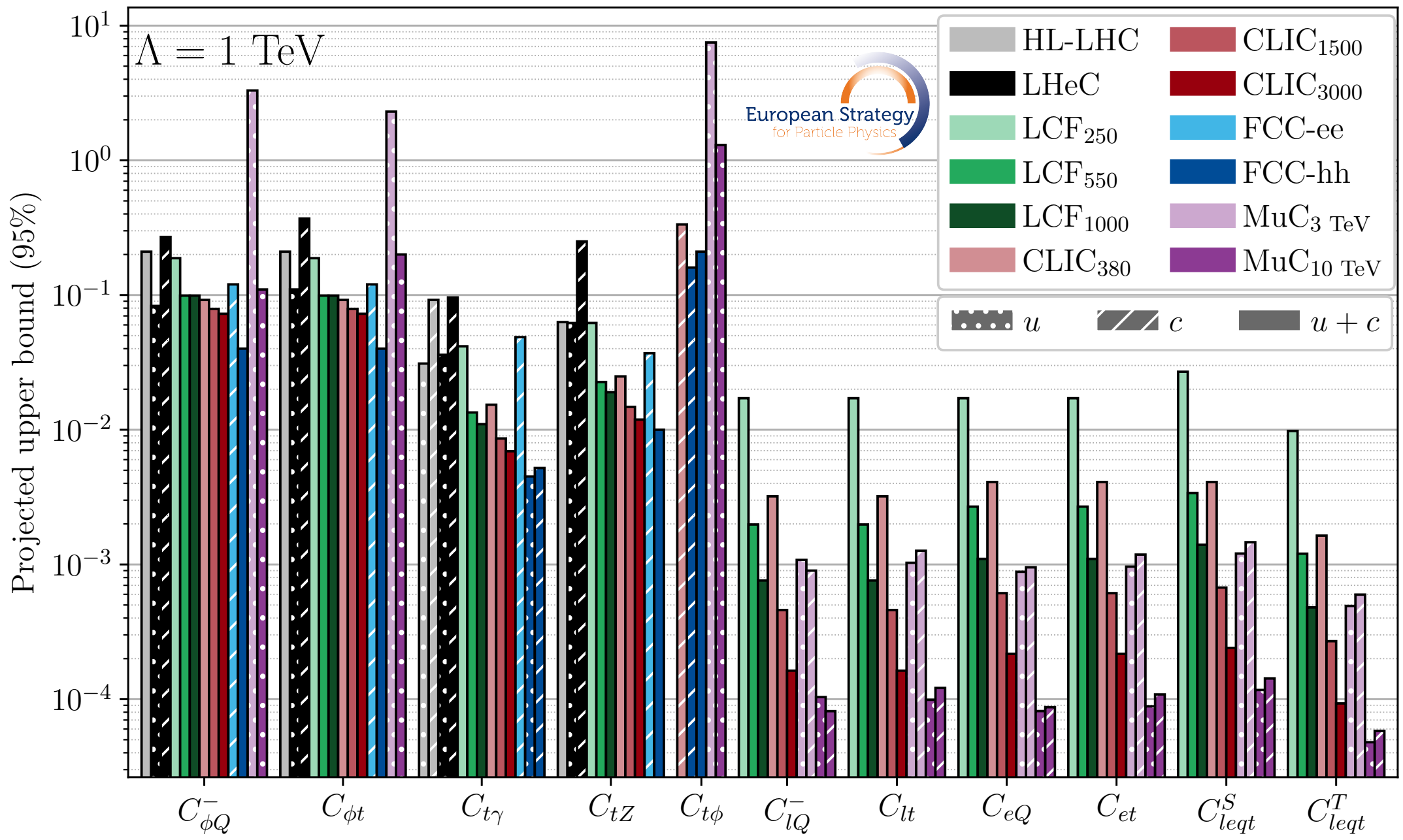

Fig. 3.2: Sensitivity of SMEFT operators to FCNCs in the top-quark sector for different collider options. Transitions between $t \to c$ and $t \to u$ and both are indicted by the different shaped areas. Inputs for some of the operators were not available for all projects.

Running at the $Z$-pole energy, the $Z$ boson mass and width and its couplings to fermions can be determined. For the $Z$ mass, the dominant systematic uncertainties are the absolute beam energy calibration (with resonant depolarization) for FCC-ee and LEP3 and the absolute momentum scale for linear colliders. For the linear collider estimates [ID140], the proposed calibration method is challenging and therefore the estimated precision on the mass should be considered as an optimistic estimate.

Measurements of the branching fractions provide the coupling strength of quarks and leptons to the $Z$ boson and can be sensitive to new physics. In these measurements, performed relative to the total hadronic branching fraction, the uncertainties due to the production cross-sections and luminosity cancel. With recent developments in flavour tagging, major improvements in the measurements of the ratios of branching fractions to bottom ($R_b$) and charm quarks to all hadrons are expected and also include first estimates to measure the ratio of the branching fraction to $s$-quarks over that for all hadrons. The projected uncertainties on a selection of these ratios are shown in Table 3.1. Minimizing these uncertainties places strong requirements on the design of the detectors.

Table 3.1: Current and projected uncertainties on a selection of EWPOs at the FCC-ee, the LCF and LEP3. The current uncertainties are taken from Ref. [13, 15]. When a single number is quoted, it refers to the total uncertainty, otherwise the statistical error is quoted and the experimental systematic error is given in the parentheses. $\Delta$ ($\delta$) stands for absolute (relative) uncertainty.

| Observable | Current | FCC-ee | LCF | LEP3 |
|---|---|---|---|---|
| $\Delta \mathrm{m_Z}$ (keV) | 2000 | 4 (100) | 200 | 7.5 (100) |
| $\Delta \Gamma_\mathrm{Z}$ (keV) | 2300 | 4 (12) | 125 | 7.5 (23) |
| $\delta \mathrm{R}_\mu$ $(\times 10^{-6})$ $\mathrm{R}_\mu \equiv \frac{\Gamma_\mathrm{had}}{\Gamma_\mu}$ | 1600 | 2.4 (2.3) | 90 (90) | 4.5 (2.3) |
| $\delta R_b$ $(\times 10^{-6})$ $\mathrm{R_b} \equiv \frac{\Gamma_\mathrm{b}}{\Gamma_\mathrm{had}}$ | 3300 | 1.2 (1.6) | 70 (60) | 2.2 (3.0) |
| $\Delta \sin^2\theta_W$ $(\times 10^6)$ | 130 | 0.4 (0.5) | 2.7 (2.3) | 0.75 (0.95) |
| $\Delta\alpha(m_Z)^{-1}$ $(\times 10^3)$ | 14 | 0.8, 3.8 | – | 1.4, 7.3 |
| $\Delta \mathrm{m_W}$ (keV) | 9900 | 180 (160) | 500 (1600) | 430 (700) |
| $\Delta \Gamma_\mathrm{W}$ (keV) | 42000 | 270 (200) | 2000 | 650 (500) |

The value of the electroweak parameter $\sin^2\theta_W$ is determined at $e^+e^-$colliders via measurements of different types of asymmetries. For linear colliders with longitudinal beam polarization, the asymmetry between the electron left- and right-handed couplings to the $Z$ boson can be measured directly via the quantity $A_{LR}$ [ID140]. For circular colliders, $\sin^2\theta_W$ can be determined via the combination of several forward-backward asymmetries ($A_{FB}^{f,0}$) and from the forward-backward asymmetry of the $\tau$-polarisation ($A_{FB}^{\mathrm{pol}(\tau)}$) [ID217]. Lower-energy $e^+e^-$colliders, $ep$ colliders and neutrino scattering facilities can also probe $\sin^2\theta_W$. For example, $ep$ colliders such as EIC and LHeC would provide information about the running of $\sin^2\theta_W$ at scales of a few tens of GeV and enable the separate, high-precision determination of the electroweak couplings of the up and down quarks. In addition, in the low-momentum range of $10^{-3} - 10^{-1}$ GeV, low-energy experiments like P2 [16], MOLLER [17] and CONUS+ [ID191] can provide a precision on $\sin^2\theta_W$ of 0.13%, 0.12% and 5%, respectively.

The current precision of the fine-structure constant $\alpha(m_Z)$ [15] will not be sufficient for future electroweak precision tests and therefore improved determinations of it are vital. At the FCC-ee and LEP3, $\alpha(m_Z)$ can be determined via off-peak measurements at the $Z$-pole of the forward-backward asymmetry [18]. More recently, a novel approach based on the comparison of the differential distributions of electrons, muons and positrons produced on the $Z$-peak has been proposed [19]. This study has the potential to greatly reduce the statistical uncertainties on the measurement of $\alpha(m_Z)$, though more work is needed to correctly assess the associated systematic uncertainties. At linear colliders, the luminosity is not sufficient to perform such a measurement via this methodology and therefore rely on predictions from Lattice QCD (see Sect. 3.4). There are several experiments devoted to probing QED in the strong-field limit, including the AWAKE plasma wakefield accelerator at CERN [20], the European XFEL in Germany [21], or the FACET facility at SLAC [22]. The proposed new experiments will probe QED in the critical field regime, which is of relevance for instance for astrophysical phenomena (for instance magnetars [23]), atoms with $Z > 137$ [24] and for high energy $e^+e^-$ colliders [25, 26].

The mass, $m_W$, and width, $\Gamma_W$, of the $W$ boson can be determined via a threshold scan at the $WW$ energy threshold or at higher energies, e.g., at 240 GeV. While there are multiple methodologies to measure the $W$ boson mass at $e^+e^-$colliders, the results are often dominated by systematic uncertainties, such as the beam energy calibration at the threshold scan and modelling of hadronisation at higher centre-of-mass energies. The LHeC enables a determination of the $W$ boson mass with a precision of 3 MeV using HL-LHC data (see more details in Sect. 4.1.2). Runs at high centre-of-mass energies also enable high-precision tests of lepton universality.

Multi-boson interactions, which give rise to final states with two or three bosons, test for BSM physics via deviations in the rate or the kinematic distributions. While the experiments at the LHC have performed a wide-range of diboson and triboson measurements, no official projections for the foreseen precision at the HL-LHC are available. For future colliders, projections for measurements of $e^+e^- \to W^+W^-$ are included in the EFT fits in Sect. 3.5, using the statistical optimal observables formalism defined in Ref. [27]. The sensitivity to new physics in diboson processes at very high energies has also been studied at muon colliders [28] and preliminary projections for similar processes at the FCC-hh are also considered [ID227]. The expected sensitivities of these measurements to EFT operators are presented in Sect. 3.5.

Figure 3.3 (top) shows the expected precision on the top-quark and $W$ masses for the HL-LHC, FCC-ee and LCF. The indirect sensitivity to $m_W$ and $m_t$, shown by the filled ellipses, are obtained via a fit of SM theory predictions to projected EWPOs. For comparison, the direct measurement precision is also shown via the dotted/solid lines for HL-LHC/$e^+e^-$ colliders. The bottom of Fig. 3.3 shows oblique parameters S and T [29] for the different $e^+e^-$ collider options. These parameters help account for possible modifications of the EWPOs and quantify universal BSM corrections to the photon, $W$ and $Z$ self-energies (e.g. they do not depend on the lepton and quark flavours). T measures the difference between the new physics contributions of neutral and charged current processes at low energies and S describes new physics contributions to neutral current processes at different energy scales. The best precision is obtained by the FCC-ee, followed by LEP3. The effects of theory uncertainties are discussed in Sect. 3.4 and shown in Fig. 3.4.

## 3.4 Theory requirements and uncertainties

Theory input is critical for the physics programme of any future collider. It is needed for the extraction of interesting electroweak and Higgs boson quantities from the data (e.g. Monte-Carlo (MC) simulations, PDFs, theory calculations for backgrounds), as well as for the interpretation of these target quantities in terms of the SM or BSM theories. Theory input is also needed for the determination of physical quantities from data, e.g. the value of a mass or a coupling, that are used in the calculations. For Higgs physics at the HL-LHC, the precision in many cases is limited by the theory [ID170], [30]. The required precision on the theory inputs is even more demanding for precision physics at future high-luminosity $e^+e^-$ and high-energy hadron colliders, but to a lesser extent for high-energy lepton colliders.

These inputs have several sources of purely theoretical uncertainties, including:

a) Missing Higher-Order Uncertainties (MHOUs) in perturbative calculations.

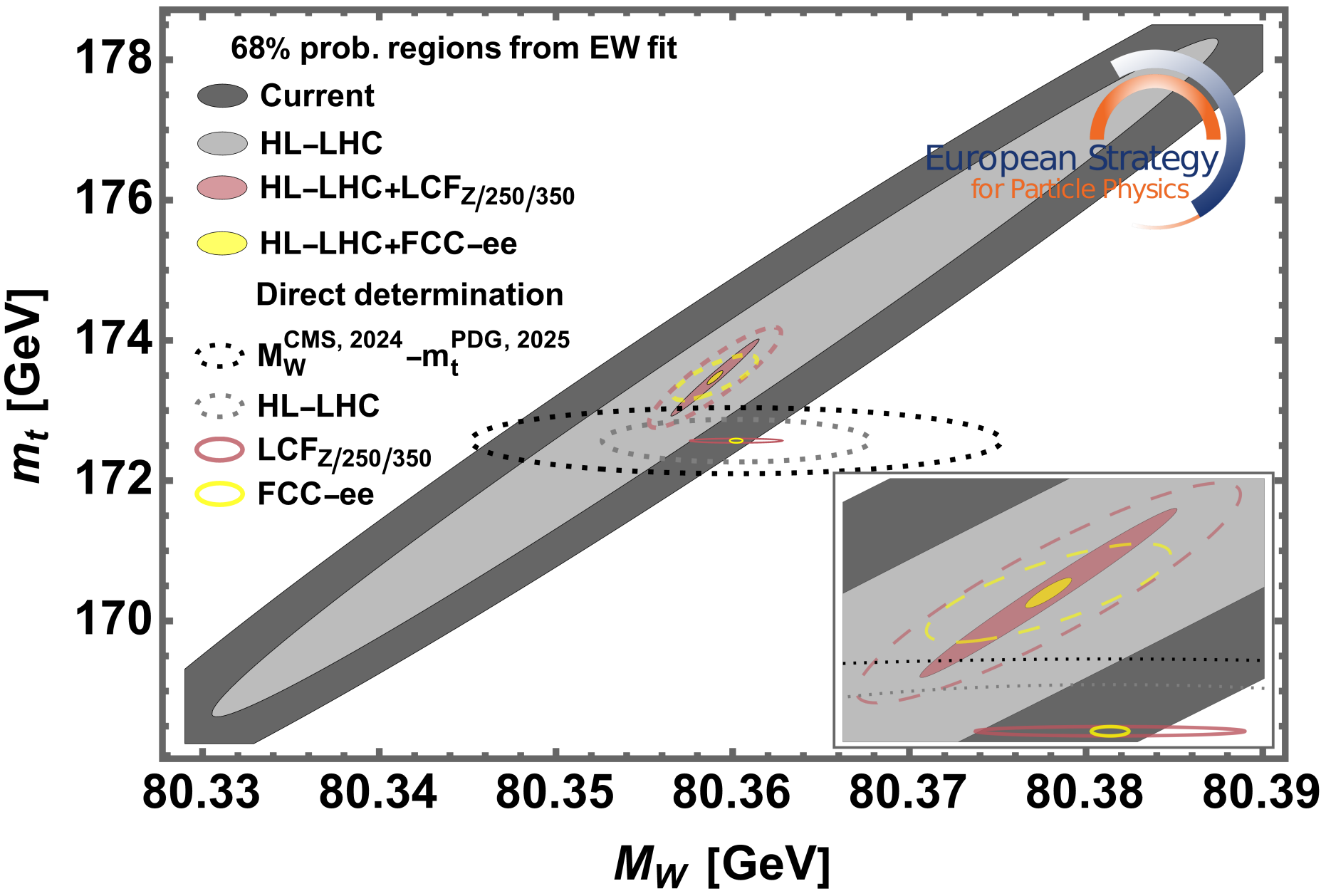

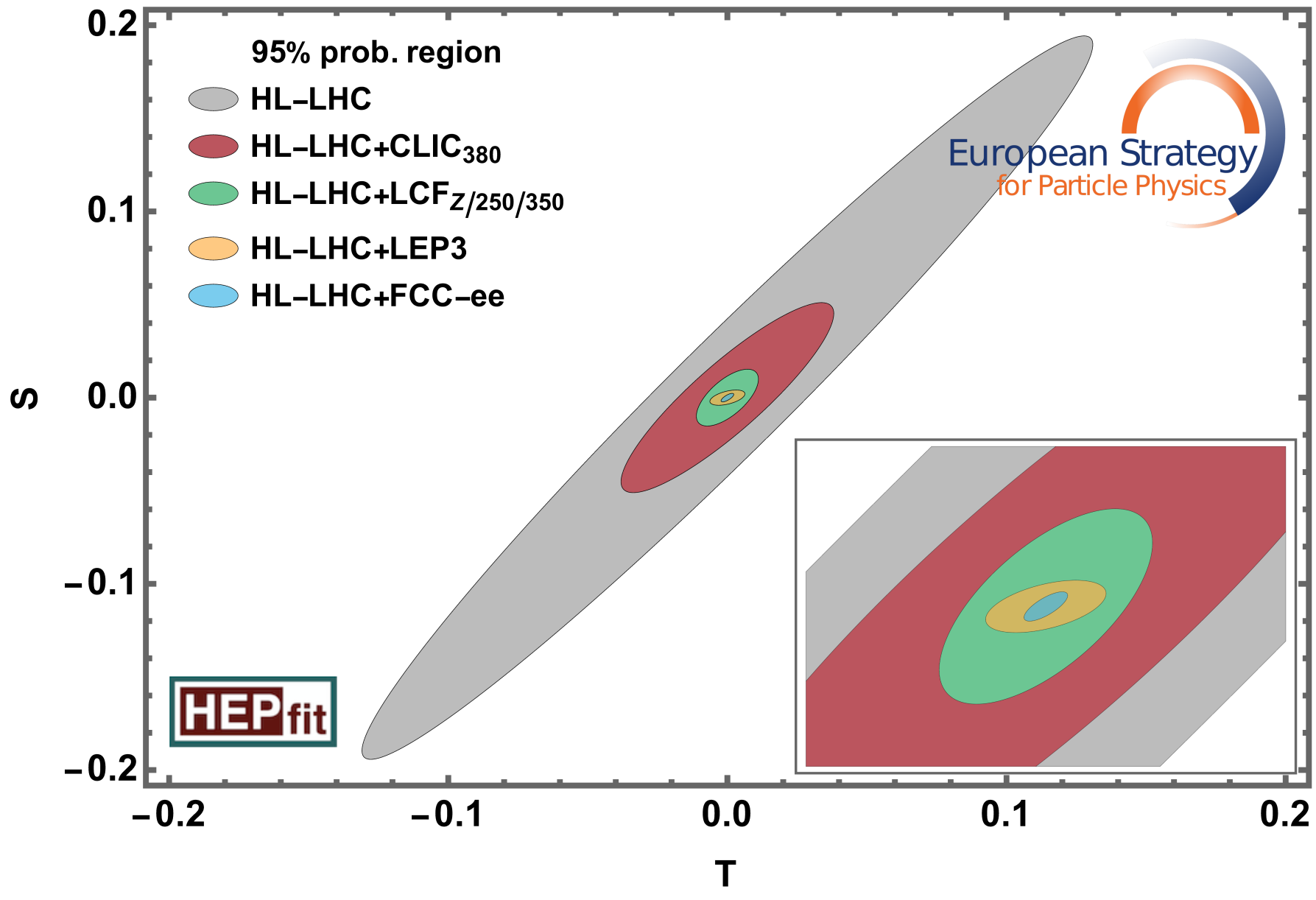

Fig. 3.3: Top: Comparison of the indirect and direct mass determination of the $W$ and top quark for different collider options. (For the EW fit results, the solid ellipses indicate the results obtained assuming the aggressive scenario for the theory uncertainties discussed in the next section. For the lepton colliders the dashed contours show the results for the conservative scenario, also discussed in the next section.) Bottom: Comparison of the S and T oblique parameters.

b) PDF uncertainties: these are due to the experimental and theoretical uncertainties for the observables used in PDF fits, and modelling uncertainties such as the parameterisation, evolution kernels, and on heavy-quark evolution of the PDFs themselves.

c) Non-perturbative (NP) effects, e.g. in running couplings or hadronisation: these are usually implemented via physics-inspired modelling and require input from data.

There is no rigorous method to quantify theory uncertainties, and any estimate relies on ad-hoc assumptions. Moreover, most of the methods do not provide intervals with a meaningfully interpretable statistical meaning.

Ideally, theory uncertainties should be reduced to a level where they are sub-dominant to statistical or experimental systematic uncertainties. However, achieving this goal requires significant theory development.

In this section, theory uncertainties estimates from Refs. [31, 32] are reviewed, updated and extended to account for recent developments. The focus is on $e^+e^-$ colliders, but some discussion of hadron and muon collider physics is also given. The aim is to estimate what improvements are needed to reach a certain precision, and not to forecast what theory improvements are likely to be achieved nor to determine any fundamental limits to possible future improvements.

Several different methods for estimating theory uncertainties have been proposed [33]. This section mostly focuses on improvements to MHOUs, using order-by-order scaling factors determined from either typical coupling and combinatorial prefactors, or from the ratio of existing theory results.

### 3.4.1 $e^+e^-$ colliders

At $e^+e^-$ colliders, the EWPOs and Higgs boson precision observables (HPOs) pose the highest demands on precision theory input. Although the current report provides estimates for a larger range of error sources than in previous works, it is nevertheless not comprehensive[2]. More details of the uncertainty evaluation can be found in Ref. [33]. Several scenarios are considered:

i) Current status (i.e. evaluation of uncertainties of existing calculations);

ii) A "conservative" future scenario, assuming theory improvements which are likely to be achieved by building on and extending existing computational methods;

iii) An "aggressive" future scenario, which would require more fundamental advances in techniques and tools.

In addition, one may consider an "ideal" scenario, in which theory uncertainties are subdominant for all observables and thus negligible.

The current framework of EWPOs may not be sufficient for the experimental precision at future $e^+e^-$ colliders, and the discussion of a more extensive parameterisation [34, 35] or new approaches inspired by lessons learned at the LHC [36], is starting within the community. For the purpose of the current report, a structure based on EWPOs is still adequate to assess the impact of theoretical uncertainties on the physics reach of the colliders, even if a new theoretical framework is needed for the actual analysis of future experimental data.

#### *Extraction of pseudo-observables*

EWPOs and HPOs are pseudo-observables (POs), which means that the extraction of these quantities from data requires theory input on a wide range of topics, including background calculations, simulation of QED/QCD radiation, modelling of hadronisation, underlying event,

[2] In particular, we do not evaluate theory uncertainties in the determination of the beam energy and luminosity [6].

and other non-perturbative effects.

**Z pole:** The extraction of the $Z$-boson peak cross-section, the total width, the branching fractions, and the various asymmetries requires the subtraction of "backgrounds", e.g. from photon exchange and box diagrams, and of ISR/FSR QED/QCD contributions [32, 34, 37]. The former are currently known at NLO, while in the future they are envisioned to be extended to NNLO. ISR/FSR is typically evaluated using MC tools, which currently include NLO corrections and resummation of leading soft and/or collinear contributions.

The conservative future scenario assumes the inclusion of fixed-order NNLO corrections and the parton-shower-like production of fermion pairs, while the aggressive scenario also assumes N3LO QCD corrections. Both assume an improvement of non-perturbative uncertainties from hadronisation by a factor 5–10 and 50, respectively. The current and projected uncertainties are listed in Table 3.2.

Table 3.2: Current and projected theory uncertainties for the extraction of $Z$-Pole POs at $e^+e^-$colliders, due to background subtraction and ISR/FSR. A dash indicates that projections for the aggressive scenario are currently not available and the uncertainly will be considered negligible for the fits in section 3.5.

| Observable | Current | Conservative | Aggressive |
|---|---|---|---|
| $\Gamma_Z$ (MeV) | 0.23 | 0.035 | — |
| $m_Z$ (MeV) | 0.3 | 0.03 | — |
| $R_\ell$ ($10^{-3}$) | 12 | 0.4 | — |
| $R_b$ ($10^{-4}$) | 4.4 | 0.44 | 0.09 |
| $R_c$ ($10^{-4}$) | 17 | 1.7 | 0.34 |
| $\sigma_{\text{had}}$ (pb) | 25 | 1.7 | — |
| $A^\ell_{\text{FB}}$ ($10^{-4}$) | 6 | 0.43 | — |
| $A^b_{\text{FB}}$ ($10^{-4}$) | 1.5 | 0.32 | 0.028 |
| $A^c_{\text{FB}}$ ($10^{-4}$) | 1.1 | 0.23 | 0.021 |

### *Predictions for pseudo-observables*

Accurate SM predictions for POs are needed for global fits and the derivation of limits on BSM physics. We describe next the projections for the uncertainties of the different types of observables due to missing higher-order calculations. There are also parametric uncertainties from the uncertainty of direct measurements of the SM input parameters. These can also be affected by the precision of theory calculations, e.g., for the top-quark mass and the electromagnetic and strong coupling constant. All these parametric uncertainties are taken into account in the fits presented in the next section.

**Higgs boson decays:** The uncertainties for Higgs boson decay calculations are evaluated using the methodology of Refs. [31, 38]. The current state-of-the-art includes NLO calculations for all decay channels, higher-order QCD calculations for hadronic final states, and partial NNLO corrections enhanced by powers of $m_t$ for the tree-level decays ($b\bar{b}$, $c\bar{c}$, $\tau^+\tau^-$, $\mu^+\mu^-$, $W^+W^-$ and $ZZ$). The conservative scenario assumes full NNLO corrections for the tree-level decays and $O(\alpha_s^4)$ QCD corrections for $H \to gg$, while the aggressive scenario considers $O(\alpha_s^5)$ corrections for $H \to bb/cc/gg$ and also mixed EW-QCD NNLO (3-loop) corrections for the $H \to gg/\gamma\gamma$. The resulting projected uncertainties are listed in Table 3.3. For several decay channels, the conservative scenario projections are already subdominant compared to the foreseen experimental uncertainties at $e^+e^-$colliders.

Table 3.3: Current and projected theory uncertainties (in percent) for the SM predictions of partial Higgs boson decay widths and production cross-sections. A dash indicates that there is no projection for this uncertainty and it will be considered negligible.

| Process | Current | Conservative | Aggressive |
|---|---|---|---|
| $H \to bb/cc$ (%) | $< 0.4$ | 0.2 | 0.1 |
| $H \to \tau\tau/\mu\mu$ (%) | $< 0.3$ | $< 0.1$ | — |
| $H \to WW^*/ZZ^*$ (%) | 0.5 | 0.3 | — |
| $H \to gg$ (%) | 3.2 | 1.0 | 0.5 |
| $H \to \gamma\gamma$ (%) | $< 1.0$ | $< 1.0$ | 0.4 |
| $H \to Z\gamma$ (%) | 1.5 | 1.5 | — |
| $e^+e^- \to ZH$ (%) | 0.3 | $< 0.1$ | — |
| $e^+e^- \to \nu\bar{\nu}H$ (%) | $\sim 1$ | $\sim 0.1$ | — |

**Higgs boson production:** For $e^+e^- \to ZH$ production, NLO and partial NNLO SM corrections with closed fermion loops are known, whereas the VBF channel, $e^+e^- \to \nu\bar{\nu}H$, is limited to NLO. For the conservative future scenario, full NNLO corrections are assumed for both processes. Since the estimated uncertainties (see Table 3.3) of this scenario are already subdominant compared to the experimental precision goals, no aggressive scenario estimates are needed.

**Z pole:** The current state-of-the-art SM predictions include full NNLO corrections and partial higher-order corrections enhanced by powers of $m_t$. For the conservative (aggressive) scenario we assume that these calculations are approximately extended by one (two) loop order(s), but only considering contributions that are enhanced by gluons and/or multiple fermion lines in the loop. The estimated uncertainties are listed in Table 3.4.

Figure 3.4 shows the effects of theory uncertainties on the oblique parameters S and T, for the FCC-ee and the LCF. In both cases, the precision on the parameters is significantly affected by the assumptions on the size of theory uncertainties, and even the most aggressive scenario may not be sufficient to fully exploit the precision of the experimental data. The impact of these uncertainties is more prominent in the case of the FCC-ee, due to the much higher experimental precision on EWPOs.

**W-boson physics:** The cross-section for $e^+e^- \to W^+W^-$ and the $W$ decay widths to leptons and hadrons are currently known to NLO and including higher-order final-state corrections. The estimated uncertainties are small, but are assumed to be further reducible with future NNLO and dominant 3rd order corrections, which would render them negligible compared to the achievable experimental precision, see Table 3.5.

### *Mass and gauge-coupling parameters*

The extraction of the masses of the top-quark and the $W$ boson from pair production near threshold require predictions of the production and decay processes in an integrated effective field theory framework. The estimates in Table 3.6 are based on Refs. [31, 39]. Estimates for the aggressive future theory scenario are currently not available, but further progress will require fundamental advances in computational techniques. The strong and electromagnetic couplings at the $m_Z$ scale are mostly affected by non-perturbative QCD uncertainties, which could be reduced with advances in lattice calculations (see also Sect. 4.1.2). The electromagnetic coupling could also be directly determined at a circular $e^+e^-$collider [18, 19], though more studies are needed for the evaluation of theory uncertainties. More discussions of these issues can be found in Ref. [33].

Table 3.4: Current and projected theory uncertainties for the SM predictions of $Z$-pole POs.

| Observable | Current | Conservative | Aggressive |
|---|---|---|---|
| $\Gamma_Z$ (MeV) | 0.4 | 0.08 | 0.016 |
| $R_\ell$ ($10^{-3}$) | 6.0 | 1.2 | 0.2 |
| $R_b$ ($10^{-4}$) | 1.0 | 0.2 | 0.035 |
| $R_c$ ($10^{-4}$) | 0.5 | 0.1 | 0.02 |
| $\sigma_{\text{had}}$ (pb) | 6 | 1.6 | 0.3 |
| $\sin^2\theta_{\text{eff}}$ ($10^{-5}$) | 4.5 | 0.7 | 0.06 |
| $m_W$ (MeV) | 4.0 | 1.0 | 0.1 |

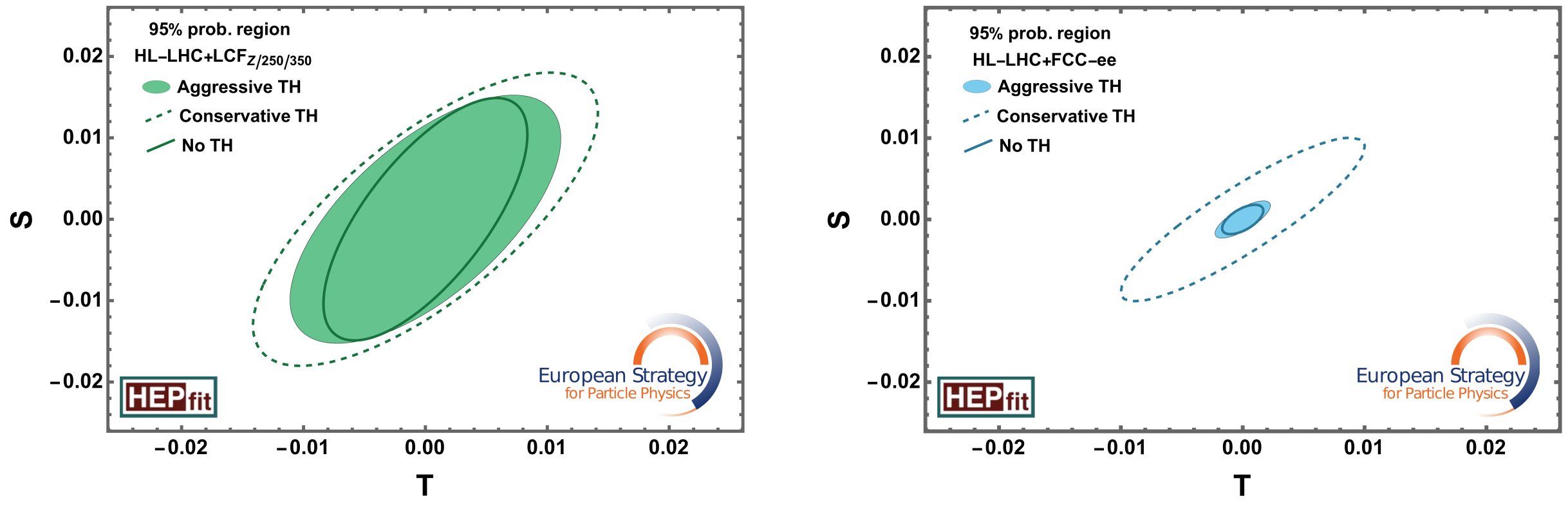

Fig. 3.4: The oblique parameters S and T shown with various theory comparisons for the LCF (left panel) and FCC-ee (right panel).

Table 3.5: Current and projected theory uncertainties (in percent) for the SM predictions of partial $W$-boson decay widths, ratios, and production cross-sections. A dash indicates that there is no projection for this uncertainty and it will be considered negligible.

| Quantity | Current | Conservative | Aggressive |
|---|---|---|---|
| $\Gamma_W^{\text{lep}}$ (%) | 0.1 | 0.013 | — |
| $\Gamma_W^{\text{had}}$ (%) | 0.1 | 0.015 | — |
| $\Gamma_W^{\text{had}}/\Gamma_W^{\text{lep}}$ (%) | 0.015 | $< 0.01$ | — |
| $\sigma[e^+e^- \to W^+W^-]$ (%) | 0.4 | 0.07 | — |

Table 3.6: Impact of theory uncertainties on the determination of various SM parameters. A dash indicates that there is no projection for this uncertainty and it will be considered negligible.

| Quantity | Current | Conservative | Aggressive |
|---|---|---|---|
| $m_t$ (MeV) | 35 | 25 | — |
| $m_W$ (MeV) | 3 | 0.6 | — |
| $\alpha_s(m_Z)$ $(10^{-3})$ | 1 | 0.3 | 0.1 |
| $\alpha(m_Z)/\alpha(0)$ $(10^{-4})$ | $< 1$ | $< 0.5$ | 0.1 - 0.3 |

### 3.4.2 Hadron colliders

Future projections for HL-LHC assume a reduction of theory uncertainties by a factor two for most precision measurements [30,40]. This goal, while relatively modest, will require improvements on many fronts, including perturbative calculations, MC simulations, non-perturbative modelling and PDFs. There are unique theory challenges for the determination of the $W$ mass (target precision 5 MeV [41]) and top-quark mass (target precision $< 0.2\,\text{GeV}$). For $m_W$, this will require theory improvements for the prediction of the differential cross-sections and for the PDFs. For $m_t$, there is a theoretical ambiguity in the definition of the top-quark "pole" mass of about 0.5 GeV [42–45]. (In particular, part of this ambiguity corresponds to an irreducible uncertainty of the order of $\Lambda_{\text{QCD}}$, which sets the ultimate precision achievable for the "pole" mass at the level of 100 MeV.) To circumvent this, improvements in MC generators and analysis techniques, using different mass definitions and observables, are needed.

For FCC-hh, QCD and EW theory uncertainties are in general significantly larger than for the LHC, except for certain final-state ratios, see Sect. 3.1.4. Dedicated studies are necessary for Higgs-boson and gauge-boson pair production processes. In particular, massive boson "radiation" will be ubiquitous at FCC-hh (see e.g. [46,47]) and it is currently unclear if this effect can be better simulated with EW parton showers or with explicit matrix elements. Given that a robust estimate of FCC-hh theory systematics is presently not feasible, the FCC-hh projections adopt two benchmark scenarios: a conservative one that assumes the HL-LHC theory uncertainty values, and an optimistic one in which all theory uncertainties are reduced by a factor of two.

### 3.4.3 Muon colliders

Similar to the FCC-hh, dedicated studies are needed for the evaluation of theory uncertainties for muon collider physics. Given that the target precision for many observables is at the percent level, it is expected that theory uncertainties are not a limiting factor [48], with a few exceptions, such as $m_W$ and $m_t$ from threshold scans. Due to enhancements from electroweak Sudakov logarithms, even percent-level precision will require corrections beyond NLO, either at a fixed order, through EW PDFs beyond leading-log approximation, or suitable matching strategies between the two [49].

## 3.5 Comparisons in the Standard Model Effective Field Theory interpretation

The input from various collider projects on Higgs, top-quark, and Electroweak physics is interpreted within the framework of the SMEFT (see Appendix A for details). The SMEFT fit employed here assumes that new physics respects a $U(2)^5$ flavour symmetry, where the third generation is aligned with the UP basis in flavour space. This assumption enables a meaningful test of new physics models that aim to address SM issues related to the EW scale, while not conflicting with the typically much stronger constraints from flavour-changing processes, discussed in Chapter 5.

These flavour assumptions are imposed at the cut-off scale $\Lambda$, above which the EFT description is assumed to break down. For this study, a benchmark value of $\Lambda = 10\,\text{TeV}$ is chosen as it provides a representative reach for the observables accessible at the high-energy collider options of the 10 TeV muon collider and the 85 TeV FCC-hh, which can perform measurements at energies beyond the cut-off value.

The analysis is restricted to dimension-6 operators, which provide the leading new physics effects in the EFT expansion. Under the $U(2)^5$ flavour assumptions and imposing CP-conservation, this leaves a total of 124 operators. [3] Taking an agnostic approach on the origin of the SMEFT operators, all interactions allowed by the imposed symmetry are assumed to be generated by new physics at the scale $\Lambda$ and are treated as free parameters in the fits — alongside the main SM input parameters of the EW sector. The SMEFT new physics corrections are also assumed to be perturbative. This theory assumption constrains the size of the new interactions even in those cases where there are more degrees of freedom than what can be constrained by the data available.

Using the 1-loop renormalization group equations, the Wilson coefficients at the scale $\Lambda$ are evolved down to the characteristic scales where the different observables are measured. The SMEFT corrections are then computed at these scales at leading order in perturbation theory. In most cases, this is done at the tree level. For the most precise observables, such as the EWPOs or HPOs, the 1-loop SMEFT finite corrections are also included [4]. Note also that the bounds on the operators depend on the choice of the renormalization scale. In the comparison of the results presented in terms of the Wilson coefficients of the operators, these are evaluated at 10 TeV.

As discussed in Sect. 3.4, theoretical uncertainties must not be neglected. For the purpose of the comparison presented here, a baseline scenario is assumed in which theoretical uncertainties under the aggressive scenario in Tables 3.2-3.6 are used. For entries in those tables that are marked with a dash, the theory uncertainty is considered negligible[5]. However, where relevant, comparisons when using the conservative scenario are shown. Finally, parametric uncertainties from the experimental determinations of the SM input parameters either from the collider project or from external determinations, e.g. future projections from lattice QCD, are included.

***Limits from electroweak precision measurements***

EWPOs place strong bounds on SMEFT interactions modifying $Zff$ and $Wff$ vertices at the tree level and also, thanks to large increase in precision, strong constrains on SMEFT oper-ators affecting these observables via quantum corrections. Measurements of 2-to-2 fermion processes above the $Z$ resonance, which provide additional complementary constraints to those from EWPO, are also included in the fits but are discussed separately at the end of this section. Electroweak precision tests in diboson processes also provide extra constraints. At energies above the $W^+W^-$ production threshold, diboson measurements test for the presence of anomalous triple gauge couplings (aTGC): $\delta g_{1,Z}$, $\delta\kappa_\gamma$ and $\lambda_Z$ [50]. In the SMEFT approach, $\delta g_{1,Z}$ and $\delta\kappa_\gamma$ are induced by operators that also modify the Higgs boson couplings to vectors bosons and are discussed together with the Higgs boson interactions. At high energies, diboson measurements bring strong sensitivity to certain operators that also contribute to EWPOs. The benefits of high-energy measurements in the EW physics programme will be illustrated here and also further discussed at the end of this section.

[3] In practice, the number of operators is reduced to 84 by ignoring some interactions that mix weakly with those entering in the precision observables considered in this study. We also note that some of the neglected operators, e.g. four-light-quark operators, are constrained by the LHC data, but no projections for HL-LHC are currently available.

[4] While these are only technically consistent after including 2-loop renormalization group effects, these are not available in the literature, and therefore not included here, and are another example highlighting the need for higher-order BSM calculations, in this case in the SMEFT, to be able to exploit the future experimental precision.

[5] This does not imply that the theory uncertainty is known to be subdominant in all these cases, but more theory work will be needed to arrive at a meaningful estimate.

The top panel of Fig. 3.5 compares the sensitivity to the combinations of SMEFT operators correcting the effective electroweak couplings $g^f_{L,R}$, defined from the SM expressions of the partial $Z$ decay widths $\Gamma(Z \to f\bar{f})$ and the left-right asymmetry parameters $A_f$. Overall, the FCC integrated project provides the best sensitivity to modifications of the fermionic EW interactions. However, the obtainable precision from EWPOs depends strongly on the assumptions taken for the theory uncertainties. This is illustrated in the middle panel of Fig. 3.5, which shows the ratio $\Delta^{A,C}_{TH} = \delta g^{A,C}_i/\delta g^0_i$, where $\delta g^{A,C}$ are the precisions obtained under the aggressive (A) and conservative (C) assumptions, and $\delta g^0$ those from the ideal scenario where theory uncertainties are neglected. The FCC-ee is most strongly affected by theory uncertainties as its results are the most precise, although its overall precision is still the best even when considering different theory uncertainty scenarios.

The EW interactions $g^f_{L,R}$ receive universal modifications from the bosonic operators $\mathcal{O}_{\phi WB}$ and $\mathcal{O}_{\phi D}$ contributing to the $S$ and $T$ parameters, respectively, described in Sect. 3.3, and fermion-specific corrections given by the operators of the type $\mathcal{O}_{\phi f}$ in Table A.1 in Appendix A. The bottom panel of Fig. 3.5 shows the bounds on the interaction scale associated to these operators. At the leading order, the parameters $g^f_{L,R}$ correspond to 12 combinations of the 15 operators in the bottom panel, which results in large correlations from the precision of EW observables. The situation is further complicated by NLO effects. To illustrate these large correlations, the bounds obtained for each operator individually are also shown.

The bottom panel of Fig. 3.5 also highlights the synergetic value provided by high-energy machines, like the 10 TeV muon collider or linear $e^+e^-$ colliders at TeV centre-of-mass energies. The stronger $Z$-pole programme at circular colliders offers better sensitivity than the linear machines or the muon collider for a larger number of the EFT combinations that enter in EWPOs. However, the high-energy options have strong sensitivity to the operators $\mathcal{O}^{(1),(3)}_{\phi l}$ and $\mathcal{O}_{\phi e}$ via diboson measurements which provides complementary information and, in some cases, a comparable reach to that of the high-statistics of the Tera-$Z$ run at circular machines. These independent probes of the previous operators do, in turn, break part of the above-mentioned correlations. Something similar is observed, to a lesser extent, for the FCC-hh in quark operators, which are generally less constrained by EWPOs than the leptonic counterparts. In general, the best sensitivity is provided by the highest precision for EWPOs in combination with the access to TeV energies, which provides additional (and complementary) handles to improve the bounds on certain operators. This is illustrated not only by the 1 TeV LCF or the FCC examples but also by the combination of the 10 TeV muon collider with LEP3.

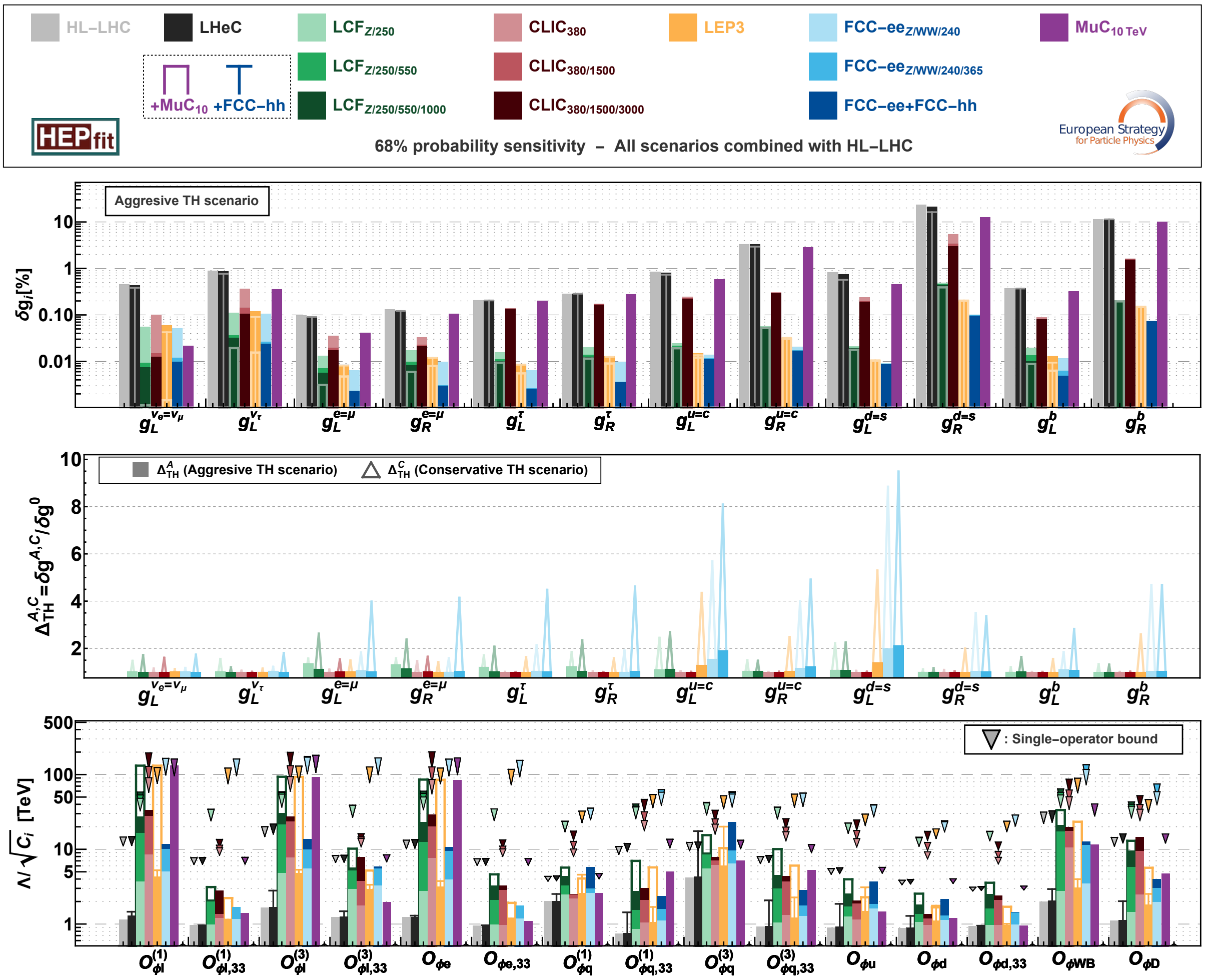

Fig. 3.5: (Top panel) 68% probability sensitivity to the combinations of operators modifying the EW couplings in the dimension-six $U(2)^5$-symmetric SMEFT framework. (Middle panel) Differences in the bounds obtained using the aggressive (bars) and conservative (triangles) theory scenarios. (Bottom panel) 68% probability reach to the interaction scale associated with operators modifying each EW fermion coupling. The last two operators induce fermion-universal effects in EWPOs. The inverted triangles in this panel indicate the single operator limits for each collider. In the top and bottom panels, the empty boxes and "T" bars indicate the results from a combination of some of the colliders with the 10 TeV muon collider and FCC-hh, respectively. See text for details.

### *Higgs boson interactions*

For comparisons of the Higgs boson interactions, effects modifying the on-shell effective Higgs couplings are considered and defined via the corresponding partial decay widths, i.e. $\delta g_{Hxx} \approx \frac{1}{2}\delta\Gamma(H \to xx)$, where $\delta$ refers to relative modifications with respect to the SM value: $\delta\Gamma(H \to xx) = \Delta\Gamma(H \to xx)/\Gamma_{\rm SM}(H \to xx)$. Figure 3.6 shows the 68% probability bounds on $\delta g_{Hxx}$ for each SM final state. The interactions with the top quark will be discussed separately. Note that, due to the $U(2)^5$ flavour symmetry, sizeable modifications of the Higgs couplings to the two light fermion families are not possible, and thus are not shown. Bounds on the aTGC, mentioned above, which receive constraints both from Higgs and $W^+W^-$ measurements, are also shown in the figure.

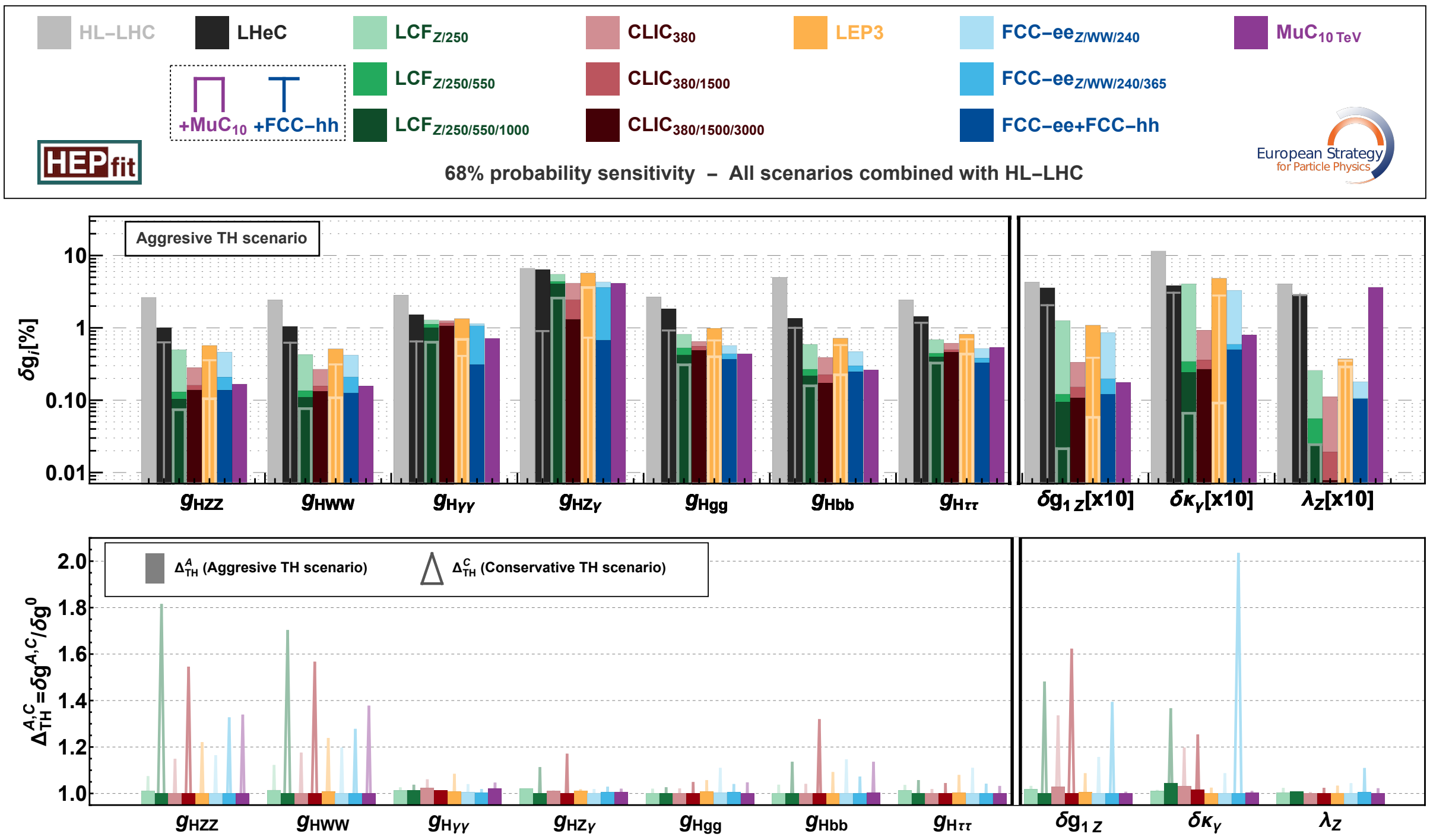

Fig. 3.6: (Top panel) 68% probability sensitivity to modifications of the effective Higgs couplings and aTGCs in the dimension-six $U(2)^5$-symmetric SMEFT framework. In this panel, the empty boxes and "T" bars indicate the results from a combination of some of the colliders with the 10 TeV muon collider and FCC-hh, respectively. (Bottom panel) Impact in the fit when using the aggressive (bars) and conservative (triangles) theory scenarios.

Similarly to the discussion of Sect. 3.1, all $e^+e^-$ and high-energy colliders yield significant improvements with respect to the HL-LHC. Unlike in the $\kappa$ framework, where the $\kappa_W$ and $\kappa_Z$ parameters are completely independent, the obtained precision on the $HZZ$ and $HWW$ couplings in the SMEFT approach is similar for a given project, as these couplings are connected to each other up to custodial symmetry breaking effects, which are themselves constrained by EWPOs. One or two permille precision can be obtained for all $e^+e^-$ colliders when including multiple energy stages. Running at only energies around 230 GeV as in the case of LEP3 yields comparatively worse results. Similar conclusions can be drawn for other couplings, e.g. to $b$ quarks or tau leptons, where high-energy stages bring strong improvement to the precision. The higher-energy stages of the LC are needed to match the high-statistical precision obtainable at the FCC-ee. For the aTCGs, a strong improvement is seen in particular when adding in higher-energy stages of the LC, due also in part to correlations with $HWW$ and $HZZ$. The effect of theory uncertainties is smaller here compared to EWPOs, as indicated in the lower panel of the figure. High-energy lepton colliders, where measurements of VBF processes are used, are more strongly affected by these uncertainties.

A high-statistics $Z$-pole run helps constrain operators that enter both in the EWPO and in the $ZH$ production, where the effect on $ZH$ production grows with increasing energy. For the FCC-ee, which projects to have the most precise $ZH$ measurement for all $e^+e^-$ options, a $Z$-pole run with several $10^9$ $Z$ bosons, much less than the anticipated statistics of the Tera-Z run, is needed to achieve sufficient precision to constrain those operators more tightly than $ZH$ production [51]. For the LC, polarisation is a valuable tool to probe contributions from s-channel photon-exchange diagrams, sensitive to SMEFT $HZ\gamma$ interactions. As the same operators generating these effects also contribute to the Higgs couplings to vector bosons, polarisation at energies of 250 GeV helps to improve the precision on $HZZ$ with respect to the case of non-polarised beams.

The HL-LHC expects to obtain a precision on the Higgs cubic interaction, $\lambda_3$, of 27%. For $e^+e^-$ projects running below a centre-of-mass energy of roughly 450 GeV, $\lambda_3$ can be determined via single-Higgs boson measurements, where it contributes via loop-level corrections; at energies above this threshold, it can be determined via $HH$ production threshold, where it contributes at tree level. The former method requires precision Higgs bosons measurements obtained at two different centre-of-mass energies, to break degeneracies. Furthermore, a wide range of measured observables, especially those involving top quarks, are needed to constrain other, non-$\lambda_3$, loop-level contributions. Being sensitive to different types of effects aside from $\lambda_3$ brings an interesting complementarity between $HH$ and single-Higgs determinations, where a non-SM signal observed in one process and not confirmed by the other would be evidence of additional BSM corrections.

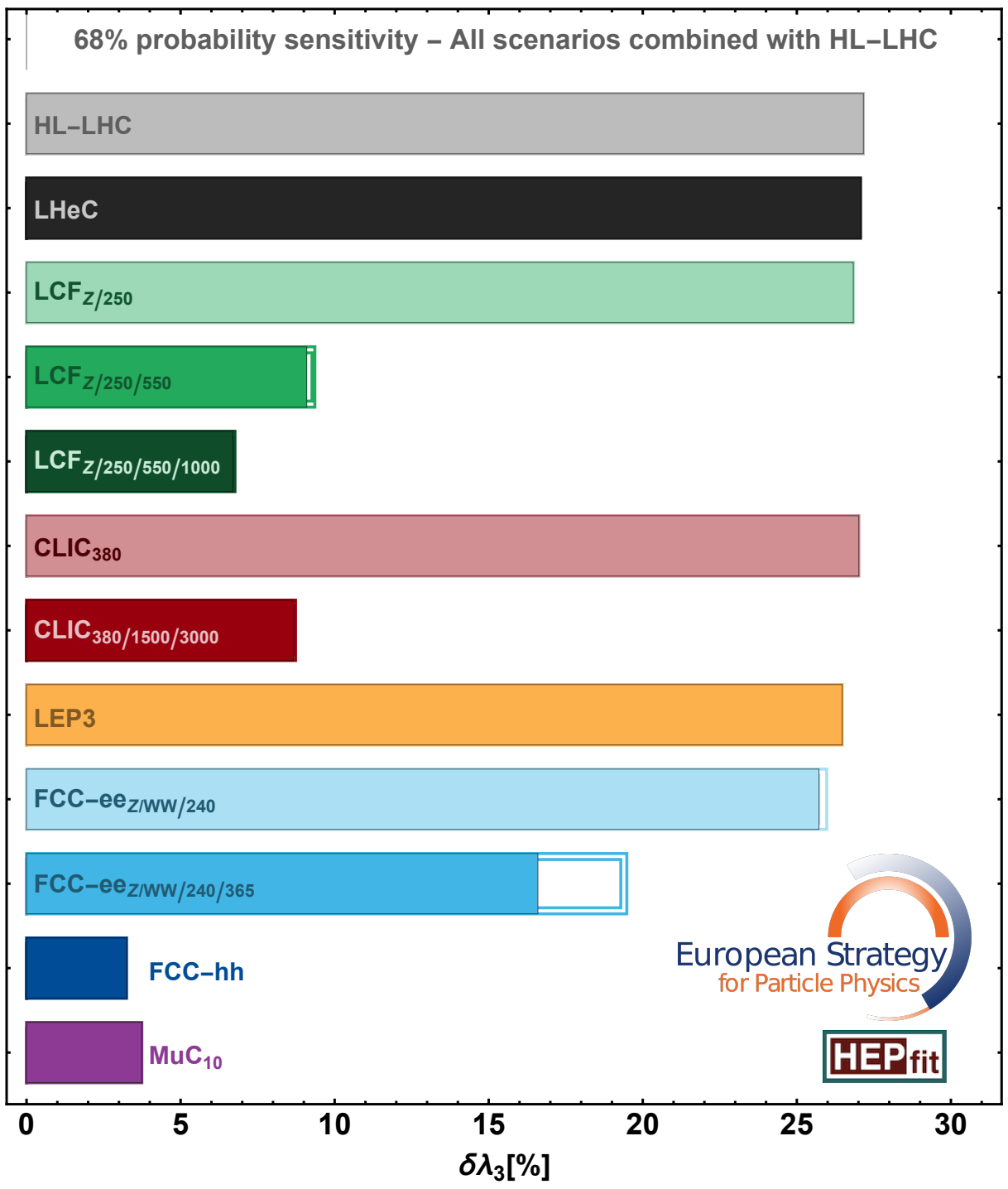

Fig. 3.7: 68% probability sensitivity to modifications of the Higgs trilinear coupling in the dimension-six $U(2)^5$-symmetric SMEFT framework. The uncertainty band in the figure reflects the result obtained from single-Higgs measurements under the aggressive/conservative scenarios for the theory uncertainties considered in this study.

Figure 3.7 shows the projected precision on $\lambda_3$ for the different collider options. As clearly seen in the figure, $e^+e^-$ colliders running at only an energy around 230 GeV, e.g. LEP3, LC at 250 GeV, FCC-ee 240 GeV, yield no significant improvement in $\lambda_3$ compared to the HL-LHC. The combination of the HL-LHC and the FCC-ee, which includes a second energy point below the $HH$ threshold, would yield a precision ~17-20%. When running at an energy above the $HH$ threshold, the precision on $\lambda_3$ improves. At linear colliders with energies of 550 GeV and 3 TeV a precision of 7-11% could be obtainable [52]. (The precision reported in the figure takes into account both the $HH$ determination and the one from single-Higgs measurements.) However, high-energy machines like a hadron collider or a muon collider are needed to reach the few percent-level precision. Both machines could obtain a precision on the self-coupling 3-4% for the SM value.

The value of $\lambda_3$, which is assumed in this fit to be the SM value, affects the cross-sections of the different $HH$ processes (see Sect. 4.4. of Ref. [ID141]). At the HL-LHC, due to the negative interference between the $\lambda_3$ and box diagrams, the absolute precision deteriorates for $1.5 < \lambda_3/\lambda_3^{\mathrm{SM}} < 4.5$ (although the relative precision still improves) [ID170]. At the LC above 550 GeV, for $\lambda_3 > \lambda_3^{\mathrm{SM}}$ the pattern of interference in $ZHH$ is positive and the cross section increases. When combining the information from $ZHH$ and $\nu\nu HH$ processes, the absolute uncertainty on $\lambda_3$ remains roughly constant over a wide range of values of $\lambda_3$. In the case of determinations using single-Higgs-boson measurements, at energies below 500 GeV, the self-coupling effects also increase the $ZH$ cross section for $\lambda_3 > \lambda_3^{\text{SM}}$. The absolute precision of $\lambda_3$ from loop effects changes slowly for moderate deviations with respect to the SM.

### *Constraints on top-quark operators*

The sensitivity to $e^+e^-t\bar{t}$ interactions can be compared across the different $e^+e^-$colliders to quantify the benefits of different centre-of-mass energies and polarisation. In a leading-order SMEFT analysis of top-quark observables, the contributions from $e^+e^-t\bar{t}$ operators cannot be disentangled from vertex corrections. However, these two types of corrections scale differently with energy and therefore can be separated by using top-quark measurements at two separate energy points. FCC-ee compensates for the absence of the high-energy lever arm via the higher precision measurements of $e^+e^- \to f\bar{f}$ processes, on- and off- the $Z$ resonance, where the top-quark operators contribute at the one-loop level. As can be seen in the top panel in Figure 3.8, the high-energy measurements obtainable at LCs would set significantly stronger bounds on $e^+e^-t\bar{t}$, in some cases by a factor of ten better than circular colliders. However, the muon collider has the strongest reach with significant improvements over other collider options. Note that the bounds on these operators can have a strong dependence on the choice of renormalisation scale. The use of a common scale for all colliders still provides a valid comparison, but some specific features may require a closer inspection of the correlations of the fit.[6] The presence of large correlations for the top-quark operators is apparent in Figure 3.8 by comparing the results of the global fit with the limits that would be obtained in a single-operator analysis.

Direct comparisons of top-quark measurements between lepton and hadron colliders is challenging due to their complementarity, in particular due to the contributions from four-fermion operators mentioned in the previous paragraph. While lepton colliders receive contributions from $\ell^+\ell^-t\bar{t}$ operators, hadron colliders measurements would be sensitive to $q\bar{q}t\bar{t}$ interactions. There are, however, some interactions in common: $C^-_{\phi Q}$, $C_{\phi t}$ and $C^{(3)}_{\phi Q}$, which correct the neutral and charged current interactions of the top quark; $C_{t\phi}$, which contributes to the top-quark Yukawa coupling; $C_{tZ}$ and $C_{tW}$, which describe dipole interactions with the electroweak gauge bosons (using the notation of the LHC Top WG [53]). Since any interaction with the top quark mixes with other dimension six interactions proportionally to the largest SM couplings, the effects of these operators can also be tested via loop-corrections to other precision measurements, such as the EWPOs. Some are also tested at the leading order in the Higgs-boson radiative processes. For these operators, shown in the bottom panel in Figure 3.8, the FCC still provides strong limits in several cases. However, a combination of measurements up to 1 TeV (3 TeV) at the LCF (CLIC), or at a 10 TeV muon collider, yields better sensitivity for most operators. An important exception is the operator $\mathcal{O}_{t\phi}$, which modifies the top-quark Yukawa coupling, where the best precision if obtained at the FCC-hh.

[6]Note that some operators enter in the observables included in the study only via loop corrections and remain weakly constrained in the global fit. This implies that bounds on operators mixing with such interactions can differ in a noticeable manner when evaluated at the scales where the measurements are performed compared to those at the chosen reference scale of 10 TeV.

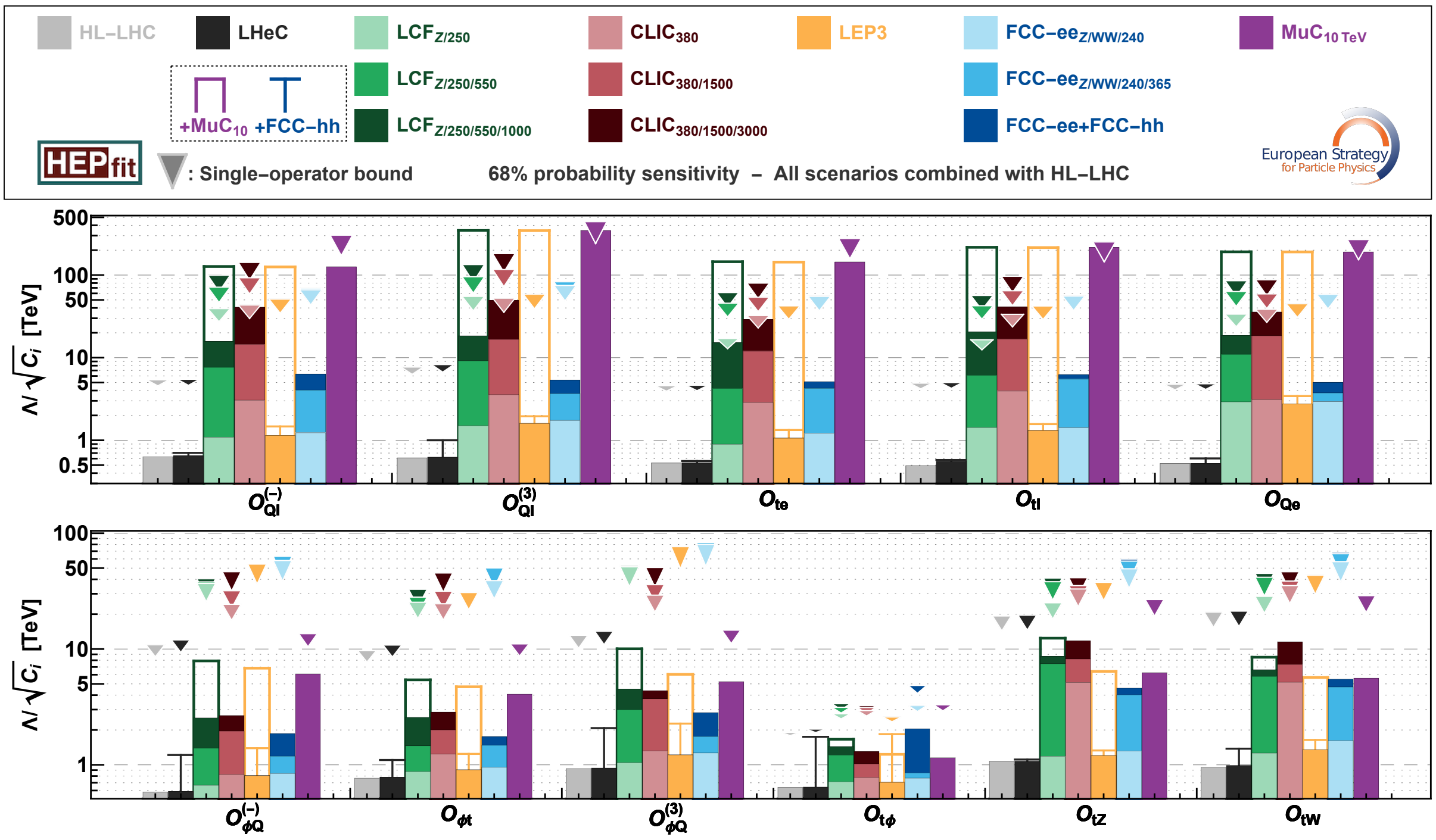

Fig. 3.8: 68% probability limits on $\ell^+\ell^- t\bar{t}$ four-fermion operators (top panel) and on operators modifying the EW and Higgs top-quark interactions (bottom panel). The empty boxes and "T" bars indicate the results from a combination of some of the colliders with the 10 TeV muon collider and FCC-hh, respectively. The inverted triangles indicate the single operator limits for each collider.

### *Precision from Energy*

For every dimension-six operator in SMEFT, there exists at least one process where its relative contribution to the SM prediction increases quadratically with energy. When such processes are experimentally accessible, higher collision energies lead to greater sensitivity to potential new physics, which is responsible for the corresponding contact interactions.

Here, high-energy two-to-two fermion scattering and diboson production processes are used to probe the sensitivity to four-fermion and two-fermion/two-boson operators. Although the specific interactions mediating these processes differ between lepton and hadron colliders, for the sake of this last comparison, it is assumed that the new physics generates universal contributions testable at all colliders, and the interpretation is performed in terms of the EFT Lagrangian in Eq. (A.4) in Appendix A. (This is also relevant for the BSM interpretation presented in Chapter 8.) In Fig. 3.9 the bounds on the Wilson coefficients $c_{2B}$, $c_{2W}$, $c_{2G}$, $c_W$, and $c_B$ are shown. $c_{2B}$, $c_{2W}$ and $c_{2G}$ are testable via two-to-two fermion scattering while $c_W$, and $c_B$ describe effects in diboson production that grow with the energy, similar to what was observed for some of the $\mathcal{O}_{\phi f}$ operators in electroweak precision measurements section. Combining with either the FCC-hh or the 10 TeV muon collider significantly improves the sensitivity. In many cases, the muon collider provides the greatest reach. On the other hand, the FCC-hh has the highest sensitivity to new physics coupled to the QCD current, as shown for the $c_{2G}$ limit.

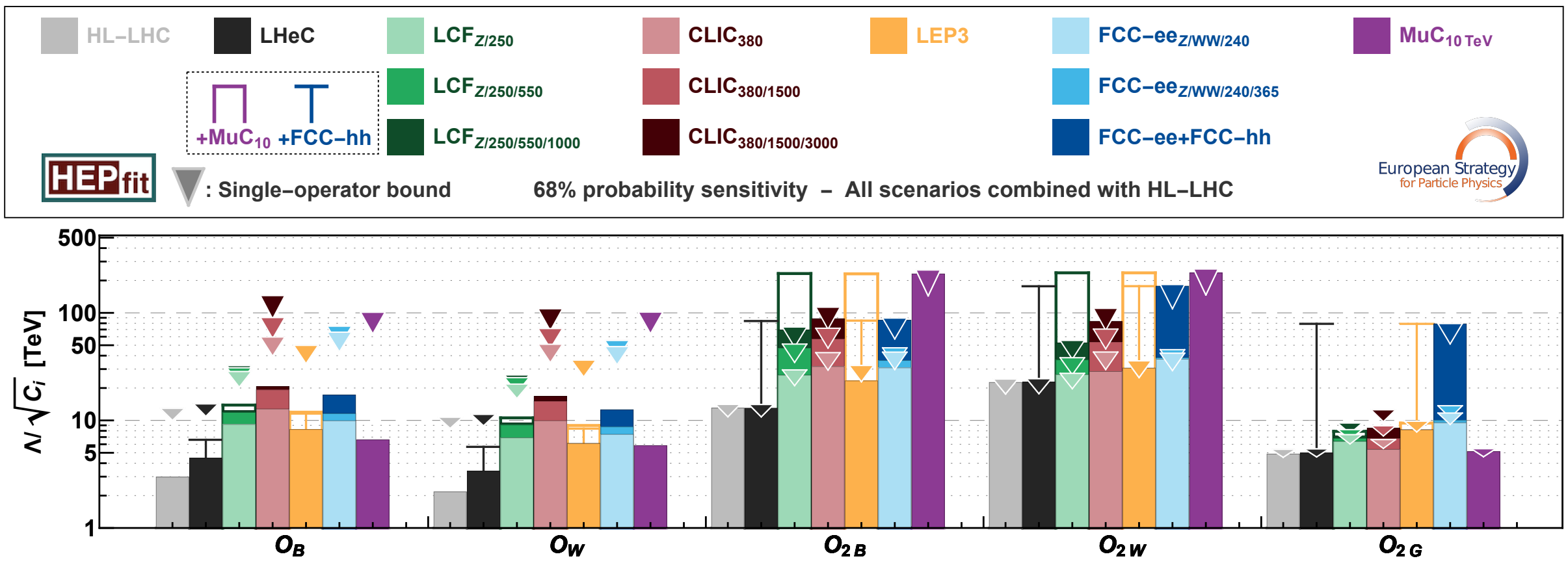

Fig. 3.9: 68% probability limits on universal two-fermion/two-boson and four-fermion contact interactions. The inverted triangles indicate the single operator limits for each collider. The empty boxes and "T" bars indicate the combination of some of the colliders with the 10 TeV muon collider and FCC-hh, respectively.

## 3.6 Conclusions

Future colliders are essential for fully probing the EW sector and understanding the fundamental mechanism of mass generation in the Standard Model and beyond. While the Higgs boson has been discovered, much remains to better understand this unique particle and its role in the EW sector. Precise and comprehensive measurements of its properties—alongside comprehensive measurements related to the top quark and EWPOs—require a diverse experimental programme that only next-generation colliders can provide.

No single collider can address all aspects of the EW sector. Electron-positron colliders offer a clean experimental environment and yield precise, model-independent results such as the total Higgs boson width. High-energy colliders are essential to explore rare processes and achieve ultimate precision in specific Higgs couplings that are inaccessible or statistically limited at lower energies.

In the Higgs sector, all proposed future options build upon the HL-LHC results and offer a significant extension of its sensitivity. These future machines provide precision measurements of the couplings including those of 2$^{\text{nd}}$ generation charged fermions. Among $e^+e^-$colliders operating near 250 GeV, the FCC-ee provides the highest precision due to its large luminosity. Higher energy collisions – such as $e^+e^-$colliders above 550 GeV, muon colliders, or hadron colliders – bring additional and significant improvements to many coupling parameters. If operated directly after the HL-LHC, the FCC-hh can substantially improve, under certain model assumptions, the precision on Higgs couplings to the photon, top quark, and muons, going beyond the reach of first-stage $e^+e^-$machines. However, for hadronic couplings and gauge-boson couplings, $e^+e^-$colliders provide superior precision, independent of model assumptions. The LHeC provides a precise determination of EW charged-current Higgs boson production. For the Higgs boson self-coupling, the best precision is reached at high energies with the FCC-hh or a 10 TeV muon collider. The FCC-ee and LC can also obtain improved results compared to those of the HL-LHC when running with at least two energy stages.

For top-quark physics, $e^+e^-$colliders are sensitive to operators modifying the electroweak-top-quark interactions via runs at the top-quark threshold or via loop corrections to precision observables. High-energy LCs with multiple high-energy points provide the strongest constraints on most top-related operators compared to other $e^+e^-$ machines. The FCC-hh provides complementary potential in the top-quark sector in particular for 4-quark operators as well as the best measurement of the top-quark Yukawa coupling. Comparatively, the muon collider provides the strongest limits on $\mu^+\mu^- t\bar{t}$ operators, which are complementary to the bounds on $e^+e^- t\bar{t}$ set by $e^+e^-$ colliders. The combination of a muon collider with a multiple energy stage LC yields overall the strongest results for $\ell^+\ell^- t\bar{t}$ interactions.

Measurements of EWPOs are in general measured more precisely at circular $e^+e^-$ machines compared to LCs, due to higher luminosities. The FCC-ee provides unprecedented precision for the EWPOs compared to any other option. However, the muon collider and high energy LCs also provide precise information on EW interactions but via diboson production at high energies.

Low-energy $e^+e^-$ colliders (e.g. LEP3, the FCC-ee up to 240 GeV only, or the LCF up to 250 GeV only) offer an inherently limited physics program. To unlock the full physics potential, especially for top-quark physics and Higgs boson self-coupling, multiple $e^+e^-$ energy stages combined with TeV-scale collision energies are required.

The best results for the Higgs, EW and top-quark sector are obtained with a precision machine followed by a high-energy option. The integrated FCC program (FCC-ee and FCC-hh) delivers the strongest overall performance in broad range of areas, thanks to the precision measurements for the FCC-ee and the energy reach for the FCC-hh. For many top-quark observables and for a selected number of Higgs-boson couplings, a high-energy muon collider in combination with an $e^+e^-$ machine offers better performance. LEP3, when paired with FCC-hh, also provides precise EWPO and Higgs-boson measurements. However, its overall performance remains lower than the FCC-ee +FCC-hh. The LHeC combined with an FCC-hh has strengths in precision measurements of the Higgs coupling to $W$ bosons and single-top-quark production. Moreover, its ability to deliver precise PDFs would directly support the FCC-hh program. A staged LC program or LEP3, followed by a 10 TeV muon collider could provide results comparable to those of the full FCC program in some areas.

Across all these collider scenarios, theory plays an indispensable role. In many cases, theoretical uncertainties could be the limiting factor in realizing the full potential of experimental measurements, unless radical advances in theory calculations and simulation tools are achieved. The theory uncertainty scenarios listed in this chapter and included in the SMEFT fits serve as guideposts for the effort that would be required, but they should not be interpreted as predictions of the improvements. Significant and sustained investment in theoretical work is necessary to match the expected experimental precision. This includes not only developing new methods and tools but also a significant investment in human and computational resources.

# Chapter 4

# Strong Interaction Physics

The strong interaction is at the heart of the visible physical world. It is responsible for binding quarks and gluons into hadrons, and ultimately for forming atomic nuclei, that constitute the vast majority of the visible matter. Its effects impact a very broad range of physical phenomena, from high-energy particle collisions to the interiors of neutron stars. The core of the strong interaction description is based on an elegant and so far unchallenged quantum chromodynamics theory (QCD), for which the unique input, besides the quark masses, is the strong coupling. While a general solution in all physics regimes is yet to be discovered, the strong interaction theory is established and fully integrated into the Standard Model (SM) of particle physics. However, the complex behaviour of the strong force across energy scales and environments remains a significant challenge for both its theoretical understanding and experimental investigation, and finally for a full understanding of the SM itself.

Among the four known fundamental interactions, the strong force remains the least-precisely characterised in several regimes. Its running coupling, confinement mechanism, and transition between perturbative and non-perturbative domains give rise to a variety of phenomena that demand distinct theoretical and experimental approaches for their study. In high-energy settings, precision predictions rely on perturbative techniques and factorisation frameworks. At low energies or high densities, non-perturbative methods such as lattice simulations are required. This interplay between regimes makes the strong force a uniquely rich subject of study. Beyond high-energy experiments, strong interaction physics connects to several domains of contemporary science. Its influence extends to hadronic structure, nuclear binding, the thermodynamic and collective properties of high-temperature strongly-interacting matter, and processes relevant to astrophysical observations and cosmic-ray measurements. These connections make it a cross-disciplinary field, with implications for both theoretical development and interpretation of experimental observations.

The contemporary research in the field of the strong interaction covers two complementary but interrelated aspects: i) the ultimate nature of the hadronic matter and interactions in all conditions, and ii) the precise estimate of the hadronic contributions to other research fields and processes, such as searches for new physics and precision electroweak measurements. The open questions and future prospects in the physics of the strong interaction have been mapped to four interrelated areas, each focusing on a specific set of benchmark phenomena and measurements.

– **Precision QCD.** This area addresses the accurate determination of the strong coupling con-

Table 4.1: Future projects (upgrades, new experiments, new facilities) with direct impact on the physics of the strong interaction.

| **Facility, Experiments** | **Colliding systems, $\sqrt{s}$** | **Timeline** | **Precision QCD** | **Partonic structure** | **Hot and dense QCD** | **QCD connections** |
|---|---|---|---|---|---|---|
| HL-LHC: ALICE 3, ATLAS, CMS p2 Upg, LHCb Upg II, LHC-spin | *pp* 14 TeV<br>*AA* 5.5 TeV<br>*pA* 8.8 TeV | 2030-2041 (ALICE 3, LHCb Upg II > 2035) | $\alpha_s(m_Z^2)$, $\alpha_s(Q^2)$, $m_t$, $m_W$ | (n)PDF, TMD, small *x*, intrinsic charm | Precision charm, beauty, hard and e.m. probes, $\mu_B \approx 0$; time evolution | Hadr. inter., (hyper/charm)nuclei, Exotics, Cosmic antinuclei, Neutron Star EoS |
| HL-LHC: FPF | LHC collisions, $\nu$-nucleon | > 2031 | | (n)PDF, small *x*, intrinsic charm | | Cosmic rays ($\nu$, modeling primary interaction) |
| SPS: NA60+, NA61 | *pA*, *AA*, 5-17 GeV | > 2030 (NA60+) | | nPDF, medium/large *x* | Charm, dileptons, critical point?, $\mu_B$=200-450MeV | Cosmic antinuclei, $\nu$ |
| FAIR SIS-100: CBM | *pA*, *AA*, 2.5-5 GeV | > 2028 | | GPD | Hadrons, dilept., critical point?, $\mu_B$ =500-700MeV | (Hyper)nuclei |
| SPS: AMBER pII | $\mu, \pi, K, p$(250 GeV)-*N* | > 2030 | | | | $K$, $\pi$ properties, spectroscopy, Cosmic antinuclei |
| MUonE | $\mu$(160 GeV)-*e* | > 2030 | $g-2$ (hadronic) | | | |
| HIE-ISOLDE Upg | Radioactive ion beams | > 2029 | | | | Nucl. phys. Inputs NS EoS |
| KEK: Belle II Upg. | *ee* 10 GeV | > 2035 | $\alpha_s(m_\tau^2)$ | | | Exotics (c,b) |
| STCF | *ee* 2-7 GeV | > 2033 | $\alpha_s(m_\tau^2)$ | | | Exotics (c) |
| EIC | *ep*, *eA* 28-140 GeV | > 2036 | $\alpha_s(m_Z^2)$, $\alpha_s(Q^2)$ | (n)PDF inclusive/diffractive, TMD, GPD, medium/large *x* | | Exotics (c,b) |
| LHeC | *ep* 1.2 TeV<br>*eA* 0.8 TeV | > 2043 | $\alpha_s(m_Z^2)$, $\alpha_s(Q^2)$, $m_W$ | (n)PDF inclusive/diffractive, TMD, GPD, small to large *x* | | Cosmic rays (small-*x* PDF), Exotics (c,b) |
| FCC | *ee* 90-365 GeV<br>*pp* 84 TeV<br>*AA* 33 TeV<br>*pA* 52.8 TeV | > 2047<br>> 2074 | $\alpha_s(m_Z^2)$, $\alpha_s(Q^2)$, $m_t$, $\Gamma_t$, $m_W$ | (n)PDF, TMD, small to large *x* | New probes of time evolution, early times | Cosmic rays ($\nu$, modeling primary interaction) |
| LCF CLIC | *ee* 0.25-1 TeV<br>*ee* 0.38-1.5 TeV | > 2050 | $\alpha_s(m_Z^2)$, $\alpha_s(Q^2)$, $m_t$, $\Gamma_t$, $m_W$ | | | |
| LEP3 | *ee* 91-230 GeV | > 2047 | $\alpha_s(m_Z^2)$ | | | |
| Muon Collider | $\mu\mu$ 3-10 TeV | > 2050 | $\alpha_s(m_Z^2)$, $\alpha_s(Q^2)$, $m_t$, $\Gamma_t$ | PDF using sec. neutrino beam | | Cosmic rays ($\nu$) |

stant $\alpha_s$ and its dependence on the momentum transfer $Q^2$. It also examines the impact of strong interaction effects on the extraction of fundamental SM parameters such as the masses of the top quark and the $W$ boson.

- **Partonic structure of protons and nuclei.** This area focuses on the determination of parton distribution functions (PDFs) in both longitudinal and transverse dimensions. Efforts are directed at improving our knowledge of the momentum and spatial distribution of partons inside protons and nuclei as functions of momentum fraction $x$ and $Q^2$.

– **Hot and dense QCD.** Investigations in this area target the understanding and the properties of strongly-interacting matter at high temperature, the quark-gluon plasma (QGP) formed in heavy-ion collisions. Benchmark measurements include heavy-flavour probes and thermal radiation of the QGP.
– **Connections of QCD with hadronic, nuclear, and astroparticle physics.** This area explores the broader implications of QCD for adjacent fields. Key topics include constraints on exotic hadron states derived from spectroscopy and hadron–hadron correlations, as well as the study of production and interaction of anti-nuclei, relevant for interpreting cosmic-ray data.

Studying the strong interaction is relevant for all future facilities (new or upgraded experiments, new accelerators), and many of them consider this research axis as their central scientific goal. About 40 Input Documents[1] to the ESPP2026 have connections to strong interaction physics. Table 4.1 maps future projects on the four physics areas and lists the main specific benchmarks considered in this document. Topics outside these areas—such as diffraction, heavy-flavour production, photon structure, etc.—which are also important, are expected to be studied in many of the proposed projects. However, they are not considered in this document, as explained in the introduction. The following sections cover the four physics areas in terms of main open questions and future experimental and theoretical prospects, as well as expected performance for the benchmark measurements.

## 4.1 Precision QCD

### 4.1.1 Future physics goals and experimental opportunities

The direct and indirect BSM physics reach of current and future colliders relies on searches for tiny deviations in the data from the SM predictions. A systematic study and precise understanding of the strong interaction is thus crucial to fully exploit a broad range of measurements and associated calculations. More precisely:

– A determination of the QCD coupling $\alpha_s(m_Z^2)$ with small uncertainties is a prerequisite for accurate and precise higher-order calculations of all particle production cross-sections and decay rates, which parametrically depend on it.
– Accurate perturbative QCD (pQCD) calculations, both at fixed (next-to)$^n$-leading (N$^n$LO) and resummed (next-to)$^n$-leading-log (N$^n$LL) orders, are fundamental to describe hadronic final states and jet dynamics, and precisely extract multiple SM quantities from the data.
– Heavy-quark, light-quark, and gluon separation (through jet-flavour tagging techniques) is key for multiple SM measurements and BSM searches, such as measuring Higgs Yukawa couplings to light-quarks and the loop-induced Higgs decay to gluons [ID141].
– Non-perturbative (NP) dynamics, such as hadronisation and colour reconnection, impacts all hadronic final states, playing a role, e.g., in $e^+e^- \to H, W^+W^-, t\bar{t} \to$jets processes.

High-energy $e^+e^-$colliders have been proposed as post-LHC facilities aiming at precisely measuring the properties of the four heaviest SM particles: the Higgs boson, $Z$ and $W$ bosons, and the top quark. Unlike hadron colliders, where PDFs, multiparton interactions, and "beam remnants" complicate the understanding of colour dynamics, $e^+e^-$ annihilation occurs at a fixed centre-of-mass (CM) energy with precisely known elementary kinematics and with a colourless initial state. In addition, events typically produce final states with well-separated jets and well-defined parton flavours, enabling detailed studies of light-quark, heavy-quark, and gluon showering and hadronisation in a controlled setup. All such aspects allow precise experimental measurements, accurate theoretical predictions, and direct tests of pQCD and NP dynamics, of relevance for the full physics programme of future colliders. Future $ep$ colliders, such as the LHeC, can probe the proton structure with unparalleled accuracy and provide complementary observables, but for common QCD observables they feature somewhat worse precision than proposed $e^+e^-$ machines. Hadron colliders at very high energies, such as the FCC-hh, will probe QCD in the uncharted multi-TeV regime, but many precision measurements will require the improved QCD knowledge provided by $ep$ or $e^+e^-$ facilities, as well as higher-accuracy theory calculations.

[1]The ESPP contributions related to strong interaction physics originate from: experimental collaborations (ID6, ID19, ID23, ID68, ID81, ID82, ID131, ID170, ID171, ID205, ID213, ID231, ID245), future colliders proposals (ID114, ID141, ID152, ID188, ID207, ID209, ID214, ID227, ID233, ID241, ID247, ID275), QCD-specific community contributions (ID2, ID7, ID29, ID55, ID89, ID103, ID201, ID224, ID235) and QCD theory collaborations (ID33, ID35, ID113, ID174).

### 4.1.2 Physics benchmark measurements

The QCD coupling $\alpha_s$, along with strong interaction effects on measurements of the top quark and $W$ boson masses in hadronic final states, serve as complementary key parameters to benchmark the ultimate achievable QCD precision at future colliders, as discussed next. Hereafter 'LC' indicates both LCF [ID40, ID140] and CLIC [ID78], under the assumption that both machines would include a $10^9$ $Z$ run and beam energy scans around $e^+e^- \to WW$ and $t\bar{t}$ thresholds.

**Strong coupling and its running.** The strength of the interaction between quarks and gluons at energy scale $Q$ is encoded in the QCD coupling, $\alpha_s(Q^2)$. Its reference value is given at the scale $Q = m_Z$, and its strength at any other energy can be predicted by the QCD renormalization group equations. The current world average has a $\pm0.8\%$ uncertainty, $\alpha_s(m_Z^2) = 0.1180 \pm 0.0009$ [15]. Knowing $\alpha_s(m_Z^2)$ precisely is crucial in particle physics, e.g., to reduce parametric uncertainties in calculations of real and virtual QCD contributions to all processes (cross-sections as well as particle decays) at the LHC and at future colliders. In addition, measuring the running of $\alpha_s(Q^2)$ at very high energies is key to explore the stability of the EW vacuum stability, as well as in searches for BSM physics (e.g., new coloured sectors, or tests of grand unification theories). The world average is derived through a combination of several determinations, accurate at least to $N^2LO$ in QCD, using observables in a broad range of scales from the tau mass ($m_\tau$) to a few TeV from jet measurements at the LHC. Values extracted from lattice calculations [54] are among the most precise included in the average, and their precision is expected to improve by a factor of two within the next decade [55]. Some of these $\alpha_s(m_Z^2)$ determinations exhibit tensions within the quoted uncertainties, underpinning the importance for multiple improved measurements to achieve a consolidated view [56].

The current uncertainty of the PDG-2024 $\alpha_s(m_Z^2)$ world-average is compared to the projected precision for different extractions at future facilities in Fig. 4.1. Blurred entries indicate that uncertainties (well) below 1% can be expected but have not been explicitly quantified to date. For instance, the blurred second to rightmost entry in Fig. 4.1 refers to hadronic $\tau$ measurements at low-energy $e^+e^-$ machines, such as the Belle-II Upgrade [ID205] and STCF [ID231]. Measurements of various DIS observables at the EIC [ID114] will match the precision of the current world average. At LHeC, PDF fits to inclusive cross sections in combination with jet cross-section measurements [ID214] will yield a determination in the 0.1% precision range, provided the theoretical uncertainties can be controlled at a similar level. The electroweak fit based on the data collected at LC would reach a $\pm0.6\%$ uncertainty [57]. At FCC-ee, mul-tiple observables can be used for precise $\alpha_s(m_Z^2)$ extractions [ID209], including low-energy collisions at $\sqrt{s} \approx 20$–$80\,\text{GeV}$ [58]. Also, per mille precision on $\alpha_s(m_Z^2)$ can be reached by exploiting hadronic $Z$- and $W$-boson decays, thanks to the very large integrated luminosities at the $Z$ pole and $W^+W^-$ threshold and the very small systematic and parametric uncertainties expected [59]. Beyond the $Z$ pole, event shapes and jet-rates at high-energy $e^+e^-$ machines can provide a complementary and theoretically-accurate test of the $\alpha_s$ running. The FCC-hh stands out for its ability to measure the running of $\alpha_s$ at scales of up to $\sim$40 TeV from jet events [ID209, ID227]. This represents an order-of-magnitude larger energy scale than probed today.

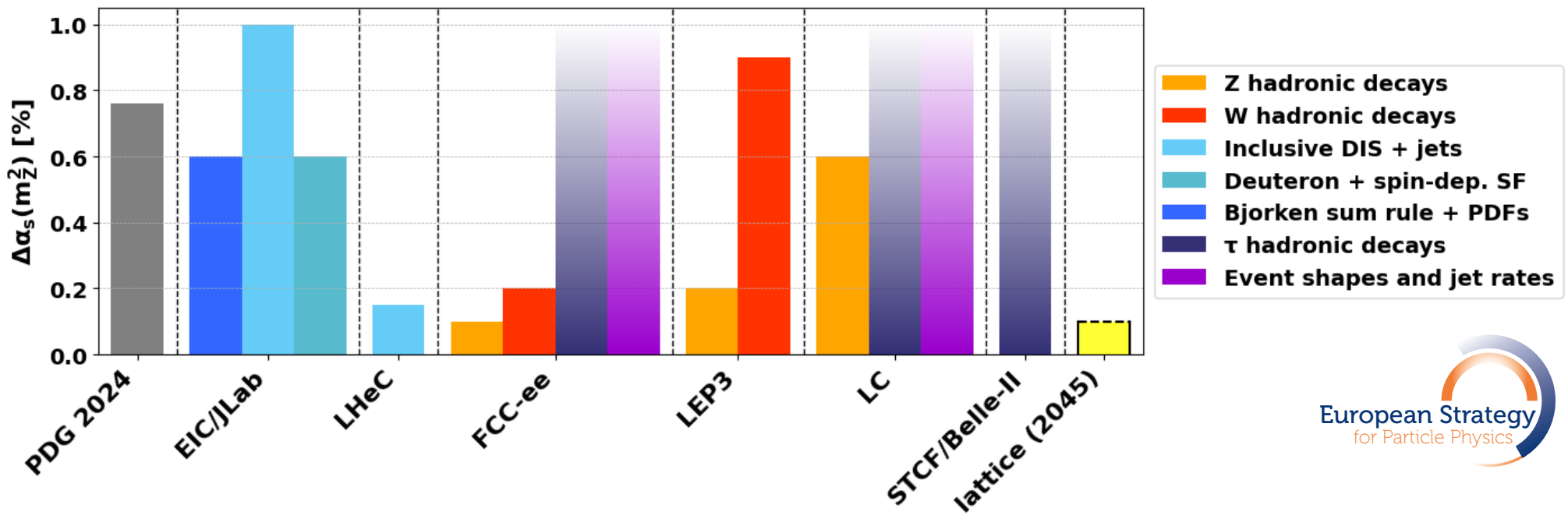

Fig. 4.1: Schematic representation of the status and projections relative to the strong coupling determination benchmark. Current precision of the world-average $\alpha_s(m_Z^2)$ value (leftmost entry) compared to the expected uncertainties at different future facilities from various observables, and expected statistical uncertainty from lattice-QCD in 20 years from now (rightmost entry).

**QCD effects on the $W$ boson mass.** At hadron colliders, the $W$ boson mass is extracted using the leptonic decays $W \to \ell\nu$ via fits of the transverse momentum $p_T^\ell$ (and transverse mass $m_T^{\ell\nu}$, at low pileup) distributions. The present uncertainty on $m_W$ at the LHC is about $\pm 10\,\text{MeV}$ [13], of which about half is due to QCD-related effects: knowledge of PDFs and of the low-$p_T$ boson distribution (dominated by intrinsic parton $k_T$ and soft gluon dynamics). Such uncertainty is expected to decrease to $\pm 5\,\text{MeV}$ at the end of the HL-LHC phase [ID170], with QCD-related uncertainties amounting to $\pm 3\,\text{MeV}$ (accounting for PDF uncertainties alone, and assuming that the low-$p_T$ $W$ boson distribution is described with much higher precision). At the LHeC, with improved determinations of PDFs and $\alpha_s(m_Z^2)$, an overall $\Delta m_W \approx 3\,\text{MeV}$ precision could be achieved [ID214]. The hadron collider $m_W$ uncertainties, including the QCD-related ones, are shown in the lower part of Fig. 4.2 (left). At future $e^+e^-$ facilities, the $W$ boson mass and width ($\Gamma_W$) can be measured through different methods. First, a threshold scan over $\sqrt{s} = 157$–$163\,\text{GeV}$ produces $WW$ pairs (nearly) at rest, and a fit of the $\sigma_{WW}(\sqrt{s})$ theory predictions to the measured lineshape provides a precise extraction of $m_W$ and $\Gamma_W$ [60]. In addition to accurate theory predictions, this method requires an excellent control of the $\sqrt{s}$ value (in the hundreds of keV range, reachable via resonant depolarization at FCC-ee). The leptonic final state, $e^+e^- \to W^+W^- \to \ell^+\nu\ell^-\bar{\nu}$, provides the most precise measurement thanks to lower backgrounds and absence of hadronic uncertainties. The forecast experimental uncertainty amounts to $\pm 0.3\,\text{MeV}$ at FCC-ee, while the present theory prediction precision for $\sigma_{WW}(\sqrt{s})$ is 3–5 MeV [61, 62]. Improving the theory prediction by more than an order of magnitude is a major challenge [ID209] that requires calculations with N$^2$LO EW accuracy, mixed EW-QCD corrections [63] and significantly improved MC modeling of QED initial-state radiation.

Second, the kinematic reconstruction of the invariant mass distribution of $W \to jj$ dijet decays in $e^+e^- \to WW$ collisions over $\sqrt{s} = 157$–$365$ GeV can also be used to determine $m_W$. The $WW$ events with semileptonic ($q\bar{q}\ell\nu$) and fully hadronic ($q\bar{q}q\bar{q}$) final states can be exploited by applying the constraint on the total four-momentum in each event, with energy equal to $\sqrt{s}$ and zero momentum, as done at LEP2. At the $WW$ threshold, the $W$ bosons are produced at rest back-to-back, and the constrained $W \to jj$ invariant mass fits are relatively free from QCD uncertainties in the semileptonic final states, reaching $\Delta m_W \approx 0.25$ MeV at FCC-ee [64]. In the all-hadronic $WW$ final states, colour reconnection (CR) affects hadrons with large separation with respect to the jet (parton) direction by pulling them towards or away from the jet. Their impact on invariant mass fits translates into $\Delta m_W \approx 1$ MeV at FCC-ee [64]. CR effects can be studied via the particle flow in the region between jets, and constrained in-situ from changes in the jet direction (or in the reconstructed $W$ mass) by varying the parameters of the jet algorithms, and by comparing $\ell+$jets to fully-hadronic events.

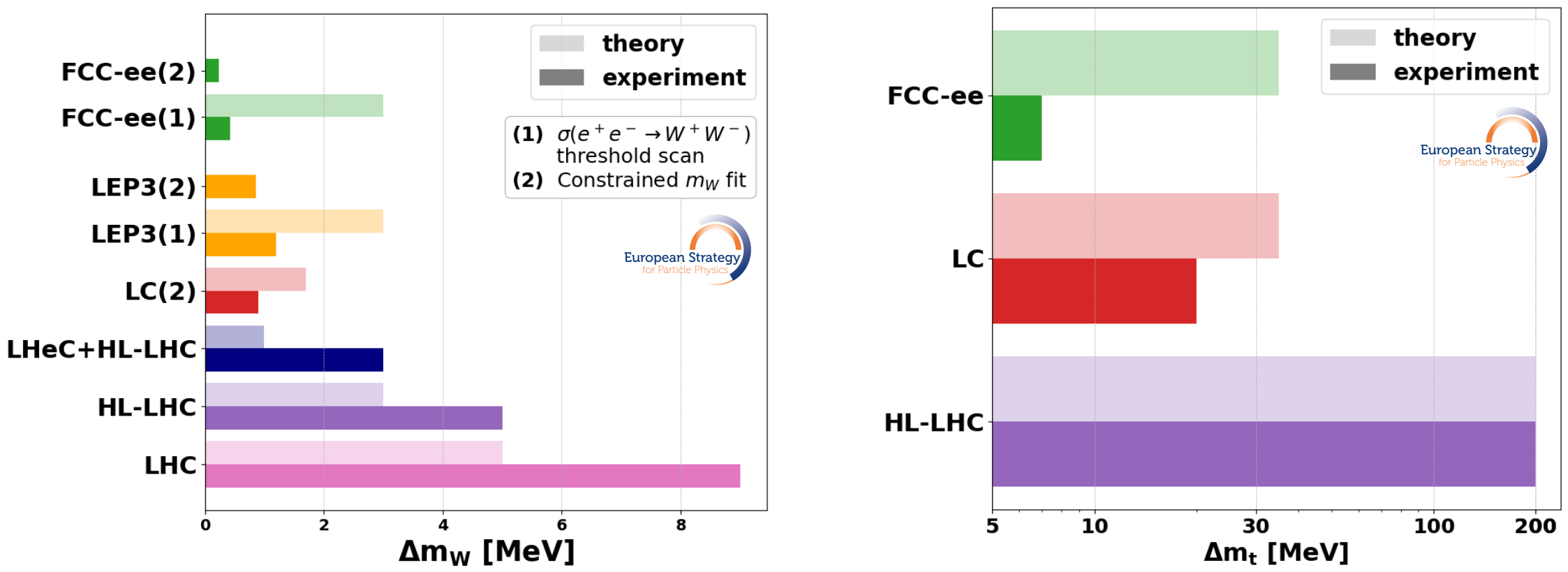

Fig. 4.2: Experimental (solid entries) and theoretical (lighter entries) precision of benchmark $m_W$ (left) and $m_t$ (right) determinations at the LHC and expected at the HL-LHC and at various future colliders. The theory precision is dominated by QCD uncertainties, except for the threshold scan measurements at FCC-ee and LEP3, for which electroweak uncertainties are more important.

The $W \to jj$ decays produced in $e^+e^- \to WW$ events at higher $\sqrt{s}$ values, such as those expected at the LC facility, are boosted, which further adds some dependence on the modelling of hadronization (in particular on the baryon and strange composition as a function of angle). Hadronization can also affect the measured jet energy, resolution and jet mass via the charged/neutral ratio fraction of very low and high momentum tracks and the shower shape. The impact of such effects on $m_W$ measurements at LC is estimated at $\Delta m_W \approx 0.9$ MeV.

The upper entries of Fig. 4.2 (left) show the expected experimental and theoretical precision on $m_W$ at different $e^+e^-$colliders from the two methods above, stressing the importance of a good control of QCD effects.

**QCD effects on the top quark mass.** The top-quark mass ($m_t$) is a key SM parameter that will be measured at the HL-LHC with a precision of $\Delta m_t \approx \Lambda_{\rm QCD} \approx 200$ MeV [ID170]. Achieving a significantly lower uncertainty requires dedicated runs at future $e^+e^-$colliders in a threshold scan over $\sqrt{s} \approx 340$–$365$ GeV, where the mass $m_t$ and width ($\Gamma_t$) can be extracted through a fit of the theoretical predictions to the measured $\sigma_{t\bar{t}}(\sqrt{s})$ lineshape[2]. Suitable renormalization schemes for $m_t$, such as short-distance mass schemes unaffected by infrared ambiguities [65–

[2]Top-threshold energy scans can also be carried out at a muon collider [ID207], if sufficient data can be collected in dedicated low-energy runs.

68], must be used for the $\sigma_{t\bar{t}}$ vs. $\sqrt{s}$ theoretical prediction. Simulation studies of the $t\bar{t}$ threshold data samples at the FCC-ee [39] —assuming that the value of $\alpha_s(m_Z^2)$ will be known within $10^{-4}$ from the $Z$ pole run (Fig. 4.1) [59], while the value of top Yukawa coupling ($y_t$) will be constrained within 3% according the expected HL-LHC precision [69]— lead to a final foreseen experimental precision on $m_t$ and $\Gamma_t$ of about 7 and 13 MeV [ID217], respectively, dominated by statistical uncertainties (Fig. 4.2, right, shows the available previsions[3] for $m_t$). Such a tiny experimental uncertainty is a challenge for the theoretical predictions. Although fixed-order QCD calculations up to N$^3$LO are available [70], the $\sigma_{t\bar{t}}$ cross-section near threshold is dominated by Coulomb interactions, since the $t\bar{t}$ pair is produced (nearly) at rest. This kinematic regime, where the top-quark velocity is of order $\alpha_s$, is described by non-relativistic QCD [71–73], which offers a framework to include QCD effects up to N$^3$LO and the resummation of Coulomb corrections [74–76]. EW effects, currently known to N$^2$LO accuracy [77], can be described within an analogous EFT framework, and the description of the $W^+W^-b\bar{b}+X$ final state also requires including nonresonant channels (without onshell top quarks) [78, 79]. For the potential-subtracted $m_t$ [67], the combined theoretical uncertainties from missing higher-order corrections lead to an uncertainty of about 35 MeV on the extracted $m_t$ value [39, 80], which are a factor of five larger than the experimental ones (Fig. 4.2, right). For observables other than the $\sqrt{s}$ lineshape, more theoretical work is still needed, in particular regarding the calibration of the top-quark mass parameter used in MC event generators, dubbed the MC mass, which corresponds most closely to a short-distance mass (such as the MSR mass [81] at a low scale), not directly to the pole mass, with an associated theoretical uncertainty of about $\pm(0.5-1.0)$ GeV [43].

### 4.1.3 Progress in theoretical predictions

Table 4.2: Wish-list for calculations of missing higher-order perturbative QCD $\mathcal{O}(\alpha_s^n)$ and/or EW $\mathcal{O}(\alpha^n)$ to match the expected experimental uncertainty at future $e^+e^-$ and $e\,p$ colliders.

| Observable | Missing higher-order & power-suppressed corrections |
|---|---|
| Hadronic $Z$ width | $\mathcal{O}(\alpha_s^5)$, $\mathcal{O}(\alpha_s^6)$, $\mathcal{O}(\alpha^3)$, $\mathcal{O}(\alpha_s\alpha^3)$, $\mathcal{O}(\alpha_s^2\alpha^2)$ |
| Hadronic $W$ width | $\mathcal{O}(\alpha_s^5)$, $\mathcal{O}(\alpha^2)$, $\mathcal{O}(\alpha_s^2\alpha)$ |
| Hadronic $\tau$ width | $\mathcal{O}(\alpha_s^5)$ |
| Hadronic event shapes ($Z$, $W$, $H$ decays) | N$^3$LO differential, N$^{3,4}$LL resummation, power corrections |
| Inclusive jet rates | 3-jet cross-sections at N$^3$LO, 4-jets at N$^2$LO, 5-jets at NLO |
| Lattice QCD results | $\mathcal{O}(\alpha_s^6)$ $\beta$-function; $\mathcal{O}(\alpha_s^5)$ heavy quark decoupling; $\mathcal{O}(\alpha_s^4)$ static potential |
| ($\alpha_s$ extr.; quark masses $m_c, m_b$) | $\mathcal{O}(\alpha_s^3)$ lattice perturbation theory matching (lattice coupling to $\alpha_s^{\overline{MS}}$ etc.) |
| $\sigma(e^+e^-\to W^+W^-)$ vs. $\sqrt{s}$ | EW N$^2$LO: $\mathcal{O}(\alpha^2)$, Mixed EW-QCD: $\mathcal{O}(\alpha_s\alpha^2)$, $\mathcal{O}(\alpha_s^2\alpha)$ |
| $\sigma(e^+e^-\to t\bar{t})$ vs. $\sqrt{s}$ | NRQCD: $\mathcal{O}(\alpha_s^5)$, Non-resonant: $\mathcal{O}(\alpha_s^5)$, $\mathcal{O}(\alpha_s^3)$ differential; QED: $\mathcal{O}(\alpha^3)$ at NNLL |
| $H\to b\bar{b}$ width | N$^4$LO ($m_b\neq 0$); N$^4$LO differential ($m_b=0$) |
| $H\to gg$ width | N$^5$LO (heavy-top limit), N$^4$LO ($m_t\neq 0$); N$^4$LO differential, N$^3$LO differential ($m_t\neq 0$) |
| MC simulations for $e^+e^-\to X$ processes | N$^{2,3}$LO matched to N$^{2,3}$LL PS. Permille control of non-pQCD effects (hadronization, CR, ...) |
| $ep\to$ hadrons (PDF and $\alpha_s$ determ.) | N$^{3,4}$LO evolution equations and inclusive cross-sections |
| $ep\to$jets ($\alpha_s$ determ.) | N$^3$LO cross-sections |

Achieving the theoretical precision that matches experimental uncertainties for future collider measurements is a formidable challenge for precision QCD calculations, far exceeding the current state of the art. For the benchmark observables, the required improvements at higher order in QCD and EW perturbation theory are summarized in Table 4.2. Progress towards those goals relies heavily on computer algebra, a crucial tool for efficiently handling and simplifying the complex mathematical expressions that arise in these higher loop calculations (Sect. 4.5). Furthermore, developments of MC generators combining fixed-order $N^2LO$ QCD perturbation theory with parton showers (PS) at $N^2LL$ accuracy, and beyond, will be essential to support the analysis and interpretation of experimental data for differential observables. Finally, the bottom rows of Table 4.2 indicate the required theoretical developments for future *ep*-collider options (EIC, LHeC, etc.), where calculations of QCD evolution equations and predictions for DIS structure functions need also to be pushed to the next quantum level. In summary, coordinated efforts to enhance the theoretical precision in pQCD are crucial to fully realize the potential of the benchmark measurements and experimental advancements at all proposed future facilities.

[3]The LC experimental precision on the top mass is expected to be around 20 MeV [ID140] with a similar theoretical uncertainty of 35 MeV.

## 4.2 Internal structure of protons and nuclei

The structure of nucleons and nuclei is of fundamental importance, serving both as an essential input to predictions and measurements in EW and BSM physics[4], and as a frontier in the strong-coupling, complex regime between particle and nuclear physics. Here we describe the state-of-the-art [86], future goals and experimental facilities. We discuss collinear parton distribution functions of protons $p$ (PDFs) [87], and nuclei $A$ (nPDFs) [88], transverse-momentum dependent (TMD) PDFs [89, 90] and generalised parton distributions (GPDs) [91] and, finally, the high-energy or small-$x$ behaviour of QCD [92]. We focus on information coming from unpolarised collisions; other aspects, e.g. spin or PDFs of unstable hadrons, are not discussed.

### 4.2.1 State-of-the-art and future physics goals

**Collinear proton/nucleon PDFs** are expected to be universal functions, with their $Q^2$ scale-dependence obtained through perturbative evolution of low-scale non-perturbative forms in Björken-$x$ and fitted to data through convolution with precision matrix-elements. The most recent global fits [93–96] include $\mathcal{O}(5000)$ experimental data points, and use a complex fit- and error-analysis methodology, including estimation of theoretical uncertainties from missing higher orders in perturbation theory (MHOUs). The highest QCD perturbative precision currently employed in the matrix elements is NNLO, but approximate $N^3LO$ is available [97–99]. The low-scale functional forms are currently minimally constrained through sum rules and parameterisation, but lattice QCD [100, 101] will potentially revolutionise PDF extraction by providing stronger *a priori* constraints. Much pre-LHC data from fixed-target and HERA *ep* experiments remain crucial to PDF fits. The LHC has provided further constraints, particularly at high-$Q^2$ and low-$x$ and in the gluon PDF through, e.g., $t\bar{t}$ production. While PDF calculations and fitting machinery are working well, methodological challenges from tensions and inconsistencies between measurements exist. For high-scale SM physics and BSM searches, the crucial requirements come from constraining the diverging relative uncertainties at high-$x$ where new physics will enter, without accidentally absorbing that physics into the fits. Methods to co-fit PDFs and new-physics effective field-theory parameters [102–104] are required, together with the inclusion of EW corrections, to match the increasing precision in QCD.

The forward sensitivity of the LHCb experiment has provided further input to global PDF

[4] Precise knowledge of PDFs and TMDs is required for an accurate extraction of parameters at high-energy colliders, e.g., $\alpha_s$, $M_W$ and $\sin^2\theta_{\text{eff}}$ at the LHC. Further uncertainties come from the existence of new dynamics at small $x$ (e.g., [82]), or from the transverse structure of hadrons and nuclei in event generators [83] and beyond: they limit our understanding of the small-system problem [84] and the extraction of QGP parameters (e.g., [85]).

fits, and proposed forward-physics facilities such as the CERN FPF [ID19] (and beyond at the LHeC [ID214] or the FCC [ID209]) can offer even greater reach into the small-$x$ region and the high-energy QCD behaviour. This is the region where a qualitative shift in internal-structure dynamics is expected, as standard (DGLAP) evolution should fail and parton saturation is predicted [92, 105]. From the theory point of view, NLO accuracy in both evolution and matrix-elements, as well as energy-suppressed corrections, are available (see Refs. [106–108] and references therein). From the experimental point of view, the essential requirement for an unambiguous observation of saturation is access to unprecedented small $x$. A large $Q^2$ coverage for small $x$ in both protons and nuclei is crucial ($pp$ and $pA$, $ep$ and $eA$), as saturation is density-driven and density increases with decreasing $x$ or increasing mass number $A$.

For **nPDFs**, efforts parallel their proton counterparts although in a less advanced stage. Several recent global sets exist at NLO [109, 110] and NNLO [111–113] accuracy, including $\mathcal{O}(2000)$ experimental data points, with similar methodology and error analysis like for protons. Data from fixed-target deep-inelastic scattering (DIS) and Drell-Yan (DY) experiments, and from $p$Pb collisions at the LHC, are complementary, with information from the LHC covering smaller $x$ [88] than for protons due to the use of $D$- and $B$-meson data. Parameterisations at low initial scale include $A$-dependence as present data are too scarce for fits to a single nucleus.

**TMDs and GPDs** [89–91] offer the next step in understanding nucleon structure, going beyond collinear factorization by considering higher-dimension structures on the road to general Wigner-function characterisation. Thus, they offer a three-dimensional view of protons and nuclei. For each parton species, there are $8+8$ different TMDs and GPDs. Thus, the amount of information to be extracted is vast, with the present knowledge mainly covering quarks and with limited flavour decomposition and range in $x$. The extension from proton to nuclei proceeds through simple modelling or, for TMDs, using collinear nPDFs. TMDs, which appear in two-scale processes, have been computed to large perturbative accuracy (up to N$^4$LL). Several global fits exist, based on semi-inclusive DIS data (SIDIS, requiring fragmentation functions, FFs) from fixed-target experiments and on DY data from the LHC, $\mathcal{O}(2000)$ data points [114, 115]. They require the knowledge of their collinear counterparts and also non-perturbative modelling, part of which can be extracted from lattice (see, e.g., [116]). GPDs provide, when Fourier-transformed, the transverse distribution of the parton content of the hadron. Their available perturbative accuracy is more limited, with some processes known at NNLO but full NLO evolution not yet available. Studied in exclusive processes, $2 \to 2$ (exclusive vector meson production, (D)DVCS, TCS) and $2 \to 3$ processes ($\gamma$–$\gamma$, $\gamma$–meson sensitive to transition GPDs and Distribution Amplitudes), their extraction method is less firmly established than for TMDs.

The present situation for collinear (n)PDFs is summarised in Fig. 4.3. Discrepancies among sets indicate either methodological differences or the need of further experimental data. Future goals include: the increase in perturbative accuracy (completion of N$^3$LO for collinear PDFs, and NLO for small-$x$ calculations and GPD evolution); the estimation of MHOUs and missing theoretical ingredients (e.g., unified evolution equations and factorisations across different $x$-domains); understanding the differences induced by methodologies (and establishing a firm one for GPD extraction); solving incompatibilities in reported data, which represent a critical issue for global fits; considering physics beyond that currently included, be it BSM or new QCD regimes; new methods to extract the numerous TMDs and GPDs through simultaneous fits of PDFs, TMDs and FFs to observables from different experiments; and procedures to employ lattice input [100, 101].

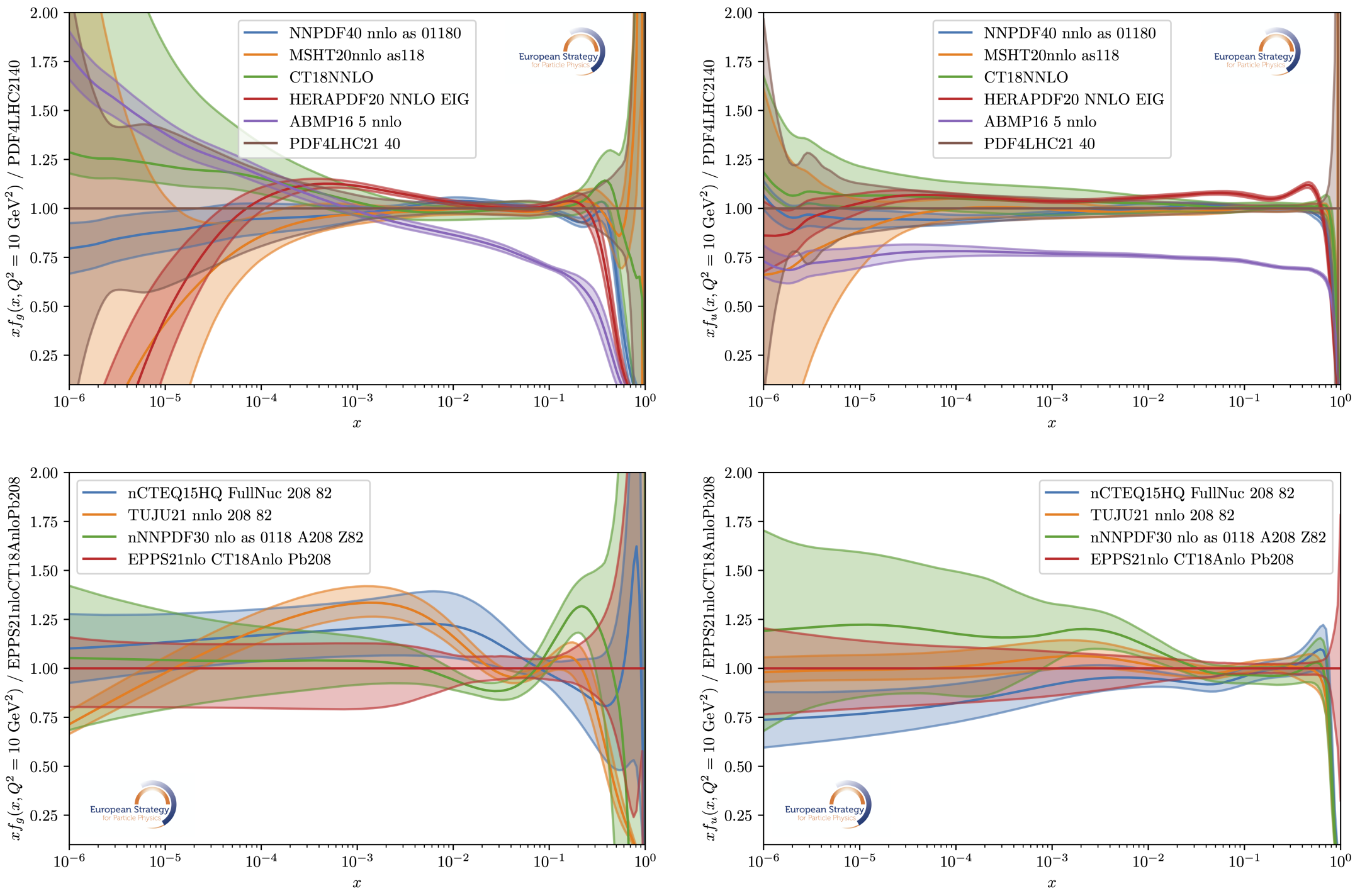

Fig. 4.3: *Top*: Comparisons of the uncertainties of gluon (left) and $u$ (right) PDFs at $Q^2 = 10\,\mathrm{GeV}^2$ from NNPDF4.0 [95], MSHT20 [94], CT18 [93], HERAPDF2.0 [117], ABMP16 [118] and PDF4LHC21 [119]. *Bottom*: Idem for nPDFs, from nCTEQ15HQ [110], TUJU21 [111], nNNPDF3.0 [113] and EPPS21 [109]. Results of a given PDF set have been normalised to those of PDF4LHC21 (top) and EPPS21 (bottom). LHAPDF naming scheme is used [120].

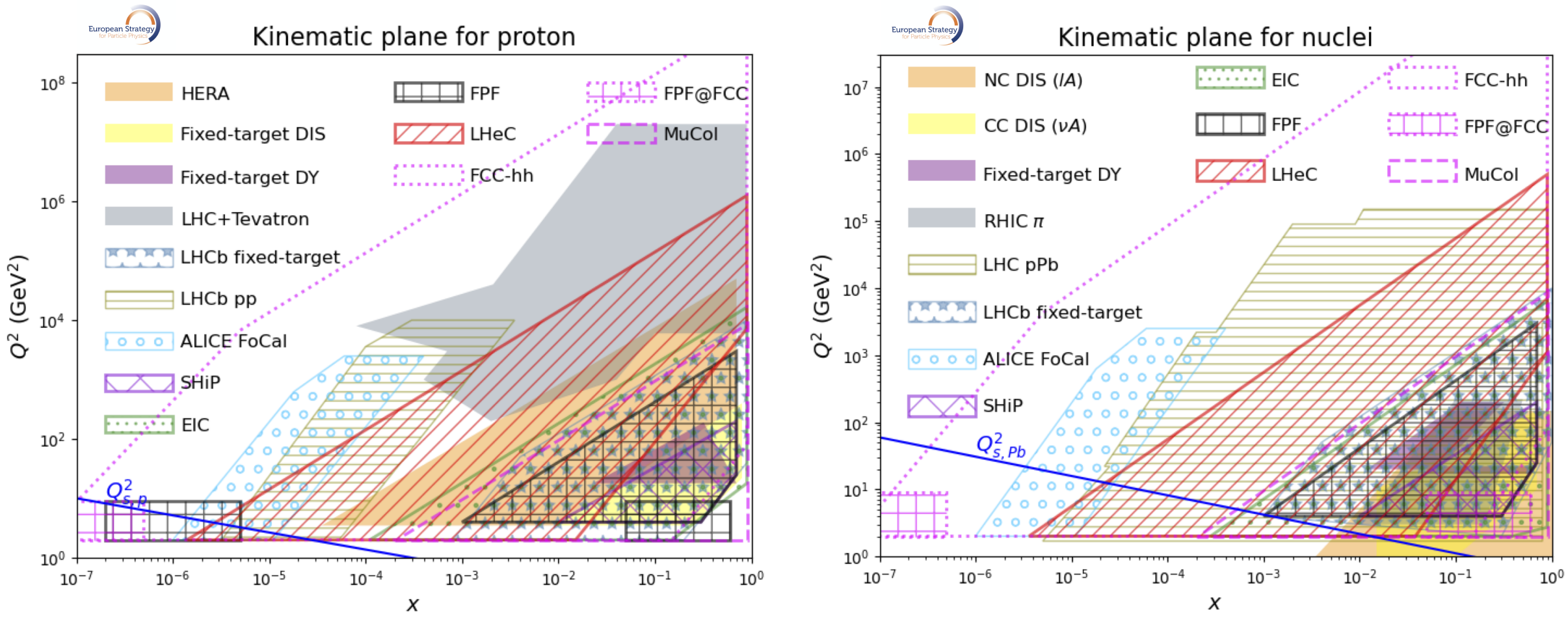

Fig. 4.4: Kinematic planes for protons (left) and nuclei (right), showing the reaches and sensitivities of present, approved and projected experiments.

### 4.2.2 Future experimental opportunities

The kinematic coverage of several future experimental facilities, both confirmed and proposed, is shown in the area overlays of Fig. 4.4. Concerning approved experiments, LHCb in both fixed-target and collider modes [ID82] (measuring $D$ and $B$ mesons, quarkonia and light hadrons), the EIC [ID114, ID216] (NC, CC and jets in DIS, light- and heavy-flavour hadrons) and SHiP [ID145] (DIS of $\nu$ from $c$ decays on fixed target; sensitivity to the poorly known $F_{4,5}$ structure functions) will cover kinematic regions overlapping with existing ones for $p$. In addition, LHCb (for $p$) and ALICE FoCal [121] [ID68] (prompt photons, neutral mesons, $J/\psi$ and jets, for $p$ and $A$) will extend our knowledge towards small $x \lesssim 10^{-5}$, in particular by exploiting high-luminosity $pA$ collisions [ID224], and the EIC will provide nPDF fits to a single nucleus and offer huge opportunities for TMDs and GPDs thanks to its PID [ID17]. Next, LHCb Upgrade II [ID81] will extend our knowledge of parton structure towards small $x$, and proposed forward-physics facilities such as FPF [ID19] ($\nu$ flux from charm for protons) would additionally offer information at moderate-to-large $x$ ($\nu$ NC and CC DIS for protons and nuclei). Beyond the HL-LHC era, the proposed LHeC [ID214] (NC, CC and jets in DIS, heavy-flavour ID) would greatly extend the kinematic plane in DIS for both proton and nuclei, with the possibilities of constraining all parton species and of PDF fits to single nuclei, e.g. for the gluon distribution at low scales of order 10 (20)% at $x = 10^{-4}$ ($x = 0.6$). Further ahead, the FCC-hh [ID209, ID247] ($W$, $Z$, DY, jets and dijets, single $t$ and $t\bar{t}$, isolated photons, $D$ and $B$ mesons, quarkonia, light hadrons) with the possibility of the FPF@FCC [122], would largely extend the kinematic plane both towards large $Q^2 > 10^6$ GeV$^2$ and small $x < 10^{-6}$, while MuCol [ID207] (NC and CC DIS of $\nu$ from muon decays on fixed $p$ or $A$ target) would cover medium-to-large $x$ with tiny statistical uncertainty.

Table 4.3: Schematic comparison of the impact and complementarities between facilities in accessing selected physics benchmarks. Dark/light/white shading indicates highest/significant/negligible contributions of the experiment to the quantities. Comments in each cell provide simplified explanations of the assessment (see the text and Fig. 4.4).

| **Quantity of interest** | **LHC: fixed target mode** $D$, $B$, quarkonium, light hadrons | **(HL-)LHC: collider mode** $D$, $B$, quarkonium, light hadrons, UPCs, DY | **ALICE FoCal** Photons, pions, quarkonium, jets, UPCs | **SHiP** DIS of $\nu$ from $c$ decays on fixed target | **EIC** NC, CC and jets in DIS, light and heavy flavour ID, excl. diffraction |
|---|---|---|---|---|---|
| **PDFs** | Most info. avail. | Most info. avail. | Simultaneous fit of proton and nuclei | Simultaneous fit of proton and nuclei; $F_4$, $F_5$ | Covered by HERA |
| **nPDFs** | Most info. avail. | Most info. avail. | Complementary e.m. probes; region overlapping with current $p$Pb | Simultaneous fit of proton and nuclei; $F_4$, $F_5$ | Cleanliness |
| **TMDs** | DY, jets | DY, jets | Limited PID | | Particle ID |
| **GPDs** | Currently UPCs | Currently UPCs | | | Particle ID |
| **Small-$x$ dynamics** | Indirect, from pre- -cision at large-$x$ | Large kinematic extent | Large kinematic extent | Indirect, from precision at large-$x$ | Kinematic reach |

| **Quantity of interest** | **FPF** $\nu$ DIS of $\nu$ from $c$ decays on fixed target | **LHeC** NC, CC and jets in DIS, heavy flavour ID, excl. diffraction | **FCC-ee/LEP3/LC** FFs of light and heavy quarks, jets, $\gamma$–$\gamma$ | **MuCol** DIS of $\nu$ from $\mu$ decays on fixed target | **FCC-hh** $D$, $B$, quarkonium, light hadrons, UPCs, DY |
|---|---|---|---|---|---|
| **PDFs** | Simultaneous fit of proton and nuclei; $F_4$, $F_5$ | Cleanliness; precision | | Simultaneous fit of proton and nuclei | Kinematic reach |
| **nPDFs** | Simultaneous fit of proton and nuclei; $F_4$, $F_5$ | Cleanliness; precision | | Simultaneous fit of proton and nuclei | Kinematic reach |
| **TMDs** | | Limited PID in detector design | FFs needed for PDFs, and TMDs in jets | | DY, jets |
| **GPDs** | | Kinematic reach | $\gamma$–$\gamma$ (transition GPDs) | | Currently UPCs |
| **Small-$x$ dynamics** | Kinematic reach | Kinematic reach in $ep$ and $eA$ | $\gamma$–$\gamma$ processes | Indirect, from precision at large-$x$ | Kinematic reach in $pp$ and $pA$ |

A full comparison of the impact of the different facilities on the several parton distributions, PDFs, TMDs and GPDs (see Fig. 4.3 for PDFs), is impossible in this summary format. Here we restrict ourselves to qualitative considerations on the different capacities of the proposed experiments and facilities, presented in Table 4.3. Note that while hadronic colliders have the largest kinematic reach, the cleanliness in the extraction of PDFs at DIS through the scattering of a point-like object with fully-constrained kinematics, cannot be matched [ID69]. Besides, DIS offers a much simpler experimental environment (no pileup, absence of colour reconnections between projectile and target, etc.) and theoretical setup for calculation of different types of corrections.

## 4.3 Hot and dense QCD

### 4.3.1 State-of-the-art and future physics goals

Over the past fifteen years, the LHC heavy-ion programme has yielded groundbreaking results (see e.g. Refs. [124,125]). Among them, it is worth mentioning the comprehensive characterisation of the nearly perfect fluidity of the QGP through light-flavoured hadron spectra, momentum anisotropies, and various multi-particle angular correlation measurements. Further outstanding results include the evidence of a novel in-medium hadronisation mechanism via hadrochemical abundances and soft charmonium production; detailed insights into quarkonium dissociation and heavy-quark energy loss in the QGP; novel insights in the dynamics of jet quenching up to transverse momenta of several hundred GeV. These results have quantified the general picture of the QGP as a strongly-coupled and rapidly-expanding fluid, and they have utilised highly-sensitive probes to characterise the physical properties of the various stages of the system evolution. A summary of this quantitative characterisation is reported in Fig. 4.5, with numerical values determined from LHC measurements, either directly or using model calculations combined with Bayesian parameter estimation [123]. Additionally, the LHC heavy-ion programme has led to unexpected discoveries, such as the observation of collectivity across all system sizes, including those in proton–nucleus and proton–proton collisions (see e.g. Ref. [126] and references therein).

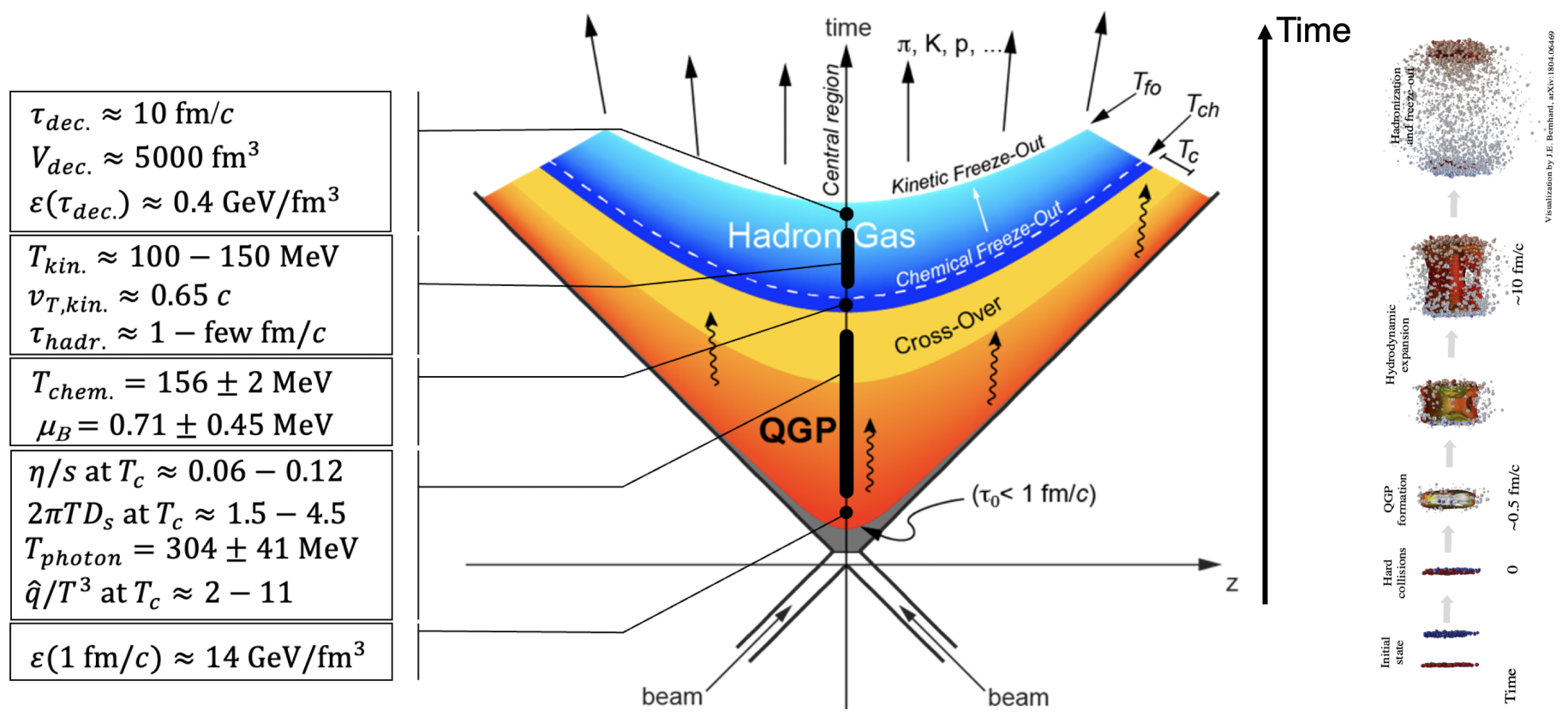

Fig. 4.5: Time evolution of the QGP formed in heavy-ion collisions, with numerical values estimated on the basis of LHC measurements or their model description. Figure from [123].

Despite significant progress, important questions remain unanswered, and the scientific opportunities for QGP studies at HL-LHC are not yet fully exploited. For example, the transport process and the interactions of heavy quarks in a nearly perfect non-Abelian fluid can be calculated from first principles in QFT, yet the corresponding QGP transport properties are only loosely constrained by the present and near-term $D$-meson measurements. A full experimental study requires precise measurements at low $p_{\mathrm{T}}$ of $D$ and $B$ meson flow, and of $D\overline{D}$ azimuthal correlations. Differential measurements of virtual thermal photons using dileptons $\gamma^* \to \ell^+\ell^-$ are necessary to estimate the temperature of the QGP with an accuracy much better than the $\pm$(10–20)% level expected in LHC Runs 3–4, and to study its time dependence during the QGP evolution. These measurements are also a crucial step toward experimentally detecting direct signatures of chiral symmetry restoration in the QCD high-temperature phase through the $\rho^0 \to \ell^+\ell^-$ spectral shape. Characterizing the QGP through medium modifications of rare hard probes, such as high transverse momentum $Z$- and photon-tagged jets, will require the integrated luminosity from HL-LHC Runs 4 and 5 [ID7, ID103].

At lower centre-of-mass energies, a strongly-interacting system with high baryon density and chemical potential $\mu_{\mathrm{B}}$ is produced. Historically, experiments exploring centre-of-mass energies at the CERN SPS at around $\sqrt{s_{\mathrm{NN}}} = 17$ GeV provided the first evidence for the production of a new state of matter with many of the predicted features of a QGP. The low energy regime of a few to ten GeV, recently explored by HADES at SIS-18, NA61/SHINE at SPS and STAR at RHIC, has important complementary interest because the onset of QGP formation and the predicted existence of a QCD critical point can be studied.

### 4.3.2 Future experimental opportunities

The future research directions develop in two branches (NuPECC [ID103], QGP@CERN Town Meeting [ID7]):

- **High-energy collisions at the HL-LHC** (baryochemical potential close to zero), where the main physics goals are the study of the time-evolution of the QGP as a many-body QCD system (linking elementary QCD interactions to equilibration at the macroscopic level) and a systematic and precise exploration of the QGP properties.
- **Low-energy collisions at SPS and FAIR** (baryochemical potential of a few hundred MeV), where the main physics goals are establishing the QCD caloric curve (temperature vs. collision energy) and phase structure (including searching for the critical point), and studying high-density QCD matter.

Figure 4.6 maps the future detectors on the QCD phase diagram (left) and on a chart of interaction rate vs. collision energy (right). In the further future, **heavy-ion collisions at FCC-hh** would give access to a hotter and longer-lived QGP and new harder rare probes. The new detectors at HL-LHC will strongly advance our understanding of fundamental questions in the field by giving access to the 4D structure (transverse, longitudinal, time), microscopic dynamics, and substructure of the quark-gluon plasma. The complementary strengths of the four large experiments provide an optimal, multi-pronged, approach for the most comprehensive characterisation of the QCD high-temperature phase by the end of HL-LHC Run 5 [ID103, ID7]:

- ALICE 3 [ID68, ID70] is a new high-energy nuclear physics experiment, based on innovative detector concepts, with particle identification and unprecedented pointing resolution over large acceptance in rapidity and transverse momentum. It offers unique opportunities in HL-LHC Run 5, in particular via multi-differential low-$p_T$ measurements of (multi-)heavy-flavour hadrons, electromagnetic radiation, and particle correlations.

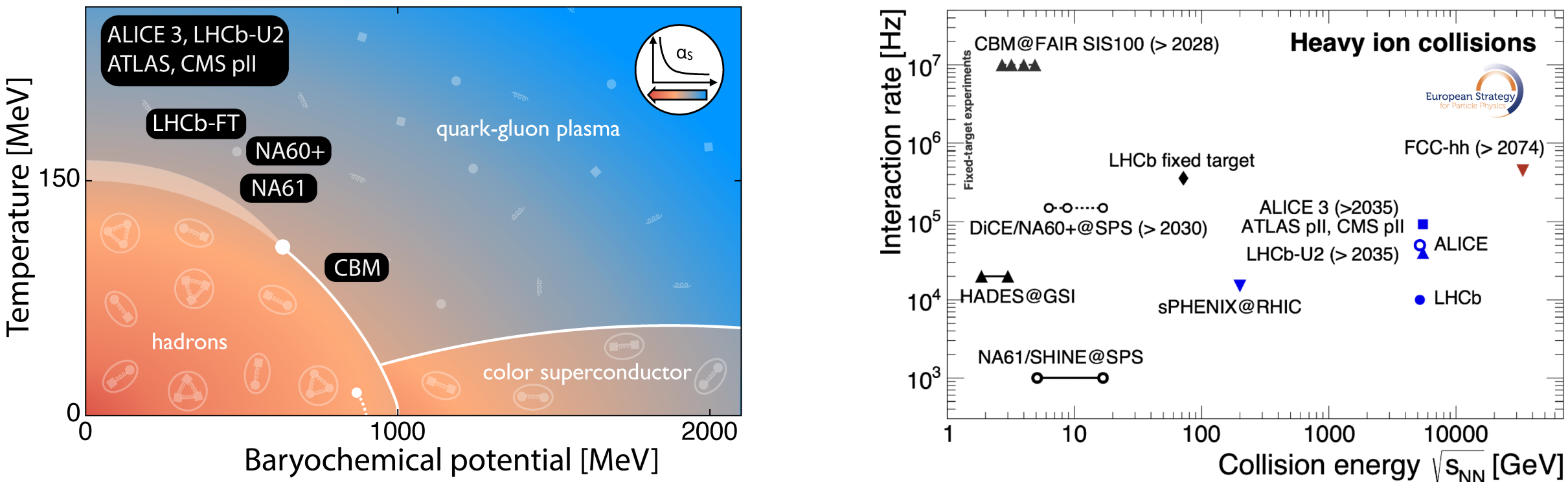

Fig. 4.6: Left: QCD phase diagram (baryochemical potential $\mu_B$, temperature $T$) with the experiments that will operate at HL-LHC, SPS, FAIR (courtesy F. Rennecke). Right: heavy-ion interaction rate and $\sqrt{s_{NN}}$ of the future proposed detectors (updated from Ref. [127]).

- LHCb Upgrade II [ID82, ID148], motivated mainly by the LHCb flavour physics programme, will also offer unique capabilities in Run 5, in particular with measurements of heavy-flavour hadrons and the initial stages in collider mode at forward rapidity and in fixed-target mode.
- ATLAS and CMS Phase-2 Upgrade detectors [ID189, ID170] feature increased pseudorapidity coverage and high-rate capability. They will significantly advance QGP research by precisely characterizing high-$Q^2$ transfer and photonuclear processes that require high statistics.

At the CERN SPS ($\sqrt{s_{NN}}$ up to $\approx 17$ GeV) there are ample opportunities for a new generation of precision measurements that address central questions about the QCD phase diagram, exploiting recent advances in silicon sensors developed by the LHC collaborations (in particular by ALICE for monolithic pixels) and extending towards lower collision energy to bridge with the FAIR facility at GSI.

- NA60+/DiCE [ID131] is a novel detector proposed at the CERN SPS that would cover the 6–17 GeV collision energy range, targeting precise measurements in the electromagnetic (thermal dilepton production around $T_c$, signals of chiral symmetry restoration) and charm sector (onset of charmonium suppression, charm hadronisation in a high-$\mu_B$ system) with unprecedented interaction rates (Fig. 4.6).
- NA61/SHINE [ID171] at the CERN SPS will continue exploring QGP-related signatures in light-ion systems through comprehensive, large-acceptance hadron measurements.
- CBM at FAIR SIS-100 [ID183] will enable the exploration of high-baryon density at lower centre-of-mass energies (2–5 GeV) through hadron and, for the first time at these energies, dilepton observables.

To fully exploit the physics opportunities offered by nuclear beams, achieving higher integrated luminosities in heavy-ion collisions at HL-LHC during Runs 4 and 5 is essential. For instance, the ALICE 3 Pb–Pb integrated luminosity target in Run 5 is 33 nb$^{-1}$ [128], which is more than twice the target till the end of Run 4. Additionally, a range of scenarios would benefit from further exploring the flexibility of CERN's accelerator complex for nuclear beam delivery to both fixed-target and collider experiments. For example, colliding lighter ion species than high-luminosity proton–nucleus runs as part of the programme [ID224] (see also Sect. 4.2) and through dedicated short runs with light ions, e.g. $^{16}$O or $^{40}$Ar [ID2].

Table 4.4: Heavy flavour and QGP benchmarks.

| Experiment | Measurement | Physics insight |
|---|---|---|
| **ALICE 3 @HL-LHC** | Beauty meson and baryon spectra and flow at $p_T > 0$ | Degree of thermalisation of heavy quarks in the QGP |
| | Multi-charm baryons ($\Xi_{cc}$, $\Omega_{cc}$) $\times$100 enhancement in Pb–Pb | Kinetic equilibration of deconfined charm quarks |
| | $D\bar{D}$ azimuthal correlations down to $p_T > 2$ GeV | Direct access to charm diffusion regime in QGP; Brownian motion of deconfined $c$ quarks |
| | $\chi_{c1,2}$ suppression patterns; low-$p_T$, mid-$y$ $\chi_{c1}(3872)$ | Access to rare $P$-wave charmonia in heavy ions; insight into internal structure of $\chi_{c1}(3872)$: molecular vs compact tetraquark |
| **LHCb Upg II @HL-LHC** | Beauty meson and baryon spectra and flow at $p_T > 0$, forward $y$ | Degree of thermalisation of heavy quarks at large $y$ |
| | Rare charmonia: $\chi_{c1}(3872)$ | Low-$p_T$, forward-$y$ $\chi_{c1}(3872)$: molecular vs compact tetraquark structure |
| **ATLAS and CMS Phase-2 Upg** | Beauty and charm hadrons at intermediate–high $p_T$ | Energy loss and hadronisation of $b$ and $c$ quarks |
| **@HL-LHC** | $t\bar{t}$ events in Pb–Pb | Jet quenching effects on $t \to W \to q\bar{q}$; explore QGP time-dependence via $p_T^{top}$ |
| **NA60+/DiCE @SPS** | Charmonium in beam-energy scan | Onset of suppression correlated with temperature measured via thermal dimuons, to identify charmonium melting threshold |
| | Intrinsic charm at mid-$y$ studying $J/\psi$ in $p$Pb | Enhanced $J/\psi$ production as signal of intrinsic charm content |
| | Charm baryon and meson cross-sections, azimuthal distributions | Short-lived medium impact on charm thermalisation; hadronisation mechanisms at high $\mu_B$ |
| **CBM @SIS-100** | Subthreshold and intrinsic charm production | Enhanced $D$ and $J/\psi$ production as a signal of intrinsic charm content |
| **FCC-hh** | $t\bar{t}$ in Pb–Pb: full exploitation | Access to time-dependence of jet quenching and QGP opacity via $p_T^{top}$ above $\sim$1 TeV |

Table 4.5: QGP temperature benchmarks.

| Experiment | Measurement | Physics insight |
|---|---|---|
| **ALICE 3 @HL-LHC** | Invariant mass spectrum of thermal dielectrons, differential in mass, $p_T$, azimuthal angle | Temperature of QGP with accuracy of $\pm(1-2)\%$ (no backgrounds), compared to $\pm10\%$ ALICE 2. Temperature as a function of time during QGP evolution |
| **NA60+/DiCE @SPS** | Caloric curve with temperatures extracted from the dimuon invariant mass spectra ($1.5 < M < 2.5$ GeV) | Study phase transition, searching for flattening of $T$ vs. $\sqrt{s}$, signature of a first-order phase transition. Experimental precision on $T$ of few MeV, i.e. $\pm(1-2)\%$ |
| **CBM @SIS-100** | Dimuon and dielectron invariant mass spectra ($1.5 < M < 2.5$ GeV) in the range $3 < \sqrt{s_{NN}} < 5$ GeV | Study phase transition, searching for a flattening of $T$ vs. $\sqrt{s} < 5$ GeV, signature of a first-order phase transition. Experimental precision on $T$ of a few MeV |
| **FCC-hh** | Quarkonium yields | Sensitivity to initial QGP temperature using $\Upsilon(1,2,3S)$ melting and regeneration patterns |
| | Total charm hadron yields and yield increase wrt $pp$ scaling with $N_{coll}$ | Thermal charm production from $gg \to c\bar{c}$ in QGP when $T \sim 1$ GeV, novel way to access initial QGP $T$ |

$^{208}$Pb (such as $^{115}$In or $^{129}$Xe) or using a 25 ns filling scheme for Pb beams could yield higher nucleon–nucleon luminosity, although with smaller QGP effects [ID7]. HL-LHC is also well-positioned to build on recent discoveries and emerging opportunities by regularly including

A heavy-ion physics programme at FCC-hh [ID209, ID7] had been explored in preparation for the FCC CDR and the 2020 ESPPU [129, 130]. It was established that the ∼6–7× larger $\sqrt{s_{\mathrm{NN}}}$ and ∼10× larger $\mathscr{L}_{\mathrm{int}}$ than HL-LHC would offer unique opportunities for QGP studies in the sector of hard probes and that the QGP created at FCC-hh energies may be the only one in which phenomena related to thermal charm production can be studied. The design choices for FCC-hh and its injectors should allow for the possibility to inject heavy ions, so that questions only accessible at higher centre of mass energies or only with future detector technologies can finally be studied as part of the FCC physics programme.

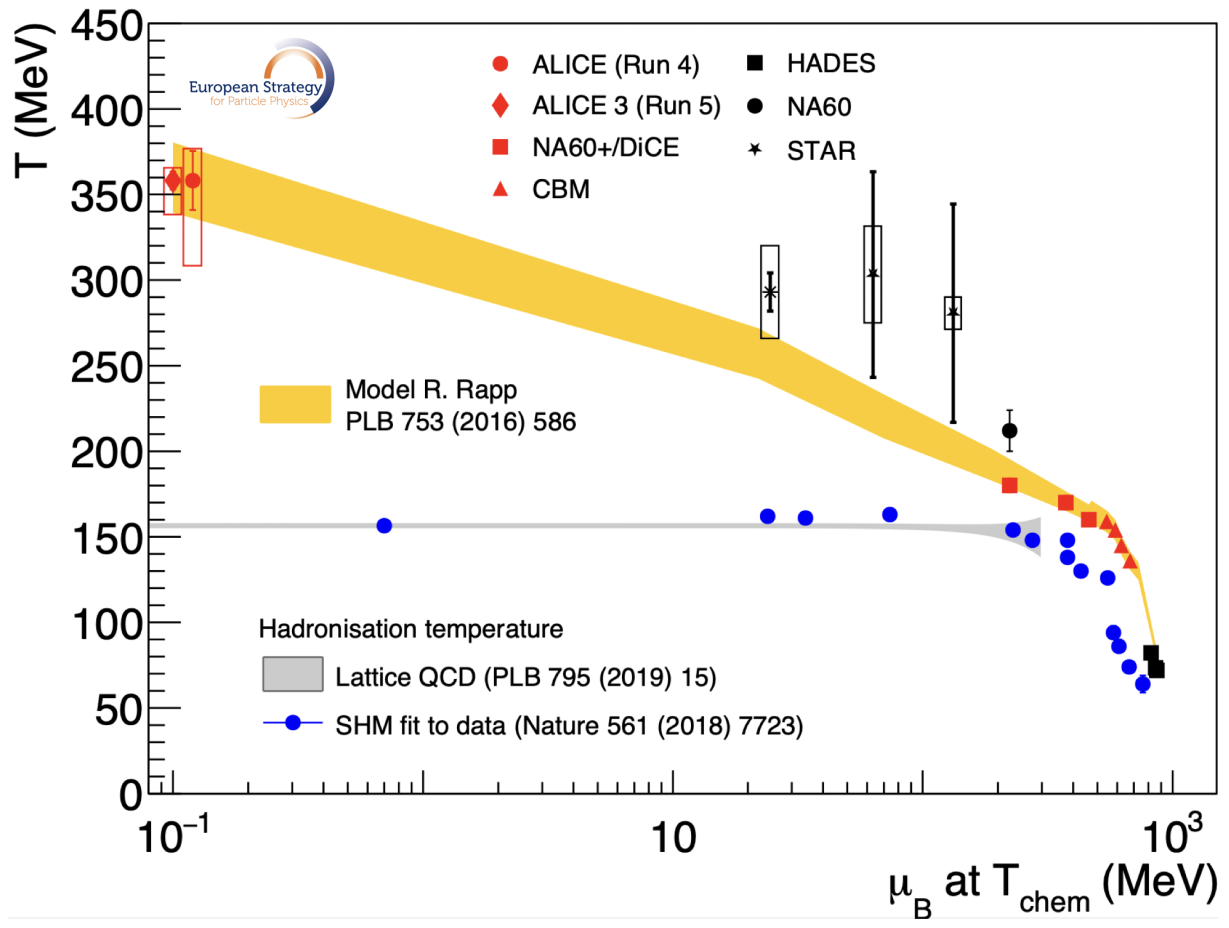

Fig. 4.7: Published (black) and projected (red) measurements of temperature from dileptons vs. baryochemical potential (courtesy of T. Galatyuk, A. Kalweit, G. Usai, I. Vorobyev). Boxes show the statistics-independent systematic uncertainties.

Selected benchmarks that document the breadth with which the future programmes can advance fundamental knowledge of QCD at extreme temperature and density are listed in Tables 4.4 and 4.5. They refer to the heavy flavour and quarkonium hadron production, electromagnetic processes sensitive to QGP radiation and rare probes, studied from the low to the high baryochemical potential $\mu_{\mathrm{B}}$ region. As an example, to highlight the large $\mu_{\mathrm{B}}$ range that will be accessible considering all the complementary heavy-ion experiments and the high precision reachable thanks to the detector upgrades, the QGP temperature is shown in Fig. 4.7 as a function of $\mu_{\mathrm{B}}$.

## 4.4 QCD connections with astro(particle) and hadron physics

### 4.4.1 State-of-the-art and future physics goals

#### *Connection to astro(particle) physics*

In connection to astro(particle) physics the following three main areas are identified.

**Equation of state (EoS) of neutron stars (NSs).** The EoS describes how the pressure of a system behaves as a function of its density $\rho$. Its dynamics is driven by the degrees of freedom that can appear as the density increases and by the strong interaction among them. The calculation of the ranges of masses $M$ and radii $R$ that allow for stable NS configuration requires the knowledge of the EoS and should reproduce the astrophysical constraints ($M_{\mathrm{max}} \gtrsim 2M_{\odot}$, $10 < R < 16\,\mathrm{km}$) [131–136]. In such compact objects, densities can exceed by far the nuclear saturation density $\rho_0 = 0.16\,\mathrm{fm}^{-3}$, and understanding the EoS under these extreme conditions remains one of most challenging aspects to be addressed from theoretical and experimental perspectives [ID89, ID106, ID235, ID126, ID76, ID183, ID117].

Constraining the properties of the nucleon-only EoS around $\rho_0$ requires a broad and complementary experimental effort: from measurements of stable/exotic nuclei properties, neutron skin thickness and nuclear collective excitations, to dedicated flow and strangeness measurements in low-energy heavy-ion collisions [137, 138]. At the high-densities achieved in the core of NSs, the production of hyperons becomes energetically favourable, but leads to a softening of the EoS and thereby to NS masses incompatible with the measured ones. This so-called “hyperon puzzle” arises from the poor knowledge of how hyperons (Y) interact at two- and many-body level with nucleons (N) and with other hyperons.

State-of-the-art results on the EoS with high-precision constraints on YN and YY interactions obtained from femtoscopy measurements by ALICE [139, 140] confirm the impossibility to achieve heavy NS masses (see green bands in Fig. 4.8). Repulsion is needed and typically advocated via three-body YNN interactions. The latter are still rather unknown, particularly when multi-strange baryons ($\Xi$) are present. The main future goal in this sector is placing tight constraints on the two- and many-body strong forces involving $\Lambda$, $\Sigma$ and $\Xi$.

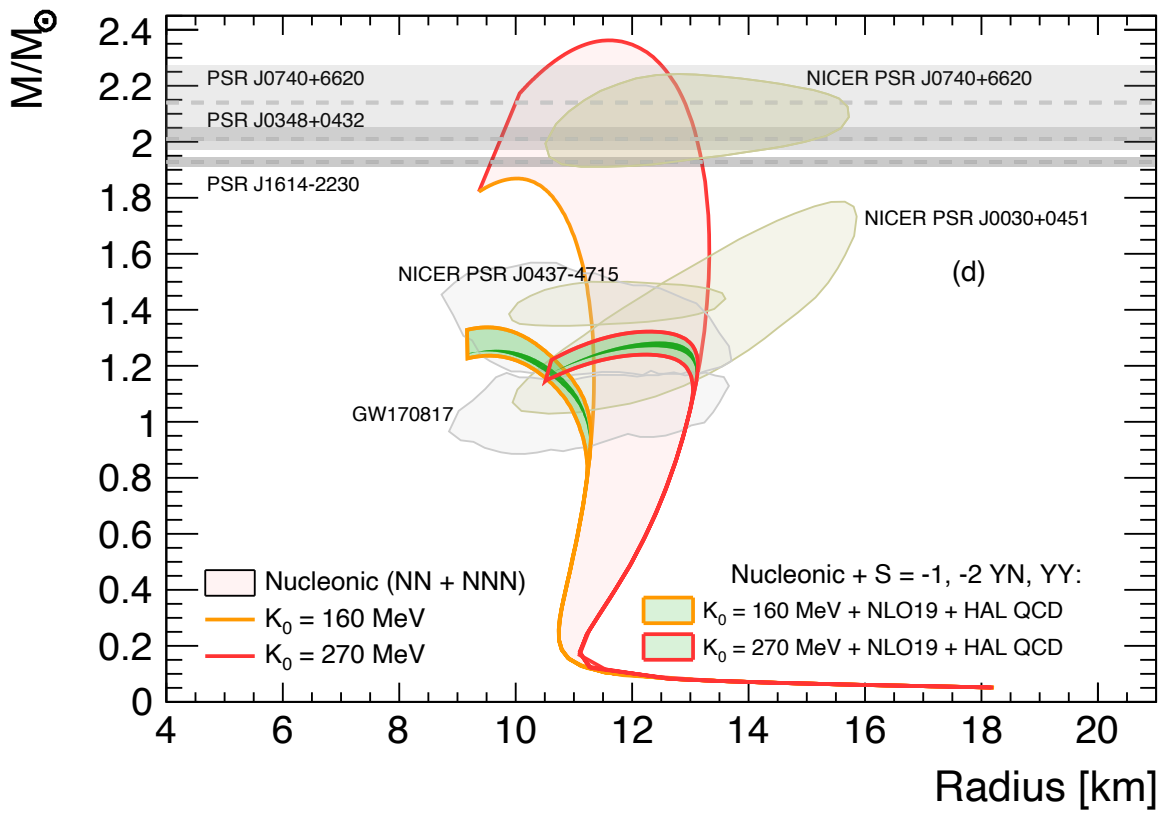

Fig. 4.8: Masses and radii of NS for a pure nucleonic EoS (red shaded area) and with the inclusion of two-body YN and YY interactions (green bands). Grey regions mark available astrophysical measurements. Figure from [139].

**Cosmic rays (CRs) and dark matter (DM).** Measurements of antimatter fluxes ($\overline{\mathrm{d}}$, $\overline{\mathrm{He}}$) in galactic CRs (GCRs) are extremely promising signatures of indirect evidence of DM annihilation (see also Sect. 7.3). Measurements at satellite and air-borne experiments have already entered the precision era, reaching uncertainties below $\pm 10\%$ for $\overline{p}$ [141]. Observation of light antinuclei ($\overline{\mathrm{d}}$, $\overline{\mathrm{He}}$) has not been confirmed, but results are expected thanks to future upgrades of CR detectors. The interpretation of antinuclei yields in GCRs is hence limited by the precision, currently at the $\pm(10-20)\%$ level, at which production and inelastic cross-sections are known [ID89]. The strength of the GCRs antinuclei flux depends on (i) how these composed objects are formed, and (ii) how they interact with the interstellar medium (ISM). Improved coalescence models, using data-driven input from measured femtoscopic radii in $pp$ collisions by ALICE, are currently under development [142, 143], along with pioneering measurements of $\overline{\mathrm{d}}$ and $\overline{\mathrm{He}}$ inelastic cross-sections [144, 145]. Results on the $\overline{\mathrm{d}}$ and $\overline{\mathrm{He}}$ production at lower energies in $pA$ collisions have been delivered by the LHCb collaboration [ID82]. The latter also set the first limits on the $\overline{\mathrm{He}}$ production in $\Lambda_b$ decays, an additional proposed source of such antinuclei in GCRs [146] to be further explored in the ALICE 3 experiment [ID68]. Future experimental efforts must focus on providing a better understanding of the production mechanism of antinuclei starting from nucleons and on improving the precision on inelastic and nuclear fragmentation cross-sections. The ultra-high energy part of CRs spectra (UHECRs) is the main focus of the physics programme of the recently upgraded Pierre Auger observatory [ID201]. Its composition is still uncertain and input from accelerator experiments is required [147], particularly in the context of the so-called “$\mu$ puzzle”, namely a muon excess (roughly $\pm(20-30)\%$ more than expected) measured in UHECRs air showers. Such a discrepancy points to either new physics or to an inaccurate modelling of hadronic interactions in Monte Carlo simulations used in the energy regime beyond colliders reach. Future precise measurements of forward-hadron production are feasible with the proposed FPF [ID19], in particular with proton–light-ion collisions [ID224], and with possible additional very-forward instrumentation at future colliders with hadron beams.

**Neutrino multi-messenger astronomy**. High-energy $\nu$ physics (see Sect. 7.2 and Sect. 7.5) could also benefit from an improved QCD input since the modelling of astrophysical neutrinos is largely limited by the knowledge of the $\nu$-nucleon (-nucleus) interaction at such energies and by uncertainties on the prompt atmospheric $\nu$ background dominated by the decay of charmed mesons [148]. For instance, significant improvements could be driven by nPDF measurements at EIC and possibly at other experiments, exploiting CC DIS scattering (see also Sect. 4.2.2).

### *Hadron spectroscopy*

Although there is some evidence suggesting the existence of exotic hadrons composed of four or more light quarks (such as the supernumerary of scalar mesons), identifying which of them are truly exotic has proven to be a challenging task. The discovery of two new unexpected mesons, $D_{s0}^*(2317)^+$ [149] and $\chi_{c1}(3872)$ [150] (also known as $X(3872)$) along with the observation of the first charged charmonium-like meson, $T_{c\bar{c}1}(4430)^+$ [151] (also known as $Z(4430)^+$), has shifted the spotlight to heavy hadrons. Over the following years, a large number of exotic hadrons has been discovered and a dedicated naming scheme has been introduced [15]. The LHC experiments have established the existence of explicitly exotic hadrons such as: pentaquarks $P_c^+ \to J/\psi p$ [152], doubly charmed tetraquark $T_{cc}(3875)^+ \to D^0 D^0 \pi^+$ [153], and fully charmed tetraquarks $T_{c\bar{c}c\bar{c}} \to J/\psi J/\psi$ [154–156].

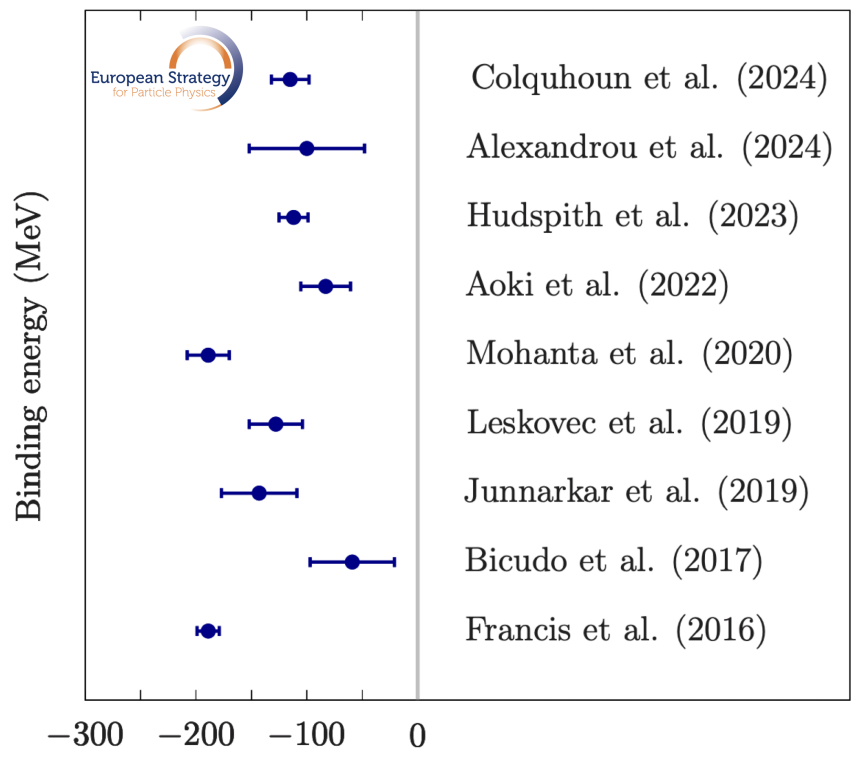

Fig. 4.9: Lattice results for the binding energy of the isospin-zero tetraquark $T_{bb}$.

Despite major experimental and theoretical progress, several fundamental questions about exotic hadrons remain unanswered. Among the most pressing ones are: How are quarks bound together within exotic hadrons (compact tetraquark vs. molecule-like meson–meson states)? Why do many exotic hadrons lack flavour or isospin partners? Do hybrids and glueballs exist?

Lattice QCD offers the possibility of computing strong-interaction physics with quantifiable uncertainties via numerical simulations. While masses of stable hadrons, lying well below hadronic thresholds, can be extracted easily, studying near-threshold or strongly-decaying states remains a relevant challenge, despite the significant progress achieved over the past decade, see e.g. [157]. As an example, doubly-charmed tetraquarks, as $T_{cc}(3875)^+$, offer the opportunity to bring together experiments, lattice simulations and phenomenology. The binding energy of the doubly-bottom tetraquark, $T_{bb}$, computed using lattice simulations (Fig. 4.9), is predicted to be large such that the hadron is stable under the strong interaction and can only decay weakly. The tetraquark $T_{bc}$ may decay weakly as well and be accessible experimentally. Further theoretical progress requires both advances in the underlying formalism and dedicated studies to quantify and control systematic uncertainties. This includes the study of extrapolations to the continuum as well as to the physical light-quark masses. An alternative approach for hadron spectroscopy is based on functional methods [158], which, together with lattice calculations, can provide important insights on the hadron substructure of exotic hadrons.

### 4.4.2 Future experimental opportunities

A summary of the most relevant experiments and facilities that will provide inputs to astroparticle and hadronic physics in the sector of the strong interaction is reported in the following, and the corresponding physics benchmark measurements are summarised in Tables 4.6 and 4.7.

The High Intensity and Energy (HIE) upgrade of **ISOLDE** [ID6], with an increased energy of the accelerated protons from 1.4 to 2 GeV, will enhance and extend the production of isotope yields. Access to a large amount of $n$-rich hypernuclei will be possible with the future proposed HyperPuma experiment. Similar studies on hypernuclei are feasible as well at the future FAIR facility [ID106, ID126, ID183].

Table 4.6: Strong-interaction benchmark measurements related to astro(particle) physics.

| Area | Topic | Measurements | Facilities / Experiments |
|---|---|---|---|
| **EoS of NS** | **Nucleonic EoS** | 1. Properties of exotic nuclei<br>2. Flow/strangeness in low-energy HICs | 1. ISOLDE HIE-Upgrade<br>2. HADES@GSI, CBM@FAIR, J-PARC-HI |
| | **"Hyperon puzzle" in NS** | 1. Correlations with hyperons $(\Lambda,\Sigma,\Xi)$<br>2. Hypernuclei ($n$-rich)<br>3. Scattering data on $\Lambda$N–$\Sigma$N, cusp spectroscopy ($\Sigma$N) | 1. ALICE 3, NA61/SHINE, J-PARC-HI, CBM<br>2. J-PARC HEF-EX, ALICE 3, HyperPuma, Panda@FAIR<br>3. J-PARC HEF-EX |
| **CRs and DM** | **Indirect DM evidence in CRs** | 1. d ($\overline{\mathrm{d}}$), He ($\overline{\mathrm{He}}$) yields<br>2. Source size via femtoscopy<br>3. Production cross-section for $\overline{\mathrm{d}}$, $\overline{\mathrm{He}}$<br>4. BR($\bar{\Lambda}_b \to {}^3\overline{\mathrm{He}} + X$) | 1. NA61/SHINE<br>2. NA61/SHINE, ALICE 3, CBM<br>3. LHCb (fixed target), AMBER, NA61/SHINE<br>4. ALICE 3, LHCb |
| | **Composition of EAS and UHECRs** | 1. Hadron production in $pp$, $pA$ and $AA$<br>2. Hadron production in forward region | 1. $pA$@LHC, $pA$@FCC-hh<br>2. FPF |
| **Neutrino astronomy** | **HE $\nu$ production, $\nu$-hadron interaction** | 1. $\nu$ spectra and $\nu$-$N$ cross-sections<br>2. Forward-$y$ charm<br>3. DIS at small $x$ | 1. FPF<br>2. FPF, NA61/SHINE<br>3. LHeC |

Table 4.7: Strong-interaction benchmark measurements related to hadron spectroscopy.

| Topic | Measurements | Facilities / Experiments |
|---|---|---|
| **Multi-heavy baryons** $\Xi_{cc}, \Xi_{bc}, \Omega_{cc}, \Omega_{ccc}$ | Excl. reconstruction | LHCb Upgrade II, ALICE 3 |
| **Compact tetraquark vs molecular model** | $DD^*$ correlation/femtoscopy, prompt/$b$-hadron/low $p_{\mathrm{T}}$ production | LHCb Upgrade II, ALICE 3, CMS, ATLAS, Belle II |
| **Flavour multiplets of exotic hadrons** | Search for in prompt and $b$-hadron decays | LHCb Upgrade II, Belle II |
| **Weakly decaying tetraquark $T_{bc}$** | Excl. reconstruction, detached $B_c^+$ production | LHCb Upgrade II |

**Low-energy experiments at SPS, FAIR, J-PARC, HIAF**. The NA61/SHINE fixed target experiment operating at SPS energies [ID161] can provide input on measurements of light (anti)nuclei spectra and yields close to the peak energy of $\overline{\mathrm{d}}$ cosmic production. Measurements of production cross-sections for $\overline{\text{d}}$ and $\overline{\text{He}}$, of nuclear fragmentation cross-section to reduce uncertainties in the CRs propagation term and femtoscopy measurements, aiming at characterizing the source size for coalescence models are in reach of NA61/SHINE and CBM. AMBER can complement with energy-dependence measurements of antiproton production rates. Regarding the EoS, low-energy heavy-ion experiments as CBM at FAIR and the heavy-ion J-PARC-HI project and CEE+ at HIAF can contribute to pinning down the properties of nuclear matter above $\rho_0$. Correlation measurements, both for coalescence studies and for constraining YN and YY interactions, are in the physics programme of NA61/SHINE and can be performed as well at the FAIR, J-PARC-HI and HIAF facilities. Scattering experiments and cusp-spectroscopy [ID35] methods within the current J-PARC infrastructure can provide additional information on the NΣ interaction and the future J-PARC-HEF-EX extension is expected to increase significantly the precision in hypernuclear spectroscopy measurements [160].

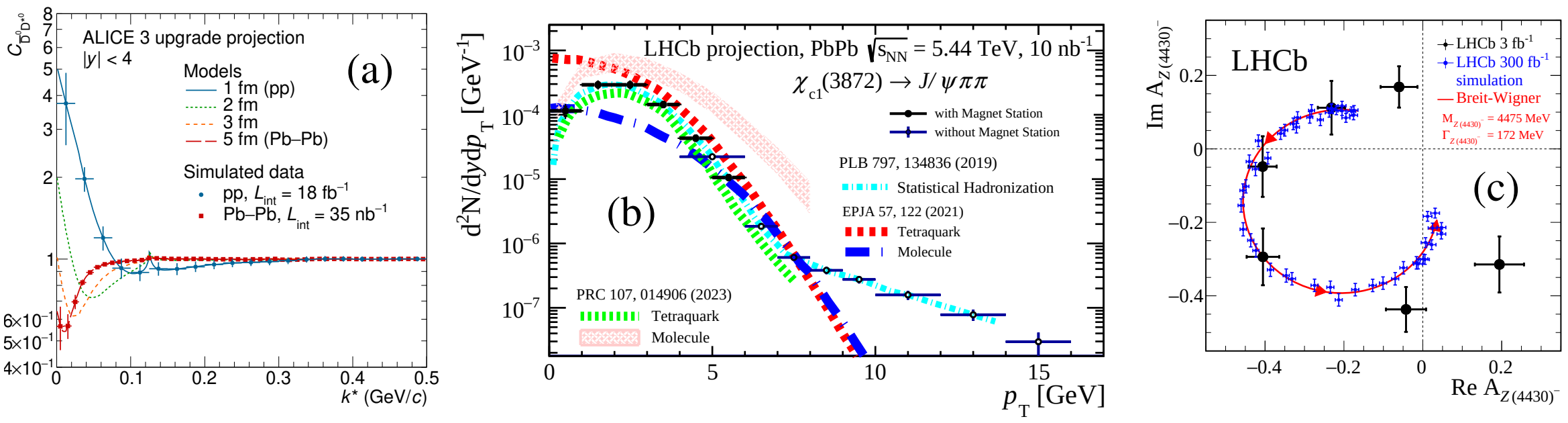

Fig. 4.10: (a) $\overline{D}^0 D^{*0}$ correlation functions at the ALICE 3 experiment in different collision systems, where the $\chi_{c1}(3872)$ meson is modelled as a molecular state. Figure from [128]. (b) Production of the $\chi_{c1}(3872)$ meson in Pb–Pb collisions at the LHCb Upgrade II. Figure from [ID82]. (c) Argand diagram of the $Z(4430)^-$ amplitude in the $B^0 \to \psi(2S)\pi^- K^+$ decays at the LHCb Upgrade II. Figure from [159].

**High-energy experiments (HL-LHC and beyond).** ALICE 3 [ID68] will be leading the femtoscopy measurements in HL-LHC Run 5. Femtoscopy studies, involving charm hadrons in the final state, will probe the molecular nature of exotic hadrons, such as $\chi_{c1}(3872)$ (Fig. 4.10-a) and $T_{cc}(3875)^+$. Measurements of the decay $\overline{\Lambda}_b^0 \to {}^3\overline{\text{He}} + X$ are among the physics goals of ALICE 3 to further explore this DM production channel for CRs. The proposed FPF at HL-LHC has the potential to explore a broad range of physics topics, among which neutrino physics plays a significant role [ID19]. Such a facility could also probe the hadron production in the forward region, important for CRs composition studies. The latter will benefit as well from measurements of particle production in different colliding systems, including $pA$ collisions in the HL-LHC experiments. Measurements of DIS processes at small $x$ in $ep$ and $eA$ at LHeC would also provide important constraints for the study of high-energy neutrinos in CRs. The LHCb Upgrade II will fully exploit the flavour-physics opportunities of the HL-LHC and will provide unique opportunities to uncover new hadronic states [ID81, ID82], such as doubly heavy baryons (e.g. $\Xi_{bc}, \Omega_{cc}$) and the first weakly decaying tetraquark $T_{bc}$. Moreover, studies of exotic hadrons in Pb–Pb collisions (Fig. 4.10-b) and $b$-hadron decays (Fig. 4.10-c) will provide insight on the internal structure of exotic hadrons (compact tetraquarks vs. molecule-like meson–meson bound states). ATLAS and CMS will contribute to the spectroscopy of the heavy tetraquarks $T_{c\bar{c}c\bar{c}}$ thanks to the large integrated luminosity collected during HL-LHC. The Belle II experiment, aiming to collect a dataset much larger than the previous $B$-factories, will be crucial to study exotic hadrons decaying to final states with photons and neutral pions. Looking back at the history of hadron spectroscopy, it is likely that upcoming projects, probing unprecedented energy regimes, will lead to the discovery of unexpected exotic hadrons.

## 4.5 Resources for theoretical developments

The research programme described in the previous sections depends critically on theoretical calculations that must meet the precision of current and upcoming experimental measurements. As discussed throughout this chapter, theoretical uncertainties often dominate the total error budget, highlighting the need for both conceptual and computational advances. Addressing such challenges require dedicated theory activities in parallel with major experimental projects, enforcing the interpretation of theoretical efforts as integral components of the overall programme. Reducing theory uncertainties will require long-term investment and coordination across multiple fronts. On the computational side, further development and maintenance of widely used open-source tools, e.g. the FORM computer algebra system, which has enabled the vast majority of precision calculations up to now [ID33], remains essential. Similarly, enhanced access to high-performance computing resources, including for lattice QCD [ID29] and artificial intelligence applications, is necessary to overcome current limitations and support precision theory efforts. Strengthening interdisciplinary collaboration between theorists and computational scientists is key to enabling progress. This includes promoting large-scale theory initiatives that align with experimental goals. In this context, it is also important to attract researchers with expertise not traditionally found within the theory community; for instance, computer scientists with strong backgrounds in algorithm development, data structures, high-performance computing, and artificial intelligence. Finally, facilitating collaboration and knowledge transfer is essential. As making experimental data open should become a standard practice, theoretical calculations should be made openly accessible, preserved for long-term use, and developed in collaborative frameworks. Enabling such openness, ideally supported by funding agencies and coordinated with experimental collaborations, will broaden participation and enhance the impact of theoretical work on the overall programme.

## 4.6 Conclusions and main messages

The strong force stands as the least understood among the four fundamental forces and its systematic investigation at all proposed future machines remains a priority. Moreover, the actual fulfilment of the unique Higgs, EW, top, flavour, and BSM programme accessible at LHC and at all future facilities relies on a significantly improved understanding of perturbative and non-perturbative QCD effects. As an emblematic example, the precision of the strong coupling $\alpha_s(m_Z^2)$ impacts parametrically all higher-order calculations of cross-sections and decay rates. Improving by an order of magnitude the $\alpha_s$ precision from the current $\mathscr{O}(1\%)$ to the $\mathscr{O}(0.1\%)$ level should be among the goals of the next decades. Such a goal is achievable at FCC-ee by exploiting the $10^{12}$ $Z$ bosons samples. Lattice QCD methodologies and measurements at LHeC provide credible alternatives, while the smaller data samples available at the linear $e^+e^-$colliders, or reduced versions of circular colliders such as LEP3, limit the foreseeable precision. QCD effects impact also the expected precision achievable on the extractions of two further key SM precision observables, $m_t$ and $m_W$. The experimental accuracy of future colliders demands a substantial improvement of pQCD calculations (up to N$^5$LO for some observables), complemented by sub-leading parton shower effects (up to N$^3$LL) and improved hadronization models.

Understanding the partonic structure of the proton remains a fundamental scientific question, significantly impacting SM measurements and searches at hadron colliders. Deep Inelastic Scattering (DIS) machines, such as the EIC and LHeC, have the potential to refine these measurements and address so far unexplored territories. In particular, the LHeC programme encompasses a large panel of measurements at the forefront of the QCD precision, including the strong coupling and the PDFs. The phase space will be further investigated through various processes and experiments, including those involving neutrinos. The strong force serves as a unique laboratory for studying the interface between perturbative and non-perturbative phenomena, as well as gluon dynamics at very low $x$ and the eventual onset of saturation, and nuclear Parton Distribution Functions (nPDFs). This rich phenomenology will be addressed with complementary approaches based on ion beams, using dedicated new forward instrumentation at existing machines (HL-LHC in particular, with proton–nucleus and photon–nucleus collisions) and new experiments at nuclear DIS facilities such as EIC and potentially LHeC.

Heavy-ion collisions are the only experimental laboratory giving access to high-temperature and high density QCD matter and the quark-gluon plasma, in which the phase structure of QCD matter and the emergence of macroscopic collective effects from elementary QCD interactions can be studied. The priorities in hot and dense QCD include: a precise determination of the QCD transport parameters with heavy quarks and jets; direct measurements of thermal radiation and the temperature evolution; the mechanisms of heavy-quark equilibration and of their hadronisation in a high-density QCD system; the phenomenon of chiral symmetry restoration; the search for the QCD critical point at large net-baryon density. These questions can be addressed, in the high-energy regime (close to zero net-baryon density), with the full exploitation of the HL-LHC in Run 5 with new detectors based on frontier sensor technologies (ALICE 3, LHCb Upgrade II, in addition to the ATLAS and CMS phase II Upgrades), for ultimate performance in heavy flavour, dileptons and jet correlations. It is also important to achieve significantly-higher luminosity with heavy ions than in previous runs and to provide the versatility for dedicated short lighter-ion runs. On the other hand, new and upgraded fixed-target experiments at SPS (NA60+/DiCE, in addition to NA61/SHINE) and FAIR (CBM) can exploit high collision rates, versatile beam configurations and the 'LHC-driven' sensor technologies to complement the LHC studies at high net-baryon density and explore the phase structure of QCD matter. New fixed-target studies could presumably be continued in the post-LHC era. On the long term, a heavy-ion programme at FCC-hh would give access to novel studies at very-high temperature, maximising the science case and community support for FCC.

Measurements in the sector of strong interactions at accelerator experiments are mandatory to address several open questions in astrophysics and in the understanding of the nature of hadronic interactions and exotic states. Providing these measurements requires broad and complementary experimental efforts across different facilities at different energy regimes. The synergy between low-energy and high-energy experiments (hadron and lepton colliders, fixed target and photoproduction experiments) is crucial. Such synergy extends to the different experimental techniques, from femtoscopy to spectroscopy measurements, from antinuclei production cross sections to hadron production in different colliding systems. The full exploitation of the HL-LHC with upgraded and new detectors, along with several upgrades and new low-energy facilities outside CERN, can provide such a broad experimental landscape.

The theoretical developments in QCD and their synergy with experimental research are crucial for the understanding of the strong interaction as well as its effects for EW and new physics searches at future facilities.

# Chapter 5

# Flavour Physics

## 5.1 The big questions

One of the most profound open questions in particle physics is why there are three generations of quarks and leptons that share identical gauge interactions but have vastly different masses. Besides the intrinsic interest of this question, baryogenesis in the early Universe may have been strongly influenced by the dynamics behind the origin of these masses, as well as by the masses of the neutrinos. Furthermore, the observed matter–antimatter asymmetry in the Universe may also be tied to the origin of quark and lepton masses.

The two key questions that flavour physics aims to address are: i) Can the observed pattern of quark- and lepton-mass matrices be predicted based on new fundamental principles or symmetries? ii) How can these underlying laws be experimentally identified, and more generally, can we observe family-non-universal interactions beyond those that result in the different fermion masses?

The quest for a deeper understanding of the pattern of fermion masses and mixings is closely linked to the nature of the Higgs field, and more broadly, to the ultraviolet (UV) completion of the Standard Model (SM). Within the SM, fermion masses are described by the Yukawa interaction between the fermion and Higgs fields. However, unlike the gauge interactions, which are uniquely determined by the requirement of local symmetry and characterized by a single coupling constant each, the Yukawa sector is different: it is not dictated by any symmetry principle; rather, it is simply compatible with the existing symmetries of the SM. In fact, the Yukawa interactions explicitly break most of the global symmetries respected by the rest of the theory and introduce a large number of free parameters.

Experimental measurements indicate that quark- and lepton-mass matrices have highly non-trivial structures, with eigenvalues spanning several orders of magnitude and markedly different mixing patterns in the quark and lepton sectors. These features suggest a deeper explanation in terms of some yet-unidentified UV dynamics. At low energies, such dynamics can manifest themselves only via the Yukawa interactions, which are the lowest-dimensional operators breaking the flavour degeneracy. However, it is natural to expect additional effects in the form of new contact interactions, explicitly suppressed by the scale characterising such dynamics. These new interactions can be probed through precision experiments: the overarching goal of flavour physics is to uncover such effects.

So far, no statistically significant deviations have been observed in flavour-changing or CP-violating (CPV) processes. This lack of evidence is traditionally expressed through stringent lower bounds on the effective scale of New Physics (NP), which exceed the TeV scale by several orders of magnitude for generic $O(1)$ couplings. However, these bounds can be partially misleading. Two plausible scenarios exist: either NP strongly breaks flavour symmetries, requiring it to lie at very high scales—though this worsens the hierarchy problem of the Higgs sector and leaves the origin of Yukawa structures unexplained—or the NP scale is relatively low (around a few TeV), but its couplings to light fermions are suppressed or aligned in a flavour-specific way. These scenarios are not mutually exclusive: a realistic UV completion may well involve multiple energy layers, each with a different flavour structure, with the lowest layer potentially tied to a solution of the hierarchy problem.

In both frameworks, improving experimental precision in flavour-changing processes is essential. In the high-scale NP case, flavour physics offers one of the few available windows into otherwise inaccessible energy regimes. In the low-scale NP scenario, precise measurements may uncover subtle imprints of flavour non-universality and shed light on the nature of the Higgs field. Thus, sustained theoretical and experimental progress in the flavour sector is crucial, not only to clarify the origin of fermion masses but also to indirectly probe and decode the new dynamics stabilizing the electroweak scale. A comprehensive strategy to uncover NP requires the complementarity of the flavour sector with respect to direct searches at the energy frontier and precision measurements in electroweak, Higgs and top physics.

## 5.2 Addressing the big questions

Regardless of whether NP lies at the TeV scale or well above it, its effects in low-energy flavour-changing processes can be systematically described as contact interactions, and the SMEFT provides a systematic tool to classify them (see Appendix A). The inclusion of flavour non-universality significantly increases the number of independent couplings controlling dimension-six SMEFT operators. While this proliferation of parameters may appear daunting, it is in fact a strength: to a good extent, each operator corresponds to a distinct observable that can be experimentally probed, at least in principle. In complete UV models these effects are correlated through only a few independent parameters and uncovering such correlations allows for identification of the underlying model. The challenge lies in identifying theoretically clean and experimentally accessible observables that are particularly sensitive to NP effects. The observables most relevant for constraining NP can be classified into the following main categories:

**Null tests of the SM:** Here, any observed signal would provide unambiguous NP evidence. The cleanest examples are lepton flavour violating decays of charged leptons (cLFV), such as $\tau \to \mu\gamma$ and $\mu \to e\gamma$, and particle electric dipole moments (EDMs). Strictly speaking, the latter are flavour-conserving processes; however their CPV nature and their extreme suppression within the SM make them ideal probes of new sources of CP violation.

**Neutral meson mixing and flavour-changing neutral currents (FCNCs):** These processes are forbidden within the SM at tree level and can often be predicted (or bounded) with high accuracy. The control of both theoretical and experimental uncertainties is what determines their ultimate sensitivity to short-distance dynamics.

**Tests of lepton flavour universality (LFU):** Ratios of quark flavour-changing transitions with identical hadronic structure and different lepton currents test the equality of lepton cou-

| Flavour Transition | Transition type $\gamma$ | quarks | $\mu$ | $e$ | $\tau$ | $\nu$ |
|---|---|---|---|---|---|---|
| $\mathbf{3_q \to 2_q}$ | $b \to s\gamma\,[{\color{red}*}]$ | $B_s \leftrightarrow \bar{B}_s\,[{\color{red}*}]$ | $b \to s\,\bar{\mu}\mu\,[{\color{red}*}]$<br>$b \to c\,\mu\bar{\nu}$ | $R_{bs}(e/\mu)\,[{\color{red}*}]$<br>$R_{bc}(e/\mu)\,[{\color{blue}\bullet}]$ | $b \to s\,\bar{\tau}\tau\,[{\color{green}\bullet}]$<br>$b \to c\,\tau\bar{\nu}\,[{\color{blue}\bullet}]$ | $b \to s\,\nu\bar{\nu}\,[{\color{green}\bullet}]$ |
| $\mathbf{3_q \to 1_q}$ | $b \to d\gamma\,[{\color{red}*}]$ | $B_d \leftrightarrow \bar{B}_d\,[{\color{red}*}]$ | $b \to d\,\bar{\mu}\mu\,[{\color{red}*}]$<br>$b \to u\,\mu\bar{\nu}$ | $R_{bd}(e/\mu)\,[{\color{red}*}]$<br>$R_{bu}(e/\mu)\,[{\color{blue}\bullet}]$ | $b \to d\,\bar{\tau}\tau\,[{\color{green}\bullet}]$<br>$b \to u\,\tau\bar{\nu}\,[{\color{blue}\bullet}]$ | $b \to d\,\nu\bar{\nu}\,[{\color{green}\bullet}]$ |
| $\mathbf{2_q \to}$ $\mathbf{2_q}$<br>$\mathbf{1_q}$ | $c \to u\gamma$ | $D \leftrightarrow \bar{D}\,[{\color{red}*}]$ | $c \to u\,\bar{\mu}\mu$<br>$c \to s\,\bar{\mu}\nu$<br>$c \to d\,\bar{\mu}\nu$ | $R_{cu}(e/\mu)\,[{\color{red}*}]$<br>$R_{cs}(e/\mu)\,[{\color{blue}\bullet}]$<br>$R_{cd}(e/\mu)\,[{\color{blue}\bullet}]$ | | $c \to u\nu\bar{\nu}\,[{\color{green}\bullet}]$ |
| $\mathbf{3_\ell \to}$ $\mathbf{2_\ell}$<br>$\mathbf{1_\ell}$ | $\tau \to \mu\gamma\,[{\color{green}\bullet}]$<br>$\tau \to e\gamma\,[{\color{green}\bullet}]$ | $\tau \to \mu\,\bar{q}q\,[{\color{green}\bullet}]$<br>$\tau \to e\,\bar{q}q\,[{\color{green}\bullet}]$ | $\tau \to \mu\,\bar{\mu}\mu\,[{\color{green}\bullet}]$<br>$\tau \to e\,\bar{\mu}\mu\,[{\color{green}\bullet}]$ | $\tau \to \mu\,\bar{e}e\,[{\color{green}\bullet}]$<br>$\tau \to e\,\bar{e}e\,[{\color{green}\bullet}]$ | | $\tau \to \mu\,\nu\bar{\nu}\,[{\color{green}\bullet}]$<br>$\tau \to e\,\nu\bar{\nu}\,[{\color{green}\bullet}]$ |

$[{\color{red}*}]$ = major progress @ HL-LHC, limited impact of $e^+e^-$ colliders
$[{\color{blue}\bullet}]$ = major progress @ HL-LHC & Belle II + significant further progress @ FCC-ee
$[{\color{green}\bullet}]$ = FCC-ee allows one order of magnitude precision improvement, or more, after HL-LHC & Belle II

| | | | | | | |
|---|---|---|---|---|---|---|
| $\mathbf{2_q}$<br>$\mathbf{1_q}$ $\mathbf{\to 1_q}$ | $s \to d\gamma$ | $K^0 \leftrightarrow \bar{K}^0$ | $s \to d\,\bar{\mu}\mu$<br>$s \to u\,\mu\bar{\nu}$<br>$d \to u\,\mu\bar{\nu}$ | $R_{sd}(e/\mu)$<br>$R_{su}(e/\mu)$<br>$R_{du}(e/\mu)$ | | $s \to d\,\nu\bar{\nu}$ |
| $\mathbf{2_\ell \to 1_\ell}$ | $\mu \to e\gamma$ | $\mu N \to eN$ | | $\mu \to e\,\bar{e}e$ | | $\mu \to e\,\nu\bar{\nu}$ |

Fig. 5.1: List of flavour-changing processes in the quark and lepton sector most relevant for constraining physics beyond the SM. The first column indicates the generations of quarks ($n_q$) and leptons ($n_\ell$) involved in the specific processes listed in the corresponding row. Transitions in the upper part of the table can be accessed in large-scale multi-purpose experiments, while those in the last two lines can best be probed in dedicated facilities ($K$, $\pi$, $\mu$ beams). The quantity $R_{ab}(e/\mu)$ denotes the electron/muon LFU ratio in the transition $a \to b$, while $M \leftrightarrow \bar{M}$ denotes the meson–anti-meson mixing amplitude of the meson $M$.

plings predicted by the SM. These ratios are predicted with high precision in the SM and are sensitive to NP effects with non-universal couplings between the different (lepton) generations.

**Ultimate determination of CKM elements:** While not directly sensitive to NP, a precise determination of the Cabibbo-Kobayashi-Maskawa (CKM) mixing matrix is a pre-requisite to make precise SM predictions for rare processes, in particular meson mixing and FCNCs.

Together, these classes of processes provide a comprehensive and highly sensitive test of the flavour structure of NP. The full set of relevant charged-current and neutral-current processes, in both the quark and lepton sectors, is summarised in Fig. 5.1. Note that each process is characterised by different quantities (e.g. exclusive and inclusive rates, angular observables, etc.). While not all of these observables are experimentally accessible and/or theoretically clean, there are more than a hundred which are relevant.

In Fig. 5.1 we also provide, via the coloured symbols, a rough indication of the expected improvement in precision for some of the processes at existing and future large-scale facilities (see classification below). This is only a qualitative indication, that will be substantiated in detail in Sect. 5.4. The impact of this improvement for NP searches, together with that expected from small-scale facilities (cf. Sect. 5.3), is discussed in Sect. 5.5.

### 5.2.1 A diverse experimental field

In flavour physics, the experimental landscape is characterised by a diverse range of facilities, which can be broadly categorised as small-, mid-, and large-scale experiments based on their scope and infrastructure. Small-scale experiments, such as those searching for EDMs and cLFV, or very rare pion decays, typically involve highly specialised, precision-focused setups designed to probe extremely rare processes with minimal backgrounds. These experiments are often laboratory-scale and even, in some very specific cases, table-top experiments. Mid-scale experiments, such as those studying rare kaon decays, require more substantial infrastructure. In both cases, they operate at dedicated beam lines or accelerator facilities. The size of collaborations in these experiments varies accordingly, ranging from small groups of a few dozen in small-scale efforts to several hundred in medium-scale projects.

Finally, a few large-scale experiments are currently running, planning for future upgrades, and being designed. They are or will be housed at major international particle physics centres, such as CERN, and the collaborations are correspondingly of the order of thousands of researchers.

The main facilities considered in this chapter are:

- The Belle II experiment, which is taking data at the SuperKEKB $e^+e^-$collider at a centre-of-mass energy of 10.58 GeV corresponding to the $\Upsilon(4S)$ resonance [ID205].

- The LHC, where the large $b\bar{b}$ and $c\bar{c}$ cross sections are used by the ATLAS, CMS and LHCb experiments for flavour physics [ID81, ID223]. The results are dominated by those of the LHCb experiment but the general-purpose experiments ATLAS and CMS, are also providing results in very specific cases.

- Proposed future $e^+e^-$high-luminosity facilities, running at the $Z$ pole and $W^+W^-$ threshold. Experiments there would take advantage of the 15% (resp. 12% and 3.4%) branching fraction of $Z$ decays into $b\bar{b}$ (resp. $c\bar{c}$ and $\tau^+\tau^-$) pairs [ID140, ID153, ID196, ID233].

The Belle II and LHC experiments have ongoing data collection and well-defined upgrade programmes. These different facilities provide complementary approaches to studying heavy-flavour phenomena. Large samples of $c$-hadrons and $\tau$ leptons could also be accessible at $e^+e^-$colliders with centre-of-mass energies of the order of 3 to 4 GeV at the Super Tau Charm Facility (STCF) [ID231].

A complementary approach to probe flavour-violating interactions is through high-energy processes, e.g. searching for $\mu\bar{\mu} \to b\bar{s}$ (rather than $B_s \to \mu\bar{\mu}$) or $pp \to \mu\bar{e}$ (rather than $\mu \to e$ conversion) at suitable high-energy colliders. The advantage over the low-energy probes lies in the ability to explore energies well above the electroweak scale, thereby enhancing the sensitivity to high-scale NP. As discussed in [ID207], a muon collider with 10 TeV centre-of-mass energy would provide significant improvements compared to the projected reach of the HL-LHC experiments and Belle II, for selected flavour-violating contact interactions. Given the long time scale required for such hypothetical high-energy measurements, and the significant improvements anticipated from low-energy probes, both on a shorter time scale and across a broader range of interactions, this class of measurements will not be discussed in this chapter.

### 5.2.2 Controlling and improving SM theory predictions

As the only systematically improvable first-principles approach to nonperturbative QCD, Lattice Quantum Chromodynamics (LQCD) provides the theoretical accuracy for SM predictions required to interpret flavour observables. LQCD denotes the set of quantum field theories defined in a finite discretised space-time volume that approach QCD in the infinite-volume and continuum limit. As long as the volume is large enough and the lattice spacing is small enough to accommodate all relevant energy scales for a given observable, systematic effects can be controlled. In practice, the number of degrees of freedom in the finite and discretised volume is still huge, such that solving the equations of motion requires access to high-performance parallel-computing resources [ID29, ID168]. Progress in the field has been impressive, in particular over the past decade, culminating recently in the seminal results that overhauled our understanding of the physics behind $(g-2)_\mu$ [161]. For more than a decade now, the Flavour Lattice Averaging Group (FLAG) [54] is giving testimony of LQCD output with relevance for flavour physics, monitoring continued improvement in precision, accuracy and novel quantities being computed.

In the near term, LQCD is essential for maximizing the scientific output of LHCb and Belle II. A new generation of high-precision lattice computations—focused on decay constants, form factors and meson-mixing matrix elements—are crucial to fully exploit experimental sensitivities to rare decays, LFU tests and CPV observables. Several flavour anomalies (e.g. in semileptonic $B$ decays) are currently limited by theory uncertainties where lattice QCD can deliver decisive improvements. The global CKM unitarity tests and searches for new physics via indirect constraints fundamentally depend on accurate lattice inputs. While FLAG considers many of the required predictions to be sufficiently mature to allow for reliable estimation of the systematic uncertainties, work is ongoing to resolve the technical challenges involved when computing more complicated quantities related to the processes in Table 5.1. Noteworthy in this context is the progress on quantities like $\varepsilon'/\varepsilon$ [162], $\varepsilon_K$, $\Delta m_K$ [163, 164], rare kaon and hyperon decay [165, 166], radiative meson decay [167–170], long-distance contributions to charm-mixing [171] and QCD+QED for semileptonic kaon decay [172].

Central to progress on tree- and rare-process SM tests is improved determination of the CKM matrix element $|V_{cb}|$. While a lattice prediction for the exclusive determination is technically challenging but now doable [173–175], novel ideas allow for the first time to compute inclusive $B_{(s)}$- and $D_{(s)}$-meson semileptonic decays in LQCD [176–179]. This provides a promising avenue for resolving the decades-old inclusive-exclusive tension, in this way strengthening the currently theory-limited $\varepsilon_K$ and other rare-process SM tests (cf. Sect. 5.4.3).

In the mid-to-longer term, the expected statistical power of Belle II and LHCb Upgrade II will increase very significantly, demanding LQCD uncertainties at or below the percent level across a wide range of processes. Achieving this requires, in addition to sustained investment in computational resources and algorithmic innovation [180], the inclusion of QED corrections and strong-isospin (i.e. up/down quark-mass difference) in simulations. For instance, a photon radiated from a quark in leptonic meson decay can couple to the final-state charged lepton, which requires a solid theoretical framework for dealing with infrared divergences. Substantial efforts go into developing the formal framework for including these effects in a non-perturbative fashion [181]. First results for leptonic kaon and pion decay [182–185] hold promise that other processes and observables will follow, provided there is progress on theory.

Increasingly, LQCD is a key scientific driver in the global quark-flavour programme. Its continued development, and thus ability to realise the full discovery potential of the next decades in flavour, will be dependent on access to high performance computing resources, large-scale data storage facilities, and support for training and software engineers.

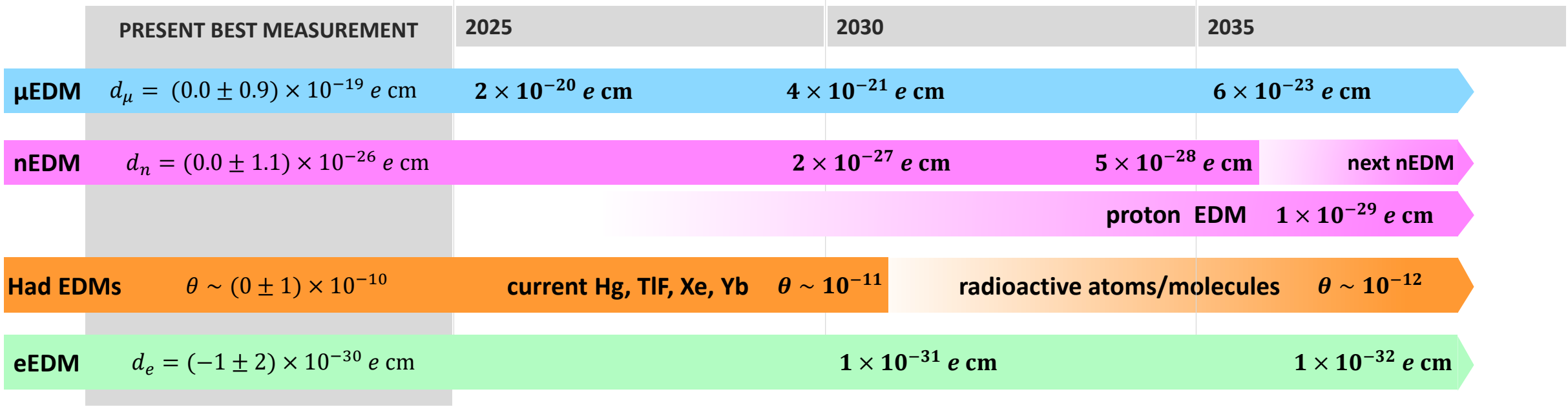

Fig. 5.2: Present best measurements and anticipated sensitivities for the EDMs of the muon, neutron, proton, diamagnetic atoms (expressed as $\theta$) and paramagnetic molecules (expressed as electron EDM). See [ID158] for more details.

## 5.3 Small and mid-size experiments

### 5.3.1 Electric Dipole Moments

The permanent EDM $d$ of a spin-1/2 particle such as the neutron or electron quantifies a possible P- and T-violating coupling of the spin to an external electric field $\vec{E}$: $H = -d\vec{\sigma} \cdot \vec{E}$. EDMs are precision observables because quasi-stable particles can be exposed to electric fields for a long time, allowing the EDM signal to accumulate. Sensitivity is typically characterized by $TE\sqrt{N}$, where $T$ is the exposure time to the electric field $E$, and $N$ the number of detected particles. In addition, the experiments rely on advanced magnetic shielding and quantum magnetometry to control the magnetic field. EDMs are expressed in units of charge times distance; the sensitivity can be appreciated by the best present bounds (95 % C.L.): $d_n < 2 \times 10^{-26} e\,\mathrm{cm}$ for the neutron [186] and $d_e < 5 \times 10^{-30} e\,\mathrm{cm}$ for the electron [187]. Several systems are being investigated [ID158], broadly categorized as: (i) neutron EDM (ii) EDMs of closed-shell (diamagnetic) atoms and molecules, which are sensitive to hadronic CP violation and provide indirect constraints on $d_n$ and $d_p$ (iii) EDMs of open-shell (paramagnetic) systems, sensitive to the electron EDM and electron-nucleon CPV; and (iv) EDMs of charged particles in storage rings (muon, proton, light nuclei) or at colliders (tau lepton, charm baryons).

Non-vanishing EDMs are induced by CPV couplings in the underlying theory. The contribution induced by the SM Yukawa couplings is extremely suppressed [188], hence EDMs are excellent SM null tests: any non-zero value would imply NP (unlike, for example, the muon $g-2$). Beyond the SM Yukawa coupling, two distinct sources of CP violation may lead to observable EDMs: the QCD $\theta$-term, as well as CPV phases from heavy NP encoded in SMEFT operators. The experimental constraints on the EDMs provide the strongest indication that the $\theta$-term may be dynamically cancelled, e.g. by the axion [189]. However, a tiny but non-vanishing $\theta$-term remains an interesting possibility [190]. As far as high-scale NP is concerned, current EDM bounds lead to bounds on effective scales of $2.5 \times 10^6$ TeV (electron) and $10^4$ TeV (neutron) for $O(1)$ couplings (see Sect. 5.5). Even assuming a chiral suppression in the couplings, these bounds exceed several hundreds of TeV, providing a unique probe of new CPV sources in the deep UV, with large margin of improvement given future prospects. Interpreting

EDM measurements in terms of fundamental couplings is a non-trivial task that involves a ladder of effective field theories, from the NP scale down to hadronic, nuclear, and atomic levels. A strong European theory effort is underway to reduce these uncertainties [ID158].

In the short term (by $\sim 2035$), significant progress is expected on the neutron EDM. Four next-generation experiments are under construction or commissioning: n2EDM at PSI [ID21], panEDM at the ILL [ID158], and two in North America (LANL [191] and TUCAN at TRIUMF [192]). All employ stored ultracold neutrons (UCNs) in double-chamber apparatuses at room temperature aiming at sensitivities in the $10^{-27}\,e\,\text{cm}$ range. Beyond 2035, several avenues could push the sensitivity further. These include more intense UCN sources, reconsideration of beam experiments at pulsed spallation sources, and the realization of the cryoEDM concept where UCNs are directly produced in the EDM cell filled with superfluid helium. This latter approach, supported by two decades of R&D in the US, appears especially promising to reach sensitivities below $10^{-28}\,e\,\text{cm}$. Long-term access to high-flux neutron sources will be essential. In this context, the European Spallation Source (ESS) has drawn strong community interest for a future programme in fundamental neutron physics [ID190], although no dedicated beamline has yet been decided.

Storage-ring experiments targeting the proton or light nuclei could potentially reach even higher sensitivities on the nucleon EDMs. Over the past two decades, extensive design studies have led to conceptual designs of a $\sim$1 km circumference electrostatic ring that aims to reach sensitivities of order $10^{-29}\,e\,\text{cm}$. The next major milestone is the completion of a full technical design. Efforts are also underway to explore EDMs of flavoured systems at PSI [ID158] and JPARC [193] dedicated experiments are under development for the muon EDM; at the LHC, the ALADDIN project targets charm baryons; and future colliders could improve the sensitivity to the EDM of the tau lepton.

Atomic and molecular (AMO) EDM searches provide another powerful and complementary approach. These table-top experiments exploit advanced AMO techniques and quantum technologies. Paramagnetic systems (ThO, $\text{ThF}^+$, or YbF) are sensitive to the electron EDM and electron-nucleon CPV, while diamagnetic systems (Hg, TlF, Xe, Yb) probe hadronic sources. Ongoing experiments aim at an order-of-magnitude gain by 2030. Looking further ahead, several avenues are being explored to boost the figure of merit $ET\sqrt{N}$ or use systems intrinsically more sensitive to CPV sources. Laser cooling and trapping techniques can increase the exposure time $T$; cryogenic matrix isolation can enhance the number of molecules $N$; and the use of species with higher $Z$ (e.g., RaF) benefit from enhanced internal electric fields or octupole-deformed nuclei. These higher-$Z$ species are by nature radioactive and facilities producing such isotopes will become increasingly relevant for future EDM searches. Overall, while projections remain uncertain, another order-of-magnitude gain in sensitivity by 2040 appears feasible. An overview of current and future EDM searches is shown in Fig. 5.2.

### 5.3.2 Charged lepton flavour violation

The conservation of lepton flavour is an accidental symmetry of the SM. The discovery of neutrino oscillations already demonstrates that this symmetry is not exact; however, neutrino masses only imply negligible LFV rates for charged leptons. Evidence of cLFV at current and future experimental sensitivities would therefore provide an unambiguous indication of new degrees of freedom. Indeed, current bounds provide stringent constraints on several NP models, see e.g. [194]. Thanks to the availability of intense muon beams, some of the most stringent limits on cLFV are derived by the so-called muon-golden channels: $\mu \to e\gamma$, $\mu \to 3e$ and $\mu N \to eN$. The combination of results from these modes will be essential to clarify the NP interpretation in the event of a positive signal. Fig 5.3 shows the time line of current and future projects with the expected sensitivities.

MEG II aims to achieve a final sensitivity of $6 \times 10^{-14}$ for the $\mu^+ \to e^+\gamma$ decay, one order of magnitude better than its predecessor, MEG. The recent result $\mathrm{BR}(\mu \to e\gamma) < 1.5 \times 10^{-13}$ [195] is based on the 2021 and 2022 data set. Data collection will continue until 2026, before the long shutdown at PSI in 2027–2028. A second phase will require a higher-intensity muon beam, supported by the High Intensity Muon Beam (HIMB) project at PSI, aiming at delivering up to $10^{10}$ $\mu^+/s$, two orders of magnitude more than the current available beam intensity [196]. A new concept for searching for $\mu^+ \to e^+\gamma$ is being discussed with the aim of enhancing the sensitivity by at least one more order of magnitude with respect to MEG II [197].

The Mu3e experiment aims to improve the sensitivity of the search for $\mu^+ \to e^+e^+e^-$. It seeks to push the limit from the current $10^{-12}$ value [198] down to $1 \times 10^{-15}$ in a first phase [199]. The HIMB project will enable Mu3e to reach its planned final sensitivity of $10^{-16}$, running at $10^9$ $\mu^+/s$.

The search for the neutrinoless coherent conversion of a muon to an electron in the field of a nucleus, $\mu^- N \to e^- N$, is carried out by the Mu2e experiment at FNAL and the COMET experiment at J-PARC. Both experiments share similar features and use an aluminum target to stop the muons. Their main goal is to increase sensitivity by four orders of magnitude compared to previous limits. COMET, which is expected to start data taking in 2026, foresees a two-stage approach: it will start the measurement with a shorter muon beam line to demonstrate the curved-solenoid technique along with physics data-taking, and will then upgrade the facility to the final configuration (Phase-II) by incorporating improvements identified in Phase-I. The target sensitivities of Phase I and II are $3 \times 10^{-15}$ and $3 \times 10^{-17}$ (or lower), respectively.

Mu2e [200] is designed to reach a single-event sensitivity of $3 \times 10^{-17}$. Data collection will take place in two phases with the same detector layout. Run I, which will collect 10% of the full statistics before the FNAL long shutdown in 2028, and Run II, which will complete the data collection starting in 2030. Run I will reach a sensitivity up to 1000 times better than the current limits and will also serve as a test phase to inform detector tuning/optimization, data reconstruction, and analysis strategies for Run II. It will also contribute to preparing an advanced proposal for the design of the experiment upgrade (Mu2e II), which will operate with the new beam from the PIP-2 linac, increasing statistics by another factor of 10 but with more demanding requirements for the detector, target design, and data handling. A new muon facility, AMF [201], which would allow a broad physics program for both $\mu \to e$ conversion and muon decay experiments, has also been suggested for FNAL's PIP-2.

### 5.3.3 Pion physics

Precise low-energy measurements of observables that can be calculated with high accuracy within the SM are powerful probes of NP. In the pion sector, two observables are of particular interest: (i) the ratio $R^{\pi}_{e/\mu} = \Gamma[\pi^+ \to e^+\nu(\gamma)]/\Gamma[\pi^+ \to \mu^+\nu(\gamma)]$, which offers a pristine test of LFU; (ii) the branching ratio of the pion beta decay, $\pi^+ \to \pi^0 e^+ \nu(\gamma)$, which enables a clean extraction of the CKM matrix element $|V_{ud}|$.

In the SM, $R^{\pi}_{e/\mu}$ is among the most precisely predicted quantities involving quarks, with a relative theoretical uncertainty of $1.2 \times 10^{-4}$ [202]. However, the current experimental uncertainty is roughly 15 times larger. Achieving measurements at the 0.01% precision level could probe NP effects in motivated SM extensions [203]. Meanwhile, the current $3\sigma$ tension [15] in the first row CKM unitarity test underscores the need for more precise determinations of $|V_{ud}|$. Achieving per-mille level precision on the pion beta decay would enable a determination of $|V_{ud}|$ with a relative uncertainty at the $10^{-4}$ level, matching the precision obtained with other experimental techniques. These improvements would further enhance the robustness of CKM unitarity tests based exclusively on decays of fundamental particles. The recently approved PIONEER experiment [204], plans to tackle these ambitious measurements using the intense pion beams available at PSI [ID134]. In the near term, PIONEER aims to improve the precision on $R^{\pi}_{e/\mu}$ reducing the uncertainty to the level of the theoretical prediction [ID115]. Over the medium and long terms, the experiment plans to measure the branching fraction for pion beta decay with a target precision of 0.1%, comparable to that of super-allowed beta transitions and anticipated future results from neutron decay measurements [ID190] (see Fig. 5.4).

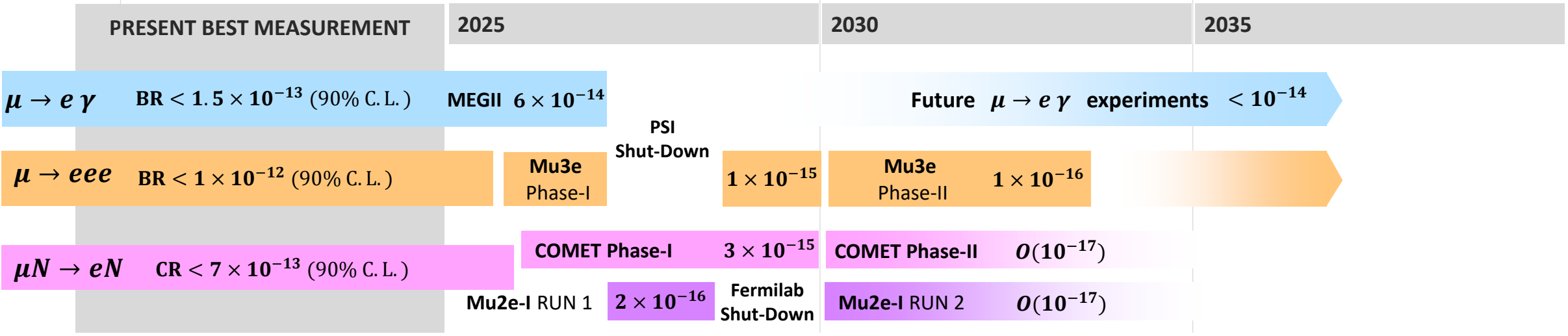

Fig. 5.3: Present best measurements and anticipated sensitivities for the golden cLFV channels of with muons.

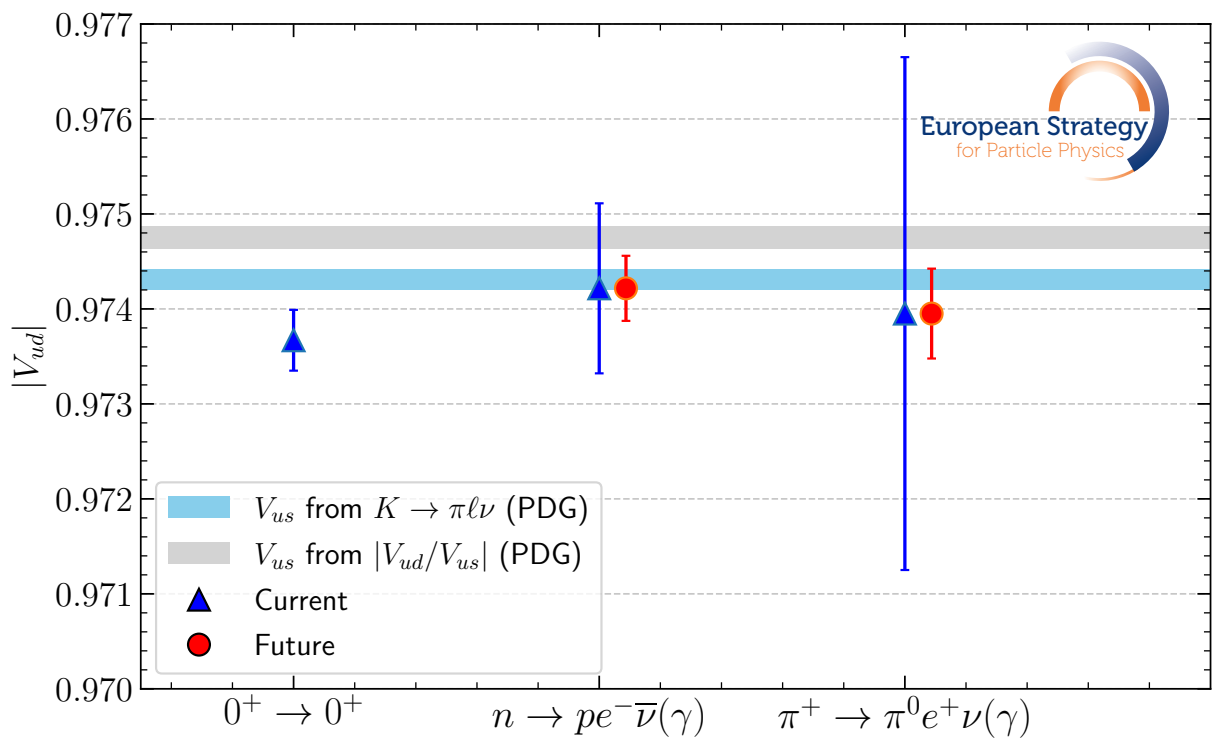

Fig. 5.4: Comparison of the current world-leading measurements of $|V_{ud}|$ with estimates from CKM first-row unitarity tests, along with projected precisions from different experimental techniques.

### 5.3.4 Kaon physics

Rare $s \to d$ FCNC transitions are indispensable for probing the flavour structure of the quark sector. Kaon decays provide the only available experimental access to the processes $s \to d\nu\bar{\nu}$ and $s \to d\ell^+\ell^-(\ell = e,\mu)$. Among these, a few gold-plated observables, dominated by short-distance (SD) dynamics, form the cornerstone of the kaon-physics program: (i) $\mathscr{B}(K^+ \to \pi^+\nu\bar{\nu})$; (ii) $\mathscr{B}(K_L \to \pi^0\nu\bar{\nu})$; (iii) the $K_{L,S} \to \mu^+\mu^-$ interference [ID55]. All of these processes are extremely suppressed and precisely predicted within the SM, making them exceptionally sensitive NP probes. Numerous well-motivated NP scenarios, consistent with current collider constraints, predict sizeable deviations from SM expectations in these channels (see [ID55] and references therein). Observables (i) and (ii) are especially compelling due to their sensitivity to interactions between light quarks ($s$ and $d$) and third-generation leptons ($\nu_\tau$). This unique feature could provide insights into the origin of flavour hierarchies and connects naturally to anomalies in heavy-flavour transitions such as $b \to c\tau\nu$ and $b \to s\nu\bar{\nu}$ (see Fig. 5.5, left). The two rare modes with missing energy are also very stringent probes on possible light dark sectors [205]. Furthermore, kaon physics allows an independent determination of three out of the four CKM parameters, without relying on $B$-physics input [206], providing another powerful test of the CKM paradigm based exclusively on loop-induced kaon processes (see Fig. 5.5, right). The observables (ii) and (iii) violate CP and thus are particularly sensitive to new CPV sources [207].

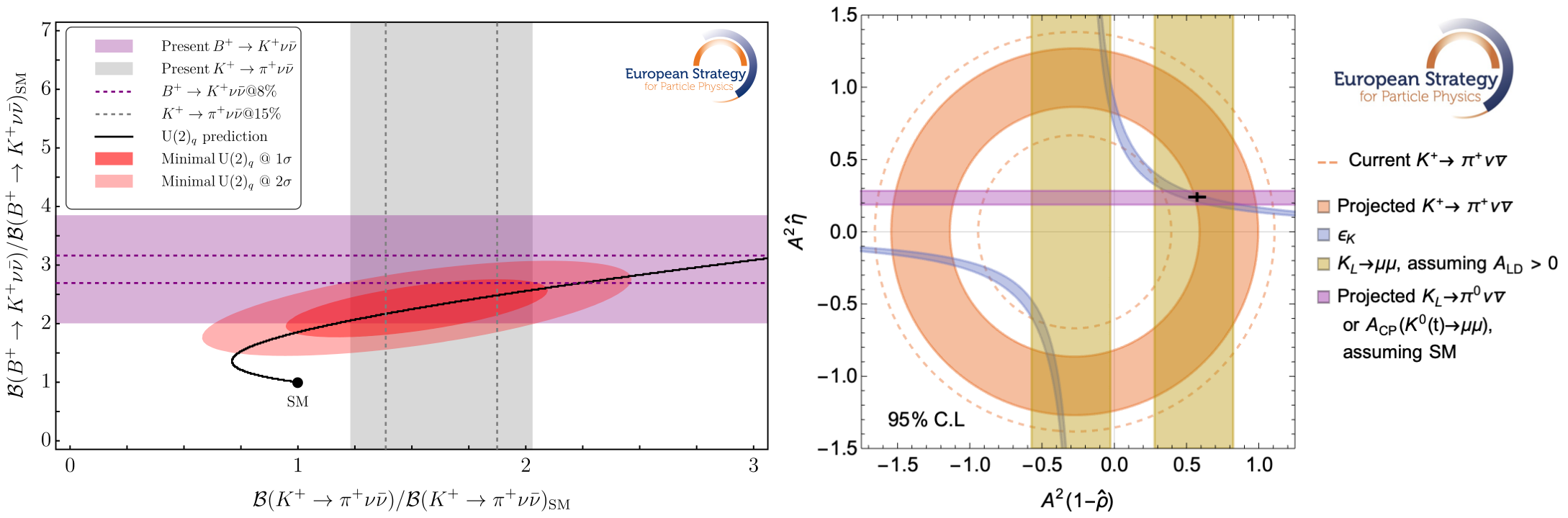

Fig. 5.5: Left: Correlation between $\mathscr{B}(B^+ \to K^+ \nu\bar{\nu})$ and $\mathscr{B}(K^+ \to \pi^+ \nu\bar{\nu})$, normalized to their SM predictions. The red contours denote the regions favoured at $1\sigma$ and $2\sigma$ from a global fit of current data in the limit of minimal $U(2)_q$ breaking [208]; the black line indicates the correlation between the two modes implied by this flavour hypothesis. The grey and purple bands indicate the current experimental constraints; dashed lines highlight future projections from Belle II and NA62. Right: Present and future constraints from kaon observables of the apex of the unitarity triangle defined in the $\{A^2(1-\bar{\rho}),\ A^2\bar{\eta}\}$ plane ($A \equiv |V_{cb}|/|V_{us}|^2$, $\bar{\rho}$ and $\bar{\eta}$ defined as in 5.4.3). Plots adapted from [208] (left) and [206] (right) with the help of the authors.

The above observables have eluded experimental detection for a long time. After decades of searches [209, 210], recently the NA62 experiment at CERN achieved the first observation of $K^+ \to \pi^+\nu\bar{\nu}$, reporting a branching fraction of $13.0^{+3.3}_{-3.0} \times 10^{-11}$ [211], consistent with the SM prediction [212]. With its full dataset, NA62 aims for a $O(15\%)$ precision of $\mathscr{B}(K^+ \to \pi^+\nu\bar{\nu})$ that would significantly constraint various NP scenarios (see Fig. 5.5). Although this is a key result, this level of precision remains well above the SM theory uncertainty of 3%. A next-generation experiment will be essential to match the theoretical precision and fully exploit the discovery potential of this channel. For the $K_L \to \pi^0\nu\bar{\nu}$ mode, whose SM rate is $\approx 3 \times 10^{-11}$ [212], the KOTO experiment at J-PARC has recently pushed its sensitivity down to $2.2 \times 10^{-9}$ [213] and aims to reach a sensitivity better than $10^{-10}$ in the late 2020s. The planned KOTO II aims to perform a 25% measurement at the SM rate in the late 2030s [ID155]. This effort also opens promising opportunities for studying other rare $K_L$ decays in the future, such as $K_L \to \pi^0\ell^+\ell^-(\ell = e, \mu)$ [ID55]. Finally, the $K_{L,S} \to \mu^+\mu^-$ interference can be accessed through measurements of the CP asymmetry in $K \to \mu^+\mu^-$ decays. Within the SM, this asymmetry is directly linked to the CPV phase of the CKM matrix. The LHCb experiment is ideally suited for this challenging measurement, thanks to the large production cross-section of neutral kaons and the capability to implement kaon tagging. A tagged analysis using the full 300 $\mathrm{fb}^{-1}$ dataset expected from LHCb Upgrade II would allow a precision investigation of these interference effects, with projected sensitivity on $|\mathrm{Im}(V_{ts}V_{td}^{*})|$ reaching below 50%, offering a unique test of CPV in the kaon sector [214].

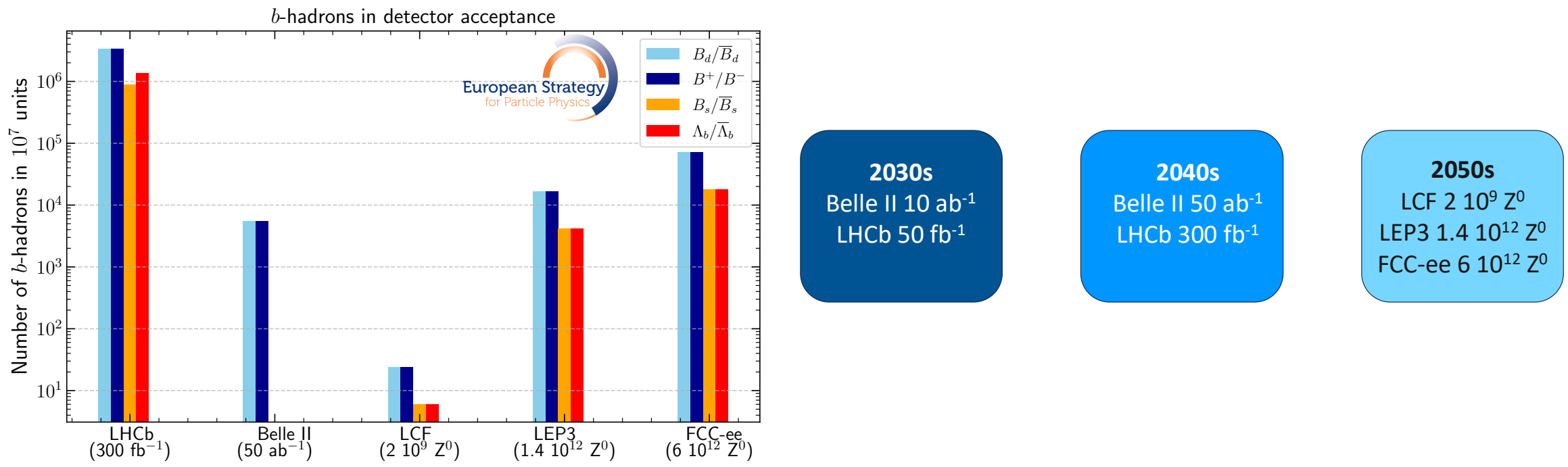

Fig. 5.6: Left: number of $b$-hadron species produced within the detector acceptance for various facilities (note the logarithmic scale). Right: approximate periods for future predictions and their definitions.

Kaon observables provide powerful tests of the SM, offering exceptional sensitivity to the flavour structure of NP through precision measurements of rare decays and CKM parameters. These measurements are a vital component of the global flavour physics programme.

## 5.4 Large-scale facilities

The number of different types of $b$-hadron species produced at the various facilities described in Sect. 5.2.1 is shown in Fig. 5.6 (left). While $e^+e^-$ colliders record the full set of events, only a subset passes the trigger requirements at the LHC. In the following, future projections will be provided for several approximate periods defined in Fig. 5.6 (right).

It is important to note that, in order to fully explore the spectrum of flavour physics at the $Z$ pole, the detectors must satisfy a set of stringent performance requirements [ID141, ID196]. Precise reconstruction of all decay vertices—especially those of the primary interaction point, the $b$-hadron, $c$-hadron and $\tau$ lepton—is essential. Achieving this requires not only the use of high-resolution silicon vertex detectors but also the use of ultra-light detector materials to minimise multiple scattering and bremsstrahlung emission. Excellent tracking and high hermeticity are essential for reducing background levels and ensuring the best achievable reconstruction of decay modes involving missing energy. Effective charged-hadron identification ($p/K/\pi$ separation), along with reliable electron and muon separation, over the full momentum range is critical. Additionally, the electromagnetic calorimeter must have not only excellent energy resolution but also good transverse and longitudinal segmentation.

Some extrapolations exhibit greater uncertainty than others due to their stronger dependence on detector design parameters, which are not yet well defined. However, it should be noted that historically the field has often outperformed its own projections as new analysis and data-processing techniques are developed, allowing the projected sensitivities to be achieved or even exceeded.

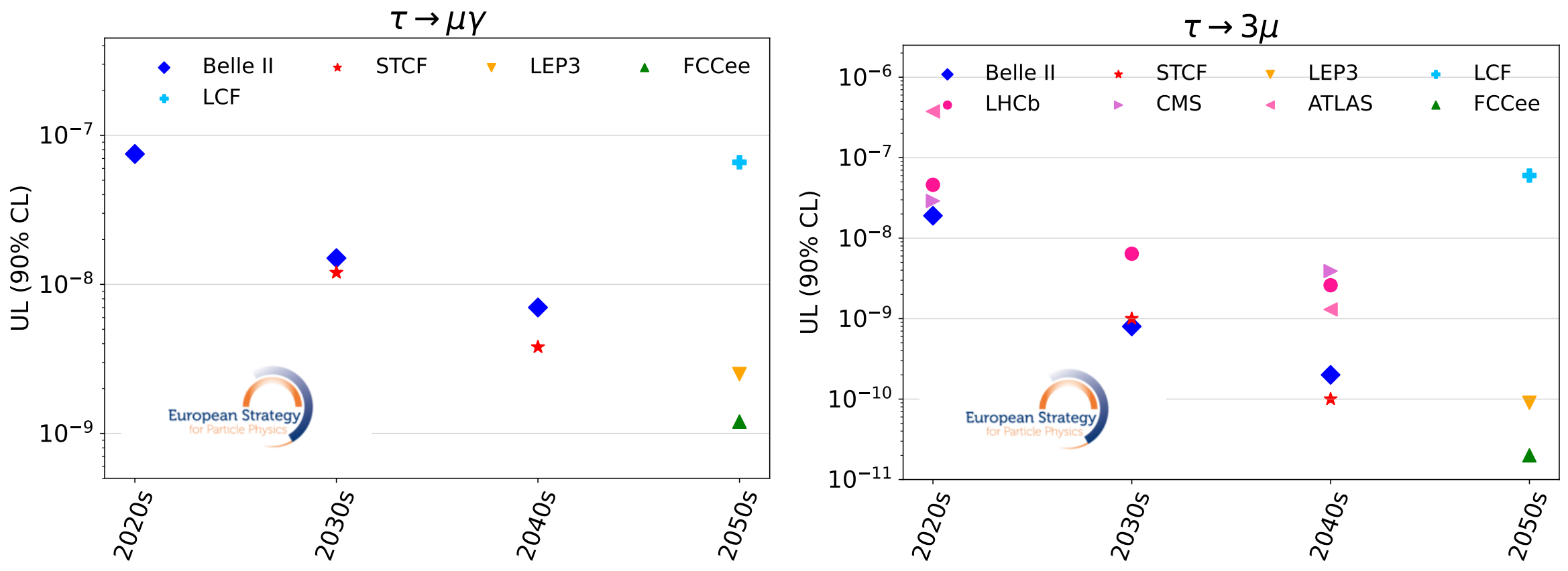

Fig. 5.7: Expected upper limits at 90% on $\mathscr{B}(\tau \to \mu\gamma)$ and $\mathscr{B}(\tau \to \mu\mu\mu)$.

### 5.4.1 $\tau$-lepton physics

The physics of the $\tau$ lepton is a key ingredient in NP searches. It covers a broad range of topics, from forbidden processes to precision measurements. In this section we focus on cLFV searches via the $\tau \to \mu\gamma$ and $\tau \to 3\mu$ channels and on LFU tests in leptonic decays.

Searches for $\tau$ LFV decays are possible at many facilities since, thanks to the absence of a neutrino in the final state, a clear signal can be expected in the invariant mass of the reconstructed decay products. While the $\tau \to \mu\gamma$ decay necessitates good spatial and energy resolution of the calorimeter, which is a limiting factor for the LHC experiments, the $\tau \to 3\mu$ channel can be sought effectively at the LHC as well. In addition, the latter mode is almost background-free, which is not the case for the former that is polluted by $\tau \to \mu\nu\bar{\nu}\gamma$ events. The expected limits at 90% confidence level can be seen in Fig. 5.7. Here the projections are taken from the submissions [ID196, ID223, ID231], except for the LEP3 and LCF facilities that are scaled from the FCC-ee projections according to $\mathscr{L}$ ($\sqrt(\mathscr{L})$) for the $\tau \to 3\mu$ ($\tau \to \mu\gamma$) decay.

LFU tests can be performed by comparing the measured branching fractions of the leptonic $\tau$ decays and their SM predictions, which are functions of the $\tau$ lifetime [215]. The precision of the SM prediction being limited by the knowledge of the $\tau$-lepton mass, the main ingredients to perform LFU tests are thus the $\tau$ lifetime, its mass, $\mathscr{B}(\tau \to e\nu_e\nu_\tau)$ and $\mathscr{B}(\tau \to \mu\nu_\mu\nu_\tau)$. The precisions on those quantities are already, or will soon be, limited by systematic uncertainties such as beam energy scale, alignment and material budget. Consequently, the extrapolations are less straightforward. The projections at future facilities can be seen in Fig. 5.8, for Belle II with 10 ab$^{-1}$ [216] and FCC-ee [ID196], compared with the current status from HFLAV [215]. In particular, Belle II will be able to improve the lifetime and mass measurements, while the precision on the absolute branching fractions will mainly come from $e^+e^-$ colliders running at the $Z$ pole. It should be noted that STCF will also be able to improve the $\tau$-lepton-mass determination thanks to the energy-scan method. Another complementary test of LFU can be performed measuring ratios of branching fractions, such as $R_\mu = \frac{\mathscr{B}(\tau \to \mu\nu_\mu\nu_\tau)}{\mathscr{B}(\tau \to e\nu_e\nu_\tau)}$, that constrains the ratio of the $e$ and $\mu$ effective couplings in leptonic charged currents. With 10 ab$^{-1}$, the Belle II experiment is expected to improve the current measurement by a factor two [216].

The Belle II and STCF experiments will provide the best limits on $\tau$ LFV decays until the Tera-$Z$ experiments enter the game. Those will be able to improve further the constraints by up to one order of magnitude, down to few $10^{-11}$ for $\mathscr{B}(\tau \to \mu\mu\mu)$ and few $10^{-9}$ for $\mathscr{B}(\tau \to \mu\gamma)$. For tests of LFU, with 10 ab$^{-1}$ Belle II will improve the precision of the $\tau$ lifetime by one order of magnitude, down to 0.06 fs, and the $\tau$ mass and $R_\mu$ measurements by a factor 2 to 2.5, down to a precision of 0.06 MeV and 0.2% respectively. The STCF proposal has also a similar projection for the mass, reaching a precision of few tens of keV. The experiments at $e^+e^-$ colliders running at the $Z$ pole will be able to gain at least one order of magnitude on the lifetime and absolute leptonic branching fractions, and a factor $\sim 3$ on the $\tau$ mass. The precision on those measurements will be limited by systematic uncertainties.

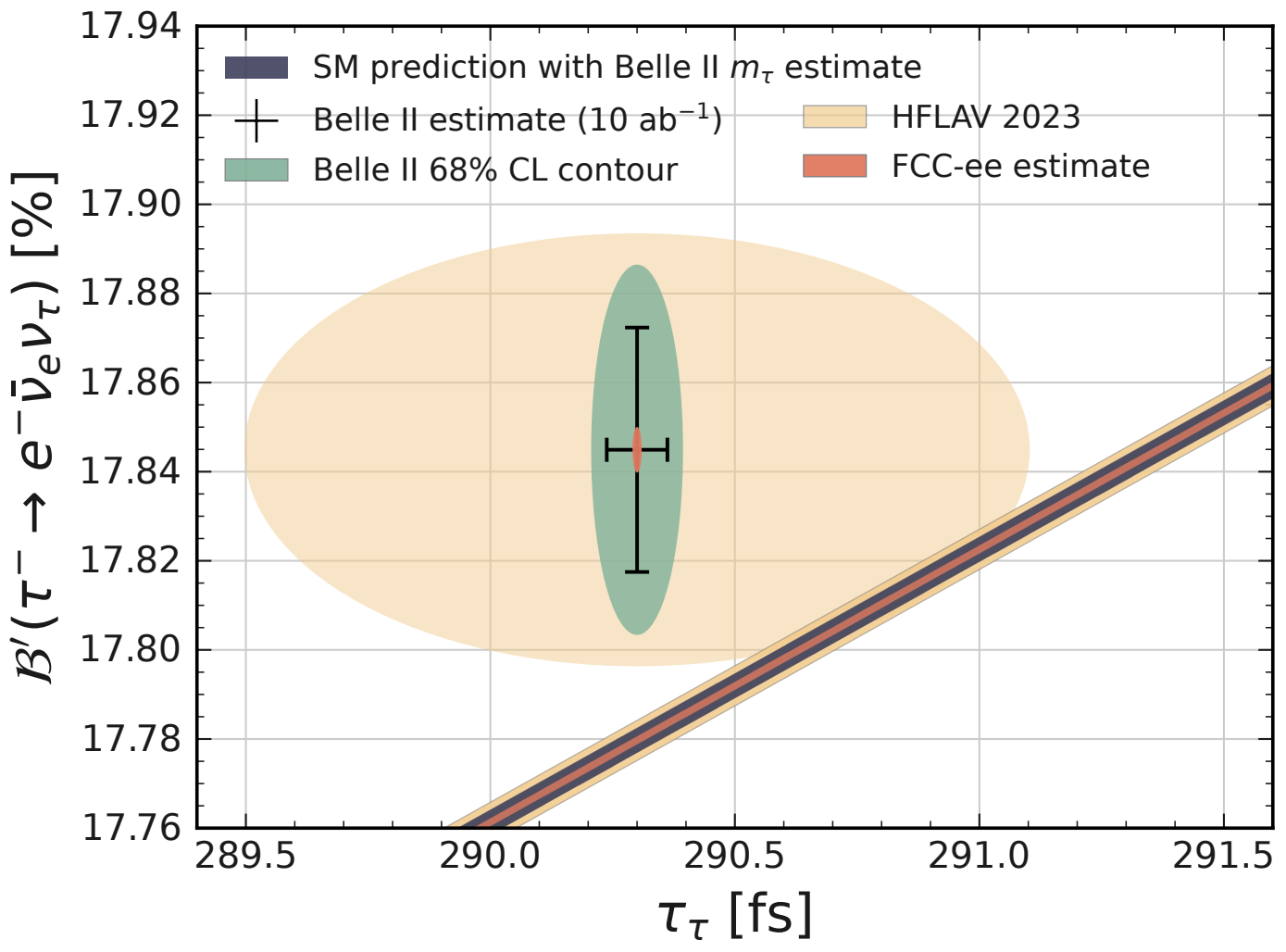

Fig. 5.8: Comparison of the measured $\tau$ leptonic branching fraction and lifetime with the SM prediction, using Belle II expectations for 10 ab$^{-1}$, averaged with other measurements from the PDG [15]. Here, $\mathscr{B}'$ is the average of the direct $\mathscr{B}(\tau \to e\nu_e\nu_\tau)$ measurement with $\mathscr{B}(\tau \to \mu\nu_\mu\nu_\tau)$ assuming electron and muon have similar weak couplings. The latest HFLAV results [215] as well as the projections of FCC-ee are also shown. Figure taken from Ref. [216].

### 5.4.2 Charm physics

Charm physics offers a unique laboratory to test the SM, predominantly due to two distinctive features: the GIM mechanism [217] is extremely effective for charmed hadron decays, and the relevant CKM matrix elements are almost real. This implies that in the SM there is very little room for CPV in the charm system, enabling several powerful null tests. At current facilities, LHCb with its planned Upgrade II [159], [ID81] and Belle II [218], [ID205] show the best prospects for studies of CPV in charm; as for the proposed future facilities, they are FCC-ee [ID196], LEP3 [ID188] and STCF [219], [ID231].

A key question yet to be settled is whether the observation of CPV in $D^0$ decays [220] can be accommodated within the SM or is a sign of NP. Due to difficulties in precisely assessing the size of hadronic effects, no consensus has been achieved yet [221–228]. In addition to future theoretical improvements, insights on this question can be obtained testing approximate SM symmetries such as isospin, see e.g. [229]. The latter requires precise measurements of CP asymmetries in isospin-partner modes, such as $D^0 \to \pi^+\pi^-$, $D^0 \to \pi^0\pi^0$, and $D^+ \to \pi^+\pi^0$. LHCb Upgrade II is expected to measure $A_{CP}(D^0 \to \pi^+\pi^-)$ with an uncertainty of $3.3\times10^{-5}$ [ID223]. For $A_{CP}(D^+ \to \pi^+\pi^0)$, similar sensitivities are expected from LHCb Upgrade II and Belle II, with an uncertainty of $1\times10^{-3}$ and $1.3\times10^{-3}$, respectively [ID223]. The decay $D^0 \to \pi^0\pi^0$ can be fully exploited with an $e^+e^-$ facility: a precision on $A_{CP}(D^0 \to \pi^0\pi^0)$ of $7\times10^{-4}$ is expected from Belle II [ID223], while an additional improvement by a factor 5–10 can be foreseen at FCC-ee [ID196].

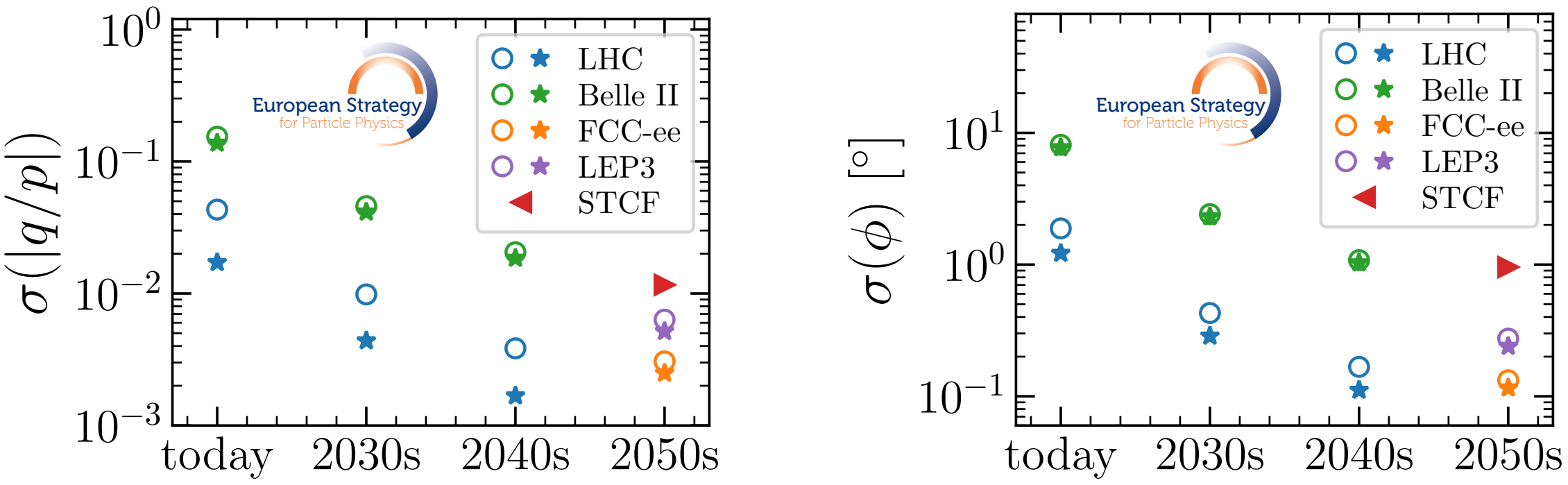

Fig. 5.9: Comparison of current and expected sensitivities on the charm mixing CPV parameters $|q/p|$ (left) and $\phi$ (right). For LHC, Belle II, FCC-ee and LEP3, the precision expected using $D^0 \to K_S\pi^+\pi^-(\pi^0)$ is shown with a coloured dot and the further inclusion of $D^0 \to h^+h^-$ is shown with a coloured star. The red triangle indicates the STCF sensitivities obtained from the combination of $D^0 \to K^-\pi^+\pi^0$, $D^0 \to K_S^0\pi^+\pi^-$, and $D^0 \to K^-\pi^+\pi^-\pi^+$ decays with 5 ab$^{-1}$.

A promising candle for NP searches is the study of CPV in $D^0$–$\overline{D}^0$ mixing: this amplitude is highly sensitive to FCNC couplings in the up-quark sector, and the related CPV observables are powerful null tests of the SM [230,231]. Moreover, the study of charm mixing is fundamental to complement corresponding studies in the neutral $K$ and $B$ systems, as the $D^0$ is the only meson in the up-type quark sector which undergoes mixing. CPV in charm mixing is encoded in the parameters $|q/p|$, which dictates the admixture of flavour states, and $\phi \equiv \arg[q\bar{\mathscr{A}}_f/(p\mathscr{A}_f)]$, the phase of decay-mixing interference in decays to a common final state $f$.

A comparison of the expected sensitivities to the mixing parameters $|q/p|$ and $\phi$, at present and future facilities, is shown in Fig. 5.9. Here, the dots refer to the projections obtained using $D^0 \to K_S^0\pi^+\pi^-(\pi^0)$ modes, for different facilities, whereas the stars account for the further inclusion of $D^0 \to h^+h^-$ decays. The triangle indicates the projections for STCF, obtained from the combination of $D^0 \to K^-\pi^+\pi^0$, $D^0 \to K_S^0\pi^+\pi^-$, and $D^0 \to K^-\pi^+\pi^-\pi^+$ decays, and assuming 5 ab$^{-1}$, corresponding to 5 years of running [ID231]. Currently, the decay $D^{*+} \to (D^0 \to K_S^0\pi^+\pi^-)\pi^+$ yields the highest sensitivity via a decay-time dependent Dalitz analysis. For LHCb and Belle II, today's values correspond to the measurements [232, 233] and the expected sensitivities with 300 fb$^{-1}$ and 50 ab$^{-1}$ are derived from Ref. [ID223]. The projections for FCC-ee and LEP3 have been estimated using the expected number of $D^{*+} \to (D^0 \to K_S^0\pi^+\pi^-)\pi^+$ decays [ID196], and rescaling the LHCb sensitivities accordingly. For Belle II, FCC-ee and LEP3, it is assumed that the decay $D^0 \to K_S^0\pi^+\pi^-\pi^0$ can be used in addition, yielding the same precision as $D^0 \to K_S^0\pi^+\pi^-$ for the same number of reconstructed decays. Including decay-time-dependent analyses of $D^{*+}(D^0 \to h^+h^-)\pi^+$ decays leads to a further constraint on $|q/p|$ and $\phi$. These analyses have been performed both at LHCb [234] and Belle [235], and projections are made based on Refs. [ID223, ID196].

### 5.4.3 CKM and neutral-meson mixing

The CKM matrix encapsulates the structure of quark-flavour mixing and CPV in the SM. Stringent tests of CKM unitarity provide powerful consistency checks that can reveal or constrain physics beyond the SM. Furthermore, many rare processes, which are sensitive NP probes, rely on CKM parameters as inputs, making their continued refinement a long-term priority for the field. The CKM paradigm can be tested using the well-known unitarity triangle (UT) [15]. The apex $(\bar{\rho}, \bar{\eta})$, quantifies the total amount of CPV in the quark sector of the SM and can be independently determined by several different types of measurement.

Measuring the apex with tree-level-dominated processes provides constraints on the CKM angle $\gamma \equiv \arg[-(V_{ud}V_{ub}^*)/(V_{cd}V_{cb}^*)]$ and $|V_{ub}|/|V_{cb}|$. The extraction of $\gamma$ is theoretically very clean and LHCb is expected to set the precision benchmark $\sigma(\gamma) \sim 0.3^\circ$ [ID223]. Sensitivity to the length of the corresponding side, $R_u = |(V_{ud}V_{ub}^*)/(V_{cd}V_{cb}^*)|$, is dominated by measurements of $|V_{ub}|$ which are all measurable with similar $\sim 1\%$, precision at HL-LHC, Belle II and FCC-ee [ID223, ID196]. LQCD will provide the required nonperturbative input with matching precision, see Sect. 5.2.2.

Whilst tree-level measurements set the SM benchmark for the $(\bar{\rho}, \bar{\eta})$ apex, NP can be explored using loop-level dominated processes. Long term prospects for measurements of $\beta \equiv \arg[-(V_{cd}V_{cb}^*)/(V_{td}V_{tb}^*)]$, are driven by the HL-LHC, with comparable precision possible at FCC-ee [ID223, ID196]. Measurements of $\alpha \equiv \arg[-(V_{td}V_{tb}^*)/(V_{ud}V_{ub}^*)]$ requires isospin relations with neutral modes and is thus the remit of the $e^+e^-$ facilities. Belle II could reach sub-degree precision, whilst prospects at FCC-ee are for sensitivities of $\mathcal{O}(0.2^\circ)$ depending on the design of the electromagnetic calorimeter [236]. At this level of accuracy, the precision on $\beta$ becomes limited by penguin pollution [237], while that on $\alpha$ by isospin-breaking effects.

Additional constraints on the apex come from measurements of the $B^0_{(s)}$-$\bar{B}^0_{(s)}$ mixing frequency, $\Delta m_{d,s}$, as well as CPV in $K^0$-$\bar{K}^0$ mixing encoded in the parameter $\varepsilon_K$. Measurements of these quantities are already very precise. For $B$-meson mixing the limiting factor currently comes from LQCD calculations used to interpret the measurements in terms of CKM elements. In the case of $\varepsilon_K$ the error in $|V_{cb}|$ dominates the uncertainty, as discussed further below.

The current status and future prospects for UT measurements are summarised in Fig. 5.10. Nonetheless, the CKM unitarity fit does not convey the entire picture. It is still possible, even likely, that NP would arise at loop level but that due to phase cancellations in ratios does not appear as an inconsistency in the $(\bar{\rho}, \bar{\eta})$ plane. Fig. 5.11 (left) shows future possible constraints on the magnitude of generic NP amplitudes in $B^0_{(s)}$ mixing diagrams [238]. The $B^0_s$ mixing phase $\phi_s$, is often considered a distinctive probe of NP models, due to its small size and precise prediction in the SM. The HL-LHC will improve the current precision by a factor of 5 with prospects for FCC-ee giving an additional factor of 2, see Fig. 5.12 (left). As for the $\beta$ angle, penguin pollution could become the limiting factor [237].

The semileptonic CP asymmetries in $B^0$ and $B^0_s$ mixing, referred to as $a_d^{sl}$ and $a_s^{sl}$, are expected to be very small in the SM and are predicted with high accuracy [239], well below the current experimental sensitivities as shown in Fig. 5.12. Provided that uncertainties related to the charge detection asymmetries are kept under control, LHCb will be able to improve the current sensitivity by an order of magnitude. An $e^+e^-$ collider at the $Z$-pole has two main advantages: the absence of production asymmetry (particularly important for $a_d^{sl}$), and reduced charge detection asymmetries. Ranges of the estimates for the sensitivity at FCC-ee are shown in 5.12 as a region spanning the most [238] and least [130] optimistic scenario. In the case of $a_d^{sl}$, the expected experimental sensitivity will be at the same level as the current theory prediction, and could be enough to observe CPV in $B^0$-$\bar{B}^0$ mixing at the SM level.

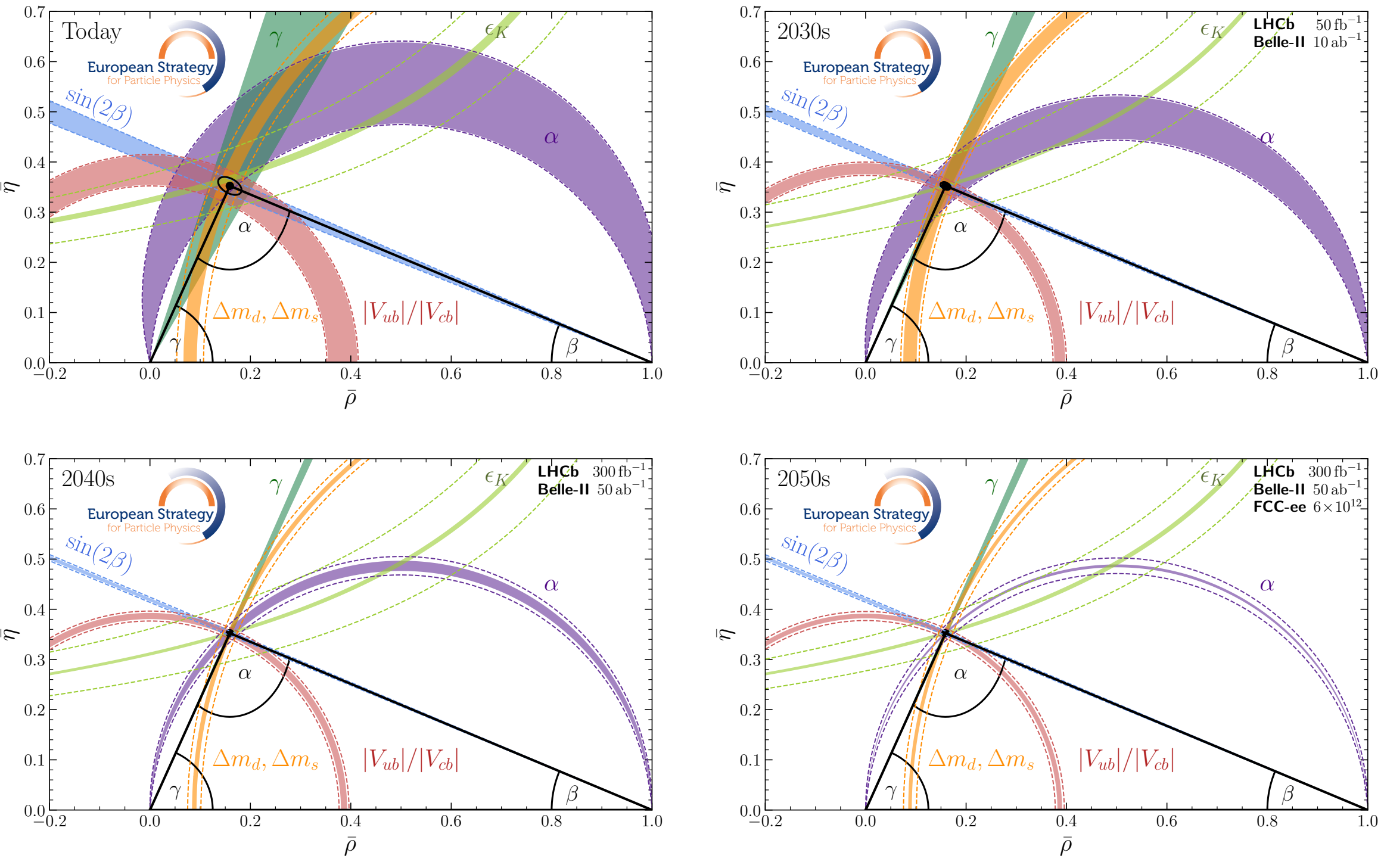

Fig. 5.10: CKM unitarity triangle today and for projections in 2030s, 2040s and 2050s. The central values are selected to correspond to those obtained from the current SM fits. The coloured regions take into account the precision of the experimental measurements only. The dashed lines also account for theory related uncertainties that arise when interpreting the experimental measurements in terms of the UT parameters.

The CKM matrix element $|V_{cb}|$ enters the UT in the sides $R_u$ (via the ratio $|V_{ub}|/|V_{cb}|$) and $R_t = |(V_{td}V_{tb}^*)/(V_{cd}V_{cb}^*)|$ (via the extraction of CKM parameters from $B$ mixing), and in the constraint from $\varepsilon_K$. Presently it is determined from semileptonic $b \to c\ell^-\bar{\nu}$ processes. There is a significant discrepancy between the inclusive and exclusive (typically $B \to D\ell^-\bar{\nu}$) determinations, which leads to an uncertainty above 5%. The issue may be resolved with future developments in LQCD predictions, see Sect. 5.2.2.

A more direct way to determine $|V_{cb}|$ is via on-shell $W$ decays to charmed and beauty jets. The anticipated $W$ boson samples at future $e^+e^-$colliders will achieve statistical uncertainties on $|V_{cb}|$ well below 1%. The critical item however is the uncertainty on the jet flavour-tagging efficiency [240, 241]. Fig. 5.11 shows the uncertainty on $|V_{cb}|$ versus that of the tagging efficiency. The jet tagging efficiency will be calibrated at the $Z$ pole, which in all facilities provides samples at least an order of magnitude larger than for the $W$ boson. Even with relatively poor understanding of the tagging efficiency, a significant improvement on $|V_{cb}|$ is to be expected, while in optimistic scenarios precisions around 0.2% can be achieved. The size of the $W$-boson sample plays a subdominant role; for tagging-efficiency uncertainties beyond 0.5% the systematic uncertainty dominates. The same argument applies to $|V_{cs}|$, whose present uncertainty of 0.6% dominates the uncertainty in the unitarity relation with processes involving strange quarks [15]. In this case ultimate precisions around $10^{-4}$ can be anticipated [240].

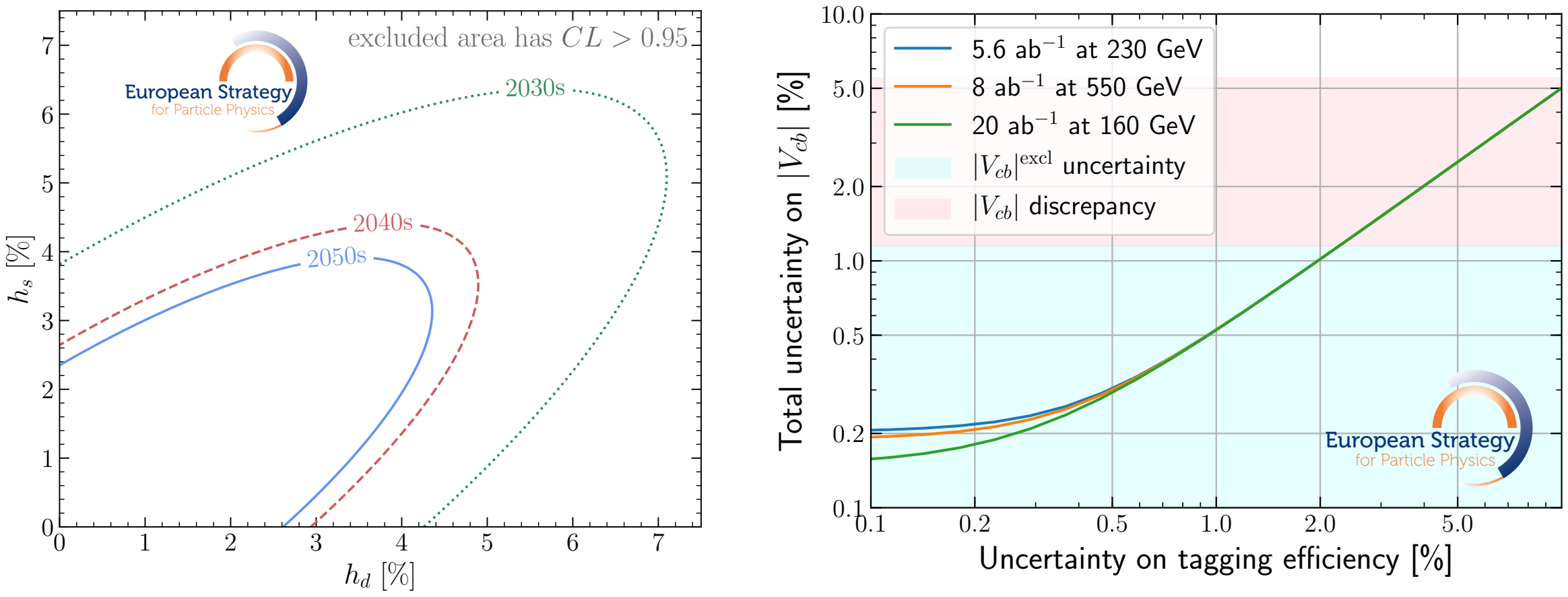

Fig. 5.11: Left: Prospects for constraints on generic NP contributions in $B^0$ and $B_s^0$ meson mixing. Here $h_{s,d}$ denote the modifications of the mixing amplitudes, in modulo, normalised to the corresponding SM values (the SM limit is recovered for $h_{s,d} = 0$). Figure adapted from Ref. [238] (with updated inputs). Right: Uncertainty on $|V_{cb}|$ versus uncertainty on jet flavour tagging efficiency for selected running scenarios [ID188, ID140, ID233]. The coloured bands correspond to the present uncertainties from exclusive semileptonic decays, and the discrepancy with the inclusive determination [54]. Figure adapted from Ref. [240].

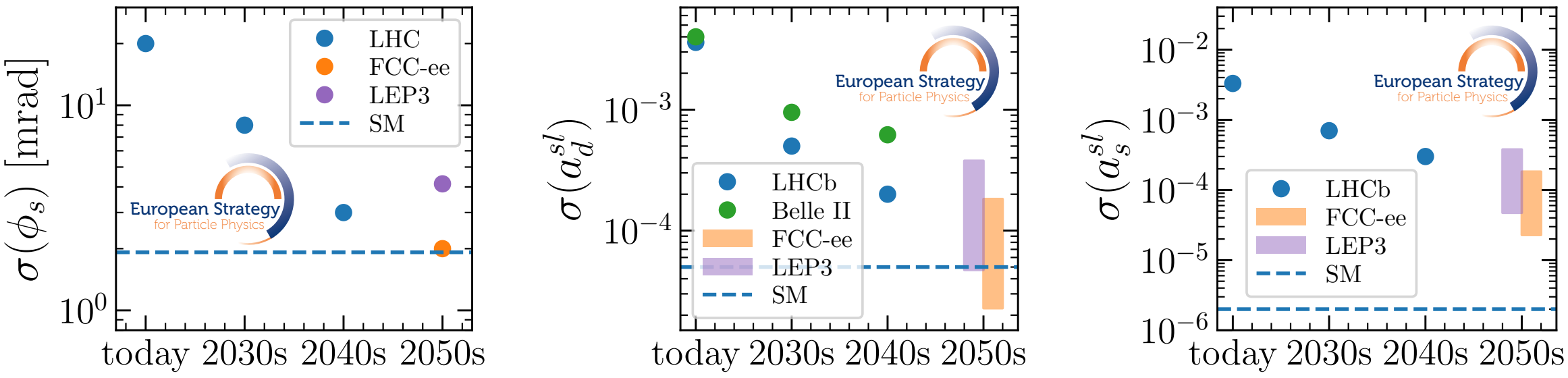

Fig. 5.12: Current and expected precision on the CPV phase $\phi_s$ (left) and the semileptonic asymmetries $a_{sl}^d$ (middle) and $a_{sl}^s$ (right) which characterise CPV in $B^0$ and $B_s^0$ mixing.

At an $e^+e^-$ collider with energies beyond the top threshold, one can determine $|V_{ts}|$ directly from $t \to sW^-$ decays. With 3 ab$^{-1}$ at the top threshold, one expects to obtain $|V_{ts}|$ with 3.1% relative precision [242], [ID233], which would correspond to 2.6% with 4.3 ab$^{-1}$ [ID78], or 2% for 8 ab$^{-1}$ at $\sqrt{s} = 550\,$GeV [ID140].

A further complementary strategy to perform precise measurements of CKM elements is via a dedicated neutrino deep-inelastic experiment at the muon collider [ID207]. This strategy could lead, in particular, to a measurement of $|V_{ub}|$ with sub-percent precision.

### 5.4.4 Tests of LFU and FCNC studies using $B$ decays

Rare decays of beauty and charm hadrons cover many decay modes accessing a wide range of observables. These include several stringent tests of the SM, with precise theoretical predic-tions, as well as null SM tests. In addition (semi)-leptonic decays to leptons of different flavours can probe lepton universality. Current measurements are largely limited by statistics. The yields to be collected in future from both current and future flavour projects will enhance sensitivity to key observables, enable additional measurements, and support new analysis strategies.

Decay modes with $\tau$ leptons in the final state are particularly interesting as they can probe SM extensions with enhanced couplings to the third generation. In this case the LFU test variable, called $R(H_c)$, is defined as the ratio of the branching ratios $R(H_c) \equiv \frac{\mathscr{B}(H_b \to H_c \tau^+ \nu_\tau)}{\mathscr{B}(H_b \to H_c \ell^+ \nu_\ell)}$ where $H_b$ ($H_c$) is a beauty (charmed) hadron and $\ell$ represents a light lepton. Experimental and theoretical uncertainties cancel in such ratios to a large extent. Averages of $R(D)$ and $R(D^*)$ from different experiments, measured in several $B$ and $D$ decays modes exceed the SM predictions by about 3.8 $\sigma$ [215]. Current measurements have a precision around 6–14%. In the next 15 years an improvement by a factor about 10 is foreseen from LHCb Upgrade II and Belle II [ID223], as shown in the left side of Fig. 5.13. Studies performed for FCC-ee considering $B_s^0 \to D_s^{(*)-} \ell^+ \nu_\ell$, using hadronic tau decays and $D_s^- \to \phi(K^+K^-)\pi^-$ final states, predict the uncertainty on $R(D_s^{(*)})$ at the permille level [243]. Thanks to progress from LQCD (cf. Sect. 5.2.2) it is envisaged that the uncertainty on the SM prediction can reach the same level of precision in the 2050s. New possibilities for LFU tests will be opened in $b \to u$ charged-current decays by high statistics samples of leptonic decays to be collected at Belle II, they will complement $b \to c$ tests with sensitivity to different couplings. The ratio $\mathscr{B}(B^+ \to \tau^+ \nu_\tau)/\mathscr{B}(B^+ \to \mu^+ \nu_\mu)$ will be determined with a precision below 10% [ID223]. Future $e^+e^-$ colliders can improve further the precision, in particular for measurements using hadronic $\tau$ decays.

FCNC transitions, allowed only at loop level in the SM, are sensitive to virtual contributions of new mediators up to energy scales much higher than those reached by direct production. In $b \to s\ell^+\ell^-$ decays, LFU between the first and second lepton generation is probed by the ratio $R_H$, defined as $R_H \equiv \int \frac{d\Gamma(H_b \to H_s \mu^+ \mu^-)}{dq^2} dq^2 / \int \frac{d\Gamma(H_b \to H_s e^+ e^-)}{dq^2} dq^2$. Here $H_s$ indicates a strange hadron, and $q^2$ is the di-lepton invariant mass squared. Current measurements are in agreement with the SM, but the expected strong improvements in precision, as shown in Fig. 5.13, will allow much more stringent tests [244], [ID223]. The current SM prediction on $R_K$ has an uncertainty of $\approx 1\%$ [245, 246]; however, it can be expected to decrease to O(0.1%) by 2040.

Various results on rates and angular observables $b \to s\ell^+\ell^-$ decays are in tension with SM predictions (see [247–251]), making improved understanding of high importance; however, the comparison with theory predictions is currently hindered by the evaluation of QCD contributions (charm loops). With high statistics it will be possible to perform new measurements to sort out the tensions, such as a full angular analysis in narrow $q^2$ bins with the determination of all angular parameters, or directly of the Wilson Coefficients. At HL-LHC and $e^+e^-$ colliders comparisons between different channels will be performed, including $B_s^0$ and $\Lambda_b^0$ decays, and studies of time-dependent CPV, for example in $B_s^0 \to \phi\mu^+\mu^-$ decays [252]. Improved theory predictions are also expected in the coming years, including from LQCD (see Sec. 5.2.2).

Decays involving tau pairs such as the $B^{+,0} \to K^{+,*0}\tau^+\tau^-$ decays have not been observed yet. They are of particular interest since they allow to test a FCNC process involving third-generation leptons, complementing the present measurements in light-leptons modes. The current branching ratio limit is $\mathscr{B}(B^0 \to K^{*0}\tau^+\tau^-) < 1.8 \times 10^{-3}$ at 90% CL [253]. An improvement in sensitivity of a factor larger than 10 is expected from Belle II with 50 ab$^{-1}$ [ID223] and up to another two orders of magnitude at future $e^+e^-$ colliders. Achieving the precision neces-sary to observe this decay at the SM level ($\mathscr{B} \sim 10^{-7}$) would require approximately $6 \times 10^{12}$ $Z$ bosons, with the exact sensitivity being highly dependent on the detector performance [ID196].

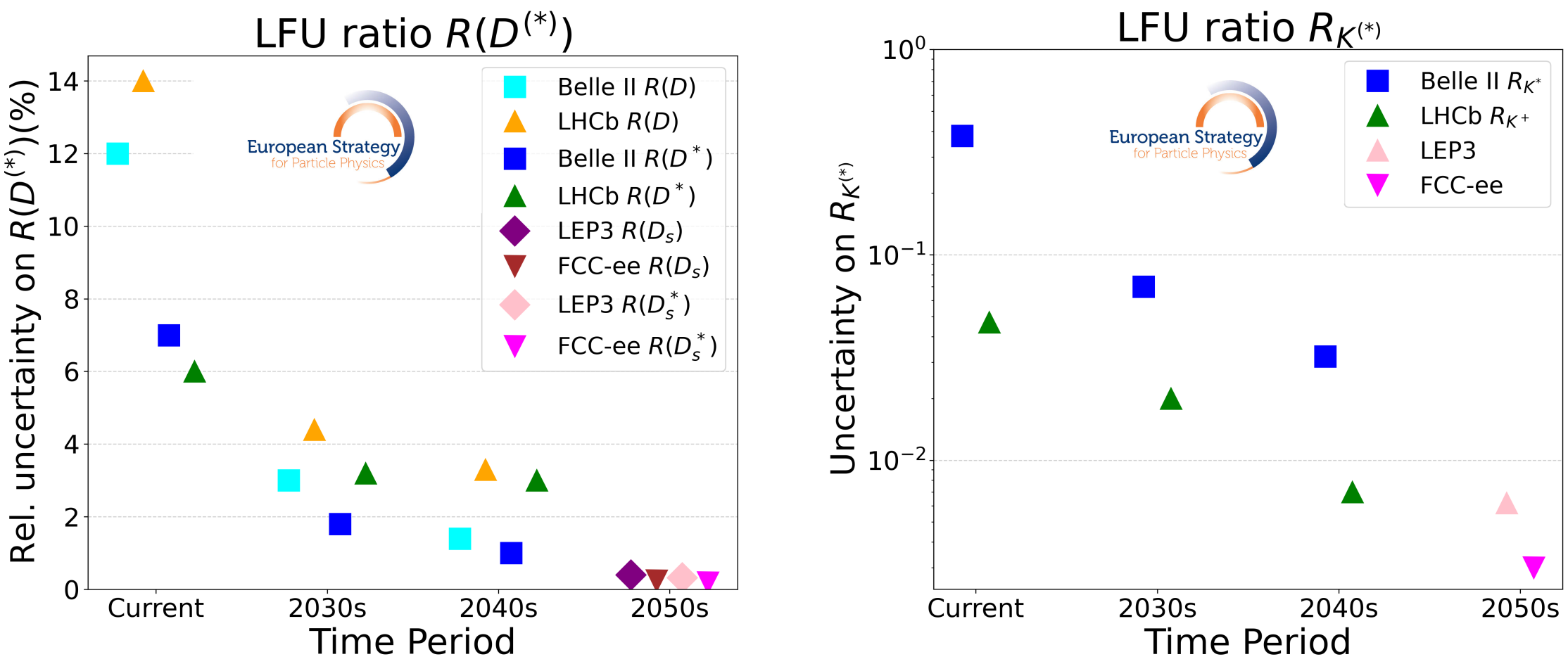

Fig. 5.13: Current and expected relative uncertainty on the LFU ratios (left) $R(D^*)$ and $R(D)$, and (right) $R_{K^{*0}}$ and $R_{K^+}$, in the $1.1 < q^2 < 6\ \mathrm{GeV}^2$ range, for different experiments.

Future data will open new opportunities for probing NP in FCNC processes with $b \to s\nu\bar{\nu}$ modes. These decay modes are of particular interest as they are free from long-distance charm-loop contributions and are predicted with high precision within the Standard Model. The first evidence of this decay has been obtained recently by Belle II as $\mathscr{B}(B^+ \to K^+\nu\bar{\nu}) = (2.3 \pm 0.5^{+0.5}_{-0.4}) \times 10^{-5}$, which is $2.7\sigma$ above the SM prediction [254]. In the longer term, the best precision for measurements at $e^+e^-$ colliders is expected with the $K^{*0}\nu\bar{\nu}$ mode. The $B^0 \to K^{*0}$ form factors will be determined from LQCD with an uncertainty at the percent level. The trend on the expected uncertainty on these decays is shown in Fig. 5.14-right. Other modes such as $B^0 \to K_S\nu\bar{\nu}$, $B_s^0 \to \phi\nu\bar{\nu}$ and $\Lambda_b^0 \to \Lambda\nu\bar{\nu}$ will also be accessible and will complement the exploration of $b \to s\nu\bar{\nu}$ transition [255].

One of the most theoretically clean and sensitive probes of NP are the $B^0_{(s)} \to \mu^+\mu^-$ decays. Being CKM, loop- and helicity-suppressed, they are predicted to be very rare in the SM: $\mathscr{B}(B_s^0 \to \mu^+\mu^-) = (3.66 \pm 0.14) \times 10^{-9}$ and $\mathscr{B}(B^0 \to \mu^+\mu^-) = (1.03 \pm 0.05) \times 10^{-10}$ [256]. While the $B_s^0$ decay is measured to be in agreement with SM with a precision of the order of 8% [15] the $B^0$ mode has not yet been observed. Measurements at LHC are performed by the ATLAS, CMS and LHCb experiments, with an uncertainty that is expected to be 4% and 12% for $B_s^0$ and $B^0$, respectively, at the end of HL-LHC [ID223]. Future $e^+e^-$ colliders will not surpass these precisions, due to the lower number of $B$ mesons produced [ID196]. The foreseen precision for the measurement of the $B^0$ branching fraction is shown in Fig. 5.14 (left). Additional variables related to the $B_s^0 \to \mu^+\mu^-$ decay will also become precisely measured and will provide more comprehensive tests. The effective lifetime [257] allows the degeneracy between any possible contribution from new scalar and pseudoscalar mediators to be broken. The current precision of 23–30% will decrease to 5% [ID223]. At future $e^+e^-$ colliders, high flavour tagging efficiency (of order of 25%), together with excellent mass resolution, will help the time-dependent measurement of CPV. The precision on the measurement of mixing-induced CPV is expected to be comparable, and of the order of 20%, at the end of the full LHCb program

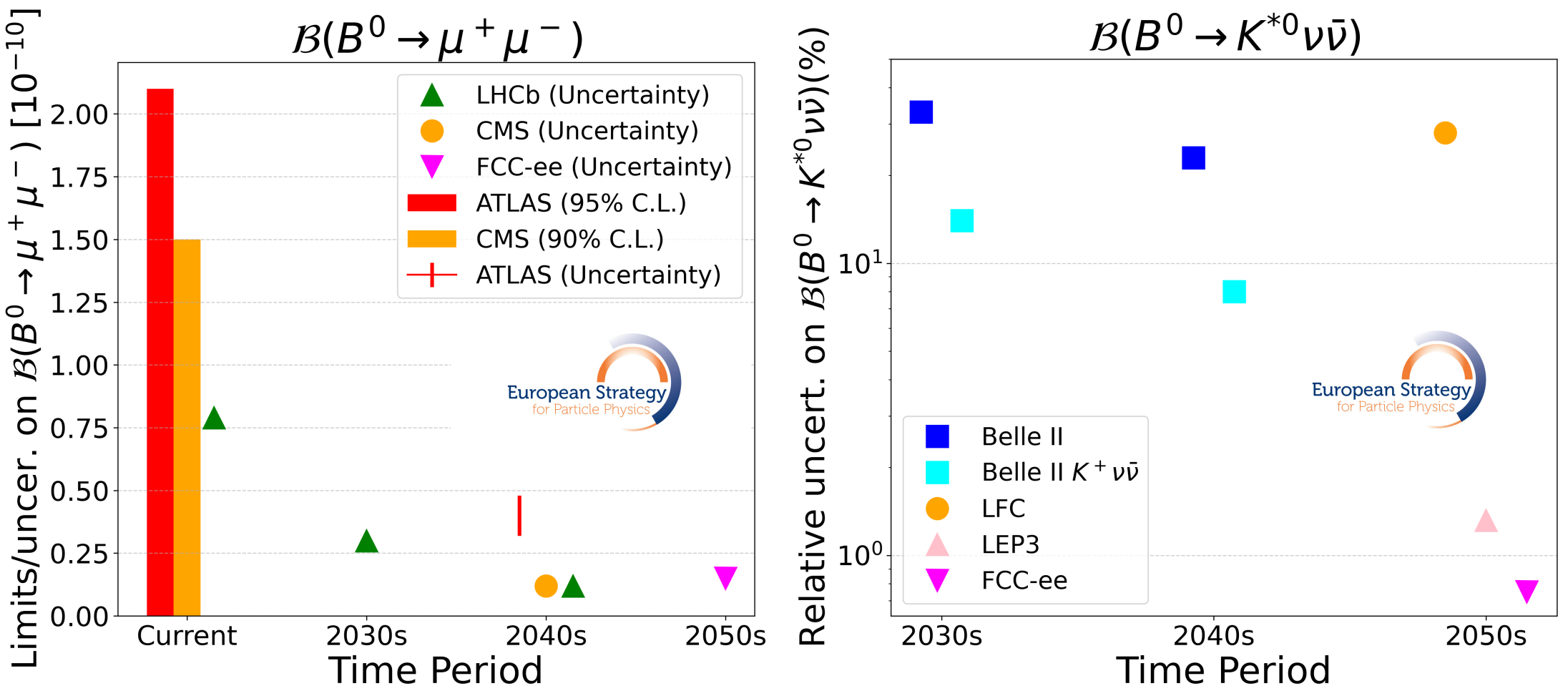

Fig. 5.14: Left: current and expected uncertainty on $\mathscr{B}(B^0 \to \mu^+\mu^-)$, for different experiments [ID223, ID196]. Right: expected relative uncertainty on $\mathscr{B}(B^0 \to K^{*0}\nu\bar{\nu})$.

and the FCC-ee $Z$ pole running.

Radiative FCNC processes offer further complementary SM tests. Due to the V−A nature of the weak interactions, the photon emitted in $b \to s\gamma$ decays is predominantly left-handed in $b$ decays (and right-handed in $\bar{b}$ decays). The polarisation of the photon can be probed in several complementary ways, with different observables [ID223]. Precise information is coming from LHCb analyses of the angular distributions of $B^0 \to K^{*0}e^+e^-$ decays at very low $q^2$, and mixing-induced CPV in $B_s^0 \to \phi\gamma$ and Belle II analysis of $B^0 \to K_S\pi^0\gamma$ decays. The current measurements are in agreement with the SM predictions with a precision of the order of 5%. In the 2030s, experimental precision is expected to reach the percent level, which corresponds to the level of accuracy currently achieved by theoretical predictions [258–260]. Since the measurements in the 2030s will be limited by statistical uncertainties, the full Belle II and LHCb datasets are expected to yield further improvements in precision, provided that theoretical uncertainties are correspondingly reduced. Similarly, the clean environment of future $e^+e^-$ colliders at the $Z$ pole will be beneficial to the precise understanding of radiative FCNC decays.

## 5.5 Impact of future programmes

As discussed in the previous sections, future projects could lead to major improvements, often exceeding one order of magnitude, in many of the observables probing the different processes in Table 5.1, as well as on the electron and neutron EDMs. Analysing the potential impact of these measurements requires making hypotheses about the UV completion of the SM. Two main scenarios are discussed below: high-scale NP with generic $O(1)$ flavour-violating couplings and SM extensions with approximate flavour symmetries and TeV-scale dynamics.

### 5.5.1 General bounds on high-scale NP

The expected bounds on the effective NP scale from selected observables are shown in Fig. 5.15. For each observable the bound on the coefficient $C_i$ of the dimension-six operator most con-strained by that observable is reported,[1] assuming $C_i = 1/\Lambda_i^2$. This functional form of the couplings is purely conventional, but it provides a qualitative indication of the high-scale sensitivity of these observables. Particularly noteworthy are the expected improvements in EDMs and LFV measurements, where the scales probed by future projects will exceed the current ones by one order of magnitude or more.

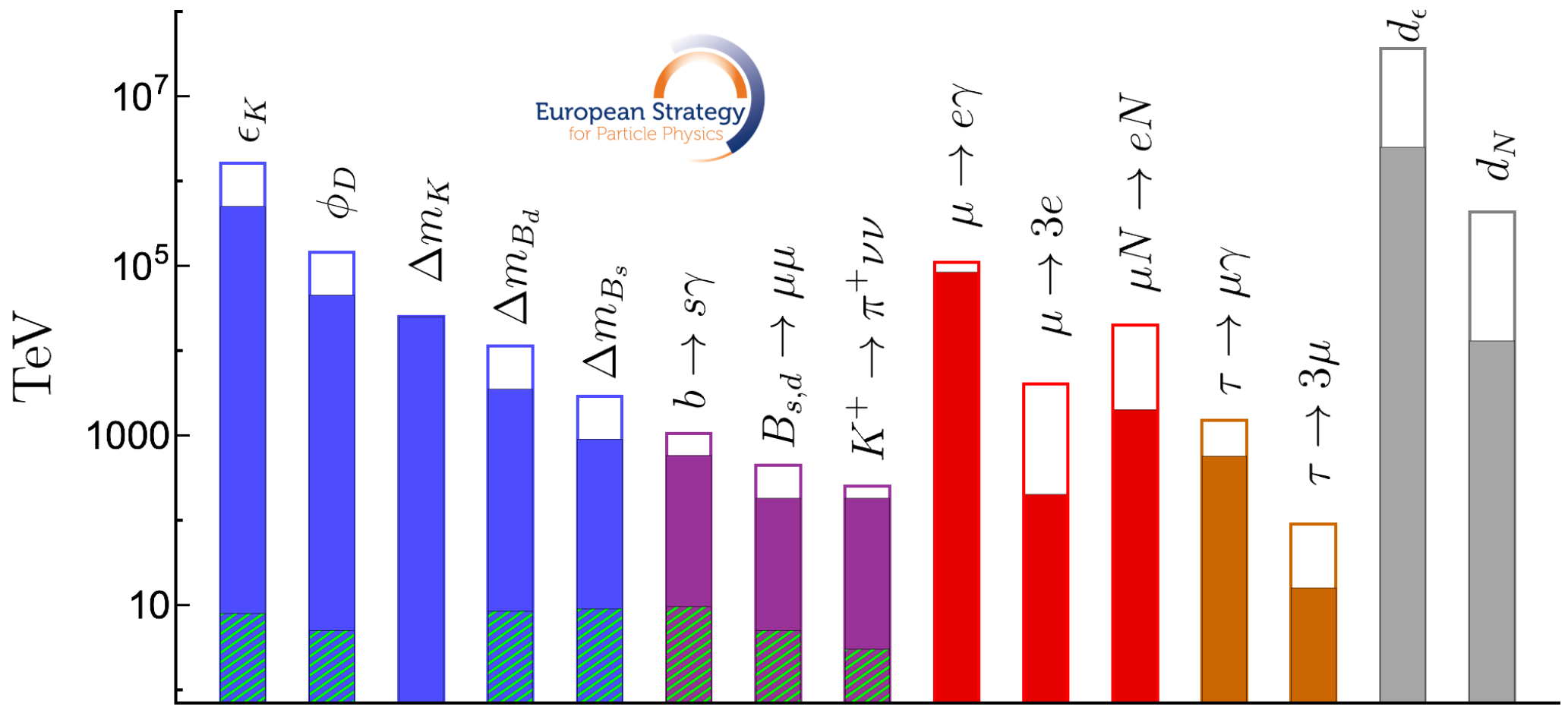

Fig. 5.15: Bounds on the coefficients $C_i$ of dimension-six operators constrained by selected flavour observables, expressed as bounds on the effective scale ($C_i = 1/\Lambda_i^2$). The coloured bars denote current limits, while the empty bars the expected improvements in the best scenario (assuming, in particular, FCC-ee for $b$ and $\tau$ observables). For quark-violating observables the hatched bars indicate current bounds assuming a MFV structure (see text).

The drawback of EDMs and LFV is that it is difficult to define a minimum value for the coupling. The situation is different in the quark sector, where the Yukawa couplings do provide an irreducible source of flavour violation in any consistent extension of the SM. The bounds obtained under the Minimal Flavour Violation (MFV) hypothesis [261–263], i.e. assuming the Yukawa couplings are the only accessible sources of flavour violation, are shown with hatched lines in Fig. 5.15. As can be seen, these bounds are as high as a few TeV, reaching in a few cases up to 10 TeV: this demonstrates that quark FCNCs currently provide constraints competitive with direct searches even in NP models with minimal flavour-violating couplings.

Concerning FCNCs and meson mixing, future improvements depend, in large part, also on the expected improvements in determining the CKM matrix elements. The best example in this respect is $\varepsilon_K$, whose current SM uncertainty, around 20%, is dominated by the parametric error on $|V_{cb}|$ (mainly due the existing tension between exclusive determinations of $|V_{cb}|$). Thanks to the precision on $|V_{cb}|$ expected from $W$ decays (see Sect. 5.4.3), as well as progress from LQCD (see Sect. 5.2.2), this could drop below 3%, with a gain of a factor 3 in the scale of NP probed.

### 5.5.2 Impact in NP models with motivated flavour structure

A realistic example of TeV-scale dynamics is provided by NP that couples mainly to the third-generation and is invariant under a $U(2)^5$ flavour symmetry acting on the light generations [265, 266]. The bounds on all the SMEFT operators containing third-generation quark and lepton doublets within this general setup—considering separately flavour, electroweak (EW), and high-energy observables (Drell-Yan processes and heavy-quark production in $pp$ collisions)—are shown in Fig. 5.16 in the two reference cases of up or down alignment. As can be seen, the completion of the HL-LHC and Belle-II programs would imply major improvements of the flavour bounds, and a further major improvement is possible later on with FCC-ee. Note also that in some cases flavour bounds exceed the electroweak ones, while in other cases it is the opposite: a general feature that holds in many realistic SM extensions with new TeV-scale dynamics [266, 267].

[1] In the cases of $\mu \to 3e$ and $\tau \to 3\mu$, the most constrained operators are the corresponding dipole LFV terms. However, since these operators are already strongly limited by $\mu \to e\gamma$ and $\tau \to \mu\gamma$, we instead report the bounds on the corresponding four-lepton LFV contact interactions.

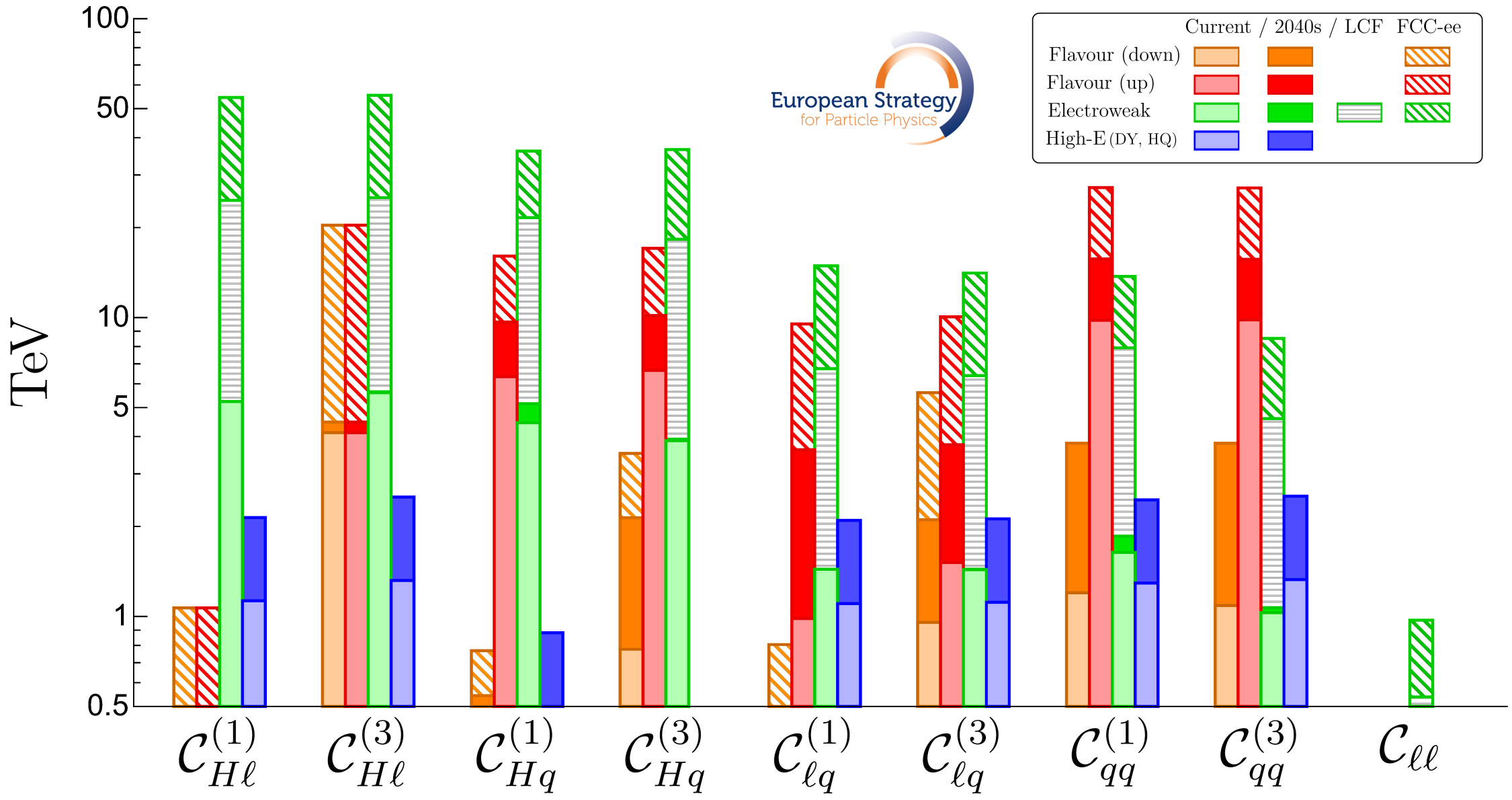

Fig. 5.16: Bounds on coefficients of selected dimension-six SMEFT operators from flavour, EW, and high-energy observables (Drell-Yan and heavy-quark production). The operator notation follows the Warsaw basis [264] with implicit flavour structure consisting only of 3rd generation fields (with up or down alignment in the quark case). The bounds are obtained considering one operator at a time and analysing separately the three different sets of observables. Projections for all the bounds at the end of HL-LHC and Belle-II are shown. For flavour and EW observables only, two different future scenarios (LCF and FCC-ee) are considered.

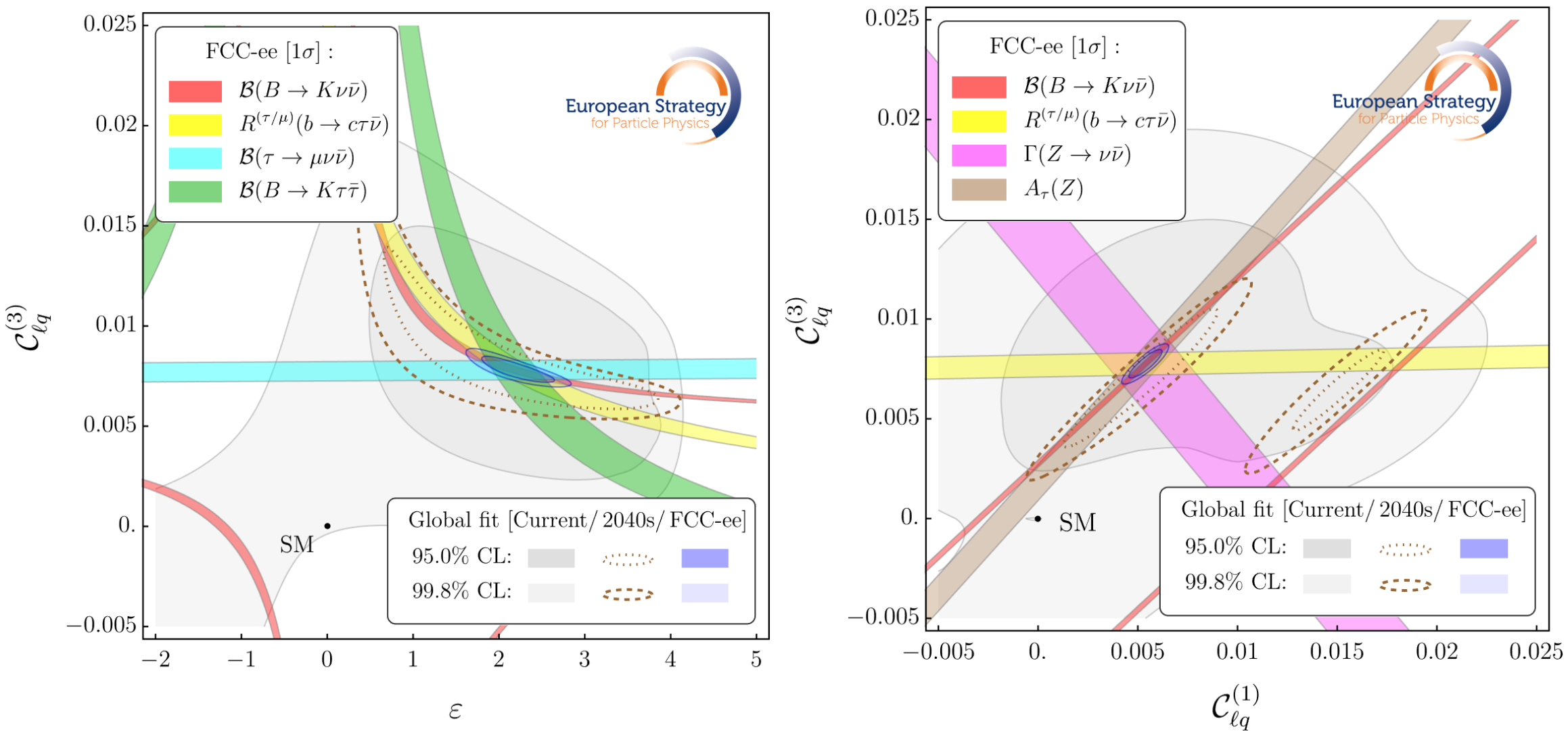

Fig. 5.17: Combined fit of $R_{D^{(*)}}$ and other flavour, EW, and collider observables sensitive to the semileptonic operators $O^{(1)}_{\ell q}$ and $O^{(3)}_{\ell q}$, with third-generation fields and flavour-mixing parameter $\varepsilon$ (see text). The gray areas denote the preferred region for the model parameters from the current global fit. Future projections for flavour and EW observables at the end of HL-LHC and Belle-II (2040), and after FCC-ee, assuming central values for the observables from the current best fit point and projected uncertainties, are also shown. The individual constraints from the most relevant observables at FCC-ee are indicated with coloured bands.

The best way to illustrate the interplay of flavour and EW observables in indirect searches, is to analyse them assuming a non-vanishing NP signal. The current tension between data and SM predictions in $R(D^{(*)})$ provides a useful testing ground in this respect. This tension can be described within a minimal setup consisting in the two third-generation semileptonic operators $O_{\ell q}^{(1,3)}$ discussed in the context of Fig. 5.16. To fit current data it is also necessary to introduce a parameter, $\varepsilon$, to describe the misaligment of the quark doublet in these operators ($q_3$) from pure down-alignment ($q_3 = q_b + \varepsilon |V_{ts}| q_s$, where $q_{b(s)}$ are the $SU(2)_L$ fields associated to $b(s)$ quarks). The setup is thus described by three parameters: the two Wilson coefficients and $\varepsilon$ (see Refs. [208, 268] for more details). The current preferred region for the model parameters and future prospects (from HL-LHC and Belle-II, and FCC-ee) are shown in Fig. 5.17. Future prospects are obtained assuming central values for flavour and EW observables corresponding to the current best-fit values of the model (corresponding to an effective NP scale around 2 TeV) and relative uncertainties as in the expected projections. The main message emerging from this exercise is the importance of using different observables both to validate a possible NP signal and to decode its interpretation. In this respect, FCC-ee represents the best opportunity given the expected improvements in both flavour and EW observables.

## 5.6 Conclusion

Flavour physics is a highly diversified area of research, characterised by its unique and wide-ranging sensitivity to physics beyond the SM. It is only through the combined exploration of flavour, Higgs and electroweak domains that the structure of physics beyond the SM can be uncovered via indirect searches. Within the flavour domain, the complementarity of the different measurements is of paramount importance. Consequently, a multifaceted programme involving small-, medium- and large-scale experiments is an indispensable part of the future strategy for particle physics.

Current bounds on particle EDMs and LFV in $\mu \to e$ transitions provide the most stringent constraints on the scale of NP in models with generic flavour structure. The large portfolio of future projects planned in these two unique areas has an outstanding discovery potential. Unique information about NP could also be obtained from kaon physics along three main directions: the rare $K_L$ decay program at KOTO II, dedicated studies of $K_{S,L} \to \mu^+\mu^-$ interference (possible at LHCb Upgrade II), and exploring opportunities for a next-generation $K^+ \to \pi^+ \nu\bar{\nu}$ experiment. High-intensity facilities for pions and neutrons, at PSI and the European Spallation Source, have the potential of significantly enhance the overall reach of SM consistency tests via improved determination of $|V_{ud}|$ and related measurements.

In the domain of $b$-hadron, $c$-hadron and $\tau$-lepton physics, major progress is anticipated through the full exploitation of existing experimental facilities up to the 2040s, particularly of LHCb Upgrade II and Belle II experiments. Thanks to these programmes, this is one of the areas of particle physics where the largest gain, in terms reduced statistical and systematic uncertainties, is expected in the next 20 years.

Despite this major progress, a significant fraction of interesting observables, particularly those involving neutrals and missing energy, would remain under-explored, significantly limiting the overall discovery potential of the flavour programme. The critical factor for this programme is the total number of $Z$ bosons produced at a future $e^+e^-$ collider, provided that the detectors are suitably designed for heavy-flavour physics. A dataset of only a few $10^9$ $Z$ bosons would fall short of delivering significant improvements beyond what is projected from the full LHCb Upgrade II and Belle II programmes. In contrast, a facility such as the FCC-ee, capable of producing $6 \times 10^{12}$ $Z$ bosons, would have a major impact. Given that a large fraction of the measurements would not be dominated by experimental systematic uncertainties, even more $Z$ bosons would be beneficial. Data taken at energies beyond the $WW$ and $t\bar{t}$ thresholds also permit measurements relevant to the flavour physics programme.

In parallel to these experimental efforts, the essential role of theoretical developments should also be emphasized. Improvements in the precision of SM predictions, which are within the reach of future LQCD efforts, are needed to match the anticipated experimental accuracy.

Chapter 6

# Neutrinos

## 6.1 Introduction: Open questions in neutrino physics

Experiments over the past decades have demonstrated that neutrinos are massive, and that a lepton flavour sector, at least as complex as the quark one, is needed in the Standard Model (SM). Massive neutrinos require the existence of new Higgs-Lepton couplings, but the extension of the SM this entails might be more complex, bringing new perspectives into other open questions in particle physics, notably the observed matter-antimatter asymmetry in the Universe. Our current knowledge of neutrino masses remains incomplete and requires significant refinement to enable meaningful connections to potential new physics.

The main phenomenological implication of massive neutrinos that has been tested so far is neutrino-flavour mixing in charged-current interactions. This is parameterized by the PMNS mixing matrix which relates the flavour neutrino states, $\nu_\alpha$, to the neutrino mass states, $\nu_i$:

$$\nu_\alpha = \sum_i (U_{\text{PMNS}})_{\alpha i}\, \nu_i. \tag{6.1}$$

The PMNS matrix depends on three mixing angles and at least one CP violating phase, $\delta$. Two very different neutrino mass differences have been measured, but it is not known if the ordering of the masses is normal (NO), with the two more degenerate states being the lightest, or inverted if they are the heaviest (IO). Global analyses of neutrino oscillation data in both scenarios show that most mass differences and mixings parameters are already known to the few percent level (see Table 6.1). The angle $\theta_{23}$ is close to maximal but suffers from an octant degeneracy, $\theta_{23} \leftrightarrow \frac{\pi}{4} - \theta_{23}$. At the $3\sigma$ level, both octants are allowed, and this affects the sensitivity to the CP phase which is largely unknown. Next-generation oscillation experiments aim at pinning down the neutrino mass ordering, discovering leptonic CP violation, and improving significantly the precision on the remaining parameters, allowing strong tests of the PMNS paradigm, as further discussed in Sect. 6.2.

For the absolute neutrino mass scale, the best sensitivity in laboratory measurements is obtained from the study of the end-point $\beta$-decay spectrum of $^3$H, which depends on all neutrino masses, through the combination $m^2_{\nu_e} \equiv \sum_i |U_{ei}|^2 m^2_{\nu_i}$. KATRIN has recently set the upper bound $m_{\nu_e} \leq 0.45\,\text{eV}(90\%\ \text{CL})$ [270], and will soon reach its target sensitivity of 0.3 eV. Neutrino masses also modify the imprint of relic neutrinos in the Cosmic Microwave Background (CMB) perturbations and in the galaxy distribution at large scales (LSS). These observables depend on the sum of all neutrino masses, $\sum_i m_{\nu_i}$. Very recent results from CMB measurements [271] have set a new bound $\sum m_{\nu_i} \leq 0.17$ eV (90% CL), within the standard cosmological model ($\Lambda$CDM), while the combination of CMB and the latest LSS measurements of DESI [272] gives the impressive bound [1] $\sum m_{\nu_i} \leq 0.048$ eV [271] at 90% CL. For a more detailed discussion see Sect. 7.6. The results of oscillation measurements on the other hand imply that the values of $m_{\nu_e}$ and the sum of neutrino masses can lie in the narrow bands shown in the left plot of Fig 6.1, in particular $\sum m_{\nu_i} \geq 0.058$ eV (90% CL). Clearly, the importance of having a direct measurement of neutrino masses from a laboratory experiment cannot be overstated. The perspectives in this area are described in Sect. 6.4.

| | PDG24 $1\sigma$ | NuFIT6.0 $3\sigma$ | Precision | | PDG24 $1\sigma$ | NuFIT6.0 $3\sigma$ | Precision |
|---|---|---|---|---|---|---|---|
| $\sin^2\theta_{12}$ | $0.307^{+0.013}_{-0.012}$ | $0.275 \to 0.345$ | $\sim 4\%$ | $\frac{\Delta m^2_{21}}{10^{-5}\text{eV}^2}$ | $7.53 \pm 0.18$ | $6.92 \to 8.05$ | $\sim 2.5\%$ |
| $\sin^2\theta_{23}$ | $0.558^{+0.015}_{-0.021}$<br>$0.553^{+0.016}_{-0.024}$ | $0.430 \to 0.596$<br>$0.437 \to 0.597$ | $\sim 5\%$ | $\frac{\Delta m^2_{3l}}{10^{-3}\text{eV}^2}$ | $+2.455 \pm 0.028$<br>$-2.529 \pm 0.029$ | $+2.46 \to +2.61$<br>$-2.58 \to -2.44$ | $\sim 1.1\%$ |
| $\frac{\sin^2\theta_{13}}{10^{-1}}$ | $0.219^{+0.007}_{-0.007}$ | $0.202 \to 0.238$<br>$0.205 \to 0.240$ | $\sim 3\%$ | $\frac{\delta(\text{rad})}{\pi}$ | $1.19 \pm 0.22$ | $0.54 \to 2.34$<br>$1.11 \to 1.93$ | |

Table 6.1: Three-flavour oscillation parameters from the PDG24 and the global fit NuFit6.0 [269] (without IC24 SK atmospheric data) for NO (upper), $l = 1$, and IO (lower), $l = 2$.

Even with the somewhat limited precision achieved so far, compared to the quark sector, it is clear that massive neutrinos provide a new perspective into the SM flavour puzzle. Lepton and quark mixing matrices are strikingly different, and neutrino masses are more than six orders of magnitude below those of the other fermions. The only compelling explanation we have of these facts is that neutrino masses might be connected to a new physics scale, $\Lambda_{\text{NP}}$, i.e. $m_\nu \propto \frac{m_f^2}{\Lambda_{\text{NP}}}$. Interestingly, such a new scale could provide an explanation for the existence of baryons in the Universe, which must have originated in a tiny matter-antimatter asymmetry at early times. The two missing ingredients in the Standard Model (SM) needed to generate the observed asymmetry—sufficiently strong CP violation and non-equilibrium dynamics above the electroweak phase transition—are generic features of models of neutrino masses that involve such a scale.

On the experimental front, neutrino detectors are designed to observe the most elusive particles: they often involve huge detectors that operate in low background conditions and/or are located near the most intense neutrino sources. Consequently, they are also well equipped to search for rare processes predicted in many new physics scenarios, whether or not these are related to neutrino masses. In this context, two new regimes of exploration with relevance for new physics searches have recently opened. The detection of coherent neutrino scattering (CE$\nu$NS), in which a neutrino recoils off a whole nucleus, has been possible thanks to the technological advances in detecting keV nuclear recoils. At much higher energies, hundreds of neutrino events with TeV energies from LHC collisions have also been recorded by the FASER and SND experiments. Furthermore, the measurement of neutrino interactions across scales is also providing new ways to test the SM, to probe nucleon and nuclear structure and constitutes

[1] This tight limit arises when combining CMB results from Planck, ACT, and SPT, datasets that show a large degree of consistency.

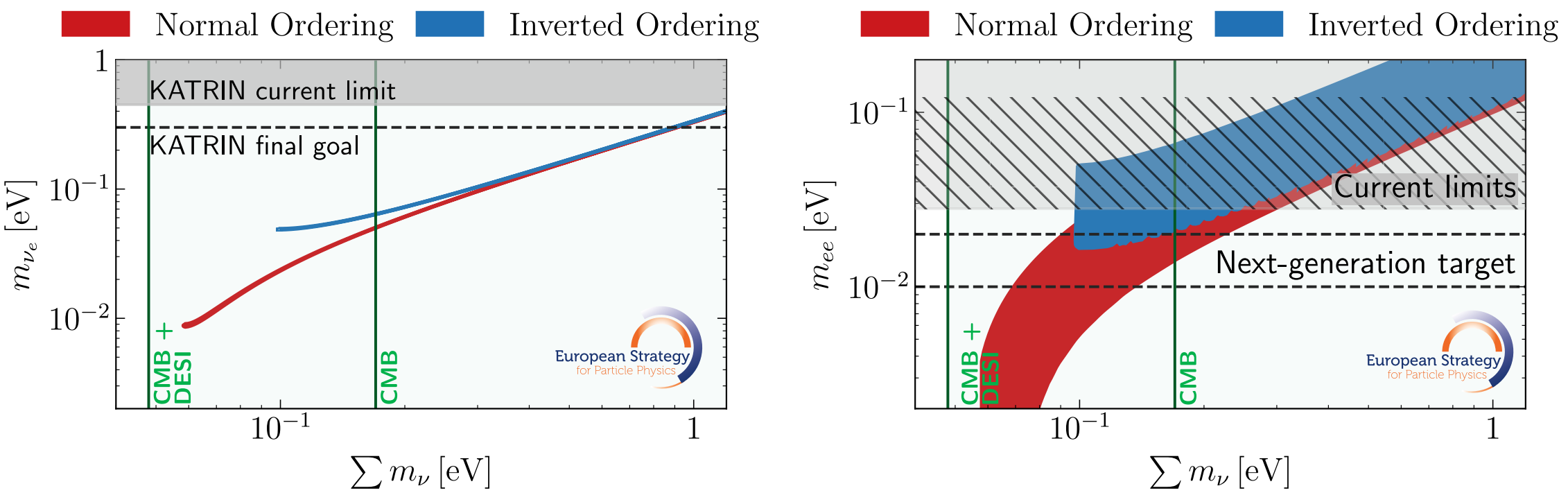

Fig. 6.1: Left: Allowed regions in the $m_{\nu_e} \equiv \sqrt{\sum_i |U_{ei}|^2 m_{\nu_i}^2}$ versus $\sum_i m_{\nu_i}$ plane for normal (red) or inverted (blue) orderings in the three-neutrino model. The current experimental limits on both parameters from KATRIN and cosmological measurements are shown. Right: same for $m_{\rm ee} \equiv \sum_i U_{ei}^2 m_{\nu_i}$ versus $\sum_i m_{\nu_i}$.

an essential input to neutrino astrophysics, see Sect. 6.3.

Both the theoretical expectations and the experimental opportunities indicate that neutrino experiments are highly relevant in the quest for new physics, whether it involves heavy sectors, or lighter and more feably-coupled ones. In the former case, a good description is provided by Standard Model Effective Field Theoty (SMEFT), where the new interactions are ranked in dimension, $d$, or in their relevance at low energies, see Appendix A. The most relevant interaction is that at $d = 5$, which corresponds to the famous Weinberg operator, Eq. (A.1), that induces Majorana neutrino masses, violates total lepton number and mediates neutrinoless-double beta decay ($0\nu\beta\beta$). The lifetime of this process is inversely proportional to the combination $m_{\rm ee} \equiv \sum_i U_{ei}^2 m_{\nu_i}$. The value of this quantity from oscillation data for both mass orderings is shown on the right plot of Fig. 6.1 [2]. A significant fraction of the relevant parameter space has already been explored by current experiments and the prospects are described in Sect. 6.5.

Next in relevance are operators at $d = 6$. There are 2499 combinations under no symmetry assumption. Among them some violate baryon number, others induce non-standard neutrino interactions (NSI) or non-unitarity in the PMNS matrix. In fact, neutrino experiments, notably Super-Kamiokande, have set the most stringent constraints on any $d = 6$ coefficient probed so far through proton decay searches [273]. Furthermore, neutrino experiments have leading sensitivity to some $d = 6$ operators inducing non-standard neutrino interactions, particularly those involving $\nu_\tau$, as discussed in Sect. 6.6.

Regarding new physics at low scales, the phenomenology of these scenarios is highly model dependent. Nevertheless, minimal models, based on the SM portals (extensions with one singlet field that couples to the SM), capture generic features of these scenarios. There are only three possibilities: a scalar, a vector and a neutrino. The neutrino portal is also the simplest and well-known model of massive neutrinos: the Type I see-saw model (at low scale). It predicts neutrino masses, the existence of extra neutrino species, which, depending on their mass, could be light sterile neutrinos (LSN) participating in neutrino oscillations, or heavy neutral leptons (HNL) if heavier. LSNs have been the standard explanation of anomalies in neutrino oscillation experiments (eg. LSND/Miniboone, reactor and gallium), while HNLs have become a standard benchmark in BSM searches for light dark sectors. Significant progress has been achieved in recent years in both areas as will be discussed in Sect. 6.6. The neutrino portal also provides a robust explanation of the matter-antimatter asymmetry of the Universe in regions of parameter space that can be tested.

[2]The bands for the two orderings assume there is no new physics beyond Majorana neutrino masses.

## 6.2 Neutrino oscillations: mixing, masses and Charge-Parity violation

Present knowledge of neutrino mixing parameters and mass differences relies on the measurement of neutrino oscillations. Neutrinos produced at various sources (accelerators, reactors, cosmic rays interacting in the atmosphere, the Sun) are detected some macroscopic distance away ($L$, baseline). By choosing the source of neutrinos and the baseline, $E_\nu/L$ can be tuned to one or the other neutrino mass difference ($\Delta m^2_{23}$ atmospheric frequency, or $\Delta m^2_{12}$ solar frequency) in such a way that the oscillation is primarily determined by a subset of the PMNS parameters. Thus, a combination of results from different experiments is needed to completely characterise neutrino oscillations.

### 6.2.1 Oscillation experiments with accelerator, atmospheric and reactor neutrinos

Experiments measuring oscillations of neutrinos produced by accelerators have two unique features. They produce a clean sample of $\nu_\mu$ or $\bar{\nu}_\mu$, depending on the polarity of the focusing optics of the parent beam. Secondly, they include near detectors to measure the neutrino rate as a function of energy and flavour before oscillation. This provides crucial control of the significant uncertainties associated with the limited knowledge of the flux and neutrino interaction cross-sections (see Sect. 6.3). This setup enables very precise measurements of $|\Delta m^2_{23}|$ and $\sin^2\theta_{23}$, clean sensitivity to Mass Ordering (MO) through matter-effects and the unique capability to probe the CP-symmetry (parametrized by the $\delta_{CP}$ phase) by comparing the oscillation of neutrinos and antineutrinos. The present generation of accelerator experiments (T2K [ID116] and NOvA [274]) will provide sensitivity at about the 2 to 3$\sigma$ level on CP violation (CPV) and MO, and is developing new techniques to suppress the impact of systematic uncertainties, that will be crucial with increasing statistics. The next-generation of experiments (Hyper-Kamiokande [ID238], Hyper-K in the following, and DUNE [ID118]) will record an unprecedented number of oscillated neutrinos having thus the potential to definitively establish MO and to discover CPV for a large fraction of possible values of $\delta_{CP}$. This programme will require an unprecedented control of the systematic uncertainties, notably neutrino flux and cross-section, to ensure accuracy of the results. In this respect, the strong complementarity of Hyper-K and DUNE will be a crucial asset. Hyper-K is based on the approach (technology, energy and baseline) of T2K and Super-Kamiokande, boosted by an increase in the size of the far detector by a factor eight and an upgrade of the beam-power to be 2.5 times larger than that used in the latest T2K analyses (515 kW [275]). Hyper-K will thus produce a very large number of (anti-)neutrinos events at the first oscillation maximum, enhacing the CPV sensitivity, while the MO sensitivity relies on the atmospheric neutrino sample. DUNE's approach aims at measuring both CP violation and the MO with beam neutrinos. To achieve this it employs a neutrino flux with wider energy range and at larger energy, and thus a longer baseline. These characteristics will be crucial to disentangle different oscillation parameters and to provide new measurements of the oscillation shape on a wider $L/E_\nu$ range, reaching up to the second oscillation maximum. To exploit this rich shape information in the oscillated spectra, DUNE will employ large liquid argon detectors, enabling an unprecedented neutrino energy resolution. Table 6.2 reports the number of events expected in these experiments and Fig. 6.2 shows their coverage in $L/E_\nu$ and their approximate energy resolution.

The construction of the Hyper-K and DUNE experiments is in progress, notably with strong support from the CERN Neutrino Platform and large European involvement. Active R&D and prototyping is underway for a range of liquid-argon and non-argon technologies for DUNE far and near detectors [ID119] as part of the Phase-II program to improve charge and light readout schemes and broaden the physics program along with improved systematic constraints. Moreover, further detector development is ongoing aiming at a second upgrade of the Hyper-K near detector (ND280++).

| sample | T2K | NOvA | Hyper-K | DUNE |
|---|---|---|---|---|
| $\nu_e$ | 94 | 181 | 2475 | 3289 |
| $\bar{\nu}_e$ | 16 | 32 | 1543 | 882 |
| $\nu_\mu$ | 318 | 384 | 8845 | 14953 |
| $\bar{\nu}_\mu$ | 137 | 106 | 12027 | 8149 |

Table 6.2: Number of events selected at the far detector in present accelerator neutrino-oscillation experiments [275, 276] and expected in the next-generation experiments after 10 years running, with $\delta_{CP} = -\pi/2$, NO and the official planned run of neutrino/antineutrino beam modes [277, 278]. For DUNE (T2K and Hyper-K) charged current events (without pions) are considered.

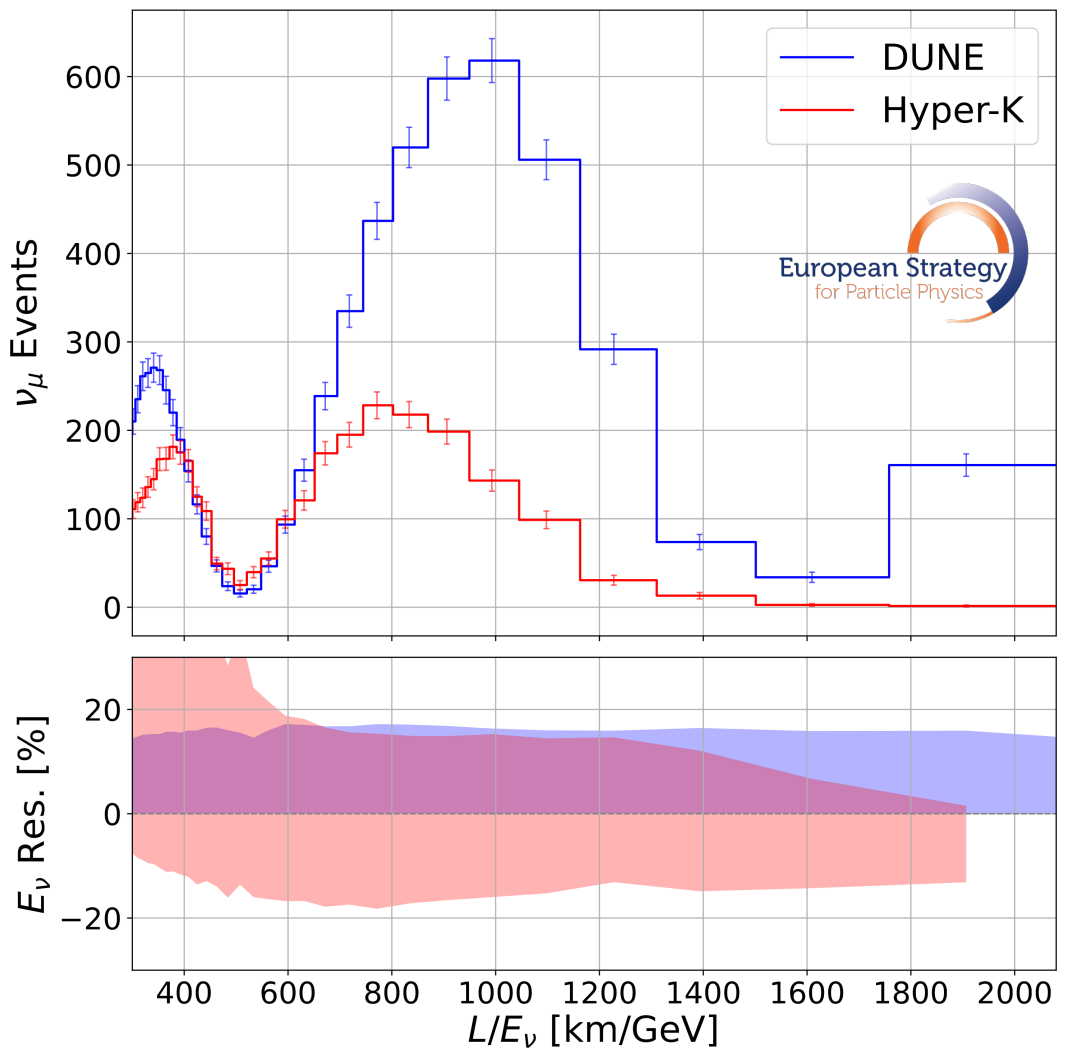

Fig. 6.2: Oscillated muon neutrino spectrum in Hyper-K and DUNE after 10 years running (top). An estimation of their energy resolution, considering only nuclear effects using, respectively, NEUT 5.8.0 and GENIE V3 (bottom). For DUNE (Hyper-K) charged current events (without pions) are considered.

Atmospheric neutrinos are produced in the upper atmosphere and span a large range of energies (MeV-TeV) and baselines ($10-10^4$ km propagating through the Earth). The measurement of atmospheric neutrino oscillations provides precise measurements of the $\Delta m^2_{23}$ and $\sin^2\theta_{23}$ parameters and a competitive sensitivity to MO. Indeed, neutrino propagation in Earth matter leads to the MSW resonance at GeV energies in the neutrino or antineutrino channel depending on the MO. On the other hand, the flux of atmospheric neutrinos is a mixture of $\nu_\mu, \nu_e, \bar{\nu}_\mu, \bar{\nu}_e$ making the oscillation measurements more challenging.

The oscillations of atmospheric neutrinos are measured by large underground detectors (such as Super-Kamiokande [279] and far detectors of future accelerator experiments, Hyper-K and DUNE) and by neutrino telescopes (underwater, as KM3NeT [ID249] or in ice, as IceCube [ID236]). While the astrophysics programme of the latter experiments is focused on very-high energy neutrinos, they also have more densely instrumented regions to enable the measurement of neutrinos with energies down to few GeV: KM3NeT-ORCA is being deployed in the Mediterranean sea and DeepCore is running in Antarctica with its further upgrade under construction. Deep-water and deep-ice experiments have gigantic target masses at the level of multiple megatons, though with limited resolution and particle identification capabilities. The present generation of neutrino atmospheric experiments (Super-Kamiokande and IceCube) is providing first hints on MO at the $2\sigma$ level and competitive measurements of $\Delta m^2_{23}$ and $\sin^2\theta_{23}$. A major challenge in atmospheric neutrino studies is the $\sim$10% uncertainty on their flux. Progress on this front requires a detailed understanding of the primary cosmic ray flux, accurate theoretical modelling of primary cosmic ray interactions, decay chains and secondary interactions.

Nuclear reactors are extremely intense sources of $\bar{\nu}_e$ in the MeV range. The previous generation of reactor neutrino experiments (Daya Bay [280], Double-Chooz [281] and RENO [282]) are tuned to the atmospheric frequency with baselines of $O$(1km). Their measurement of the disappearance of reactor neutrinos gave the most precise measurement of the smallest mixing angle, $\theta_{13}$ (see Table 6.1). Reactor neutrino experiments at baselines of $O$(100) km, such as KamLAND [283], are tuned to the solar frequency and provide the most precise measurement of $\Delta m^2_{12}$, while the angle $\theta_{12}$ is best measured by solar neutrino experiments. The next-generation of reactor experiments is JUNO [ID36] that has started data taking in August 2025. JUNO operates a large liquid scintillator detector at a baseline of 50 km and can reach sub-percent precision in both the solar parameters. While the leading oscillation in JUNO is controlled by these solar parameters, there is a subleading modulation of the signal from the faster atmospheric oscillation frequency, which is sensitive to $\theta_{13}$ and $\Delta m^2_{31}$, including the MO. In particular, JUNO aims to reach sub-percent precision on $|\Delta m^2_{31}|$. In order to exploit this feature, a very precise control of the energy scale and resolution is needed over the whole JUNO volume. At MeV energies, neutrinos interact mainly with hydrogen nuclei with well-known cross-section. On the other hand, the rate and spectrum of reactor neutrinos is rather uncertain. Significant progress in the modelling of reactor fluxes has been achieved in recent years, notably thanks to dedicated measurements from Daya Bay [284] and the short baseline reactor experiment STEREO [285], yet uncertainties remain. The JUNO near detector (TAO) will make further progress in measuring the neutrino spectrum, minimizing the effects of those uncertainties on JUNO physics programme.

### 6.2.2 A global programme: sensitivity and interplay of the different experiments

Studies of neutrino oscillations will benefit in the next decade from an unprecedented avalanche of new data, most importantly, from different neutrino sources and from highly complementary experiments, which will enable a much improved characterization of the PMNS paradigm and increased sensitivity to non-standard oscillation effects (see Sect. 6.6). Figs. 6.3 and 6.4 present the projected precision from individual experiments.[3] These separate results are affected by important degeneracies between different oscillation parameters, as well as, between oscillation parameters and systematic uncertainties. The combination of measurements from different experiments has the potential to significantly reduce these degeneracies, boosting the physics reach well beyond their plain statistical sum. The second phase of DUNE data-taking will also be particularly powerful for lifting these degeneracies.

Significant synergies exist by combining reactor, accelerator and atmospheric data on $\Delta m^2_{23}$. In particular, the combination of JUNO with $\nu_\mu \to \nu_\mu$ disappearance measurements of present and future long-baseline and/or atmospheric experiments can provide significant sensitivity to the MO. Sensitivity at the 2-3$\sigma$ level on the MO can be provided with this approach in the next few years. Beyond this, the comparison of $\nu_\mu \to \nu_e$ and $\bar{\nu}_\mu \to \bar{\nu}_e$ oscillations in presence of matter effects is the main road to provide sensitivity to MO. The achievable MO sensitivity depends on the value of $\sin\theta_{23}$. The first individual experiment which will provide unambiguous and definitive ($> 5\sigma$) determination of the MO is expected to be DUNE. JUNO will also determine the MO, albeit with lower sensitivity ($\sim 3\sigma$), independently of matter effects. Comparing the results of DUNE and JUNO will be a stringent test of the PMNS paradigm.

[3]The sensitivities are those provided in the inputs of the individual collaborations and do not necessarily use the same statistical methods. Furthermore, there is a sizeable dependence of the sensitivity on the true values of the parameters, particularly the octant of $\theta_{23}$ and the MO for low exposures. Not all experiments have provided information on this dependence.

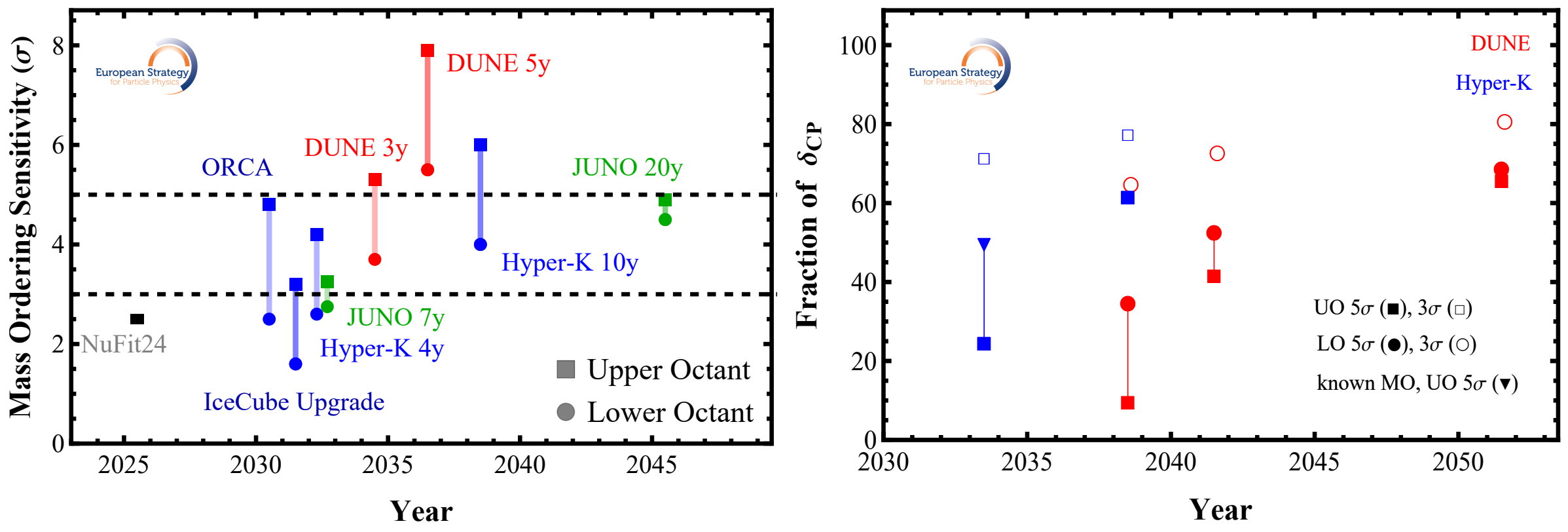

Fig. 6.3: Sensitivity achievable by the various neutrino oscillation experiments on MO (left) and CP violation (right), according to the present schedule. On the right, the points shown correspond to 5*y* and 10*y* of Hyper-K, and 7*y*, 10*y* and 20*y* of DUNE. True NO is assumed in all cases. The dependence on the true octant is shown for the MO plot. It is also shown for the DUNE $\delta_{CP}$ sensitivity, while the HyperK CPV reach assumes upper octant ($\sin^2\theta_{23} = 0.528$). The estimation of Hyper-K 5y sensitivity to $\delta_{CP}$ is available with beam data only, so we show the result if the MO is known, and if it is not. This range encompasses any dependence on the true octant, which is estimated to be few percent for known MO.

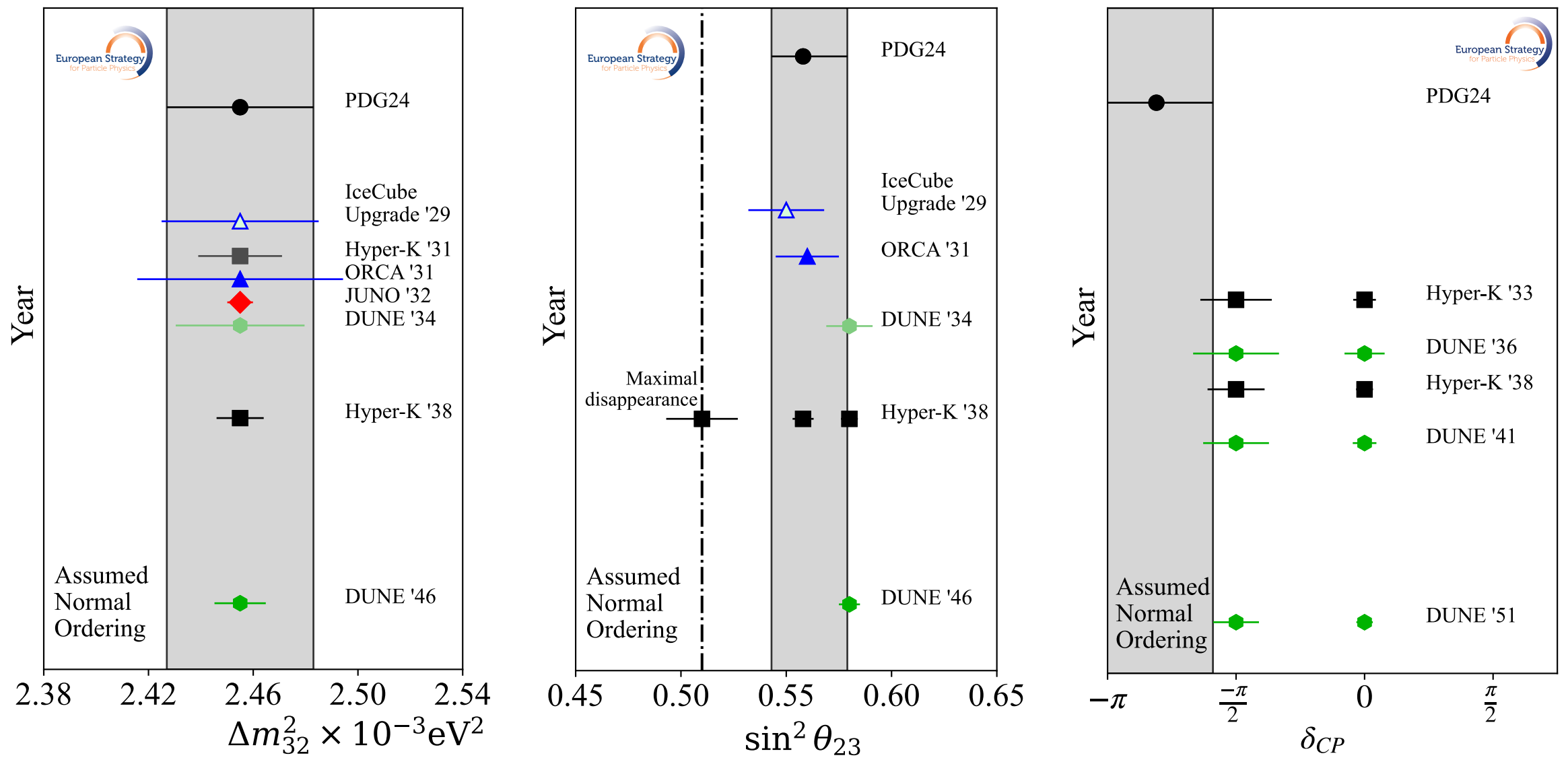

Fig. 6.4: Precision achievable by the various neutrino oscillation experiments on $|\Delta m^2_{23}|$, $\sin^2\theta_{23}$, $\delta_{CP}$ (from left to right). The grey area indicates the present precision (from PDG2024). The precision achievable on $\sin^2\theta_{23}$ depends on its value (as reported for Hyper-K, for example). For this parameter, the central point of the $3\sigma$ region is shown as the central value, and 1/6th of the $3\sigma$ range is shown as uncertainty.

The discovery of CPV is the domain of accelerator experiments, where clean and separate samples of $\nu_\mu \to \nu_e$ and $\bar{\nu}_\mu \to \bar{\nu}_e$ oscillations can be directly compared. While the measurement of $\sin^2\theta_{13}$ from reactor experiments may help reach CPV sensitivity sooner by relying on the $\nu_\mu \to \nu_e$ channel alone, the next-generation of accelerator experiments will ultimately collect sufficient samples in both $\nu$ and $\bar{\nu}$ modes to enable a direct probe of CPV and a new competitive measurement of $\sin^2\theta_{13}$, providing a new test of the PMNS paradigm. DUNE can measure separately MO and $\delta_{CP}$ with beam neutrinos for all possible values of $\delta_{CP}$. The Hyper-K measurement with beam neutrinos is affected by degeneracies between MO and CPV effects for half of the possible values ($\delta_{CP} > 0$ in NO, $\delta_{CP} < 0$ in IO). Nevertheless, for CPV close to maximal in the non-degenerate region, Hyper-K can discover CPV in 2 to 4 years of data-taking. Degeneracies can also be resolved in Hyper-K when combining with atmospheric data. With ultimate exposure the two experiments will be able to discover CPV for about 60% of possible $\delta_{CP}$ value. The precision on $\delta_{CP}$ will span, for DUNE and Hyper-K, between 6 to 7 degrees (in case of CP conservation) to about 18 to 20 degrees (in case of maximal CPV). In the first case the achievable sensitivity is limited by rate uncertainties in the $\nu_e$ and $\bar{\nu}_e$ appearance samples, while in the latter case the limiting systematic uncertainties are those affecting the energy shape of the $\nu_e$ oscillated sample.

To push the physics reach beyond what is described above, various feasibility studies are being conducted for new experiments in Europe: a new gigantic reactor experiment at Chooz (SuperChooz [286]) and a new neutrino beam to be generated using the 5 MW proton linac of the European Spallation Source (ESSnuSB [ID151]). ESSnuSB proposes to exploit the second oscillation maximum, where the CPV effect is enhanced, with the prospect to improve the error of $\delta_{\mathrm{CP}}$. It will also have a dedicated programme to measure sub-GeV cross-sections. Feasibility studies are also being conducted for a possible new 100 kiloton-scale detector based on the dual readout (Cherenkov plus scintillation light detection technology, THEIA [ID263]) that would enable improved reach in the physics programme of low-energy neutrinos.

### 6.2.3 Improving systematic uncertainties

In accelerator experiments, a highly capable set of near detectors is being developed in order to constrain flux and cross-section systematic uncertainties. The goal of this new generation of near detectors is the comprehensive and precise reconstruction of particles in the final state from neutrino-nucleus interactions. The T2K experiment has recently upgraded its near detector (ND280), lowering the threshold for particle reconstruction and enabling, for the first time, the direct measurement of the energy of neutrons from (anti-)neutrino interactions on an event-by-event basis. This detector will also serve Hyper-K. A new Intermediate Water Cherenkov Detector (IWCD) is also being built for Hyper-K. A further upgrade of ND280 is being studied to be possibly deployed during Hyper-K data taking. The DUNE near detector liquid argon TPC will provide precise and exclusive measurements of neutrino interactions. The addition of a high-pressure gas argon TPC is planned for the second phase of data taking. The IWCD in Hyper-K and the near detector argon TPCs in DUNE will use the so-called PRISM technique: these detectors will be moved off-axis in order to measure neutrino-nucleus cross sections at different average energies. This approach will lift degeneracies between flux and cross section modelling, reducing the scope for biases in neutrino oscillation analyses which may be induced by an incorrect cross-section model. Magnetized near detectors (such as ND280 and DUNE near detectors) also provide unique control of the neutrino background in the antineutrino beam mode, a crucial input for the CPV search and the measurement of $\delta_{CP}$.

Other experiments are measuring neutrino-nucleus cross-sections in the relevant energy range (0.1–few GeV) and on multiple targets (Ar, C, CH, $H_2O$, He,Fe, Pb). MINER$\nu$A [ID149] has measured neutrino interactions at energies relevant for DUNE, including nuclear effects, and plans to release all their data. Pioneered by the ICARUS collaboration in LNGS, the unique capabilities of liquid argon TPCs to measure precisely the final state of neutrino interactions have been recently demonstrated by the MicroBooNE experiment [287]. SBND [ID232] will measure neutrino-argon interactions with unprecedented statistical precision ($>10^6$ events) exploiting the PRISM technique, whilst ICARUS [ID226] will perform measurements closer to the energy of DUNE's first oscillation maximum.

Nonetheless, the DUNE near detectors will be the first to measure neutrino interactions on argon in the higher end of the DUNE energy regime. In particular, the ultimate precision in terms of exclusive final states will be achieved by the high-pressure gas argon TPC, with a significantly lower momentum threshold for charged hadrons.

The exploitation of near detector measurements to predict the neutrino energy spectra at the far detector has intrinsic limitations. The near and far detectors have different acceptances, and different neutrino energy distributions, because of oscillations. In some cases an extrapolation between different target materials is needed and the number of electron neutrino events is limited at the near detector, before oscillation. Most importantly, the near-detector measurements cannot determine flux and cross-section separately and have limited neutrino energy resolution. Thus the accuracy of oscillation measurements cannot be unambiguously ensured in the absence of a realistic, robust and comprehensive theoretical model of the flux and the cross-section. It is therefore crucial to keep developing models of hadron production (as in GEANT4 [288] and FLUKA [289]) for neutrino fluxes, and models of nuclear physics for neutrino cross-sections (championed by many European initiatives alongside the NuSTEC collaboration [ID135]). Such endeavours will require sustained efforts from the nuclear theory community, in close collaboration with experimentalists. The concurrent development and update of dedicated Monte Carlo tools (GENIE [290], GiBuu [291], NEUT [292], NuWro [293], ACHILLES [294]) is crucial to synthesise available nuclear physics models into a complete simulation of neutrino nucleus interactions and to allow their use in experimental data analysis.

Specific measurements from various dedicated experiments have demonstrated their pivotal role in tuning the parameters encoding the uncertainties of flux and cross-section models. Regarding the flux, NA61/SHINE [ID171] is the only experiment that can provide the data needed for sufficiently accurate neutrino flux predictions. In particular, the CPV sensitivity and the ultimate precision on the CP phase will be limited by the uncertainty in the neutrino contamination in the beam of antineutrinos. This pollution is produced by wrong-charge hadrons which escape de-focusing due, for instance, to their re-scattering in the beam line material. The NA61/SHINE measurement of low energy hadrons is essential to accurately estimate this contamination. To this aim the collaboration has proposed, with the support of the next-generation of accelerator neutrino experiments, a new movable low-energy proton beam line at the SPS. The project would also halve the uncertainties on the atmospheric neutrino flux.

The main limitations of the neutrino-nucleus cross-section measurements using near detectors and other dedicated experiments in standard neutrino beams are the low number of events available of $\nu_e$ and $\bar{\nu}_e$, the impact of the flux uncertainty, and the limited neutrino energy resolution, which does not allow for the separate measurement of the different nuclear effects that contribute to the cross-section at different energies (see Sect. 6.3). Various projects have been proposed to overcome these limitations. Kaon decay at rest, as in JSNS$^2$ [295], can be used to measure the neutrino cross-section at monochromatic energy of 235.5 MeV. Within the muon-collider roadmap the nuSTORM project [ID200], proposes to exploit the decay of muons in a storage ring to provide a large samples of electron and muon (anti-)neutrinos in the entire relevant energy region for DUNE and Hyper-K (300 MeV to 5.5 GeV). The overall flux rate is known at an unprecedented level ($< 1\%$) and various neutrino energy regions (of $> 1$ GeV width) could be scanned by tuning the muon beam energy. Finally, the nuSCOPE proposal [ID101] has been put forward at the Neutrino Platform to construct an instrumented proton beam line based on slow extraction to enable a monitored neutrino beam, where the lower instantaneous intensity allows to directly measure the parent $K,\pi$ decays to precisely determine the $\nu_\mu$, and $\nu_e$ neutrino fluxes. Furthermore, in the case of two-body decays ($\nu_\mu$), the detected neutrinos can be *tagged*, i.e., associated with the parent meson decay, and their energy reconstructed on an event-by-event basis. nuSCOPE is proposed to operate at the CERN SPS beam, collecting about $10^6$ tagged ($10^4$ monitored) charged-current $\nu_\mu(\nu_e)$ interactions, using about $2.5 \times 10^{18}$ protons on target per year. The necessary detector technology for beam instrumentation has either been demonstrated with prototypes or is being developed. Both the overall flux rate and the energy of muon neutrinos event-by-event can be known with < 1% precision. Knowing the neutrino energy with such precision, independently on any cross-section and flux model, allows a direct constraints on the map between energy seen in a detector and the true neutrino energy, resolving a pivotal source of uncertainty for neutrino oscillation measurements at accelerators.

Finally, flux and cross-section uncertainties introduce particular challenges when combining results from multiple experiments, as the important sources of uncertainties due to neutrino flux (for both beam and atmospheric neutrinos) and to neutrino-nucleus cross-section may be strongly correlated. A comprehensive model of these uncertainties must be built and shared among the different collaborations. Thus a robust combination of results will require significant theoretical and experimental efforts over the coming years.

## 6.3 Neutrino cross-section across energy scales

A rich variety of neutrino sources is available in Nature from sub-eV up to multi-PeV energies (see Fig. 7.1) with the neutrino scattering cross-section on matter spanning from $10^{-31}$ to $10^{-1}$ mb. Precise knowledge of the cross section is necessary to characterize these sources. Moreover, interactions of neutrinos from natural and artificial sources can be used to probe properties of the target—notably of nuclear matter—and neutrinos, within the SM and beyond.

Neutrinos with energies $\lesssim \mathcal{O}(10\,\text{MeV})$ offer a unique physics opportunity because their interaction with nuclei is dominantly coherent (called coherent neutrino-nucleus scattering, CE$\nu$NS) and thus strongly enhanced. Consequently, detectors of $\mathcal{O}(10)$ kg are massive enough to measure CE$\nu$NS. However, its signature is very challenging to detect: the only observable is a few-keV energy deposit from the nuclear recoil. Various detection technologies can be used: a summary of previous measurements and proposed experiments is presented Fig. 6.3 and Table 6.3, respectively. Two artificial neutrino sources have been exploited for this measurement: pion decay at rest ($\pi$-DAR) at neutron spallation sources, for neutrino energies of tens of MeV, and nuclear reactors, for neutrino energies of a few MeV. $\pi$–DAR gives unique access to both electron and muon neutrinos with a well-defined ratio and time delay. Reactor neutrinos enable complementary measurements at lower neutrino energies, but the detection is even more challenging: a very good knowledge of the response to very low energy deposits is needed. Furthermore, this could have interesting applications in reactor monitoring.

Neutrinos of these energies are also copiously produced in astrophysical sources. CE$\nu$NS is the dominant interaction process of solar neutrinos in low-threshold detectors, constituting the so-called "neutrino-fog" in direct DM searches. The process is also very important in supernova explosions and in the detection of neutrinos produced in such explosions. CE$\nu$NS gives also access to interesting nuclear physics, as the loss of coherence provides a unique measurement of the neutron form factor. Conversely, in the fully-coherent region the cross-section is known well. There, CE$\nu$NS provides a measurement either of the Weinberg angle ($\sin^2\theta_W$) at 0.01–0.1 MeV energy or of the neutrino charge radius. CE$\nu$NS measurements are also unique probes of BSM neutrino electromagnetic properties, such as millicharges or magnetic moments. Finally, measurements at CE$\nu$NS experiments provide opportunities for other BSM searches, as discussed in Sect. 6.6.

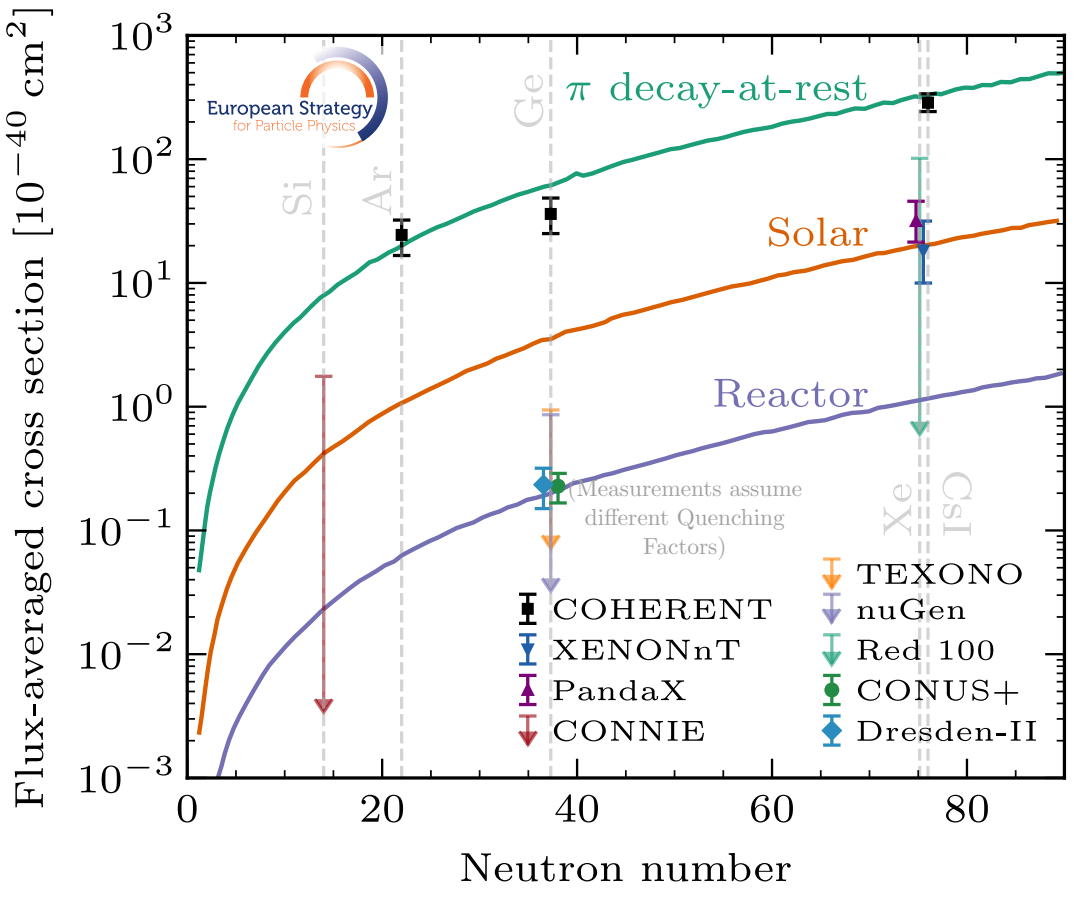

Fig. 6.5: Current CEvNS measurements

| Experiment | Neutrino source | Detector material |
|---|---|---|
| COHERENT | $\pi - DAR$ | NaI, 3500 kg |
| COHERENT | $\pi - DAR$ | Liquid Ar, 750 kg |
| COHERENT | $\pi - DAR$ | Cryogenic CsI, 10 kg (low threshold) |
| COHERENT | $\pi - DAR$ | Ne, 20 kg |
| Captain Mills | $\pi - DAR$ | Ar, 10 ton |
| ESS | $\pi - DAR$ | CsI[Na], 130 kg; Si, 1 kg; Xe, 20 kg; Ge, 7 kg; Ar, 10 kg; $C_3F_8$, 10 kg (large $\nu$ flux) |
| SuperCDMS | Solar | Si, 3.7 kg; Ge, 25 kg |
| RES_NOVA | Solar | Pb, 170 kg; 2.4–465 ton |
| CYGNUS | Solar | He+$CF_4$ / He+$SF_6$, 15 kg–1500 kg |
| CONUS100 | Reactor | Ge, 100 kg |
| CONNIE upgrade | Reactor | Si, 1 kg–10 kg (low threshold) |
| MINER | Reactor | Ge, 1.5 kg; Si, 10 g |
| NEON | Reactor | NaI, 13.3 kg |
| NUCLEUS | Reactor | $CaWO_4$, Ge, $Al_2O_3$,10 g–1 kg (low threshold) |
| Ricochet | Reactor | Ge, kg; Si, Zn, Al, Sn, 300g (low threshold) |
| SBC | Reactor | Ar, 10 kg |
| PALEOCCENE | Reactor | Different materials, 10 g–1 kg (exploits defect formation, low threshold) |
| RELICS | Reactor | Xe, 10–100 kg |

Table 6.3: Future CEvNS experiments (ongoing and proposed).

Interactions of neutrinos with energies in the range of 10 MeV to few GeV involve complex nuclear physics effects. At the lower end of this range, recent measurements from the JSNS$^2$ experiment [295] use mono-energetic neutrinos at 235.5 MeV from kaon-decay-at-rest to study nuclear ground state effects. At higher energies, this region has pivotal importance for the measurement of oscillations of accelerator and atmospheric neutrinos, for which these nuclear effects are a major source of systematic uncertainty (see Sect. 6.2.3). A rich set of measurements is being performed both by using the near detectors of accelerator oscillation experiments and other dedicated accelerator experiments. These experiments, if complemented by equally strong theoretical efforts, provide an opportunity to improve our understanding of the hadronic and nuclear physics involved in neutrino-nucleus/-nucleon interactions.

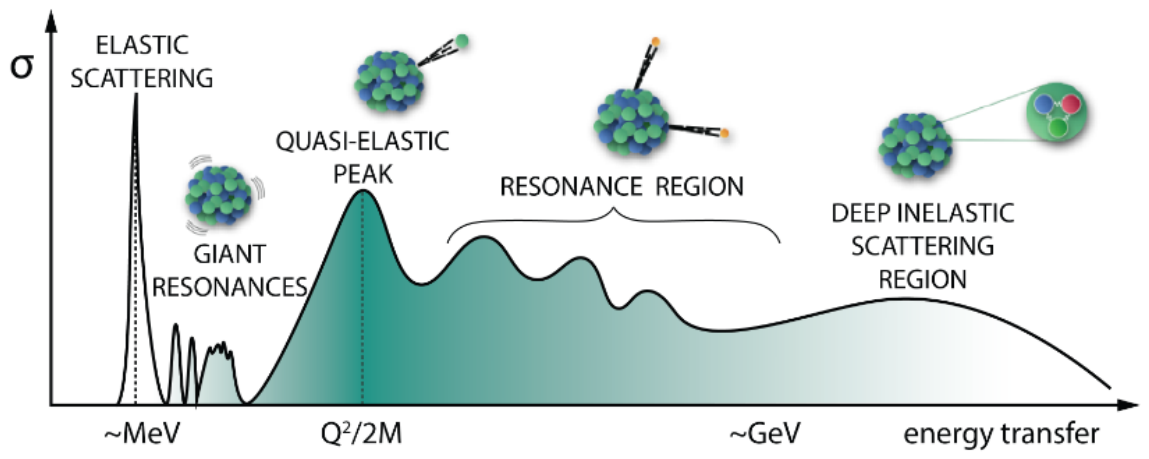

Fig. 6.6: A qualitative overview of the evolution of the neutrino-nucleus inclusive interaction cross-section as a function of energy transfer for a fixed scattering angle (from [ID135]).

This energy region has also been historically explored with electron-nucleus and electron-nucleon scattering experiments. Recent results from the CLAS experiment [296] have highlighted clear deficiencies in the nuclear modelling of neutrino interactions. Nevertheless, such measurements are sensitive only to vector interactions, leaving the axial-vector components, which are crucial for neutrino interactions, largely unconstrained.

Concerning neutrino-nucleon cross-section measurements, sparse data on hydrogen or deuterium targets is available from old bubble-chamber experiments. More recently, the axial vector form factor for quasi-elastic interactions has been directly calculated from Lattice QCD (LQCD) [297] and efforts are ongoing to further expand LQCD calculations to other interaction channels and to light nuclear matter. Simultaneously, methods to measure neutrino scattering on hydrogen within hydrocarbon in scintillator detectors have been suggested and recently investigated by the MINERvA experiment [298]. More precise measurements from the T2K [299] near detector are also expected in the coming years.

Moving to neutrino-nucleus interactions, Fig. 6.6 shows a qualitative overview of several nuclear effects at play at different energy transfers. In the 10 MeV to 100 MeV region, the cross-section is affected by giant collective nuclear resonances and long-range intranucleon correlations are relevant. The quasi-elastic peak (QE) is arguably the best-known region, but still its exact position and width are affected by nuclear binding energy and Fermi motion. Slightly above the QE peak, there is a poorly-known contribution from short-range and multi-nucleon correlations. At even higher energy, the cross-section is characterized by poorly understood nuclear resonances and their interference. Above the resonance region is the complex transition region where the relevant degrees of freedom shift from hadrons to quarks and are described, within large uncertainties, by quark-hadron duality. Finally, above these energies is the Deep Inelastic Scattering region where uncertainties are dominated by PDF modifications in nuclear matter. The landscape is further complicated by hadronization and final-state interactions (FSI) of the outgoing hadrons, that, among other effects, distort the kinematics of the process. Measurements of proton-nucleus and pion-nucleus interactions (e.g. in protoDUNE [300]) can constrain some aspects of FSI modelling.

Building comprehensive nuclear and hadronic models of neutrino-nucleus interactions, with well-parameterised uncertainties and capable of covering all these features, has proven to be an extremely challenging endeavor. Despite the wealth of available neutrino-nucleus scattering measurements, different effects cannot be experimentally disentangled because the initial neutrino energy is unknown. As a result, different models are typically able to cover only restricted energy regions, and still differ by more than 50% in specific kinematic configurations. Progress in this field relies on supporting both new measurements (notably using tagged neutrino beams or with mono-energetic electron and hadron beams) and a joint experiment-theory effort to develop more accurate nuclear models. European groups are spearheading many current efforts alongside the NuSTEC collaboration [ID135].

Neutrinos at higher energy, from a few GeV up to multiple TeV, are observed in atmospheric neutrino detectors and in neutrino telescopes. A new approach in this high-energy domain has been demonstrated by the FASER [ID23] and SND@LHC[ID63] experiments. These experiments detect the neutrinos produced in proton-proton collisions at the LHC in the very forward direction of the ATLAS experiment, benefiting from a relatively well-known neutrino flux compared to astrophysical neutrinos. Updates of these experiments in the HL-LHC era should collect very large neutrino samples of all flavours. SND@LHC also aims to demonstrate detector technology that could be implemented in the future SHiP beam-dump experiment, which will collect unprecedented number of neutrino interactions (notably of tau neutrinos) in the 10–100 GeV region. Further expansions of this programme are being proposed, including a new large-scale infrastructure at the LHC (FPF [ID19]), hosting several multi-ton detectors to collect millions of TeV neutrino events. Similar facilities are being studied for future hadron colliders, while a muon collider could provide an exceptional laboratory for precision neutrino measurements.

Experiments detecting neutrinos from colliders provide new measurements of the neutrino-nucleus interaction cross-section in the high-energy DIS region. Among other BSM physics such as NSI or HNLs, they will deliver new tests of lepton flavour universality. Within the SM, these experiments are a powerful probe of forward hadron production in proton-proton collisions, by detecting the neutrinos produced in their decays. As an example, neutrinos from charm-hadron decays give access to the poorly constrained PDFs of gluons at low $x$ and of intrinsic charm at high $x$. Such determinations are important for present and future colliders and for astroparticle physics (to address, for instance, the so-called muon puzzle, an excess on cosmic muons from cosmic-ray showers, which may be explained by an enhanced strange quark PDF).

The cross-section of neutrinos at even higher energies, 100 TeV and above, can be measured by neutrino telescopes by looking at the angular distortion of the neutrino flux due to the absorption by the Earth. As demonstrated by IceCube [ID236], this method provides a cross-section determination that is free of flux uncertainties.

## 6.4 Neutrino mass scale in the lab

The KATRIN experiment [ID132] has been leading the field of neutrino mass measurement. It is approaching its final run in 2025 and is expected to reach a final sensitivity to $m_{\nu_e}$ of 300 meV. The development of the next-generation of $\beta$-decay and electron-capture neutrino mass experiments is currently underway. In order to push beyond the 300 meV mass scale, improvements in energy resolution, background reduction—as well as operating a well-understood radioactive source—are mandatory.

Two experiments, HOLMES [ID181] and ECHo [ID258], have pioneered the measurement of neutrino mass from the analysis of the end-point region of the $^{163}$Ho electron capture spectrum. The excitation energy of the daughter atom is measured for each event using multiplexed cryogenic bolometers, leading to a *differential* spectrum. In the case of HOLMES, Transition Edge Sensors (TES) are used, while ECHo relies on Metallic Magnetic Calorimeters (MMC). Recently, small prototypes of both projects have demonstrated good energy resolution and linearity with self-calibrating peaks, as well as excellent background levels as shown in

Table 6.4. Upper limits of $m_{\nu_e} < 19$ eV and $m_{\nu_e} < 27$ eV have been achieved by the ECHo and HOLMES experiments, respectively. The expected sensitivity targets for the upcoming phase of these projects are shown in Fig. 6.7. Scaling to the sub-eV scale, however, presents a number of significant experimental challenges. A minimum of $10^6$ detectors would need to be operated in order to achieve the necessary statistical accuracy. In addition, there is currently no sufficiently precise knowledge of the electron capture spectrum. Advances on both these technical and theoretical fronts are necessary.

On the tritium beta decay front, there are four experimental efforts currently underway: Project-8 [ID225], QTNM [301], KATRIN++ [ID132], and PTOLEMY [ID28]. Each of these experiments make use of a differential spectral measurement (in contrast with the integrated signal of KATRIN) in order to ascertain the neutrino mass, thus providing a statistical boost to enhance sensitivity. In addition, these experiments are planning to use atomic tritium as source. An atomic source removes an irreducible systematic uncertainty that comes from using molecular tritium, mainly related to the rotational and vibrational final states following the tritium beta decay. PTOLEMY uses atomic tritium embedded on graphene, which allows for surface loading of tritium, but final-state effects due to delocalization are expected. The development of an atomic tritium source requires production, cooling and trapping with extreme high purity levels. Leveraging on the Karlsruhe Tritium Laboratory (TLK), Project-8, QTNM and KATRIN++ collaborations are joining efforts in the global consortium Atomic Pathfinder at TLK, to address this common challenge. The planned sensitivity to the neutrino mass of the upcoming phase of the tritium experiments is also shown in Fig. 6.7.

| | $\sigma_E$ | Bg |
|---|---|---|
| $^3$H | | |
| KATRIN (I) | 1eV | 0.15 cps |
| Project-8 (D) | 1.7eV | $3 \cdot 10^{-10}$ cps/eV |
| $^{163}$Ho | | |
| HOLMES (D) | 7eV | $2 \cdot 10^{-9}$ cps/eV |
| ECHo (D) | 8.3eV | $4 \cdot 10^{-11}$ cps/eV |

Table 6.4: Performance figures of running differential (D) and integrating (I) neutrino-mass detectors: energy resolution, $\sigma_E$, and background level in counts-per-second (cps), and per eV in the case of differential measurements.

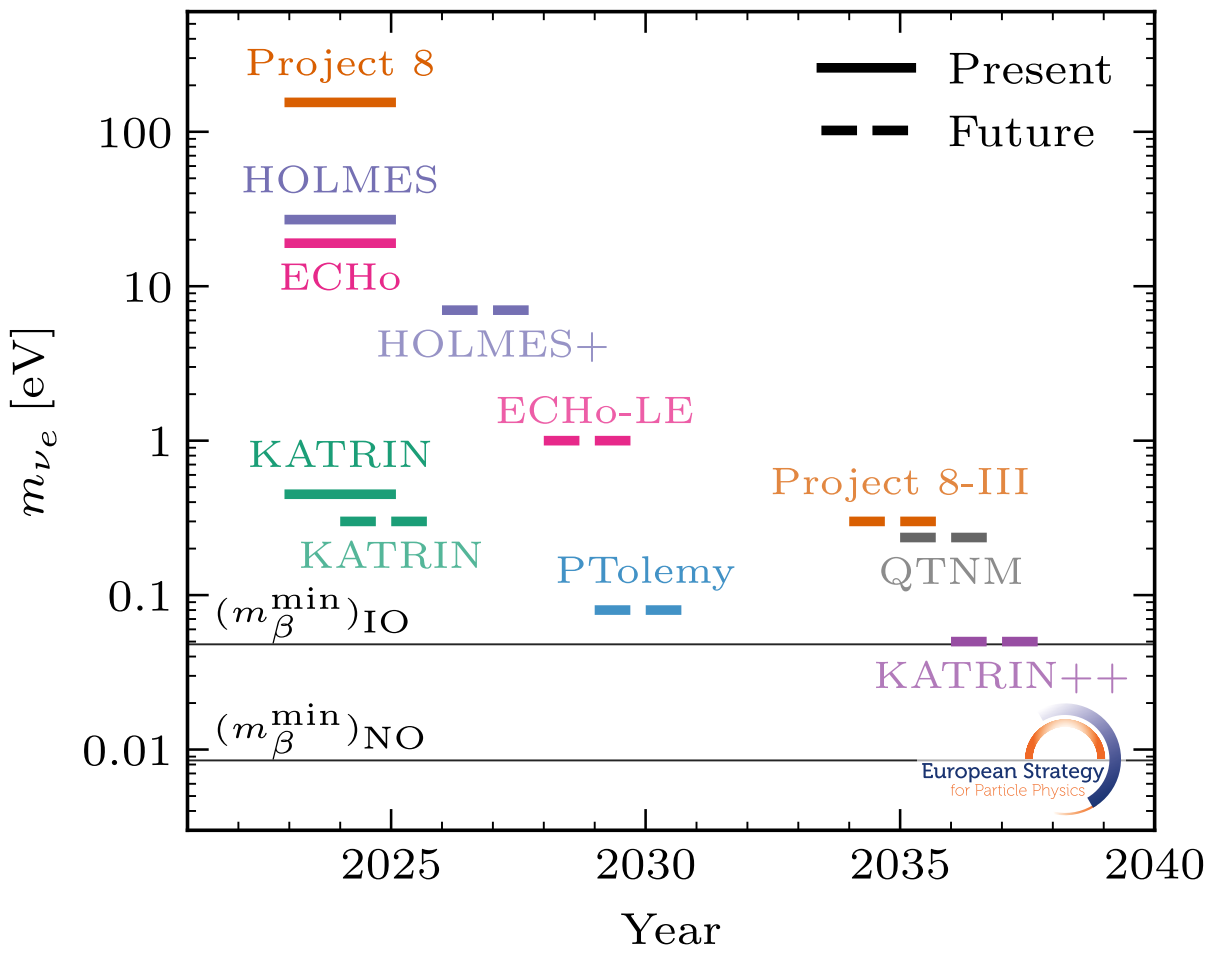

Fig. 6.7: Upper bounds on $m_{\nu_e} \equiv \sqrt{\sum_i |U_{ei}|^2 m_{\nu_i}^2}$ of running experiments (solid) and expected sensitivities by future upgrades/projects (dashed).

Regarding energy measurements, Project-8 has recently pioneered a new technique to measure the electron energy using cyclotron emission from radiating electrons (CRES). It allows for a high precision differential measurement resulting in an excellent background level. The technology has been demonstrated in a small prototype, $\mathcal{O}(\text{mm}^3)$, using molecular tritium and a limit of $m_{\nu_e} \leq 155$ eV has been obtained. Despite the excellent energy resolution inherent in the technique, the experiment still faces significant experimental challenges, including developing a complex magnetic field environment for electron and atomic trapping, as well as the need for ultra-low-frequency noise measurements across large volumes. KATRIN++ is also exploring various ideas for a differential spectrometer with tagging based on CRES or superconducting quantum sensors.

Finally, the ultimate goal of the PTOLEMY project [ID28] is to measure the relic cosmological neutrino background from neutrino-capture in a gram-level tritium source embedded in graphene. To demonstrate the technology, a measurement of the $\beta$-decay end-point spectrum is underway using 1–100 $\mu$g. Challenges are significant even for this more modest goal. PTOLEMY has recently shown 90% tritium loading efficiency on graphene, but the detector concept aims at high energy resolution using an in-flight RF tagger and an EB drift filter that needs to be experimentally demonstrated.

In summary, significant R&D efforts need to be carried out in the next few years to demonstrate the novel technologies that are needed to improve the state-of-the-art laboratory measurements of the neutrino mass. The interplay of laboratory and cosmological measurements is entering a very interesting era, and a direct measurement in the laboratory is of the highest priority.

## 6.5 Neutrino-less double beta decay

The most promising way for establishing the existence of a new physics scale connected to neutrino masses is the measurement of $0\nu\beta\beta$. Under the assumption that the only source of lepton number violation is a neutrino Majorana mass, the half-life of this process is predicted to be

$$(T^{0\nu}_{1/2})^{-1} = G^{0\nu}|M^{0\nu}|^2|m_{ee}|^2 \,, \tag{6.2}$$

where $G^{0\nu}$ is a well known phase factor, $M^{0\nu}$ are poorly known nuclear matrix elements and $m_{ee} \equiv \sum_i U^2_{ei} m_i$. The search for this very rare process (lifetimes $T^{0\nu}_{1/2} \gtrsim 10^{26}$ years) is an extremely challenging experimental endeavour. The nuclear transition creates two electrons depositing all the available energy, $Q_{\beta\beta}$, in the range of a few MeV. It is therefore a monochromatic peak at the end point of the standard double-beta ($2\nu\beta\beta$) spectrum, an energy region which is also populated by several ubiquitous natural radioactivity backgrounds. A large $Q_{\beta\beta}$ is important to maximize the number of events since $G^{0\nu} \propto Q^6_{\beta\beta}$, and to ensure that the signal peaks at energies above the largest natural radioactive backgrounds. Only some nuclear isotopes feature a sufficiently high $Q_{\beta\beta}$ to be viable, and in most cases, their natural abundances are small, so enrichment is mandatory to achieve large target masses, a process that remains an important challenge for some isotopes. The control of backgrounds in the region of interest is also extremely challenging. Operating underground with strict requirements on radiopurity, combined with an excellent energy resolution, $\Delta E$, is crucial for minimising backgrounds. Under conditions of background subtraction in the region of interest, the sensitivity to the half-life scales as $T^{0\nu}_{1/2} \propto \sqrt{\frac{Mt}{b\Delta E}}$, with $Mt(\mathrm{kg \cdot y})$ the exposure[4], $b(\mathrm{cts/kg \cdot y \cdot keV})$ the background and $\Delta E(\mathrm{keV})$ the energy resolution.

Various technological approaches, listed in Table 6.5, meet the experimental challenges mentioned above. The best limit of $T^{0\nu}_{1/2} > 3.8 \times 10^{26}$ y (90%CL) on $0\nu\beta\beta$ has been reached by the KamLAND-Zen experiment [302], which exploits gaseous $^{136}$Xe dissolved in organic liquid scintillator. This approach enables an extremely large exposure at the expense of a limited energy resolution, as shown in Table 6.5. This is problematic given the relatively low value of $Q_{\beta\beta}$, 10 keV above a radioactive line from the radon decay chain. In order to decrease the background level, an upgrade is being deployed (KamLANDd2-Zen) with improved resolution and radiopurity. A different strategy is followed in experiments based on xenon Time Projection Chambers (TPC), which feature better intrinsic resolution and can profit from additional handles to suppress the background. The EXO collaboration [303] uses a liquid (enriched) Xe TPC with double read-out (charge and light): the anti-correlation between the ionization and scintillation signals ensure good energy resolution, while the ratio between the two channels, combined with the fiducial self-shielding, enables a good background control. This approach will be further pursued with improved radiopurity and larger mass by the nEXO [304] and XLZD [ID175] experiments. The latter, which is optimized for dark matter searches, plans to use a very large mass of natural Xe, that contains a $\sim 9\%$ fraction of $^{136}$Xe. Backgrounds have been estimated in [305]. Finally, the NEXT project [ID87] uses $^{136}$Xe in high-pressure gaseous TPCs, which achieves better energy resolution and enables topological discrimination of signal and backgrounds thanks to the reconstruction of the two electron tracks, as shown by the NEXT-White demonstrator. The presently running NEXT-100 experiment will confirm the scalability, and optimize the design (e.g. optimal pressure, compromising between the largest possible mass and the effect of charge diffusion) that will be finally deployed in the ton-scale experiment NEXT-HD. The AXEL project [306] is also pursuing this technology, implementing a novel amplification system, that has recently achieved $< 1\%$ FWHM energy resolution.

[4] The exposure in this formula should be understood as the effective exposure that takes into account the fiducial volume, the enrichment fraction, etc.

Another excellent isotope is $^{76}$Ge which has a relatively low $Q_{\beta\beta}$ but can be exploited in high-purity, highly-enriched germanium semi-conductor detectors, which feature unrivaled energy resolution. The technology has demonstrated excellent performance in GERDA [307] and MAJORANA [308], and is being exploited on a larger scale by LEGEND-200 [ID270]. The combined limit on the lifetime has reached $T^{0\nu}_{1/2} \geq 2.8 \times 10^{26}$ y (90% CL) [309]. LEGEND-1000 [ID270] aims to push the technology to the ton-scale with improved suppression of external backgrounds thanks to shielding with underground argon. The main challenge is the scalability, which requires the combination (assembly and read-out) of numerous sub-detectors.

A different approach, pioneered by the CUORE experiment, uses bolometers with the $^{130}$Te isotope [310]. A major breakthrough in this technology has been the use of the $^{100}$Mo isotope, combining the traditional heat channel with the scintillation light signal. This approach reduces very effectively the $\alpha$ background, provides excellent resolution and features the largest $Q_{\beta\beta}$ so far. The major challenge is ensuring crystal radiopurity and scalability, due to the complex industrial process of production of the bolometric crystals. The technology enabling $\alpha$-particle rejection has been demonstrated by the CUPID-Mo pathfinder [311]; while the infrastructure of CUORE, that will be reused by CUPID [312], has proven well-characterized performance and radioactive background levels. A similar technology is being pursued by the AMoRE experiment [ID197].

Finally, $^{130}$Te is an interesting isotope, being the only one with relatively large (34%) natural abundance. After CUORE, this isotope will be further exploited by the SNO+ experiment [ID125], which consists of a large volume of liquid scintillator loaded with Te, that relies on self-shielding through fiducialization. While it has limited resolution, similar to Kamland-Zen, it can reach, affordably, very large isotope mass (ton scale). The SNO+ project has a broader physics programme which includes the observation of geo- and solar neutrinos. If the approach of Te-loaded liquid scintillator is successful, this could become an interesting option for other present and future very large liquid scintillator detectors.

| Experiment | Isotope | $Q_{\beta\beta}$ [keV] | $T_{1/2}^{0\nu}(0.1\text{eV})$ $\times[10^{26}y]$ | $b$ [cts/kg/keV/y] | $\Delta E$ [keV] FWHM | Isotope Mass [kg] |
|---|---|---|---|---|---|---|
| GERDA | $^{76}$Ge | 2039.06 | 2.8-15.6 | $5.2\times10^{-4}$ | 2.6 | 41 |
| MAJORANA | $^{76}$Ge | 2039.06 | 2.8-15.6 | $6\times10^{-3}$ | 2.5 | 24 |
| LEGEND-200 | $^{76}$Ge | 2039.06 | 2.8–15.6 | $5\times10^{-4}$ | 2.5 | 96 |
| LEGEND-1000 | $^{76}$Ge | 2039.06 | 2.8–15.6 | $10^{-5}$ | 2.5 | 900 |
| EXO-200 | $^{136}$Xe | 2457.83 | 0.8–14.5 | $4\times10^{-3}$ | 69 | 80 |
| nEXO | $^{136}$Xe | 2457.83 | 0.8–14.5 | $7\times10^{-5}$ | 58 | 1800 |
| KamLAND-Zen | $^{136}$Xe | 2457.83 | 0.8–14.5 | $1.1\times10^{-4}$ | 240 | 670 |
| KamLAND2-Zen | $^{136}$Xe | 2457.83 | 0.8–14.5 | $10^{-5}$ | 120 | 720 |
| NEXT-White | $^{136}$Xe | 2457.83 | 0.8–14.5 | $4\times10^{-3}$ | 25 | 5 |
| NEXT-100 | $^{136}$Xe | 2457.83 | 0.8–14.5 | $4\times10^{-4}$ | 25 | 87 |
| NEXT-HD | $^{136}$Xe | 2457.83 | 0.8–14.5 | $4\times10^{-6}$ | 12 | 900 |
| XLZD | $^{136}$Xe | 2457.83 | 0.8–14.5 | $6\times10^{-6}$ | 38 | 5316 |
| CUORE | $^{130}$Te | 2526.97 | 0.4–9.8 | $10^{-2}$ | 8 | 202 |
| SNO+ Phase-I | $^{130}$Te | 2526.97 | 0.4–9.8 | $10^{-4}$ | 230 | 1330 |
| CUPID-Mo | $^{100}$Mo | 3034.4 | 0.4–3.3 | $5\times10^{-3}$ | 7.4 | 2.3 |
| CUPID | $^{100}$Mo | 3034.4 | 0.4–3.3 | $10^{-4}$ | 5 | 240 |
| AMoRE-I | $^{100}$Mo | 3034.4 | 0.4–3.3 | $2.5\times10^{-2}$ | 13 | 3 |
| AMoRE-II | $^{100}$Mo | 3034.4 | 0.4–3.3 | $10^{-4}$ | 10 | 85 |

Table 6.5: Key performance figures for various present and future $0\nu\beta\beta$ experiments: isotope, $Q_{\beta\beta}$ in keV, predicted lifetime range for $m_{ee} \equiv \sum_i U_{ei}^2 m_{\nu_i} = 0.1$ eV [313] (varying the nuclear matrix elements), background level in counts per kg-keV-year, FWHM energy resolution in keV, and the isotope mass used/planned. The backgrounds for the future experiments are estimates. For XLZD, the cosmogenic $^{137}$Xe background is not included since the experimental site has yet to be determined.

In summary, there are multiple experimental techniques able to improve the search of $0\nu\beta\beta$ towards the $10^{27}$ y limit. A comparison of the technologies in the form of the exposure[5] (in arbitrary units) needed to reach a fixed sensitivity to $m_{ee}$ at the 0.1 eV level is shown in Fig. 6.8. Given the relevance for fundamental physics of this process, and in view of the complex nature of the radioactive backgrounds affecting this search, the implementation of multiple, complementary technologies emerges as one of the priorities of the European particle physics community, that has very strong involvement in CUPID, LEGEND, NEXT and XLZD. This endeavour crucially relies on fully instrumented underground laboratories.

The interpretation of the measured half-time in terms of $m_{ee}$ or some other BSM physics requires modelling of the nuclear matrix elements. Theoretical models differ in their predictions by large factors resulting in the bands in Fig. 6.8. Recently, significant theoretical progress has been made with the identification of new short-distance contributions [314] that were previously neglected, and the emergence of *ab-initio* calculations (for a review see [315]). These improvements are already showing better convergence in the predictions and higher confidence in the quoted uncertainties. Precise knowledge of matrix elements is also crucial for exploiting the complementarity of measurements with different isotopes. Finally, $0\nu\beta\beta$ experiments are also providing precise measurements of the $2\nu\beta\beta$ spectrum, which is also of high interest to nuclear physics. The theoretical efforts to improve the matrix element calculations for double-beta processes are strongly encouraged.

[5] In the case of nEXO, their multivariate analysis estimation [304], which shows that the half-life sensitivity scales parametrically not as $(b\Delta E)^{-1/2}$ but as $b^{-0.22}\Delta E^{-1}$, is reported.

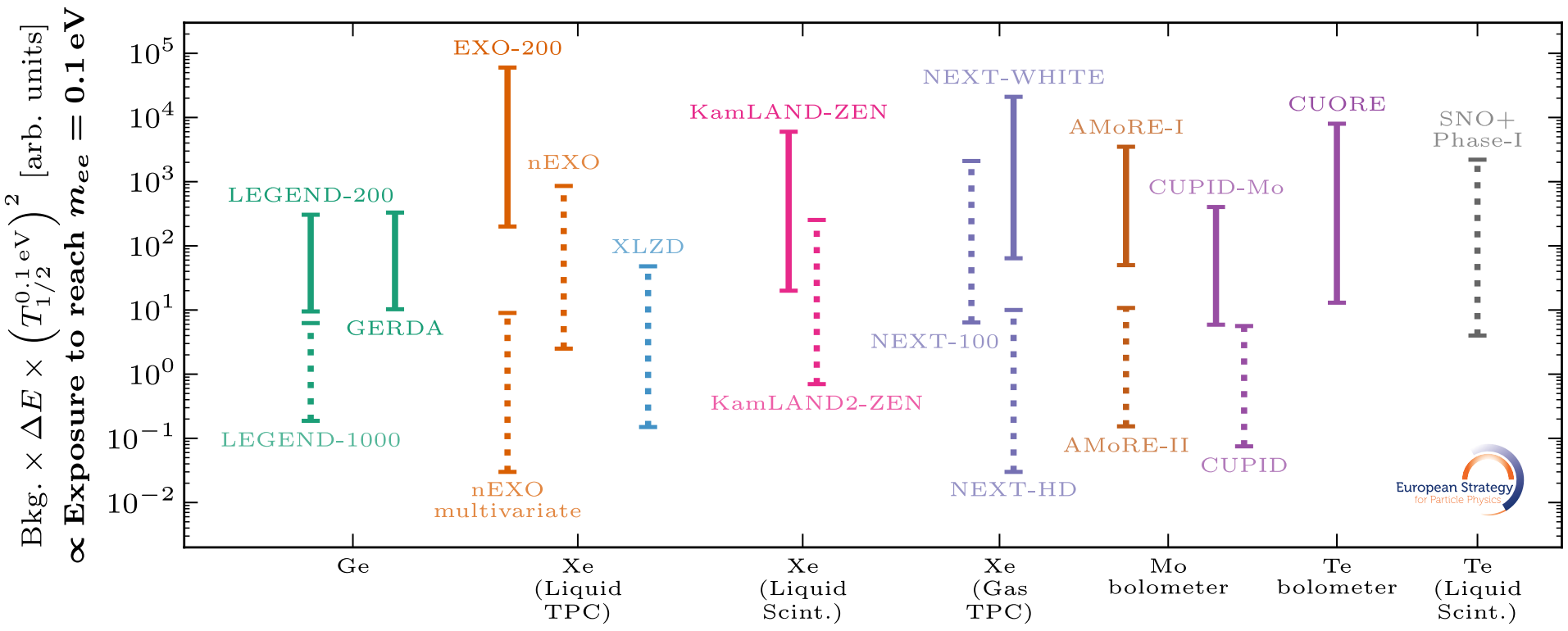

Fig. 6.8: Exposure in arbitrary units required to reach 0.1 eV sensitivity in $m_{ee}$ by the different detector technologies/isotopes, estimated as $(T^{0\nu}_{1/2}(0.1\text{eV}))^2 \times b\Delta E$, from the numbers in Table 6.5. Solid lines are for experiments that have already presented $0\nu\beta\beta$ limits, while dashed lines are for experiments at different stages of development or operation. The range shown represents the uncertainty in the matrix elements.

On a longer time scale, new ideas for further enhancing the $0\nu\beta\beta$ sensitivity are being explored. Tagging Ba ions in xenon detectors would provide a background-free approach [ID87]. The feasibility of using isotopes at even larger $Q_{\beta\beta}$ in bolometric crystals such as $^{96}$Zr, $^{150}$Nd, $^{116}$Cd, and $^{48}$Ca (with the highest end-point energy of all isotopes at $\sim 4.3$ MeV), are also being investigated. In this case enrichment is also a challenge to be addressed.

## 6.6 On BSM searches in neutrino experiments

As explained above, the tiny values of the neutrino masses are suggestive of being the effect of a heavy mediator. If the mass of the mediator (and of all the new states) is sufficiently large, a good description at low energy of this new physics is provided by the SMEFT. The leading operator at dimension $d = 5$ can be probed with $0\nu\beta\beta$, but at $d = 6$ the list of possible new interactions is very long if no additional symmetry assumption are made. In this context, neutrino physics provides very important information.

On the one hand, neutrino experiments set the most stringent bounds on baryon number (B) violating operators. Table 6.6 contains the present and future limits on a selected set of B-violating modes. Improvements by factors of $3-6$ are expected in the lifetimes of the golden proton channels, $p \to e^+\pi_0$ and $p \to \bar{\nu}K^+$, covering a range favoured by generic GUT models [316]. Improvements by orders of magnitude can be achieved in some neutron decay channels, and in neutron-antineutron oscillation searches.

Regarding other operators that modify neutrino observables, a very important class is that of the four-fermion operators including two quarks and two leptons. They induce non-standard neutrino interactions (NSI) usually parameterized phenomenologically as

$$\delta\mathscr{L}_{\text{NSI}} = -2\sqrt{2}G_F\left(\varepsilon^{q,P}_{\alpha\beta}\,\bar{\nu}_\alpha\gamma_\mu\nu_\beta\;\bar{q}\gamma_\mu Pq + \varepsilon^{qq',P}_{\alpha\beta}\,\bar{l}_\alpha\gamma_\mu\nu_\beta\;\bar{q}\gamma_\mu Pq'\right). \tag{6.3}$$

NSIs modify matter effects in neutrino propagation as well as neutrino production and detection, impacting significantly in neutrino oscillations observables.

| Mode | PDG24 | Future | |
|---|---|---|---|
| | ($\times 10^{34}$ years) | ($\times 10^{34}$ years) | |
| $p \to e^+\pi^0$ | $> 2.4$ | $> 7.8$ | Hyper-K |
| | | $> 3.8$ | THEIA |
| $p \to \bar{\nu}K^+$ | $> 0.59$ | $> 1.3$ | DUNE |
| | | $> 3.2$ | Hyper-K |
| | | $> 1.9$ | JUNO |
| $n \to e^+K^-$ | $> 3.2 \cdot 10^{-3}$ | $> 1.1$ | DUNE |
| | ($\times 10^{8}$ seconds) | ($\times 10^{8}$ seconds) | |
| $n - \bar{n}$ | $> 4.7$ | $> 5.5$ | DUNE |
| | | $> 10 - 10^3$ | HIBEAM/NNBAR (ESS) |

Table 6.6: Present limits from the PDG2024 on partial lifetimes at 90%CL of selected proton and neutron $B$-violating decay modes and neutron-antineutron oscillations. The expected limits from future projects are shown assuming the exposures: 1900 Mton·y (Hyper-K), 400 kton·y (DUNE), 200 kton·y (JUNO) and 800 kton·y (THEIA). ESS will have the unique capability to look at $n - \bar{n}$ oscillation in vacuum with the HIBEAM (NNBAR) proposal [ID179].

In the SMEFT, neutrinos are part of the lepton doublet and as a result NSI are accompanied by new interactions of charged leptons, which are often better constrained. Nevertheless, recent analyses [317, 318] have shown that some of the operators that induce NSI are unconstrained or poorly constrained from electroweak and charged flavour observables, while oscillation data and CE$\nu$NS measurements can significantly improve the limits. Neutrino data are therefore relevant and need to be included in global fits of the general SMEFT.

NSIs can also arise from the exchange of mediators below the electroweak scale. At energies below the mass of the mediator, the Weak Effective Field Theory (WEFT), where the massive EW gauge bosons are also integrated out is adequate. It includes NSI as in Eq. (6.3), not necessarily correlated to similar interactions involving charged leptons. Even if correlations with charged lepton interactions are absent, a self-consistent global analysis of NSI is highly non-trivial [319]. For example NSI can change the commonly assumed adiabaticity condition in solar neutrino propagation. Although a global analysis including future constraints has not been carried out at the same level of rigour as with present data, significant improvements on present constrains are expected in future neutrino oscillation experiments [320], and should be pursued. On the other hand, recent analyses have also demonstrated the importance of the forward physics programme at LHC in this context. For instance, the constraints on some of the flavour-changing $\varepsilon$ parameters can be improved in FASER-nu and SND@HL-LHC even beyond the very stringent limits from other flavour observables [321].

It is important to stress that the presence of new states below the EW scale generically induces other effects beyond NSIs, since these states can be produced in particle collisions. The phenomenology is, however, model dependent and it is difficult to draw general conclusions. Minimal models, such as the SM portals, provide a good starting point to analyze these low-scale new physics scenarios. Neutrino experiments are specially relevant in the context of the neutrino portal, which is equivalent to a low-scale Type-I see-saw model with $n$ extra singlet fermions, $N$:

$$\mathscr{L}_{\text{Neutrino Portal}} = \mathscr{L}_{\text{Type-I}} = -i\bar{L}Y\sigma_2 H^* N - \bar{N}MN^c. \tag{6.4}$$

The model gives generically $3+n$ Majorana neutrino states, three of which are light , $m_{\text{light}} \propto$

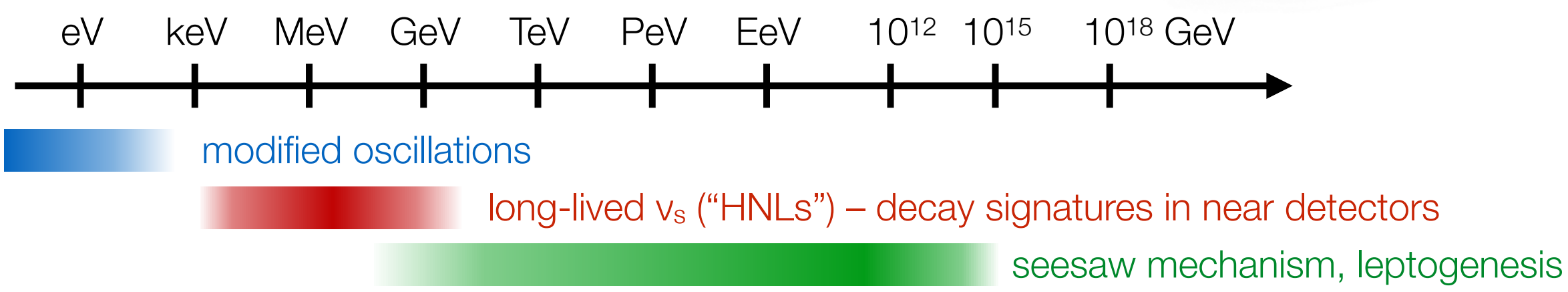

Fig. 6.9: Majorana mass scale in the neutrino portal or Type-I see-saw model.

$Y^T \frac{v^2}{M} Y$ (with $v$ the Higgs vev), and $n$ are heavy with masses $m_{\text{heavy}} \propto M$. The light neutrino mixing matrix is only approximately unitary up to corrections of $\mathcal{O}(\frac{m_{\text{light}}}{m_{\text{heavy}}})$. The known neutrino masses do not allow for disentangling $Y$ and $M$ and therefore the scale of the heavy states is presently unconstrained, but different ranges give rise to interesting phenomenology, see Fig. 6.9.

If the scale $M$ is in the $\mathcal{O}(eV)$ range, the heavy neutrinos can participate in oscillations, that is they are light sterile neutrinos, and may help resolve some of the anomalies found in neutrino oscillation experiments. An intense scrutiny of these anomalies has been carried out in recent years with several dedicated experiments. The MiniBoone anomaly [322] has not been confirmed by MicroBooNE [323], and a tension between the two experiments is emerging. This will be further explored in the SBN programme at Fermilab. The reactor anomaly [324, 325] seems to be explained by systematic uncertainties in the neutrino flux [326], while several experiments NEOS [327], STEREO [285], PROSPECT [328], DANSS [329], SOLID [330] and Neutrino4 [331] have explicitly searched for an $L/E$ dependence of reactor neutrinos. All except one experiment have excluded an $L/E$ variation compatible with the reactor anomaly. Finally, while the Gallium anomaly has been reinforced by the recent results of the BEST experiment [332], the possible explanation based on light sterile neutrinos has been excluded with high significance by various experiments, notably KATRIN [333]. The presence of light sterile neutrinos is nevertheless best tested through neutrino oscillation experiments, either exploring an $L/E$ dependence or searching for non-unitarity effects in the mixing matrix.

If the scale of $M$ is larger, there exist heavy neutral leptons that couple via charged and neutral currents to the three lepton flavours. Depending on their mass range, constraints on their mixings to the $e, \mu$ or $\tau$, from $\beta$-decay, reactor, $0\nu\beta\beta$, rare meson decays, gauge boson and Higgs boson decays in colliders or electroweak precision tests can be obtained. Very strong constraints can also be derived from cosmological observables for masses below GeV. The parameter space for masses $\gtrsim$ GeV is broadly consistent with successful baryogenesis, and will be further probed in future experiments. Future limits from various neutrino experiments (e.g. KATRIN/Tristan, DUNE, Hyper-K, etc.) will be competitive in the future. A compilation of future constraints in the large mass range can be found in Fig. 8.19 in Chapter 8.

Beyond the neutrino portal, neutrino experiments, particularly near detectors in neutrino oscillation experiments that are located near intense beam dumps, or neutrino experiments close to colliders, will significantly contribute to searches for long-lived particles and general feebly-interacting or dark sectors. Major contributions are expected from several facilities in the near future: the short-baseline neutrino programme at Fermilab (SBND ID232 and ICARUSID226), the beam dump experiments at the SPS at CERN [ID145], the forward physics programme at the LHC [ID19, ID23, ID63], coherent neutrino scattering measurements, and the near detector complex at DUNE [ID118] and Hyper-K [ID238] with 1-2 MW proton beams. Nevertheless, the dark sectors could well be more complex than the minimal SM portals, leading to a richer phenomenology [ID250]. Experimental searches for dark sectors should be as model independent as possible, focusing as much as possible on rare topologies and avoiding model-specific correlations. Examples include searches for long-lived particles where lifetime and production rate are kept independent of each other, even if they are correlated in most models, or looking for semi-visible decays of dark states. Other examples of richer portals involve new gauge symmetries, such as left-right models, that can be strongly constrained with $0\nu\beta\beta$ searches [334].

## 6.7 Conclusions

The origin of neutrino masses remains one of the most important open question in particle physics, and holds profound implications for the matter–antimatter asymmetry, and the evolution of the Universe. Upcoming neutrino oscillation experiments—using accelerator, reactor, and atmospheric neutrino sources—are expected to make major progress in the determination of neutrino masses and mixings in the next decade. They can determine the neutrino mass ordering and can discover leptonic CP violation across a large region of parameter space. Fully realizing this potential requires significant improvements in our understanding of neutrino fluxes and cross-section systematic uncertainties. Measurements from highly capable near detectors, hadron-production experiments and possibly dedicated beams, complemented by advances in nuclear theory, will be pivotal in this effort. European contributions to DUNE and Hyper-K are anchored in the CERN Neutrino Platform and strongly rely on this facility for both commissioning and physics exploitation. The forward physics program at the LHC and neutrino telescopes enables measurements of neutrino interactions at TeV–PeV energies, with direct implications for proton parton distribution functions and cosmic-ray modelling, while the emerging field of coherent neutrino scattering offers novel opportunities to probe neutrino properties. Substantial progress is also anticipated in determining the absolute neutrino mass scale. Complementary approaches include cosmological observations and measurements of the end-point spectra of beta decay and electron-capture processes. Several projects aiming to improve upon KATRIN's sensitivity are underway, and the novel detector concepts involved will require sustained R&D. The discovery of $0\nu\beta\beta$ would demonstrate that neutrino masses are linked to a new physics scale. Multiple isotopes and technological strategies are being pursued at various underground laboratories. Given the scale of the challenge and the profound implications of such a discovery, independent confirmation from several experiments, together with progress in determining nuclear matrix elements, will be essential. Neutrino experiments and their infrastructures also provide unique opportunities to search for new physics, including proton decay, light sterile neutrinos, heavy neutral leptons, and non-standard neutrino interactions. The theoretical community has played a vital role in the interpretation of neutrino data, in the prediction of neutrino cross sections, and in identifying opportunities for new physics searches. Close collaboration between experiment and theory will remain indispensable to fully exploit the scientific potential of future neutrino experiments.

# Chapter 7

# Cosmic Messengers

Both the hot plasma of the early Universe and astrophysical environments in and around stars and their remnants provide extreme conditions and thus unique laboratories to test our understanding of the Standard Model of particle physics and possible extensions. Most likely, the energy densities in the primordial plasma exceeded the energies achievable in human laboratories, and have left us with cosmic relics, such as the observed matter-antimatter asymmetry and dark matter, which require extensions to the Standard Model. Astrophysical environments in turn can act as cosmic accelerators, producing charged particles at extremely high energies, which can be trapped in galactic magnetic fields for astronomical times before these particles or their decay products reach the Earth. These processes are natural probes of rare processes, providing an avenue for probing very weakly coupled new physics. A major challenge in exploiting these cosmological laboratories is the indirect nature of the probes, which renders the task of accurately understanding and modelling the complex astrophysical environments involved very complicated. First-principle methods often reach their limits and need to be supplemented by data-driven approaches, relying on observations in different frequency bands, observations of different types of cosmic messengers or complementary laboratory measurements.

Recent years have seen remarkable progress in our ability to make precision observations of the Universe in different ways. The James Webb Space Telescope (JWST) [335] delivered its first deep field images in 2022, pushing the understanding of the earliest forming galaxies and black holes while the Euclid satellite [336] saw first light in 2023, and is currently in the process of establishing the most precise 3D map of the matter structures in the Universe, complemented by the ground-based observatories DESI (early data released in 2024) and Vera Rubin (first light in 2025). Neutrino telescopes such as IceCube, KM3NeT and GVD [ID 236, 249] are mapping the galactic and extragalactic neutrino sky for the first time, and the hunt for the most energetic cosmic neutrinos has only just begun. Direct cosmic ray detectors have reached percent-level accuracy in measurements of cosmic antimatter fluxes, searching for relics of dark matter decays. The gravitational detector network LIGO-Virgo-Kagra (LVK) [ID122] has by now observed over a hundred mergers of black holes in the frequency range around 100 Hz, while at much lower frequencies pulsar timing arrays are seeing indications of a Gravitational Wave (GW) signal from supermassive black hole binaries, formed during the collision of galaxies. Measurements of the cosmic microwave background (CMB) and galaxy surveys have reached the precision to probe particle physics properties such as the neutrino mass and the number of effective degrees of freedom. Moreover, their precision measurements of the ΛCDM parameters at different scales are now providing a stress-test for the cosmological standard model.

Building on this success, upgrades of current observatories as well as new observatories are expected to come online in the coming years, with scientific questions, methods and technologies which are deeply intertwined with particle physics. In this chapter the cosmic messengers and observatories most relevant for this Strategy Update are reviewed, before concluding with some considerations on the synergies with particle physics. Roadmaps for astroparticle physics in Europe have been outlined in the APPEC [ID276] and EuCAPT [ID215] inputs.

## 7.1 High-energy gamma rays

Electromagnetic radiation has been the backbone of astronomy for centuries and continues to drive our exploration of the Universe. In this process, it has proven crucial to exploit the full bandwidth of electromagnetic radiation, since both the sources and the absorption of the radiation, when propagating through the intergalactic and interstellar medium as well as the Earth's atmosphere, vary significantly with frequency. In this section the focus lies on high-energy gamma-rays, emitted in extreme astrophysical environments, while CMB radiation and galaxy surveys are discussed below. A key science target is the search for dark matter, or more generally searches for light BSM sectors, which are discussed in Chapter 9.

The detection strategy is based on the imaging of the electromagnetic showers produced by gamma rays. Current instruments include the Large Area Telescope [337] on the Fermi satellite, the water Cherenkov telescopes HAWC [338] and LHAASO [339] as well as the atmospheric Cherenkov telescopes HESS [340], MAGIC [341] and VERITAS [342]. Building on this technology, the Cherenkov Telescope Array Observatory (CTAO) [343] has been established as a European Research Infrastructure Consortium in preparation of a next-generation atmospheric observatory at two sites in the northern and southern hemispheres, which could be complemented by the water Cherenkov telescope SWGO [344] in the Atacama dessert. Notably, CTAO is expected to detect or close the parameter space for the remaining simplest WIMP dark matter models, see Fig. 9.7 in Chapter 9.

## 7.2 Neutrinos

Neutrinos are ubiquitous in the Universe. Some of these neutrinos point back to astrophysical objects ranging from stellar bodies like the sun to the most extreme environments in the Universe, such as Supernova (SN) explosions. Atmospheric neutrinos are produced in the damping of cosmic rays on the atmosphere, while nuclear reactors and the Earth itself are also powerful sources. In the lowest energy range, the cosmological neutrino background is a relic from the Big Bang, frozen at the time of neutrino decoupling when the temperature of the Universe was of the order of MeV. At the highest energies, a flux of cosmogenic neutrinos from proton collisions with the background radiation is expected. These predicted steady neutrino sources are shown in Fig. 7.1. Several of these cosmic neutrino sources have been observed. Solar and atmospheric neutrinos have been instrumental in the discovery of neutrino masses and mixings. Solar neutrinos still provide the best sensitivity to one of the mixing angles, and atmospheric neutrino experiments have excellent sensitivity to the mass ordering. Reactor and accelerator neutrino experiments have superseded the precision in most of the remaining oscillation parameters. Nonetheless, solar neutrinos still contribute significantly to understanding and modelling the Sun, while geo-neutrinos are providing useful information on the Earth's interior.

Some $O(20)$ neutrinos from the supernova (SN) explosion SN1987A were detected by the old generation of neutrino experiments, notably Kamiokande. In spite of the limited number of events, this observation has helped constrain many non-standard properties of neutrinos as well as models of supernova explosions, that are not yet well understood. The future observation of neutrinos from a close-by SN will hopefully help establishing a "standard supernova model", which is a crucial step to fully exploit their excellent potential in constraining new physics.

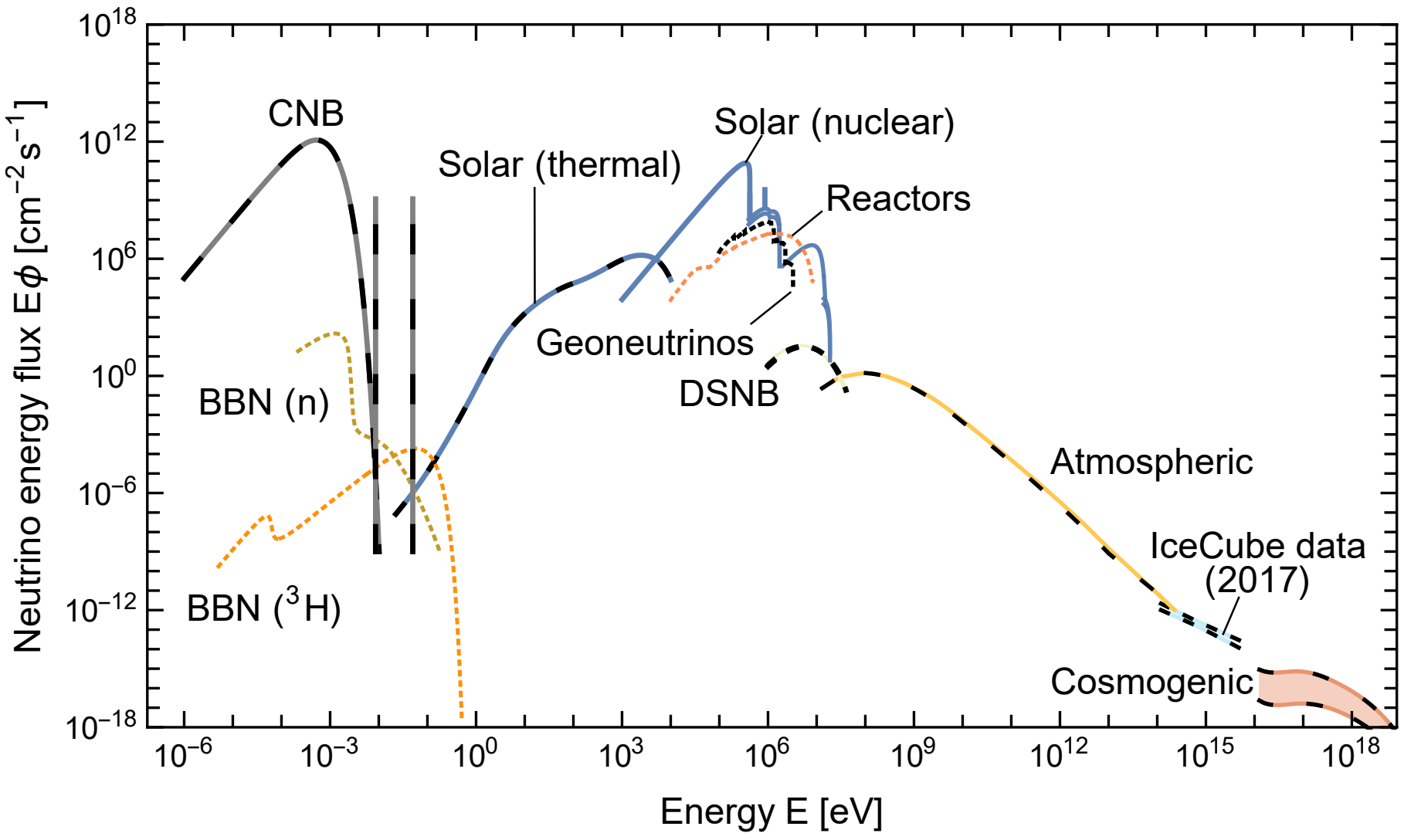

Fig. 7.1: Fluxes of neutrinos (solid), antineutrinos (dashed or dotted) or both (superimposed solid+dashed) integrated over directions and summed over flavours from different known sources: cosmic, solar, Earth, reactors, atmospheric, astrophysical and cosmogenic. Figure taken from Ref. [345].

Much larger neutrino signals from a galactic SN are expected in the next generation neutrino experiments: DUNE, Hyper-K and JUNO. Hyper-K alone could detect $O(10^4)$ neutrino events from a SN at 10 kpc, and will provide a pointing precision of $1^\circ$. DUNE will be able to gather a clean sample of $\nu_e$, while JUNO will have the best energy resolution and lowest threshold. IceCube and KM3NeT will also provide timing and contribute to pointing by triangulating the timing from different detectors. An alert system, SNEWS [346], is in place to make sure that the SN signal is not missed by any observatory on Earth. These observations will accurately measure the time and energy dependence of the SN neutrino flux, a crucial benchmark for SN simulations.

A diffuse supernova background from all the past supernovas is also expected. The gadolinium-loaded Super-Kamiokande detector is starting to explore the favoured target region. Intriguingly, a small excess has been building up in recent analyses [347], so this discovery could be around the corner. If such an excess becomes a signal, 5-10 events per year are expected in Hyper-K and in JUNO.

The cosmic neutrino background has not been directly observed, although efforts are being made to develop the required technology, e.g. PTolemy. On the other hand, abundant indirect evidence of this relic neutrino background exists from the abundances of light elements in Big Bang Nucleosynthesis, the perturbations in the cosmic microwave background, or the large scale distribution of galaxies. At the highest energies, IceCube has discovered a high energy neutrino flux, whose origin is still largely unknown. A very small fraction of this flux has been associated to two known sources [349, 350], while evidence of a diffuse background from the galactic plane has also been found [351]. KM3NeT on the other hand has observed [348] the highest energy neutrino event ever recorded with an estimated energy of $E = 220^{+570}_{-100}$ PeV (see Fig. 7.2). The flavour composition of the neutrino flux is one of the golden measurements for particle physics. It can help discriminate between the main production mechanisms but also provide stringent tests of non-standard neutrino properties such as neutrino decay or Lorentz violation. Along with the discovery and study of additional astrophysical sources of neutrinos, a detailed spectral and flavour composition of the neutrino flux is thus amongst the top priorities. Improvements of existing instruments are underway, with the imminent upgrade of IceCube [ID263] and the completion of the KM3NeT [ID249] and GVD [352] projects. The European KM3NeT experiment will have an enhanced angular resolution and a good view of the galactic centre, and will thus be particularly suited for the discovery of new neutrino sources. The P-ONE [ID53] project in Canada and three projects in China (Hunt [353], Trident [354] and Neon [355]) are in the R&D phase, with demonstrators being deployed and aiming at instrumenting multi-km$^3$ fiducial volumes. Neutrino astronomy is still statistically limited, and these new experiments are complementary, providing different views of the sky. Several of these observatories (Antares, IceCube/IceCube-Gen2, KM3NeT, Baikal/GVD, P-ONE and RNO-G [356]) are working together in a Global Neutrino Network (GNN).

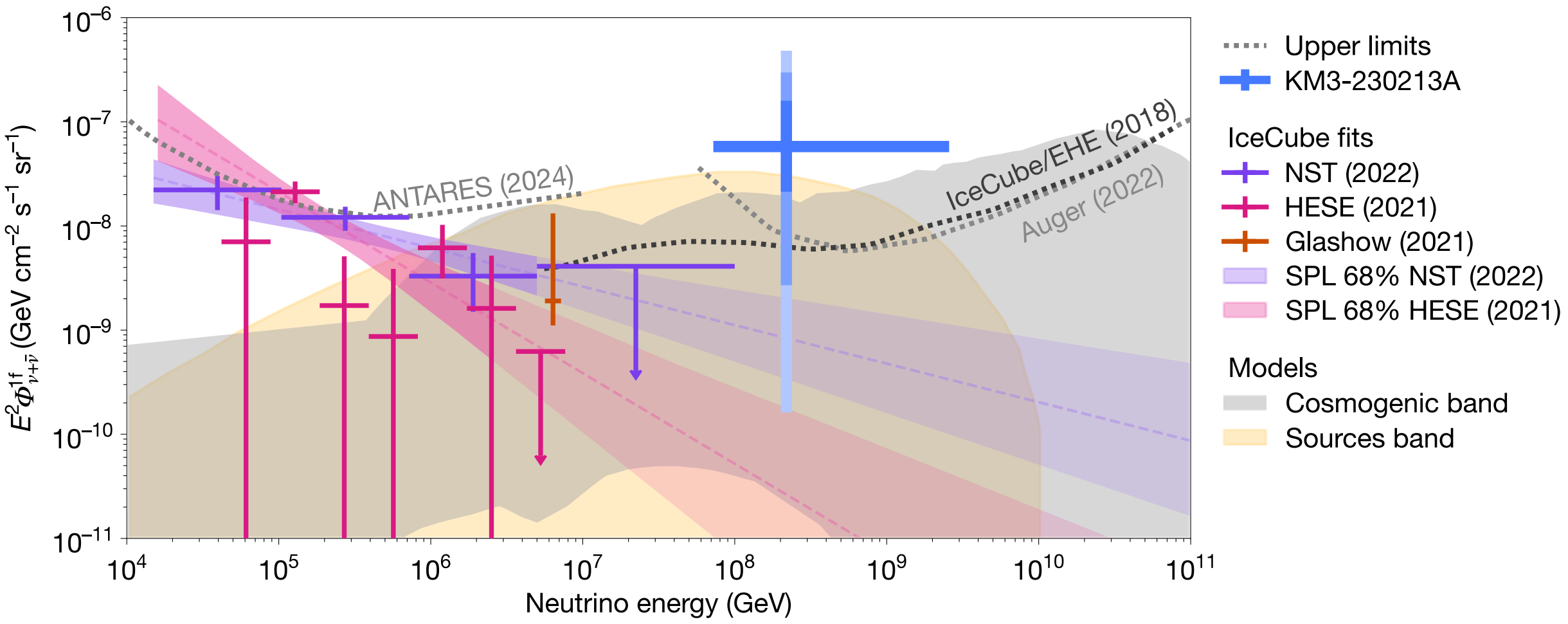

Fig. 7.2: Astrophysical neutrino fluxes measured by IceCube and KM3NeT. Limits from ANTARES, IceCube and Pierre Auger are also shown. The shaded regions cover the expected fluxes of astrophysical and cosmogenic origin. Figure adapted from Ref. [348].

Many new initiatives are being planned to extend the searches to higher energies (EeV-ZeV), where new detection strategies will be needed. The best constraints in the highest energy range are those from IceCube, Auger and ANITA. Plans exist to improve by at least two orders of magnitude these limits, including measuring ground-based air showers (GRAND [357], Trinity [358], Tambo [ID272], BEACON [359]), radar in ice (RET-N [360]), radio in ice (RNO-G, IceCube-Gen2), air showers in balloons or in space (ANITA [361], EUSO-SPB2 [362], PUEO [363], POEMMA [364]), or radio signals from the moon (SKA [365]). The detection of cosmogenic neutrinos in the next generation of experiments looks promising.

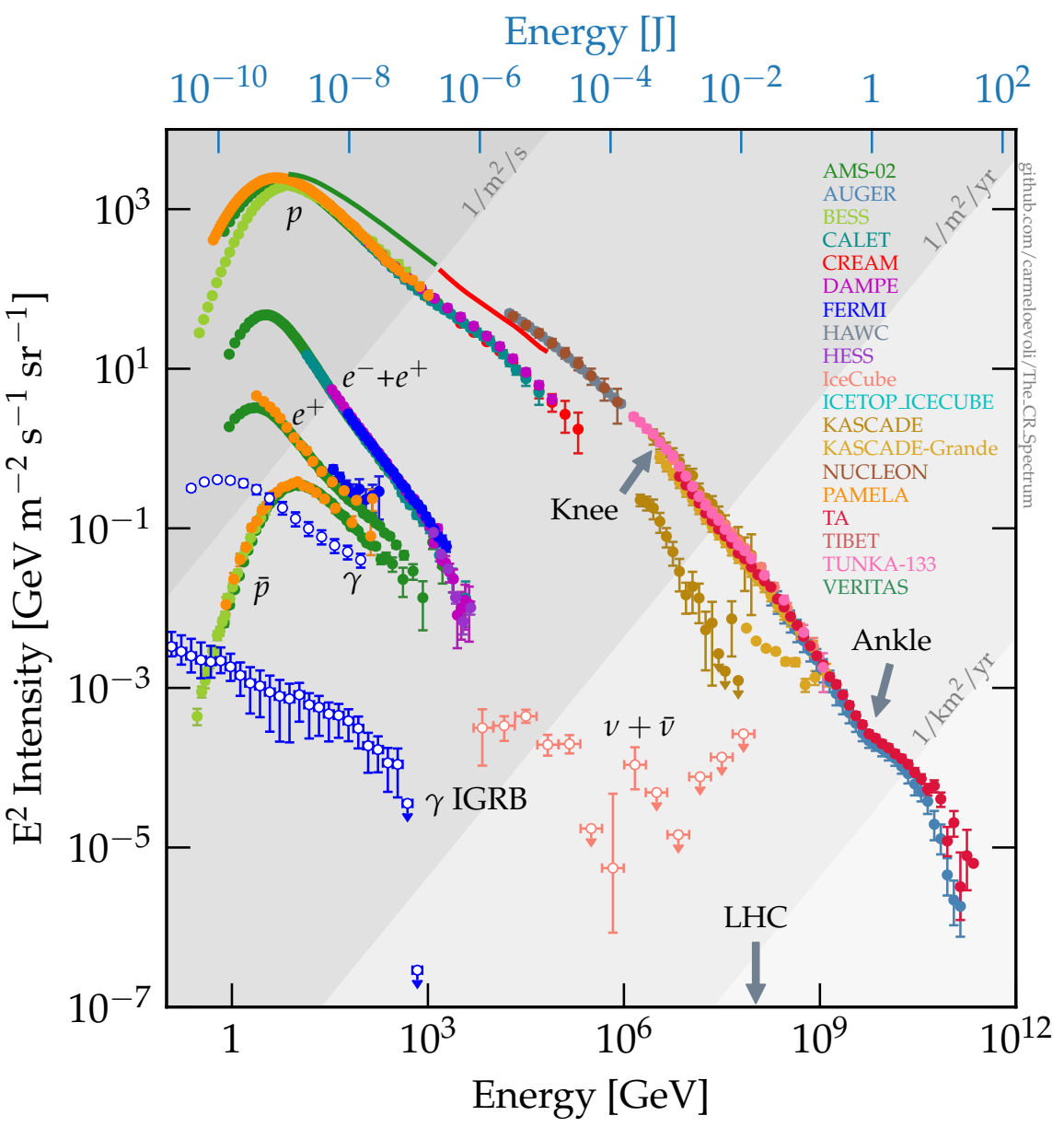

Fig. 7.3: Flux and energy of different cosmic ray constituents and $\gamma$-rays, colour coded by detector. Cosmic rays with energies below about $10^6$ GeV are typically of galactic origin and are probed by direct detection above the Earth's atmosphere, whereas higher energies indicate extragalactic origin lead to air showers that are measured by ground-based observatories. Figure from Ref. [366].

## 7.3 Cosmic rays

Cosmic rays are charged particles that are produced and accelerated in galactic or extra-galactic astrophysical environments and are observed over many decades in energy, see Fig. 7.3. As charged particles, they can spend astronomical time-scales trapped by magnetic fields, making them excellent probes of rare interactions that alter their spectrum in a characteristic way. Their production mechanisms and their interactions with the interstellar medium probe particle physics at both high energies and at weak couplings to potential new physics. In particular, an excess of antiparticles in cosmic rays could arise as a result of dark matter decay.

Direct cosmic ray detectors are particle physics detectors operating above the Earth's atmosphere, either from a balloon (e.g. HELIX [367], GAPS [368]), from a space station (AMS [369], CALET [370]) or from a satellite (DAMPE [371]), targeting cosmic ray particles up to about $10^6$ GeV. These instruments will be taking data until about 2030; proposals for successors include TIGER-ISS [372] and HERD [373] on the Chinese space station. A key challenge in the interpretation of these datasets is the accuracy with which cross-sections of cosmic ray particles are known. Cosmic-ray detectors measure the sum of all processes leading to a given final-state particle and thus knowledge of all processes (during production, propagation and detection) is required. For example, AMS has measured the antiproton flux at 10 GeV at the percent level, whereas cross-section uncertainties exceed 10% [366]. The interpretation in terms of, e.g., dark matter decay, is hence currently limited by the latter. Fixed-target experiments such as LHCb-SMOG [374], AMBER (COMPASS) [375], NA61/SHINE (NA49) [376] and nTOF [377] have provided key input to reduce cross-section uncertainties in galactic CR propagation (spallation), and will be critical in improving these searches [ID89] also in the near future, see also Sect. 4.4.

At energies above about $10^6$ GeV, cosmic rays induce hadronic air showers in the atmosphere. These can be observed with ground-based observatories such as the Pierre Auger Observatory in Argentina [ID201] and the Telescope Array [378] in the US. These instruments are studying the arrival directions and mass composition of the highest-energy cosmic rays. The latter has implications for cosmogenic neutrino production (see Sect. 7.2), while the former could reveal a dominant close-by source of ultra-high energy cosmic rays. Prospects for instruments beyond 2030 include the Giant Radio array for Neutrino Detection (GRAND) [357] and the Global cosmic ray observatory (GCOS) [379]. An accurate interpretation of their data will require precise modelling of hadronic interactions in air showers, driven by Monte-Carlo simulations and accelerator data, in particular LHC runs with oxygen and other light ions, as well as data from forward detectors such as TOTEM, FASER, SND and the proposed Forward Physics Facility, see also Sect. 4.4.

## 7.4 Gravitational waves

The first direct detection of GWs by LIGO a decade ago has opened up a new window for the exploration of the Universe. Predicted by general relativity, gravitational waves (GWs) are probes of the most massive and energetic events in our cosmic history: mergers of black holes and other compact objects as well as violent processes in the hot plasma of the early Universe. As such, they are powerful probes of our understanding of gravity, in particular in the strong field regime. Moreover, GWs interact so weakly with ordinary matter that they can traverse the entire Universe essentially unperturbed. As such, GWs carry information about events on cosmic time and distance scales. Crucially, this includes times before the decoupling of the CMB, when the Universe was not yet transparent to photons, and even as early as cosmic inflation. This property has earned GWs the reputation of a "smoking gun"signature of many BSM scenarios involving e.g. first-order phase transitions or the formation of topological defects, see Fig. 7.4.

Currently, the network of ground-based interferometers shaped by LIGO, Virgo and KAGRA (LVK, [ID122]) is in its fourth data taking run. With a peak sensitivity at around 100 Hz, primary targets are mergers of neutron stars (NSs) and black holes (BHs) of stellar origin. To date, more than a hundred BH mergers have been detected, drastically increasing our knowledge of stellar and black hole populations. The one multimessenger NS observation achieved to date has led to new constraints on the QCD equation of state [380], on the mass of the graviton [381], and to an independent measurement of the Hubble parameter [382]. The next generation of these experiments, the Einstein Telescope (ET) [ID198] in Europe and the Cosmic Explorer in the US, are currently under active development. They are expected to significantly extend the science reach of ground-based interferometers, e.g. detecting $10^4$ NS mergers up to a redshift of $z \sim 3$ and detecting all stellar-origin BH mergers in the observable Universe, i.e. a sensitivity reaching the dark ages before the formation of stars. Increasing the current sensitivity to GWs by two orders of magnitude, corresponding to a gain of four orders of magnitude in the energy density of stochastic gravitational wave backgrounds (SGWBs), these detectors are expected to make precision measurements of GW waveforms while simultaneously being discovery machines for exotic astrophysical objects and BSM physics, such as new light bosons or cosmological SGWB sources.

At much lower frequencies, pulsar timing arrays (PTAs) have recently reported evidence of a GW detection at nHz frequencies, likely associated with binary supermassive BHs located in the centres of galaxies. PTAs use radio telescopes to analyse the light emitted by millisec-ond pulsars, which act as cosmic light houses. A GW crossing the line of sight induces a change in the time of arrival of the pulses. This effort is driven by a range of international collaborations: The European and Indian (EPTA + InPTA) [383], Parkes (PPTA) [384], and Chinese (CPTA) [385] pulsar timing arrays as well as the US-led PTA Nanograv [386] and Meerkat [387] (a precursor of the next-generation Square Kilometer Array (SKA) [388]), which collaborate through the International Pulsar Timing Array (IPTA) [389]. Telescope upgrades, increased observation time and cadence dedicated to existing and newly discovered pulsars as well as improved noise modelling are expected to reveal the origin of the currently observed signal and push the exploration of the low-frequency GW sky.

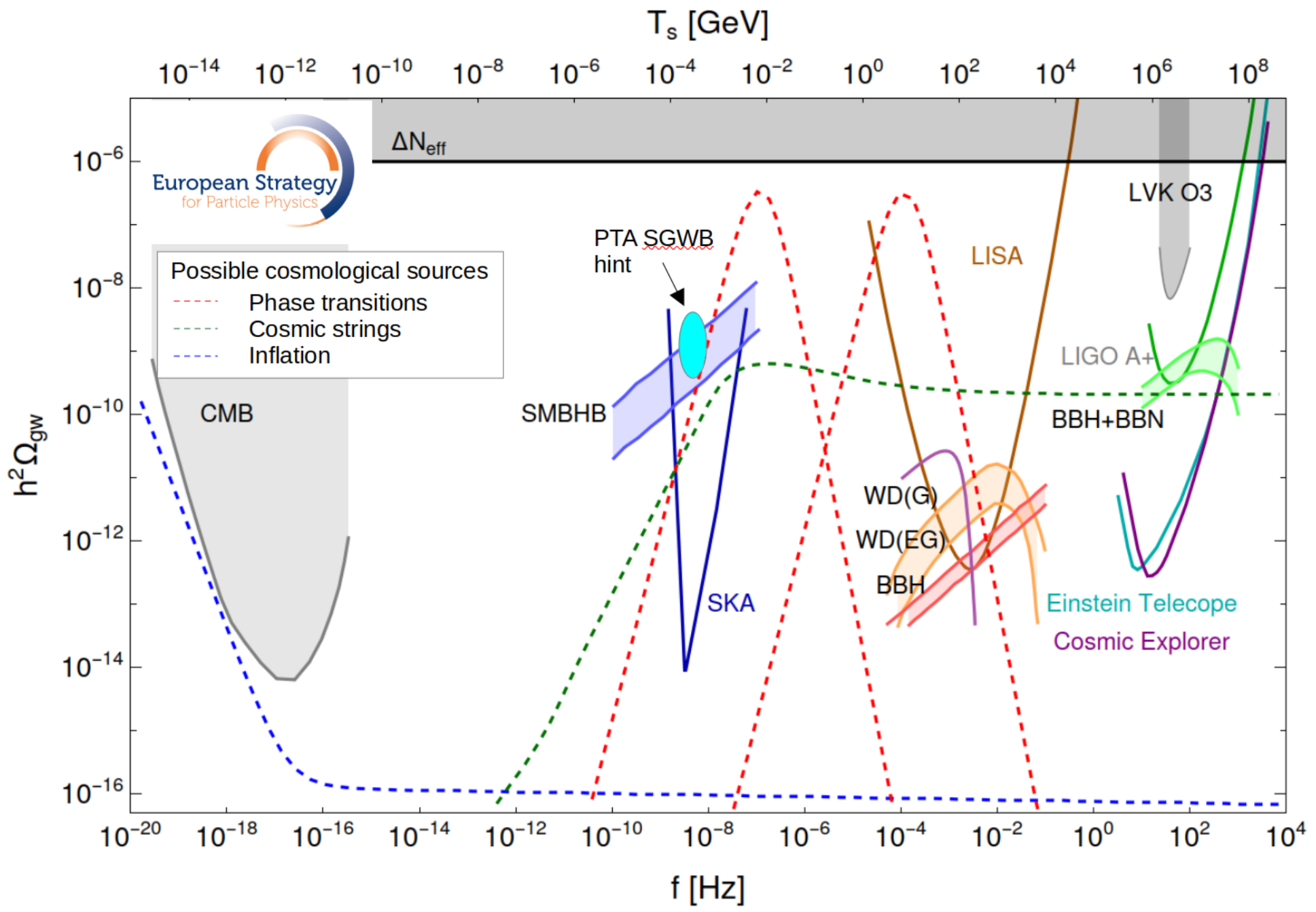

Fig. 7.4: Prospects for SGWB detection. Grey shaded regions indicate existing constraints, while solid curves are 1-year power law integrated sensitivity curves of the next generation of experiments. Shaded regions of matching colour indicate expected astrophysical SGWBs, due to supermassive black hole binaries (SMBHB), binaries of galactic (G) and extragalactic (EG) white dwarfs (WDs), black hole binaries (BBH) and neutron star binaries (BBN). These backgrounds are instrument specific, as increased sensitivity reduces the number of unresolved sources. Dashed lines show some examples of BSM scenarios leading to large SGWBs; shown are first-order weak and QCD phase transitions at temperatures $T_s$, and cosmic strings associated with a symmetry breaking process around $10^{13}$ GeV.

At intermediate frequencies, the space-based interferometer LISA [ID164] is scheduled for launch in the early 2030s. Primary targets are mergers of the heaviest supermassive BHs, as well as binaries of lighter compact objects, including BHs, NSs and white dwarfs (WDs). Reaching up to redshift $z \sim 7$, LISA will be able to probe our understanding of cosmology, performing, among others, an independent percent-level measurement of the Hubble parameter. It is also expected to extend the sensitivity to the energy density in GWs in this frequency range by seven orders of magnitude and is a remarkable discovery tool for BSM scenarios that predict SGWBs. Notably, the frequency range of LISA coincides with the Hubble horizon at the electroweak scale, making it a probe of extended scalar sectors that could render the electroweak phrase transition in the SM a first-order transition.

As for all other cosmic messengers, the exploration of different frequency bands yields complementary information, motivating further research into closing the existing gaps in the observational programme. Ultimately, the goal is to map the GW sky from the smallest frequencies imprinted in the CMB to the highest frequencies associated with processes in the very early Universe, using technologies ranging from radio telescopes to atom interferometers [ID37] and quantum sensing [ID260].

## 7.5 Multimessenger astronomy

Starting from the galactic supernova 1987A (SN1987A), which was observed in optical light and associated with the emission of $\mathcal{O}(20)$ neutrinos, multimessenger astronomy has proven to be a powerful tool in understanding astrophysical sources, breaking degeneracies in the modelling of source and transport, and setting limits by comparing the fluxes in different cosmic messengers. SN1987A has been fundamental to our understanding of core-collapse supernovae, but has also been instrumental in understanding neutrino properties and setting limits on light dark sectors, providing some of the strongest limits still to this date [390]. Its neutron star remnant is in fact still under investigation today, most recently by the JWST [391].

The simultaneous emission of neutrinos and photons remains at the core of the quest of determining the origin of extragalactic neutrinos, with so far only two sources identified as active galactic nuclei (AGNs) [350, 392]. Upcoming neutrino telescopes, see Sect. 7.2, promise a better understanding of such AGNs, which are considered prime candidates for the most powerful cosmic accelerators and have also, by comparing primary and secondary electromagnetic emission of blazars, delivered hints of cosmic magnetic fields [393]. The angular resolution of electromagnetic and neutrino telescopes will allow a localization and characterization of these sources, which is in turn key to understanding the origin and production mechanisms of high-energy charged particles, i.e., cosmic rays. Once their production and propagation are sufficiently well modelled, they can probe rare decay and interaction processes, key for dark sector searches (see Sect. 7.3)

Another recent example of multimessenger astronomy is the observation of the neutron star merger GW170817 [394], observed in gravitational waves by LVK and, due to an early-alert system, in a plethora of electromagnetic telescopes, from the first gamma-ray burst in Fermi to the afterglow weeks after the merger. As mentioned above, this single event has led to leading constraints on fundamental physics such as the QCD phase diagram and the speed of GWs. Increasing the number of detectors in the network (Kagra, LIGO India) and thus the capability of sky-localization from GWs, as well as increasing the sensitivity (Einstein Telescope, Cosmic Explorer) will enable the systematic observation of more such events, pushing our understanding of these fundamental physics questions.

Besides neutron stars, also mergers of supermassive black holes surrounded by gas clouds are expected to produce an optical counter part. These 'bright sirens' allow a simultaneous measurement of the luminosity distance (through GWs) and the redshift (through the electromagnetic emission), and thus a direct measurement of the expansion rate of the Universe, i.e. the Hubble parameter. LISA [ID164] expects to achieve a precision of better than 10%, thus providing a fully independent measurement which may help resolve the long-standing Hubble tension. Moreover, so-called 'dark sirens', for which the optical counterpart is not observed, can also be used for cosmological measurements by using cross-correlations with galaxy catalogues to determine possible host galaxies.

This non-exhaustive list of examples demonstrates the rich physics which can be unlocked with multimessenger astronomy, but also the challenges that need to be overcome, in particular collaboration between different observatories including early-alert systems, open exchange of data and accurate modelling of complex astrophysical environments using information gathered in multiple different ways.

## 7.6 Cosmic relics

The hot big bang model predicts a range of cosmic relics which decoupled from the cooling thermal plasma of the Universe at different stages of its evolution. These include isotropic components such as the cosmic microwave background (CMB), the cosmic neutrino background (C$\nu$B) and GW backgrounds, the dark matter abundance, the primordial light element abundance and the average matter-antimatter asymmetry. It also predicts that small anisotropies present in the primordial plasma result not only in the observed CMB anisotropies but also seed structure formation, leading to the observed distribution of both dark and visible matter today. Over the past decades, the observation of these cosmic relics, notably through CMB observations and galaxy surveys, has led to the emergence of $\Lambda$CDM, the standard model of cosmology. The current generation of experiments has reached for the first time the sensitivity required to stress-test this cosmological model across different scales, while reaching a precision to cosmological observables that entails sensitivity to particle physics parameters, notably the neutrino mass, the number or relativistic species and the scale of cosmic inflation. These observations are thus informing the particle physics landscape.

The most precise CMB observations to date stem from the legacy of the ESA satellite mission PLANCK [395], together with the ground-based observatories such as ACT [396], SPT [271], Bicep/Keck [397] and the Simons Observatory [398]. The next generation of ground-based observatories are being developed under the umbrella of the US-led initiative CMB S4 [399], while the Japanese space agency JAXA is preparing to launch the next-generation satellite mission LiteBIRD [400]. Proposals for a spectral distortion measurement, successor of the satellite-based instrument FIRAS operating in the early 1990s, highlight a gap in the current plans for future CMB observatories. The CMB S4 + LiteBIRD programme aims to test a compelling class of inflation models including Higgs inflation, to detect or rule out any light relativistic particles that decoupled after the start of the QCD phase transition, and to measure or severely constrain the impact of the neutrino mass in cosmology with an uncertainty comparable to the minimal neutrino mass inferred from neutrino oscillations [ID252].

Recent years have seen a dramatic increase in the precision of galaxy surveys. The two satellite missions Euclid (ESA) and SPHEREx (NASA) were launched in 2023 and 2025, respectively, with the first Euclid data release expected in 2026. On the ground, the Dark Energy Spectroscopic Instrument (DESI) released its second data set in early 2025, measuring for the first time the scale of baryon acoustic oscillations across a wide redshift range. This measurement rivals the accuracy of CMB experiments on cosmological parameters, measured in an independent way at very different scales. While roughly agreeing, a detailed analysis of the remaining discrepancies, possible systematic uncertainties and possible indications for new physics is currently under investigation. This applies particularly to the upper bound on the sum of the neutrino masses, which is starting to display some tension with neutrino oscillation experiments. While it is too early to draw definitive conclusions, these measurements have demonstrated that the instruments have reached the sensitivity and the statistics required to perform precision cosmology of relevance to particle physics. In the mean time, new instruments are coming online, with the Vera Rubin Observatory receiving first light in 2025 and the Roman Telescope (NASA) scheduled to be launched in 2027. The picture is completed by other tracers of the large-scale structure, including the Lyman-$\alpha$ forest, 21 cm intensity mapping, supernovae, weak lensing and distance ladder measurements of the Hubble parameter.

## 7.7 Synergies with high-energy particle physics

Many of the questions targeted by astroparticle physics and cosmology are also fundamental questions of HEP. Does the SM accurately describe nature in the extreme conditions in the hot thermal plasma of the early Universe, in the dense environments of stars and their remnants and at the ultra-high energies reached in cosmic accelerators? Are there dark or secluded sectors weakly coupled to the SM? What is the nature of dark matter? How does gravity operate at large scales and in the strong field regime, is there a microscopic explanation of the cosmological constant? What were the key steps in the evolution of the Universe, how did the matter-antimatter asymmetry arise, how was dark matter produced and were there symmetry breaking processes beyond those known in the SM? Does an extended scalar sector exist which could render the electroweak phase transition first order?

Addressing these questions requires theoretical and computational tools deeply rooted in both fields. This includes simulation and reconstruction software used for particle showers both in the atmosphere and in particle detectors, quantum field theory tools extended to compute waveforms of GWs and properties of large scale structure, high-dimensional and rapid data analysis required for analysing collisions at the LHC and in complex astrophysical data sets as well as setting limits and interpreting possible anomalies in a consistent theoretical framework. The European Consortium for Astroparticle Theory (EuCAPT, [ID215]), whose central hub is currently hosted at CERN, was founded in the last Strategy Update with the purpose of supporting this research and facilitating exchange.

Moreover, specific measurements in HEP facilities can drive progress in astroparticle searches and vice versa (see also Sect. 4.4). For example, measurements of the forward neutrino detectors at the LHC are key to understanding the cross-section for neutrino production in astrophysical environments, fixed-target experiments have provided (and are hoped to continue to provide) key input to reduce cross-section uncertainties in galactic cosmic ray propagation, the light-ion runs at the LHC are informative for cosmic ray air showers, and a neutrino mass measurement in the lab would provide a key verification point for cosmological models. Vice versa, any indication of BSM physics in cosmology or astroparticle physics would provide a target for laboratory experiments.

Finally, astroparticle and particle-physics experiments share many technology and infrastructural needs. This includes civil engineering and vacuum systems required for colliders and the next generation of ground-based GW experiments; instrumentation, trigger and data acquisition systems used in collider experiments as well as in cosmic ray and neutrino detectors; design and development of sensor and readout technology operating at low temperatures and exposed to radiation, required both for particle-physics experiments and CMB satellite missions; as well as expertise in the management of large international collaborations.

Exploiting these synergies at all levels, from theoretical tools over coordinated measurement campaigns to shared R&D efforts, has the potential to leverage efforts towards common goals.

## 7.8 Summary

Over the past decades, astroparticle physics and cosmology have established themselves as powerful and complementary approaches to study the fundamental questions of particle physics alongside laboratory experiments. The field has rapidly grown in diversity, maturity and precision, achieving key milestones such as stunning precision measurements of the CMB, the first measurement of extragalactic neutrinos, percent-level accuracy in the measurement of antimatter fluxes and the discovery of gravitational waves. Across all cosmic messengers, surveys are being upgraded and new instruments are coming online, which are expected to provide us with a wealth of data in the upcoming years. To stay on this successful trajectory, interdisciplinary and international collaboration is crucial. This includes combining information and sharing alerts across different cosmic messengers to fully exploit the power of multimessenger astronomy, as well as close interactions covering areas from cosmology over astroparticle physics to particle physics and including observatories, laboratory experiments and theory. Moreover, several next-generation projects are currently being developed, such as the next generation of ground and space based CMB observations, large-scale neutrino telescopes, an the next generation of gravitational wave experiments. Continued support from the European community is crucial to the success of these projects.

Chapter 8

# Beyond the Standard Model Physics

## 8.1 Introduction

The Standard Model (SM) represents a triumph of reductionist philosophy applied to some of the most fundamental questions in physics. A combination of experimental ingenuity and theoretical insight has led to a simple picture of a quantum field theory that describes the dynamics and interactions of the known fundamental particles. Yet, facing this enormous success—whose capstone was the experimental discovery of the Higgs boson by ATLAS and CMS in 2012—we are still left with a number of unanswered questions, many of which pertain to the origins of our own existence.

The discovery of the Higgs boson raises as many questions as it answers: what deeper theory explains electroweak symmetry breaking (EWSB) and the nature of the Higgs boson itself? Next-generation colliders will provide crucial insights. The Higgs boson also lies at the heart of the SM flavour structure, where generic extensions typically induce large corrections. This points to a possible link between the microscopic dynamics of the Higgs sector and the unexplained patterns and hierarchies of flavour. These patterns highlight the special role of the third generation, especially the top quark. New physics coupling preferentially to third generation particles, still largely unexplored, can be effectively probed at a future $e^+e^-$ collider, in particular through Tera-$Z$ and $t\bar{t}$ threshold runs.

Moving beyond the terrestrial, further gaps remain in our understanding of the Universe. Observations of the cosmos have led to the discovery of dark matter (DM) via its gravitational interactions, demanding a particle-physics explanation. Future colliders, in combination with direct detection experiments and astrophysical observations, can help pin down the nature of DM (see Chapter 9). We also lack knowledge of the earliest times in the Universe, when electroweak symmetry might have been restored. Theories addressing these questions often lead to signatures that can be probed at a future electroweak scale factory, or in upcoming gravitational wave experiments. Some answers may reside in a hidden sector, connected to the SM by portal particles potentially discoverable at the HL-LHC, future colliders, or complementary facilities, including neutrino beams, whose ultimate goal is to unveil the origin and the scale of neutrino mass generation (see Chapter 6).

In the modern era, we have come to interpret the SM as a bottom-up effective framework. This perspective enables tests across a wide range of scales, from flavour and electroweak precision observables, to measurements of Higgs boson couplings, to kinematic distributions in many processes. Together, these complementary probes can map the landscape of possible SM extensions, and as emphasized throughout this chapter, provide remarkably strong indirect sensitivity. Extracting bounds from these measurements requires precise predictions for both signal and background processes. Perturbative and non-perturbative effects can impact their interpretation, highlighting the need for a strong precision theory programme to support the experimental BSM searches.

In this chapter, we provide representative benchmarks illustrating the experimental reach of proposed future facilities. These benchmarks are not intended to cover every possibility for how nature may reveal itself, but rather to provide a motivated sample of simple models addressing the above questions and yielding diverse collider phenomena. Where possible, we complement plots for specific benchmarks with general lessons on the potential of future experiments for similar physics.

Our organising principle is that most new physics relevant to the open problems of the SM remains linked to the electroweak scale and cannot be arbitrarily decoupled. This leads to a clear message: a programme that combines unprecedented precision with direct exploration at energies around the 10 TeV scale would cover much of the parameter space where answers are most likely to emerge, from the origin of EWSB to the nature of DM. In charting this terrain, the observation of even one single robust deviation from the SM would mark the first step of a revolution in particle physics; conversely, the absence of any deviation would sharply constrain ideas about the microscopic origin of the SM.

The chapter is organized as follows. In Sect. 8.2 we address the question most directly tied to the weak scale: the microscopic origin of the Higgs sector. We consider whether the Higgs could be composite, part of a larger supersymmetric framework, or linked to new states neutral under some SM forces. Section 8.3 examines the structure of the Higgs sector itself and its implications for the early Universe. New forces, including resonances and interactions tied to the SM flavour structure, are discussed in Sect. 8.4. Finally, Sect. 8.5 presents four benchmark scenarios introducing new portals that may provide the first experimental hints of a hidden sector, testable through complementary methodologies. We conclude in Sect. 8.6. Throughout, we highlight connections with other chapters, particularly those on Electroweak, Flavour, and Dark Matter/Dark Sector physics.

## 8.2 Origins of the weak scale

The discovery of the Higgs boson provided a triumphant indication that the Brout-Englert-Higgs (BEH) mechanism accounts for the masses of fundamental particles in the SM. But this discovery now poses deeper questions about the microscopic origin of the Higgs potential and the phenomenon of EWSB itself.

The Higgs potential governs EWSB, but its defining parameters—including the sign of the quadratic term—cannot be predicted from first principles in the SM. In this respect, the BEH mechanism for EWSB is strongly reminiscent of the Ginzburg-Landau effective theory governing the superconducting phase transition. Ultimately, the Ginzburg-Landau model was understood to be a phenomenological parameterization of a more complete microscopic description, the BCS theory of superconductivity, from which the parameters of the Ginzburg-Landau model can be derived in terms of more fundamental constants. We now face the same challenge for the BEH mechanism: can we uncover the microscopic theory of EWSB which predicts the sign of the mass parameter along with the shape of the Higgs potential?

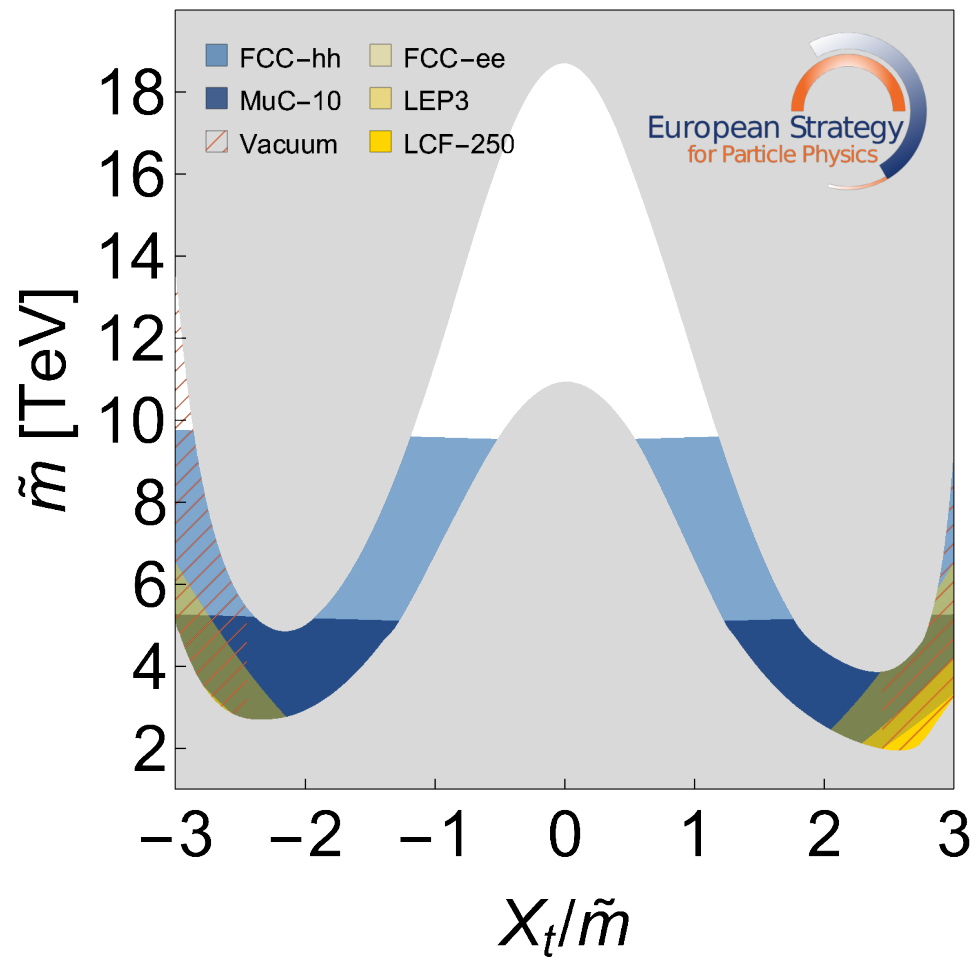

Fig. 8.1: Parameter space compatible with measured Higgs mass $m_h = 125$ GeV as a function of average stop soft mass $\tilde{m} \equiv (\tilde{m}_{Q_3}\tilde{m}_{U_3})^{1/2}$ and mixing parameter $X_t \equiv A_t - \mu \cot\beta$ for $\tilde{m}_{Q_3} = \tilde{m}_{U_3}$. Here $\tan\beta = 20$, $\mu = 1$ TeV; all other soft masses are 5 TeV, and the Higgs mass is computed using `FeynHiggs 2.19.0` [401] at two loops with NNLL resummation. The band between gray regions indicates compatibility with the measured Higgs mass within the `FeynHiggs` $2\sigma$ uncertainty [402]. The blue coloured regions indicate the projected 95% exclusion limit by a given collider on the mass of the lightest stop eigenstate, assuming large mass splitting to the LSP. The yellow regions indicate the projected 95% exclusion limit from the measurement of the $T$ parameter at FCC-ee, LEP3, and LCF-250. The hashed red region is disfavoured by vacuum instability.

There are many candidate theories that could explain the origins of the Higgs potential. Ultimately, experimental exploration at colliders with energies higher than the electroweak scale is the only way to determine the path chosen by nature. Among the leading candidates are SM extensions with additional global or spacetime symmetries, which make the Higgs potential calculable via new particle masses and couplings. In Composite Higgs models, the SM Higgs doublet emerges as a pseudo-Nambu-Goldstone boson (pNGB) of a global symmetry spontaneously broken by new strong dynamics. In this scenario the Higgs potential can be calculated in terms of the scales of explicit and spontaneous global symmetry breaking. In supersymmetric (SUSY) models, an enhanced spacetime symmetry relates SM particles to superpartners whose spin differs by 1/2. The theory remains weakly coupled at all scales, naturally giving rise to a SM-like Higgs. SUSY makes the Higgs potential calculable in terms of the SM couplings and masses of SUSY partner particles.

In both scenarios, as the scale of new physics is raised above the weak scale, it becomes increasingly difficult to understand how the observed parameters of the Higgs potential are generic predictions of the microscopic theory. In other words, the low energy predictions become extremely sensitive to small variations of the fundamental parameters. This notion of *naturalness*, validated in effective theories tested across scales ranging from quarks to the cosmos, suggests that new physics may appear around the TeV scale. Some explanations for the Higgs potential are so predictive that they provide a strong target for experimental searches without leaning on expectation from naturalness. For example, in the Minimal Supersymmetric Standard Model (MSSM), the Higgs mass is sharply predicted from the superpartner spectrum, particularly the stop, with only mild dependence on other superpartners. In particular, the light Higgs quartic coupling is fixed at tree level by electroweak gauge couplings and grows logarithmically with stop masses. Consequently, the observed Higgs mass favors stops above the TeV scale, while naturalness—requiring the measured Higgs vacuum expectation value as a generic outcome of the spectrum—prefers them below it. Thus, the measured value $m_h \simeq 125\,\text{GeV}$ points to stops likely lying beyond HL-LHC reach, in tension with perfect naturalness, but accessible to future colliders. Figure 8.1 shows the stop masses $\tilde{m}_3$ and mixing $X_t$ compatible with the Higgs mass (band reflects current SM parameter uncertainties) compared with collider reaches. Compatibility near the TeV scale occurs only with large stop mixing. Moreover, the outstanding precision electroweak programme of FCC-ee [ID233, ID242] probes the least fine-tuned region of this parameter space. This provides a concrete demonstration of the power of an integrated programme, combining precision and energy frontiers, in advancing our understanding of the origin of the weak scale. In the following subsections, we introduce benchmarks illustrating the main phenomenological consequences of these scenarios.

### 8.2.1 Compositeness

The low-energy phenomenology of a new composite sector can be broadly characterized by the mass scale of new composite resonances $m_*$ and the coupling constant $g_*$, which together govern the scale of spontaneous global symmetry breaking, $f \sim m_*/g_*$. Compositeness can be tested through two complementary strategies: direct searches for new states at the scale $m_*$ and indirect searches for new effective interactions.

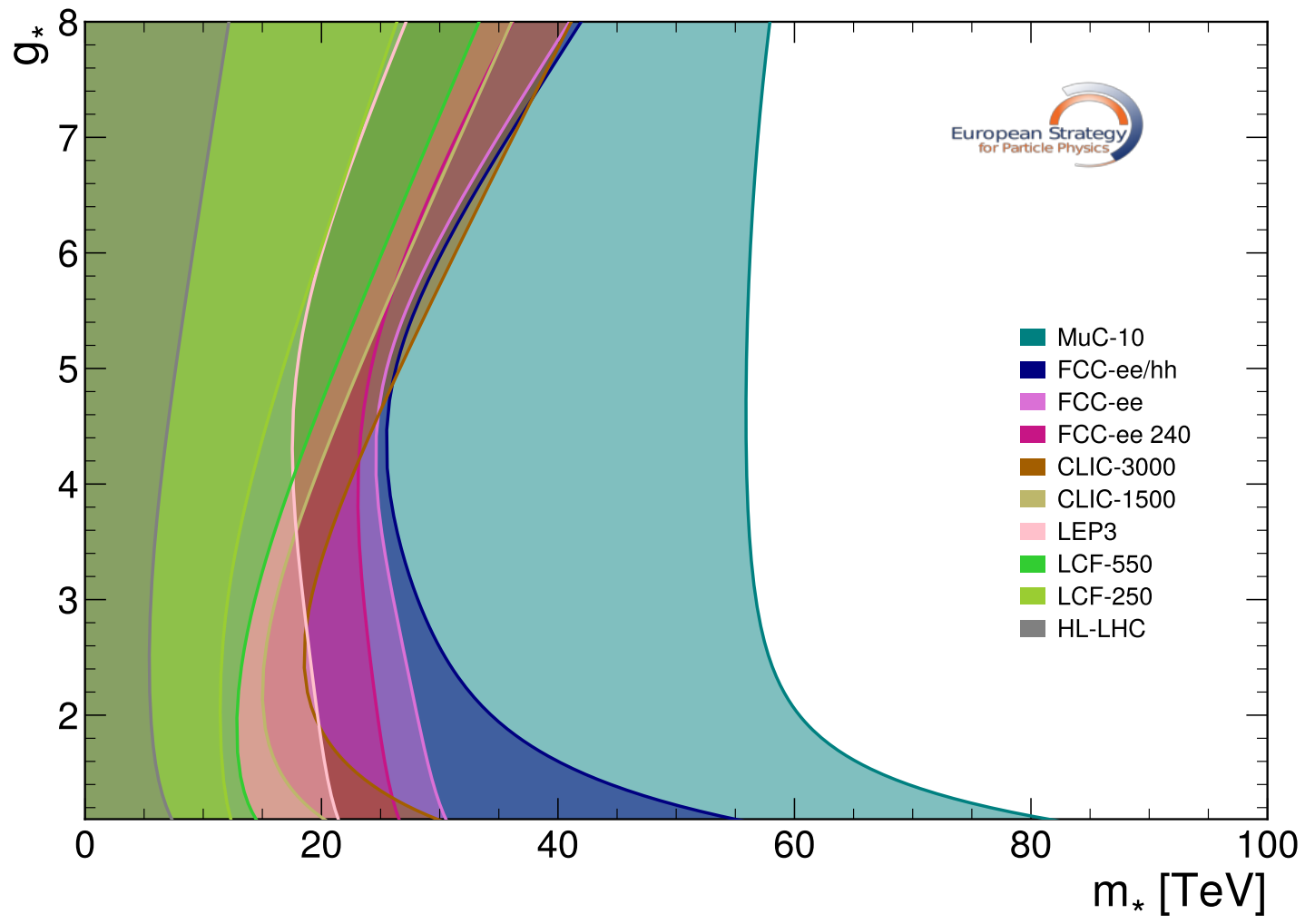

Fig. 8.2: Exclusion reach on the flavor universal Composite Higgs model.

When the scale of the new strong sector is much higher than the weak scale, it leaves an imprint at low energies that can be captured using the SMEFT, following the framework of the "Strongly Interacting Light Higgs" [403], that organizes the expectations for SMEFT operators in terms of $m_*$ and $g_*$. Figure 8.2 shows the projected reach for such indirect effects in the $(m_*, g_*)$ plane for various collider options, under a RG-evolved EFT interpretation.

The figure illustrates the reach of low-energy $e^+e^-$ precision (e.g. FCC-ee [ID233, ID242]) and high-energy reach (e.g. 10-TeV muon collider [ID207]) in constraining operators like $C_H$ and $C_W$, probing compositeness scales well above 10 TeV. This analysis shows that future colliders could probe such scales with deviations of order $m_H^2/m_*^2 \sim g_{\text{SM}}^2/(16\pi^2)$, which is a compelling theoretical target for such models; see also Sect. 8.2.3. Top quark compositeness can also be tested using the same strategy, making the reach even stronger. The improvement on the reach is significantly model dependent, but sizable for each machine.

A complementary probe is to directly search for the new vector boson resonances, analogous to the $\rho$ of QCD, with mass near the scale $m_*$ and coupling strength (to the longitudinal modes) set by $g_*$. Current LHC bounds already place these new resonances beyond the direct reach of future $e^+e^-$ colliders. However, it may be possible to produce them resonantly at future hadron or muon colliders, highlighting the complementarity between machines: indirect hints from $e^+e^-$ experiments could be confirmed through direct production at higher-energy facilities.

References [404] and [ID242] present studies of such spin-1 resonances in composite Higgs models. Figure 8.3 shows the result for Model E as defined in these works. Hadron colliders are sensitive to spin-1 resonances in various channels, typically reaching up to ∼0.5 of the $pp$ center-of-mass energy, a conclusion supported by current LHC results [405, 406]. Projections for HL-LHC and FCC-hh [ID227, ID247] suggest a reach of up to 5–7 TeV and 30–50 TeV, respectively, depending on the model. Again, we see that future colliders could probe the parametrically interesting regime where $m_*$ is larger than the weak scale by more than a loop factor. The region probed directly by FCC-hh overlaps with that accessible to lower-energy $e^+e^-$ machines, enabling direct confirmation of indirect hints at lepton colliders. Moreover, FCC-hh has indirect sensitivity to spin-1 states through the $W$ and $Y$ parameters, providing additional insight into the nature of new physics. This indirect reach also helps cover gaps of $e^+e^-$ machines, which are mainly sensitive via the $S$ parameter and can be blind in scenarios such as Model D of Ref. [404] and [ID242], particularly if hadronic $Z$-pole observables underperform expectations.

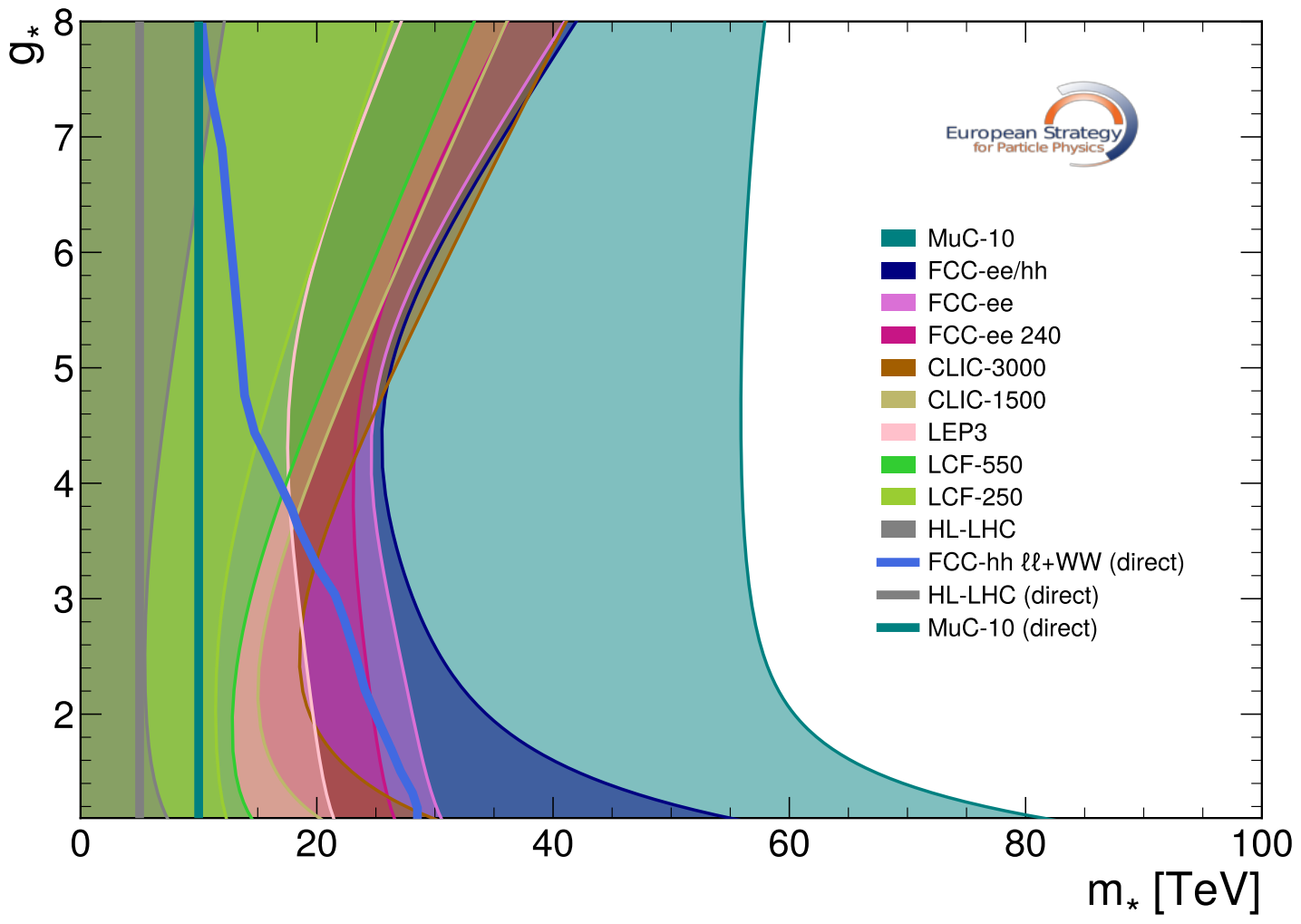

Fig. 8.3: Indirect and direct reach on resonances of Model E, corresponding to a triplet $(\mathbf{1},\mathbf{3})$ of $SU(2)_L \times SU(2)_R$ of resonances, in the $SO(5) \to SO(4)$ composite Higgs model defined in Refs. [404] and [ID242]. The vertical line for MuC-10 is from direct production of the resonance, the curved line corresponds to the indirect sensitivity. FCC-hh has direct sensitivity in the $VV$ and the $\ell\ell$ final states as well as indirect sensitivity from high-$p_T$ Drell-Yan from the $W$ and $Y$ parameters. LEP3 and FCC-ee are indirect bounds and benefit from the combination of all foreseen energy runs.

Finally, a generic prediction of composite Higgs models is the presence of fermionic top partners near the weak scale. These states are essential to the top-quark coupling and play a central role in generating the Higgs mass. For concreteness, we consider a benchmark with an $SU(2)_L$-singlet top partner and examine the complementary reach of the HL-LHC, FCC-hh, and a 10-TeV muon collider (MuC). We assume a common coupling $\kappa$ between the top partner and the electroweak bosons ($Z$, $W$, $H$). Top partners can be probed through QCD-driven pair production, independent of $\kappa$, and through single $T$ production via $bW$ fusion, which depends on $\kappa$. Pair-production reaches for the HL-LHC and FCC-hh are obtained with the Collider Reach tool [407], using mass exclusions from [408, 409], while for a 10 TeV MuC the reach is taken as half the centre-of-mass energy. Single-production projections use [410] for the HL-LHC (scaled with luminosity) and are extrapolated to the FCC-hh with higher luminosity and $\kappa$-dependent cross-sections at 85 TeV. The MuC reach follows [411]. Figure 8.4 presents the projected reach of each collider. The grey region indicates where the top-partner width approaches its mass, invalidating a perturbative description. All machines substantially extend the accessible parameter space: the 10 TeV MuC excludes masses below $\sim$8 TeV, while the FCC-hh reaches up to $\sim$15 TeV. Such direct searches will strongly inform the origin of the weak scale, since probing top partners at 10 TeV requires tuning UV parameters below the percent level [412, 413].

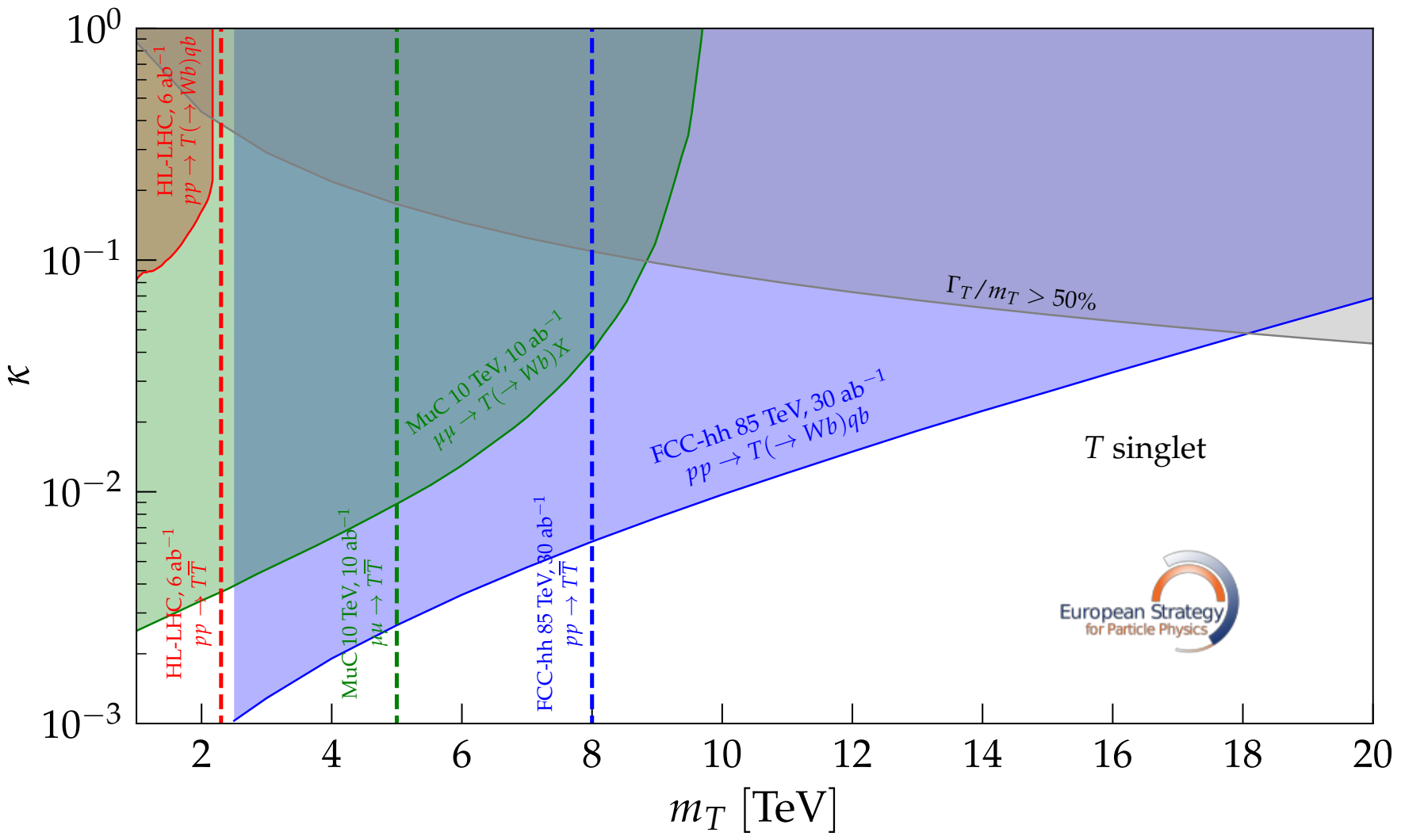

Fig. 8.4: Projected 95% exclusion limits for a fermionic top partner $T$ (an $SU(2)_L$ singlet vector-like quark), in the $\kappa$ versus $m_T$ plane, where $\kappa$ parametrizes the coupling between the $T$ quark and the SM gauge bosons from mixing with the SM top quark. The projections shown correspond to the direct reach of single $T$ production and pair production for three collider scenarios: the HL-LHC, FCC-hh, and MuC 10 TeV. The region where the $T$ width is more than half its mass, and thus perturbation theory breaks down, is shown as a shaded region.

Combining the reach for spin-1 and spin-1/2 states with the indirect sensitivity to $m_*$ and $g_*$ reveals a rich landscape. High-precision EW programmes at $e^+e^-$ colliders can probe $m_* \sim 20$–50 TeV, potentially pointing to direct signals accessible at proton colliders, while muon colliders extend the indirect reach to $\sim$60 TeV. Proton and muon machines also enhance the direct reach for fermionic top partners, which are most closely tied to the Higgs potential and may evade both indirect probes and spin-1 searches. Even if a future $e^+e^-$ programme found full agreement with the SM—down to $10^{-3}$ in Higgs couplings and $10^{-4}$–$10^{-5}$ in key EW

observables—compositeness would remain viable well beyond indirect sensitivity. FCC-hh could probe even deeper, directly accessing resonances and SMEFT operators with coefficients scaling as $1/m_*^2$ at sensitivities surpassing $10^{-4}$. Thus, even without indirect evidence at lepton colliders, exploring the energy frontier to multi-tens of TeV remains essential to establish whether the Higgs is elementary or a composite pNGB. Direct searches for fermionic top partners could be decisive, and ultimately an integrated programme covering precision and energy, such as that of the FCC-ee+FCC-hh, will offer the best path to discover compositeness and extend our understanding deep into the tens of TeV regime.

### 8.2.2 Supersymmetry

SUSY remains one of the most sound explanations of the electroweak scale: it renders the Higgs potential calculable from SM couplings and superpartner masses; predicts an elementary, SM-like Higgs consistent with data; offers viable DM candidates (including the thermal higgsino) when assuming $R$-parity conservation; and enhances prospects for gauge coupling unification. While evidence for superpartners near the weak scale has not yet emerged at the LHC, there is still strong motivation to search for SUSY at future colliders: in the MSSM, the observed Higgs mass suggests that top-quark superpartners lie beyond the reach of the LHC but are possibly within that of its successors.

Precision and energy play complementary roles in the experimental pursuit of evidence for supersymmetric extensions of the SM. On one hand, these extensions typically violate the accidental global symmetries of the SM, such as flavor and custodial symmetry, and necessarily imply corrections to Higgs properties at both tree- and loop-level. On the other hand, they are weakly coupled and so their contributions to low-energy observables rapidly decouple, making direct searches at high energy the most decisive tests. This is particularly apparent in the paradigm of "natural supersymmetry", in which only the superparticles most directly relevant to the Higgs potential (such as the partners of the Higgs boson and top quark) are light, while states most susceptible to SM precision tests are heavy.

Supersymmetric extensions of the SM warrant additional attention above and beyond their potential relevance to the weak scale: they are extraordinary signature generators for physics beyond the SM. Supersymmetric models provided the first motivation for searches that now form the backbone of the modern experimental physics programme at colliders, such as missing energy, disappearing tracks, and long-lived particles. In this respect, the prospects for SUSY at future colliders illustrates the potential for discovering BSM physics more broadly.

To assess the prospects for SUSY at future colliders, we consider a suite of simplified models. These typically include a given superpartner and the lightest stable superpartner (LSP), assumed to escape detection. While not capturing the full signature space of complete theories, simplified models provide a practical framework for connecting SUSY scenarios to searches. In such models, pair production cross-sections are fixed by SM quantum numbers, leaving only the sparticle and LSP masses as free parameters. At hadron colliders, the sensitivity depends strongly on the sparticle–LSP mass splitting, which controls the amount of missing energy in the final states. To capture this dependence, we consider benchmark scenarios with both large and small splittings. Lepton colliders, with lower backgrounds, are less sensitive to the splitting and can typically probe masses up to the pair-production threshold ($\sqrt{s}/2$) for sparticles that carry electroweak charges.

The direct reach of future colliders, HL-LHC [ID170], FCC-hh [ID227, ID247], LCF 550 GeV [ID140], CLIC 3 TeV [ID78], muon collider at 10 TeV [ID207], are compared to the current LHC limits, for several SUSY simplified models and illustrated in the plots of Fig. 8.5. Limits for hadron colliders are obtained with luminosity scaling using the COLLIDER REACH tool [407], while limits for lepton colliders correspond to half of the center-of-mass energy for distinctive final states.

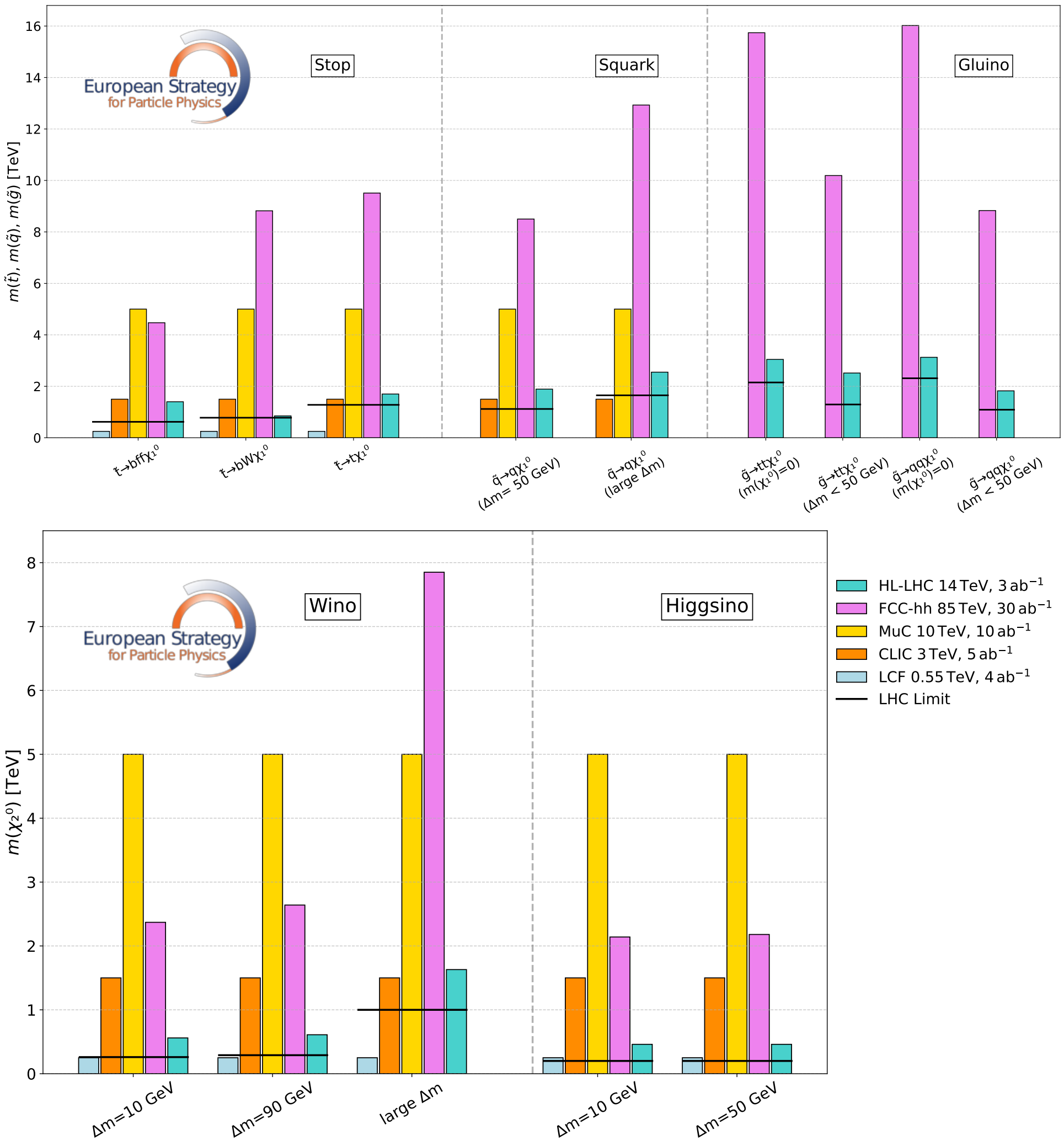

Fig. 8.5: Projected 95% exclusion limit for strongly interacting (top) and weakly-only interacting sparticles (bottom) as a function of their mass in a LSP simplified model for representative decay modes. For stops: (i) Top quark+LSP with large stop-LSP mass splitting $\Delta m$. (ii) Bottom quark + $W$ boson + LSP with $m_W < \Delta m < m_t$. (iii) Bottom quark + off-shell $W$ boson + LSP with $\Delta m < m_W$. For light-flavour squarks: Quark + LSP, corresponding to (i) large squark-LSP mass splitting and (ii) with $\Delta m = 50$ GeV. For gluinos: (i) Top quarks + LSP, corresponding to large gluino-LSP mass splitting $\Delta m$. (ii) Off-shell top quarks + LSP with a compressed spectrum, $\Delta m = 50$ GeV. (iii) Light-flavor quarks + LSP, with large $\Delta m$. (iv) Light-flavor quarks + LSP with $\Delta m = 50$ GeV. For winos, decays into bino + electroweak bosons in three cases, compressed ($\Delta m = 10$ GeV), moderately compressed ($\Delta m = 50$ GeV) and large wino-bino mass splittings. For the higgsinos, decays of neutral and charged higgsino to lightest neutral higgsino, corresponding to compressed mass splittings $\Delta m = 10$ and 50 GeV.

The left panel of the upper plot in Fig. 8.5 shows the reach for the lightest stop, the scalar superpartner of the top quark. We assume the decay interaction couples the stop, top quark, and LSP, with the final state determined by the available phase space. We consider three decay modes corresponding to different mass splittings between the stop and the LSP: top quark plus LSP for $\Delta m \equiv m_{\tilde{t}} - m_{LSP} > m_t$, bottom quark, $W$ boson, and LSP for $m_W < \Delta m < m_t$, and bottom quark, quark-antiquark pair, and LSP for $\Delta m < m_W$. In the middle panel the reach for light-flavour squarks, the scalar superpartners of first- and second-generation quarks is shown. We assume a dominant decay into a light quark and LSP, and consider two representative squark–LSP mass splittings to highlight the impact of missing energy on collider sensitivity. Finally, the right panel displays the reach for the gluino, the fermionic superpartner of the gluon. The gluino decays via an off-shell squark (assumed heavy and decoupled) into a quark and the LSP. The dominant final state depends on both the available phase space and the mass hierarchy among squarks. Accordingly, we consider four benchmarks, combining off-shell squarks or stops with either large or small gluino–LSP mass splittings.

The lower plot of Fig. 8.5 shows the reach for winos and binos, fermionic superpartners of the electroweak gauge bosons and for higgsinos, the fermionic superpartners of the two Higgs doublets in the MSSM. The winos are pair-produced via electroweak interactions and decay into the bino LSP plus electroweak gauge bosons. We study three benchmarks with large, moderate (on the order of $m_Z$), and small mass splittings, summarized in the left panel. In the case of the Higgsinos, mixing with heavier neutralinos induces small mass splittings between the charged components and the lightest neutral state. We consider two benchmarks with small splittings, summarized in the right panel. Alternatively, if the only splitting arises from SM loops, the corresponding phenomenology is discussed in the DM section (see Chapter 9).

Taken together, these projections demonstrate that future colliders offer a striking improvement in sensitivity to SUSY relative to the HL-LHC, with complementary strengths between the highest-energy machines, i.e. the FCC-hh and a 10-TeV muon collider. Together they can provide significant coverage of natural parameter space for EWSB and meaningful coverage of the targets indicated by the Higgs mass. In parallel, precise measurements at weak scale lepton colliders and in particular at the FCC-ee may offer early hints of superpartners consistent with the observed Higgs mass, as seen in Fig. 8.1. More broadly, precision tests of SM symmetries will powerfully complement direct searches at the energy frontier. Finally, as representative benchmarks for BSM physics, supersymmetric simplified models illustrate an order-of-magnitude gain in mass reach across a rich collection of experimental signatures.

### 8.2.3 Beyond minimal naturalness

We have shown that a broad class of observables, from indirect effects in precision electroweak and Higgs measurements to a multitude of possible final states associated with the direct production of new particles both resonantly and in pairs, is associated with the two canonical symmetry-based approaches to making the Higgs mass calculable. Both hypotheses, compositeness and/or SUSY, offer discovery opportunities at HL-LHC and future colliders. Here, we briefly comment on the possibility that a more complex theoretical structure could be realised in Nature, potentially obscuring these signatures.

One irreducible prediction of any model that explains the origins of the weak scale is that there must exist some kind of "top partner" states that cancel the top-quark loop contributions to the Higgs mass-squared parameter. These are the stops in SUSY and fermionic top partners for composite Higgs models discussed above. In these standard approaches, the top partners have the same SM charges as the top quark. The premise of "neutral naturalness" is to change the nature of the top partners, so that they are neutral under some or all SM interactions. The two canonical examples are the Twin Higgs [414] and Folded SUSY [415] models, both of which exhibit these features.

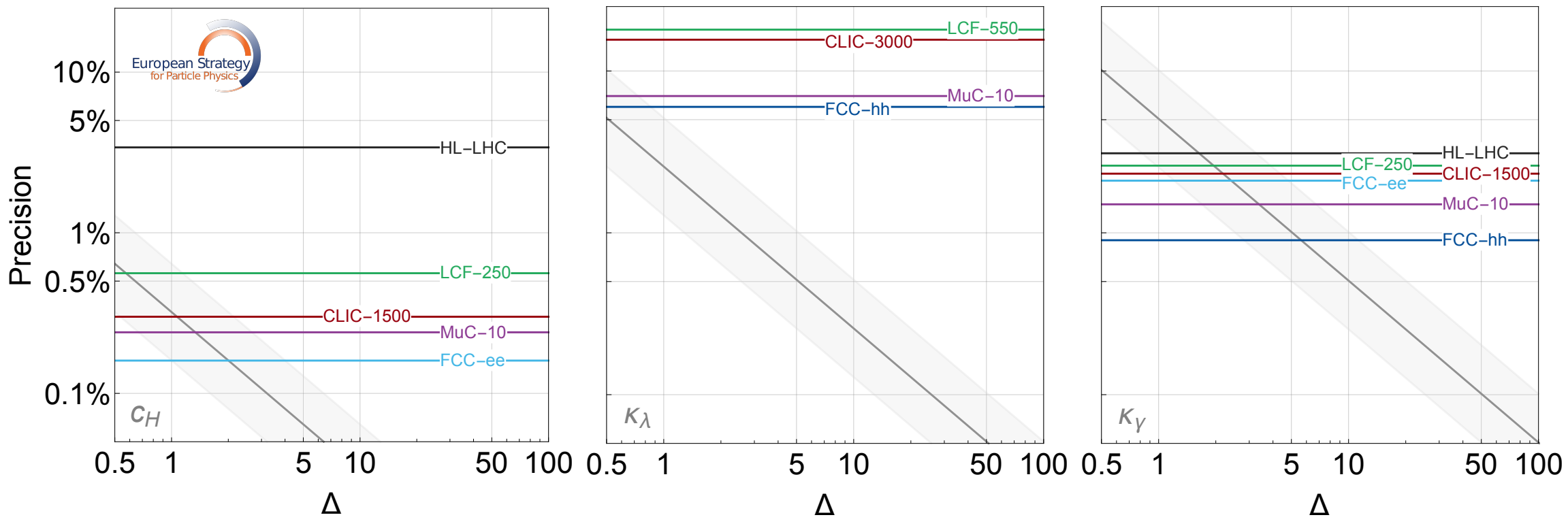

Fig. 8.6: Schematic relation between fine-tuning $\Delta$ and minimal deviations in Higgs couplings (universal shift $c_H$ and deviations $\delta\kappa/\kappa$ in the $hhh$ and $h\gamma\gamma$ couplings). The band indicates the model dependence in mapping a UV theory to these observables. Horizontal lines show projected precisions (95% CL) of various future collider scenarios . This highlights how Higgs precision measurements inform expectations for tuning in neutral naturalness scenarios.

The paradigm of neutral naturalness has two major implications for the correlation between the tuning in the model and the limits on the search for new physics effects. The first is that direct search limits are weaker because the top partners in neutral naturalness models are un-coloured and in principle can evade direct discovery at a collider, though searches for long-lived particles (LLPs) could provide sensitivity to many models [416, 417]. The second is that the calculable contribution to the Higgs mass due to the combination of the top and top partner loops scales as[1]

$$\delta m_H^2 \sim \frac{y_t^2}{16\pi^2} m_{\rm NP}^2\,, \tag{8.1}$$

where $m_H^2$ is the Higgs mass squared parameter, $m_{\rm NP}$ is the mass of the new top partner states, and $y_t$ is the top Yukawa coupling. Overall, this implies that the tuning $\Delta \equiv \delta m_H^2/m_H^2$ in such models naively scales as $\Delta \sim m_{\rm NP}^2$. Assuming neutral top partners evade direct discovery, their leading effects may appear in precision electroweak observables. Dimensional analysis suggests that these contributions to Higgs observables scale as $1/m_{\rm NP}^2$. Including naive coupling factors yields the estimates in Fig. 8.6, illustrating the potential impact of precision Higgs measurements on probing these models.

## 8.3 Scalar sector

The Higgs boson is the first known scalar particle that appears to be elementary. This invites the search for additional scalars: given that the $J = 1/2$ and $J = 1$ sectors of the Universe are rich in multiplicity, why not the $J = 0$ sector as well? Additional scalars are closely tied to the

[1]Note that for the MSSM there is also a tree-level contribution $\Delta m_H^2 \sim m_{\rm NP}^2$, where $m_{\rm NP}$ is the mass of the Higgsinos, which are harder to discover due to the smaller production rate and the intrinsic small splitting.

Higgs boson itself, either because they play the role of additional Higgs bosons or communicate with the SM through the Higgs portal. Indeed, $|H|^2$ is the lowest-dimensional gauge-invariant operator in the SM and allows for marginal couplings to additional scalars regardless of their SM quantum numbers. Such interactions could have dramatic ramifications for the Higgs potential, EWSB, and the electroweak phase transition. The nature of the electroweak phase transition is a particularly sharp target, because a strongly first-order phase transition requires non-decoupling physics and hence a bounded target space in precision and energy.

Additional scalars neutral under QCD, especially complete SM singlets, remain weakly constrained due to small production rates and subtle final states. Future colliders offer enormous discovery potential through both precision Higgs measurements and direct searches. The expected precision of Higgs factories is sufficient to probe even loop-level effects from scalar singlets, granting unique sensitivity to subtle SM extensions. Additional states at the electroweak scale also lead to rich phenomenology, potentially accessible at $e^+e^-$ colliders with centre-of-mass energies below 1 TeV.

Although there are many possible scalar extensions of the SM, here we exemplify the physics case using the simplest possible benchmark: a real scalar singlet coupled through the Higgs portal. The relevant potential terms for a real scalar singlet $S$ are

$$V(H,S) \supset b_1 S - \frac{\mu_S^2}{2} S^2 + \frac{b_4}{4} S^4 + \frac{b_3}{3} S^3 + \frac{a_1}{2} |H|^2 S + \frac{a_2}{2} |H|^2 S^2 \,. \tag{8.2}$$

In general, $S$ acquires a vacuum expectation value (either from its own potential or via Higgs couplings after EWSB) and mixes with the Higgs. This leads to tree-level modifications of Higgs couplings and resonant single production of the new scalar with SM decays. If the scalar has an unbroken $\mathbb{Z}_2$ symmetry, then $b_1 = b_3 = a_1 = 0$; Higgs couplings are modified only at loop level, scalars are pair produced via the Higgs, and final states are typically invisible. Although the $\mathbb{Z}_2$-symmetric case is very difficult to detect (the "nightmare scenario"), future colliders could probe the most motivated parameter space. More complex scalar extensions yield a richer signal space, but the singlet scalar already captures many core features of the physics case.

Although the real scalar singlet has several free parameters, the parametric behaviour simplifies in the limit $\mu_S^2 \gg a_1 v^2, \mu_H^2$ where the model enjoys an approximate custodial symmetry that implies $BR(S \to ZZ) = BR(S \to hh) = BR(S \to WW)/2$. In this limit, the phenomenologically relevant parameters become the mixing angle $\theta$ between the singlet and the Higgs boson and the mass $m_S$ of the mostly-singlet mass eigenstate. The degree of correlation between these parameters depends on whether or not $S$ acquires a vev from its own potential or via EWSB. In the former case, one expects $\sin\theta \sim m_h/m_S$, while in the latter case one expects $\sin\theta \sim m_h^2/m_S^2$. The reach of direct collider searches and indirect constraints is shown in Fig. 8.7 which collects input from HL-LHC [ID170], FCC-ee [ID233, ID242] FCC-hh [ID227, ID247], LCF 250 GeV [ID140], CLIC 1.5 TeV [ID78], muon collider at 3 and 10 TeV [ID207]. The FCC-hh direct search reach comes from a study of $S \to ZZ$, while the muon collider reach comes from a study of $S \to hh$ [ID207].

Additional scalars expand the potential landscape, which becomes a function not only of the Higgs vev but also the vevs of all other scalars. Even in the simplest case where no new scalars acquire a vev, they can be considered as modifying the Higgs potential in the effective SM-like theory where the new states have been integrated out. This is illustrated in Fig. 8.8, which shows a Petrossian-Byrne plot of our current and future experimental knowledge of the

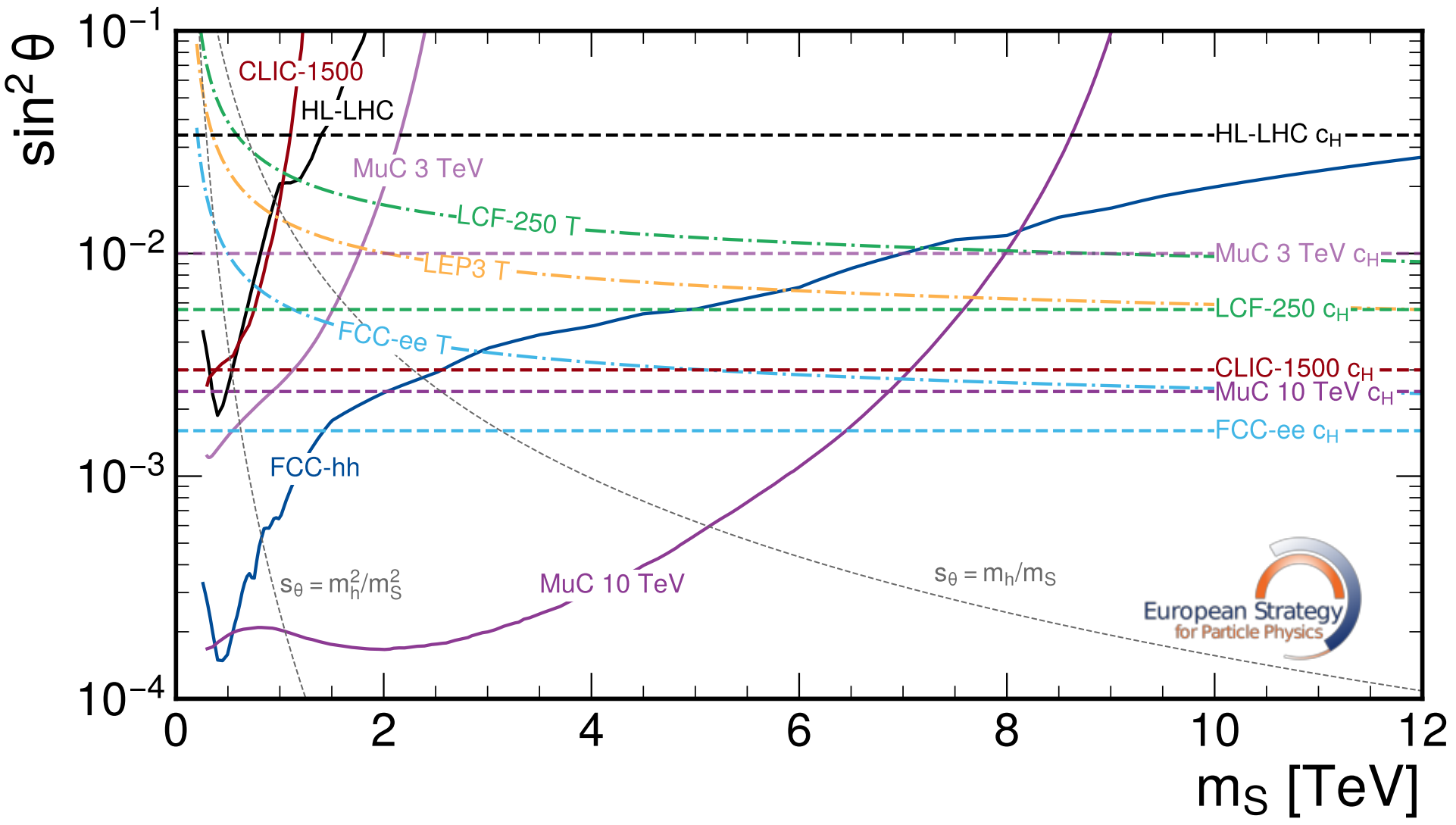

Fig. 8.7: Direct and indirect limits on a singlet scalar mixing with the Higgs boson as a function of singlet mass $m_S$ and squared mixing angle $\sin^2\theta$ in the "custodial limit" $BR(S \to hh) = BR(S \to ZZ) = BR(S \to WW)/2$. Solid curves indicate direct search limits, dot-dashed curves indicate indirect limits on the $T$-parameter, and dashed horizontal lines indicate indirect limits on the universal shift to all Higgs couplings ($c_H$), at 95% CL. Dotted grey curves indicate two characteristic scalings of the mixing angle as a function of the singlet mass corresponding to an independent singlet vacuum expectation value and an EWSB-induced singlet vacuum expectation value, respectively.

Higgs potential inferred from Higgs self-coupling measurements, relative to the changes in the shape of the potential induced by a representative $\mathbb{Z}_2$-symmetric singlet scalar benchmark. A similar approach based on an EFT interpretation of the constraints on the trilinear coupling expected at the HL-LHC, point to the exclusion of a first-order phase transition at $2\sigma$ (for $\kappa_3 = 1$) [ID170].

### 8.3.1 Electroweak phase transition

One of the most interesting consequences of an extended scalar sector is the possibility of featuring a sufficiently discontinuous, i.e. strong, first-order electroweak phase transition (FOPT). In the SM, a Higgs mass $m_h \sim 125$ GeV implies that the electroweak phase transition is a smooth crossover, but this can be altered by additional light degrees of freedom that modify the Higgs potential. Scalar extensions of the Standard Model can lead to a first-order phase transition consistent with current data. For instance, Ref. [ID140] studies generic and inert two-Higgs-doublet models (2HDMs) that can induce a first-order electroweak phase transition by modifying the Higgs self-coupling, including unexpectedly large two-loop effects, while leaving other couplings largely unchanged. A strong FOPT in the early Universe would have remarkable implications, including potentially observable stochastic gravitational waves. Moreover, it could provide the out-of-equilibrium conditions required for baryogenesis, which would explain the observed matter-antimatter asymmetry. This departure from equilibrium is one of the three Sakharov conditions for baryogenesis, alongside baryon number violation and the violation of charge (C) and charge-parity (CP) symmetries—for constraints on CP violation from low-energy observables, such as electric dipole moments (EDMs), see Chapter 5.

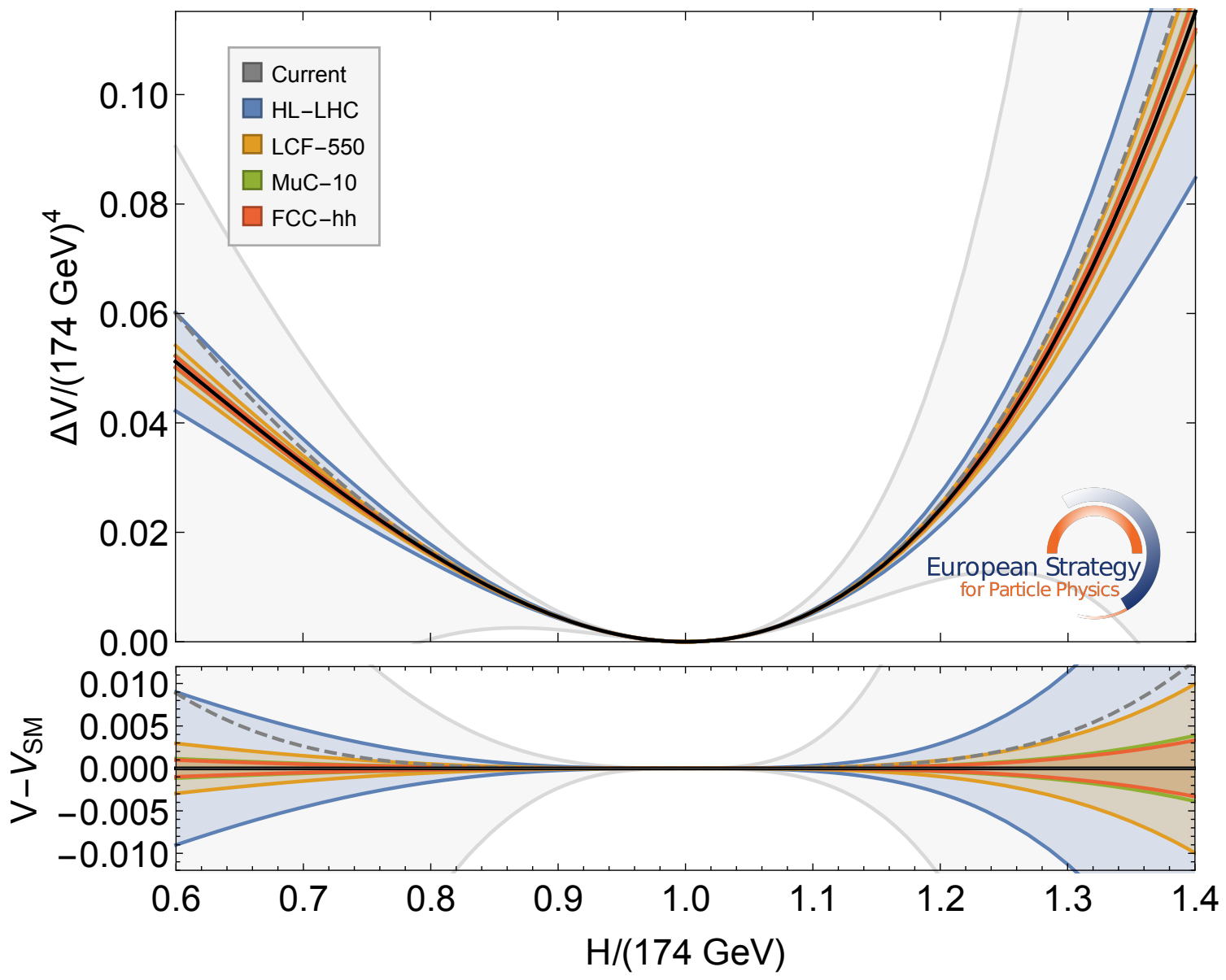

Fig. 8.8: Illustration of the effective understanding of the Higgs potential provided by the 95% CL uncertainties on $\kappa_3$ obtained by the CMS experiment at the LHC and expected at the HL-LHC [ID170], LCF [ID140], FCC-hh [ID227, ID247], and a 10-TeV muon collider [ID207]. Shaded regions show the 95% uncertainty on $\kappa_3$, interpreted as variations in the Higgs potential within the SMEFT-6 parameterization, where trilinear coupling effects dominate. The solid black line is the SM prediction, while the dashed grey line represents the effective potential of a $\mathbb{Z}_2$-symmetric singlet scalar with quartic coupling $a_2 = 8$ and mass $\mu_S = 4m_t$.

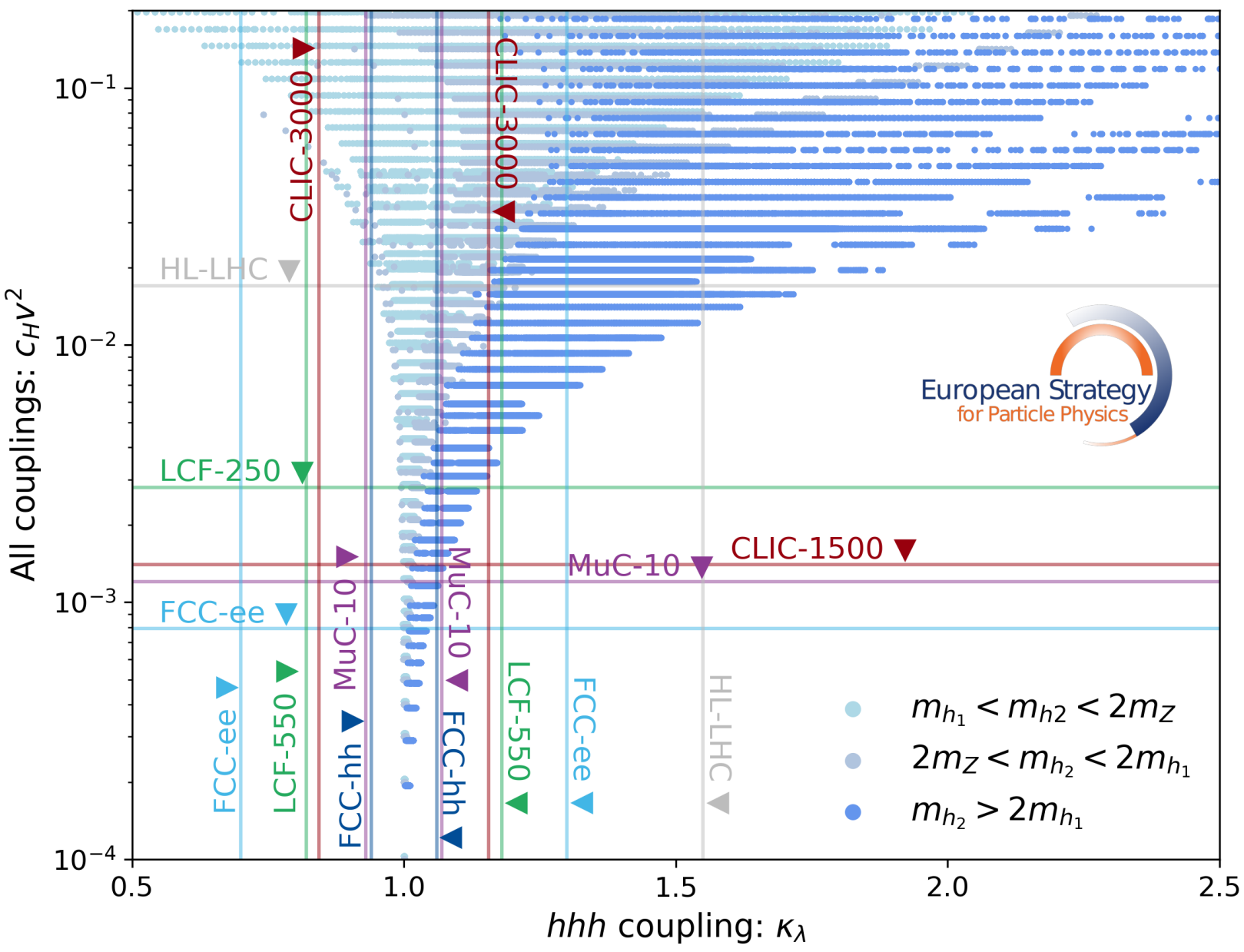

Fig. 8.9: Parameter space of the singlet scalar leading to a first-order electroweak phase transition as a function of the universal shift to all Higgs couplings ($c_H$) and corrections to the Higgs self-coupling, at 95% CL.

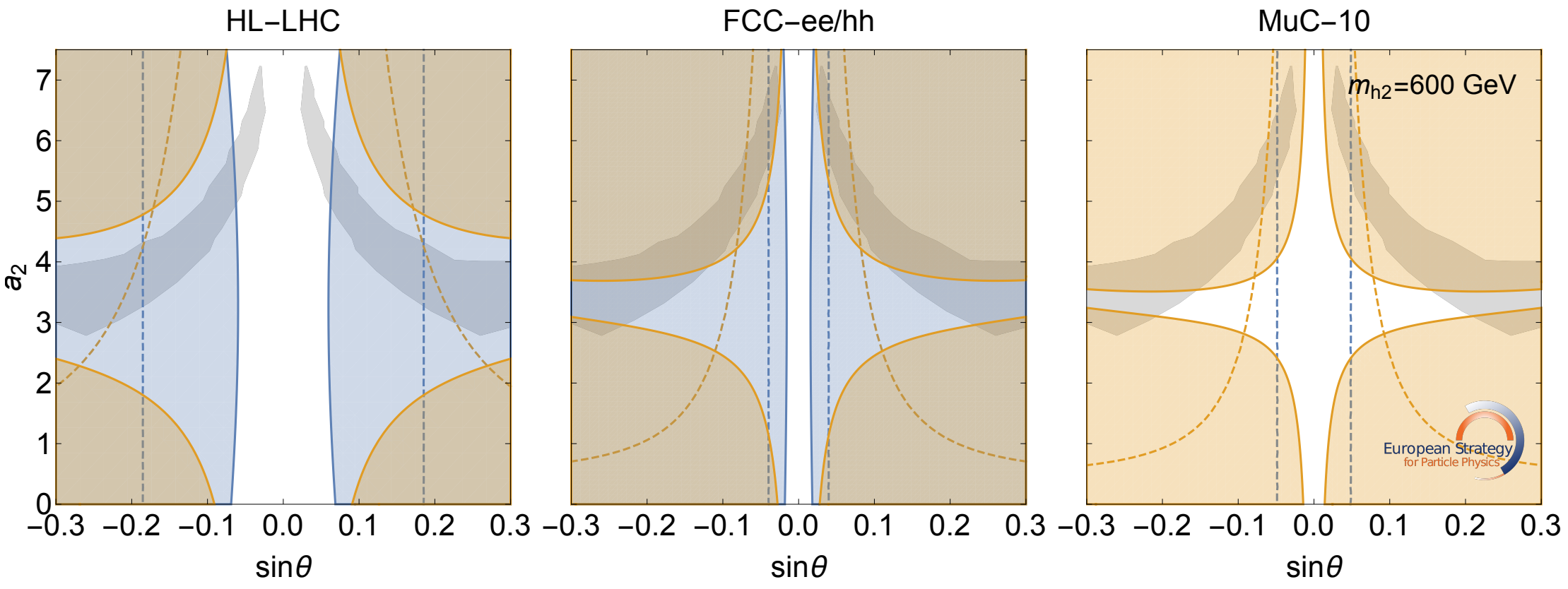

Fig. 8.10: Direct and indirect limits on a representative singlet scalar benchmark with $m_{h_2} = 600\,\text{GeV}$ as a function of the mixing angle $\theta$ and cross-quartic $a_2$. The grey shaded region indicates the parameter space for a strong FOPT. The blue solid region indicates the projected 95% CL limit from direct searches for $S \to ZZ$, while the orange solid region indicates the projected 95% CL limit from direct searches for $S \to hh$. Blue and orange dashed lines indicate the projected 95% CL limit from constraints on the universal shift to all couplings ($c_H$) and the Higgs self-coupling, respectively. In the middle panel, the $c_H$ limit comes from FCC-ee and the $h^3$ limit from FCC-hh.

In the case of a real scalar singlet, a strong FOPT typically requires $m_S \lesssim 1$ TeV and large couplings to the Higgs, implying corrections to Higgs couplings and the Higgs self-coupling that are poorly constrained by current measurements but accessible at future colliders. This is illustrated in Fig. 8.9, which shows model points from a parameter scan of the real singlet scalar that yield a strong FOPT, relative to changes in the Higgs self-coupling and the universal shift to all couplings.

The precision attainable at future colliders covers the vast set of parameter points yielding a strong FOPT. Direct searches also have an important role to play. This is illustrated in Fig. 8.10, which shows the coverage of direct searches for $S \to ZZ$ and $S \to hh$ alongside $hhh$ and universal limits for a singlet scalar benchmark with $m_S = 600\,\text{GeV}$, where the region giving a strong FOPT is illustrated in the plane of $\sin\theta$ and the cross-quartic $a_2$.

The power of both direct and indirect searches in this parameter space leverages the consequences of Higgs-singlet mixing, which is eliminated when the singlet enjoys a $\mathbb{Z}_2$ symmetry (the above mentioned "nightmare scenario"). This limit is nonetheless accessible thanks to the precision in Higgs couplings and the Higgs self-coupling anticipated at future colliders. This is illustrated in Fig. 8.11, which shows the coverage of these measurements on the parameter space of the $\mathbb{Z}_2$-symmetric singlet space compatible with a strong FOPT [418]. This parameterization also allows a straightforward understanding of the experimental prospects for $\mathbb{Z}_2$-symmetric scalars with electroweak quantum numbers; coverage of the real triplet scalar and inert doublet scalar are also shown in Fig. 8.11.

In summary, we have examined a simple singlet extension of the SM scalar sector that, by exhibiting rich collider phenomenology, can serve as a proxy to identify the main features of extended scalar sectors, such as 2HDMs, including regions of parameter space leading to a strong FOPT. Taking into account both direct and indirect constraints, which complement each other in a non-trivial way, we find that, in the minimal case of a singlet extension of SM scalar sector, the integrated FCC-ee +FCC-hh programme and/or a 10-TeV muon collider could explore the entire allowed parameter space.

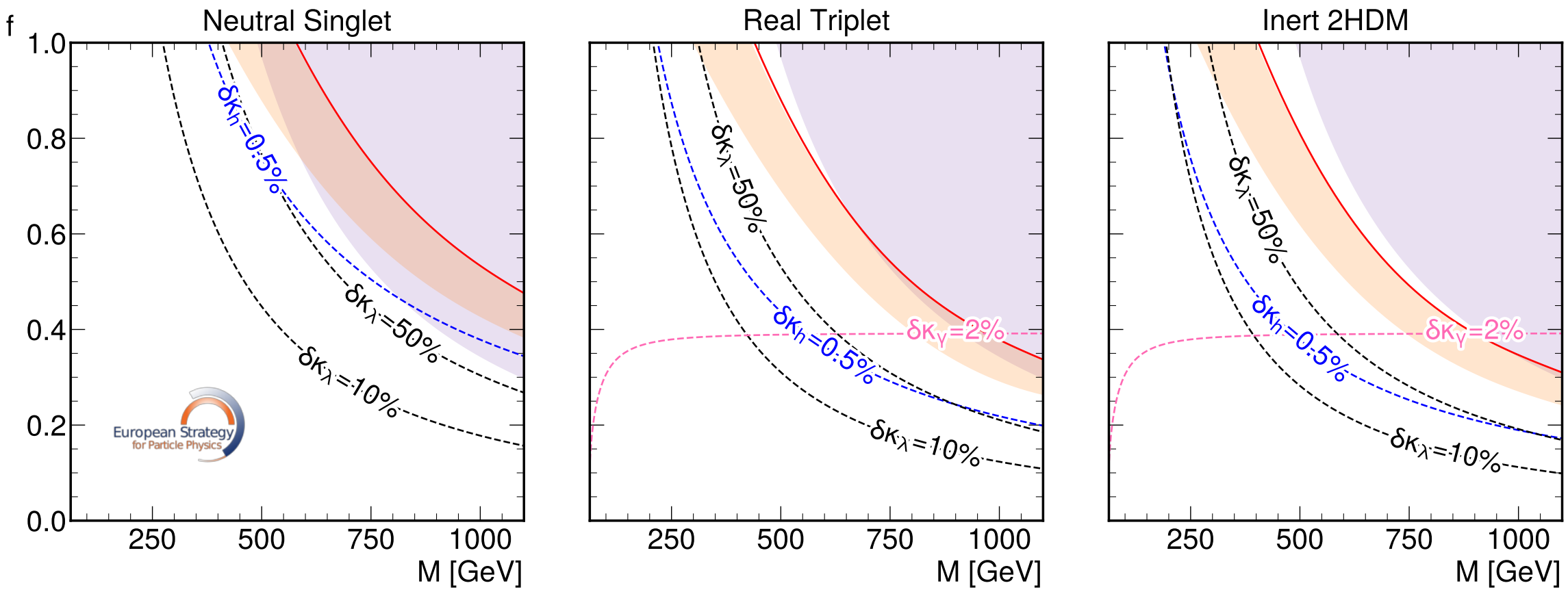

Fig. 8.11: Parameter space of custodially-symmetric $\mathbb{Z}_2$-symmetric scalars with various electroweak charges (singlet, real triplet, and inert doublet) coupled to the Higgs boson as a function of the physical scalar mass and the fraction $f$ of mass obtained from EWSB. The red shaded region indicates a FO phase transition, while the red line indicates gravitational waves observable by LISA. The purple region indicates the breakdown of perturbative unitarity, while the region above each dashed line is excluded at 95% CL by the corresponding Higgs measurement.

## 8.4 New forces

Many models of new physics include larger or additional gauge groups extending the $SU(3)_c \times SU(2)_L \times U(1)_Y$ of the SM. The spontaneous breaking of these groups down to the SM implies gauge bosons mediating new forces, which may provide the first hints of larger frameworks, such as Grand Unified Theories (GUTs), theories of flavour, and theories addressing the hierarchy problem.

The collider phenomenology of these new gauge bosons depends on their couplings to SM fields. For spin-1 resonances with large couplings to gauge bosons we refer to the study in Sect. 8.2.1. Here we focus on vectors with significant fermion couplings. The requirement of anomaly cancellation restricts the fermion couplings of a flavour-universal neutral $Z'$ to only a few possibilities [419]. Among them, the "Y-universal" $Z'$ boson (denoted $Y'$ in the following), which couples to the SM fields according to hypercharge, is the only non-trivial option that remains anomaly-free independent of the number of right-handed neutrinos. This model serves as a useful experimental benchmark, as it couples to all SM fermions.

Figure 8.12 (Left) shows projected 95% CL limits on the $Z'$ coupling $g_{Z'}$ vs mass $M_{Z'}$ for such a $Y'$ boson across various colliders. At collision energies lower than $M_{Z'}$, the model matches directly onto the effective operator corresponding to the oblique $Y$ parameter [420], such that $g_{Z'}/M_{Z'} = g_1\sqrt{Y}/m_W$. Limits from HL-LHC [421] and FCC-hh [ID227, ID247] include both direct resonant searches and indirect constraints from four-fermion operators affecting high-energy Drell-Yan production (shown as diagonal lines). For lepton colliders, only indirect bounds are shown [ID233, ID242]. Except at low masses ($m_{Y'} \sim 5-35$ TeV), where FCC-hh direct searches dominate, the strongest constraints come from four-fermion operators at the highest partonic energies. A 10-TeV muon collider stands out as the most powerful indirect

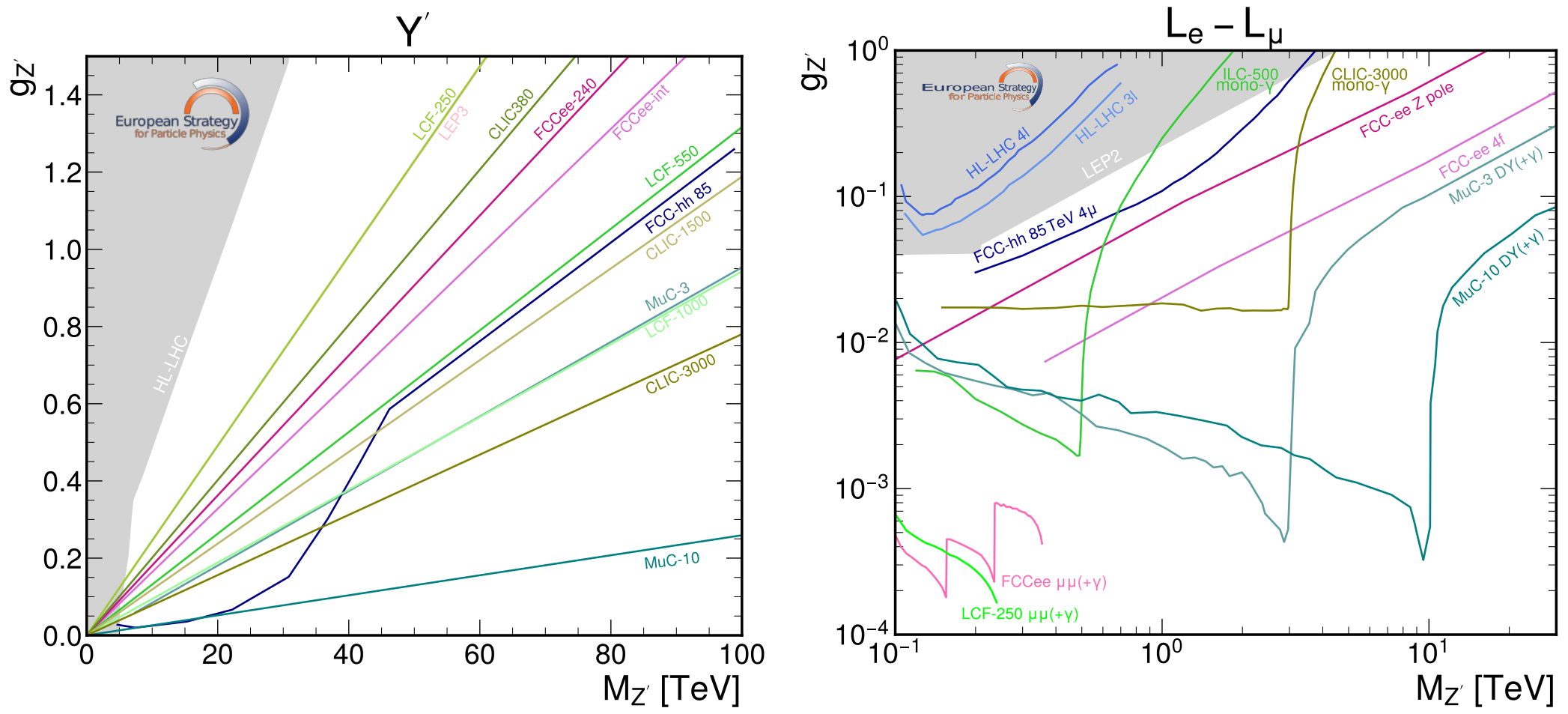

Fig. 8.12: Projected 95% CL exclusions on the coupling vs. mass plane for (Left) a $Y$-universal $Z'$ and (Right) an $L_e - L_\mu$ $Z'$ boson.

probe of such gauge bosons, with the added benefit of enabling model discrimination [422].

If a new $Z'$ boson has flavour-dependent couplings, there are more anomaly-free options for the charges. Simple examples include gauged $L_i - L_j$ or $B_i - L_i$, where subscripts $i, j$ denote fermion families. Fig. 8.12 (Right) shows projected exclusions for an $L_e - L_\mu$ $Z'$ boson. This scenario is challenging for hadron colliders, where searches rely on associated production channels such as $pp \to \mu\nu Z'(\to \mu\mu)$ and $pp \to \mu\mu Z'(\to \mu\mu)$ [ID227, ID247]. Lepton colliders are better suited since the $Z'$ couples directly to electron or muon beams, offering several probes: indirect effects at the $Z$ pole [ID233, ID242], four-fermion operators at high energies [ID207, ID241, 247], and direct searches via $\mu\mu(+\gamma)$ and mono-$\gamma$ signatures at $e^+e^-$ colliders [ID78, ID140, ID233, ID242], as well as $s$- and $t$-channel $\mu\mu(+\gamma)$ production at a muon collider [ID207]. For LCF-250 and FCC-ee we show $\mu\mu + \gamma$, while for cases not yet studied, e.g. LCF-550 and CLIC 3 TeV, we show mono-$\gamma$ results as an indication, noting that stronger bounds are expected from $\mu\mu(+\gamma)$. Strong complementarity arises: $e^+e^-$ colliders dominate for $M_{Z'} < 350$ GeV, while a muon collider extends sensitivity well into the multi-TeV regime depending on energy.

New gauge forces in BSM theories with a flavour mechanism often have flavour non-universal couplings to the SM. Models that address the flavour puzzle while predicting such forces include *flavour deconstruction models* [ID141, ID241, ID242]. Here, third-generation fermions and the Higgs are charged under one gauge group, while lighter generations are charged under another. These groups break to the SM at the TeV scale or above, yielding new gauge bosons with flavour-dependent interactions. Simple versions that generate realistic Yukawa textures involve deconstructing parts of the electroweak group, i.e. $U(1)_Y$ or $SU(2)_L$ (for an overview of these models see [423]). The deconstructed $U(1)_Y$ model predicts a neutral $Z'$, while the deconstructed $SU(2)_L$ model yields a triplet $W'$ [2].

Figure 8.13 shows current constraints and future sensitivities on the parameter space of these gauge bosons. The left plot shows the parameter space of the $Z'$ of the deconstructed $U(1)_Y$ model, while the right plot shows the parameter space of the $W'$ of the deconstructed $SU(2)_L$ model. The light grey region on both plots is disfavored by naturalness, with Higgs mass corrections $\delta m_h^2 > 1\,\text{TeV}^2$. For small $\theta$, couplings to the third generation and the Higgs are enhanced, making electroweak precision tests at the $Z$ pole most relevant. For large $\theta$, couplings to light generations dominate, so Drell-Yan and $e^+e^- \to f\bar{f}$ processes are more sensitive. The gauge basis is aligned with the up-quark mass basis. In the $SU(2)_L$ model, Tera-$Z$ precision at FCC-ee could probe all of the natural region at 95% CL. For the $U(1)_Y$ model, Higgs mass corrections are smaller and projected constraints weaker due to $g_Y < g_L$. Tera-$Z$ projections assume the "aggressive" theory uncertainties as presented in the Electroweak, Top, and Higgs Chapter 3. The region to the left of the black line can be reached by direct production at FCC-hh [424].

[2] Their fermion couplings are $g_\psi = g_{SM}\,\mathrm{diag}(-\tan\theta, -\tan\theta, \cot\theta)$, and their Higgs coupling is $g_H = g_{SM}\cot\theta$, where $g_{SM}$ is the SM coupling: $g_{SM} = g_1 Y_{\{\psi,H\}}$ for the deconstructed $U(1)_Y$ model and $g_{SM} = g_2$ for the $SU(2)_L$ model.

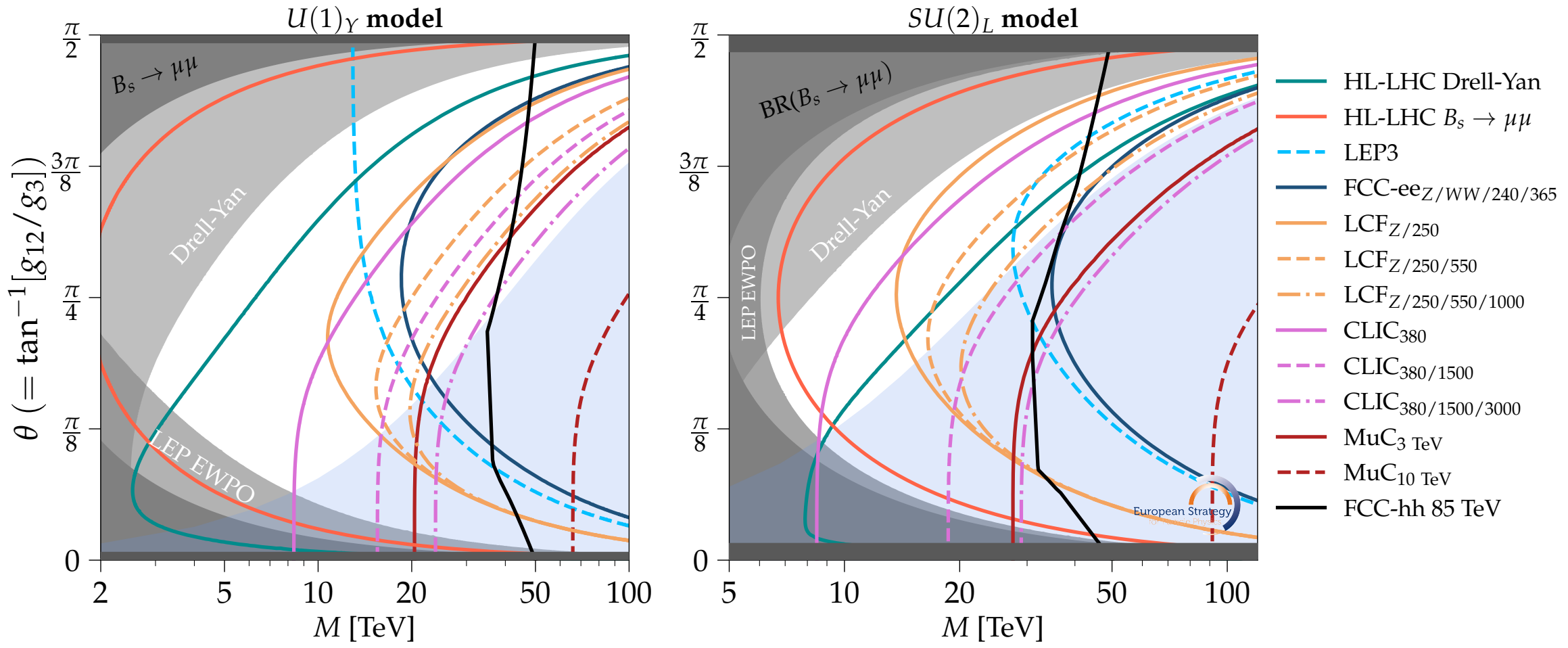

Fig. 8.13: Current constraints and projected 95% CL sensitivities on new flavoured vector bosons from a deconstructed gauge model. Left: Parameter space for a $Z'$ singlet from a deconstructed $U(1)_Y$ symmetry. Right: Parameter space for a $W'$ triplet from a deconstructed $SU(2)_L$ symmetry. The pale blue shaded region is disfavored by Higgs naturalness, while the solid grey horizontal regions at very large and very small $\theta$ indicate a coupling becoming non-perturbative.

Vector leptoquarks with dominant couplings to third-generation quarks and leptons can explain observed anomalies in the $R(D^{(*)})$ observables, which test lepton universality in semileptonic $B$ decays. They may also arise as heavy gauge bosons in flavour-deconstructed or flavour-specific Pati-Salam models [425, 426]. We consider a simple model with couplings only to left handed fermions, described by the Lagrangian:

$$\mathscr{L}_{U_1} \supset \frac{g_U}{\sqrt{2}} \beta_{ij} \bar{Q}_L^i \gamma_\alpha L_L^j U_1^\alpha + h.c.. \tag{8.3}$$

Figure 8.14 shows current constraints and future sensitivities for this model under two leptonic coupling scenarios. In the left panel, non-zero couplings are set to $\beta_{\tau b} = 1$, $\beta_{\tau s} = 0.1$. Current LHC exclusions [427] are shown in grey, and the green band marks the $2\sigma$ region favoured by $R(D^{(*)})$ measurements. Direct pair production and $t$-channel exchange at HL-LHC, Muon Collider, and FCC-hh [ID227, ID247] [428] will probe higher masses, while indirect constraints from electroweak precision measurements at FCC-ee, LEP3 or LCF offer crucial complementary reach [429]. Notably, current bounds on $b \to s\tau\tau$ are far above SM predictions, but FCC-ee [ID233, ID242] could measure these decays to ~20% precision (see Chapter 5), greatly enhancing sensitivity to third-generation couplings. In the right plot, the non-zero couplings are $\beta_{\mu b} = 1$, $\beta_{\mu s} = 0.1$. Current LHC exclusions [430–432] are shown in grey. Here, current $B_s \to \mu\mu$ limits already impose strong constraints, and future improvements are limited. However, direct searches at HL-LHC, Muon Collider, and FCC-hh remain sensitive to higher masses, and $\mu\mu \to jj$ at the Muon Collider offers additional indirect reach. Additionally, Muon Collider searches for flavour changing contact interactions can provide complementary sensitivity to the $bs\mu\mu$ interaction. At linear colliders and FCC-ee, leptoquarks coupling to electrons can be tested indirectly through $e^+e^- \to q\bar{q}$ processes (see e.g. [433]).

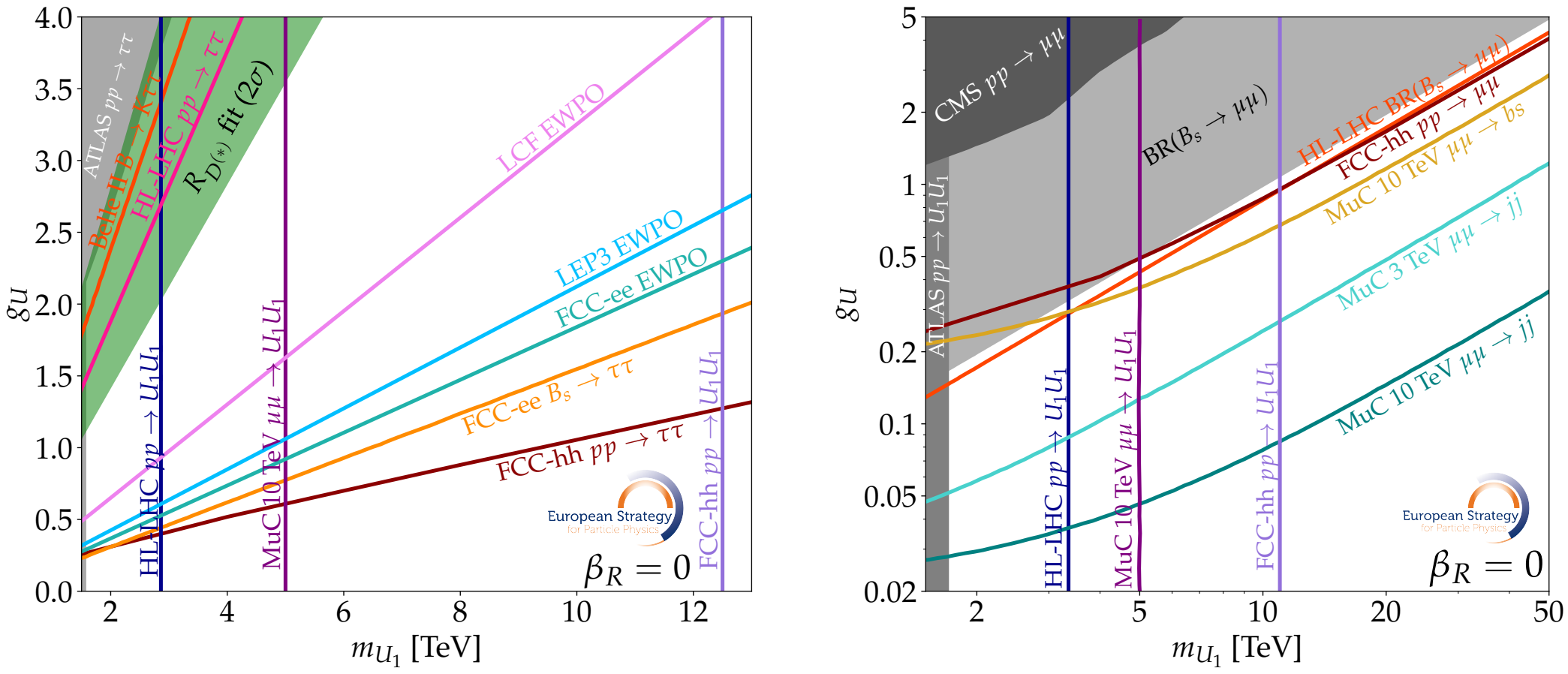

Fig. 8.14: Current constraints and projected sensitivities at 95% CL on the vector $U_1$ leptoquark coupling predominantly to third-generation quarks. Left: Leptonic couplings only to the third generation. The green shaded region indicates the $2\sigma$ preferred region to explain current $R(D^{(*)})$ measurements. Right: Leptonic couplings only to the second generation.

To summarise the experimental messages of this section, new gauge bosons with couplings to fermions can be well tested indirectly through their fermionic contact interactions. Significant tree level effects occur in $f\bar{f} \to f'\bar{f}'$ processes if the boson couples directly to electrons (for $e^+e^-$) machines, quarks (for hadron machines) or muons (for a muon collider). In these cases, sensitivity depends mostly on the energy of the collider. If the boson couples to a Higgs current, then it will modify $Z$ boson properties and can be sensitively probed by a precision Tera-$Z$ programme. Bosons with flavour non-universal couplings to quarks will in general mediate flavour changing neutral currents, which can also be studied at a Tera-$Z$ programme, which will in particular cover new ground for bosons which couple predominantly to third generation leptons.

## 8.5 Portals

A motivated class of scenarios connecting the SM to a broader framework involves *portals*: renormalisable or higher-dimensional interactions linking SM fields to new sectors, often tied to DM, neutrino masses, or other cosmological puzzles. These sectors may be neutral under SM forces, making portals the only experimental window into their dynamics. Despite their simplicity, portals yield rich phenomenology, from modified Higgs and electroweak properties to missing-energy or LLP signatures. Here we focus on four representative cases, dark photons, axion-like particles, dark scalars, and heavy neutral leptons, illustrating the complementarity between precision and direct searches in collider and non-collider experiments [ID235]. These scenarios serve as benchmarks for signature-driven studies rather than complete models; as highlighted in recent work [ID250], realistic constructions often lead to qualitatively different phenomenology. For example, in all cases discussed below DM particles are assumed out of reach, either too heavy or disconnected, so the coupling to the dark sector is typically taken to be negligible.

### 8.5.1 Dark photons

In the vector portal considered here, the interaction arises due to kinetic mixing between a dark and a visible Abelian gauge boson. The dark gauge boson, commonly referred to as dark photon $A'$, is a new vector boson associated with an additional $U(1)$ symmetry. It interacts with the visible SM sector via kinetic mixing with the SM photon, while remaining neutral under SM gauge interactions. The effective interaction term in the Lagrangian is:

$$\mathscr{L} = \frac{\varepsilon}{2\cos\theta_W} F'_{\mu\nu} B^{\mu\nu}, \tag{8.4}$$

where $\varepsilon$ is the kinetic mixing parameter between the dark and ordinary photon, $B_{\mu\nu}$ is the $U(1)_Y$ field strength, and $F'_{\mu\nu}$ is the dark photon field strength. Such light vector particles naturally arise in many BSM theories featuring hidden sectors, and may be linked to DM. These models can be tested experimentally both at collider and beam-dump experiments with constraints also arising from astrophysical and cosmological observations.

The relevant parameters describing this class of models are $\varepsilon$ and the dark photon mass $m_{A'}$. The studies considered here assume mass in the range $m_{A'} > 1$ MeV, suitable for accelerator searches. $m_{A'}$ is assumed to be smaller than $2m_\chi$, thus allowing the dark photon to decay back to SM particles (visible decays). These studies generally focus on minimal dark photon scenarios, featuring a single new state $A'$.

Current constraints as well as sensitivity projections for future experiments are shown in Fig. 8.15. In addition to cosmological constraints [434], searches targeting $e^+e^-$ and $\mu^+\mu^-$ final states, as well as reinterpretations of data from fixed-target and neutrino experiments at low masses (below 1 GeV), constrain the parameter space for $\varepsilon \gtrsim 10^{-3}$. The strongest constraints come from the LHCb, NA48/2, A1, BaBar, and KLOE experiments [435]. In the 0.01–10 GeV mass range, LHCb extends the sensitivity to $\varepsilon$ values as low as $10^{-3}$. Additional constraints from electron beam-dump experiments such as SLAC's E141 and E137, Fermilab's E774, and NA64 [436], as well as from the proton beam-dump experiment NA62 [437], further probe the $\varepsilon$ range down to $10^{-7}$. The past proton beam-dump experiments, such as CHARM and NuCal [438], also provide valuable constraints in this parameter space.

Future collider and beam-dump experiments offer essential complementarity in expanding current constraints: collider experiments are sensitive to high-mass, large-coupling scenarios, while beam-dump experiments are ideal for probing the low-mass, extremely weakly coupled regime. The sensitivity of each experiment depends on the dominant production mechanisms, which vary according to the experimental setup. Beam-dump experiments and forward experiments primarily probe dark photon production via bremsstrahlung, meson decays, and meson mixing [438–440]. In contrast, lepton colliders are sensitive to production through radiative return, associated production, and $Z$-boson decays. Hadron colliders, such as the LHC, can access dark photons mainly through Drell-Yan processes and decays of heavy mesons. NA62 in dump mode [441], as well as SHiP [ID145], FASER [ID23], and the Forward Physics Facilities (FPFs) [442], are expected to be sensitive in the low-mass (0.01–1 GeV) and very small kinetic mixing ($\varepsilon \lesssim 10^{-6}$) region. The LHCb Upgrade [443], Belle II [ID205], NA64 [ID50], and the LHeC [444] can bridge the intermediate region ($10^{-5} < \varepsilon < 10^{-3}$) for masses between 0.02 and 0.5 GeV. SHIFT [445] can extend the coverage of this mass range to intermediate couplings ($10^{-6} < \varepsilon < 10^{-5}$). Care must be taken at low mass, as this region is affected by large theoretical uncertainties, which are treated differently depending on the experiment [438, 440, 446]. Future high-energy collider experiments such as HL-LHC, FCC-ee [447][ID233, ID242], FCC-hh [ID227, ID247], the Linear Collider Facility (LCF) [ID140], and the Muon Collider [448] will provide sensitivity for high-masses (>10 GeV), probing down to $\varepsilon \sim 10^{-4}$. HL-LHC and FCC-hh projections have been derived from current CMS scouting and offline searches for resonant muon pairs, and scaled to expected luminosity and cross-section. The LCF projections presented here are an update of those in [ID141], incorporating the latest luminosity estimates from the LCF Collaboration. In addition, a likelihood ratio weighting has been applied to account for the two polarization configurations. Projections for muon colliders, however, are very conservative, as existing studies consider only the associated production mode, to which this type of collider has limited sensitivity; extended sensitivity studies might be conducted in the future.

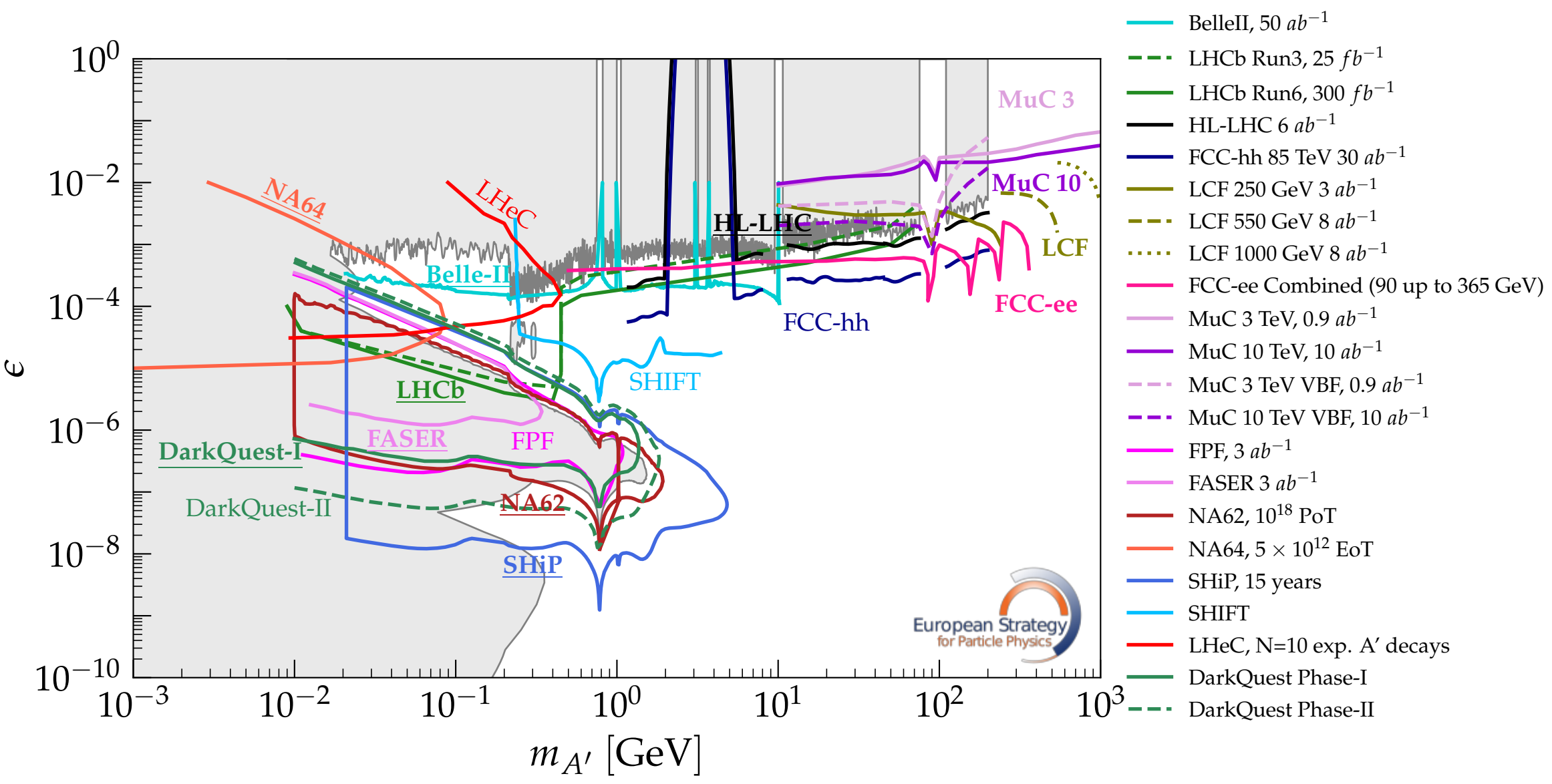

Fig. 8.15: Sensitivity projections for dark photons in the plane mixing parameter $\varepsilon$ versus dark photon mass $m_{A'}$. All curves correspond to 90% CL exclusion limits, except for FCC-ee, FCC-hh, and LCF (95% CL exclusion limits). The LHeC curve corresponds to $N = 10$ expected signal events and zero background. The dashed MuC curves correspond to the VBF production.

### 8.5.2 Axion-like particles

Axion-Like Particles (ALPs) are light pseudoscalar bosons, $a$, that can serve as mediators between the SM and hidden sectors. Theoretically motivated by the QCD axion, ALPs are less constrained and arise naturally in many SM extensions addressing its limitations, including DM.

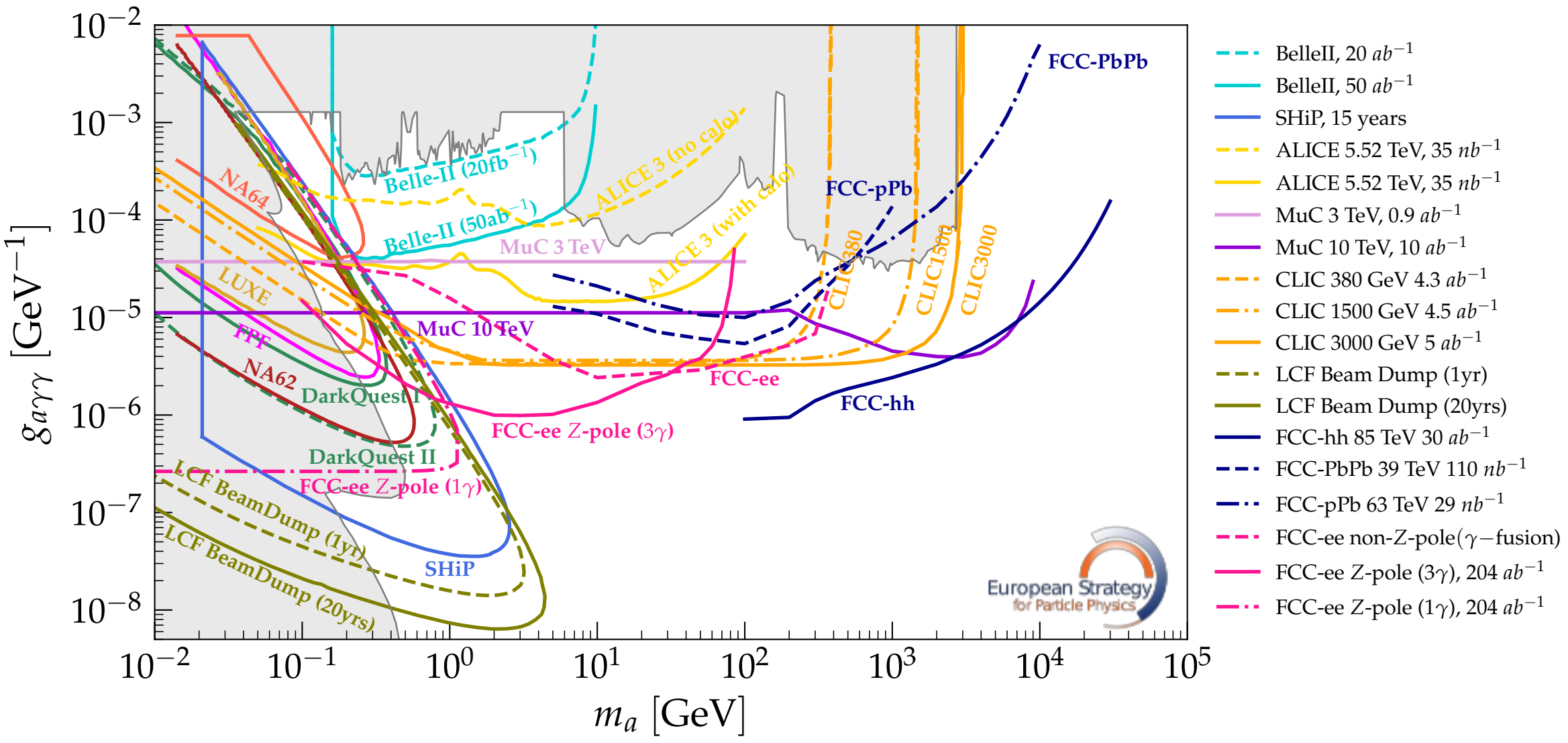

Fig. 8.16: Sensitivity projections for ALPs coupling to photons as a function of the ALP mass. All curves correspond to 90% CL exclusion limits, except for FCC-ee, FCC-hh, and LCF (95% CL exclusion limits).

In the simplest scenario, ALPs couple to photons through the dimension-5 interaction

$$\mathscr{L} = \frac{g_{a\gamma\gamma}}{4\cos^2\theta_W} a B_{\mu\nu}\widetilde{B}^{\mu\nu}. \tag{8.5}$$

Figure 8.16 illustrates the sensitivity of various experimental approaches (fixed-target and beam-dump experiments, and high-energy colliders) to ALPs.

ALPs with mass $m_a \lesssim 90$ GeV may be efficiently probed via displaced decays at beam-dump experiments, prompt $a \to \gamma\gamma$ resonant decays, or via missing energy searches, depending on the coupling range. Notably, at the beam-dump experiments, such as SHiP [449] [ID145] and proposed lepton colliders in dump mode (e.g., LCF-250 [ID140]), ALPs are mainly produced via the Primakoff process or through decays of light mesons [450]. NA64 will probe the domain of masses $m_a \lesssim 300$ MeV and couplings $g_a \gtrsim 10^{-5}$ GeV$^{-1}$; SHiP and LCF-250 will probe tiny couplings down to $g_{a\gamma\gamma} \sim 10^{-8}$ GeV$^{-1}$ in the mass range $m_a \lesssim 5$ GeV. FCC-ee is expected to be sensitive in this same mass range using the beamstrahlung photon dumps. The prompt decay signature mainly probes significantly larger couplings, but in the significantly broader mass range, extended to $m_a \simeq m_Z$ and beyond. It is efficient at both lepton and hadron colliders. The former include the currently running Belle II [ID205], and future FCC-ee, CLIC 3 TeV (with the ALP production processes $e^+e^- \to \gamma a$ and the photon fusion $e^+e^- \to e^+e^-a$); and at ALICE3 at the HL-LHC [ID68], where ALPs are produced via the quasi-coherent scattering $\mathrm{Pb}+\mathrm{Pb} \to \mathrm{Pb}^{(*)} + \mathrm{Pb}^{(*)} + a$. Finally, the missing energy searches may be very efficient at FCC-ee and at MuC [451–454].

At higher masses ($m_a > 90$ GeV), ALPs are produced via associated processes like $e^+e^-$ or $\mu^+\mu^- \to a\gamma$ or photon fusion ($\gamma\gamma \to a$). These channels can be explored at hadron (FCC-hh) and lepton (FCC-ee, CLIC, MuC [451–454]) colliders (note that CLIC projections assume 4 signal events). In this regime, FCC-hh could extend the mass reach up to $m_a \sim 10^4$ GeV.

We note that ALPs are also expected to couple to fermions and gluons, providing other production mechanisms and decay modes [455, 456]. Some studies of the future prospects of these exist in the literature [454, 457, 458] [ID141].

### 8.5.3 Dark scalars

DM particles could get some of their mass by coupling to a complex scalar field, in the same way as SM particles. And in a similar way as the Higgs field, this new field could spontaneously break a new gauge symmetry, leading to the appearance of a new scalar boson referred to as dark scalar or dark Higgs boson $S$. The phenomenology of this new particle depends on its mass and is similar to that of the SM Higgs boson, as described in Sect. 8.3. The dark scalar couples linearly to other fields in the theory, and because of this, it is unstable. Typically, it is considered the lightest dark sector particle, so it decays only into visible SM particles, as the decay width to the SM neutrinos is negligible. Its decay pattern follows that of a light Higgs boson, preferring the heaviest pair of decay products kinematically allowed. In particular, its decay width to the SM neutrinos is vanishingly small, so all the decays are visible.

The high scalar mass case $m_S > m_h/2$ is discussed in Sect. 8.3, while here the focus is on the lower masses. Such dark scalars have been sought at various accelerator experiments [437, 457–459], thanks to the efficient production in the SM Higgs boson decays or in rare hadron decays. The dominant decay modes depend on the mass $m_S$, and range from a pair of electrons at the masses $m_a \sim 2m_e$, to a pair of $b$ quarks for the masses above 10 GeV.

After EWSB, the effective Lagrangian for the dark scalar has the form

$$\mathscr{L}_{\text{eff}} = \sin\theta\, m_h^2 hS + \frac{\alpha}{2} hS^2\,, \tag{8.6}$$

where $\theta$ denotes the mixing between the SM Higgs boson and the dark scalar, and $\alpha$ is the trilinear coupling. We present sensitivities for the parameter choices adopted in [457] to allow for direct comparison of the different facilities and experiments. Namely, $\alpha$ is fixed at a given value of $m_S$ by setting $\mathscr{B}(h \to SS) = \alpha^2 \cdot f(m_S^2)$ to 0 or 0.01, ($f(m_S)$ is a kinematical factor).

In the first scenario ($\mathscr{B}(H \to SS) = 0$), the number of model parameters is reduced to $\theta$ and $m_S$. Figure 8.17 (Top) shows the excluded parameter space and projected sensitivities for future experiments. The sensitivity in this scenario is dominated by $K$ and $B$ meson decay searches, and therefore, the mass reach for the dark scalar is limited to ∼5 GeV. Complementary reach is ensured by including prompt and displaced signatures at LHCb [443][ID81]; performing searches at the dedicated MAPP and MAPP2 experiments [ID235] [460]; the FASER2 experiment [ID19, ID235] at HL-LHC; and the NA62 experiment in the beam dump mode [441]. The SHiP experiment [ID145] extends the reach of other experiments in the domain of very small mixing angles $\theta^2 \simeq 10^{-11}$. Dedicated experiments at FCC-hh [ID227, ID247], like FOREHUNT, can give access to much lower coupling $\theta$ values exploiting the large number of $B$ mesons produced [ID44].

In the second scenario ($\mathscr{B}(h \to SS) = 0.01$), additional scalar production mode opens up – the Higgs boson decays into a pair of dark scalars. It can dominate the production at the experiments with large enough energies to produce $h$ bosons, extending the dark scalar mass reach to $m_h/2$ [461]. Figure 8.17 (Bottom) shows the sensitivity for this scenario.

Dedicated transverse detectors at the LHC, such as ANUBIS [ID273], CODEX-b [ID199] and MATHUSLA [ID61], improve the projected HL-LHC reach of the main detectors for hadronically decaying LLPs above 10 GeV, making them well suited to probe larger dark-scalar masses. The main collider experiments can access this region as well [461], but their sensitivities are typically presented in the $c\tau$–$\text{Br}_{h\to SS}$ plane rather than in the $m_S$–$\sin^2\theta$ plane, see

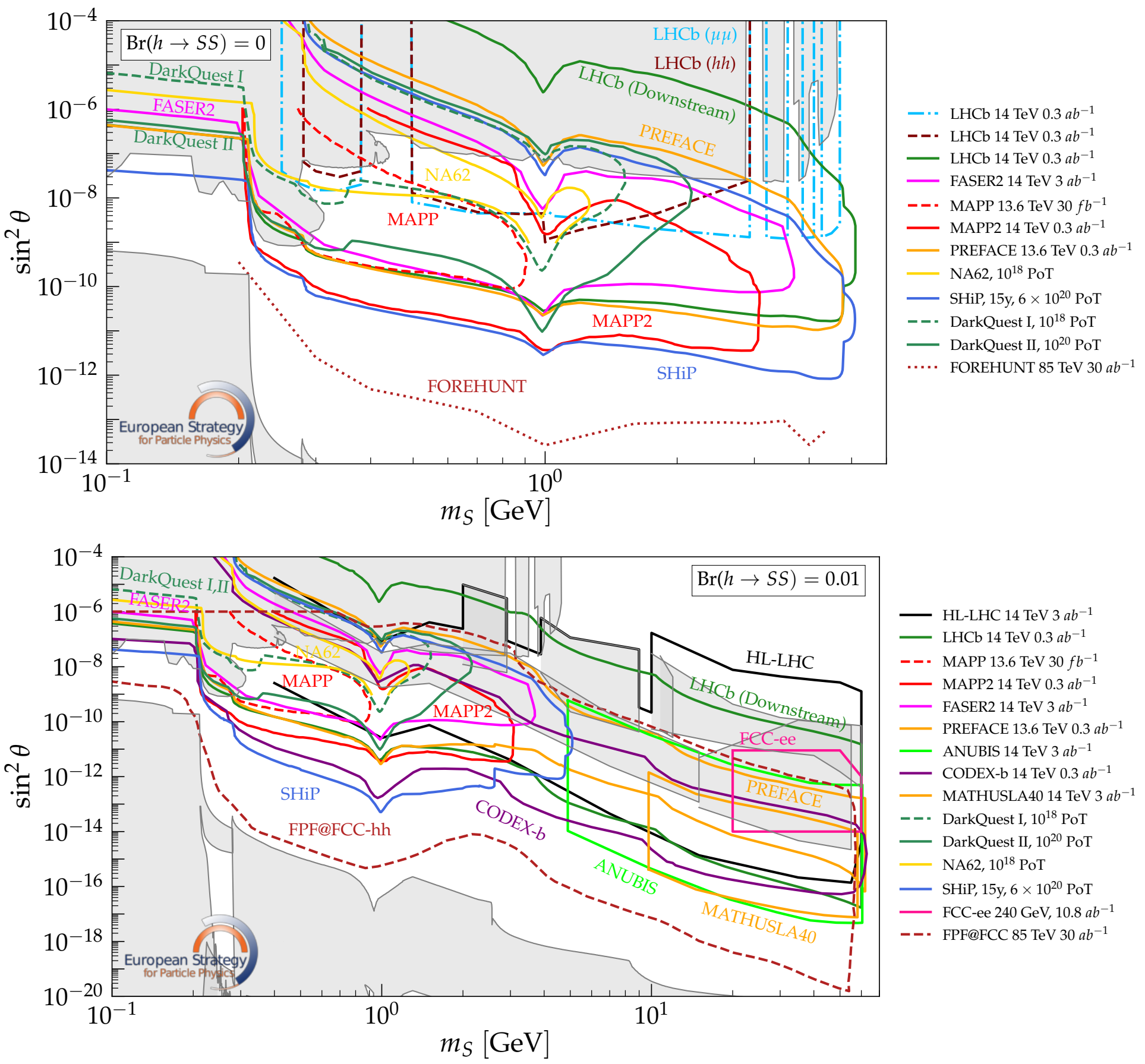

Fig. 8.17: Exclusion limits for a dark scalar mixing with the Higgs boson $\sin^2\theta$ as a function of its mass $m_S$, assuming that (Top) the Higgs boson does not decay to a pair of dark scalars, and (Bottom) the Higgs boson decays to a pair of dark scalars with a 1% branching fraction. Filled areas denote excluded regions, while curves show the projected sensitivities.

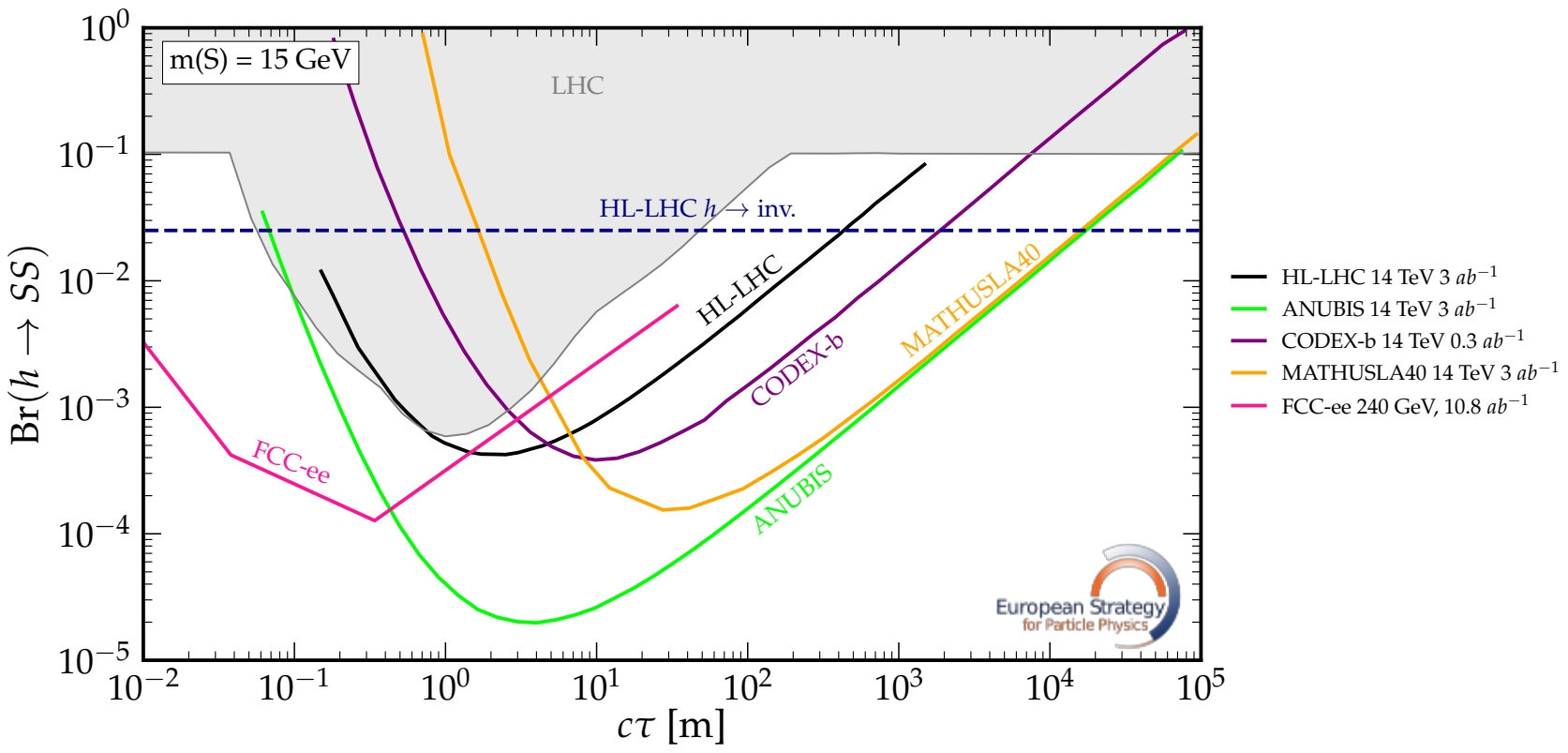

Fig. 8.18: Simplified model in which $S$ is a LLP produced in Higgs boson decays. Sensitivities of collider experiments are shown in the plane of $c\tau$ and BR($h \to SS$).

Fig. 8.18. A dedicated FPF-like facility at the FCC-hh would improve HL-LHC sensitivities by two orders of magnitude in the value of $\sin^2\theta$ [122][ID241, ID247]. Sensitivity from an FCC-ee study [462] rescaled to $\mathscr{B}(h \to SS) = 0.01$) is also shown. FCC-ee experiments will be sensitive to much lower branching fractions, of the order of $10^{-4}$.

In addition, it shows the need to probe Higgs boson branching fractions several orders of magnitude below the projected sensitivity of the measurement of invisible Higgs boson decays. The dominant decay mode in this case, a pair of $b$ quarks, is particularly challenging at hadron colliders in the domain of small scalar lifetimes. The FCC-ee experiments can provide additional sensitivity to this parameter space, thanks to the clean $e^+e^-$ collision environment [ID233, ID242].

### 8.5.4 Heavy neutral leptons

Heavy neutral leptons (HNLs) appear in minimal extensions of the SM, which aim to explain the observed light neutrino masses via a see-saw mechanism. At least two HNLs are needed for realistic neutrino masses.

In simple HNL models, the couplings required to explain neutrino masses have a lower bound set by the "see-saw line". This bound is too small for accelerators; however, the couplings may be much larger even in the minimal model with just two HNLs. This would be expected if there is an approximate lepton number symmetry, in which the two HNLs form a quasi-Dirac pair [463], manifesting themselves via oscillation in the number of LFV events as a function of the HNL decay length. It has been shown that this minimal model can robustly account for the observed matter-antimatter asymmetry.

The $\nu$MSM model [464] involves three HNLs: two more massive forming a quasi-Dirac pair to explain leptogenesis, and the third at the keV scale to explain DM.

HNLs only interact with SM particles by mixing with the active neutrinos, denoted by a mixing matrix $U$. Consequently, the HNL production mechanisms are the same as those for active neutrinos, suppressed by the value of $|U_\ell|^2$, where $\ell$ is the flavour of the corresponding active neutrino, with which HNL mixes. The HNL decay modes are also governed by the interactions of the active neutrinos, and include both neutral $Z^{(*)}\nu$ and charged $W^{(*)}\ell$ currents. Depending on the HNL mass and coupling strength to active neutrinos $U_\ell$, they can decay promptly or be long-lived on the collider experiment scale.

HNLs have a well-characterised parameter space, constrained by cosmological observations [465, 466] and with a target parameter range suitable to accommodate the see-saw mechanism for the measured active neutrino masses. Therefore, either the observation of HNLs or the exclusion of theoretically motivated HNL parameters would fundamentally impact particle physics. To compare the reach of various proposed facilities, simplified scenarios with a single-flavour mixing dominance are considered, with the HNL mixing with only one type of active neutrino: electron, muon, or tau. While realistic models generally involve all flavours [467], we avoid committing to a specific framework and aim for illustrative results here. Current constraints and future projections are shown in Fig. 8.19 for the muon (Top) and electron (Bottom) dominance scenarios.

Low-mass HNLs (below $\sim$5 GeV) are best explored by beam-dump experiments like SHiP [ID145] or LCF [ID140], and by off-axis, shielded detectors such as FASER2 [ID19, ID235], CODEX-b [ID199] or ANUBIS [ID273]. Though disfavoured in the minimal $\Lambda_{\text{CDM}}$ scenario, HNLs below the kaon mass may still be viable cosmologically; they can be probed at neutrino facilities like DUNE [ID118], SBND [ID234] or Hyper-K [ID238].

Above these masses and below the mass of the $Z$ boson, searches at FCC-ee [ID233, ID242] at the $Z$-pole dominate the sensitivity, in both prompt and displaced signatures. Hadron collider experiments like ATLAS, CMS, LHCb [ID81], as well as at the LHeC [ID214] in the case of electron-coupling dominance, probe masses above the $b$-quark threshold, though their ultimate sensitivities are significantly below the FCC-ee reach. FCC-hh [ID227, ID247] capabilities remain to be fully explored, including proposals for dedicated detectors, such as the mid-$\eta$@FCC-hh [468]. HNLs above 100 GeV become accessible at high-energy colliders such as LCF [ID140], FCC-hh, or a muon collider [ID207]. Moreover, experiments like SHiP [ID145] and FCC-ee [ID242] offer observation sensitivity to over an order of magnitude improvement in coupling reach beyond HL-LHC exclusion power, and unique opportunities to study HNL properties if hints emerge during HL-LHC operation.

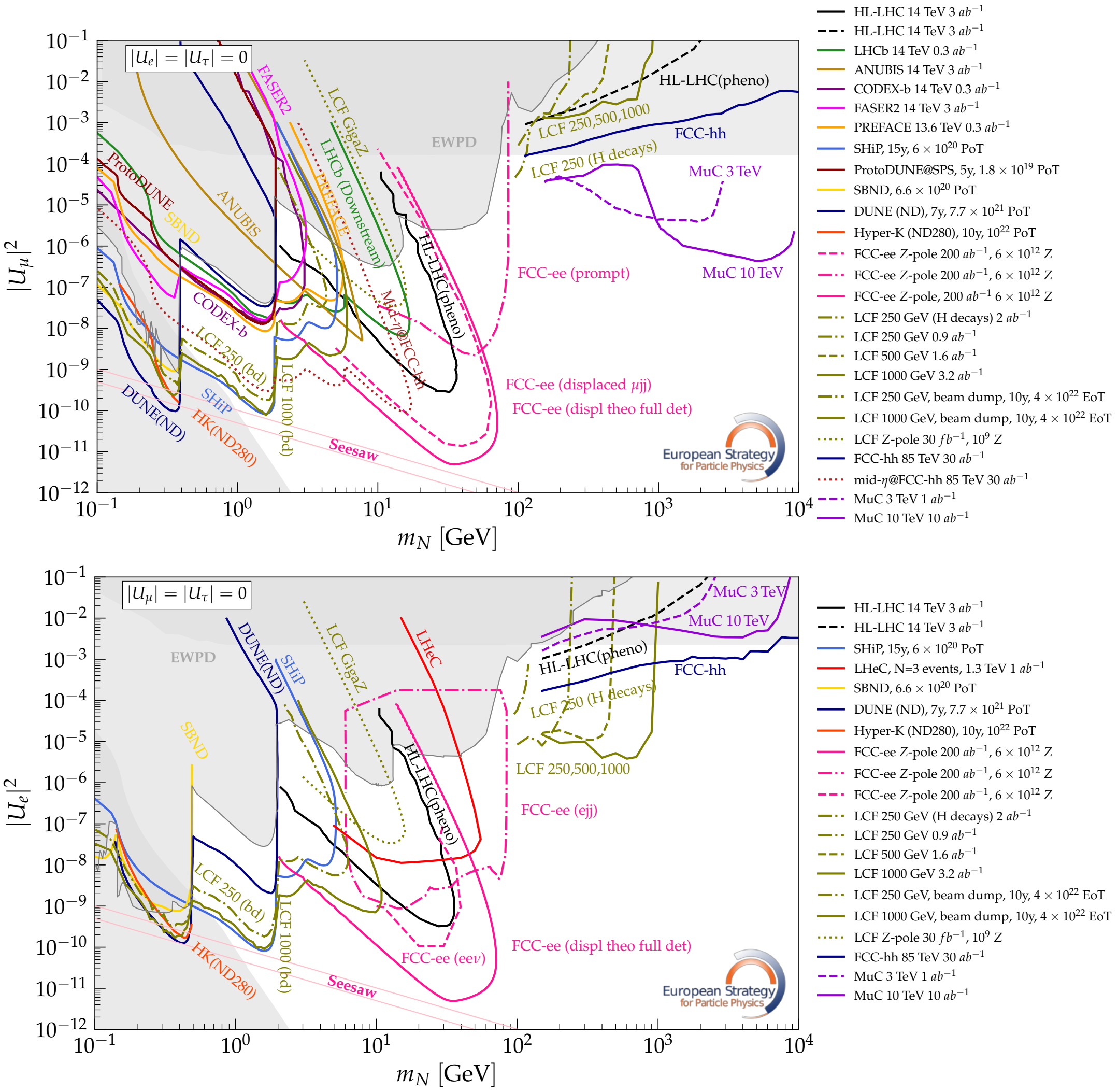

Fig. 8.19: Exclusion limits for HNLs mixing with muon (Top) and electron (Bottom) neutrinos.

Finally, electroweak precision observables provide indirect constraints on $|U_\ell|^2$ at the level of $10^{-4} - 10^{-2}$, depending on the $\ell$ flavour and HNL mass, up to masses far beyond the reach of any current or future collider [469, 470]. These constraints are expected to improve by another two orders of magnitude in $|U_\ell|^2$ in the FCC-ee era [471–473].

## 8.6 Conclusions

In this chapter, we have painted a picture of the many ways that future exploration beyond the SM could yield discoveries and deepen our understanding of fundamental interactions. We do not yet know how Nature resolves the outstanding puzzles of the SM, and so a broad and comprehensive experimental programme is essential to uncover new physics and study its properties.

Perhaps the most pressing of these questions concerns the origin of the weak scale. The experimental confirmation of the Brout-Englert-Higgs mechanism shows that an effective parameter in the SM is inherently sensitive to the presence of arbitrarily short distance scales that must exist in nature. As discussed in this chapter, there are several possible pathways to discovering the particles underlying the weak scale, whether the Higgs boson is composite, supersymmetry exists at the TeV scale, or something more exotic. A multi-pronged strategy is critical: the interplay of indirect and direct probes will be key to constraining each of these scenarios. For supersymmetry, the particles most relevant to the weak scale can be probed at unprecedented levels: a high-energy hadron collider would be sensitive to gluinos up to $\sim$15 TeV and stops up to $\sim$10 TeV. For lepton colliders (e.g., LCF, CLIC, MuC), the generic reach extends to $\sqrt{s}/2$, with the highest reach among the options discussed provided by a 10-TeV muon collider, capable of discovering Higgsinos up to $\sim$5 TeV. The challenging scenarios of colourless top partners can be tested indirectly through Higgs-coupling measurements, probing the most natural parameter space.

Composite Higgs models can be probed through three complementary avenues: indirect SMEFT effects, direct/indirect searches for vector resonances, and the direct search of fermionic top partners. Precision measurements at future $e^+e^-$ colliders can constrain SMEFT operators associated with the new strong sector, probing compositeness scales at 10–20 TeV, with a 10-TeV muon collider extending the indirect reach to about 60 TeV. Direct searches target vector resonances near the compositeness scale, with hadron colliders such as the HL-LHC and FCC-hh extending sensitivity to 5–7 TeV and 30–50 TeV, respectively. Finally, top partners, central to Higgs mass generation, are accessible through both pair and single production: the HL-LHC will push current bounds, while FCC-hh and a 10-TeV muon collider can extend sensitivity up to $\sim$15 TeV and $\sim$8 TeV, respectively.

In broad classes of BSM scenarios, additional scalar particles are expected, with wide-ranging phenomenological consequences best explored through a complementary programme of precision measurements and direct searches. An especially intriguing aspect of extended scalar sectors is their potential impact on the electroweak phase transition and, consequently, on the origin of the matter/antimatter asymmetry. The minimal extension studied here, with rich phenomenology but parameter space constrained by the requirement of a FOPT, shows that the integrated FCC-ee +FCC-hh programme and/or a lepton-lepton collider at the highest energies, such as a 10-TeV muon collider, could fully test it. More generally, future high-energy colliders probing the Higgs potential, together with complementary EDM experiments, offer a path to uncover the origin of this asymmetry, either through the discovery of new states or by achieving percent-level precision on the Higgs self-coupling.

Many theories predict additional forces beyond those already present in the SM. Searching for the associated resonances is a hallmark of collider physics, and highlights the interplay of direct and indirect searches. This class of models can be well tested by $\ell^+\ell^- \to f\bar{f}$ and electroweak precision tests at lepton colliders, depending on the flavour structure and couplings to the Higgs doublet. For flavour-universal cases and a benchmark coupling $g_{Z'} = 0.5$, a 250 GeV $e^+e^-$ collider could probe ∼20 TeV, an 85 TeV hadron collider up to ∼40 TeV, and a 10-TeV muon collider beyond ∼100 TeV. Flavoured gauge forces tied to the origin of SM Yukawas can be probed beyond the regions favoured by naturalness, while forces coupling only to the third generation may first appear in flavour observables or electroweak precision tests at a Tera-$Z$ run.

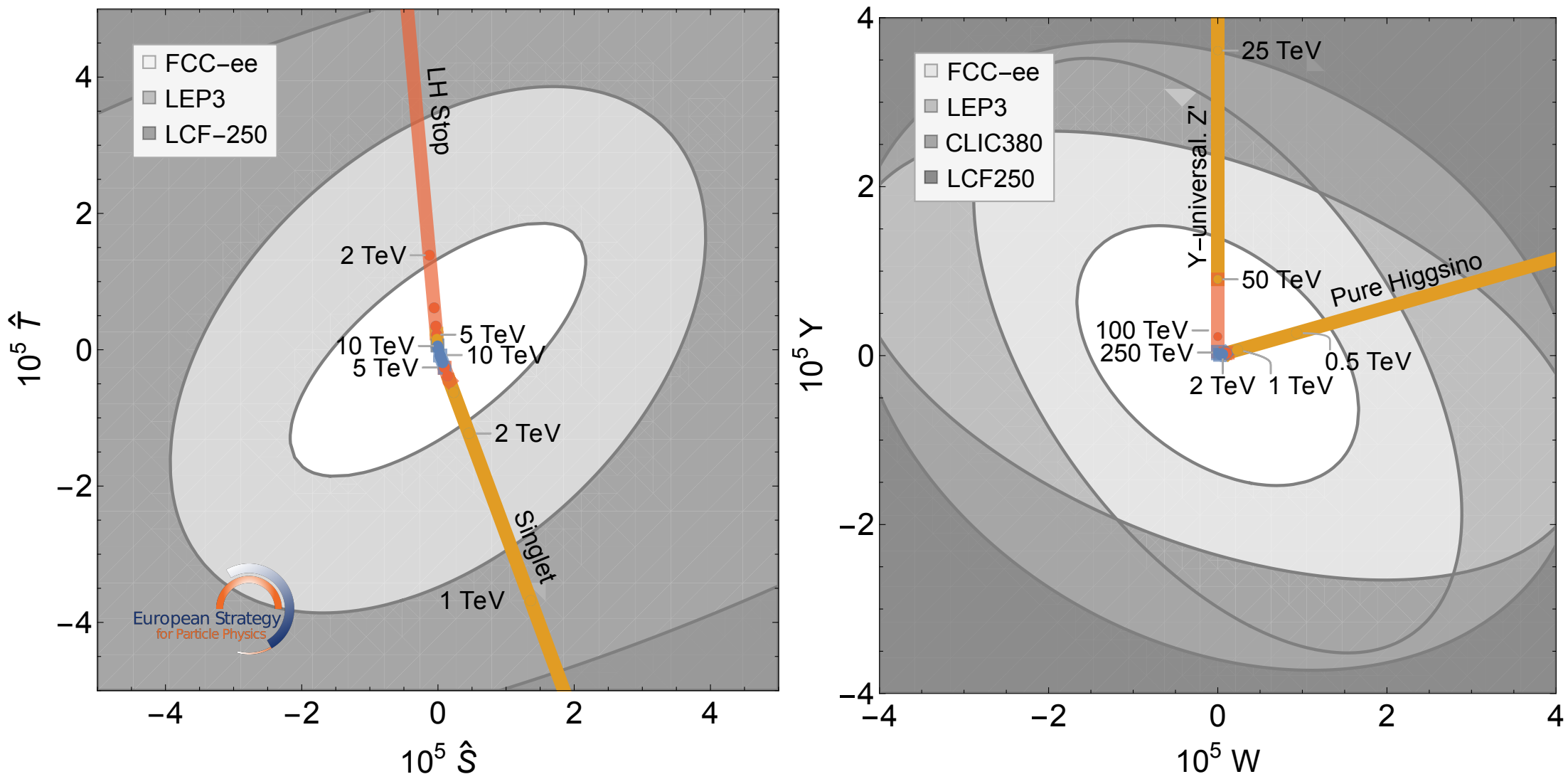

Fig. 8.20: Reach of energy frontier colliders in representative BSM scenarios relative to electroweak precision observables at low-energy $e^+e^-$ colliders. Left: $(\hat{S},\hat{T})$ 95% CL ellipses relative to contributions from a left-handed stop squark and a singlet scalar (with $\sin\theta = m_h/m_S$) as a function of particle mass. Orange and red shading of the stop and singlet lines indicates masses within reach of FCC-hh and a 10-TeV muon collider, respectively. Masses covered by both colliders are shaded by the collider with the lower overall reach, while blue shading indicates contributions outside the reach of both colliders. Right: $(W,Y)$ 95% CL ellipses relative to contributions from a pure Higgsino and a $Y$-universal $Z'$ (with $g_{Z'} = g$), shaded according to coverage by FCC-hh and a 10-TeV muon collider.

Portal interactions allow colliders to probe hidden sectors that are otherwise inaccessible. In this chapter, we studied four portal scenarios, all emphasizing the complementarity between a variety of experimental probes. These simple light extensions of the SM display a rich phenomenology, highlighting the need for diverse and complementary strategies to search for new physics. Hadron and muon colliders probe high-mass, strongly coupled scenarios, while beam-dump experiments target low-mass particles with very weak couplings. The Tera-$Z$ run of FCC-ee could reach $g_{a\gamma\gamma} \sim 10^{-6}$ GeV$^{-1}$ for axion-like-particles, branching ratios of $10^{-1}$ for Higgs decays into dark scalars, and sensitivity near the see-saw line for heavy neutral leptons.

By exploring the implications for a set of selected models, we have demonstrated the broad strengths and weaknesses of different collider options. The main conclusion of this chapter is that combining the best achievable precision with the highest attainable energies offers the potential to probe a wide range of BSM scenarios at scales of order 10 TeV and above. In this respect, FCC-ee (and, to a luminosity-proportional lesser extent, LEP3), followed by FCC-hh and/or a 10-TeV muon collider, are found to be the most powerful options. In the scenarios presented here, an LCF with energy runs between 250 and 1500 GeV reaching the planned luminosities (e.g., a giga-Z run) [ID140] provides a significantly reduced reach with respect to an integrated ee-pp programme or a 10-TeV muon collider.

Figure 8.20 quantitatively summarizes our findings. Model-independent precision constraints on the EWPO observables $\hat{S}, \hat{T}, W, Y$ are shown in shades of grey, together with the corresponding scales in the models studied in this chapter. The colour coding of the lines indicates whether direct searches can access these scales. The examples in the figure show that precision measurements and direct searches should generally be regarded as independent probes. For example, some models would provide early signals in EWPO at FCC-ee or LEP3 (e.g. compositeness), while others (e.g. SUSY) would not, but either could be discovered later at FCC-hh or at a 10-TeV muon collider. Considering also that high-statistics $Z$-pole runs will provide (in several cases unique) sensitiveness to low-mass weakly interacting new particles, the examples discussed in this chapter underscore not only the essential complementarity of precision and energy-frontier programmes, but also their necessity in our quest to uncover the dynamics behind the weak scale.

# Chapter 9

# Dark Matter and Dark Sector

## 9.1 Introduction

The evidence for the existence of dark matter (DM) is multifaceted and the scientific consensus is that the existence of some form of DM is incontrovertible. Identifying its fundamental nature is a task of paramount importance in particle physics. A broad range of experimental techniques pursue this goal; we broadly categorize these in this chapter as cosmological constraints, direct detection, indirect detection, collider production, and accelerator-adjacent experiments, and report here on the current status and prospects of each towards identifying the nature of DM.

The primary milestone would be to achieve a detection of properties which go beyond the suite of gravitational phenomena observed to date. To search for such a non-gravitational interaction requires a strategy for where and how to look experimentally, and in this respect the perspective of the particle physics community has evolved significantly in the last decade.

The visible sector comprises $\sim 18\%$ of all the matter in the Universe, and DM comprises the other $\sim 82\%$, accounting by cosmologically-stable matter budget. Within that small $\sim 18\%$ we observe enormous phenomenological diversity across a wide range of scales, despite the fact that, by matter budget accounting, most of the visible matter in the Universe is in the form of hydrogen ions. Clearly a physical picture of the visible Universe based on a matter budget accounting misrepresents the phenomenological landscape. Thus while it is plausible that the majority of DM is one type of relatively weakly-interacting field, the physical phenomena engendered by it could be much richer. It is this realisation which has led to a broadening of the conceptual landscape over the past decade, from DM itself to the possibility of dark sectors more generally.

As in the visible sector, the dark sector may very well contain additional states which are unstable, yet give rise to phenomena which could open the shutters to the dark sector, possibly without even involving the dominant DM state. One need only recall the phenomena associated with the photon, electron, muon, pions, neutrinos and elements heavier than helium to see why searches allowing for phenomenological diversity are well-motivated. How may we access the cornucopia of possible dark sector phenomena whilst retaining a targeted approach to accessing the dark sector experimentally? A first step is to organise phenomena according to the scale on which they arise. That is the logic behind the structure of this chapter, to organise by length scale (mass) rather than by interaction type.

A subsidiary organising concept is that of portals, as introduced in Chapter 8. In attempting to explore the dark sector, we consider the 'portal' interactions allowed between the Standard Model (SM) and new gauge-neutral states. Depending on the particle spin, we distinguish the following lowest-dimension, gauge-invariant portals: scalar, pseudoscalar, neutrino, and vector [457]. They add, correspondingly, a Higgs-like scalar, an axion-like particle (ALP), a heavy neutral lepton (HNL), and a mediator coupled to the combination of the baryon, lepton, and the electromagnetic current (like a dark photon). We emphasize that this portal approach may capture dark sector particles which comprise the DM, but also intentionally has scope beyond just the DM itself.

Considerations of DM production mechanism offer guidance on the possible mass range of the dark sector. For decades, the dominant paradigm has been thermal freeze-out, in which the DM is in thermal equilibrium with the SM bath at $T \gg m$, until the two decouple when the annihilation rate becomes smaller than the expansion rate of the Universe, and the DM abundance is thus frozen, diluted only by cosmological expansion. Weakly Interacting Massive Particle (WIMP) DM is the prototypical example. However, other production mechanisms are possible. Alternatively, DM could start with negligible abundance and be populated by very rare interactions with the SM bath, until decoupling. This is the so-called freeze-in mechanism [474, 475]. Non-thermal production mechanisms are required for ultra-light DM to behave as cold DM, consistent with astrophysical observations. Axion DM, first posited to solve the strong CP problem [476–478], is the classic example.

The mass range of a DM particle candidate dictates the experimental search strategies, optimally with strong complementarity between approaches. Ultralight DM ($m \lesssim$ eV) searches employ atomic and optical interferometry, haloscopes, and gravitational probes such as pulsar timing arrays. Light DM (keV $\lesssim m \lesssim$ GeV) searches for stable DM employ direct detection, looking for DM particles to interact with atomic nuclei or electrons, while unstable DM within the keV-MeV range may be probed using X-ray searches, CMB distortions, and other astrophysical and cosmological observations. Accelerator experiments may efficiently explore the parameter space in both the light and heavy DM mass ranges, utilizing signatures such as missing energy or additional decays of dark sector states accompanying the DM particle. Heavy DM (GeV $\lesssim m \lesssim$ 10 TeV) searches concern colliders, direct, and indirect detection, which aim to detect the by-products of DM scattering, decay, annihilation, and in the case of indirect detection additionally the capture and accretion of DM in astrophysical sources. Ultra-heavy DM ($m \gg$ 10 TeV) searches employ direct and indirect detection. Cosmology and astrophysical searches provide important constraints on e.g. primordial black hole DM candidates. Together, this wide range of experimental approaches provides valuable and complementary constraints on models of DM and physics beyond the Standard Model.

## 9.2 Ultralight Dark Sectors

In ultralight DM (ULDM) searches a vibrant experimental landscape has emerged, with great diversification over the past decade, spanning from innovative concepts to technically mature designs. This section draws on community input documents [ID54, ID37, ID146, ID164, ID198, ID204, ID228, ID252, ID260] and reports [ID13, ID27, ID112, ID235] to highlight ongoing progress and future opportunities across laboratory, astrophysical, and cosmological probes.

Benchmark models considered here for sub-eV DM candidates include QCD axions and ALPs as the primary focus, as well as dark photons and dark scalars. The MeV–GeV mass win-dow for ALPs and dark photons is accessible at accelerator experiments, discussed in Sect. 9.3.

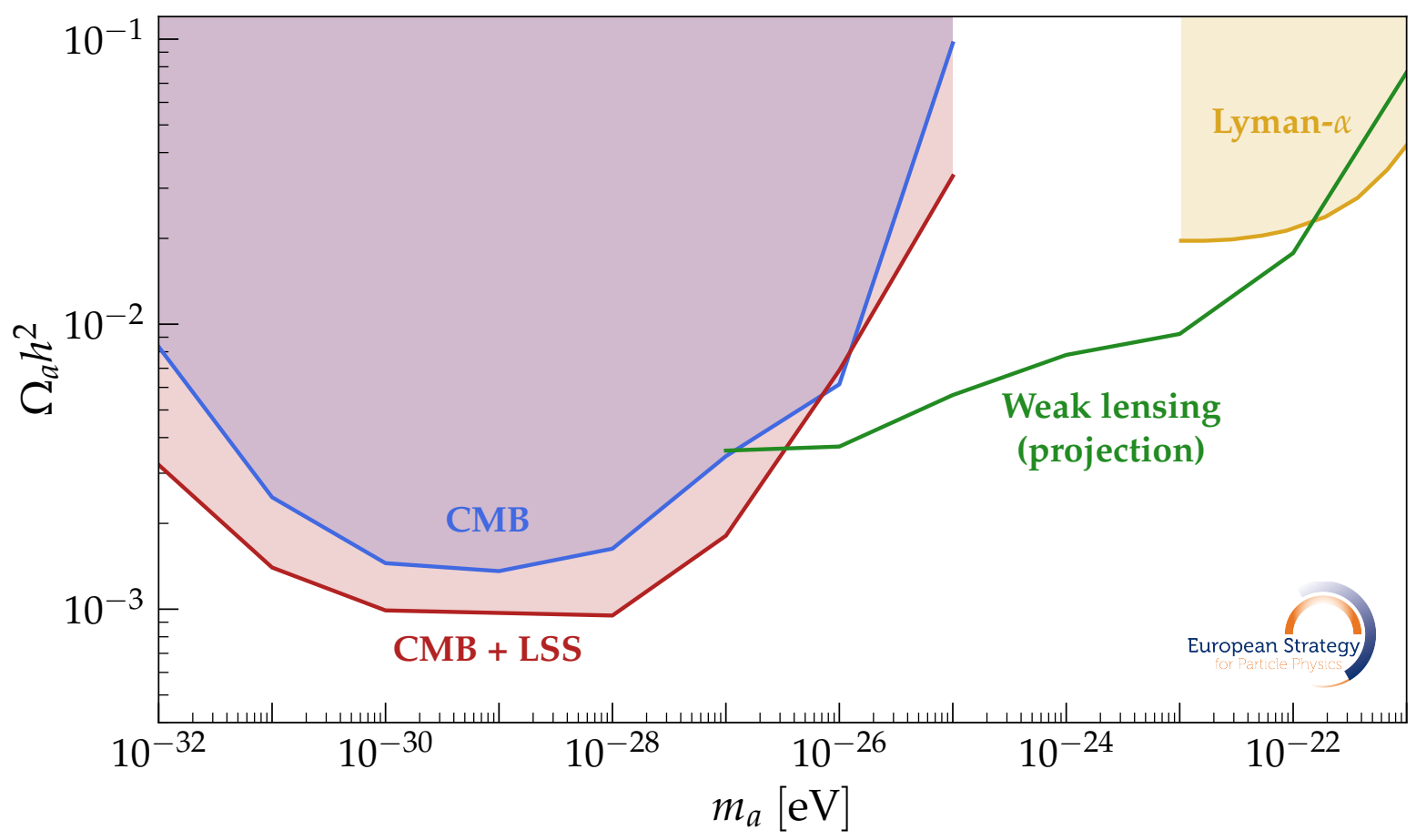

Fig. 9.1: The allowed parameter space for the energy density in ULDM as a function of mass. Shaded regions correspond to current bounds, while the lines display projections for datasets that will come online within the next decade.

Axions naturally emerge as pNGBs from the PQ solution to the strong CP problem with mass $m_a \sim m_\pi f_\pi / f_a$, with $f_a$ the axion decay constant [479–481]. The dimensionless couplings to photons are typically assumed to be of order unity, as predicted e.g. by the KSVZ ($g_\gamma = -0.97$) [482, 483] and DFSZ ($g_\gamma = 0.36$) [484, 485] model benchmarks. High magnetic fields can provide high energy density scattering targets, thus most experiments target the axion-two-photon coupling $g_{a\gamma\gamma} = (g_\gamma \alpha/\pi)/f_a$. ALPs are generically predicted in high-scale theories like string theory, and allow a wider parameter space. Couplings around $g_{a\gamma\gamma} \sim 10^{-11}$ GeV$^{-1}$ are particularly interesting as they may explain astrophysical anomalies [486, 487]. Ultralight dark photons can arise from extra U(1) sectors, and kinetic mixing of the dark photon with the visible photon has signal strength characterized by the mixing angle (and is independent of magnetic field). In such models, the dark photons may be accompanied by a dark millicharged sector, which can be detected due to the small mixing angle.

Constraints on ULDM candidates come from cosmological and astrophysical observations (Sect. 9.2.1) and dedicated axion conversion or production experiments (Sect. 9.2.2). We note the importance of pursuing multiple larger and smaller-scale search efforts to cover all viable axion parameter space spanning over 10 orders of magnitude in mass; together these offer a strong prospect of testing most of the currently favoured QCD axion model band region in the coming decade and beyond.

### 9.2.1 Cosmological and Astrophysical Constraints

Cosmology and astrophysics are sensitive to ULDM when the de-Broglie wavelength of the ULDM is so large that it affects the gravitational dynamics of astrophysical objects, and the Large Scale Structure (LSS) of the Universe more generally [488]. Current data place strong constraints on the fraction of DM comprised of ultra-light particles, as shown in Fig. 9.1. At the low-mass end, ULDM becomes indistinguishable from a cosmological constant, then the constraints are progressively dominated by measurements at increasingly smaller scales: the CMB and the LSS first [489], followed by weak lensing [490], then astrophysical probes like the Lyman-$\alpha$ forest, strong lensing and counts of Milky-Way satellites [491–493]. Beyond the cosmological bounds, constraints for $10^{-18}\,\mathrm{eV} \lesssim m_a \lesssim 10^{-12}\,\mathrm{eV}$ can in principle be obtained using observations of dwarf galaxies, Pulsar Timing Arrays, and population studies of black hole spin-down triggered by superradiance effects induced by a light scalar field [494–499]. Moving towards higher masses, cosmology also places bounds on a possible population of hot, i.e. thermally produced, axions with $m_a \sim \mathcal{O}(\mathrm{eV})$, as they unavoidably contribute to the energy density in relativistic species [500], offering complementarity to direct detection searches in this mass range [501]. Light relics e.g. $\mathcal{O}(\mathrm{eV})$ light gravitinos, could also contribute to a fraction of the DM today, with strong constraints from CMB and LSS data [502].

Solar observations, including helioseismology, solar age considerations and neutrino fluxes, limit the axion-photon coupling to $g_{a\gamma\gamma} \lesssim 6 \times 10^{-10}\,\mathrm{GeV}^{-1}$ [503]. Similar considerations can be used to constrain dark photons [504–506]. Globular cluster stars, particularly horizontal branch stars, constrain photon couplings to similar values ($g_{a\gamma\gamma} \lesssim 0.65 \times 10^{-10}\,\mathrm{GeV}^{-1}$ at 95% CL) through modifications in stellar evolution [507], and X-ray, gamma-ray and radio analyses set constraints $g_{a\gamma\gamma} \lesssim$ a few $\times 10^{-12}\,\mathrm{GeV}^{-1}$ for $m_a$ below $10^{-5}\,\mathrm{eV}$ in mass [508–511]. White dwarf cooling [486, 512] and red giant branch stars [513, 514] restrict the CP-odd axion-electron Yukawa coupling to $g_{ae} \lesssim$ a few $\times 10^{-13}$ [503]. The cooling of neutron stars (NS) can also provide constraints: for young NS (e.g. [515, 516]) a bound can be derived on the coupling of axions to neutrons, while for old NS (e.g. [517]), where neutrino emission becomes inefficient and surface photon emission starts to dominate, the coupling of axions to protons can be constrained. Beyond this, neutrino data from Supernova SN1987A limit nucleon couplings, implying axion masses $m_a \lesssim$ few $\times 10^{-2}\,\mathrm{eV}$ for QCD axions [518]. Stellar and astrophysical bounds have recently been summarized in Refs. [503, 519].

### 9.2.2 Experimental Landscape for ULDM Searches

Current search strategies for axions and ALPs can be broadly categorized based on their production source: (1) Haloscopes [520] aim to directly detect axions that make up the local DM halo, (2) Helioscopes [521] search for axions generated in the Sun, leveraging solar axion production mechanisms without relying on the assumption that axions are the DM, and (3) Laboratory-based experiments would produce and detect axions entirely within controlled environments, offering searches that are independent of astrophysical or cosmological assumptions.

Current experimental limits on axions and ALPs, along with the projected reach of next-generation searches, are presented in Fig. 9.2 and discussed in the following. The global search for ULDM particles such as the axion has gained substantial momentum, with major European initiatives like IAXO aiming to extend the sensitivity to solar axions, haloscope experiments such as MADMAX targeting the direct detection of axion DM, and complementing large light-shining-through-a-wall experiments like ALPS II providing competitive, model-independent laboratory tests. These flagship efforts, together with a diverse range of smaller-scale experiments, will require sustained support to comprehensively explore the axion parameter space. This broad experimental landscape could eventually converge toward a large-scale ‘Ultimate Axion Facility’, offering shared infrastructure such as cryogenics and high-field magnets, while fostering collaboration and knowledge exchange across different approaches. The overlap in technologies, design challenges, and scientific goals among these efforts is reviewed in [522].

***Haloscopes.*** Resonant cavity haloscopes search for axion DM, particularly within the QCD axion band, via the axion-to-photon conversion in a strong magnetic field inside a resonant microwave cavity tuned to match the axion mass. ADMX [524] has pioneered this approach and continues to lead in sensitivity at low masses. Other efforts such as QUAX and similar experiments [525–528] are exploring complementary ranges and implementing innovative technologies, such as quantum-limited amplifiers and squeezed states. A new generation of resonant haloscopes is under development including RADES [529], GrAHal [530, 531], SUPAX [532], CADEX [533], ORGAN [534], TASEH [535], and PXS [536], which aim to extend frequency coverage, increase scanning speed, and improve sensitivity.

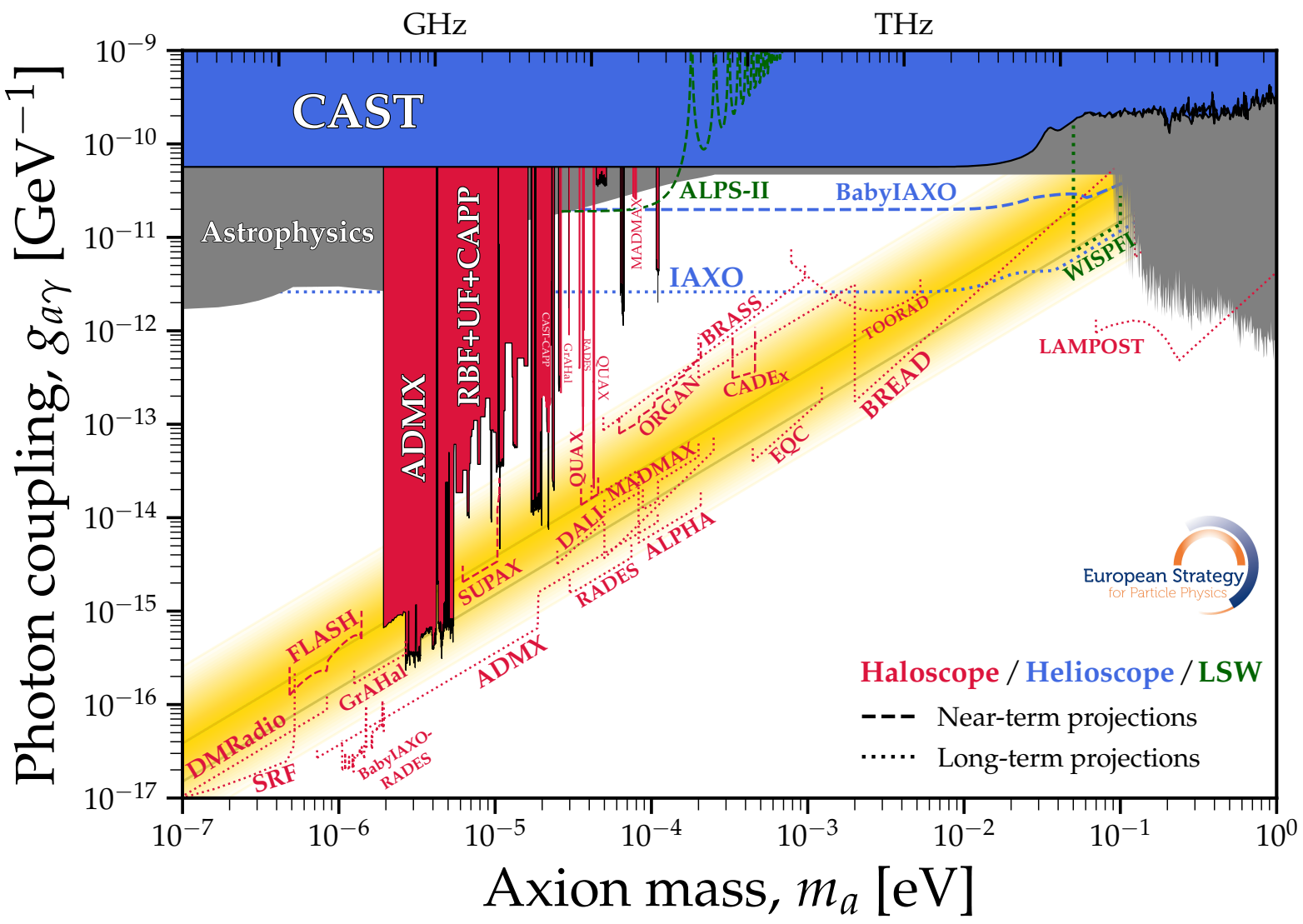

Fig. 9.2: Experimental prospects and current bounds on the coupling of axions and ALPs to photons in the sub-eV mass-scale for different types of axion searches [523]. Solid lines indicate existing experimental (red: haloscopes, blue: helioscopes, green: pure laboratory searches) and observational (grey) bounds, dashed and dotted lines for near-term and long-term prospects. The yellow shaded region indicates the typical prediction of QCD axion models. Couplings to particles other than photons can be found in [523].

To probe higher axion masses above the reach of conventional cavities, new concepts are being pursued. Dielectric haloscopes, such as MADMAX [537], DALI [538], LAMPOST [539], and EQC [540], use layered dielectric interfaces to coherently enhance axion-photon conversion at higher frequencies. Dish antenna experiments like BRASS [541] and BREAD [542] aim to detect broadband photon signals from axion conversion on reflecting surfaces. Plasma haloscopes, including ALPHA [543], manipulate the dispersion relation of photons in plasma to enable resonant detection at higher masses. TOORAD [544] explores using magnetic topological insulators as frequency-tunable, solid-state resonant targets.

Several efforts target lower mass ($<\mu$eV) axions. FLASH [545] and BabyIAXO-RADES [546] are exploring novel magnet and cavity concepts in this low-frequency regime. Lumped-element detectors, such as DMRadio [547] and WISPLC [548], replace resonant cavities with circuit components like inductors and capacitors, enabling searches at lower masses while maintaining high sensitivity through cryogenic and quantum sensing techniques. SRF-m$^3$ [549] proposes a novel heterodyne approach to access low frequency signals via resonant conversion of photons between two neighboring high-frequency modes of a single high-Q cavity.

***Helioscopes.*** Axion helioscopes are designed to detect axions produced in the core of the Sun, where they could be generated via the Primakoff effect [550] in the strong electromagnetic fields of the plasma. On Earth, these solar axions can reconvert into X-ray photons when passing a strong, transverse magnetic field. No assumptions requiring axions to be the DM are included.

The CAST experiment at CERN has led this effort, delivering the most stringent limits to date on the axion-photon coupling in the sub-meV mass range [551, 552]. With CAST now concluded, the International Axion Observatory (IAXO, [553]) is being developed as the next-generation helioscope, aiming to improve sensitivity to axion-photon couplings by roughly 1.5 orders of magnitude. IAXO will access the unexplored parameter space of realistic QCD axion models in the meV mass regime, and investigate persistent astrophysical anomalies, such as anomalous stellar cooling and unexpected gamma-ray transparency. In addition to probing photon couplings, IAXO is expected to be sensitive to axion-electron and axion-nucleon interactions, enabling it to constrain or even determine the axion mass in case of a discovery [501]. As a key intermediate step, BabyIAXO [554] is entering its construction phase at DESY with support from CERN. It serves a dual role: reduction of technical risk by validation of crucial components and integration strategies, and operation as a scientifically capable pathfinder with the potential to deliver first physics results within the next five years. In parallel, the implementation of a haloscope setup (BabyIAXO-RADES) is anticipated, which would allow the IAXO infrastructure to also search for dark-matter axions in the microwave regime.

Beyond terrestrial helioscopes, novel strategies have been proposed to use X-ray space telescopes as axion detectors. The magnetic fields in the outer solar atmosphere could serve as natural reconversion zones for axions, allowing satellites such as NuSTAR [555, 556] or future missions to detect excess solar X-rays as potential axion signals.

**Pure Laboratory Experiments.** Laboratory axion searches using the light shining through a wall (LSW) principle offer a fully controlled and model-independent approach to probing axion-photon interactions. In these experiments, laser photons would be converted into axions in a magnetic field, pass through an opaque wall, and are regenerated as photons in a second strong magnetic field on the other side. This would provide a clean measurement of the axion-photon coupling independent of assumptions of the local DM density. ALPS II [557] at DESY is the most advanced LSW experiment to date. With high-finesse optical cavities and superconducting magnets, it is currently running and on track to reach its design sensitivity to explore new parameter space, particularly for sub-meV axion-like particles. Looking ahead, novel setups, like for example WISPFI [558], a fibre interferometer concept, are being developed to target higher axion masses in the meV to 100 meV range. Future projects like HyperLSW [559] could deliver targeted high-sensitivity verification of a potential axion discovery within a laboratory setting.

**Beyond Axion-Photon Couplings.** While many experiments focus on the generic axion-photon coupling, a growing set of laboratory searches target other axion interactions, such as couplings to gluons, nucleons, and electrons. These experiments open complementary pathways into the axion parameter space by probing distinct physical effects. For example, CASPEr [560] uses Nuclear Magnetic Resonance (NMR) techniques to search for axion-induced spin precession, which would signal couplings to nucleons or gluons. Similarly, QUAX [561] investigates axion-induced spin precession in a ferrimagnetic crystal (YIG) to target the axion-electron coupling, while ARIADNE [562] aims to detect axion-mediated spin-dependent forces. GNOME [563] constitutes a network of optical magnetometers with sensitivity to spin-dependent couplings of ultralight bosonic fields. In addition, the NASDUCK experiment [564, 565], based on hot vapour magnetometers, probe ALP DM nucleon coupling for masses below $2 \times 10^{-10}$ eV.

**Future Directions with Quantum Sensors.** A growth area since the last European Strategy Update concerns the leveraging of advances in quantum sensing technologies to target

ULDM, and to open synergies in detecting DM and gravitational waves (GW). Quantum-limited amplifiers are being deployed by most experiments targeting signals in the ∼GHz range. At higher frequencies where blackbody emission is exponentially suppressed, much better noise performance can be attained with qubit-based single microwave photon detection, as pursued by [566–568]. Of note is the recent demonstration of a 2-qubit quantum non-demolition detector with sufficiently low dark rate to cover the predicted QCD axion coupling in the ∼10 GHz band [569, 570]. In the infrared-optical band, the world-leading sensor is the superconducting nanowire single photon detector which has achieved dark count rates as low as 1 count/day [571]. It may also find use in high-occupancy collider detectors due to its picosecond risetime. Even lower dark rates are targeted by [572], with an optical heterodyne detector.

Scalar DM with photon and/or gluon coupling can be efficiently probed by atomic clock comparison [573–576] and by $^{229}$Th nuclear clocks [577–580]. Ultralight scalar DM is targeted in atom interferometry experiments [581–584] by probing differential signals in the 0.1–1 Hz band, overlapping with GW science. Solid-state and cryogenic detectors aim to detect absorption signals, while storage ring experiments [585] search for oscillating EDMs. A related recent development is the use of axion detection techniques to search for gravitational waves. These have similar 2-photon couplings as axions. Efforts are underway to develop cavity detectors targeting gravitational wave signals [586, 587] [ID260].

### 9.3 Light Dark Sectors

Searches for DM candidates with mass between keV and GeV, and for DS with new particles in a similar mass range, have been boosted considerably in recent years, with experiments not even at the proposal stage at the time of the past European Strategy setting world-leading constraints now. This area has been a strong driver for innovation in detector development and has become a major player in the LHC-extended and non-LHC science programme at CERN and beyond, as illustrated in this section, based on community input documents [ID19, ID23, ID44, ID46, ID50, ID61, ID81, ID92, ID109, ID118, ID126, ID145, ID175, ID199, ID235, ID268, ID273, ID275, ID280].

Benchmark models considered here include portal particles which may be cold or warm at freeze-out; and DM coupled via mediators. Portal particles as light DM include $Z'$ and massive dark photons, ALPs and light scalars, and sterile neutrino DM. The former can be in the light DM range depending on the model choices; for example one can have $m_{Z'} \sim 10-100$ keV as a DM candidate with production via freeze-in. ALPs were introduced in Sect. 9.2, and may populate the light DM mass range as well as ultralight. Sterile neutrinos, $\nu_s$, are a possible warm DM candidate in the keV to MeV mass range, which is light enough to be sufficiently long-lived yet heavy enough to not be in tension with structure formation and X-ray observational constraints, in the context of a modified cosmological setup and/or additional $\nu_s$ interactions (see e.g [588] and subsequent work).

Alternatively, portal particles may not be the DM but rather mediators coupling the visible to the dark sectors. In such a case, the DM particle is typically stable and its mass may range from MeV up to the electroweak scale and more. In the example of a dark photon mediator $A'$, after spontaneous dark symmetry breaking, one is left with two physical Majorana states $\chi_i = \{\chi, \chi'\}$ and their interactions with $A'$ and the dark Higgs boson $S$. The phenomenology of the model and experimental probes is controlled by the couplings determining the mixing between the SM photon and the dark photon, as well as the masses $m_{A'}, m_S, m_\chi, \Delta \equiv (m_{\chi'} -$ $m_\chi)/m_\chi$ [457, 589]. In the simplified scenario when (a) the dark scalar sector is unobservable, and (b) only the off-diagonal vectorial coupling $\chi'\chi$ is present, three cases are considered: (i) elastic DM, $\Delta = 0$; (ii) quasi-elastic DM, where $\Delta$ is tiny yet non-zero; and, (iii) inelastic DM, where $\Delta$ is sizeable. Depending on the value of $\Delta$, the strongest constraints come from direct detection or accelerator searches.

Light and feebly coupled sectors are efficiently explored at accelerator-adjacent experiments. Conversely, collider experiments can probe mass ranges and couplings in hidden sectors not accessible there, particularly in the $m \gg m_B$ and large-mixing regime. Sensitivities of current and future experiments are briefly summarised in Sect. 9.4.

Constraints on light DM candidates and dark sectors come from cosmological and astrophysical observations (Sect. 9.3.1), direct detection experiments (Sect. 9.3.2) and accelerator-adjacent and collider searches (Sect. 9.3.3). We note that leading constraints today in this mass range come from experiments across a range of scales (time, cost, complexity), and this diversity of approaches is vital to realise the strong prospect for testing most of the freeze-in and neutrino bound benchmarks in the coming decade and beyond.

### 9.3.1 Cosmological and Astrophysical Constraints

Cosmological observations place stringent limits on thermal sub-GeV DM with $s$-wave annihilation into SM particles other than neutrinos. The energy injection during the CMB epoch leads to spectral distortions and an altered ionization history, excluding thermal relics with masses below 10 GeV [395, 590, 591]. However, models with velocity-suppressed $p$-wave or $d$-wave annihilation are viable, as the annihilation rate at recombination becomes negligible. Alternatively, freeze-in scenarios evade BBN, CMB, and 21 cm constraints, as they lack significant late-time energy injection. Elastic interactions with baryons and electrons are constrained by CMB observations. Such interactions induce a drag force between DM and the tightly coupled photon-baryon fluid, modifying the damping tail and phase shift of acoustic peaks [592–594]. These constraints potentially disfavour significant portions of the sub-GeV DM mass range for models involving light mediators, milli-charged particles, or atomic DM, among other possibilities. Finally, BBN measurements provide powerful constraints on unstable DS particles with lifetimes above $\tau \gtrsim 0.02$ s and masses up to the TeV scale [595–597].

For sub-MeV DM strong cosmological constraints apply if the DS reaches thermal equilibrium with the SM, as only electrons, photons and neutrinos are present in the SM bath at these temperatures. In this case a thermal DM would unavoidably produce an increase in the energy density in relativistic particles well beyond the current constraints from BBN and CMB, providing a *lower* bound on the DM mass of $\mathcal{O}(10)$ MeV [598–600]. While this bound can be somewhat relaxed in modified cosmological scenarios, strong constraints still apply as this can alter the primordial abundances and the physics of recombination. Even in an almost completely secluded DS, any particle, e.g. a light mediator, that achieves thermal equilibrium with the SM up to the QCD crossover would produce an increase in the relativistic energy density that can be probed by the next generation of CMB experiments [398, 601, 602].

Warm DM (WDM) relates to a scenario in which the particles associated with it possess significant free-streaming lengths, $\lambda_{\rm fs} \sim$ Mpc for keV-scale particles. They owe this property to being created semi-relativistically at a time when their cosmic abundance is being set in the early Universe. The free streaming effectively suppresses structure formation [603, 604] below halo mass scales $M \approx 3 \times 10^8 M_\odot \, (m_{\rm WDM}/{\rm keV})^{-3.33}$ where $m_{\rm WDM}$ is a thermal-relic equivalent particle mass [605]. Mapping $m_{\rm WDM}$ to specific WDM models depends on the WDM production scenario. Strong constraints emerge from Lyman-alpha forest data, imposing lower mass bounds $m_{\rm WDM} \gtrsim$ few keV [606, 607], dwarf galaxy abundances, and phase-space density analyses [608, 609]. Upcoming/operational observational facilities like Rubin and Euclid promise stringent tests of the WDM scenario [610–613].

Sterile neutrinos $\nu_s$, gauge-singlet right-handed fermions coupled to the $\bar{L}i\sigma_2 H^*$ operator, constitute a prominent WDM candidate with mass in the keV-MeV range [614] Alternative WDM candidates that predate more recent developments include keV-scale gravitinos in supersymmetric models [615], axino (fermionic superpartners of axions [616]), and, more generally, thermal relics with masses of $\sim$ keV–tens of keV [604]. These candidates similarly suppress small-scale structures, observable through galaxy counts, dwarf galaxy kinematics, and Lyman-$\alpha$ forest analyses.

Additional astrophysical constraints below the GeV scale come from indirect detection searches for DM annihilation, decay or capture products. keV-scale sterile neutrinos or ALPs could decay into photons, leading to monochromatic X-ray or soft gamma-ray signals. The long-debated 3.5 keV X-ray line [617, 618], once interpreted as a signal of decaying DM, has been reassessed [619] and will be further tested in future missions *XRISM* [611] and *Athena* [620]. Instruments like *XMM-Newton*, *eROSITA*, and *INTEGRAL* set limits on radiative decays and DM annihilation into two photons from the Galactic halo [621], and the Galactic center [622, 623]. Secondary emission via inverse Compton scattering by DM-produced electrons and positrons can be constrained in the MeV–GeV range through the diffuse Galactic emission in the X-ray and soft gamma-ray band, including the 511 keV line [624–626]. Astrophysical bounds on sub-GeV DM are very competitive below 1 MeV and above ~200 MeV, and future MeV telescopes will be in the reach of $p$-wave DM models [627, 628]. Cosmic rays provide complementary constraints, and Voyager-1 and -2, having exited the heliosphere, are positioned to detect sub-GeV charged particles unaffected by solar modulation [629].

DM capture in compact stars, such as white dwarfs and neutron stars, offers an alternative window on light DM. Annihilation can lead to observable heating [630, 631], or may trigger explosive phenomena such as supernova explosions or collapse into black holes [632, 633]. These probes offer constraints on nucleon and electron scattering cross-sections competitive with direct detection and extend down to 100 keV, albeit with astrophysical uncertainties.

Light mediators in the DS, with $m \lesssim 100$ MeV, are constrained through decay and cooling signatures in astrophysical environments. Examples include bounds from gamma-ray signatures [634, 635], various effects on SN1987 [434, 457, 636–639], and neutron star mergers [640, 641].

### 9.3.2 Direct Detection Searches

Noble liquid time-projection chambers (TPCs) (XENON [642], LZ [643], PandaX [644], DarkSide50 [645]) leverage single-electron sensitivity, achieving energy thresholds of 10–20 eV and probing sub-GeV DM via nucleon scattering. The sensitivity of the forthcoming DarkSide-20k experiment, for example, goes beyond current exclusion limits by up to 2 orders of magnitude [646], [ID268] as shown in Figure 9.4. Simultaneously, solid-state detectors like CRESST [647] and SuperCDMS [648] aim at lowering thresholds below 10 eV, targeting 100 MeV DM masses through nuclear recoils, while R&D explores new targets such as gaseous or superfluid helium (DarkSphere [649], DELight [650], HeRALD [651], QUEST-DMC [652]), and hydrogen-doped xenon (HydroX [653]) to improve kinematic matching for light DM interac-tions. Further progress derives from including in the signal model smaller, irreducible inelastic channels, like the Migdal effect [654], inducing atomic ionization in nuclear recoils in noble liquids and semiconductors [655], which extends sensitivity to tens of MeV for DM-nucleon interactions inaccessible to conventional searches.

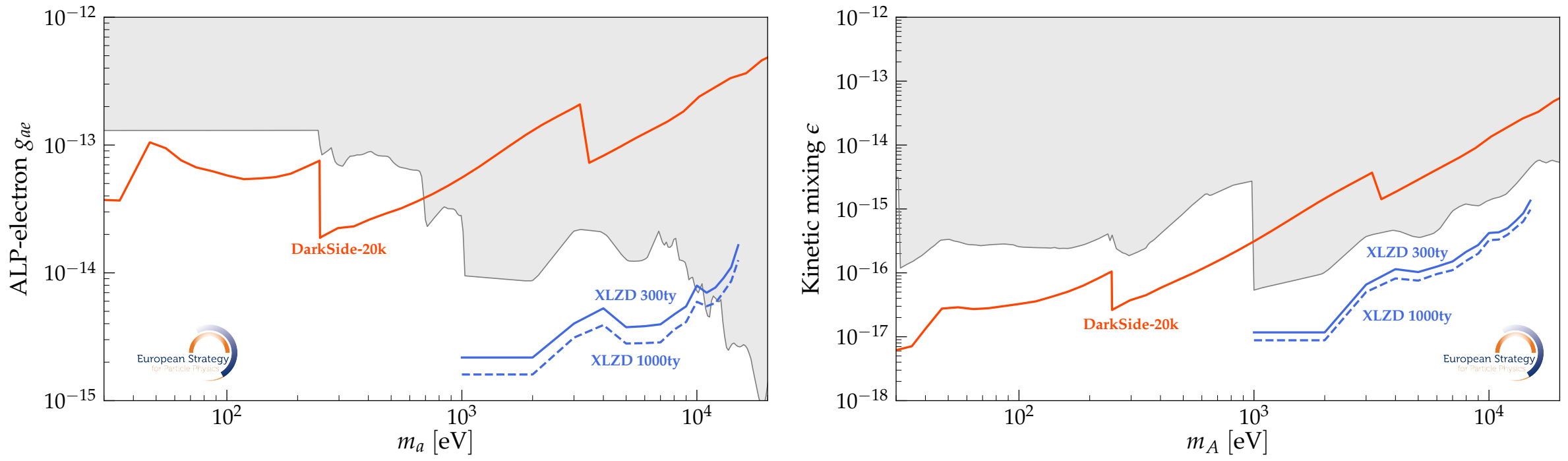

Fig. 9.3: Sensitivity of noble liquid TPCs to ALP-electron (left) and Dark Photon-electron interactions (right), for the case where the ALP or Dark Photon represent all of the galactic DM. The notebook to generate the figure is available at [660].

For DM masses below 100 MeV, where nucleon scattering becomes kinematically disfavoured, experiments exploit electron-recoil channels. The skipper-CCD programmes (DAMIC [656], SENSEI [657], and the future OSCURA [658]) probe both freeze-in and freeze-out benchmarks [659] in the 10–100 MeV range via single-electron threshold (1.1 eV) and sub-eV readout noise. We highlight that these constraints can be re-interpreted as constraints on elastic DM models discussed in the preceding, with direct detection searches setting the current most stringent constraints in the MeV range in the context of these models (Fig. 9.4). Interactions with electron final states are also interpreted in models with absorption of dark photons and axion-like particles (ALPs), in which case direct detection searches cover keV–MeV scales (shown in Fig. 9.3) and offer compelling complementarity to astrophysics and searches with helioscopes.

If part of the DM in the Galactic halo is accelerated at high speed (e.g. boosted DM) enhanced signatures are expected in direct detection experiments. Mechanisms include solar reflection [661], cosmic-ray up-scattering [662–664], acceleration and cooling in astrophysical jets and star-burst galaxies [665–667], or alteration of high-energy cosmic particle fluxes. These mechanisms can enable even more stringent constraints on light DM from conventional direct detection experiments, for scattering with both nucleons and electrons.

The interest in exploring the parameter space at even lower DM masses, where modest exposures can provide competitive sensitivity, is driving intense research and development towards novel detector technologies and materials capable of reaching ultra-low energy thresholds. R&D-scale searches for DM interactions in quantum sensors like QROCODILE can probe DM with mass as low as 30 keV [668]. Projects growing beyond the R&D scale (TESSERACT [669], SPLENDOR [670]), aim to address key challenges including ultra-low-energy backgrounds, both radiogenic and non-radiogenic, and scaling detectors to larger masses while preserving sub-eV resolution.

### 9.3.3 Collider and Accelerator-adjacent Searches

Portal particles may be efficiently sought using both collider and accelerator-adjacent experiments. In our schema, collider probes include searches at the LHC experiments (ATLAS, CMS, LHCb, ALICE and their respective upgrades at the HL-LHC) and Belle II [216], and future projections from FCC-hh, FCC-ee, ILC, CLIC, LHeC, and muon collider as well as the Electron-Ion Collider (EIC) [671]. Accelerator-adjacent experiments are intended as those with the decay volume separated by a macroscopic distance from the collider collision point, and beam dump setups. These include the ongoing NA62, NA64, SND@HL-LHC, FASER, and SBND experiments [437,441,672,672–675], [ID232], LUXE-NPOD under construction [676], and the recently approved DUNE, SHiP, DarkQuest, and the proposed Lohengrin, FPF, PREFACE, MATHUSLA, and ANUBIS [677–685], [ID46]. The motivation for such searches goes beyond explaining DM: dark sector (DS) scenarios with non-DM states include e.g. minimal SM extensions adding gauge singlet particles [457,458,686], or unstable particles appearing in more complex models, including inelastic DM [589], and dark QCD [687].

A major evolution within the DS field since the past strategy has been the demonstration of searches for dark sector particles using forward detectors at the LHC and the approval of the SHiP programme to move forward to the Technical Design Report (TDR) phase [ID145]—representing an important investment by CERN in future accelerator-adjacent DM and DS searches.

***Hidden Sector Portal Models*** Figs. 8.15-8.19 show the parameter space that ongoing and future experiments could explore for the minimal models of dark photons, ALPs, Heavy Neutral Leptons, and visibly decaying dark scalars introduced in Sect. 8.5. Given the sensitivity reach, a discovery could potentially observe thousands of events, enabling identification of the nature and properties of the underlying new physics model [688–691]. Caveats to these comparisons of sensitivities are that the methods used are not uniform across experiments [692], and carry substantial theoretical uncertainties [436, 438, 440, 455, 456, 693–695] not shown. Details on constraints in the high-mass, large-coupling regions, mainly from collider experiments, are depicted in Sect. 9.4.

**Dark Photons** Fig. 8.15 shows constraints and projected sensitivities to dark photons from various experiments in the mass-coupling plane. In the high-mass region ($m_{A'} \gtrsim 10\,\mathrm{GeV}$), sensitivity is dominated by collider searches, discussed in Sect. 9.4.3. In the mass range $m_{A'} \lesssim 10$ GeV, the large couplings $\varepsilon \gtrsim 10^{-5}$ will be probed by LHCb and accelerator searches [218, 443, 445, 672, 696]. The LHeC can probe dark photons in the GeV mass range with very low backgrounds [444]. The smaller couplings $\varepsilon \lesssim 10^{-5}$ are better probed with displaced searches at currently running [441, 697], upcoming [678, 680], and proposed projects [682].

**Axion-like particles (ALPs) coupled to photons** Sensitivities to ALPs in the mass-coupling plane are presented in Fig. 8.16. The accelerator-adjacent experiments, utilizing the displaced decay signature, may efficiently explore ALP masses $m_a \lesssim 5$ GeV and couplings down to $g_{a\gamma\gamma} \simeq 10^{-8}\ \mathrm{GeV}^{-1}$. The sub-GeV mass range will be well-probed with currently running [441, 672] and upcoming [678, 680, 698] searches, extending sensitivities up to a few GeV.

**Heavy Neutral Leptons** The parameter space of HNLs, assuming their Majorana nature, is shown in Fig. 8.19[1]. In the mass range $m_N \lesssim m_B$, the main production modes of HNLs are decays of flavoured mesons $K, B, D \to N$, which are copiously produced at meson factories [701, 702], neutrino beams [703, 704], and at SHiP [678], with the prospect of probing couplings of HNLs down to the see-saw line in the mass range $m_N \lesssim 1.5$ GeV.[2] In the domain $m_N \gtrsim m_B$, the main probes come from searches at colliders.

[1]In the minimal model with one HNL, having the mixing angle above the so-called see-saw limit $U^2 \simeq 5 \cdot 10^{-11}\ 1\ \mathrm{GeV}/m_N$ is inconsistent with neutrino oscillations. This may be avoided without spoiling laboratory

**Dark scalars** Fig. 8.17 shows the parameter space of Higgs-like scalars in the plane ($m_S$ vs. $\sin^2\theta$) assuming two values of $\mathrm{Br}(h \to SS)$: 0 and 1%. These correspond to the widely considered benchmarks "BC4" and "BC5" [457, 458, 692]. In the first case, the production modes are decays of $B, K$ mesons and proton bremsstrahlung. The main probes come from $B$ and $K$ factories [701, 702], SHiP [678], and MAPP [692]. Assuming $\mathrm{Br}(h \to SS) = 0.01$, more channels open up [705], increasing sensitivity.

***Elastic, Quasi-Elastic and Inelastic Dark Matter*** The parameter space of portal DM in the context of models with dark photons coupling to DM, for the elastic and quasi-elastic cases, is shown in Fig. 9.4. The main detection signatures are DM scattering $\chi + e/p \to \chi' + e/p$ and missing energy. The elastic case may be efficiently sought by direct detection experiments. In the quasi-elastic case, accelerator-adjacent experiments may produce these states and detect their scattering or missing energy. In the inelastic DM case, current constraints and projections from future searches at the LHC and accelerator-adjacent experiments for the reference set of parameters $\Delta = 0.1, \alpha_D = 0.1, m_\chi/m_V = 1/3$ are shown in Fig. 9.5; future lepton colliders, such as FCC-ee and CEPC, may explore this model as well [706]. The main signatures are missing energy and decays of the heavier $\chi$ state, $\chi' \to \chi + \mathrm{SM}$, depending on the mass splitting and the $\chi'$'s decay length.

Finally, in the scenario where the dark photon mass is tiny, $m_{A'} \ll 1$ eV one has millicharged particles [710], for which there are sensitive accelerator-based probes, including milliQan, MAPP [ID145] and the proposed FORMOSA detector [ID109]. This scenario may be complementarily explored with direct detection and ion trap experiments [711] (see also [712]).

## 9.4 Heavy Dark Sectors

Heavy DM searches have coalesced since the last European Strategy Update into a relatively small number of large programmes, driving to reach the so-called 'neutrino fog' benchmark in the case of direct detection for DM above the 100 GeV scale, to reach comparable coupling sensitivity below 100 GeV at colliders, and to probe the full allowed mass range for the thermal relic benchmark in the case of indirect detection. This section draws on community input documents [ID112, ID175, ID268] and reports [ID36, ID53, ID236, ID238] to highlight ongoing progress and future opportunities in direct and indirect detection, and on documents [ID65, ID74, ID77, ID132, ID133, ID144, ID159, ID176, ID195, ID201, ID204, ID214, ID228, ID229] reporting accelerator-adjacent and collider searches, with the relevance of complementary DM and DS searches underlined in [ID106] and [ID260].

In the context of the thermal freeze-out mechanism, the observed DM relic density suggests particles with masses in the range from multi-keV to about 100 TeV, and couplings to SM particles at or below electroweak strength. Thus heavy DM has been the most explored target to date in the search for DM. Benchmark models considered here for heavy DM candidates include the canonical WIMP candidates that emerge from extensions of the SM invoking supersymmetry among others, in the context of simplified models with minimal thermal relics, probes, e.g., by introducing more HNLs [699, 700]. This is implicitly assumed in the benchmark scenarios here.

[2]Future prospects for NA62 are also promising, though not yet publicly available.

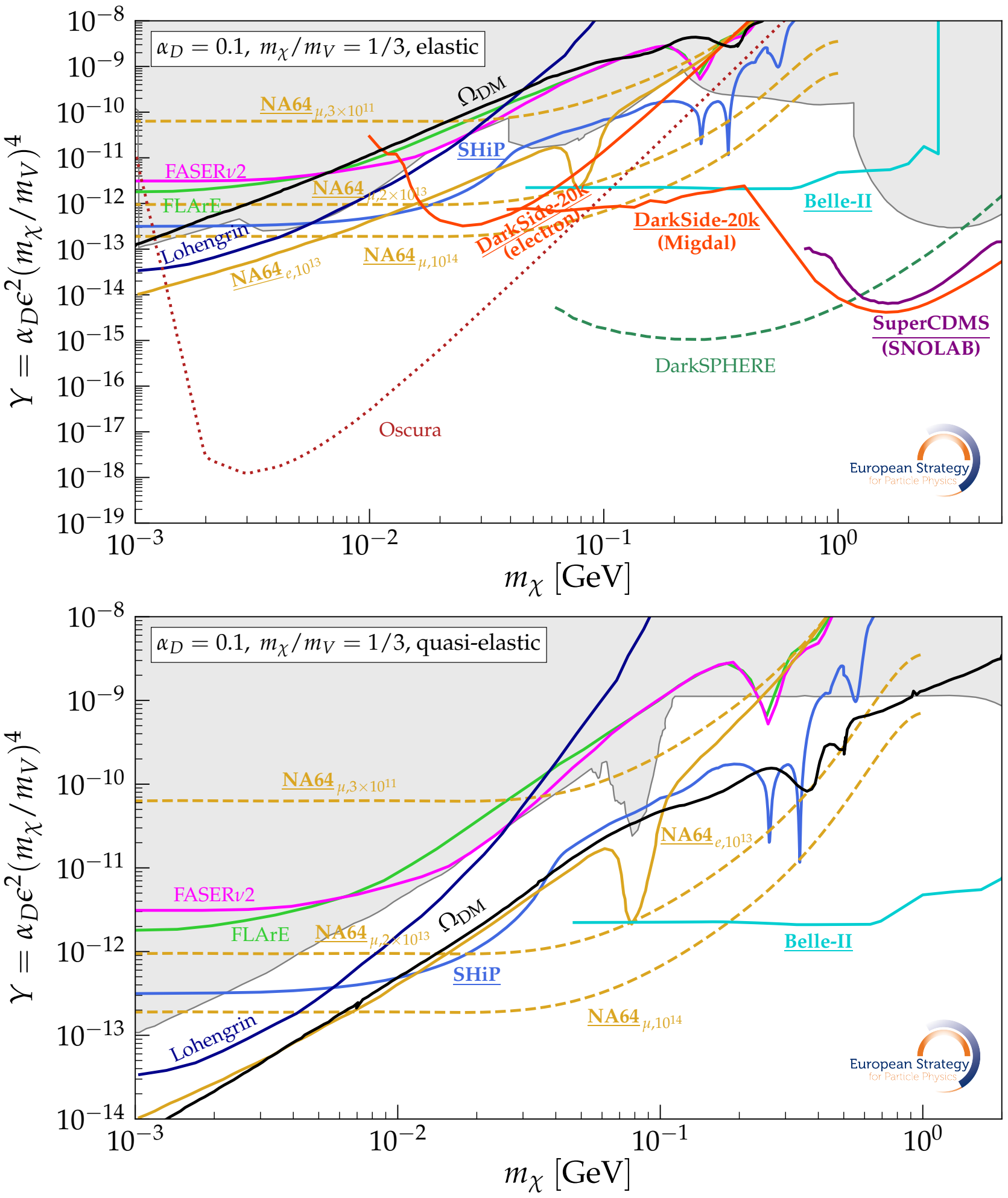

Fig. 9.4: Parameter space of elastic (top) and quasi-elastic (bottom) DM models with the dark photon mediator, in the plane $m_\chi - Y$, with $Y = \varepsilon^2 \alpha_D (m_\chi/m_V)^4$. The model parameters are $\alpha_D = 0.1, m_\chi/m_V = 1/3$ and in the quasi-elastic case the splitting is small, but sufficient to suppress direct detection signatures. The thick black line "$\Omega_{DM}$" corresponds to the parameter space where the model explains the DM abundance in the standard scenario. The grey shaded constraints come from the combination of accelerator and DM direct detection probes [645,656, 672,707]. NA64 sensitivity [672] comes from the invisible dark photon decay signature. The SBND experiment also has sensitivity to these models, but existing projections [708] require knowing background status for interpreting them.

i.e. wino and Higgsino scenarios, axial-vector and scalar mediators, Higgs portal DM, and hidden sector portal particles, the latter reported in Sect. 9.3. WIMP candidates are expected to have couplings at the weak scale or below, thus experiments searching for WIMP interactions with atoms in sensitive terrestrial detectors require large target mass, long exposure, and ultra-low backgrounds, as discussed in Sect.9.4.1. Indirect detection searches have complementary sensitivity via annihilation, with the prospect to reach exclusion of minimal DM relic models, described in Sect.9.4.2. Collider searches probe a broad range of DM candidates, with strong interplay with direct and indirect detection experiments, discussed in Sect.9.4.3, and a wide range of hidden sector portal models, as introduced in Sect.9.3.3.

In summary, a complementary mixture of larger, high-sensitivity searches across direct,

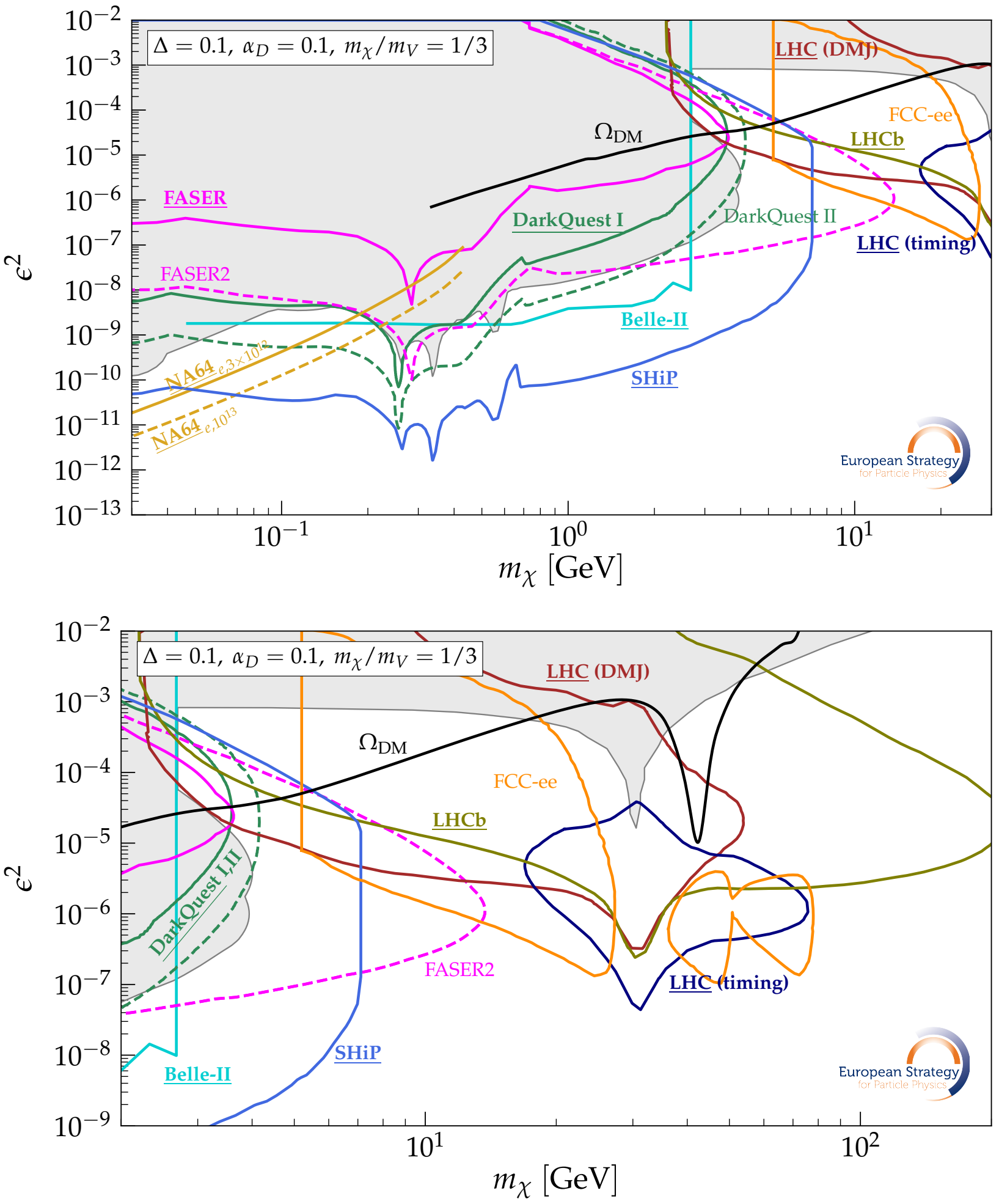

Fig. 9.5: The parameter space of the inelastic DM model in the plane $m_\chi - \varepsilon^2$. Top: with focus on small masses $m_\chi \lesssim 30$ GeV. Bottom: with focus on larger masses. The choices of the other parameters are $\Delta = 0.1, \alpha_D = 0.1, m_\chi/m_V = 1/3$. Depending on the parameters, the reach of different experiments may significantly change; this is caused by the scaling of the decay length $l_{\text{decay}} \propto \alpha_D^{-1}(m_\chi/m_V)^{-4}\Delta^{-5}$. Similarly to Fig. 9.4, the thick black line shows the DM abundance assuming the standard cosmological history. Shaded constraints come from the combination of accelerator-adjacent probes, as well as from LEP [589]. NA64 sensitivity [672] comes from the invisible dark photon decay signature. Some experiments also have sensitivity but are not shown in the figure because there are no prepared projections for the considered benchmark scenario (e.g. SBND [709]).

indirect and collider detection is essential to comprehensively probe the heavy DM parameter space; with important complementarity and interplay of techniques, with the prospect of direct and indirect detection probing new physics up to mass scales beyond the reach of even the most ambitious future colliders envisioned today.

### 9.4.1 Direct Detection Searches

Direct detection of heavy DM particles continues to be dominated by the technology of noble liquid time projection chambers (TPCs), a technology now so advanced that the sensitivities of these detectors will soon be limited only by irreducible backgrounds from astrophysical neutrinos, the so called neutrino fog, as shown in Fig. 9.6. These detectors are designed to observe nuclear recoils at recoil energies below 100 keV, and the main challenge has been to reduce backgrounds and lower the energy threshold, while increasing detector target volume.

Noble liquid TPCs are sensitive to a variety of physics channels through interactions with both nuclei and shell electrons, from light DM candidates (Sect. 9.3.2) through superheavy DM. In addition to spin-independent coherent scattering on the whole nucleus, direct detection experiments can also access spin-dependent DM-nucleon interactions with sensitivities approximately a factor of $1 \times 10^6$ weaker than the spin-independent limits, depending on the spin content of the target. Indirect detection offers complementarity with spin-dependent DM-nucleon cross-section for masses above $\sim 1$ GeV, as discussed below. Xenon detectors as well aim to study neutrinos through neutrino-less double beta decay [713]. Reaching the neutrino fog, while a background to the DM search, also presents opportunities to study cosmic neutrino sources.

Within Europe, one such detector is currently operating: The XENONnT experiment at Gran Sasso National Laboratory (LNGS) in Italy [714]. Internationally, three more detectors of similar scale exist: PandaX-4t in China [715], LZ in the US [716], and DEAP-3600 in Canada [717]. With no credible signal observed to date, these projects have excluded large areas of parameter space for heavy WIMPs, shown as the grey shaded region in Fig. 9.6.

The next detector to come into operation is DarkSide-20k at LNGS (liquid argon) [ID268] [718], which has realised major technological synergies with CERN through its membrane cryostat delivered by the Neutrino Platform, and its SiPM array readout building upon decades of silicon detector development for colliders. Future projects include PandaX-xT in China under construction (liquid xenon) [719], and XLZD in Europe or North America [ID175, ID176] (liquid xenon) [713] and Argo in Canada (liquid argon) [ID268] at the proposal and planning stages. Together, they aim to probe the WIMP parameter space into the neutrino fog. The argon and xenon projects are highly complementary, having independent systematic uncertainties, offering the prospect of independent confirmation of a candidate signal, and, if a signal is detected in both an argon and a xenon-based detector, allowing for improved reconstruction of the DM particle properties through joint analysis. Direct detection complements collider searches by accessing masses that are out of reach to colliders (see Fig. 9.6), and can identify the origin of a signal as from the galactic DM via modulation signatures.

European countries contribute significantly to these flagship projects, as well as to smaller-scale direct detection experiments [ID112]. Europe is well-positioned for global leadership in direct DM detection, with deep-underground laboratories (Boulby lab in the UK, LNGS in Italy, LSC in Spain, LSM in France) hosting the current leading detectors and R&D facilities.

### 9.4.2 Indirect Detection Searches

Signals from DM decay or annihilation arise at energies corresponding to the centre-of-mass energy of the process, around the DM particle mass. Heavy DM signatures are probed by both space-based and ground-based detectors. The challenge is to discriminate faint DM signals from more abundant astrophysical backgrounds, induced by standard processes which may not be well understood. Below the TeV scale, the current most stringent bounds on DM annihilations come from: (1) combined gamma-ray data from the direction of dwarf spheroidal galaxies in the Milky Way [720]; (2) radio observations of the Large Magellanic Cloud looking for synchrotron emission from DM produced electron-positron pairs [721]; and (3) anti-proton flux measurements [722–725]. Decaying DM is strongly constrained [721, 726], excluding lifetimes shorter than $10^{28}$s for masses above 20 GeV. Above 1 TeV, Galactic centre observations set strong constraints, reaching within a factor of 2 of the thermal relic cross-section benchmark for both continuum [727] and line-like signals [728]. The thermal DM wino is excluded by gamma-ray data up to 3 TeV, even allowing for large DM cores [729]. Lower mass winos, as subcomponent DM, are also excluded down to $\sim 1.5$ TeV. In contrast, gamma-ray data exclude thermally produced Higgsinos up to 270 GeV at 95% CL [730], thus the thermal Higgsino (1.1 TeV) remains a viable DM candidate.

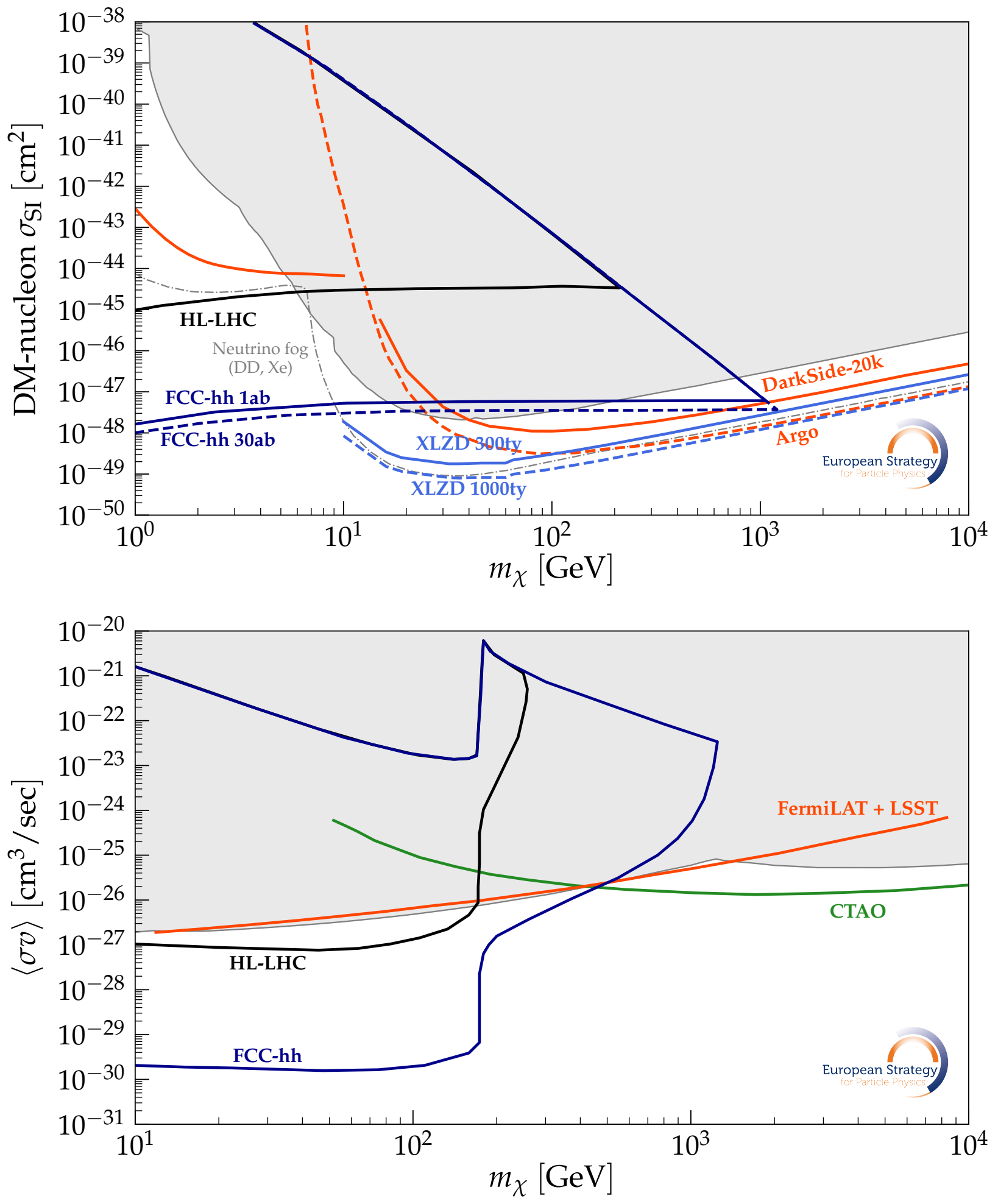

Fig. 9.6: Collider limits for (top) scalar and (bottom) pseudo-scalar models compared to direct and indirect constraints, respectively—see text for references. The grey area represents current constraints from direct detection experiments. Future projections for XLZD [713] and DarkSide [718] are overlaid. The notebook to generate the figures is available at [660].

Neutrino telescope searches targeting the galactic halo and centre are exploited to set competitive constraints on DM annihilation/decay for DM masses above a few TeV [731, 732]. Capture of DM into the Solar core and its subsequent annihilation set the most stringent bounds on DM-nucleon spin-dependent cross-sections for DM masses above 200 GeV [733, 734], and for annihilation into neutrinos IceCube sets the strongest constraints above a few GeV DM mass.

Excesses and anomalies in the multi-wavelength sky are often found, and regularly interpreted as BSM physics, however, no unambiguous detection of a DM signal has yet emerged. A notable example is the *Fermi* GeV excess [735], an intriguing anomaly that continues to spark debate. Alternative astrophysical explanations exist and will be tested by upcoming radio observations targeting millisecond pulsars [736]. Another yet-unexplained anomaly is the rise of the cosmic-ray positron fraction seen by several experiments [737]. Despite being technically possible to fit the excess using multi-TeV DM, this interpretation faces severe challenges [738]. On the other hand, a few nearby pulsars may naturally dominate the positron emission [739].

For the future, in addition to proposed satellite telescopes in the GeV–TeV energy domain, like HERD and VLAST [740], *Fermi*-LAT and LSST data sets are anticipated to discover hundreds of additional dSph galaxies, giving a factor of 2-3 increase in DM annihilation sensitivity [613]. The flagship Cherenkov Telescope Array Observatory (CTAO) is projected to improve upon current constraints by roughly an order of magnitude for both continuum and line-like DM signals [741, 742]. In parallel, the Southern Wide-field Gamma-ray Observatory (SWGO) will also provide compelling insights on TeV DM [743]. Both CTAO and SWGO aim to confirm or exclude Higgsino DM in the first 5–10 years of operation for a range of DM density profiles [744–747]. For antimatter searches, advances are expected to come from anti-deuteron (and eventually anti-helium) measurements from AMS-02, and sensitivities reaching the thermal cross-section for DM masses below 100 GeV are targeted by GAPS and GRAMS [748]. These searches benefit from the synergy with collider/accelerator cross-section measurements for reducing the uncertainties on cosmic-ray production cross-sections [366]. Neutrino telescopes [ID236, ID238, ID53, ID36] also aim to push sensitivities down by a factor of 10–100, complementary to direct detection.

### 9.4.3 Collider Searches

Collider experiments explore the nature of DM and its interactions under uniquely controlled laboratory conditions. Importantly, collider probes do not depend on the thermal history of the Universe nor local astrophysical conditions, yet are sensitive to new species which may contribute to the DM relic density.

The primary signature of DM production at colliders is an imbalance in transverse momentum or missing energy, arising from the presence of invisible stable particles in the final state. If the DM particle couples to the Higgs boson and has a mass below $m_h/2$, invisible Higgs decays provide a search channel. Disappearing charged track signatures provide a key handle for scenarios where the mass splitting between the DM candidate and heavier charged particles is very small. In other scenarios, the DM abundance is determined by interactions mediated by new particles, which may be accessible through visible decay modes at colliders and collider-adjacent experiments, with complementary sensitivities. While the discovery of such mediators would not confirm the existence of DM, the identification of associated invisible signatures would be a critical step toward identifying the associated new physics model. In hidden portal models, collider experiments can probe mass ranges and couplings not accessible to accelerator-adjacent experiments, particularly in the $m \gg m_B$ and large-mixing regime.

***Wino and Higgsino Scenarios*** These arise in supersymmetric models but can be studied independently as minimal thermal relics. Their relic abundance fixes the mass of the DM particle to approximately 1.1 TeV for Higgsinos (with a charged-neutral mass splitting of $\sim$ 350 MeV) and 2.8–3.1 TeV for winos (charged-neutral mass splitting of $\sim$ 160 MeV) [749, 750]. The prospects for discovering Higgsino and wino DM at future colliders are illustrated in Fig. 9.7, for the HL-LHC [751], FCC-hh (rescaled to the current nominal c.o.m. energy) [752], where the main collider signature is production of charged states that decay into soft SM particles and the neutral state. Searches therefore focus on large missing energy and soft decay products inside the detector. At a Muon Collider [753] the reach for Higgsinos is enhanced due to the possibility of soft track reconstruction, which allows sensitivity beyond 1.1 TeV. Indirect probes via electroweak precision observables can further extend the reach in a model-dependent way, beyond the kinematic limit. These are shown as indirect constraints in Fig. 9.7.

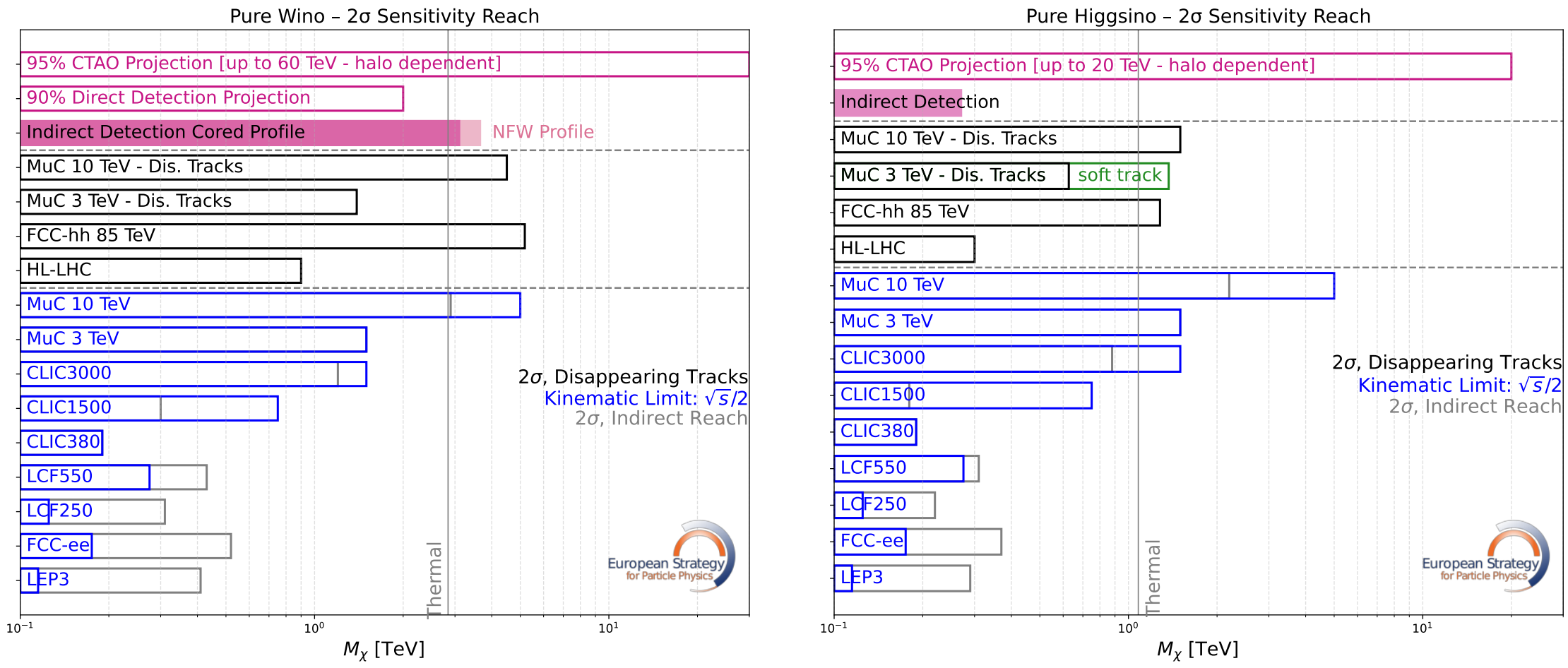

Fig. 9.7: The 2$\sigma$ sensitivity reach for the wino (Left) and the Higgsino (Right). Pink bars indicate the experimental reach assuming each candidate makes up 100% of the DM, with CTAO prospects shown [744–747]. Filled pink bars indicate present exclusion. Black bars show collider reach via disappearing track searches. Blue blocks mark the kinematic limits of various collider setups, while grey blocks indicate indirect sensitivity from precision measurements.

Direct detection experiments have limited sensitivity to pure Higgsino scenarios as the spin-independent scattering cross-section lies below the onset of the neutrino fog, while the wino cross-section is above it. Featuring a large annihilation cross section enhanced by Sommerfeld effects, winos and Higgsinos are a prime target for indirect detection through gamma rays, discussed in Sect. 9.4.2.

***Simplified Models: Axial-Vector and Scalar Mediators*** Simplified models provide a tractable framework allowing collider searches to be interpreted in terms of a small number of well-defined parameters. In the benchmark scenarios considered here, as in Ref. [3, 754], the dark sector consists of a single Dirac fermion DM particle and a single mediator. The key parameters are the masses of the mediator and DM particle, and their couplings to SM and DM fields. The mediators considered here can be scalar, pseudoscalar, or axial-vector bosons. Figure 9.8 shows the projected reach, compared with direct detection searches. The specific DM mass choices represent the points of approximate maximum sensitivity for direct detection, where sensitivity reach varies significantly with mass. Not shown, Muon Colliders also have sensitivity through mono-*W* analyses [755].

Figure 9.6 (top) shows the comparison between collider and direct detection sensitivities for a scalar mediator, where collider reach is recast as a constraint on the DM–nucleon scattering cross-section. Collider searches can reach DM masses up to the TeV scale, even when the mediator is significantly heavier than the DM particle, whilst direct detection can probe DM masses above the reach of the FCC-hh. Figure 9.6 (bottom) compares collider and indirect detection reach in terms of the DM annihilation cross-section. We note that in this particular model direct detection is spin-dependent and velocity-suppressed, such that it does not have sensitivity. In this representation, LHC or FCC-hh sensitivity to a given point in the mediator–DM mass plane is translated into an effective cross-section value, as in Ref. [3]. The comparison illustrates a complementary picture: collider searches have better sensitivity to DM candidates below the top mass threshold, while indirect detection is more powerful for higher DM masses [613, 741].

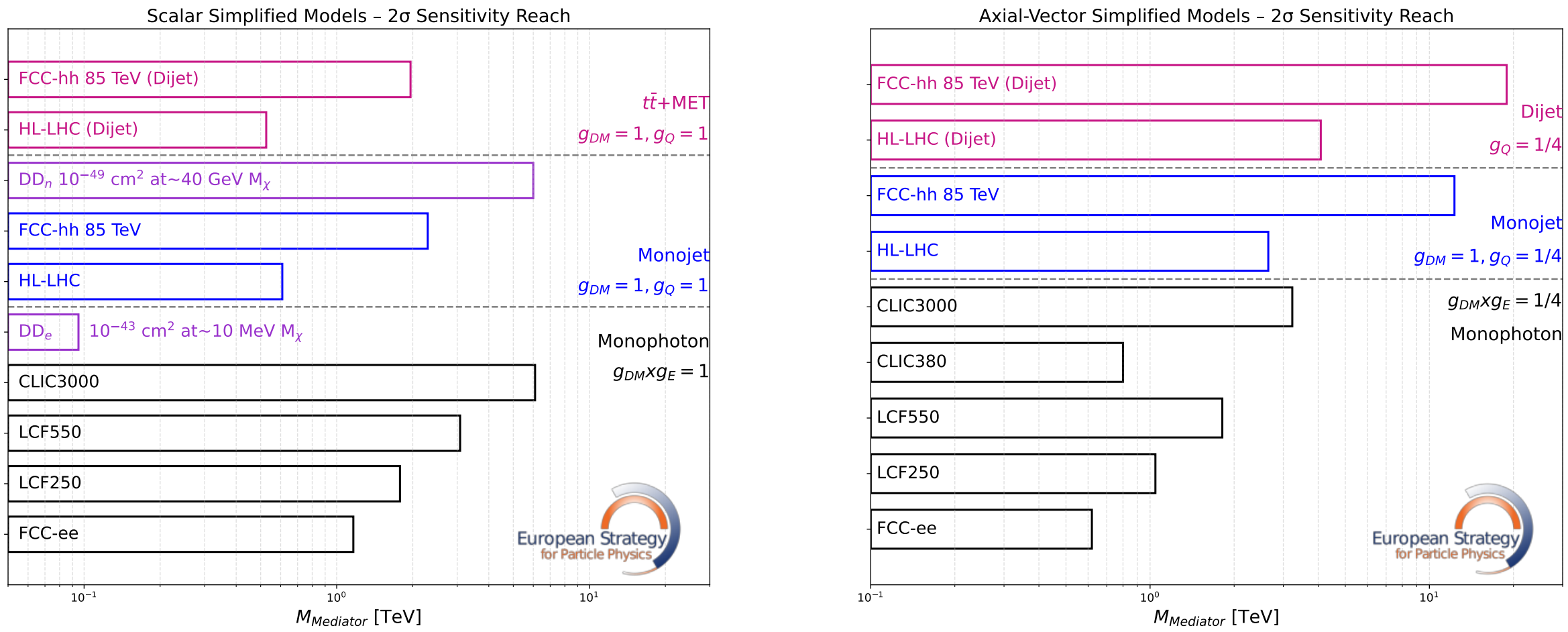

Fig. 9.8: Collider limits for (left) scalar and (right) axial-vector models compared to direct constraints – see text for details.

***Higgs Portal Dark Matter*** A minimal scenario for DM interactions is the Higgs portal model, in which the Higgs boson mediates the interaction between the Standard Model and a stable DM candidate. In the benchmarks considered here, the DM is either a real scalar or a Majorana fermion. The only free parameters are the DM mass and its coupling to the Higgs field.

Figure 9.9 shows the complementarity between future collider and direct detection sensitivities in the DM–nucleon scattering plane. Collider results are derived from the projected reach of invisible Higgs decay searches and translated into the scattering plane. Current LHC upper limits constrain the Higgs boson invisible branching ratio to below 10% at the 95% confidence level [756, 757]. HL-LHC is expected to improve this limit to around 0.9% for 3 ab$^{-1}$, whilst the FCC-hh to a BR of $2\times10^{-4}$ significantly tightens the sensitivity to Higgs-portal DM. Lepton collider projections from ILC [758] and FCC-ee [759] are also shown. For DM masses above roughly 10 GeV, future direct detection experiments are expected to provide stronger constraints than collider searches. However, collider experiments retain complementary sensitivity in the lower mass range below $m_h/2$, where the Higgs can decay invisibly to DM, offering the unique prospect of identifying galactic DM in a direct detection experiment and measuring its properties at a collider in the 10-100 GeV DM mass range.

***Hidden sector portal models*** Collider experiments probe mass ranges and couplings for hidden sector portal models complementary to accelerator-adjacent experiments. Beyond the benchmarks here, chosen to ease comparison of different facilities and experiments, other relevant scenarios include Hidden Valley Models [760] and their variants.

**Dark photons** Figure 8.15 shows projected sensitivities for visible dark photon decay searches e.g. to $e^+e^-$ or $\mu^+\mu^-$. In the high-mass region ($m_{A'} \gtrsim 10\,\text{GeV}$), sensitivity is domi-nated by prompt dilepton resonance searches: HL-LHC and FCC-hh reach is extrapolated from CMS dark photon analyses using high-rate scouting and offline triggers [761, 762]. Lepton colliders (ILC [763] and CEPC [764], FCC-ee [765], and Muon Collider [448]) exploit associated production and radiative return channels to probe lower $\varepsilon$ values with high precision.

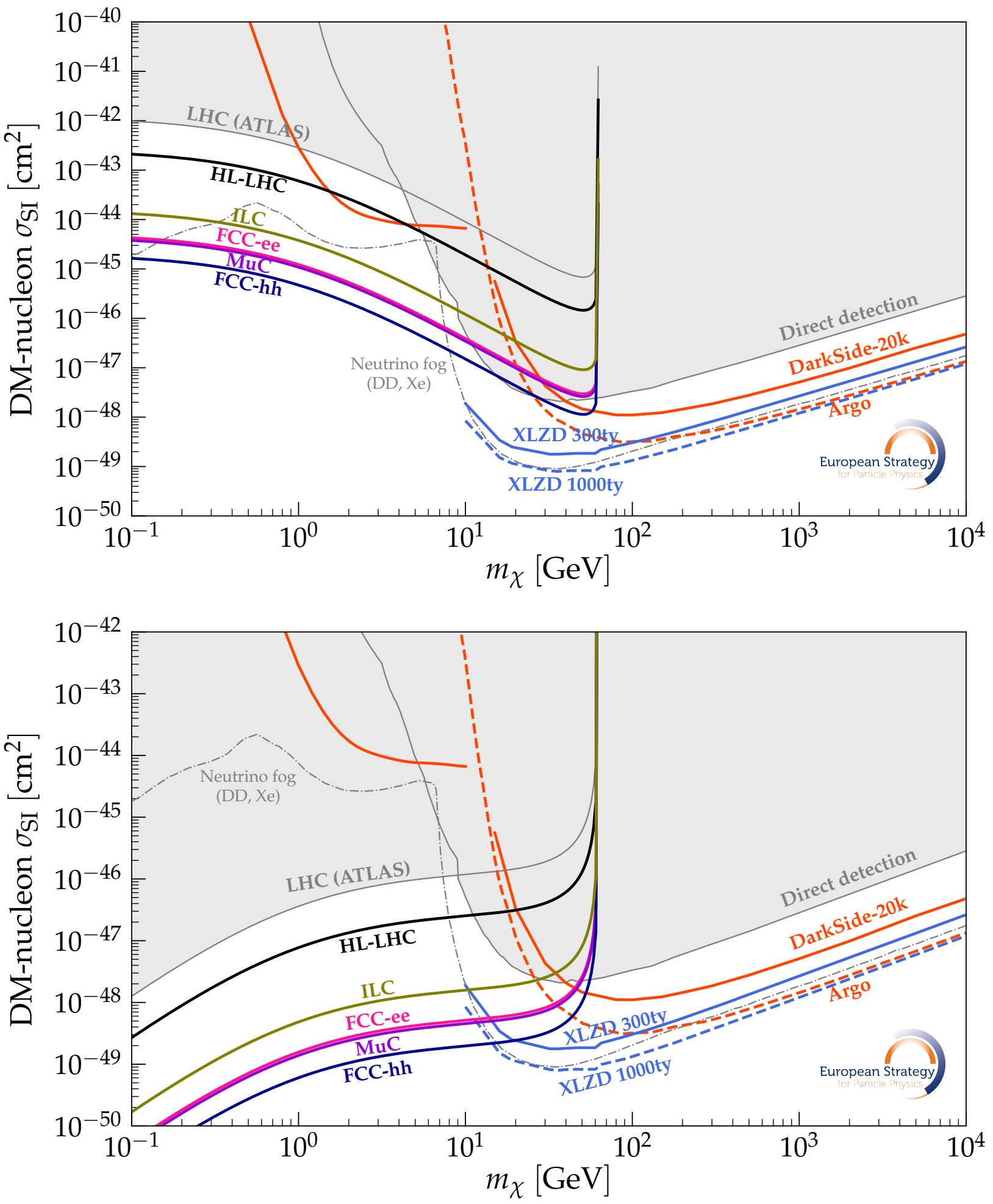

Fig. 9.9: Collider limits overlaid onto direct detection. (Top) Assuming Higgs portal scalar DM, (Bottom) Assuming Higgs portal Majorana DM. The grey area represents current constraints from direct detection experiments. Future projections for XLZD [713] and DarkSide [718] are overlaid. The notebook to generate the figure is available at [660].

**Axion-like particles (ALPs)** In collider environments, ALPs can be produced either resonantly or through exotic decays of the $Z$ or Higgs bosons, such as $Z \to a\gamma$, $h \to aZ$, or $h \to aa$. Additional production channels include vector boson fusion and associated production at $e^+e^-$ colliders. Ultra-peripheral collisions of heavy ions, where ALPs may be produced in the quasi-coherent scattering $\mathrm{Pb} + \mathrm{Pb} \to \mathrm{Pb}^{(*)} + \mathrm{Pb}^{(*)} + a$ [ID68], also provide a promising probe. Collider sensitivities are shown in Fig. 8.16, where projections typically assume $C_{WW} = 0$ and are derived from limits on $C_{\gamma\gamma}/\Lambda$ converted into $g_{a\gamma\gamma}$ as in [766]. At $e^+e^-$ colliders, FCC-ee is particularly sensitive at the $Z$ pole via the exotic decay $Z \to a\gamma$ [766, 767], considering both visible and mono-photon channels. Hadron colliders such as the FCC-hh, and heavy-ion searches in FCC-Pb and ALICE access ALP production through gluon fusion and photon fu-sion ($\gamma\gamma \to a \to \gamma\gamma$) [768, 769]. The Muon Collider offers complementary reach in the case of invisible ALP decays [753, 770]. Linear colliders like the ILC and CLIC provide further sensitivity through $e^+e^- \to a\gamma$, both in standard and beam-dump modes [771]. Sensitivity from $ep$ colliders such as LHeC [772] (or even $eA$ at EIC) is not in this figure because of differences in assumptions in their respective results.

**Heavy Neutral Leptons** At colliders, HNLs can be produced in leptonic decays of the $W$, $Z$, and Higgs bosons, with probabilities governed by their mixing with SM neutrinos. Additional production mechanisms include $t$-channel $W$ exchange processes at $e^+e^-$ and $ep$ colliders (e.g., $eq \to Nq$ and $e^+e^- \to N\nu$), and $\gamma W$ fusion at $e^+e^-$ machines ($e\gamma \to NW$). HNLs decay via $W$, $Z$, or Higgs boson emission, if kinematically allowed, leading to either prompt or displaced signatures. Projected sensitivities and exclusion limits are shown in Fig. 8.19 for muon- and electron-dominance, respectively. At FCC-ee, three scenarios have been explored, with the optimistic projection chosen [773]. Results for the ILC running scenarios are based on Ref. [774]. Beam dump and Giga-$Z$ sensitivity at the ILC was assessed using inputs based on [775]. Constraints from LHeC, covering an intermediate mass-mixing region in the electron-dominance case, are taken from [776].

**Dark Scalars** Collider searches probe both the single production of the dark scalar $S$ via mixing and exotic Higgs decays $h \to SS$. Figure 8.17 (right) shows the exclusion regions on the plane ($m_S$ vs. $\sin^2\theta$) assuming that the branching ratio BR($h \to SS$) is 1%. For scalar masses $m_S \gtrsim 10$ GeV and moderate-to-large mixings, FCC-ee is sensitive to intermediate mixings [777]. High-energy colliders like the HL-LHC, FCC-hh, as well as other lepton colliders, also have sensitivity via searches for resonant scalar decays or displaced signatures. Collider experiment sensitivity to $h \to SS$ is typically presented as a function of the scalar proper decay length $c\tau$. Displaced vertex searches at the HL-LHC provide coverage over a wide range of lifetimes, from prompt decays to those occurring at adjacent experiments [416, 692, 778–780], shown in Fig. 8.18.

## 9.5 Superheavy Dark Sectors

Superheavy DM candidates have seen increased interest since the last European Strategy Update, given the non-observation of lighter DM candidates to date. This section aims, for completeness, to briefly review superheavy candidates, and the status of current experimental efforts to explore this mass range. For particle DM, thermal relic scenarios are constrained by the unitarity bound, which limits the mass of DM candidates to roughly 100 TeV. Heavier particles—often called WIMPzillas—must be produced non-thermally to avoid over-closing the Universe without additional entropy injection. These candidates typically interact very weakly with the SM. Nonetheless, several mechanisms—such as gravitational production or inflaton-induced effects—can generate the correct relic abundance. While theoretical models exist to accommodate such scenarios, their detection remains challenging due to suppressed interaction rates and energies far beyond current collider reach. The large noble liquid direct detection searches have set the strongest bounds on non-thermal DM candidates, up to the Planck scale in the case of DEAP-3600, employing multiple-interaction signatures [781, 782]. Indirect detection offers the strongest constraint on unstable superheavy DM candidates, in particular looking for signals of radiative decays in fluxes of ultra-high-energy cosmic rays, from e.g. the Pierre Auger observatory [ID201]. Above $10^5$ GeV the strongest constraints on the DM lifetime come from multi-messenger observations of cosmic rays, neutrinos and gamma rays, including recent LHAASO observations [783–785].

Primordial black holes (PBHs) offer a compelling superheavy alternative to particle DM in this mass range, as they may have formed in the early Universe from the collapse of large density fluctuations generated during inflation. The probability of PBH formation depends sensitively on the shape and amplitude of these perturbations, linking their abundance to features in the primordial power spectrum. PBHs with masses above $\sim 10^{14}$ grams can survive to the present day, making them viable DM candidates. On cosmological scales, they behave like cold (or warm) DM, but their compact nature can lead to detectable small-scale granularity effects. A range of observational strategies—including constraints from Hawking radiation, microlensing, large-scale structure, and gravitational wave signals—has excluded most of the allowed mass range, except for the so-called asteroid-mass window between $\sim 10^{17}$ and $10^{20}$ grams, which is only mildly constrained by current data [786]. Direct detection searches offer the prospect of testing this mass range via searches for evaporating PBHs, as in the DarkSide-50 dataset [787]. For a complete review see [788].

## 9.6 Conclusions

The pursuit of DM has evolved significantly in the last decade. Experimentally and theoretically the old vision of discovering the lone DM particle has given way to a modern paradigm based on an integrated vision of plausible DS phenomena. It is acknowledged that DS phenomena may be complex and so a combined view of DM, its portals to us and potentially associated but adjunct unstable DS particles has emerged as the modern, connected, stage on which to strategise. Just as gravitons, photons and neutrinos act as envoys from the heart of some of the most dynamic visible sector phenomena, so too could dark photons, ALPs or sterile neutrinos be the first agents crossing the frontier between dark and visible sectors.

This broadening phenomenological backdrop requires keeping an open mind experimentally. One expects DS phenomena to be associated with a mass scale, be that the DM mass itself, a mediator mass, or more generally a mass scale associated with some non-trivial dynamics. Observational approaches depend on this mass scale.

Our Universe is a fertile archaeological site, with dynamics having scanned over an exponentially large range of mass scales since the time of reheating. If non-trivial DS dynamics emerged at a specific mass scale then it may have left fossils in the observable Universe at the associated redshift. Low-redshift cosmology allows for non-trivial probes of ULDM, all the way up to BBN providing a window on dynamics that could occur at the MeV scale and even earlier Universe dynamics setting the stage on which the DM abundance may have been set.

Astrophysics opens a complementary window, with astrophysical cores providing access to high luminosity and energy environs with particle scattering up to the MeV scale also. Galactic cores offer the high DM densities to enhance DM annihilation, with visible end-products, and galactic dynamics allows for the prospect of direct detection, with fast, but non-relativistic, DM particles impinging on the Earth.

However, the highest DS energies that can be accessed in a controlled environment are realised with the aid of particle accelerators. Whether operating in a high luminosity beam-dump mode or as particle colliders, accelerators offer a unique and indispensable window onto DS dynamics at mass scales above tens of MeV. As illustrated in this chapter, the future extension of the dark intensity frontier to the highest mass scales achievable would greatly extend our knowledge of dark sectors, welcoming a new era of exploration.

It should also be kept in mind that producing the observed DM abundance does not in principle require any coupling to the SM. Nonetheless, it is still possible to discover the properties of the DS via its gravitational interactions. Such secluded scenarios include the ULDM and PBH considered here. Furthermore, any dark radiation will unavoidably contribute to the energy density in relativistic particles and thus to the expansion of the Universe. For these scenarios, upcoming data from optical and infrared telescopes will substantially improve over current bounds.

Hunting in the dark requires a multi-faceted approach. A great deal of synergy exists between the various experimental approaches, including within the proposed high energy collider probes. The enormous statistical power of the FCC-ee Tera-Z programme allows for unique dark sector exploration of unprecedented depth, extending the intensity frontier from B-factory energies up to the Z-boson mass. Proposed high energy colliders such as FCC-hh or a high energy muon collider offer similar prospects for covering the classic WIMP mass range, however any comparison is model-dependent, relying on assumptions as to how the DM couples to the visible sector, such as through quarks or leptons. In any scenario, a convincing discovery of galactic DM will require non-collider experiments.

With a modern integrated view of DM and dark sectors, combined with the broadest possible future experimental efforts across all dark frontiers, including cosmology, astroparticle physics, indirect detection, direct detection and accelerator-based probes, we may be optimistic in maximising our chances of resolving one of the greatest scientific questions of our time.

# Chapter 10

# Accelerator Science and Technology

This chapter presents a summary of the Accelerator Science and Technology state-of-the-art and challenges associated with the main accelerator projects proposed at CERN and elsewhere, and reviews the respective submissions to the 2026 ESPPU. R&D progress to support the quest for higher luminosity, higher energy and/or higher intensity is outlined together with its present and potential societal applications. Environmental aspects of accelerator facilities are also briefly addressed.

## 10.1 Challenges towards high luminosity

Future colliders will have to deliver extremely high instantaneous and integrated luminosities to reach unprecedented statistical precision, notably at the $Z$-resonance and as Higgs factories, or to further push the energy frontier to at least $\mathscr{O}(10\,\text{TeV})$ parton-centre-of-momentum (pCM) energy. The luminosity $L$ of any collider is given by:

$$L = \frac{kN_b^2 f}{4\pi\sigma_x^*\sigma_y^*} F R_{\text{HG}} H_{\text{D}} = \frac{kN_b^2 f\gamma}{4\pi\sqrt{\beta_x^*\varepsilon_x^*}\sqrt{\beta_y^*\varepsilon_y^*}} F R_{\text{HG}} H_{\text{D}} \tag{10.1}$$

where $k$ is the number of colliding bunch pairs, $N_b$ the bunch population, $f$ the repetition rate (i.e. the revolution frequency for Circular Colliders (CCs)), $\sigma_x^*$ and $\sigma_y^*$ the horizontal and vertical beam sizes at the Interaction Point (IP), with $\sigma_{x,y}^* = \sqrt{\beta_{x,y}^*\varepsilon_{x,y}^*/\gamma}$, $\varepsilon_{x,y}^*$ the normalized emittance, $\gamma$ the relativistic factor and $\beta_{x,y}^*$ the betatron functions, determined by the machine optics, at the IP. $F$ is a geometric reduction factor (<1) in the presence of a crossing angle $\theta$ between the two colliding beams, $R_{\text{HG}}$ is a reduction factor (<1) accounting for the hourglass effect and $H_{\text{D}}$ is an enhancement factor (>1) for $e^+e^-$colliders discussed later [789]. The parameters of the colliding beams are assumed to be identical (with the exception of the particle charge).

### 10.1.1 $e^+e^-$colliders

Circular Colliders (CCs) have larger repetition rates ($\mathscr{O}$(kHz) for high energy machines) compared with Linear Colliders (LCs) (10 Hz to 100 Hz) but their beam current is limited by the emitted Synchrotron Radiation (SR) power $P_{\text{SR}}$ unless larger radii of curvature $\rho$ are considered. LCs are not limited by SR (except for the Beamstrahlung (BS) radiation at the IP) but their beam current is limited by the required Radio Frequency (RF) power, by the positron source and to a lesser degree by the polarized electron source considering that the beams are dumped after each collision. The beams circulate over many turns in CC and multi-turn (e.g. beam-beam) effects might limit their luminosity. More experiments can be served in parallel in CCs (though potentially increasing beam-beam effects) while luminosity can only be shared between experiments in LCs. In CCs, assuming $\sigma_y^* \ll \sigma_x^*$ (flat beams) and horizontal crossing angle, $L \approx \frac{3}{8\pi m_e c^2 r_e^2} R_{\mathrm{HG}} H_{\mathrm{D}} \xi_y \frac{\rho}{\beta_y^*} \frac{P_{\mathrm{SR}}}{\gamma^3}$ where $\xi_y$ is the so-called (vertical) beam-beam parameter which, from Large Electron Positron (collider) (LEP) and LEP2 experience, should be smaller than $\mathcal{O}(0.1)$ [790].

High-energy BS photons are generated at the IP of high-energy $e^+e^-$colliders when particles in one beam are accelerated by the intense electromagnetic fields of the opposing beam. In CCs, BS leads to an increase of both the energy spread and the bunch length of the colliding beams. To handle the large photon power and minimise experimental background, dedicated absorbers are required. BS is mitigated by constraining the minimum value of $\beta_x^*$.

The designs of the presently proposed CCs, Future Circular $e^+e^-$Collider (FCC-ee), LEP3 and Circular Electron Positron Collider (CEPC), are based on the crab-waist concept, pioneered at DAΦNE [791], and in operation at SuperKEKB [792], and the experience gained there is providing important lessons and input for the design of the new proposed colliders. In particular, high current ($\geq$1.3 A) and ultra-low $\beta_y^*$ (1 mm) operation have been demonstrated at SuperKEKB. The luminosity achieved so far by SuperKEKB is approximately a factor 12 lower than nominal [793] and it is limited by localized Sudden Beam Losses (SBLs), not captured by the collimation system, very likely due to local beam interaction with dust produced by contamination of the beam vacuum by a sealant. The occurrence of SBL events has significantly reduced in a few regions where this non-conformity has been addressed. In addition, the minimum achieved $\beta_y^*$ is approximately 3 times the nominal one because of the large emittance from the injectors and the limited dynamic aperture generating large background at the detector at injection. A series of measures has been devised and will be implemented to reduce the emittance blow-up of the beams in the injectors and in the transfer lines to the collider. Nonlinearities, misalignment and local coupling in the Interaction Region (IR) are also suspected to generate emittance blow-up when combined with beam-beam effects. The limited knowledge of the field quality of the installed IR magnets, their fringe fields, and the rather complex geometry inherited from the KEKB machine are affecting the accuracy of the simulations and the correction capabilities. These limitations could be addressed via an upgrade of the IR, which is being considered [793]. While the first limitation (SBL) is due to a non-conformity, the others appear to be specific to SuperKEKB, and are not expected to be an issue for FCC-ee or other future CCs.

The LCs luminosity is limited by power consumption constraining the power $P_{\mathrm{beam}}$ of each colliding beam. Crab cavities maximize the overlap between the colliding beams in the presence of a crossing angle ($F \approx 1$) and $\beta_y^* \approx \sigma_z$ in order to minimize the hourglass effect ($R_{\mathrm{HG}} \approx 1$). The lower repetition rate $f$, as compared to CCs, demands very small $\sigma_{x,y}^*$ to obtain acceptable luminosities $L \approx \frac{N_b P_{\mathrm{beam}}}{4\pi m_e c^2 \sqrt{\beta_x^* \varepsilon_x^* \beta_y^* \varepsilon_y^*}} H_{\mathrm{D}}$. Because of the extremely small sizes, each beam acts as a strong focussing lens pinching the opposing beam (if of opposite charge) and enhancing the luminosity by a factor $H_{\mathrm{D}}$ $\approx$1.5–2 for typical LC parameters, though the exact value can be computed only by simulations. The resulting bunch shape distortion enhances BS, requiring shielding of the experiments, and can lead to instabilities which make operation sensitive to relative beam-beam offsets. Particles emitting BS photons will collide at a lower than nominal c.o.m. energy. The minimum value of $\sigma_x^*$ is selected so that the width of the luminosity spectrum due to BS is comparable to that resulting from Initial State Radiation (ISR) [789,794]. For a given $P_{\text{beam}}$ the luminosity of LCs is approximately constant as a function of energy. The presently proposed Superconducting (SC) LCs (International Linear Collider (ILC) and Linear Collider Facility (LCF)) can deliver higher beam intensity (with larger bunch population, larger bunch spacing and longer pulses) allowing larger beam sizes to obtain a given luminosity as compared to Normal Conducting (NC) LCs (Compact LInear Collider (CLIC)). The latter, being operated at higher RF frequency, can achieve higher gradients resulting in a more compact installation for a given energy.

LCs require continuously intense $e^+$ beams, with average currents between one and two orders of magnitude larger than those necessary for CCs and than that achieved at the Stanford Linear Collider (SLC) [795, 796][ID40, ID233, ID247]. The PSI Positron Production (P3) experiment is intended to demonstrate improvements of positron yield per incoming electron, but not the full intensity required for FCC-ee [797]. The baselines of the proposed LCs include polarized $e^-$ beams. Polarized $e^+$ beams are considered as baseline for LCF and ILC and are generated from the conversion of polarized photons generated by the high energy ($E > 128\,\text{GeV}$) $e^-$ beam in a 230 m-long helical undulator before colliding at the IP. The $e^+$rate for LCs is still a significant performance concern [798,799] and R&D is being pursued [ID40].

The performance of LCs critically relies on the generation of beams with low emittance (at least one order of magnitude smaller than for CCs) and on its preservation up to the IP. Misalignment of the components can generate dispersion and wakefields in the linear accelerators (particularly for CLIC due to the small aperture of the 12 GHz structures) and transfer lines. Ad-hoc design of the accelerator and beam delivery components' support structures, rigorous pre-alignment and active stabilization are required to minimize the effect of vibrations and static misalignment and a series of complex feedforward systems (dispersion-free steering, wakefield-free steering, etc.), pioneered at SLC [796] and further developed at FFTB, ATF2, FACET and FERMI@Elettra, need to operate in parallel.

Strong focussing of the beams at the IP (particularly in the vertical plane) is needed and demands an extremely good control and correction of chromatic aberrations and other nonlinear effects. This has been studied at the Final-Focus System (FFS) test facilities FFTB and ATF2. The design normalized vertical emittance ($\gamma\sigma_y^{*2}/\beta_y^*$) targets for ILC/LCF could be approached at ATF2 (within ≈20%), for a larger than ATF2 design (×10) $\beta_x$ function and at lower bunch population ($0.07 \times 10^{10}$) as compared to the target value of $0.5 \times 10^{10}$ for ATF2 and $2 \times 10^{10}$ for ILC/LCF [800]. This limitation is partly explained by wakefield effects at ATF2 [ID78] and scaling the results to the LC designs relies on simulations. Additional studies at ATF2 and potentially at SuperKEKB are proposed [ID40, ID78]. ATF2 has provided a proof-of-principle of beam-beam offset control at the nanometre level by means of a feedback acting from one bunch to the next for ILC/LCF [801] (allowed by the bunch spacing: 366 ns to 554 ns) and within a train for CLIC (due to the smaller bunch spacing: 0.5 ns) [802].

### 10.1.2 High-Energy Hadron colliders

The successful LHC operation provides a solid base for the design of future hadron colliders at c.o.m. energies of $\mathcal{O}(100\,\text{TeV})$. For small crossing angles (e.g. with crab cavities) and for round beams (i.e. with equal horizontal/vertical $\varepsilon^*$ and $\beta^*$), their luminosity is given by $L \approx \xi \frac{I_{\text{beam}}}{r_p e \beta^*} \gamma$ where the beam-beam parameter $\xi = r_p N_b/(4\pi\varepsilon^*)$ is typically limited to $\approx \mathcal{O}(0.01)$ per IP.

Magnet field quality and reproducibility as well as a low noise environment (e.g. power converters) have proven to be critical to mitigate beam-beam effects (and the latter even beam stability) [803]. Luminosity levelling at the pile-up (or at the SR power [804]) limit might be required as SR effects will be significant. Radiation damping will cause emittance reduction during the fill and luminosity increase in the early phases of a fill when burn-off is not dominating. SR will deposit significant heat load on the beam screens. For the Future Circular hadron–hadron Collider (FCC-hh) the beam screens will have to absorb a total power of 2.4 MW [ID247] at the beginning of the fill with a dependence $P_{\mathrm{SR}} \propto I_{\mathrm{beam}}\gamma^4/\rho$. Electron cloud and impedance effects are additional potential performance limitations for high energy hadron colliders but are generally well understood and measures to mitigate their effects are available and can be implemented in the early phases of the design. Technological aspects related to High Field Magnets (HFMs), cryogenics, vacuum, beam stored energy and machine protection represent the major challenges (see Sect. 10.4).

### 10.1.3 Muon Colliders (MCs)

Because of the larger muon mass ($m_\mu \approx 200\,m_e$) circular MCs are not limited by SR even at $\mathcal{O}(10\,\mathrm{TeV})$ energies as $P_{\mathrm{SR}} \propto E^4/m^4$. BS, disruption and ISR are significantly reduced, but muons are produced via a tertiary process and have a lifetime of 2.2 µs at rest. The proposed designs therefore consider one intense single bunch for each of the colliding $\mu^+$ and $\mu^-$ beams and no crossing angle (i.e., $k = 1, F = 1, H_{\mathrm{D}} \approx 1$). The lost muons need to be replaced rapidly with a limited number of turns $n_{\mathrm{turns}}$ between consecutive injections (at collision energy) and the injectors need to provide them with a high repetition rate $f_{\mathrm{inj}}$. For round beams $L \approx \frac{\left\langle N_b^2 \right\rangle_{n_{\mathrm{turns}}} n_{\mathrm{turns}} f_{\mathrm{inj}} \gamma}{4\pi\beta^*\varepsilon^*} R_{\mathrm{HG}}$ indicating the importance of maximizing $n_{\mathrm{turns}}$ by minimizing the collider circumference with HFMs. In addition, small emittances and $\beta^*$ are necessary to compensate for the fewer number of bunches together with small $\sigma_z$ to minimize the hourglass effect. At higher energy, smaller $\sigma_z$ and $\beta^*$ can be considered, potentially leading to a favourable ($\propto \gamma$) dependence of the luminosity per unit of power consumption, provided that control of non-linearities can be guaranteed.

## 10.2 Large-Scale Collider Projects at CERN

Several large-scale collider projects have been proposed for possible realisation at CERN. These include next-generation $e^+e^-$ colliders FCC-ee [ID233], LCF [ID40], CLIC [ID78], and LEP3 [ID188], as well as a hadron collider FCC-hh [ID247], a Large Hadron-electron Collider (LHeC) [ID214], and a MC [ID207]. These are discussed in this section. Additional LC options/upgrades have been proposed [ID140], including some based on wakefield acceleration such as the Hybrid Asymmetric Linear Higgs Factory (HALHF) [ID57] and Advanced Linear collider for Very high Energy (ALiVE) [ID210]; these are discussed in Sect. 10.4.5.

### 10.2.1 Compact LInear Collider (CLIC)

CLIC is an $e^+e^-$collider based on a Two-Beam Acceleration (TBA) scheme powering a high gradient (72 MV m$^{-1}$ to 100 MV m$^{-1}$) X-band NC RF linac. The design was developed as an efficient path to multi-TeV energies. The TBA technique extracts the X-band 12 GHz RF power from a high power drive beam which is generated in a highly efficient complex operating at lower frequency (1 GHz). CLIC is foreseen to be built in stages with the first stage operating at a c.o.m. energy of 380 GeV in a 12.1 km tunnel powered with a single drive beam complex. The same drive beam complex can support a higher energy collider up to 1.5 TeV in a 29.4 km tunnel. The CLIC design is documented in Refs. [805, 806]. The updated design, described in Ref. [ID78], (see Table 10.1) includes two IRs sharing luminosity by multiplexing pulses and increased luminosity at 380 GeV by a factor of 3 due to a doubled repetition rate (100 Hz) operation and more ambitious assumptions on the spot sizes at the IP. While at 550 GeV it will be possible to operate at 100 Hz, at 1.5 TeV, the repetition rate would be decreased to 50 Hz serving a single IP, limited by power consumption, and $N_b$ would be decreased to mitigate wakefield effects [ID78].

Multiple test facilities have been constructed to support elements of the design. The CTF, CTF2 and CTF3 focused on the demonstration of the TBA concept. In particular, CTF3 achieved 50% energy extraction from a 25 A drive beam transported over less than 20 m [807], while CLIC is aiming for 90% energy extraction from a 100 A drive beam transported over ≈880 m. Drive beam loss control is critical to avoid component damage and radiation to the electronics in the tunnel. X-band acceleration facilities at SLAC, KEK and CERN demonstrated accelerator structures operating at high gradients. Starting in 2031, a 1 GeV X-band RF linac will be available within the framework of the EuPRAXIA project [808, 809] to enable further tests under operational conditions.

The CLIC design requires tight alignment tolerances and a high energy density due to the low beam emittance and high current primary beam, requiring an extensive machine protection system [806]. The main linac tunnel hosts a limited number of active RF components but, due to the tight alignment tolerances, a large number of active alignment systems with associated electronics. A few CLIC modules with active alignment have been constructed, but there is limited experience operating a system with several thousands of them [806, 810] in a radiation environment.

An 8-year preparation phase is contemplated before starting construction. This would include optimization and industrialization of components requiring large-scale production such as the accelerator structures and modules. It might also include additional development of the high-power drive beam beyond what has been achieved at CTF3, as well as other beam physics studies that would address some of the risks described above [ID78].

### 10.2.2 Future Circular Collider (FCC)

The FCC 'integrated programme' consists of an $e^+e^-$ collider FCC-ee [ID233] followed by a hadron-hadron collider FCC-hh [ID247]. Each collider would be installed in the same 90.7 km circumference tunnel; all of the FCC-ee Civil Engineering (CE), and much of the technical infrastructure, would be reused for the subsequent FCC-hh. Each collider lattice design incorporates a four-fold superperiodicity to permit the accommodation of up to four detectors. A detailed placement study has been conducted considering, among others, the territorial, geological and environmental conditions. This study has not yielded any showstopper for construction, and dialogue with the public and host-state authorities has started.

FCC-ee is engineered as a double-ring collider with four IPs and $P_{SR}$ is maintained constant at 50 MW per beam. A full energy booster with an injection energy of 20 GeV, located in the same tunnel (5.5 m diameter) as the collider, is used to steadily top-up the beam currents in the two colliding rings. A High-Energy (HE) linac accelerates the $e^-$ and $e^+$ beams, extracted from a damping ring at 2.86 GeV, to 20 GeV for injection into the full-energy booster ring.

| | CLIC | | | FCC-ee | | | | LCF | | | | LEP3 | | |
|---|---|---|---|---|---|---|---|---|---|---|---|---|---|---|
| | | | | | | | | LP | FP | | | | | |
| c.o.m. energy [GeV] | 380 | 550 | 1500 | 91.2 | 160 | 240 | 365 | 250 | 91.2 | 250 | 550 | 91.2 | 160 | 230 |
| Circumference/length collider tunnel [km] | 12.1 | 15 | 29.6 | 90.7 | | | | 33.5 | | | | 27.6 | | |
| Number of experiments (IPs) | 2 | 2 | 1 | 4 | | | | 2 | | | | 2 | | |
| SR power/beam [MW] | — | | | 50 | | | | — | | | | 50 | | |
| Longitudinal polarisation ($e^-$ / $e^+$) [%] | ±80 / 0 | | | 0 / 0[a] | | | | ±80 / ±30 | | | ±80 / ±60 | 0 / 0[a] | | |
| Number of years of operation (total) | 10 | N/A | 10 | 4 | 2 | 3 | 5 | 5 | 1 | 3 | 10 | 5 | 4 | 6 |
| Nominal years of operation (equivalent)[b] | 8 | N/A | 9 | 3 | 2 | 3 | 4.5 | 3 | 1 | 3 | 9 | 5 | 4 | 6 |
| Instantaneous luminosity per IP above 0.99 $\sqrt{s}$ (total) [$10^{34}\,cm^{-2}\,s^{-1}$] | 1.3 (2.2) | 1.6(3.2) | 1.4 (3.7) | 140 | 20 | 7.5 | 1.4 | 1 (1.35) | 0.28 (0.28) | 2 (2.7) | 2.25 (3.85) | 40 | 6.2 | 1.6 |
| Integrated luminosity above 0.99 $\sqrt{s}$ (total) over all IPs over each phase [$ab^{-1}$] | 2.56 (4.4) | N/A | 1.51 (3.96) | 205 | 19.2 | 10.8 | 3.1 | 0.72 (0.97) | 0.067 (0.067) | 1.44 (1.94) | 4.85 (8.32) | 48 | 6.0 | 2.3 |
| Peak power consumption [MW] | 166 | 210 | 287 | 251 | 276 | 297 | 381 | 143 | 123 | 182 | 322 | 200 | 226 | 250 |
| Electricity consumption per year of nominal operation [TW h/y][c] | 0.82 | 1.1 | 1.4 | 1.2 | 1.3 | 1.4 | 1.9 | 0.8 | 0.7 | 1.0 | 1.8 | 0.94 | 1.1 | 1.2 |
| Cost [BCHF] [d] | 7.24 | +30% [e] | +7.1 [e] | 13.73 | | | +1.26 | 8.29 | +0.77 | | +5.46 | 3.74 | | |

[a] Vertical polarisation of at least a few percent for ∼200 non-colliding pilot bunches, enabling precise quasi-continuous measurement of the beam energy. No longitudinal polarisation in the baseline. Residual longitudinal polarisation of colliding bunches should be controlled at the $10^{-5}$ level.

[b] This row lists the equivalent number of years of operation at nominal instantaneous luminosity, hence taking into account the luminosity ramp up.

[c] Computed from the peak power consumption and the assumptions on the operational year (see Table A.1 in Ref. [ID281])

[d] Total installation and construction cost quoted by the proponents of the projects in 2024 prices. The additional cost of each individual upgrade is indicated. It includes the cost of the technical components, materials, contracts, services, civil construction and conventional systems and associated implicit labour such as that provided by a company to produce components. It does not include labour provided by the host institution and the collaborating laboratories, contingency, any potential future inflation, the costs prior to project approval (construction and R&D), off-line computing, spares, maintenance, beam commissioning. The cost of the experiments is not included. The cost of land acquisition, site activation (e.g. external roads, water supplies, power lines) and spoil removal are not included for CLIC and LCF though they are expected to represent a minor contribution to the total cost (at the percent level). The additional cost of each individual upgrade is indicated.

[e] Cost of the upgrade from 380 GeV.

Table 10.1: Main parameters for the $e^+e^-$ colliders proposed for CERN [ID40, ID78, ID188, ID233]. The instantaneous and integrated luminosity in-between parentheses includes also the contribution from energies below 99 % of the c.o.m. energy $\sqrt{s}$. The data for CLIC at 550 GeV are not the result of a detailed estimate but an extrapolation from the data for the baseline configuration (380 GeV) and the upgrade at 1.5 TeV.

The FCC-ee collider and booster will be equipped with large Superconducting Radio Frequency (SRF) systems to compensate for the SR energy loss. They will consist of 2-cell 400 MHz Nb thin film/Cu (only for the collider) and 6-cell 800 MHz bulk-Nb cavities operated at 4.5 K and 2 K, respectively. A single-cell bare 400 MHz cavity (based on the LHC design) has been tested so far and approached, but not yet achieved, the required parameters. A bare 5-cell 800 MHz prototype cavity has exceeded the gradient specifications in a vertical test performed at JLab as part of a collaboration with CERN for the LHeC and FCC-ee initiatives [811]. An extensive R&D plan is ongoing, including the construction of a dedicated SRF facility at CERN, aiming for demonstration of the associated technologies by 2031. The RF installation is identical for the $Z$, $WW$ and $ZH$ modes of operation. Additional RF cavities will have to be installed for $t\bar{t}$ operation. Before the $t\bar{t}$ upgrade the collision energy can be flexibly changed between these three modes of operation at nominal luminosity; after the upgrade, operation at lower energy is possible but at lower luminosity. The klystron galleries are separated from the main tunnel, allowing for installation activities during machine operation. The FCC-ee main parameters are listed in Table 10.1 [ID233].

Proof-of-principle of critical FCC-ee technologies, such as crab-waist collisions, small vertical beta function at the collision point, and top-up injection, has been demonstrated at previous or presently operating colliders including LEP, SLC, KEKB, PEP–II, DAΦNE, BEPC–II and SuperKEKB. To further reduce accelerator costs and energy consumption, technology research and development efforts are advancing for ultrahigh-efficiency Continuous Wave (CW) RF power sources, twin-aperture dipole and quadrupole magnets, SRF cavities with enhanced quality factor ($Q_0$) values, etc. The project has completed a Feasibility Study [ID233] in which designs for the tunnel and subsurface structures, surface sites, technical infrastructures, injector, transfer lines, booster, and collider are presented.

The FCC-ee design incorporates significantly higher stored beam energy and BS power compared with previous circular $e^+e^-$ colliders. Robust machine protection and collimation systems to prevent damage from beam losses and SR and to limit experimental background are the subject of R&D. Mitigation measures against radiation-induced effects to guarantee high machine availability are being studied.

The FCC-hh baseline c.o.m. energy is 85 TeV, and the design calls for an integrated proton-proton luminosity of about 20 ab$^{-1}$ in each of two multi-purpose experiments during 25 years of operation. The remaining two IRs could house additional specialised detectors. FCC-hh could also accommodate ion-ion and proton-ion collisions and one interaction point would be upgradeable to electron–hadron collisions. As the second phase of the integrated programme, FCC-hh could start beam operation in the mid-2070s.

Key FCC-hh design challenges include the dipole magnet technology and the cryogenic power consumption in the presence of strong SR. To limit the latter the bunch population is reduced relative to that of HL-LHC. The baseline FCC-hh dipoles have a target field of 14 T and are based on $Nb_3Sn$ superconductor operating at 1.9 K. The SR will be intercepted by a beam-screen operated at 40 – 60 K: it represents the dominant part of the thermal load on the beam screen independent of the magnet temperature. The average heat load on the cold mass requires similar cryogenic power as the beam screen. The peak load during magnet ramps shall be buffered and cooled over a full operational cycle. Operation of $Nb_3Sn$ at 4.5 K could reduce the required cryogenic power. High Temperature Superconductors (HTSs) magnets, operating at 20 K at the baseline field, could further reduce the electricity consumption, or allow for higher field and beam energy at constant power. These are active research directions of the HFM programme [ID243] (see Sect. 10.4.1).

The FCC-hh technical timeline is largely determined by progress in arriving at a cost effective design for the arc dipole magnets [ID243]. In the baseline $Nb_3Sn$ approach, industrial magnet production could be ready to start around 2045 and could take roughly 10 years. Hence, if FCC-hh were to proceed as a standalone project, from a technical perspective, operations could start around 2055 [ID247]. Earlier involvement of industry may further reduce this timeline in very optimistic scenarios. For HTS magnets the feasibility of accelerator-quality magnets with Rare-earth Barium Copper Oxide (ReBCO) tapes needs first to be demonstrated (see Sect. 10.4.1). Assuming that accelerator requirements can be met, the timeline for HTS magnets is expected to be 10 to 20 years longer than the most optimistic one for Low Temperature Superconductor (LTS) magnets (see Table 10.3). The wide uncertainty range is due to the low Technology Readiness Level (TRL) of ReBCO accelerator magnets. The uncertainty is largest for the highest fields in the target range of 14 T to 20 T.

The capital cost for FCC-hh construction as FCC phase 2 is estimated to be 18.9 BCHF [ID247], with large uncertainty. This estimate is dominated by the collider magnets, amounting to about 10 BCHF. The main additional CE structures required are the beam-dump tunnels and transfer-line tunnels, with a total construction cost of $\sim$ 0.5 BCHF. Most of the electrical, cooling and ventilation installations are reused. The cryogenics infrastructure for the main magnet cooling drives the capital cost of the technical infrastructure which amounts to $\sim$ 4 BCHF. The cost of the injector and transfer lines is estimated at 1 BCHF; the final cost will depend on which injector option is chosen [ID247]. For a standalone project the current FCC-hh cost estimate is 28.4 BCHF for the collider and its injectors and transfer lines as well as the overall (ex-novo) CE and technical infrastructure. The estimated yearly electricity consumption of FCC-hh is 2.34 TW h at a magnet cold-bore temperature of 1.9 K, and below 2 TW h at 4.5 K.

### 10.2.3 Linear Collider Facility (LCF)

The LCF is an $e^+e^-$ LC based on the ILC design (see Sect. 10.3.2 and Table 10.2) in a 33.5 km tunnel, with two IRs sharing the pulses delivered by the collider and double the repetition rate (10 Hz) of ILC [ID40]. The width of the tunnel is smaller than that of ILC and no separation wall is present between the main linac and the RF power sources. The first phase (Low-Power (LP)) aims at 250 GeV c.o.m. energy operation with the same number of bunches per pulse (1312) as ILC, followed by operation at higher luminosity by doubling the number of bunches per train (Full-Power (FP)) with parameters listed in Table 10.1. The 33.5 km tunnel is sized to accommodate an upgrade to 550 GeV c.o.m. energy, using the same SRF technology. Upgrades to higher energies based on possible future accelerator technology developments are outlined in Ref. [ID140].

LCF relies on 1.3 GHz bulk-Nb SRF cavities with a gradient of 31.5 MV m$^{-1}$ and a $Q_0$ of $2 \times 10^{10}$ (twice the specifications for ILC). The design includes contingency measures for underperforming cavities (up to 20%), such as active RF power modulation. An average gradient of 27 MV m$^{-1}$ is acceptable if e.g. 10% more cavities are installed, with only an expected ≈3% cost increase. Lower $Q_0$ values are manageable to a certain extent, as the required cryogenics cooling power is dominated by static losses. For instance the impact of a lower $Q_0$ value of an average factor of 2 is evaluated at around 3% of the cost of the project for additional cryogenic power [ID40]. The target accelerating gradient has already been demonstrated in small production series [812] (see also Sect. 10.4.2). The EU-XFEL facility operates 640 cavities (those in the HE part of the linac) at ≈70% of the LCF nominal gradient when running at its maximum energy of 17.3 GeV [813].

The positron source [814], which is designed to produce polarised $e^+$using a helical undulator (see Sect. 10.1.1), has also undergone significant R&D, proving its fundamental principles and design features. However, no full-scale version has yet been built, and challenges such as target fatigue, high-speed rotation, and radiation resistance are the subject of dedicated ongoing R&D to confirm feasibility and reliability. The 230 m-long undulator design is also considered mature, as similar-length undulators with similar apertures operate successfully at the EU-XFEL. Studies on wakefields and SR effects show no significant performance limitations [815]. To mitigate risks, an alternative electron-driven $e^+$ source, which preserves the expected intensity but not the polarisation feature, is under consideration. Damping rings have been extensively studied in the ILC and CLIC design efforts, and, based on the performance of advanced synchrotron light sources, the target emittance levels have been achieved. The two main beam dumps, each designed to dissipate up to 17 MW of beam power, have a validated thermodynamic design based on the absorption of beam power in water. While conceptually sound and supported by experience (at around 2 MW beam power) at SLAC, they lack a full-scale engineering design, particularly of beam windows, and raise concerns such as the production of tritium. A comprehensive and resource-loaded R&D programme has been proposed in Ref. [ID275] to address, during the preparation phase of the project, all the technology gaps identified.

### 10.2.4 Large Electron Positron (collider) 3 (LEP3)

The LEP3 proposal [ID188] foresees an $e^+e^-$collider in the 27 km-long LHC [816] tunnel which has already hosted LEP [817, 818] and LEP2. The proponents estimate that dismantling of the HL-LHC and installation of LEP3 could be done in approximately 5 years. This proposal features two IPs with collision c.o.m. energies ranging from around 91.2 GeV to 230 GeV ($Z$, $WW$ and $ZH$-operation modes). The SR power is limited to 50 MW at all stages for power consumption considerations. The smaller $\rho$ as compared to FCC-ee limits proportionally the maximum circulating current and a larger RF system is required (for both the collider and booster, located in the same tunnel) at constant emitted $P_{SR}$ and beam energy, since the energy emitted per turn $U_0 \propto \gamma^4/\rho$. The maximum c.o.m. energy is limited to 230 GeV by the length of the existing LHC Long Straight Sections (LSSs) available for installation of the RF cryomodules.

The injector complex is expected to be similar to that of the FCC-ee. The HE linac injects into a booster ring, which is located on top of the two collider rings, allowing for top-up injection at the nominal energy. An injection energy of 10 GeV is considered as a result of the higher injection magnetic field of the booster due to the smaller radius. Although there is no detailed design for the injector complex, it is presumed to fit on the CERN Prévessin site.

High intensity $Z$ operation requires bulk-Nb 800 MHz 1-cell RF cavities operated at a maximum accelerating gradient of 12.9 MV m$^{-1}$ while at the two higher energy modes bulk-Nb 800 MHz 4-cell RF cavities are needed because of their larger accelerating gradients of 19.3 MV m$^{-1}$. Therefore LEP3 cannot switch flexibly between the low and the two higher energy modes at nominal luminosities. For the booster bulk-Nb 800 MHz 6-cell cavities operated at a maximum gradient of 20.3 MV m$^{-1}$ are foreseen for all operation stages. Two LSSs host the collider RF systems, while two others host the booster RF. The proposed RF cavities have a similar design as the FCC-ee RF cavities.

Since no detailed lattice or beam dynamics studies have been performed for the submitted proposal, the presented parameters are mainly obtained by scaling from FCC-ee [ID233] or from an earlier study conducted in 2017 [819,820]. They are summarized in Table 10.1. Shorter lattice cells are considered as compared with FCC-ee to provide similar emittances and, in addition, a larger main dipole radius of curvature (by ≈7%) than in Refs. [819,820] is contemplated to limit the required total RF voltage. Nested SC quadrupoles and sextupoles are needed to keep the power consumption at acceptable values; it is envisaged to install these in all Short Straight Sections (SSSs) profiting from the existing LHC cryogenic lines, however operating at 40 K.

Several technical aspects need detailed design studies before it can be considered feasible to achieve the proposed machine performance and in order to provide a bottoms-up cost estimate. These include: lattice and optics design; HTS SSSs and the associated cryogenics; integration of the booster in the same tunnel as the collider rings and the RF-cryomodules; required CE modifications; dismantling and storage or disposal of the LHC components; placement and design of the full injector complex.

### 10.2.5 Large Hadron-electron Collider (LHeC)

The LHeC is proposed to follow the HL-LHC phase, operating in standalone electron-proton (ep) and electron-ion (eA) collision modes. Due to the single-pass nature of the interaction between the $e^-$ and hadron beams, utilising a 50 GeV, 50 mA $e^-$ beam from a multi-pass Energy Recovery Linac (ERL) supports high luminosity without beam-beam limitations, potentially achieving up to 1 ab$^{-1}$ integrated luminosity in 6 years of operation. Colliding with HL-LHC's proton or ion beams, LHeC would achieve c.o.m. energies around 1.18 TeV (0.74 TeV u$^{-1}$), and luminosities up to $2.3\times10^{34}$ cm$^{-2}$ s$^{-1}$ ($0.7\times10^{33}$ cm$^{-2}$ s$^{-1}$)[1] for ep (eA) respectively. The ERL is based on 800 MHz SRF technology and it is located in a new tunnel complex tangential to the LHC ring, tentatively in IP2, where a dedicated experiment would be installed. The length of the ERL must be chosen to ensure the matching of bunch patterns of the colliding beams. 1/3 of the LHC circumference is considered in Ref. [ID214], compared to 1/5 in Ref. [821] with a corresponding reduction in energy from 60 GeV to 50 GeV, resulting in both SR and cost reduction. According to the proponents, the ERL could in principle be built in parallel with operation of the LHC.

[1]This is the nucleon luminosity assuming operation with lead ions and 1200 bunches, i.e. the luminosity multiplied by the number of nucleons: 208 for lead ions.

Achieving the required high-luminosity ep or eA collisions depends on successfully scaling up ERL technology to unprecedented beam power of 2.5 GW (approximately three orders of magnitude larger than demonstrated so far); this requires validation through the multi-pass 20 mA ERL prototype PERLE [ID56], under construction at IJCLab. Its phased implementation begins with a single-turn 89 MeV operation around 2029, progressing to a three-turn mode reaching 250 MeV and 5 MW beam power by 2031. In the longer term, it may serve as a test facility for improving energy efficiency in future $e^+e^-$ colliders [822], or for testing future SRF developments. The beam dynamics of LHeC is challenging and although demonstration of high current operation with three turns will be experimentally tested at PERLE, the impact of beam disruption after collision on losses and efficiency of energy recovery, and the effects of SR, cannot be verified at PERLE. LHeC operation in recirculation mode (no energy recovery) would still be possible, though at significantly lower current (likely by more than an order of magnitude) limited by power consumption and beam dump design, and therefore at a correspondingly lower luminosity.

A staged approach to the LHeC has been proposed in another submission [ID174] with phase one starting with an initial 20 GeV single-pass ERL phase which will not need the spreader/recombiner modules. The operation of a phase-one ERL will allow optimisation of the handling and shielding of SR and the corresponding management of high-power RF systems. The downtime required to install additional passes would need to be optimally managed. The cost savings would not be substantial as the tunnel will be the same. Both this proposal and LHeC could only commence in the post HL-LHC era.

The LHeC construction costs were estimated to be 1.6 BCHF (2018) for the ERL. In principle part of this infrastructure such as SRF cryomodules, magnets, cryogenics, and diagnostics could be reused for FCC-ee or LEP3, potentially reducing future capital investment. However, a detailed assessment will be required to understand if the SRF infrastructure will meet FCC-ee and/or LEP3 requirements. The expected maximum power consumption of LHeC with the associated experiment is 220 MW, when operated in dedicated mode. Excluding the LHC injector complex the yearly electricity consumption is expected to be approximately 1.1 TWh.

#### 10.2.6 Muon Collider (MC)

The baseline MC design is a green-field, 10 TeV c.o.m. collider whose initial concept was developed by the US Muon Accelerator Programme (MAP) until 2017 [823][ID152]. The International Muon Collider Collaboration (IMCC) [ID207, ID152] has been updating and advancing the design since 2022 [824].

A short, high-intensity proton pulse is produced by the proton driver, which hits the target and creates pions. The pions are guided by the decay channel, which uses a buncher and phase rotator system to create a beam with the resulting muons. The required brightness can be achieved only by longitudinal and transverse cooling to be achieved in a single pass by means of so-called 6D ionization cooling [825]. Ionization cooling is achieved in a $\mathcal{O}$(km) long channel composed of a sequence of low-Z wedge absorbers interleaved with high-gradient (20 MV m$^{-1}$ to 30 MV m$^{-1}$) NC RF cavities operated in strong solenoidal/dipolar magnetic fields of $\mathcal{O}(10\,\mathrm{T})$. Full 6D ionization cooling has not been tested and therefore a demonstrator facility is a necessary milestone to validate the feasibility and performance of a MC [ID207].

After cooling the muon beam must be accelerated to high energy very rapidly so as to profit from time dilation and hence reduce decay losses. The beams are accelerated to 63 GeV by a system comprising a linac and two Recirculating Linear Accelerators (RLAs). This is followed by a series of high-energy Rapid-Cycling Synchrotrons (RCSs) that reach TeV energies. Large RF voltages are required to accelerate the beams (up to 90 GV and 3000 cavities) in combination with fast-ramping NC magnets (4200 T s$^{-1}$) [824]. The last acceleration stage for the higher-energy option requires a hybrid RCS incorporating alternating fixed-field 10 T SC dipoles and pulsed 1.8 T NC dipoles to increase the energy swing, which induce large radial excursions (up to ≈2 cm) of both counter-rotating $\mu^+$ and $\mu^-$ beams during acceleration. The large RF voltage and the hybrid configuration imply new beam dynamics regimes.

Finally, the beams are injected at full energy into the collider ring, where they will circulate and collide within the detectors until they decay. In order to achieve competitive luminosities, small $\beta^*$ (down to 2 mm), and single bunches with large population ($1.8 \times 10^{12}$) and short bunch length (2 mm) are required. This implies a challenging lattice design with large chromatic aberrations and sensitivity to errors as well as significant collective effects, though the latter are mitigated by the short beam lifetime.

Initial exploration indicates that a MC with 2 IPs could be implemented at CERN [ID251] by reusing the existing SPS and LHC tunnels to accelerate the muon beam, though the feasibility of integrating the RCS in the SPS and LHC tunnels has not yet been demonstrated. This design could provide an initial stage at 3.2 TeV c.o.m. collision energy and a successive stage with a reach of up to 7.6 TeV. In a first option a single new collider ring tunnel is constructed and used for both energy stages, which is consistent with the use of 11 T $Nb_3Sn$ magnets at full energy and Nb-Ti technology at the lower energy. A second scenario contemplates two independent collider rings, with $Nb_3Sn$ dipoles for the 3.2 TeV and HTS for the 7.6 TeV stage, to be implemented in successive phases. Projected luminosities are close to $10^{34}$ cm$^{-2}$ s$^{-1}$ ($10^{35}$ cm$^{-2}$ s$^{-1}$); considering the operational scenario described in Table A.2 of [ID281] integrated luminosities of 1 ab$^{-1}$ (10 ab$^{-1}$) at 3.2 TeV (7.6 TeV, respectively) could be accumulated in less than 10 years.

A number of technological challenges are associated with the various elements of a MC complex. The variety and range of beam dynamics effects demand the development of start-to-end simulation tools to validate the overall performance. In addition the control and reduction of the ionizing radiation generated by the large neutrino flux resulting from muon decays is critical and is being studied. The IMCC suggests an extensive R&D programme requiring significant resources (well beyond those presently available) in order to assess the feasibility of a MC with the projected performance. The first phase is supposed to consist of a start-to-end facility design. Hardware development for components including the muon cooling technology demonstrator and SC magnets [ID105] is also included. This would be followed by the construction and operation of a 6D cooling demonstrator. A comprehensive and realistic project cost, performance and timeline assessment can be provided only after successful completion of the proposed R&D programme. For the accelerator this is estimated to take more than 10 years and require a material budget of at least 300 MCHF plus 1800 FTEy.

## 10.3 Large-Scale Accelerator Projects elsewhere

### 10.3.1 Circular Electron Positron Collider (CEPC)

The CEPC features a $\sim$ 100 km collider with two IPs. Its injector chain includes a 30 GeV 1.8 km-long linac, a 1.1 GeV positron damping ring and a booster synchrotron accelerating the electron and positron beams from 30 GeV to the beam collision energy. The CEPC baseline design involves three operating modes ($ZH$, $Z$, and $WW$, in chronological order) with a $P_{SR}$ of 30 MW per beam for the $ZH$ and $WW$ operating modes and 10 MW per beam for the $Z$ mode. Possible upgrades include [826] an increase of the power to 30 MW per beam for the $Z$ mode of operation and to 50 MW per beam for the $ZH$, $Z$ and $WW$ modes of operation, as well as an increase of the energy to 360 GeV for $t\bar{t}$ operation at 30 MW and 50 MW per beam (see Table 10.2). A SC proton-proton collider, the Super Proton Proton Collider (SPPC), designed to operate at a c.o.m. energy of 125 TeV with 20 T SC dipoles, could be installed at a later stage in the same tunnel. According to the proponents, this could operate simultaneously with the CEPC [826].

The integrated luminosity and electricity consumption have been estimated assuming the same operational scenario and machine availability as for the other $e^+e^-$ colliders considered at CERN for comparison purposes, though no luminosity ramp-up has been assumed by the CEPC proponents [826].The main technical requirements of CEPC are similar to those of FCC-ee and the overall technology readiness is comparable. The CEPC ambitions are to start construction in 2028 and operation in 2037 [826], [ID153].

| | CEPC | | | | ILC |
|---|---|---|---|---|---|
| c.o.m. energy [GeV] | **91** | **160** | **240** | 360 | 250 |
| Circumference/length collider tunnel [km] | 99.955 | | | | 20.5 |
| Number of experiments (IPs) | 2 | | | | 2 (1)[a] |
| Longitudinal polarisation ($e^-$ / $e^+$) | 0 / 0[b] | | | | 0.8 / 0.3 |
| Number of years of operation (total) | **2** | **1** | **10** | 5 | 10 |
| Nominal years of operation (equivalent)[c] | **2** | **1** | **10** | 5 | 8 |
| SR power per beam [MW] | **10** / 30 / 50 | **30** / 50 | **30** / 50 | 30 / 50 | – |
| Instantaneous luminosity per IP above 0.99 $\sqrt{s}$ (total) [$10^{34}$ cm$^{-2}$ s$^{-1}$] | **38** / 115 / 192 | **16** / 26.7 | **5** / 8.3 | 0.5 / 0.8 | 1 (1.35) |
| Integrated luminosity above 0.99 $\sqrt{s}$ (total) over all IPs over each phase [ab$^{-1}$] | **18.3** / 55.2 / 92.2 | **3.8** / 6.4 | **12.0** / 19.9 | 0.60 / 0.96 | 1 (1.3) |
| Peak Power consumption [MW] | **100**[d] / 203 / 287 | **225** / 299 | **262** / 339 | 358 / 432 | 111 |
| Electricity consumption per year of nominal operation [TW h/y][e] | **0.47** / 0.95 / 1.3 | **1.1** / 1.4 | **1.3** / 1.6 | 1.8 / 2.2 | 0.64 |

[a] Two experiments at a single IP ("push-pull" mode).
[b] No longitudinal polarisation is considered in the baseline, but such a possibility is being explored, especially at the $Z$ energy.
[c] Equivalent number of years of operation at nominal instantaneous luminosity, hence taking into account the luminosity ramp up.
[d] Extrapolated from Ref. [826].
[e] Computed from the peak power consumption and the assumptions on the operational year (see Table A.1 in Ref. [ID281]).

Table 10.2: High-level parameters of the ILC baseline in Japan [827][ID275] and of the CEPC for different options [826][ID153] (baseline parameters for CEPC are in **bold**), re-scaled according to the operational scenario for $e^+e^-$colliders described in Table A.1 of [ID281].

### 10.3.2 International Linear Collider (ILC) in Japan

The ILC is an $e^+e^-$collider, based on SRF technology, proposed to be built in Japan. The baseline design considers a 250 GeV c.o.m. energy [ID275] and is upgradable up to 1 TeV using the same technology, but assuming further improvements in the SRF cavity gradient and/or $Q_0$, and a longer tunnel. The main parameters are listed in Table 10.2. The International Development Team (IDT), established by ICFA in 2020, coordinates the international R&D and the strategy. R&D on SRF technology, particle sources, nano-beam stability, and final focus systems is conducted in Japan, Europe, South Korea, and US within the ILC Technology Network (ITN), created in 2023. The ILC parameters are more conservative compared with to those of LCF, with a single IP and lower repetition rate (5 Hz). In 2024, the ILC cost was re-estimated reflecting inflation, detailed design changes, and international vendor input and served as a basis for the cost estimate for the LCF accelerator.

## 10.4 Accelerator Technology

Following recommendations in the 2020 ESPPU [828] a European accelerator R&D Roadmap was established (see [ID154]). The roadmap incorporates five R&D areas with respective steering panels. Two of the corresponding R&D programmes cover technologies: HFM development for a future hadron collider, and RF systems including both RF sources and SC and NC accelerating structures. The other three R&D programmes cover advanced concepts: Plasma Wakefield Accelerators (PWFAs), MCs, and ERLs. These concepts have potential advantages over conventional accelerator approaches, but they are associated with a need for significant R&D as well as with increased complexity. The first phase of the roadmap was launched in 2022 and will conclude in 2026, after which it is expected that a next phase will implement a programme guided by the recommendations of the 2026 ESPPU. In February 2025 a comprehensive review [ID154] of the R&D status was performed on behalf of the Laboratory Directors Group (LDG) with the objective of assessing progress against initial plans, the maturity of key technologies, and future plans. The status of each of the five R&D programmes is summarised in more detail below.

### 10.4.1 High Field Magnets (HFMs)

In 2024, the main parameters of the FCC-hh with $Nb_3Sn$ main dipoles were updated. The primary change from the 2019 Conceptual Design Report (CDR) [829] is the reduction of the operational field from 16 T to 14 T [ID247]. This decision allows, for example, the reduction of the required $Nb_3Sn$ conductor baseline performance from 1500 A mm$^{-2}$ to 1200 A mm$^{-2}$ at 4.2 K and 16 T, which corresponds to the existing state-of-the-art conductor, an increase in operational margins, and a reduction of the maximum coil stress. Together, these measures establish a baseline with a high confidence level based on the HL-LHC experience [ID243].

The main near-term goal of the HFM programme [830, 831], comprising CERN and five national institutes as well as six universities, is to demonstrate short (1 m to 2 m-long) model magnets which meet the above specifications and reach routinely a maximum field of 15 – 15.5 T, thus demonstrating sufficient engineering margins for length scale-up and industrial production. In this context, the Racetrack Model Magnet 1 (RMM1) [832], designed and built at CERN, reached stable operation at 15.5 T in the aperture, and a maximum field of 16.9 T, showing that state-of-the-art conductor can reach the above specifications[2].

Beyond the baseline parameters, research on technology to further improve cost and/or performance is being pursued [ID243]. To reduce the price of $Nb_3Sn$ conductor by a factor 3 to reach the FCC-hh target, HFM is working with industries in Japan and Korea to increase the supply base. At the same time, HFM is preparing for a more competitive market for HL-LHC high-performance wire as the respective patent has recently expired. Moreover, wire performance at or above 1500 A mm$^{-2}$ at 4.2 K and 16 T could be reached by the introduction of artificial pinning centres, a technology that is developed in the US-Magnet Development Programme (US-MDP) as well as in HFM. An additional path to reduce the impact of the conductor cost is a hybrid $Nb_3Sn$/Nb-Ti coil, with a saving potential of up to 50% of $Nb_3Sn$ conductor. Operation at 12 T is also being explored as a lower cost option. The technical feasibility and cost of operating at 4.5 K, which could reduce the cryogenic power consumption by up to 33%, and the helium inventory by 52%, is also being studied. Note that the cryogenic power consumption due to SR, intercepted on the beamscreen at 40 – 60 K, is independent of the cold-mass temperature, which also applies to HTS magnets operating at or above 10 K discussed below. The technically-limited timeline for the LTS option for FCC-hh is outlined in Table 10.3 with first beam in 2050 – 2055, according to the scenarios.

[2]Note that RMM1 is a block-coil magnet built from flat racetrack coils; it, therefore, does not demonstrate all features of an accelerator magnet, which would require flared coil ends.

| Date | | LTS | Date | | HTS |
|---|---|---|---|---|---|
| Aggressive | Baseline | | Optimistic | Pessimistic | |
| end 2028 | end 2028 | Test of first short models and selection of design | 2030 | 2035 | Selection of cable technology and coil design |
| 2028-2032 | 2028-2032 | Short model programme to verify reproducibility | 2030-2035 | 2035-2043 | Achieve accelerator quality and target field (≥14 T) in short magnets |
| 2032-2036 | 2032-2036 | Scaling to 5-m-long magnets | 2035-2040 | 2043-2050 | Reproducibility and first scaling in length |
| 2036-2040 | 2036-2040 | Scaling to 15-m-long magnets | 2040-2045 | 2050-2055 | Scaling to 15-m-long magnets |
| 2038-2042 | 2040-2045 | Industrialization | 2043-2047 | 2055-2060 | Industrialization |
| 2042-2049 | 2045-2053 | Production and testing | 2047-2054 | 2060-2068 | Production and test |
| 2043-2050 | 2047-2055 | Installation and commissioning | 2048-2055 | 2062-2070 | Installation and commissioning |

Table 10.3: Expected technically-limited timelines for the LTS and HTS options for FCC-hh [ID243] for different possible scenarios. The HTS timeline is affected by large uncertainties due to the relatively low TRL and it is conditional on the demonstration of the feasibility of such accelerator-class magnets; fields above 16 T are associated with the largest uncertainty.

Two types of HTS are available from industrial processes: Bi-2212 multi-filamentary wires and ReBCO tapes. The market for ReBCO production has seen a rapid growth over the past decade with multiple suppliers worldwide, driven by the demand from commercial fusion companies. ReBCO is produced as a flat tape of 4 – 12 mm width. Screening currents, which are induced in all types of conductors during a magnet ramp, increase with the major dimension of the superconductor. Keeping ramp losses low and achieving good, stable, and reproducible field quality during injection, ramp, and physics is more challenging with ReBCO tapes than other conductors with smaller filament sizes (e.g., Nb-Ti: ≈5 µm, $Nb_3Sn$ and Bi-2212: ≈50 µm).

As of today, accelerator-readiness with ReBCO magnets has not been fully demonstrated. A high-priority research direction of the HFM Programme is therefore the exploration of different cable architectures, and their test in sub-scale magnets for an initial qualification. Several racetrack-shaped magnets with 5 T in the magnet centre have been tested. First tests of magnets with apertures for field-quality measurements are planned for 2026. Fast-turnaround R&D with sub-scale magnets will allow one to develop and validate numerical models, and use them to extrapolate parameters such as field quality and ramp losses to an HTS magnet for FCC-hh. These predictions are required also to include beam dynamics, cryogenics, powering, magnet protection, and vacuum considerations and define appropriate magnet specifications (field, operating temperature, operational margin, field quality, etc.) that balance the competing criteria of physics reach, power consumption, and magnet cost. The main field in an HTS FCC-hh dipole may range between 14 T to 20 T, with a factor 4 larger SR heat load and increased ramp losses for the 20 T/120 TeV case. For the highest fields, US-MDP is developing single-aperture hybrid Bi-2212/$Nb_3Sn$ and ReBCO/$Nb_3Sn$ options with stress management.

| Target Date | Technology Milestone |
|---|---|
| 2032 | Demonstration of HTS split solenoid, representative of a 6D cooling cell. |
| 2033 | Generate 20 T in a 1.4-m bore at an operating temperature of 20 K. |
| 2034 | Demonstrate final cooling solenoid, with 40 T in a 50 mm bore, and 150 mm length. |
| 2036 | Long demonstrator of a $Nb_3Sn$ wide-aperture dipole (short demonstrator in 2034). |
| 2045 | Long demonstrator of a ReBCO wide-aperture dipole (short demonstrator in 2041). |
| 2045 | Long demonstrator of a ReBCO IR quadrupole (short demonstrator in 2041). |

Table 10.4: Plans for technology milestones in the MC HFM R&D programme [ID105].

The technically-limited timeline for HTS magnet development for FCC-hh is given in Table 10.3. A key milestone in this timeline is the successful demonstration of a short-model dipole with accelerator quality by 2035. This is essential for a decision between LTS and HTS technology for FCC-hh. Due to the low TRL of HTS technology for accelerator magnets, the 2035 demonstrator milestone and all subsequent target dates are affected by uncertainties on the scale of +5 to +15 years, with the highest risk associated with the highest fields (given the quadratic scaling of forces and stored energy). This uncertainty range is expressed by the two scenarios in Table 10.3.

To date, Iron Based Superconductor (IBS) samples, produced in China on the scale of a meter length, perform a factor 3–4 lower at 4.2 K than industrially produced ReBCO samples of several 100 m length at 20 K. Nevertheless, HFM will evaluate the conductor's potential to meet the required critical-current performance and industrial-scale manufacturing.

Magnet technology for a MC is more varied in magnet types [ID105]. The target solenoid is a 1.4 m-aperture 20 T solenoid, exposed to significant radiation. The 6D cooling system comprises over 3000 split solenoids with bore fields ranging from 3 T to 18 T and apertures from 6 cm to 80 cm, where higher fields are associated with smaller apertures. The solenoids are placed at small axial distances and with opposing field directions, making magnetic and mechanical coupling a major design challenge to be mastered, particularly in case of a magnet quench. The final cooling solenoids feature 40 T in a 5 cm aperture and 0.7 m length. Finally, among the MC HFMs, the collider main magnets are single-aperture dipole and quadrupole magnets with wide apertures to accommodate shielding. For different collider energies, different magnet technologies apply. For dipoles, fields range from 5 T with Nb-Ti, to 11 T with $Nb_3Sn$, and 14 T with ReBCO, with apertures between 110 and 150 mm. IR quadrupole apertures may be up to 290 mm wide. Note that stored energy and forces scale with the aperture and with the field squared. All of the listed HFMs are DC-operated. The MC R&D programme proposes to build six high-field demonstrators over the coming 20 years (see Table 10.4).

### 10.4.2 Superconducting Radio Frequency (SRF) Technology

Many future HEP facilities demand high performance SRF technology capability. The respective R&D has been started or is ongoing, depending on the project or in some cases more generic research lines. SRF cavity gradient and $Q_0$ continue to improve for both bulk-Nb and thin films. For bulk-Nb, recent more environmentally-friendly surface processing advances (N-doping/infusion, HT/mid-T/2-step baking, Plasma Electro Polishing (PEP)) could enable much higher project specifications to be realized compared with large-scale facilities in operation, such as the EU-XFEL. SRF technology can fundamentally drive higher sustainability provision for future large-scale facilities: higher gradients and $Q_0$, i.e. reduced cryogenic losses, breakthrough for thin-films and Ferro-Electric Fast Reactive Tuners (FE-FRTs) integration verification. The latter are a potential game-changer for SRF cavity tuning: transient beam load-ing/microphonics can be addressed using this new method for rapid change of frequency.

Bulk-Nb cavities integrated in multi-cavity cryostats are essential subsystems of ILC/LCF and FCC-ee. The operation frequency is either following the well-established 1.3 GHz TESLA technology or lower, with cavities still of elliptical bellows-like geometry. Both ILC/LCF and FCC-ee require best performing state-of-the-art cavities, produced in large series. As a consequence, well qualified industrial vendors need to use most recent surface processing techniques, very likely with improved new infrastructure (e.g. Ultra-High Vacuum (UHV) furnace, manufacturing tooling, Quality Control (QC) etc.). The request for lowest rejection rate and highest cavity performance creates high aspiration regarding QC during production. Regarding FCC-ee, industrialisation at this highest performance level is a must for the upcoming years, according to the FCC-ee schedule until the early 2030s. A prototype module, based on Fermilab's Proton Improvement Plan - II (PIP-II) design, is in the design phase. Production, scheduled by 2031, requires a collaborative, very active partnership.

FCC-ee also requires 400 MHz Nb thin film/Cu cavities, and a 4×2-cell cryomodule for which the proven technology is targetted for 2031. The pre-prototype is under construction. Nb thin film/Cu still requires essential R&D. The recently mentioned timeline [833] expects the development of coating for multi-cell cavities around 2028 – 2032. The related cryomodule technology is not expected before 2030. FCC-ee 800 MHz bulk-Nb cavities are clearly much more conservative; the development can profit from the known large scale 1.3 GHz production (EU-XFEL, LCLS–II, PIP-II and others). The finalized technical design requires surely an industrialization period of some few years. Given other large-scale SRF-based projects, collaboration partners should be available.

ILC/LCF require the largest number of standardized high performance 1.3 GHz 9-cell cavities. The LCF design uses a factor-2 higher $Q_0$ than ILC, a choice which relates to recent R&D results. In several laboratories single-cell cavities reach gradients above 35 MV m$^{-1}$ at $Q_0 \geq 2\times10^{10}$ [834, 835]; here, a two-step baking process (mid-T followed by low-T) is applied. The transfer to 9-cell cavities is expected in due course. This important R&D extends the performance range of the EU-XFEL, where only single 9-cell cavities have reached the envisaged LCF performance, in vertical test. After successful knowledge transfer covering most recent surface preparation techniques, the challenge will be to use all available, and qualified, cavity vendors worldwide. Production on a reliable level would be key to success, and can reduce the number of extra modules integrated in the facility layout to guarantee physics reach.

The MC RF requirements are demanding in terms of number of cavities/modules, to be realized at highest performance. Many challenges exist, such as muon bunch intensity, required cavity aperture, high accelerating gradient, and last, but definitely not least, possible degradation of SRF cavities (stray magnetic fields and $\mu$ decay). A respective MC SRF R&D plan with timelines needs development.

Also, LHeC would ask for 800 MHz SRF cavities. Here the issues are high dynamic losses and high HOM excitation/need for extraction. The above-mentioned FE-FRT would address the request for fast frequency tuning.

In summary, the worldwide SRF community with strong European contributions is addressing key R&D topics. Not all project-requested results exist; more and in some cases essential work needs to be done. There is a lot of progress, bulk-Nb cavities are very advanced, and thin film cavities are strongly supported and have good prospects [836].

### 10.4.3 Normal Conducting (NC) RF Technology

NC RF technology is important in the majority of lepton collider projects under consideration. In some projects NC RF technology is a key performance and cost-driver, for example in CLIC and the Cool Copper Collider ($C^3$) [ID97] where it is used in the main linacs, in the muon capture and cooling channel of the MC project, and in HALHF, where it is used in the $e^+$ linac. In addition NC technology plays an important role in specific accelerator complexes, for example the $e^+e^-$injectors of the FCC-ee, CLIC, $C^3$ and HALHF and the drive beams of CLIC and HALHF. These complexes are in some cases quite large but only require relatively conservative performance parameters.

One important performance parameter for the CLIC and $C^3$ main linacs, the HALHF positron linac and the MC cooling channel is a high accelerating gradient. Relevant for the first three is that accelerating gradients well in excess of $100\,\mathrm{MV\,m^{-1}}$, and consistent with the 380 GeV CLIC breakdown rate specification of $2 \times 10^{-6}\,\mathrm{m^{-1}}$ per pulse, have been thoroughly demonstrated in 12 GHz CLIC prototype structure tests carried out in many laboratories [ID78]. Gradients above $70\,\mathrm{MV\,m^{-1}}$ have already been adopted and are being used in a number of smaller scale accelerators. These results mean that the high-gradient performance required for these projects have been achieved in a number of contexts. The design accelerating gradient for the MC cooling channel is lower, between 25 and $30\,\mathrm{MV\,m^{-1}}$, but it must be achieved in cavities immersed in a multi-T external magnetic field, potentially reducing the electric field holding capacity. Various tests have shown different aspects of the required performance, but the required performance showing all aspects simultaneously still requires demonstration [ID207]. For this reason, dedicated new test infrastructure is being proposed by the MC collaboration.

Dedicated tests have shown that the achievable accelerating gradient goes up by as much as 30% when copper structures are cooled to temperature below 80 K, as shown in measurements carried out in the context of the $C^3$ project [ID97]. Cryogenic temperatures also increase the $Q_0$ by a factor 2 to 3 which can be used to reduce the peak power requirement, and hence the number of power sources. Additional power is, however, needed for the cryogenic system which results in a somewhat higher overall power consumption compared with an equivalent room temperature RF system. Using cryogenic copper is now being considered in various applications so more experience is being gained. In addition, experiments are underway investigating the use of HTSs for RF, which could combine elegantly with cryogenic copper systems and result in very high efficiency.

In addition to the performances demonstrated in the tests and prototypes described above, significant progress has been made in understanding the fundamental processes that result in the main limitations to gradient [837]. The main limitations are vacuum breakdown, pulsed surface heating and field emission. This understanding is contributing to performance improvements in projects through, for example, quantitative optimization of beam and RF structure parameters and guidance for technology choices and development programmes.

Another important performance aspect for NC RF cavities is the ability to limit both long- and short-range wakefields. Very strong HOM suppression in CLIC accelerating structures has been directly demonstrated in beam tests at the ASSET and, later, FACET facilities [ID78]. Short-range wakefields are controlled through very precise structure assembly and alignment. The micron-level tolerances required for CLIC have been demonstrated, for example in X-band CLIC prototype structures and C-band SwissFEL accelerating structures [ID78].

### 10.4.4 RF Power sources

RF power sources are a key technology for all proposed future projects, linear or circular, normal or superconducting. RF sources are a cost driver for many projects, especially for the lepton colliders, and strongly influence facility efficiency.

The RF power sources considered for the HEP collider projects are all vacuum devices, mostly klystrons or similar. The vacuum devices in existing accelerator facilities have been produced by commercial suppliers for many decades. There are catalogue devices available today that would fulfil the basic requirements for most of the proposed projects. There have, however, been important developments in the field of klystrons in recent years, with the active development of higher-efficiency tubes. The main branch of this work began in the CLIC study as part of an effort to increase the efficiency of the drive-beam accelerator, and to provide improved X-band power sources for testing and a klystron-based design. The work has now expanded into a broader effort supporting many of the collider projects. In addition to the direct benefit of efficiency on reduced power consumption, investment cost is reduced through, for example, reducing modulator requirements and lowering installed mains power and cooling.

The first high-efficiency tubes of this type, a pair of 1 GHz, 21 MW tubes for the CLIC drive beam accelerator [ID78] and a 400 MHz, 380 kW tube for the HL-LHC project, have been ordered and produced in industry. Both tube types show a measured electronic efficiency[3] of 70%. Further improvement in efficiency is expected based on design studies of so-called two-stage klystrons. Designs of a two-stage klystron for the CLIC drive beam give an electronic efficiency of 80% (pulsed) and 86% (CW) for an FCC-ee klystron. A two-stage klystron for ILC would be very similar. The FCC-ee has adopted a new powering scheme which allows a single installation of the RF system, however imposing the requirement that the power source must operate over a range of powers [ID233]. This variable power requirement results in lower efficiency from klystrons which are efficient usually only in saturation. A multi-beam version of a so-called tristron (triode-klystron) is being developed for that purpose to further improve efficiency.

Higher frequency (3 GHz to 12 GHz) high-power (40 MW to 80 MW) klystrons are needed for the FCC-ee injector, the $C^3$ main linac, the HALHF positron linac, and CLIC accelerating structure conditioning stand. A high-efficiency version of a 50 MW 12 GHz klystron has been ordered by the EuPRAXIA [808, 809] project and is being prepared in industry. All of these tubes can benefit from the high-efficiency developments and they are being adopted by the non-HEP projects described elsewhere.

Solid-state RF source capability is steadily increasing and is used for many applications. Solid-state technology is steadily making inroads into accelerator applications, for example in synchrotron light sources such as SOLEIL, and the SPS. The collider projects discussed in this section all use vacuum devices with the exception of the FCC-ee booster during the $t\bar{t}$ stage, for which a solid-state system is considered [ID233]. This 800 MHz system is expected to be within industrial capability when required.

Industrial supply of RF power sources is a critical issue for all proposed future projects for a number of reasons. European laboratories currently depend on a limited industrial supplier base of three companies, only one of which is European. Such a small supplier base may limit project risk-mitigation through the use of multiple suppliers. High-power vacuum tubes have a very limited range of applications, which results in a very irregular market which makes the emergence of new suppliers unlikely, and even puts at risk the current industrial base. Large numbers of units are required for all projects, well beyond current capacity. Finally there is limited accelerator laboratory capability to design and fabricate even prototype devices.

[3] Electronic efficiency refers to the efficiency of only the klystron itself. Wall-plug-to-RF efficiency depends on many further parameters, including pulsed vs. CW operation, modulator efficiency etc.

### 10.4.5 High-gradient Wakefield Acceleration

High-gradient wakefield accelerators (HGWFAs) use dielectric structures or plasmas, driven by laser pulses or particle beams, to accelerate charged particles with gradients $\geq$GeV m$^{-1}$. In the context of HEP, plasma wakefields are currently the main focus, having demonstrated 30 – 100 GeV m$^{-1}$ gradients [838–840]. Such high gradients can enable compact, cost-effective future facilities provided beam quality and acceleration efficiency meet application requirements.

Three collider conceptual design studies are underway: i) HALHF [ID57] is the most advanced wakefield collider concept. It considers asymmetric collisions of 375 GeV $e^-$ (accelerated in multiple electron-beam driven plasma wakefield stages) and 42 GeV $e^+$ (accelerated using $C^3$ technology) to yield a c.o.m. energy of 250 GeV with a luminosity of $\mathcal{O}(10^{34}\text{cm}^{-2}\,\text{s}^{-1})$ and a possible length of $\approx$5 km. It requires significant technology R&D (estimated to take 13 years and >150 MCHF), including a demonstration of multi-stage plasma acceleration. Upgrades to 550 GeV c.o.m. energy are contemplated. ii) In ALiVE [ID210], both $e^-$ and $e^+$ are accelerated to 125 GeV in two single proton-driven plasma wakefield stages. Short proton bunches are generated by a multi-10 MW, 500 GeV RCS, currently at the pre-conceptual design stage with feasibility yet to be demonstrated. Upgrades to 10 TeV c.o.m. energy are considered. iii) The 10 TeV Initiative [ID98] is a US-led effort launched in 2025 after the Particle Physics Project Prioritization Panel (P5) report, aiming for a start-to-end design with roadmaps and resource estimates targetted for 2028.

Several critical wakefield collider components remain at the conceptual or pre-conceptual level [ID154], including: the generation of collider-quality beams with high single-stage energy and efficiency; staging of multiple plasma modules while preserving beam quality; development of compact beam delivery systems; efficient $e^+$ acceleration with maintained beam quality; stable, high-repetition-rate, long-term operation; integrated start-to-end system design and simulation. The next steps are to further develop collider concepts and demonstrate multi-stage plasma acceleration with collider-quality beams. The Advanced LinEar collider study GROup (ALEGRO) [ID159] coordinates international HGWFA efforts for HEP, identifying R&D priorities and fostering global collaboration.

In parallel to the collider studies, HGWFAs are moving towards delivering beams for applications, with several ongoing projects: EuPRAXIA [808, 809] aims to commission a compact soft X-ray FEL user facility driven by a plasma acceleration module at high repetition rate (>100 Hz) by 2031 and DESY is developing a Laser-driven Plasma Wakefield Accelerator (LWFA) as a PETRA IV injector [841]. These facilities can also demonstrate improved sustainability by reducing the facility footprint and possibly energy use compared to existing infrastructures. AWAKE [ID172] plans for multi-10 GeV beams for fixed-target or Strong Field Quantum Electrodynamics (SF-QED) experiments [ID191] in the mid-2030s [ID235]. Other proposals include test beams for detector R&D, multi-stage SF-QED experiments [842][ID191] and laser-based beam collimation (spin-off), e.g., for FCC-ee.

In summary, HGWFAs are moving from proof-of-principle experiments to delivering beams with quality tailored for applications [843–846]; first projects are expected to become operational within the next decade. Collider concepts are under active study, potentially en-abling HE upgrades of a future LCF [ID140]. Continued technology R&D will be essential for HGWFA to impact multiple areas of HEP [ID57, ID210, ID98, ID159, ID172, ID140, ID235, ID191].

## 10.5 Accelerators for Physics Beyond Colliders

High-intensity secondary and tertiary beams (e.g. muons, kaons, pions) are essential for many particle physics studies, e.g. neutrino physics and Physics Beyond Colliders (PBC). These beams are mainly generated through high-intensity proton accelerators at various facilities including CERN [ID120, ID235, ID251], ESS [ID151], FAIR, Fermilab [ID120], ISIS, J-PARC and PSI [ID134]. High-intensity proton accelerators are also considered as drivers for future colliders [ID207, ID210]. The two primary classes of limitations in high-intensity accelerators are beam losses during acceleration, which must be minimized to maintain operational safety and performance, and heat and stress on beam-intercepting devices like targets and foils, which must tolerate extreme conditions. Specific technical bottlenecks include charge-exchange injection and foil durability, target resilience, radiation damage, as well as cooling, space charge effects and wakefield-induced instabilities [847]. Numerical modelling and multi-physics simulations are key to improving accelerator performance. Tools include PyORBIT [848], IMPACT [849], OPAL [850], Warp [851] and others. More recently CERN Xsuite has been developed as a modular package for beam dynamics simulations including several single-particle and collective effects [852].

High-intensity and collider accelerators share several technologies (linacs, RF cavities, muon cooling, collimation, etc.). The applications of high intensity accelerators span HEP and non-HEP domains (e.g., chemistry, biology, materials science). In addition, accelerator technology (e.g., SRF, HFMs, cryogenics, etc.) can support the development of fundamental physics experiments for Axion-like Particles (ALPs) searches, gravitational-wave detection, atomic physics and others [ID235].

## 10.6 Environmental impact of Accelerator Facilities

Research facilities in Europe and beyond are investing in the evaluation of their environmental impact in order to align with the objectives fixed by the United Nations Sustainable Development Goals (UN-SDG) [853], and those of the Agenda 2030 for sustainable development [854]. In this context the current approach for new facilities relies on the methodology of Life Cycle Assessments (LCAs) to quantify emissions, resource use and other environmental burdens across all stages of a facility's existence: construction, operation, and decommissioning. Metrics include Greenhouse Gas (GHG) equivalent emission, particulate emissions, land and water use, and impacts on ecosystems. The analysis is performed according to international standards, and the community agreed to align in particular to the International Organization for Standardization (ISO) 14040/14044 [855,856] describing the general framework, and to European Norm (EN) 17472 [857], EN 15804+A2 [858] in case of specific applications [859].

While these metrics provide a structured and recognized framework, they are not always applied with consistent scope, boundaries or functional units. Moreover the result may vary considerably depending on the maturity of the project. For this reason it is important to adapt the scope and the level of detail to the status of the project, with the goals of improving the environmental performance of a Research Infrastructure (RI), identifying the main drivers for environmental impact and addressing those main drivers in successive phases of a project. Com-parison among projects is not straightforward for the reasons explained above, and should be performed with great care. In particular, also resorting to the objectives of the UN 2030 Agenda, particular focus has been driven on the GHGs emissions generated during construction and operation of RIs, as well as on their electrical power consumption [860][ID281]. The lessons learned from the first exercises of LCAs on large facilities (CLIC, FCC, LCF), is that CE work has a considerable impact, from 30% to 60% of the emissions of a new facility, together with accelerator HW construction. In the short term (before 2030), savings of 40% to 60% of the emissions during the construction of the RIs with respect to the current practices are considered to be within reach.

## 10.7 Societal applications

Accelerators-based RIs generate socio-economic impacts beyond scientific knowledge, across several of the 17 UN-SDGs. Future colliders need to address technology gaps and therefore drive the advancement of technologies that can lead to substantial societal impact and synergies with other fields in the form of collaborative projects, technology transfer, accessible infrastructures and the training of Early Career Researchers who will be involved in the projects [ID90]. The large majority of them will not remain in academia, and in returning to the public or private sector they will bring a capital of international culture, capacity to bridge the gap towards international markets, and with a relevant technical competence acquired during the design and construction of facilities based on new technologies and sometimes of unprecedented dimensions and complexity.

Concrete and ongoing examples for collaborations in the domain of HFMs include light sources, neutron spectroscopy, energy generation and transport (fusion, power distribution), medicine (particle therapy), and transportation (electric planes, maritime transport) which will have a tangible impact [ID243, ID105]. More efficient RF power sources and DC power converters will reduce the environmental impact of RIs in HEP but also in other sectors. Similarly, NC RF technology for HEP is used in many other projects including photon sources, industrial linacs, medical accelerators and the injectors for hadron accelerators. The knowledge and understanding gained during the high-gradient R&D programmes and the R&D towards high-efficiency RF power sources carried out by the LC studies is now being applied to a wide range of applications and conversely development and industrialization in the wide range of applications results in greatly improved knowledge and industrial base for use in future HEP projects.

## 10.8 Summary and Conclusions

The FCC-ee, CLIC, and LCF proposed designs for $e^+e^-$ colliders have similar levels of complexity and varied challenges. The LEP3 design is still at a pre-conceptual level and it is not yet validated by start-to-end simulations to confirm performance (luminosity, power consumption, etc.) and cost with a comparable level of confidence as for the others. CCs offer the possibility to serve more experiments and FCC-ee has the capability of varying the collision energy between 90 and 240 GeV seamlessly and at high luminosity; their larger repetition rate (determined by the revolution frequency in the kHz range) allows them to provide significantly higher integrated luminosities per unit of electricity consumption than LCs up to energies of at least 300 GeV (see Fig. 10.1). It is worth noting the significant improvement expected in the integrated luminosity per unit of electricity consumption for future CCs as compared to their predecessors, LEP and LEP2. Beam-beam and intensity (e.g. electron cloud) effects in CCs are challenging but there is extensive experience at CERN. SuperKEKB, and potentially DAΦNE, will provide significant experience for future $e^+e^-$ CCs in further understanding and addressing potential limitations and devising solutions.

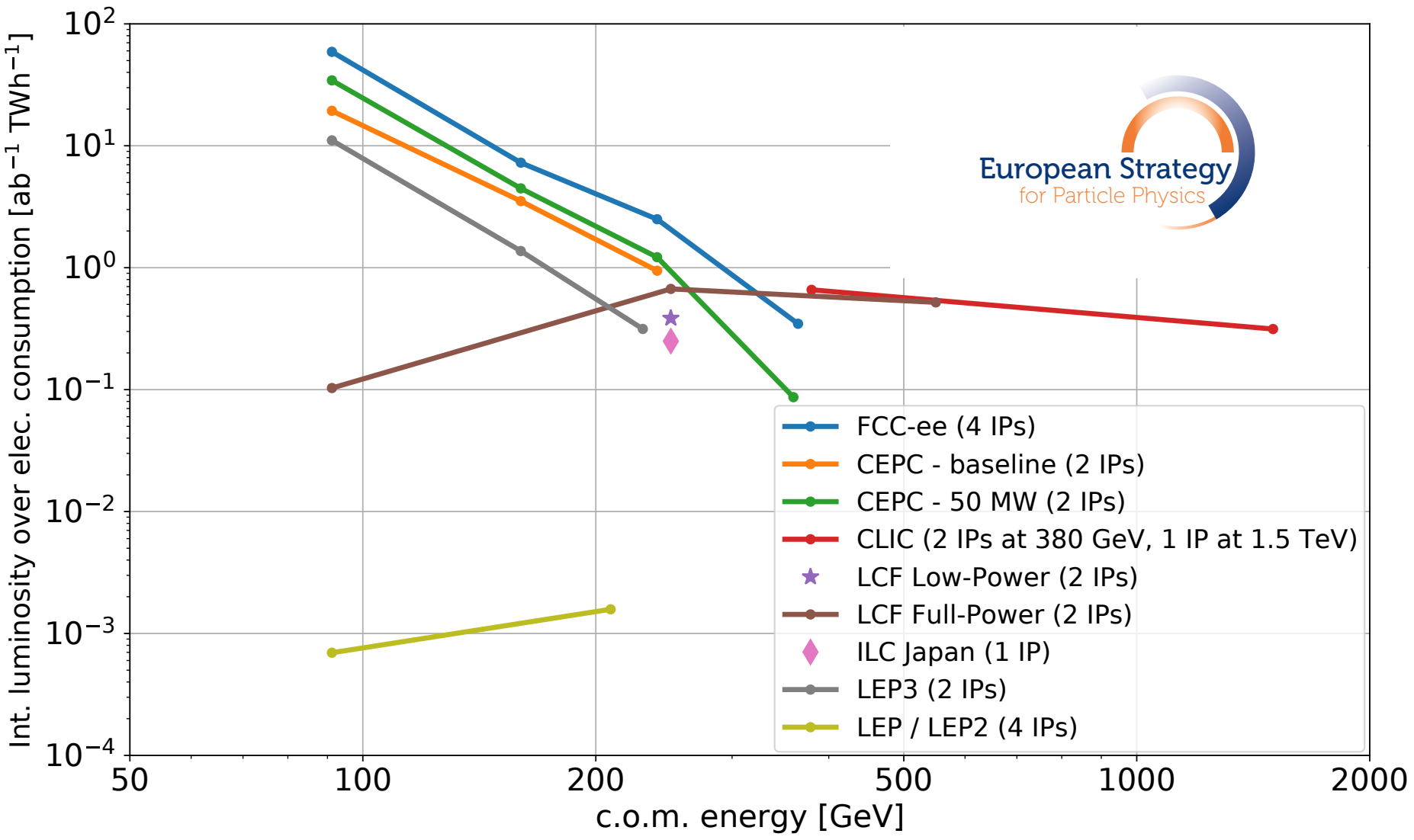

Fig. 10.1: Nominal yearly integrated luminosity over all experiments divided by the corresponding electricity consumption for future $e^+e^-$ colliders (excluding off-line computing) as reported in Tables 10.1 and 10.2. For LCs the total luminosity (including that below 99% of the nominal c.o.m. energy) is considered. LEP and LEP2 data are derived from Refs. [790, 861, 862] (Courtesy N. Mounet - CERN).

LCs can operate at higher c.o.m. energies than CCs, up to $\mathcal{O}(1\,\text{TeV})$ with conventional RF accelerating technology. They rely on sophisticated feedbacks/feedforwards conceived at SLC and require unprecedented $e^+$ production rates (particularly for LCF). The latter could pose a potential performance concern and are the subject of ongoing R&D. LCs can deliver polarized beams. Among LCs, those based on SRF technology can deliver higher beam intensity with larger bunch population and bunch spacing as well as longer pulses, which allow larger beam sizes for a given luminosity, while NC LCs, operated at higher RF gradient, allow higher energy reach for a given facility length. The ATF2 has partly reached the IP beam size target and successfully studied IP position stabilization for a single beam. However, scaling to LCs for wakefield effects needs to be addressed by simulations. The SRF technology underlying the ILC and LCF proposals benefits from the R&D, production and operational experience at EU-XFEL and LCLS–II built upon the TESLA study at DESY. The high gradient X-band NC RF structures required for CLIC have been tested successfully at CTF3 and are being used in a number of smaller scale accelerators. CTF3 has also served as a proof-of-principle of TBA though the efficiency of this process has been validated only partly and on a small scale compared with that needed for CLIC.

The LHeC performance critically relies on an ERL, operated in multi-turn mode and at unprecedented beam power, whose feasibility needs to be validated with a demonstrator (PERLE) under construction and expected to deliver the first results in the early 2030s. The beam-dynamics of LHeC is challenging and the impact of beam disruption after collision on the efficiency of the energy recovery remains a critical aspect that will be possible to address only with simulations. The development of the SRF technology underpinning ERLs is synergistic with the SRF developments for FCC-ee and LEP3.

The design of a next-generation high energy hadron collider lies on the solid foundation of the LHC and HL-LHC designs, construction and operational experience, however, it faces significant technological challenges, namely HFMs, cryogenics, vacuum and machine protection. The FCC-hh technical timeline is largely determined by progress on the development and industrialization of the arc dipole magnets. The baseline $Nb_3Sn$ magnet production could be ready to start around 2045 and could take roughly 10 years while, with current knowledge, the timeline for HTS magnets is around 5 to 15 years longer, conditional on the successful demonstration of the feasibility of an accelerator-class FCC-hh short HTS dipole. A decision point on the conductor technology is expected around 2035 at the earliest. This timeline is ambitious and subject to significant uncertainties, given the low TRL of HTS technology for accelerator-class magnets.

MCs could be an option for achieving HE lepton collisions but they have not yet reached the level of maturity of the other large-scale collider project proposals presented in Sect. 10.2. The proton-driver for the muon beam production shares the technology and beam-dynamics challenges of high intensity-proton accelerators developed for PBC and would benefit from the common R&D in that domain. Particularly critical is the demonstration of the 6D ionization cooling which requires a dedicated medium-sized test facility. A variety of technological challenges is associated with the MC cooling channel and the downstream accelerators. Start-to-end simulation tools need to be further developed to validate the overall performance of MCs. Neutrino flux mitigation remains a critical subject and is being studied.

Significant technical progress has been achieved in the domain of HGWFA and in particular in PWFAs since the 2020 ESPPU. Beam quality, stability and efficiency improved individually. However, collider applications require them simultaneously and major challenges remain towards colliders based on these technologies. While PWFA projects for applications are underway, first conceptual collider design studies have started recently; among them the most advanced is HALHF, working towards a CDR for an asymmetric collider preceded by a demonstrator including multiple stages of acceleration of beams of the required quality.

High-efficiency klystron designs have been developed as part of an effort to increase the efficiency of the CLIC L-band drive beam accelerator, and to provide improved X-band power sources for testing. This R&D now benefits not only all the proposed collider projects, as they use mostly klystrons as RF power sources, but also photon sources, industrial linacs, medical accelerators and the injectors for hadron accelerators.

The accelerator R&D for HEP has enjoyed a strengthened collaboration among European Laboratories and international partners within the framework of the LDG European Accelerator R&D Roadmap, which was set up in 2022, and it has established productive synergies with several other domains beyond collider and particle physics. It has also contributed to train the next generation of accelerator physicists, engineers, and technicians. In this framework particular attention is also being devoted to evaluate the environmental impact of accelerator facilities with the aim of identifying the main contributors to their environmental impact in the early phases of the projects and to address them in guiding the design choices.

Particle accelerators are powerful engines of discovery and innovation. They not only drive breakthroughs in fundamental physics, but also advance medical diagnostics, radiation and nuclear therapy, and contribute to the design of clean-energy technologies and, thus, more sustainable RIs. The continuation and reinforcement of a sustained and well-coordinated R&D programme will, therefore, be of utmost importance. In addition to supporting the design, construction and operation of the next generation large-scale accelerator project, and thereby securing the longer-term broad future of HEP, a strong R&D effort will also ensure beyond state-of-the-art advancements in neighbouring scientific domains, with careful consideration of environmental impact and, more generally, long-term sustainability. Achieving the unprecedented performance targeted by future accelerators will require the development of increasingly accurate accelerator-physics models and powerful multi-physics simulation frameworks, capable of capturing the full complexity of these machines. Reaching the required high availability will further demand more sophisticated models, including the creation of digital twins of both, the accelerator and the associated technical infrastructure, while incorporating experience from existing machines. The interplay between technological innovation and advanced modelling tools will be crucial to guarantee maintenance, improve commissioning and enhance operation, to create the next generation of efficient and sustainable particle accelerators.

# Chapter 11

# Detector Instrumentation

## 11.1 Introduction

This chapter covers instrumentation for future detectors for both collider and non-collider projects, including instrumentation for neutrino physics and rare event searches. For detectors at future colliders, and for detector upgrades for existing colliders, the timeline may be separated into three eras. The **first era** encompasses the high-luminosity phase of the LHC (HL-LHC) at CERN, the SuperKEKB electron-positron collider at KEK, Japan, and the upcoming Electron-Ion Collider (EIC) at BNL in the US. Major instrumentation projects planned for the HL-LHC in addition to the ATLAS and CMS Phase-2 Upgrades include ALICE 3 [ID70], [128,863] and the LHCb Upgrade II [ID148], [864, 865]. The Belle II experiment at SuperKEKB [ID205], [866] is considering upgrades on similar time scales, and the ePIC detector [ID17], [867] is planned as the first experiment at the EIC.

Future electron-positron colliders that serve as "factories" for the physics of the Higgs boson, the top quark and for electroweak physics ("HET factories", [6]) mark the **second era** of future colliders. An HET factory at CERN may be realised with an electron-positron storage ring such as the FCC-ee [ID233], [868, 869] or with a linear collider, either the Linear Collider Facility (LCF, [ID40], [870,871]) based on technology developed for the International Linear Collider (ILC, [ID275], [872–874]) or the Compact Linear Collider (CLIC, [ID78], [805,875,876]). For the second era, several detector concepts have been developed; these link detector requirements, which are driven by the physics programme, to concrete detector technologies. The detector concepts guide the research and development (R&D) efforts, while leaving freedom to combine technologies differently at a later stage. Further projects proposed for the second era that reuse parts of the existing LHC infrastructure at CERN are LEP3 [ID188], [877], a new electron-positron collider in the LHC tunnel, and the Large Hadron-electron Collider (LHeC, [ID214], [821, 878]), in which protons or ions from the LHC collide with electrons from a new accelerator.

The **third era** of future colliders includes accelerators that enable partonic centre-of-mass energies of up to 10 TeV, such as intended for extensions to the LCF, a muon collider [ID207], [753, 755] or a high-energy hadron collider like the FCC-hh [ID247], [868, 869].

**Key points:** First priority must be the success of experiments in Era 1, laying the foundation for the developments required for Eras 2 and 3. For Era 2, specific topics with long lead times must be pursued earlier.

## 11.2 The DRD Programme

The 2020 update of the European Strategy for Particle Physics (ESPP) [879] suggested that "*The community should define a global detector R&D roadmap that should be used to support proposals at the European and national levels.*" The roadmap should develop a diversified detector R&D portfolio that has the largest potential to enhance the performance of the particle physics (PP) programme in the near and long term. Thus, a roadmap process was started, consisting of a phase of structuring (May–December 2020), the collection of inputs from the community by several open symposia (January–May 2021), and the drafting of the conclusive documents (June–October 2021). The outcome was the **ECFA Detector R&D Roadmap** document [880], released in 2021, which defines the major detector R&D themes (DRDTs) and ten General Strategic Recommendations (GSRs): 1. Enhanced and broad support for R&D facilities; 2. Provide engineering support for detector R&D; 3. Maintain and update specific software for instrumentation; 4. Expand international coordination and organisation of R&D activities; 5. Establish distributed R&D activities with centralised facilities; 6. Establish long-term strategic funding programmes; 7. Provide support for "blue-sky" R&D; 8. Attract, nurture, recognise and sustain the careers of R&D experts; 9. Establish tactical industrial partnerships; 10. Embed Open Science in all instrumentation work.

During the roadmap process, the work was organised into nine Task Forces (TF), each reflecting a different detector technology domain; eight produced proposals that led to **DRD collaborations** [ID157], while TF9 became the ECFA Training Panel [ID30]. The DRDs are designed to advance promising concepts from exploratory "blue-sky" research to mid-level technological maturity, preparing for the engineering phases of experiment-specific implementation. They focus on the validation, optimisation, and system-level evaluation of detector technologies, corresponding to Technology Readiness Levels (TRLs) 3 to 6, visualised in a typical timeline of a PP experiment in Fig. 11.1. This intermediate positioning ensures that DRDs produce cross-cutting, experiment-agnostic innovations that facilitate the instrumentation needs of future collider and non-collider experiments.

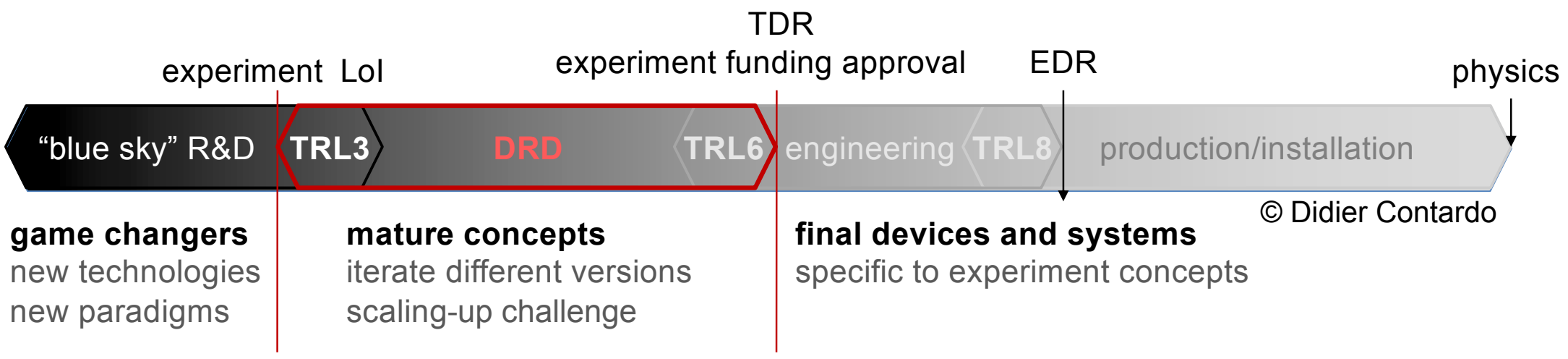

Fig. 11.1: Timeline of technology developments, showing the strategic R&D performed by the DRDs between "blue sky R&D" and deployment in a PP experiment. Strategic R&D starts with the Letter of Intent (LoI) and ends when the Technical Design Report (TDR) is being finalised. The engineering phase ends with the Engineering Design Report (EDR).

The DRD collaborations can be categorised into three groups, as listed in Table 11.1: DRD 1, 3 and 6 are each based on (several) predecessor collaborations with already established community. DRD 2, 4 and 5 are new, which implies that the first focus was on community building, establishing trust and identifying the benefit of being a "CERN-hosted" collaboration. Two collaborations on electronics (DRD7) and mechanics (DRD8) cover transversal activities, but perform their own genuine R&D, possibly with a lower TRL, and provide input to the whole field and the other DRDs in the long term. The current DRD8 focuses its work on mechanics and cooling for tracking detectors and does not cover all the topics initially mentioned in the Roadmap, since the portfolio was very diverse. Future extensions of DRD8 may include calorimeter mechanics.

Table 11.1: List of the eight established DRD collaborations with the number of interested institutes listed in each proposal document. The predecessor collaborations mentioned in the first box only refer to direct forerunners.

| | |
|---|---|
| *Based on previous R&D collaborations:* | |
| DRD1: | Gaseous detectors (based on RD51): 161 institutes [881], [ID229] |
| DRD3: | Semiconductor detectors (previously RD42, RD50): 145 institutes [882] |
| DRD6: | Calorimetry (CALICE, RD52, Crystal Clear, other collaborative R&D efforts): 135 institutes [883], [ID108] |
| *New:* | |
| DRD2: | Liquid detectors: 86 institutes [884] |
| DRD4: | Photodetectors & particle identification: 74 institutes [885] |
| DRD5: | Quantum sensors and emerging technologies: 112 involved institutes [886], [ID204] |
| *Transversal activities:* | |
| DRD7: | Electronics: 67 institutes [887] |
| DRD8: | Mechanics & integration: 38 institutes [888] |

All DRD collaborations adhere to a common organisational architecture and have reached a comparable operational state. They leverage CERN infrastructure and follow established procedural frameworks comparable to those of experiments [889]. Governance is formalised through Memoranda of Understanding (MoUs), which are signed by participating institutes and outline resource commitments, deliverables, and collaborative responsibilities. The process of negotiating and signing these MoU's is currently ongoing (status: September 2025). The internal structure of each DRD consists of resource-loaded Work Packages (WPs), aligned with the DRD R&D Themes (DRDTs) defined in the Roadmap, as well as transversal Working Groups (WGs) that address shared challenges, such as software tools, facilities for irradiation, characterisation or production, or engagement with industrial partners. Some collaborations (DRD1 and DRD3) have introduced a fixed annual membership fee (Common Fund) to support WG activities. Collaborative governance is exercised through a Collaboration Board (CB), comprising institutional representatives, and a spokesperson or a designated executive management body. Oversight is provided by the CERN-based DRD Committee (DRDC), which surveys DRD proposals, monitors progress through annual reviews and reports, and ensures alignment with evolving scientific requirements. Strategic guidance is provided by the ECFA Detector Panel (EDP), while interactions with funding agencies are facilitated through Resource Boards as stipulated in the MoUs.

The DRD collaborations are considered an established concept by many input proposals. Neighbouring fields such as nuclear and astroparticle physics reference them in the NuPPEC Long-Range Plan [890] and the APPEC Roadmap Update 2023, respectively, and emphasise large overlap and synergy. The US PP community has been engaging broadly in the process of defining and shaping the DRDs. In addition, *Detector R&D Collaborations* (RDCs), were established following recommendations of the Snowmass process [ID230] and administered by the Coordinating Panel for Advanced Detectors (CPAD) [ID93]. They are structured along similar lines as the DRDs to build strong connections [ID230]. In Japan, KEK established an *Instrumentation Technology Development Center* (ITDC), to cover strategic R&D in addition to developments in experiments like Belle II [ID205], Hyper-K [ID238] or at J-PARC [ID155].

**Key points:** The Detector R&D (DRD) collaboration structure has been created with the goal of implementing the ECFA Detector R&D Roadmap. It must be fully exploited and the activities must be adequately resourced.

## 11.3 Detector Concepts for a Higgs/Electroweak/Top Factory

Linear collider detectors [ID94, ID102] operate at low duty cycles ($\mathcal{O}(0.5\%)$ for LCF and down to $10^{-5}$ for CLIC), with stringent timing requirements due to very high instantaneous background rates. The low duty cycle allows for "power pulsing," i.e., reducing the power consumption of the front-end electronics during the 200 ms (LCF) or 20 ms (CLIC) gaps between the bunch trains [891], enabling air cooling solutions of its vertex detectors. At the FCC-ee, continuous powering of the electronics is needed, thus requiring active cooling, which heavily impacts the material budget and services. Thus, the detectors at the FCC-ee [868] are the most challenging among the proposals for Era 2. The detectors are designed to provide the best performance as required to achieve the physics goals (Table 11.2), while operating under the very demanding conditions of the accelerator [868, 892]. The high luminosity of FCC-ee will result in large backgrounds in the detectors, especially close to the beam line, which influence both the detector readout and the trigger strategy. Detector integration aspects, such as the detector cooling, power distribution and readout architectures, directly impact the detector acceptance and material budget and ultimately its performance.

Table 11.2: Main detector performance requirements at FCC-ee (adapted from Ref. [868]). The quantities $\delta M_H, \delta M_Z, \delta M_{\mathrm{HNL}}, \delta\Gamma_Z$, and $\delta\mathscr{L}$ refer to the desired precision for the Higgs and $Z$ boson and neutral heavy lepton masses, the $Z$ width, and the instantaneous luminosity.

| Performance indicator | Requirement | Physics motivation |
|---|---|---|
| Vertex hit resolution and material budget | $\approx 3\,\mu$m; $\approx 0.1\%\ X_0$ per layer | $B$-meson, Higgs-boson and $\tau$ lepton physics |
| Momentum resolution | $\sigma(p_\mathrm{T})/p_\mathrm{T} \approx 2\times 10^{-5}\cdot p_\mathrm{T}(\mathrm{GeV}) \oplus 0.2\%$ | $\delta M_H = 4\,\mathrm{MeV}$; $\delta M_Z = 15\,\mathrm{keV}$ |
| ECAL | $\sigma(E)/E \approx \mathrm{few}\%/\sqrt{E(\mathrm{GeV})}$ and high granularity | $B$-meson, $\tau$ and EW physics, $\pi^0$ and $\gamma$ reconstruction |
| Jet energy resolution | $\approx 30\%/\sqrt{E(\mathrm{GeV})}$ | Higgs and multi-jet events |
| Magnetic field stability/mapping | better than $10^{-6}$ | Point-to-point energy uncertainty in $\delta\Gamma_Z$ |
| Timing resolution | tens to 100 ps per track | Particle identification; $\delta M_{\mathrm{HNL}}$ |
| Particle identification | $> 3$ standard deviation $\pi$–$K$ separation 1 GeV to 30 GeV; standalone muon ID | $H \to s\bar{s}$ ; $b \to s\nu\bar{\nu}$; long-lived particle searches |
| Luminosity detector alignment | Position along beam line: $\delta z = 110\,\mu$m; Inner radius: $\delta R_{\mathrm{min}} = 1\,\mu$m | $\delta\mathscr{L} = 10^{-4}$ from Bhabha events at the $Z$ pole energy |
| Data acquisition and trigger | 50 MHz trigger rate and up to 25 Gbit s$^{-1}$ module readout | Processing of beam-induced backgrounds at the $Z$ pole |

Integrated designs are leveraging detector R&D through the DRDs (Sect. 11.2), with the help of full-simulation models for sub-detector and system performance on specific physics benchmarks. Several solutions are being investigated, resulting in four detector concepts, built around different calorimeter and detector solenoid choices, while the other subsystems are mostly interchangeable (vertex, tracking and muon systems): CLD [ID95], [893] and ILD [ID102], [894]——based on earlier designs for linear colliders—as well as ALLEGRO [ID211], [895], and IDEA [ID95], [896].

The design of the **interaction region** is rather complex [868, 892], placing strict constraints on the placement of the luminosity detector and the angular clearance around the beam line. Thus, it demands careful integration between the machine and the detector and makes the accurate positioning of the luminosity detector around the beam pipe challenging, posing stringent requirements on its manufacturing and surveys during operations [868]. The expected radiation levels in the detectors, on the other hand, are much smaller with respect to the LHC, with 1-MeV-neutron equivalent fluences of $1 \times 10^{13}\,\mathrm{cm}^{-2}$ and $5 \times 10^{12}\,\mathrm{cm}^{-2}$ per year in the vertex and luminosity detectors, respectively.

**Vertex detectors** will be located very close to the beam pipe, and will be based on air-cooled layers of Monolithic Active Pixel Sensors (MAPS), which allow for fully integrated electronics in thin silicon substrates, with very low power density. A carbon-fibre supported flat layout results in a material budget of about 0.25% of a radiation length, $X_0$, per layer, while an ALICE ITS3-inspired curved layout [897] may provide 0.07% $X_0$, albeit with reduced hit efficiency and coverage [892]. The impact parameter resolution is ultimately limited by the necessity of an actively cooled beam pipe, inspired by SuperKEKB [898], with a material budget of about 0.7% $X_0$ that might jeopardise the excellent vertex resolution, which initiated an R&D programme in DRD8. The air cooling of the vertex detectors is challenging due to service integration and requires reducing the induced vibrations below $\approx 1\,\mu\mathrm{m}$ to not spoil the vertex resolution. The systems developed for Era-1 experiments will be instrumental in MAPS developments for FCC-ee, and R&D lines have been proposed on MAPS, detector layout and the interaction region [ID95], [882], while ALICE is demonstrating the state-of-the-art in the current development of ITS3, which uses wafer-scale, ultra-thin MAPS that are approximately 50 µm thick and 266 mm long, seamlessly stitched and bent into truly cylindrical layers (Fig. 11.2).

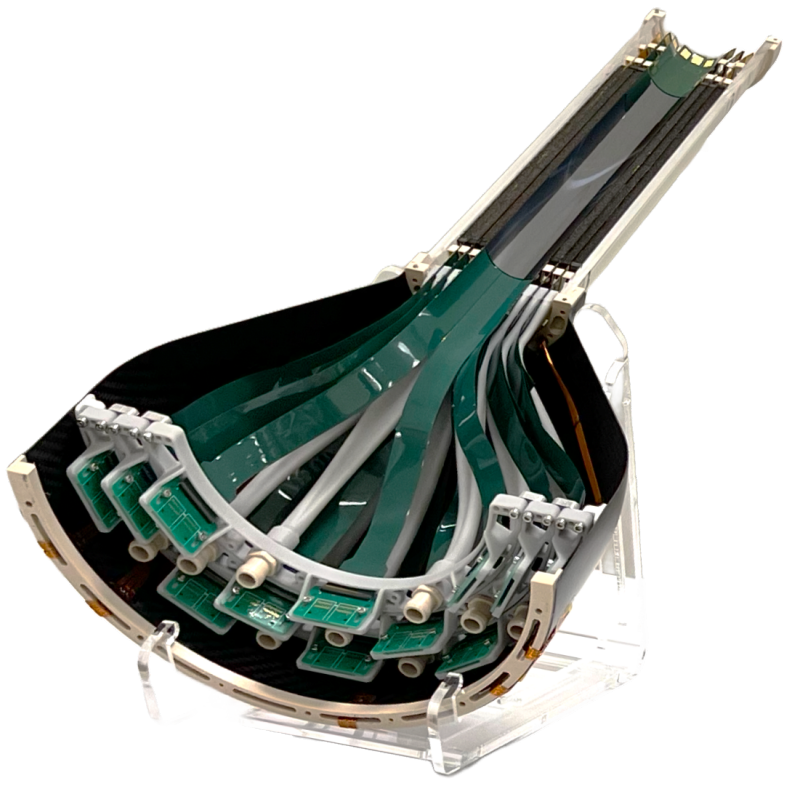

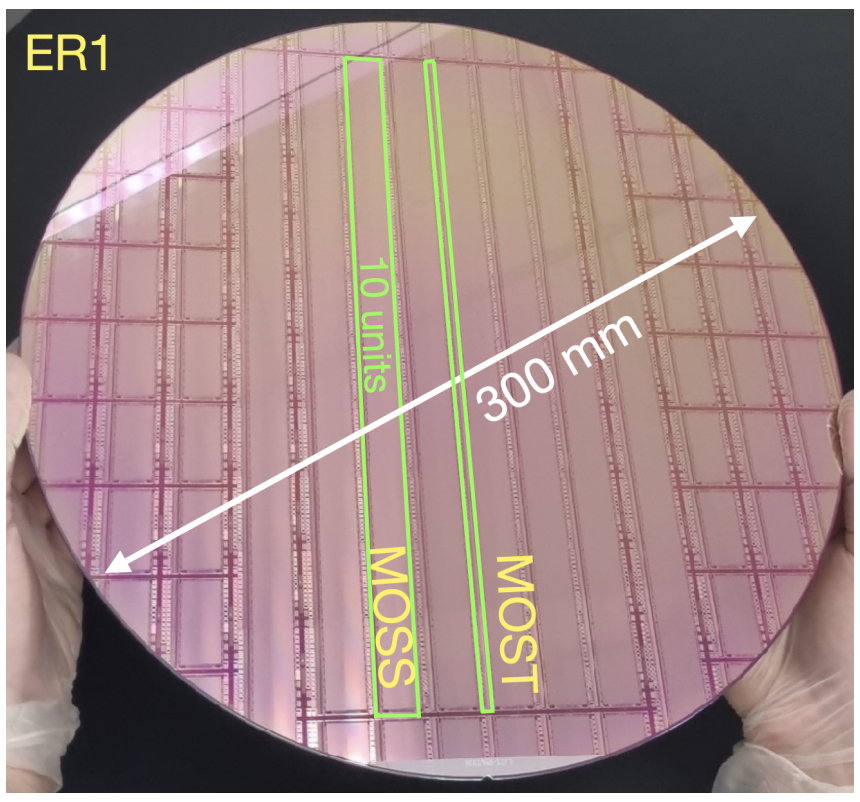

Fig. 11.2: Prototypes for the barrel-only ALICE ITS3 Vertex detector. Left: Full-scale engineering model with air-cooling and bent sensors. Right: Stitched full-wafer long MOSS and MOST monolithic active pixel sensor prototypes from Engineering Run ER1 (both pictures courtesy of ALICE collaboration).

The background from **beamstrahlung**, especially at the high luminosity running at the $Z$ pole energy, limits the beam-pipe radius and results in large occupancies in the detectors [892]. Electron-positron pairs generated by incoherent scattering of either real or virtual photons from the two interacting beams generate a large background in the vertex detector, resulting in high hit rates (up to $200\,\mathrm{MHz\,cm^{-2}}$), which are challenging for the readout (about $25\,\mathrm{Gbit\,s^{-1}}$ per detector module) and thus influence the trigger and data acquisition (TDAQ) design. Synchrotron radiation is suitably shielded before the experiments, and careful simulations of possible residual effects are being carried out.

Different technologies for the **main tracker** are envisaged. One solution employs a large number of moderate precision (about 100 µm) layers using several techniques: helium-based drift chambers for IDEA and ALLEGRO, a time projection chamber (TPC) for ILD, straw tubes or scintillating fibres for ALLEGRO. Both ALLEGRO and IDEA complement the central tracker by a precise silicon wrapper, also providing timing information. Another solution is adopted by CLD and is based on a small number of silicon detector layers with about 10 µm precision; the current implementation results in a higher material budget compared to the gaseous detectors and optimisation is ongoing. The TPC readout at the $Z$ pole for FCC-ee is particularly affected by the large rate of ions produced by the beamstrahlung background, which results in distortions in the hit reconstruction of up to 1 cm that need to be compensated. Due to the smaller interaction rate, the TPC operation is less affected at linear colliders, even at the $Z$ pole and in presence of intense beamstrahlung [894].

**Particle identification** exploits traditional energy loss ($\mathrm{d}E/\mathrm{d}x$) or cluster counting ($\mathrm{d}N/\mathrm{d}x$) methods in gaseous detectors, resulting in relative resolutions of about 4.5% and 2.5% [896], respectively. One of the challenges for cluster counting is the need to digitise the signal with very fast sampling electronics, which may result in an increased power budget for the front-end electronics. The CLD concept uses an aerogel ring-imaging Cherenkov (RICH) system to identify particles with momenta between 2 GeV and 50 GeV. The ALLEGRO and IDEA concepts use a time-of-flight (TOF) system in the silicon wrapper to cover low-momentum tracks (below about 2 GeV), based on either low-gain avalanche detectors (LGADs) or complementary metal-oxide-semiconductor (CMOS) sensor-based systems with time resolutions between few tens and 100 ps.

Different solutions for highly granular **calorimeters** have been proposed. Two detector concepts (CLD and ILD) exploit compact very high granularity sampling calorimeters, as developed in the CALICE collaboration. Instrumental input on their behaviour will come from Era 1 experiments [128, 698, 867, 899, 900]. The IDEA concept uses a dual readout technique, which disentangles electromagnetic and non-electromagnetic contributions in hadron showers by separating scintillator and Cherenkov light detection with high granularity. ALLEGRO calorimeters use fine grained noble liquids (liquid argon with lead or liquid krypton with tungsten), based on the ATLAS experience but with smaller electrodes. Dedicated cryogenic electronics and integrated cryogenics with superconducting detector solenoids are being studied.

Large area **muon detectors** are needed to instrument the outer layers, with a resolution of a few hundred micrometers [901] . To detect long-lived particles (LLPs) and other exotic particles, which are candidates for dark matter, self-triggering and stand-alone tracking capabilities are required, resulting in granular cells and multiple layers. Gas-based systems will require eco-friendly gas mixtures [881].

Controlling **trigger** efficiencies for FCC-ee $Z$-pole operation will be a challenge for precision measurements. Most of the occupancy in the sub-detectors will result from large beam-induced backgrounds rather than genuinely interesting physics signals. Thus, experiments aim for a trigger-less design with full detector readout, as described in section 11.11. However, the need for triggered readout of some sub-systems will have to be evaluated, expecting impact on the front-end electronics and therefore on the material budget. This optimisation also needs to take into account calibration, data processing and offline data storage needs.

**Detector magnets** for the FCC-ee detectors need to provide a magnetic field over a large volume (nearly $70\,\mathrm{m}^3$). Compared to the ILC, the field strength at the FCC-ee is smaller: 2 T instead of up to 5 T for SID [ID94], [902]. However, the field affects the circulating beam emittance, hence the luminosity [868]. The most recent large solenoids were built for the LHC experiments, over 20 years ago. No fully qualified vendor for superconducting cables exists today; however, ALICE is actively pursuing commercial options for Al-stabilised cables for the ALICE 3 magnet [ID68], while ePIC [ID17] will use Cu-stabilised cables [ID114]. A CERN initiative pursues alternatives, notably also for high-temperature superconductors at 20 K, developed mainly for fusion reactors [903].

**Key points:** The broad physics programme and challenging experimental conditions for the proposed HET factories demands detectors beyond the present state of the art. The operation of the FCC-ee at the $Z$ pole with high rates, but respecting low power and material budgets, requires novel approaches to TDAQ architectures.

## 11.4 Semiconductor Detector Technology

Future experiments at both electron-positron and hadron colliders, as well as in non-collider settings, will operate under vastly different conditions, each presenting unique challenges for tracking, vertexing, and timing systems. Silicon-based detectors remain the most promising technology to meet these demands, thanks to their proven performance and versatility. The challenges of Era-2 experiments have been discussed in Sect. 11.3. For any HET experiment, finer segmentation and thinner sensors are needed to achieve low-mass, high-precision vertexing, but these must be balanced against power consumption and readout constraints. At hadron colliders like FCC-hh, the main challenges stem from extremely high particle fluences and intense radiation doses. Detectors in these conditions must withstand 1-MeV-neutron equivalent radiation fluences exceeding $1 \times 10^{17}\,\mathrm{cm}^{-2}$ while maintaining acceptable spatial resolution and operational performance. Timing resolution at the level of tens of picoseconds becomes critical to mitigate pile-up from hundreds of simultaneous collisions. Meeting these diverse demands requires tailored solutions. R&D for these diverse topics is carried out within DRD3. Its programme, described in detail in Ref. [882], is structured around the following key R&D themes: Development of fine-pitch CMOS sensors with timing capabilities, advancement of 3D sensors and LGADs, investigation of radiation effects and novel materials, and development of integration and hybridisation techniques for ultra-light, high-density modules.

For ultrafast timing applications, **LGADs** [904, 905] have emerged as the leading technology. LGADs are thin silicon sensors incorporating a controlled gain layer that provides modest internal charge multiplication (gain $\sim$10–40), enabling excellent timing resolution of the order of 30 ps to 50 ps. They are already being implemented in the high-granularity timing layers of the Phase-2 ATLAS [906] and CMS [907] upgrades and considered for several Era-1 experiments and upgrades [ID17, ID70, ID148]. Ongoing R&D aims to improve their spatial granularity, fill factor, and radiation tolerance. Conventional LGADs have dead areas between pads or pixels because the gain layer cannot be implemented there. New designs such as *AC-coupled*

*LGADs (AC-LGADs)* and *Inverse-LGAD (iLGAD)*, which use a continuous gain layer combined with resistive or capacitive readout, and *trench-isolated LGADs*, which use deep isolation structures to reduce cross-talk and improve radiation hardness, are under development [908]. Work is also ongoing on gain layer defect engineered devices, such as carbon co-implanted LGADs and *compensated LGADs*, which preserve gain beyond fluences of $1 \times 10^{15}\,\mathrm{cm}^{-2}$ by optimising both p- and n-type dopants to counteract radiation-induced changes in doping concentration [909].

**Three-dimensional (3D) sensors** [910], in which electrodes are fabricated perpendicular to the wafer surface, are already used in the most radiation-hard regions of the ATLAS Insertable B-Layer [911]. Their vertical geometry shortens carrier drift paths and lowers depletion voltage, maintaining charge collection efficiency even after fluences above $1 \times 10^{16}\,\mathrm{cm}^{-2}$. Both ATLAS and CMS will use 3D sensors in the upgrades. Future developments focus on producing thinner sensors to reduce material budget and power consumption, devices with gain, improving large-scale fabrication techniques, and achieving finer pitch for higher granularity and better spatial resolution. Recent work also explores *trench 3D sensors* for enhanced fill factor and improved timing response, particularly promising for LHCb Upgrade II [912].

For ultra-low-mass vertex detectors at electron-positron colliders, **Monolithic Active Pixel Sensors (MAPS)** based on CMOS technologies are the preferred choice [913]. In these devices, the sensing and readout electronics are integrated on the same silicon wafer, eliminating the need for bump bonding and enabling extremely thin, lightweight detectors with high granularity. Small-electrode CMOS designs—such as the ALPIDE chip developed for ALICE ITS2 [914]—offer excellent noise and power performance with pixel pitches of $27\,\mu\mathrm{m} \times 29\,\mu\mathrm{m}$. The ALICE ITS3 is pioneering wafer-scale, ultra-thin MAPS with ultra-low material budget (Sect. 11.3). MAPS production is typically carried out in industrial foundries using commercial imaging sensor nodes down to 55 nm; however, process modifications tailored to high-energy physics applications are often required—for example, the modified Towerjazz/TPSCo process reduces low-field regions to enhance charge drift and thus improve collection speed and mitigate charge trapping, improving radiation hardness. To increase depletion depth and enhance charge-collection speed, *large-electrode* designs and *high-voltage* or *high-resistivity* CMOS processes are also being developed. The Mu3e experiment [915] employs high-voltage CMOS technology to achieve very low material budgets of about $0.1\%\ X_0$ per layer for studies of muon decays at rest, using chips thinned to 50 µm to 75 µm, aluminium-kapton flex circuits, helium cooling, and 25 µm carbon fibre supports. Additional R&D aims to further enhance the performance of monolithic sensors by offering ultra-low-power architectures, in-pixel gain, low noise, sub-nanosecond time resolution, and ultra-fine pitch. Monolithic technology is also being explored for strip sensors (“passive CMOS”) and LGAD detectors to cover large areas. This R&D will not only improve the performance of future vertex detectors, the main current application of MAPS, but will also expand the use case to large trackers, particle identification systems, and calorimeters.

In parallel, R&D efforts are ongoing on **alternative materials** that may outperform silicon under extreme conditions. Chemical vapour deposition (CVD) *diamond detectors*, with high radiation hardness and carrier mobility, are being investigated for timing layers, while techniques such as laser-induced 3D electrodes aim to significantly increase radiation hardness, timing and position resolution in single and polycrystalline CVD diamond [916]. *Silicon carbide (SiC)* detectors show promising characteristics under irradiation, and early-stage research on *gallium nitride (GaN)* and *graphene/SiC hybrids* is also underway, although these technologies are still at a low maturity level.

**Key points:** The development of semiconductor detectors must be driven beyond the state of the art using advances in microelectronics, with near-term emphasis on low mass, high precision and rate capability, whilst keeping long-term challenges in radiation hardness on the agenda.

## 11.5 Gaseous Detector Technology

A rich ensemble of gaseous detector technologies is currently being developed, including single and multi-stage resistive plate chambers (RPCs), straw and tube chambers, multi-wire proportional chambers (MWPCs), drift chambers, transition radiation detectors (TRD), time projection chambers (TPCs), and micro-pattern gaseous detectors (MPGDs). The ECFA Detector R&D Roadmap [880] identified four main themes for future gaseous detectors: 1. Improve time and spatial resolution for gaseous detectors with long term stability; 2. Achieve tracking with $\mathrm{d}E/\mathrm{d}x$ and $\mathrm{d}N/\mathrm{d}x$ capability in large volumes with very low material budget and different readout schemes; 3. Develop environmentally friendly gaseous detectors for very large areas with high-rate capabilities; 4. Achieve high sensitivity in both low and high-pressure TPCs. DRD1 [ID229] was formed to address these challenges, which are shared among the different gaseous detector technologies. This allows the gaseous detector community to advance cutting-edge technologies while maintaining a synergistic effort. DRD1 relies on working groups providing tools such as simulation, production techniques and test beam facilities, and on work packages addressing specific technological needs.

While existing technologies are likely sufficient to meet the performance requirements of **muon systems** in future collider experiments, **tracker systems** demand further research and development. These systems must provide high-rate capabilities of the order of $1\,\mathrm{MHz\,cm^{-2}}$ to $10\,\mathrm{MHz\,cm^{-2}}$ over operational periods spanning several decades (see, e.g., [ID95, ID133, ID141, ID142, ID211], [868]). Critical R&D directions include the development of novel resistive detector architectures, advanced materials and geometries, low-noise front-end electronics, and high-granularity readout schemes to control channel occupancy. Achieving timing resolutions at the nanosecond scale—and targeting resolutions in the range of 10 ps to 100 ps—is essential for mitigating pile-up in high-luminosity environments. Scalability and cost effectiveness will be central to future deployment, necessitating the establishment of large-scale, serial production techniques and cost-reduction strategies. Moreover, environmental sustainability must be addressed through the use of gas mixtures with low global warming potential (GWP) or the implementation of efficient recirculation systems to minimise greenhouse gas emissions [917].

Strategic R&D for **drift chambers and TPCs** is progressing toward the realisation of large-volume detectors for tracking and particle identification in future HET factories. Key areas of focus in drift chambers include mechanical optimisation of detector structures, the integration of state-of-the-art front-end electronics and the development of hydrocarbon-free gas mixtures. Alternative gas mixtures for TPCs are also being investigated to enhance $\mathrm{d}E/\mathrm{d}x$ resolution and suppress ion backflow in high-multiplicity environments. These studies are complemented by the implementation of advanced readout electronics and the development of innovative gating schemes to optimise TPC performance. Further refinement of cluster-counting techniques to enhance particle identification capabilities are important for both drift chambers and TPCs. For **straw and drift tubes**, R&D activities have been conducted with the aim of minimising the material budget while achieving a timing resolution of a few nanoseconds and a spatial resolution of a few tens of µm. The effort is centred on the development of innovative self-supporting detector structures and the integration of high-granularity readout electronics.

The widespread use of **MPGDs** is the result of the constant and cross-field R&D focusing on the advances of new amplification structures, studies of new materials and coatings (e.g., resistive, low outgassing), and selection of the appropriate gas mixtures. Operating MPGDs with stable and uniform gain in certain conditions (e.g., charging up of insulators in highly ionizing environment, highly ionizing events, variable irradiation fluxes) remains a challenge to be addressed by future developments. For **gaseous timing detector** technologies, such as RPCs and MPGDs with Cherenkov radiators, current R&D efforts are focused on the development of stable, large-area, and robust devices [918] capable of achieving picosecond-level timing resolution, minimisation of the electronic jitter and optimisation of the intrinsic time response.

**Key points:** Gaseous detector technologies for main tracking systems with particle identification capabilities need to be advanced to achieve higher rate capability and ion backflow suppression, finer granularity and improved timing precision towards the picosecond range. Performance and reliability of operation with gases with lower global warming potential and higher radiation hardness must be established both for gaseous tracking and RICH detectors, and in parallel recirculation systems must be developed that additionally prevent emissions of environmentally harmful gases.

## 11.6 Calorimeter Technology

At present, calorimeters are no longer considered crude instruments to measure only energy. The need to associate deposited energy with tracks measured in the inner tracker and to distinguish superimposed events calls for a new paradigm in calorimetry, with five-dimensional capability. Achieving precise position measurements requires increased detector granularity. However, this introduces challenges, including a higher channel count, which affects mechanical integration, demands low-power electronics, and complicates cooling. In addition to the 3D information recorded so far, timing information becomes important as well, with a further increase of the data to be collected and processed, as well as the need for faster detectors and faster electronics, which further worsens the required power consumption. On top of this, a multi-messenger approach, which is based on additional signal information, such as pulse shape, spectral information or time of propagation, is becoming more and more a game-changing element to increase calorimeter performance.

There are several proposed approaches to address the target performance described above: a reduction of fluctuations, for example, with the dual readout technique [919], and the improvement of the energy reconstruction through the use of advanced algorithms, such as the particle flow technique [920] and machine learning approaches. All these technologies are now developed in the framework of DRD6 [ID108]. The different approaches have been grouped into three main categories, which are described below, to tighten the collaboration on similar techniques and exploit synergies. This is also true for general needs such as electronics, mechanics and cooling, new material development or characterisation, photodetector use and software. The performance of calorimeters can only be completely determined with full shower containment prototypes, which are bulky objects, especially for hadronic calorimeters. Moreover, they need to move beyond component R&D to system demonstrators early, to validate the approach and prepare for scalable production solutions.

**Sandwich calorimeters with fully embedded electronics** are optimised for the particle flow approach, with a view to an integrated system composed of both electromagnetic and hadronic calorimeters, in synergy with tracking information [ID94, ID102]. Since the sen-sor and the electronics are fully embedded in the calorimeter structure, both the sensor and the system aspects need to be considered jointly. At present, different types of sensors for the active media have been considered, ranging from solid-state sensors to optical media and to gaseous gaps. The sensor development is addressed in full synergy with the corresponding DRDs (DRD3, DRD5 and DRD1, respectively), with groups joining both collaborations. Moreover, due to the key importance of readout electronics in the detector development, a close connection to the DRD7 on electronics, in particular on application-specific integrated circuits (ASICs), is also crucial here more than in other types of calorimeters. In fact, the high number of channels poses highest challenges on power and heat management for continuous readout.

**Liquefied noble gas calorimeters** have a long story of success in particle physics, such as the LAr sampling calorimeters in ATLAS [921] and D0 [922] or neutrino physics detectors (Sect. 11.8). This is also due to the typical capability of such calorimeter types to keep systematic uncertainties low. Present developments [ID211] aim at increasing the currently reached granularity, making them suitable for the particle-flow approach. This requires a new design optimisation for both the readout and the mechanics. The option of cold electronics is worth being investigated to achieve an almost noiseless readout and to simplify cable routing, though requiring careful development due to the technological challenge for the large number of channels expected.

The category of **optical calorimeters** includes all calorimeters based on optical media as active media for both homogeneous and sampling calorimeters. Following the multi-messenger approach described above, these types of calorimeters feature enhanced granularity, fast response for both the material and the photodetectors. The latter need to provide photon readout over a large dynamic range and with good linearity. The development and characterisation of new materials to support the above-mentioned characteristics is an integral part of the R&D programme. This includes, for example, new garnet-based materials, scintillating glasses, as well as new materials such as those based on nano scintillators and quantum dots (Sect. 11.10).

**Key points:** The physics requirements of a future HET factory have spurred the development of innovative approaches to calorimetry. They must be validated early enough at the system level for electromagnetic and hadronic showers due to their fundamental role in the overall detector design and the challenges in scalability and industrialisation. Interactions with sensor and electronics development need to be strengthened.

## 11.7 RICH and Photon Detector Technology

Photon detectors are foundational to particle identification (PID) systems in PP. PID subsystems such as Ring Imaging Cherenkov (RICH), Detection of Internally Reflected Cherenkov light (DIRC), Time-of-Flight (TOF), and Time-of-Propagation (TOP) detectors depend critically on fast, efficient, and radiation-hard single-photon sensors. These systems underpin precision measurements in flavour physics, e.g., LHCb [ID148], Belle II [ID205], heavy-ion studies in ALICE 3 [ID70], and lepton-hadron scattering experiments. Era-2 collider detectors [ID17, ID94, ID95, ID102, ID140, ID141, ID211] and neutrino experiments (e.g., DUNE [ID119]) will place more stringent requirements on photon detection instrumentation. Detectors must combine high photon detection efficiency (PDE), precise timing (targeting less than 50 ps to 100 ps single-photon resolution), radiation tolerance, operation in magnetic and cryogenic environments, and robustness under high event rates (exceeding 100 MHz per channel).

These requirements demand sustained innovation in both sensor technologies and their system-level integration. As photon detector granularity increases, there is a growing need to closely integrate sensors with fast, low-power readout electronics to maintain timing precision, limit power consumption, and handle high data rates. This level of integration presents additional challenges in packaging and thermal management, and must be co-developed alongside the photodetectors themselves to ensure optimal system-level performance.

Key challenges in next-generation **photosensors** involve improving performance under high rates, radiation, and extreme conditions, while ensuring scalability and integration with fast electronics. **Silicon photomultipliers** (SiPMs) are compact, highly granular, and magnetic-field tolerant solid-state detectors. Ongoing R&D targets improved cryogenic and high-rate performance, and radiation tolerance. A major challenge is uniting the integration benefits of CMOS single-photon avalanche diodes (SPADs) with the high detection efficiency and low noise of custom SPADs. Hybrid approaches using 3D integration or wafer bonding are promising but remain technologically demanding. Radiation hardness, timing jitter for larger cell sizes, and long fabrication cycles continue to limit deployment scalability. **Microchannel plate photomultiplier tubes** (MCP-PMTs) offer unmatched timing resolution (10 ps to 20 ps), but challenges include limited photocathode lifetime, magnetic field sensitivity, and scalability to large areas. High-rate environments demand trade-offs between gain and longevity. Efforts are focused on atomic layer deposition-coated MCPs for durability and new photocathodes for improved quantum efficiency, though vacuum-based production prolongs development. **Hybrid photodetectors** combine fast timing and spatial resolution but face constraints in gain, radiation tolerance, and thermal management. Integrating MCP-PMTs with ASICs like Timepix4 [923] offer precise photon tagging but present packaging and power challenges. Further R&D is needed to mature these hybrids for robust, large-scale deployment.

**Ring imaging Cherenkov detectors** are key to modern PID systems, enabling hadron identification across wide momentum ranges. A critical challenge is replacing fluorocarbon gases due to their environmental impact; green alternatives must offer suitable Cherenkov thresholds, low dispersion, and high transparency, while remaining compatible with existing optical and mechanical designs, including pressurised or proximity-focusing geometries. Their performance is critically dependent on the properties of both the radiators and the photosensors. As future collider experiments aim to reduce material budgets and improve vertex resolution, RICH detectors must evolve towards thinner, more integrated designs. This requires advances in lightweight mirror substrates, transparent support structures, and compact photon detector arrays. Efficient coupling of photons to detectors—through mirrors or lenses—must preserve angular resolution without compromising acceptance. Reducing material in tracking regions requires lightweight mirrors, transparent supports, and compact photon detectors that maintain angular resolution and acceptance. Progress depends on innovations such as low-density aerogel radiators, metamaterials, integration of fast sensors like SiPMs and MCP-PMTs, and compact readout electronics. Demonstrators and beam tests are vital for assessing performance under realistic conditions.

**Key points:** The application range of photosensors for extreme radiation or thermal environments must be expanded. System-level development of RICH and photon detectors should focus on compact and low-mass modules as needed for Era-2 collider experiments.

## 11.8 Technologies for Neutrino Experiments

The next generation of neutrino experiments (Chapter 6) aims to address fundamental PP questions such as the absolute neutrino mass scale and the Majorana or Dirac nature of neutrinos through the search for neutrinoless double beta decay ($0\nu\beta\beta$). The neutrino mass ordering and CP violation in the lepton sector will be investigated by precise measurements of neutrino oscillation parameters, where more precise neutrino cross-section measurements are crucial for accurate predictions of the neutrino flux. Detecting the coherent elastic neutrino-nucleus scattering process (CE$\nu$NS) offers a powerful tool for studying neutrino properties and potentially probe BSM physics. To achieve these ambitious physics goals, advanced detector technologies are essential which are capable of identifying subtle signals while effectively mitigating background noise.

The **neutrino detection requirements** vary significantly depending on physics goals and neutrino sources. For neutrino oscillation measurements, very large target volumes are required, often located underground to minimise background. Key requirements include the identification of neutrino flavours through different neutrino interactions, precise measurement of energy, reconstruction of final-state particles, and determination of the neutrino direction. The search for $0\nu\beta\beta$ decay demands large masses of isotopes, very low background levels, and excellent energy resolution. Neutrino mass measurements through beta decay kinematics requires excellent endpoint energy resolution, effective background removal, and good control of systematic uncertainties.

**Liquid detectors** [ID36, ID53, ID116, ID119, ID151, ID232, ID263, ID266] have played a crucial role in the field of neutrino physics and rare event searches due to their unique capabilities and scalability. The three primary types are water Cherenkov (WC), liquid scintillator (LS), and noble liquid detectors, each with distinct advantages. The Hyper-Kamiokande [924], JUNO [925], and DUNE [926, 927] experiments plan to build and operate WC, LS and liquid argon (LAr) TPCs, respectively. Key challenges include scalability to very large masses, the need for efficient and fast photodetectors (Sect. 11.7), increased light yields, and pixelated readouts. WC detectors use Gd-loaded water to enhance neutron tagging [928]. CERN has achieved a major technological breakthrough with the construction of large cryostats for LAr TPCs in collaboration with the liquefied natural gas shipping industry [929]. DUNE has developed TPC readout cryogenic electronics immersed in LAr to reduce channel capacitance and noise [929]. The main R&D directions for liquid detectors include readout technologies to enhance spatial and energy resolution, noise reduction, improved material properties of targets and detectors, and scalable technologies for large systems, with strategic R&D in the context of DRD2 [884]. New charge amplification techniques in liquid TPCs are being explored, trying to improve the 3D imaging, using lower power and lower pixel thresholds [930]. There are proposals to use multiple modality (light and charge) pixels [931]. New large-area wavelength shifters, reflectors, and collectors are being developed along with digital cryogenic SiPMs and new cryogenic facilities for vacuum ultraviolet (VUV) sensor characterisation. The operation of photodetectors in high voltage environments has been enabled through the use of Power-over-Fibre (PoF) and Signal-over-Fibre (SoF) techniques [932]. Isotope loading of target liquids is being investigated to improve light yield and detection efficiency. The use of opaque LS allows for a better localisation of the light source [933]. Solutions for target procurement, production, and purification at large scales, as well as large-area readouts, are also being explored.

**Solid-state detectors** [ID132, ID197, ID253, ID258, ID264] are crucial for experiments focused on the neutrino mass measurement, CE$\nu$NS, and the search for $0\nu\beta\beta$ decay. Cur-rent technologies include enriched high-purity germanium detectors (LEGEND [934], CONUS [935], COHERENT [936]), silicon drift detectors (KATRIN [937]), and skipper CCDs (CONNIE [938]). For phonon detection, crystal-based macro-bolometers (CUORE/CUPID [312, 939], AMoRE [940], NUCLEUS [941], RICOCHET [942]), magnetic microcalorimeters (MMCs) (ECHo [943], KATRIN++ [944], AMoRE [945]), and transition-edge sensors (TES) (HoLMES [946]) have been developed. Achieving extreme radiopurity, active background rejection, sub-keV to few-eV low energy thresholds, excellent energy resolution (0.1% at MeV to keV scales) and scalability to large numbers of detectors are among the key challenges for solid-state detectors. The main R&D directions include the development of new quantum sensors (Sect. 11.10), advanced detector materials, large-scale isotopic enrichment, enhancement of detector platforms, and development of low-vibration dilution refrigerators to achieve millikelvin temperatures for large volumes.

**Gaseous detectors** [ID87, ID119]: Gas TPCs are currently used in the T2K [947] and NEXT [948] experiments, and are planned as near detectors for long-baseline experiments, like DUNE ND-GAr [949]. The aim is to build a high-pressure magnetised gaseous-argon (GAr) TPC for which TPC amplification and readout techniques are being studied, along with efficient and low-noise light readout. In the context of $0\nu\beta\beta$, key challenges are related to the scalability of high-purity gaseous xenon TPCs and barium tagging [950].

**Tracking and emulsion detectors** [ID23, ID171] are used to measure neutrino interactions at colliders. Experiments like FASER [951] and SND [952] at the LHC, and NA61/SHINE [953] at CERN, utilise these technologies. These experiments aim at continuing their operation during the high-luminosity LHC run with upgraded detectors, including high-granularity tracking calorimeters, high-precision silicon tracking layers, and advanced emulsion-based detectors.

**Spectrometers and radiofrequency (RF) cavities** [ID132, ID225] are vital for precise measurements of neutrino mass. KATRIN [954] is the European flagship experiment using the precise spectroscopy of tritium beta decay to determine the effective electron neutrino mass. The limitations of the molecular tritium source have led to the development of new atomic tritium sources for improved precision, together with the Project 8 [955] experiment. Project 8 has successfully demonstrated Cyclotron Radiation Emission Spectroscopy (CRES) as a promising technique for precision electron spectroscopy [956]. Key challenges include the development of scalable CRES-compatible magnet systems and precise frequency analysis.

**Key points:** The next generation of neutrino experiments faces significant technological challenges related to detector performance and scalability. The unique CERN Neutrino Platform is playing a key role in developing the R&D programmes needed for future neutrino experiments.

## 11.9 Technologies for Dark Matter Detection and Rare Event Searches

Over the past decades, our view of dark matter (DM) candidates has expanded from a narrow focus expecting new physics at the TeV scale to a rich and diversified dark sector landscape (Chapter 9, see e.g. Ref. [957]). This has driven an impressive diversification of experimental strategies and detection technologies, which cover complementary regions of the DM parameter space, from keV-scale energy thresholds to ultra-light bosonic fields. This section will primarily focus on the technological landscape of **direct DM detection experiments**. DM production at accelerators (such as ongoing programmes at CERN at the SPS [672] and the HL-LHC [958–960], at ELSA (Bonn) [961], at beam dump facilities [900, 962], and many others) provides a complementary path to explore DM candidates and interactions. Instrumentation for DM searches is related to DRDs 1, 2, 4, 5, 6, and 7 and requires extensive infrastructures including underground laboratories, accelerator facilities, and highly specialised R&D platforms for cryogenics, RF technologies, and quantum sensors. The DMInfraNet initiative [ID112] is proposed to foster European collaboration in DM searches.

The search for DM using **noble liquid detectors** has advanced significantly in recent years. In Europe, the Global Argon Dark Matter Collaboration (GADMC) is pioneering large-scale dual-phase LAr TPCs, such as DarkSide-20k [963], to be installed at the LNGS underground laboratory in Italy. It is designed to achieve ultra-low backgrounds and exceptional sensitivity to weakly interacting massive particles (WIMPs) and is also sensitive to light DM and supernova neutrinos [646, 964]. Key enabling technologies include the extraction of low-radioactivity underground argon, further purification and isotopic separation with a cryogenic distillation column, and the deployment of large-area cryogenic SiPMs (Sect. 11.7) for improved photon detection. These innovations, combined with pulse shape discrimination capabilities unique to LAr, enable unprecedented rejection of electronic recoil backgrounds. In parallel, the XLZD collaboration is driving the development of a massive liquid xenon (LXe) observatory [965]. Building on the success of XENONnT and LZ, XLZD will operate a dual-phase LXe TPC with a 60-ton active target. LXe detectors benefit from high density and stopping power, allowing compact designs with excellent self-shielding. They achieve low-energy thresholds by detecting both scintillation and ionisation signals, enabling sensitivity to light dark matter and precision studies of solar and supernova neutrinos. Challenges such as achieving ultra-low radon levels and procuring large quantities of xenon are met with sophisticated purification systems and material screening programmes.

**Axion and ultralight dark matter detection:** The International Axion Observatory (IAXO) represents the next generation of axion helioscopes, aiming to detect solar axions with unprecedented sensitivity [966]. Building on experience from CAST [967], IAXO uses a large superconducting toroidal magnet inspired by CERN's ATLAS magnet technology, combined with X-ray focusing optics and ultra-low background detectors, e.g., Micromegas MPGDs and alternative technologies like transition edge sensors and MMCs (Sect. 11.8). Its modular design enables versatile searches for axion-like particles (ALPs), dark photons, and even high-frequency gravitational waves. The intermediate-scale BabyIAXO, in preparation at DESY, serves as a pathfinder to demonstrate key technologies [968]. Complementing helioscope efforts, several RF cavity experiments explore the axion dark matter parameter space by exploiting the axion-photon coupling in strong magnetic fields. Projects such as RADES [969] and SRF heterodyne [970] are developing quantum-limited techniques to enhance sensitivity and explore masses of axion-like particles (ALPs) in the μeV range. These initiatives also open pathways to probe high-frequency gravitational waves using similar setups (GravNet [971], MAGO [972]). At Grenoble, the BabyGrAHal and future GrAHal projects [973] are developing haloscope setups to search for axion dark matter in the 1 μeV to 150 μeV range. These efforts integrate advanced cryogenics and high quality-factor resonators to explore regions of parameter space inaccessible to previous experiments. Other approaches aim to search for ALPs and other ultralight bosonic fields through their coupling to nuclear spins (CASPEr [974], GNOME [975]) or through apparent variations in the fine structure constant (QSNET [976]).

**Bolometers** are central to European efforts in detecting low-mass DM particles, as they can sense tiny energy deposits at cryogenic temperatures. Key projects include CRESST [707] at LNGS, which uses cryogenic calorimeters with $CaWO_4$ scintillating crystals to detect sub-GeV WIMPs and achieve excellent background discrimination through simultaneous measurement of phonon and light signals. Recently, other targets have been explored such as Si, $LiAlO_2$, $Al_2O_3$ (sapphire) and Silicon-On-Sapphire (SOS). The COSINUS experiment [977] employs similar cryogenic calorimeters with NaI crystals to provide an independent test of the DAMA/LIBRA annual modulation signal. Additionally, TESSERACT [978], to be installed at the Modane underground laboratory (LSM) in France, is developing next-generation cryogenic sensors, such as TES, with sub-eV energy resolution and single-electron sensitivity to probe ultra-light dark matter candidates.

Thick, ultra-low-noise, fully depleted **silicon charge-coupled devices** (CCDs) enable the measurement of extremely small ionisation signals with sub-electron noise levels, making them uniquely suited to probe DM particles with masses below a few GeV. The most recent evolution, DAMIC-M [979], implements novel skipper CCDs that allow multiple non-destructive charge measurements per pixel. With noise levels as low as $\approx 0.1\,e^-$, DAMIC-M achieves single-electron resolution. Installed at the LSM, DAMIC-M is optimised for detecting nuclear and electronic recoils induced by interactions of low-mass WIMPs and hidden-sector particles with silicon nuclei.

**Key points:** Europe has a unique ecosystem of underground laboratories, accelerator facilities and RF cavity expertise. It often provides a test bed for cutting-edge detector technologies for all of PP, spanning noble liquids, cryogenic bolometers, and quantum sensors. Coordinated efforts are suggested to advance instrumentation for DM searches to take advantage of synergies with DRD5 by complementing their low-TRL activities with the development of next-generation experiments.

## 11.10 Quantum Sensors

Quantum sensors (QS) have seen very rapid development and uptake in low-energy precision measurement and search experiments, and significant potential for further development exists [980–982]. Challenges requiring shared R&D on a wide range of existing quantum sensor-based approaches and systems incorporating them include scaling up from individual sensors to large-scale devices, increasing sensitivity, readout, or modifications for operation in less extreme environments, among many others. Such activities are well under way for low-energy PP [983, 984]. At the same time, first applications of several families of quantum sensors as elements of high-energy PP detectors are starting to be considered. If the corresponding R&D indicates performance comparable to or better than traditional PP detectors, similar challenges to those of low-energy PP will still have to be addressed. DRD5 [886], [ID204] coordinates shared R&D on quantum sensing technologies for the benefit of low- and high-energy PP.

Four major **technology directions for quantum sensors** in PP can be identified among the wide range of quantum sensor approaches under development, corresponding to DRD5 working groups: 1. Low-dimensional materials: quantum dots [985], structures built up from monolayers, and generally engineering at the atomic scale [ID11, ID140, ID155]. 2. Spin-based [983] and optomechanical sensors for DM and gravitational wave detection [ID260, ID271]. 3. Atomic and nuclear systems of atoms, molecules, or highly charged ions [986], exotic bound systems [987], clocks [982, 988, 989], atom interferometry [990], [ID6, ID37, ID71]. 4. Superconducting devices and electronics, e.g., nanowires [991] [ID95, ID211], cryo-calorimetry and -spectroscopy [992, 993] [ID28, ID72, ID132, ID181, ID197, ID258], superconducting RF cavities [ID260], and cryo-electronics [ID54, ID225].

Particularly, directions 1. and 4. are relevant for near-term high-energy PP applications: narrow-band emitting quantum dots could enable longitudinally segmented "chromatic calorimeters." Superconducting devices are already in use in calorimetry (e.g., MMCs or TESs for neutrino mass measurements in KATRIN++, AMoRE, HOLMES+ or for exotic atoms, see also Secs. 11.8, 11.9) or are being considered at colliders (e.g., superconducting nanowire single photon detectors as luminometers). Applications of 2. and 3. are also being investigated, e.g., for TPCs with improved sensitivity or helicity or polarisation-sensitive detectors for high-energy PP [994], and entirely new experiments based on quantum sensing for low-energy PP.

R&D on these and other quantum technologies is both developmental and exploratory. Opportunities to develop transformative quantum sensors for PP require awareness of the quantum technology landscape. It is opportunistic, identifying specific characteristics of existing quantum sensors that would potentially enhance existing particle detectors. It may permit entirely new types of measurements (e.g., helicity or polarisation of single particles). Finally, the potential to enhance sensitivity and enable novel sensing modalities for PP through the use of squeezing, back-action evasion, and entanglement will need to be studied. Challenges such as cost, scaling up or readout, must also be addressed. DRD5 will continue fostering the exchange of expertise and ideas among disparate communities with little prior interactions towards common goals. Considering the very rapid growth in capabilities and range of quantum technologies, potential applications in neighbouring areas and synergies with, and expertise from, many domains must be explored. Dedicated resources and their efficient sharing, as well as transnational access to infrastructures, are required to evaluate the viability of new concepts quickly and coherently. R&D in QS requires a multidisciplinary workforce, familiar with quantum technologies, detectors, as well as particle, astroparticle, and nuclear physics. The shorter time scales of QS-based experiments compared to high-energy PP enable attractive combinations of R&D and physics results.

**Key points**: Quantum sensing devices for low-energy PP have the potential to significantly supplement and enhance existing high-energy PP detector techniques. It is crucial that dedicated R&D is carried out on a wide range of suitable sensor technologies and this is supported at national and international levels. Particular emphasis should be put on the development of pilot applications in PP experiments.

## 11.11 Trigger and Data Acquisition

Future collider experiments, including those in all three eras, face enormous throughput challenges in Triggering and Data Acquisition (TDAQ). Driven by the pursuit for high precision, where hardware trigger requirements reduce signal efficiency for a majority of processes, Era-1 and Era-2 experiments shift toward trigger-less designs, implying full detector readout at very high rates. As shown in Fig. 11.3 (left), LHCb will face the largest integrated data rate in Run 5 of the LHC, followed by FCC-ee experiments at the $Z$ pole [ID211, ID102, ID95]. However, full detector readout and offline-quality real-time reconstruction for Era-2 experiments and beyond might be extremely challenging. Trigger-less operation will require compression and feature extraction directly on or near the detector. TDAQ requirements also depend highly on the choice of detector technology, e.g., readout of ultra-light tracking detectors like TPCs is currently not possible at 50 MHz.

For Era 3, extreme pile-up conditions will result in data rates of $\mathcal{O}(1\,\mathrm{PB\,s^{-1}})$ for silicon trackers. Even if suitable link technology for the readout will exist at the time, infrastruc-ture constraints and radiation hardness may render full tracker readout impossible. This would necessitate hardware trigger systems with offline-quality reconstruction capabilities, such as charged particle tracking. The extreme throughput and latency constraints for such systems require highly PP-tailored solutions capable of operating at sub-microsecond latencies. FCC-hh front-ends will require a throughput one order of magnitude larger than HL-LHC front-ends and commercial applications such as Netflix, as shown in Fig 11.3 (right).

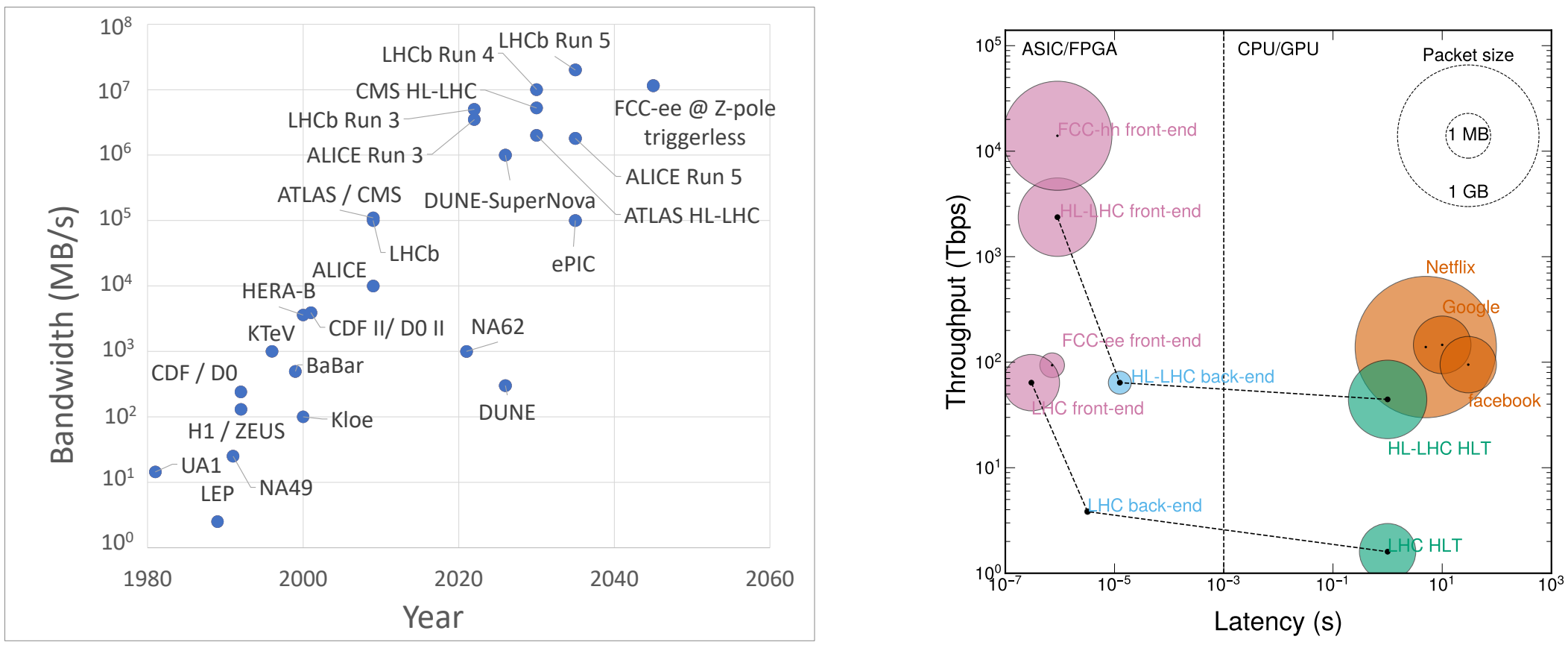

Fig. 11.3: Left: Evolution of data rates over time for large high-energy physics experiments. (Source: A. Cerri (Università di Siena & INFN Pisa), D. vom Bruch (CPPM & Aix Marseille University)) Right: Characteristic throughput versus latency for real-time processing systems. Event sizes for FCC-hh and FCC-ee are approximate, based on reference detectors [896, 995]. LHC and HL-LHC data are from CMS/ATLAS. Packet size is the unit size of transmitted data (e.g., event size in colliders; movie, image, or webpage size in industry). (Source: T. Aarrestad (ETH Zurich), S. Summers (CERN))

TDAQ is evolving along three interconnected directions: 1. Reducing data through compression at the front-end using intelligent ASICs; 2. Reducing data through filtering on hardware back-ends, such as field-programmable gate array (FPGA) farms (triggered); and 3) Reducing data through real-time reconstruction, alignment and calibration, and selections in software (trigger-less). Both trigger-less readout and hardware triggers require data to be compressed, in some cases directly at the source, to enable full-rate readout; and real-time processing in heterogeneous systems follows very similar design principles regardless of whether operating in triggered or trigger-less mode.

**On-detector data compression and transmission:** The vertex detector, subject to extremely high hit rates of $\mathscr{O}(1\,\mathrm{TB\,s^{-1}})$ from the inner layers in Era-2 experiments, faces the largest readout challenge. Efforts to reduce rates include intelligent front-end filtering to discard low-momentum tracks, or feature extraction directly at source on pixel front-end ASICs. For example, neural network designs in low-power 28 nm CMOS technology have the potential to achieve bandwidth savings of 50% to 75% [996]. Reconfigurable logic in ASIC design, through the embedded FPGA (eFPGA) framework, also offers a promising pathway to implementing machine learning directly at the source [ID95]. It can be designed using integrated open-source eFPGA frameworks such as FABulous [997]. One early implementation of this approach comes from CMS for HL-LHC, where a neural network encoder will be used to compress data from the high-granularity endcap calorimeter into a smaller format suitable for filtering in the back-end hardware trigger [998]. An orthogonal approach to addressing the data transfer bottleneck is wireless data transmission [999], [ID95]. The 60 GHz band or higher frequencies, offer the necessary bandwidth, where technologies such as the Wi-Fi standard 802.11ac/ad can provide multi-Gbit $s^{-1}$ links. By broadcasting a single signal to multiple receivers, it is possible to significantly reduce cabling and infrastructure requirements.

**Back-end processing in hardware:** If all sub-detectors cannot be read out at the full rate, a hardware trigger becomes necessary. To maximise efficiency, offline-like quality reconstruction on specialised hardware such as FPGAs is crucial. For the HL-LHC, CMS will run track reconstruction on FPGAs and has demonstrated that track seeding, building, and Kalman filtering to identify final track candidates and determine track parameters is feasible within $\mathcal{O}(1\,\mu s)$ [1000]. Machine learning solutions are being implemented to enhance signal efficiency, using HEP-designed tools for low-latency inference on specialised hardware [ID95, ID17, ID205, ID148], such as `hls4ml` [1001], `QKeras` [1002], and `conifer` [1003], which are also used in the design of the custom integrated circuits discussed in Sect. 11.12. Extremely low-latency, machine learning-based anomaly detection algorithms are already operational in the CMS [1004] and ATLAS [1005] hardware trigger systems, with several additional systems planned for deployment during the HL-LHC era [1006].

**Real-time analysis in software:** Offline-quality real-time analysis comprises reconstruction, alignment and calibration, and selections with selective persistency to minimise the information stored per event. Given the evolution of compute architectures for high-throughput processing, heterogeneous systems are necessary to tackle data rates in modern PP experiments [ID67, ID18, ID127]. ALICE has pioneered the usage of graphics processing units (GPUs) for real-time reconstruction since LHC Run 1 [1007]. Both ALICE and LHCb have deployed a trigger-less readout system for Run 3 of the LHC, where ALICE runs real-time reconstruction for data compression on GPUs [1008], while LHCb has deployed a full real-time analysis system in two stages, the first being processed entirely on GPUs, and the second on CPUs [1009]. Preparing for the HL-LHC, CMS has also deployed a heterogeneous software trigger in Run 3, where parts of the reconstruction are offloaded to GPU processors [1010].

**Key points:** To ensure the high precision and efficiency required by future experiments, TDAQ systems must evolve to deliver offline-like performance in extremely high-throughput environments. R&D efforts for TDAQ systems must include intelligent processing at both the detector front-end and back-end, as well as real-time reconstruction, alignment, and calibration. To mitigate the risk of data acquisition bottlenecks in future detector systems, the combined investment in on-detector intelligence and heterogeneous computing frameworks should be strengthened.

## 11.12 Electronics

Advances in electronics continue to be a central driver of progress in PP. Electronic systems are vital to modern experiments and are among the most critical elements of developing detectors in terms of engineering complexity and risks. They use a wide range of technologies, from nanoscale semiconductors, high current and voltage power supplies to optoelectronics, and rely on complex software and firmware. A breadth of skills in engineering, modelling, simulation and systems integration are needed to design, develop and deliver the electronic systems for future experiments. At the same time, it is increasingly difficult to secure such skills in the community, and to regroup them into teams of sufficient size to serve the ambitions of the field.

Today's electronics lack the necessary combination of performance, power efficiency,

and radiation hardness to meet the requirements imposed by next-generation facilities, and thus risk to be a limiting factor in the scientific reach of future experiments. Some aspects critical to the use in PP experiments, such as the operation in high radiation environments, in high magnetic field or at cryogenic temperatures, are often not relevant for commercial providers. This necessitates fundamental R&D to make new technologies useable for PP. DRD7 on on-detector processing and electronics addresses key open questions in the field of electronics for PP. Modern microelectronics technology is evolving rapidly, driven by developments in industry. While PP was still using up-to-date technology nodes in the late 1990s, today, the field is at least 10 to 15 years behind in technology adoption, exemplified by the current use of a 65 nm process commercially introduced in 2006. Driving factors for this development include the increasing development, tooling and fabrication cost in smaller technology nodes, and the time and resources needed for PP-specific qualifications such as radiation tolerance.

**ASICs – lessons learnt:** One area that has emerged as a particular challenge in the LHC detector upgrades are ASICs, in particular mixed-signal ASICs used in different stages of the readout and control chains of essentially all detectors. The Phase-2 upgrades of ATLAS and CMS required the development of a total of about 50 different ASICs, with some common, widely used chips for data transmission or powering, and other subsystem- and experiment-specific solutions. All require high radiation tolerance and high speed. Many of these ASICs required one or more prototyping iteration beyond what was foreseen, and were delayed by several years compared to the original planning. A dominant contributing factor was a lack of ASIC engineering experts with experience in complex projects using the full suite of tools including modern verification tools and design workflows. The long time scales of the projects also resulted in loss of personnel in participating institutes due to contracts running out, shifts to other projects and changing academic interest. In response to the delays and other problems occurring, CERN has founded the CERN-HEP IC design Platform and Service (CHIPS) initiative, which injected support by experienced personnel in critical projects, and has been instrumental in bringing the Phase-2 ASICs to the stage in which they no longer drive the critical path of the upgrades. This illustrates that the current model of ASIC development in PP, which relies on many small groups at universities and research institutes, is hitting its limits. These arise through conflicts of academic interest, which emphasises the design of new features, with project needs, such as the adherence to schedule and the mitigation of risks, and the desire for ownership, which result in duplication and fragmentation of the overall effort.

**ASICs – a path forward:** To move forward, coordination, collaboration and pooling of resources will be crucial. Common submissions of multi-project wafers enable R&D projects in a cost-effective manner. Community-wide access to professional state-of-the-art design tools, software licences and process design kits (PDKs), as provided through EUROPRACTICE for purely academic development, is essential, as is access to foundries and the supporting legal framework provided by CERN. To address the challenges in the design and the adoption of new technologies, DRD7 is working out a proposal for a hub-based structure for ASIC development in PP, with the goals of establishing and maintaining access to cutting-edge technologies, a professional approach to design, verification, prototyping and fabrication, the facilitation of collaborative work across distributed teams, and the management of risks. A further generalisation of selected developments also beyond PP can add new dynamics and has the potential to open up new funding avenues, while allowing the use of more cost-intensive technologies through higher volumes and fewer separate developments. One example of such an approach is given by the MediPix/TimePix project [ID161].

Beyond the development and fabrication of ASICs, **interconnect and packaging** technologies have a central role in the integration of individual silicon components into detector systems. Ongoing R&D projects in DRD7 and elsewhere aim at a full integration of the front-end electronics with data transmission via silicon photonics and programmable on-detector intelligence. The chiplet technology allows to combine smaller chips into larger ASICs, each potentially produced in a different technology. For some applications, this has the potential to increase the re-use of existing designs and increasing flexibility, and would allow to use the optimal technology for each different element in the larger ASIC.

**Key points:** To overcome the limitations of the current microelectronics development model, new forms and structures for collaborative designs need to be established. Novel packaging, interconnection and integration technologies must be adopted.

## 11.13 Ecosystem for Research and Development on Instrumentation

A strong and supportive ecosystem for detector R&D is a foremost requirement for the community to be able to execute the ambitious future PP programme envisioned by the ESPP. Several key aspects have been identified that are necessary to provide such an ecosystem. These are addressed in the ten GSRs of the ECFA Detector R&D Roadmap (Sect. 11.2), which still hold and are partly implemented by the DRDs and the ECFA Training Panel. A vibrant instrumentation community with a well-trained workforce working towards the realisation of physics projects is essential. This needs to be supported by state-of-the-art test facilities. Broad multi-disciplinary participation and leadership, together with industry partners, will be vital to maintain and leverage developments, as will be the integration of advanced software. Working with society and engaging in outreach is equally important [ID157]. It is important to note that at this moment we do not know where the new physics will manifest itself. Unexpected phenomena may have very subtle signatures and the task of future detectors is not to confirm the expected, but to discover the unexpected that withstands the scrutiny of our peers.

**Coordinated community and projects:** First and foremost, a coordinated community needs to exist that provides a venue for scientists to exchange ideas and identify collaborative projects. With the creation of the DRDs (Sect. 11.2), a nexus has been established for scientists and engineers with a shared interest in specific technologies. It is very important to recognise the implications of the vastly different time scales and sizes of PP projects for a healthy ecosystem for detector development. Currently many proposed projects take at least a decade to go from inception to construction and are very large in scope. Directed R&D aimed at meeting targeted performance specifications is central to the realisation of experiments, which is adequately addressed by the formation of the DRDs. It is critical to provide the opportunity for the development of completely new, novel technologies that represent a high-risk, high-gain effort. Transformative development can happen on a relatively short time scale as demonstrated by, for example, the development of the SiPM [1011], a device that has become a workhorse for the field (Sect. 11.7). The development of the LGAD (Sect. 11.4) went from concept to large-scale deployment in less than a decade [1012, 1013]. This illustrates the necessity to retain “blue sky” R&D in the portfolio. Because of the scale of future experiments it it also necessary to include research at the system level in future development programmes. Subtle effects or low-rate features can manifest themselves at the system level that would go unnoticed at the device level and could affect performance. R&D efforts are most effective when they are targeted or embedded in a science project and not stand-alone.

**Test facilities:** Beam test facilities are critical for R&D on detectors for future experiments. European projects like AIDAinnova and Euro-LABS were instrumental to provide access and develop them further, e.g., with high-resolution beam telescopes. Maintenance and upgrade activities of accelerator facilities affect their availability and it is important to coordinate these activities to provide continuous access for the community [ID244]. A schedule of all major facilities is shown in Fig. 11.4, highlighting the shortage in available beam lines during CERN Long Shutdown 3 in 2027–2028. Irradiation facilities are required for developments for future experiments that provide easy access and high particle fluences so that irradiation doses can be achieved within a reasonable time frame. Access to a general-purpose high-field magnet platform will be necessary for future experiments where detector systems can be tested inside a representative magnetic field. The need for specialised facilities will also increase. This will include facilities to measure relevant fundamental parameters with higher accuracy, calibration platforms, low-background and assay setups and general-purpose platforms to test and characterise quantum devices.

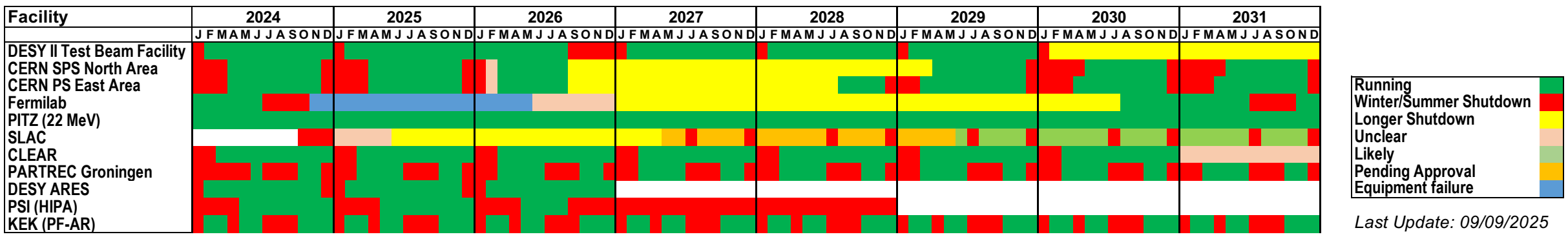

Fig. 11.4: Availability of beam test facilities over the next years. Available online at `https://cern.ch/international-facilities`

**Workforce:** Arguably the most important aspect for a vibrant future PP programme is a workforce that is fluent in building detectors of the scale and complexity proposed. Developing this workforce through training is of critical importance. Although various opportunities already exist at different levels of student competency, there is a clear demand for a comprehensive curriculum in instrumentation at multiple levels of fluency. Measures proposed by the ECFA Training Panel [ID30] to complement the university training in instrumentation to include a European master study programme and a CERN School of Instrumentation. To retain the early-career workforce, strong career opportunities need to be provided. Retention of more senior staff will require a continuous flow of challenging projects and continuous education to keep up with the increasing demands and complexities of the rapidly advancing technologies.

**Partnerships** are indispensable to provide for a continuous, challenging and attractive environment for instrumentation development [ID161]. Industrial partnerships to help with new technology development, but also to stay abreast of the latest developments, will be important. A different type of industrial partnership is the engagement with the commercial sector to licence technologies, which demonstrates the value of our research to society and continues development well beyond deployment of the detectors in a PP experiment. Collaboration and partnerships with other scientific disciplines will become mandatory. Instrumentation technologies have a broad range of applications beyond PP, ranging from muography to medical imaging. Key challenges include ensuring reliable operation, developing cost-effective and low-power detection systems, and enabling effective technology transfer to industrial partners for large-scale implementation. At the same time, other areas of science are delivering impactful advances relevant to instrumentation that could be incorporated in our detector designs and could have deep impact for the physics reach [ID204]. Broadening the community to embrace a multidisciplinary environment will be immensely beneficial. Joint DRD meetings and meetings of the DRDs with other fields could be a great forum to initiate such interactions. Partnerships and integration of advances in high performance computing, AI and ML, and quantum technologies also need to be fully integrated into the workflow of PP detector design, simulation, front-end processing and data taking (Sects. 11.10, 11.11). These transformative technologies will usher in a new paradigm and the field needs to take full advantage of them. Another most important partnership is the relationship with society that needs to be nurtured through outreach. Engagement with local schools is especially important, since it will provide a pathway to create a "brain gain" by reaching the younger generation at an early age, getting them interested in science.

**Key points:** Both early-career researchers and trained experts must be attracted and retained within PP instrumentation through enhanced training measures and career opportunities. Test infrastructures such as test-beam areas with particle identification instrumentation and a wide range of particle types and energies, as well as magnets, cryogenic environments, and irradiation facilities need to be sustained. Creating a broader multi-disciplinary instrumentation environment is crucial, including tactical partnerships with industry as well as outreach and interactions with society.

## 11.14 Conclusions and Recommendations

Advances in instrumentation are a core driver of particle physics and a key enabler of discoveries. The proposed future facilities and their physics ambitions require detector technologies beyond the present state-of-the-art. The ECFA Detector R&D Roadmap highlights major strategic R&D themes and identifies synergies between both parallel and sequential projects, the former playing the role of a stepping stone to the latter. The Detector R&D (DRD) collaboration structure, created to implement the Roadmap, must be fully exploited.

First priority must be the success of the Era-1 projects at the HL-LHC, at SuperKEKB and at the EIC. With the HL-LHC detector upgrades nearing completion, and with the identification of a future flagship collider at CERN, strategic detector R&D must be intensified at CERN, and the resources be correspondingly ramped up at CERN, at national laboratories and at universities. It is suggested to set up a DRD Office at CERN to support interplay and enhance coherence among the DRDs and overarching topics like training. An incremental update of the Detector R&D Roadmap should be organised by ECFA once a flagship project is approved and its scope and timeline established, to support further focussing of the activities. An update should be reconsidered at the latest in the context of the next ESPP Update. Detector systems with long lead times or early installation dates, such as experiment solenoids and large calorimetric systems, must be addressed before others, and decisions affecting the machine-detector interface or conventional infrastructures need to be taken early. Efforts to rebuild the expertise and industrial basis for magnet conductor technology and to explore high-temperature superconductors need to be strengthened. Simulation and engineering studies at full detector level must be pursued early to address conceptual challenges such as the general TDAQ approach, optimisation of overall dimensions and minimisation of machine-induced backgrounds.

Detector R&D needs a supportive ecosystem to prosper as spelled out in the GSRs of the ECFA Detector R&D Roadmap. Beam test facilities are central and critical, and efforts should be made to upgrade their instrumentation and to minimise time periods of limited availability due to simultaneous accelerator upgrades. Irradiation facilities must continue their services, and access to generic test platforms, e.g., with high-field magnets, must be ensured. PP must main-tain its agility and the creativity oriented culture to explore and develop novel technologies in "blue-sky" projects to complement and ultimately feed the pipeline of more targeted R&D. The field of quantum technologies is rapidly evolving, and in DRD5 new applications are opened up, in particular for rare event detectors.

Tactical partnerships with industry should be sought and nourished, both to stay abreast of latest technologies and to foster the application of PP technologies to the benefit of society at large. Partnerships with neighbouring fields should be explored to broaden the application range. Targeted outreach should be intensified to engage new communities, for example in electronics and mechanical engineering. In order to unfold the potential of advanced microelectronics for next-generation detectors and to overcome the limitations of the current electronics development model, new collaborative design structures need to be established. The tiered approach followed in DRD7 should be fully implemented with support from CERN, the LDG, ECFA and national bodies. A framework for the conception of high-throughput TDAQ architectures including computing aspects of near-detector online processing should be established and AI methods in on-detector electronics should be developed.

The R&D programme will only be possible if both early-career researchers and trained experts are attracted and retained within PP instrumentation. Training, including first-hand exposure to the latest technologies, should be supported by establishing new instrumentation schools.

# Chapter 12

# Computing

The role of Computing is of great importance in current- and next-generation High Energy Physics (HEP) initiatives. The Computing domain includes several interconnected aspects:

- the infrastructure (the planning, deployment and operation of a distributed system of dedicated computing centres, as well as the use of additional external resources)
- the software engineering (with LHC experiments having built a software base of 5-10 million lines of code each in the last two decades)
- the technology (including AI as well as other new, potentially disruptive technologies expected to become viable in the medium or long term)
- more social aspects (the positioning of HEP computing with respect to other sciences and with national initiatives in general, with a special emphasis on the problem of sustainability of systems and careers in HEP computing).

In order to guarantee the exploitation of the physics potential of HEP initiatives, computing activities need to be properly funded, planned, structured, tested, and executed, requiring adequate funding opportunities.

## 12.1 Facilities

As part of the 2026 update to the European Strategy for Particle Physics (ESPP), the Facilities Working Group (WG) was tasked with reviewing and synthesizing community inputs on the future of computing infrastructure for High Energy Physics (HEP). Out of approximately 50 submissions to the computing track, around 25% focused directly on facilities, reflecting a strong and growing interest in addressing the technological and strategic challenges associated with computing at scale.

Central to any discussion of future facilities is the Worldwide LHC Computing Grid (WLCG) [ID9], the computing infrastructure handling the data collection and analysis at the LHC. Its success as a robust, scalable and collaborative infrastructure has been key to the LHC physics programme. Future HEP projects universally assume in their inputs that the WLCG will continue to provide this foundational layer of distributed high-throughput computing infrastructure, building upon its services (going beyond the LHC-specific focus suggested by the “L” in the acronym). This continuity, however, depends on a dedicated effort to preserve and evolve WLCG operations and computing R&D through the end of the LHC programme and beyond. A key challenge will be understanding how to sustain this successful community-driven model in the long term. Such evolution is already described, among many other areas, in the "WLCG Strategy 2024-2027" document [ID9].

Across the contributions, six key themes emerged: (1) Evolution of Processor Architectures, (2) HPC Integration in HEP, (3) Sustainability and Energy Efficiency, (4) AI Facilities, (5) Specialized Analysis Infrastructure, (6) Interdisciplinary and Industry Collaborations. The following sections highlight the strategic directions and technical efforts shaping computing facilities in HEP, with a focus on enabling the scale, adaptability, and sustainability required for future experiments.

### 12.1.1 Evolution of Processor Architectures

The scientific computing landscape is shifting rapidly as industry increasingly prioritizes AI-focused technologies and specialized architectures. Advancements in general-purpose CPUs, which have served as the foundation of HEP computing for decades, are being outpaced by specialized or reconfigurable processors such as GPUs, TPUs, and FPGAs. These technologies offer immense computational potential, but also require substantial effort to adapt scientific codes and workflows that were not originally designed with such architectures in mind, as detailed in Sect. 12.2.

The HEP community has already begun responding to this challenge. Many workflows—especially in online systems—have been ported to GPU- or FPGA-based workflows, and offline ones are starting to follow suit. Several collaborative initiatives are supporting this transition, including benchmarking tools and campaigns [1014], GPU workflow developments, powerful abstraction layers such as HLS4ML [1015] and efforts to improve interoperability. CERN openlab [ID18] plays a central role in this area, offering a Heterogeneous Architecture Testbed for evaluating emerging technologies and fostering partnerships with hardware vendors.

Adaptation to these architectures is not just a technical necessity but a strategic imperative. High-precision simulation and reconstruction tasks in HEP must coexist with industry-driven developments in low-precision AI-specialized hardware. This divergence creates both a risk and an opportunity: the need to preserve scientific integrity while actively collaborating with industry to ensure that future architectures remain suitable for research-grade computation. European projects like SPECTRUM [ID180] and ODISSEE [1016], along with white papers from JENA [ID124] and national roadmaps, underscore the urgency of this work.

### 12.1.2 HPC Integration in HEP

A major transformation underway is the expansion of HEP computing beyond traditional high-throughput computing (HTC) models toward federated infrastructures that seamlessly incorporate HPC and cloud resources [ID12]. This evolution builds on the long experience in Lattice QCD, where HPC usage is well established, but now extends to a broader set of workloads including simulation, AI, and analysis, and in general data-intensive workflows, as needed by the HEP large datasets and I/O intensive processing. The transition is motivated not only by performance requirements, but also by funding landscapes and policy trends. National and European HPC initiatives—including the EuroHPC JU [1017]—are investing heavily in building large-scale HPC systems at multiple sites. These initiatives aim to maximize system utiliza-tion by promoting open, competitive access based on scientific excellence. Priority is given to scientific fields that can demonstrate both technical readiness and strategic importance. Today, Lattice QCD is a major user of these infrastructures which have been crucial for precision physics calculations [ID29]. As soon as possible, HEP as a broader domain should position itself not just as users of these facilities but to shape their development through co-design and shared planning activities.

This integration is far from trivial. Much of the HEP software ecosystem is tightly coupled to x86 architectures, which limits compatibility with many modern HPC systems. Additionally, the distributed workflows commonly used in HEP experiments depend on connectivity and interfaces that are often lacking in HPC environments. Portability—both of software and workflows—is therefore a critical priority. To address these challenges, which involve both technical hurdles and coordination across communities, several collaborative efforts are underway. These include the HEP-HPC strategy meetings [ID12], the JENA federated computing initiative [ID124], and international projects like SPECTRUM [ID180] and InPEx [1018]. Meanwhile, experiments including ATLAS [ID187], CMS [ID107], Belle II [ID205], LHCb [ID127], and DUNE [ID121] are contributing concrete use cases and infrastructure proposals.

Commercial Cloud resources have been tested as potential pledged resources and as an opportunity for RD on novel architectures. While the latter approach is promising, the former seems disfavoured by economic considerations [1019].

### 12.1.3 Sustainability and Energy Efficiency

As computing grows in scale and scope, so too does its environmental and financial cost. Sustainability is no longer a secondary consideration; it is rapidly becoming a core design principle for any large scale scientific infrastructure. In HEP, the need to reduce the energy/carbon footprint of computing has sparked a range of initiatives focused on both operational efficiency and long-term planning.

Key efforts include energy-aware scheduling and workload optimization techniques, energy/carbon tracking tools, and consistent use of metrics for measuring data centre efficiency such as Power Usage Effectiveness (PUE). Tape storage systems are being prioritized for their energy efficiency in massive data processing workflows and long-term data preservation, and emergent architectures like ARM and RISC-V are being investigated for their potential to deliver high performance with lower power consumption.

Importantly, sustainability must be considered across the full lifecycle of computing: from hardware procurement and infrastructure planning to daily operations and decommissioning. The transition to green computing is a multi-dimensional challenge that intersects with funding models, hardware choices, software design, and facility governance.

### 12.1.4 AI Facilities

The emergence of AI as a transformative tool in scientific research brings with it a new class of infrastructure requirements. The training, deployment, and inference at scale demand facilities that are optimized for data-intensive workflows, specialized accelerators, and ML-specific collaborative development frameworks.

In response, the European Union has launched an initiative to develop what are being called "AI factories"—dedicated platforms that support the full machine learning lifecycle from experimentation to inference at scale. These facilities integrate machine learning operations (MLOps) frameworks, provide access to high-throughput storage, and offer optimized computing resources for model development. Community efforts and projects such as EuCAIF [ID167, ID185] and SPECTRUM are helping to define the functional and architectural requirements of these platforms within the context of HEP. The current on-premises computing infrastructure in HEP is often insufficient for the scale and flexibility that modern AI workflows require. This has prompted calls for shared, community-wide platforms that can support cross-experiment use and interface with broader scientific initiatives.

### 12.1.5 Specialized Analysis Infrastructure

As the volume and the complexity of data from HEP experiments increase, the analysis phase of the scientific workflow becomes more demanding and more specialized. Traditional batch computing models are ill-suited to the needs of modern analysts, who require fast, interactive access to large datasets and scalable processing environments. This shift has spurred the development of specialized analysis facilities that blend interactivity with elasticity. Elastic Analysis Facilities (see [ID237] for example) offer on-demand compute capacity tailored for user-driven workflows. These platforms are being designed with a strong emphasis on usability, featuring seamless access to data, low-latency storage, and intuitive interfaces that reduce the technical overhead for users. The goal is to create environments that support both novice and expert users, enhance scientific productivity, and enable efficient sharing of infrastructure across experiments. This is particularly important in the context of future colliders and multi-experiment facilities, where interoperability and reuse will be key to maintaining sustainability. Emerging priorities also include support for data preservation, reinterpretation of legacy analyses, and reproducibility of scientific results. These considerations highlight the need for long-term planning and investment in analysis platforms that go beyond immediate operational needs.

### 12.1.6 Interdisciplinary and Industry Collaborations

The future of HEP computing should not be considered in isolation. As computing becomes increasingly interdisciplinary and cross-sectoral, collaboration with other scientific domains and with industry is essential to maximize impact and ensure long-term sustainability.

Several strategic directions are emerging in this area. Shared infrastructure planning across domains such as Lattice QCD, astroparticle physics, and radio astronomy can improve efficiency and enable new scientific synergies. Interoperable data services, promoted by initiatives like SPECTRUM and ESCAPE [1020], are enabling more seamless cross-site access. FAIR data principles (Findable, Accessible, Interoperable, Reusable) are being adopted as a baseline for data stewardship across the research ecosystem.

Moreover, the HEP community should actively explore the formation of EuroHPC Centres of Excellence (CoEs) that would provide focused support for scientific computing, including for HEP and Lattice QCD applications. Projects like EVERSE [1021], which aim to improve the quality and reproducibility of research software, further underscore the importance of a professional, well-supported software ecosystem.

These efforts rely not only on technological alignment but also on governance, policy coordination, and shared investment. Interdisciplinary and industry partnerships offer a path toward a more integrated and resilient research computing infrastructure.

The HEP domain, given its history of excellence, should strive to obtain a central role in these multi-disciplinary collaborations, and have its role recognized by policy and funding entities.

### 12.1.7 Recommendations

- **Secure the WLCG continuity and evolution:** Ensure the long-term continuity, evolution and sustainability of the WLCG through dedicated support for computing R&D and operations, preserving it as the foundational infrastructure for current and future HEP programs.

- **Invest in next-generation infrastructure:** It is imperative that HEP attains increased relevance within the domain of high-performance computing, ensuring that its unique scientific requirements and contributions are recognized and prioritized in policy and funding decisions. This needs the concerted effort of CERN, of the experiments, and of the Funding Agencies. Invest in scalable AI platforms, sustainable computing practices, and the development of a federated, heterogeneous computing ecosystem that integrates general-purpose facilities such as HPC centres or Clouds, to meet evolving scientific demands and match the expected resource landscape.

- **Expand collaborations:** Actively pursue and broaden collaborations with other scientific domains, funding agencies, and industry to leverage shared expertise, drive innovation, and address the scale of future computing challenges.

## 12.2 Software

The inputs to the computing track evidence the immense software challenges High Energy Physics (HEP) is facing in the coming decade with the HL-LHC and in the longer term with future colliders. Largely driven by the unprecedented data volumes forecast, the community—from experimental collaborations and infrastructure providers to theoretical groups—agrees that a strategic and coordinated evolution of software is essential, with a focus on creating shared, common solutions. Initiatives like the HEP Software Foundation (HSF) [ID67] and JENA Computing [ID124] can play a crucial role in fostering and coordinating international efforts in HEP software and computing towards the adoption of common solutions. Alongside this, the community requires vastly improved career support and structures for the expert workforce that develops and maintains these critical tools.

### 12.2.1 Software landscape for the next decade - current colliders

**Evolution of resources and consequences for software.** There is a consensus that HEP computing must urgently transition to a hybrid model that integrates the traditional High-Throughput Computing (HTC) sites with High-Performance Computing (HPC) centres, and more in general with heterogeneous resources. This is motivated by the $O(10)$-fold increase in computing and storage needs for the HL-LHC.

Integrating HPC sites presents software challenges due to HEP tools having been developed over decades for a different distributed computing model. Whereas HEP software is traditionally optimised for x86 CPU architectures, modern HPC sites increasingly accommodate heterogeneous systems with GPU and FPGA accelerators within new architectures noted for their energy efficiency like ARM and RISC-V. The rapidly evolving hardware landscape—largely driven by AI—means no single technology is expected to be dominant for long, in contrast to the last decades of x86 CPU dominance.

For these reasons HEP software must evolve and transition to a heterogeneous model. A collective development and investment plan is needed to modernise the millions of lines of code within legacy codebases ensuring performance across diverse hardware without a need for frequent, extensive and expensive (and therefore probably not realistic) rewrites. A key aim here is achieving "performance portability"—developing code that runs with near native-performance across different architectures/accelerators; frameworks like Kokkos and Alpaka [ID107, ID124] are tools, among others, for writing applications with "performance portability" targeting all the hardware used in major HPC platforms. By isolating vendor-specific features (as these tools do) a single codebase can be maintained and detrimental vendor lock-in avoided [ID187, ID107].

To face the real-time analysis challenge of processing integrated data rates of $\mathscr{O}(10\,\mathrm{TB\,s^{-1}})$, it will be crucial to leverage the progress in commercial off-the-shelf (COTS) technologies [ID18] and the considerable experience in effective and efficient software design gained during ongoing experiments [1022–1024]. For this, software tools that support rapid adaptation to emerging hardware technologies are essential. These tools are best designed and maintained by a team of physicists, engineers, and computer scientists [ID67]. Future reconstruction and classification challenges, including 4D reconstruction for detectors providing both space and time information, can be addressed by exploiting a mix of classical and machine-learning (ML)/artificial-intelligence (AI) algorithms. Processing alignment and calibration tasks fast enough to fit into buffer requirements will become a challenge at future experiments as well. Here, the software tools for parallel architectures developed for reconstruction can be leveraged to minimise computation times.

Collaboration between the HEP community, funding agencies, and HPC providers is needed to ensure that the HEP software evolution is aligned with the global distributed computing resource landscape.

**The need of common software solutions.** To maximize efficiency and avoid duplicated effort, the inputs submitted to the ESPP endorse the development and support for common software solutions. Shared tools like Rucio [1025], PanDa [1026] and DIRAC [1027] for data and workflow management, and A Common Tracking Software (ACTS [1028]) and Coffea [1029] for reconstruction and analysis, are provided as successful examples of cross-experiment tools.

Shared, foundational HEP software — upon which more specific applications are built — like Geant4 [288] and ROOT [1030] are undergoing significant modernization to support heterogeneous hardware and are shifting away from traditional event loops towards more declarative, columnar analysis to reduce time-to-insight. Particularly when these tools are external to HEP, collaboration with the development teams is essential. A key example is Geant4's integration of the AdePT [1031] and Celeritas [1032] tools aimed at accelerating parts of the Geant4 simulation workflow on GPUs.

Event generators are another important category of software external to the experiments, but vital to their physics exploitation. Tools like Madgraph [1033, 1034] and Sherpa [1035] are investing quite some effort on GPUs and future-proofing and MC generators support and necessary development are essential for HEP current and future initiatives; among these it should be noted the critical problem of negative weights [1036] in event generation, which risks to increase by large factors the resources needed for event generation.

**The transformative role of AI.** AI/ML will move beyond offline analysis and become deeply embedded in every stage of the data processing and analysis pipeline. It is seen as a critical solution for mitigating the overwhelming computational costs across the entire data lifecycle. Key, recent advances include ultra-fast ML trigger algorithms running in real-time and reconstruction tasks like particle tracking and clustering. In particular, simulation remains one of the largest consumers of computing resources. While the Geant4 toolkit will remain crucial for high-fidelity simulation, fast simulation options are required for the vast samples needed. Therefore, alongside the continued optimization being undertaken for Geant4, experiments are aggressively developing and deploying fast simulation techniques which heavily employ AI/ML, including generative models and end-to-end “flash simulation" (see for example [ID107], [ID127]) and that can produce analysis-level objects directly from generator input. The community recognizes the need for a structured approach to AI research, proposing dedicated R&D groups and the development of domain-specific Large “physics" Models. AI is further explored in Section 12.3.

**Evaluating analysis requirements - benchmarking.** To determine the requirements for analyses over the next decade and develop software accordingly, a comprehensive assessment of current needs is essential. Initiatives like the Analysis Grand Challenge (AGC [1037]), organized by the Institute for Research and Innovation in Software for High Energy Physics (IRIS-HEP [1038]), are crucial for this purpose. The AGC provides practical “demonstrators" in the form of realistic, end-to-end benchmark physics analysis pipelines that use publicly available Open Data from CERN experiments. This allows stress-testing of new systems (eg. platforms, facilities etc.) and provides developers with real-life benchmarks for performance comparison and software optimization, as exemplified by the development of RNTuple by the ROOT team. This ensures that future software is suitable for the immense challenges of the High-Luminosity LHC.

**The HEP software ecosystem.** High-Energy Physics relies on a diverse ecosystem of software tools and has greatly benefited from open-source, community-driven projects such as Scikit-HEP [1039] and PyTorch [1040]. To fully exploit these tools in HEP workflows and allow flexible, innovative research there are two key requirements: seamless interoperability between software tools, and package-able and portable shared software environments. The HEP Software Foundation (HSF) ESPP input [ID67] is endorsed by the four main experiments of the LHC—ALICE, ATLAS, CMS, LHCb,—as well as by Belle II, DUNE, ePIC, MCnet and the WLCG.

The HSF has played a crucial role in fostering and coordinating international efforts in HEP software and computing and defining a common roadmap. The HSF promotes the investigation of interoperable tools and frameworks, and increases the user-base of emerging languages such as Julia [1041] that boasts significant performance benefits over Python and usability benefits over C++. Technologies such as containerization and environment management tools such as Conda-Forge [1042] that allow portable and versionable environments will remain crucial for exploiting distributed resources maximally and for analysis preservation.

Software licensing models are an important aspect when workflows need to be deployed globally, on owned, leased, and opportunistic resources. The four main LHC experiments have by now turned to Open Source derived licensing (either Apache 2.0 [1043] or GPLv3 [1044]); it is important that new initiatives choose this same direction from the beginning.

**Sustainable software.** The multi-decade lifespan of HEP experiments necessitates a focus on long-term quality assurance, software sustainability, and collaborative development. This is achieved through modern software practices, including robust training programmes and a development model based on Git repositories. Within this model, continuous integration pipelines automatically put proposed software changes through rigorous checks — from auto-mated compilation and unit tests to large-scale physics workflow validation — before it can be merged. HEP should continue to adopt such industry-standard software development practices as they evolve.

**The critical role of software training.** The expertise required for modern (HEP) software now extends far beyond the scope of a typical physics degree, creating a critical need for robust and specialized training programmes. For these programmes to be effective and sustainable, HEP collaborations must provide clear incentives and formal recognition for the experts who deliver and maintain the training materials.

While individual experiments must maintain specific onboarding programmes, these should be supplemented by shared community resources and events. High-quality training programmes are maintained by organizations like the HSF [1045] and The Carpentries [1046] on a range of software topics. See Section 12.6 for additional details.

### 12.2.2 Software landscape in the longer-term - future colliders

The future collider projects of HEP are also facing significant software challenges, albeit with a slightly different perspective due to the much longer time scale involved for any follow-up project to the HL-LHC. On the one hand the community will most likely have already addressed many of the challenges as sketched out above - on the other hand there are two additional challenges the future collider community has to address: a complete and fully functional offline processing chain for all major proposed colliders (FCCee/hh [ID211], LCF [ID78,ID102], MuonCollider [ID207], ...) with detailed simulation models for the envisioned detectors is needed basically now and at the same time one has to build-in the flexibility to adapt to — potentially transformative or even revolutionary — developments of novel technologies.

**The Key4hep software stack.** The HEP community has actually addressed the need for a suitable offline processing chain through the Key4hep [1047] project that, initiated in 2019, aims at providing a complete software ecosystem for the development, study and optimisation of future collider detector concepts. The Key4hep project takes the idea of common software projects in HEP to a whole new level by providing one large common software stack that is used (and developed) by all future collider projects (CEPC, CLIC, FCCee, FCChh, ILC, MuonCollider and others) collaboratively. Key4hep itself is not re-inventing the wheel but utilizes the best-practice tools available and under development in the HEP community (see e.g. 12.2.1). At the same time there are a number of specific software tools targeted at future colliders most prominently the EDM4hep [1048] event data model that are developed by a relatively small group of people. As pointed out in [ID240] the person power for Key4hep project, that is central for all future collider projects that are currently discussed in this strategy update process, is threatening to become sub-critical. This is partly due to the general problem of career opportunities in HEP software R&D and partly due to the general difficulty of getting funding for research software engineering.

**Integration of and adaptation to novel technologies.** Looking back at the pre-LHC period, it is clear how the computing landscape has evolved. LHC preparation started with single core, 32 bit Intel processors on uniform Grid nodes, on which single thread processes were executed. Some 20 years later, WLCG is best described as a *system of heterogeneous systems* [ID9], where multi-threaded processes are scheduled on multi core CPUs, GPUs and FPGAs from dedicated (owned) centres, private/public clouds and HPC systems. We can expect a similar evolution in the next decades, with even more disruptive technologies expected to be relevant (quantum computing as a clear example). It is of paramount importance that HEP computing and in particular the future collider community stays open and flexible to adapt new technologies - as described in Section 12.5, in order to profit from the best performance at the best price level.

### 12.2.3 Recommendations

– Encourage the adoption of portable programming models and abstraction layers as needed to leverage the computing power provided by HPC centres, thus ensuring HEP software can run efficiently across diverse and rapidly evolving architectures.

– Foster collaboration and continue to develop/support common software solutions that serve multiple experiments (even from outside the HEP domain) to maximize efficiency and avoid duplicated effort across the experiment pipeline from simulation to data management.

– Strategically develop and integrate AI/ML technology with AI-driven fast simulation and ML in real-time triggers and reconstruction through structured, collaborative R&D programmes.

– Ensure long-term preservation of entire software ecosystems — code, workflows, environments, statistical models — leveraging technologies like containerization and workflow management systems to adhere to FAIR principles.

– Prioritize environmental and economic sustainability in HEP computing, developing more efficient algorithms, optimizing software for energy-efficient hardware, establishing common metrics for power consumption, and designing cost-effective computing models.

– Ensure that HEP institutes provide sufficient resources for the research software engineering on the common software tools and projects that the HEP community is relying on (e.g. Key4hep, ACTS).

– Ensure the longer term maintainability and evolution of HEP software and computing infrastructures, through structured software R&D projects similar to the DRDs for detectors, with funding and/or personpower commitments.

## 12.3 AI

Artificial Intelligence (AI) is reshaping the methodology and capabilities of fundamental physics. High Energy Physics (HEP), with its large data volumes and long-standing tradition of developing advanced analysis techniques, serves as both a testbed and a driver for AI innovation.

This section provides a strategic overview based on ESPP inputs and symposium discussions, focusing on: (1) the AI landscape for current and future colliders, (2) community needs to maximize AI's impact, (3) Strategic recommendations for ensuring leadership in this transformative field.

### 12.3.1 AI Landscape in High Energy Physics

**The Next 10 Years (2025–2035): AI at Current Colliders.** Over the past decade, AI has moved from peripheral to central in collider workflows. In ALICE, ATLAS, CMS, LHCb, and Belle II, deep learning models now outperform traditional techniques in event classification, reconstruction, simulation, and triggering. AI is already mature and indispensable in HEP:

- **Classification and Tagging:** GNNs, transformers, and end-to-end classifiers have significantly advanced jet and flavour tagging, increasing sensitivity in Higgs and rare decay analyses [1049, 1050]. Models trained on simulations and refined with Run 3 data are deployed in real-time systems.

- **Offline Reconstruction algorithms:** CNNs, transformers, GNNs and in general AI-driven algorithms are used extensively as substitutes of classical algorithms, aiming at both physics performance and reduction of computing needs . Their scope varies from the piece-wise replacement of specific target codes, to a full redesign of time critical algorithms such as tracking or calorimeter reconstruction [ID187, ID107, ID127] [1051].

- **Simulation and Generative Modelling:** GANs [1052], normalizing flows, and diffusion networks [1053], achieve 10–100x faster simulation with percent-level accuracy. These surrogates are integrated into detector subsystems (e.g., calorimeters, tracking), supporting design and unfolding tasks [1054–1056].

- **Triggers and Real-Time Systems:** low-latency AI models on FPGAs/ASICs enable jet tagging and anomaly detection within O(100 ns) [1057]. Such embedded models underpin modern triggers and pave the way for near-triggerless architectures at the HL-LHC and FCC [1004, 1058].

- **Inference and Unfolding:** simulation-based inference and AI-driven unfolding enable high-dimensional, un-binned parameter estimation and robust multidimensional corrections, improving both precision and speed [1059–1061]. End-to-End Learning – Models trained on raw detector data (e.g., calorimeter hits) preserve fine-grained features, directly optimizing physics sensitivity and supporting detector design optimization.

These applications illustrate that AI in HEP is beyond the exploratory stage, forming a foundation for more ambitious paradigms.

**Beyond 2035: AI for Next-Generation Colliders.** Looking beyond the HL-LHC era, the challenges of future facilities such as the FCC-ee and FCC-hh will likely place AI at the very heart of experimental design and operation. The expected data rates, reaching exabyte scales, will make traditional workflows of collection, storage, and offline analysis no longer be viable. Instead, machine intelligence could be embedded directly into detectors and data pipelines.

One direction under active exploration is the development of **intelligent frontends**. Here, ML-based feature extractors integrated on ASICs or smart pixels act as gatekeepers, filtering and compressing data streams at the detector level. Early prototypes, already demonstrated in CMOS technology, show that convolutional or transformer-like architectures can operate at sub-100 nanosecond latencies, suggesting that real-time AI within detectors is not only plausible but necessary.

At the algorithmic level, the emergence of **foundation models** marks a paradigm shift. Inspired by large language models in NLP, these networks would be pre-trained on massive and diverse physics datasets—jets, calorimeter images, time series—before being fine-tuned for specific tasks such as classification, regression, or simulation [1062]. Such models could provide universal representations of events, transferable across experiments or energy regimes, drastically reducing the need for experiment-specific retraining and accelerating innovation.

Complementing this vision are so-called **Large Physics Models (LPMs)**, which aim to merge neural architectures with symbolic reasoning and physical constraints. If realized, these models would not simply classify or regress; they could act as co-pilots in scientific discovery, suggesting hypotheses, designing searches, or even interacting with physicists through natural language and visualization. While still speculative, they embody a longer-term aspiration: AI systems that actively participate in knowledge generation rather than passively supporting it.

Another promising frontier is **differentiable programming**. By making entire detector simulations differentiable, the geometry and readout of detectors could be optimized through gradient descent, directly targeting physics performance. This approach [1063] has the potential to revolutionize how detectors are conceived, effectively merging design and physics optimization into a single computational loop.

Taken together, these trends suggest a future where AI is not just a tool for analysing data, but a foundational component of detector technology, data acquisition, and possibly part of the conceptual framework of discovery itself.

### 12.3.2 Needs of the HEP Community

To move from today's successes to this more ambitious future, the HEP community must invest in both **infrastructure** and people. Modern AI requires significant computational resources: the training of large-scale foundation models, or even high-fidelity generative simulations, demands access to GPU clusters, AI factories, and HPC centers. These must be adapted to the particular requirements of physics workloads, which mix large-scale simulation, low-latency inference, and distributed training. The adoption of federated learning will also be critical, allowing models to benefit from datasets that remain distributed across experiments, thereby preserving privacy and reducing duplication of effort.

Equally important is the **software ecosystem**. While tools such as hls4ml [1015], or Brevitas [1064], have enabled deployment of ML models in resource-constrained environments, the community still lacks standardized, well-maintained libraries for key tasks such as quantization, pruning, or graph-based inference. Experiment-agnostic workflows will be needed to make models portable across detectors and facilities. To ensure comparability and reproducibility, **shared benchmarking datasets and evaluation metrics** must be established, forming the methodological backbone of AI in HEP.

**Human capital** is perhaps the most critical element. Dedicated training programs, from summer schools to PhD courses, are needed to prepare the next generation of particle physicists fluent in AI. Equally, mechanisms for career development must recognize AI specialists, whether their expertise lies in hardware or software. Without clear career paths, the community risks losing talent to other fields. Cross-disciplinary collaboration with computer science and industry should be encouraged, to import methodological advances and to attract diverse expertise. At the same time, inclusivity must be ensured so that the benefits of this transformation are broadly accessible.

### 12.3.3 Recommendations

- **AI should be institutionalized as a crucial component for current and future HEP experiments**, treated on the same level as detector development or theoretical modelling, and integrated from the outset into detector design, simulation, and analysis. As such, AI research for HEP should be tackled with similar organisational structures, for example by defining AI-RD common efforts. The inputs from EuCAIF [ID167,ID185] suggest a possible direction.

– Future investment must prioritize **low-latency AI and smart instrumentation**, embedding intelligence into detectors and DAQ systems, and should also support HEP-specific foundation models, pre-trained on diverse datasets and transferable across experiments.

– A culture of **open science** is essential: models, datasets, and benchmarks should be shared through centralized repositories to ensure transparency and accelerate innovation. At the same time, AI in HEP should remain **interpretable** and **physics-informed**, embedding symmetries and conservation laws in intrinsically interpretable AI models.

## 12.4 Computing Resource Needs and Projections

This section has two primary components. First a comparison of anticipated computing needs for potential future collider experiments at CERN, based on current physics programme projections as well as inputs on the computational and storage requirements from experimental experts is made. Second, themes identified in contributions related to future computing resource needs, as well as potential requirements to be considered in resource planning beyond total resource need as is currently tracked are identified.

In recent years, the increase in computing resources has been constrained to the so-called "flat budget model", in which the yearly monetary contribution for computing stays at most constant. In the scenario, a steady increase of computing resources relies on the constant decrease of unitary costs for hardware resources[1]. While in the last $\sim$10 years the year-over-year decrease was considered of the order of 20% for storage and CPU resources, the trend has significantly slowed down, with today's best estimates of the order of 10% if not less [1066]. For illustrative purposes, a 20% decrease gives on the time scale of 10 years a "flat budget increase" of $\sim$ 6x, while a 10% only $\sim$ 2.5x.

### 12.4.1 Computing needs for potential future collider experiments

The primary focus for computing at CERN over the next decade is the HL-LHC. Inputs from the CMS [ID107] or ATLAS [ID187] experiments did not include updated HL-LHC projections for computing requirements. Thus the most recent computing projections are from 2022 [1067, 1068], and are shown in Fig. 12.1.

ALICE 3 [ID68] and LHCb Upgrade II [ID127] are planned upgrades planned for LHC Run5, still in the phase of approval. As such, the computing modelling is at an early stage, and was discussed with the two collaborations outside official ESPP inputs. In both cases, even in the presence of important increases in the instantaneous luminosities and the interaction rates (a factor $\sim$ 7x for LHCb and about 25x for pp collisions in ALICE), the current modellings show no large divergences from a flat budget scenario with reasonable year-over-year decrease parameters.

All the projections indicate that a significant R&D programme is needed to achieve the baseline physics programme within an estimated flat-budget planning scenario. Missing that, factors in the resource needs should be expected, to be solved either by increasing the funding level, or by decreasing the physics scope.

Matching the computing requirements and needs of the HL-LHC physics programme, and the R&D required to get there is one of the main priorities of the field.

[1] Specifically for CPUs, this is usually referred to as Moore's law [1065].

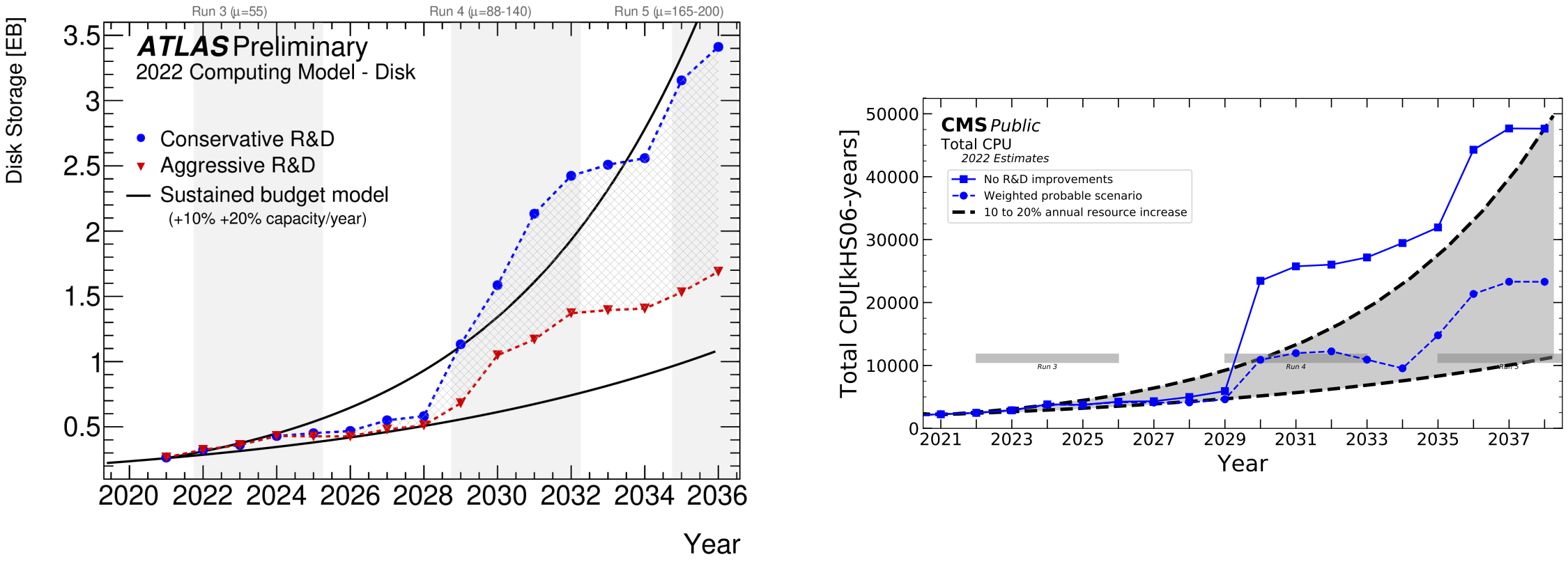

Fig. 12.1: Projected Disk needs for the ATLAS Experiment at HL-LHC, from Ref. [1068] (left). Projected CPU needs for the CMS Experiment at HL-LHC, from Ref. [1067] (right).

Computing related inputs from each of the prospective facilities beyond HL-LHC have been analysed. For the most part the analysis has been limited to summarizing and extrapolating based on inputs from proponents and thus resource needs are estimated with somewhat differing approaches from initiative to initiative; this is to be expected given the time scales involved.

The process is summarized in Table 12.1, and a number of dedicated comments expanding on the process leading the table values are listed as follows:

The **FCC-ee** [ID233] computing requirements are driven by statistics of the tera-$Z$ programme, which is currently planned for six trillion $Z$ events across four experiments; all the rest of the physics programme needs much smaller computing resources in comparison. Current projections [ID233] include storing O(1 MB)/event for both data and Monte Carlo simulations, corresponding to expected raw data, or simulation hit level, event sizes. As such they are quite conservative and can be considered as tape storage requirements. Future technology costs can then inform the most affordable alternative of storing detailed simulation results vs recomputing them if needed. Having the computational needs driven by a modest fraction of the overall programme does mean that the scale of owned compute resources strongly depends on the running programme. For example, collecting and processing a full running year of $Z$ data would require more compute than if the run was interspersed with other data taking. If compute resources are well-suited to bursts in demand (e.g., HPC, cloud), then this is less of a planning consideration.

Another driving requirement is the ratio of Monte Carlo simulated events to data events. The LEP programme had a much higher ratio (10x) than is the case for LHC. FCC-ee projections are considered for up to 10x simulation events, which can be justified by the precision physics programme requirements.

The **LEP3** resource needs are again driven by the scale of the $Z$-pole run. Projections were made using the FCC-ee project, scaling them down by the ratio of expected integrated luminosity of LEP3 vs FCC-ee at the $Z$ pole [ID188].

The **Linear collider** computing requirements are modest in comparison either FCC-ee or LEP3. A comprehensive analysis of requirements, including replication and other factors not yet considered in other estimates, was made in Ref. [1069, 1070]. The running scenarios expected have evolved somewhat since these studies [ID40]. Resource requirements have been scaled up according to the respective increases in integrated luminosity.

The **LHeC** resource needs assume a streaming readout without any hardware triggering.

This is expected to be possible due to the relatively modest event rate. An event rate to storage of 10 kHz is assumed, independently of the delivered luminosity, corresponding to being able to retain events with $Q^2 > 10$ GeV$^2$ at a luminosity of $10^{33}$ cm$^{-2}$s$^{-1}$. Based on current expectations, an event size of around 100 kB and a per event processing time of 100 HS06 $\cdot$ s are assumed [ID214].

An extensive resource estimate of the **10 TeV muon collider** can be found in Ref. [ID207]. Controlling and reducing beam induced backgrounds is critical, and modelling these backgrounds is resource intensive. A trigger design that is adequate for reducing the data rate by 1000x from the full rate is assumed to be achievable, but it is not yet reported. As with the FCC-ee estimates, storage estimates depend considerably on whether detailed simulation information should be retained or not.

The **FCC-hh** [ID247] resource projections are made scaling from HL-LHC projections after making corrections for the event rate and pile-up. The event rate from the high-level trigger is estimated to be 100 kHz, or 10 times the expected HL-LHC rates, if the current LHC trigger thresholds on physics objects are retained. The expected per bunch crossing pileup is approximately five times that of HL-LHC. Including the effects of an 85 TeV beam energy, this corresponds to approximately 1700 overlaid 14 TeV interactions (10x HL-LHC) in terms of total tracks, for an overall factor of 100 times with respect to the HL-LHC computing requirements, with a linear increase in computing with pileup assumed. Should the compute requirements scale non-linearly with pileup, new approaches to tracking and other reconstruction algorithms are necessary. Among them, gains from timing detectors are potential large since these can potentially reduce the effective pile-up seen by reconstruction algorithms, by removing hits from pileup interactions before the events are analysed offline.

#### 12.4.2 Additional inputs related to computing resource needs and modelling

Looking forward, given the evolution of analysis techniques, computing hardware and experimental timescales, several new dimensions into resource projections/modelling in the near future need to be incorporated. One important example is the evolution of the necessary scale of resources for AI/ML in production and analysis applications. Requirements for both inference and training need to be included. Estimations, e.g., benchmarks, and tools to facilitate the optimization of energy footprints and environment impact both by sites and experiments is increasingly important. I/O requirements are increasingly important given the slower evolution in I/O relative to total capacity in storage. This includes storage I/O, in particular tape bandwidth requirements, as well as network requirements (e.g., WA or transatlantic).

#### 12.4.3 Recommendations

- The scale of HL-LHC computing is not a solved problem given the expected (at best) flat budgets for computing. Thus, the short term, e.g., five years from now, is already a challenging future. **A significant R&D programme in order to exploit state-of-the-art technologies is needed to ensure the achievement of the physics goals for HL-LHC and a timely delivery of the results, while staying within an estimated flat-budget planning scenario**. Missing that, factors in the resource needs should be expected, to be solved either by increasing the funding level, or by decreasing the physics scope.

- Beyond HL-LHC, the initiatives under consideration that would take place at the end of the 2040s **cannot be differentiated based on their feasibility in terms of comput-**

| | HL-LHC (GPE) | LC | FCC-ee | LEP3 | LHeC | 10 TeV muon | FCC-hh |
|---|---|---|---|---|---|---|---|
| Programme | $ab^{-1}$, Number of physics events | | | | | | |
| $pp$ 14 TeV | 3, 1$e$12 | | | | | | |
| $pp$ 85 TeV | | | | | | | 40, 2$e$13 |
| $ee$ $Z$-mass | | 1, 3$e$10 | 205, 6$e$12 | 48, 2$e$12 | | | |
| $ee$ $WW$ | | | 19, 2.4$e$8 | 5.6, 4$e$7 | | | |
| $ee$ $ZH$ | | 2, 4$e$5 | 11, 2.2$e$6 | 2.4, 2.4$e$5 | | | |
| $ee$ $tt$ | | 0.2,– | 3, 2.5$e$6 | | | | |
| $ee$ 550 GeV | | 8,2e6 | | | | | |
| $ep$ 1.2 TeV | | | | | 1, 70$e$9 | | |
| $\mu\mu$ 10 TeV | | | | | | 10, 1$e$10 | |
| Potential start | 2030 | 2042 | 2045 | 2045 | 2042 | 2050 | 2055 (or 2070) |
| Length [Yrs] | 12 | 25 | 15 | 20 | 6 | 5 | 25 |
| Total CPU [MHS23-y] | 500 | 0.5 | 110 | 30 | 4 | 200 | 100000 |
| Ave annual CPU [MHS23-y] | 40 | 0.02 | 7 | 1.5 | 1.5 | 40 | 4000 |
| Total storage [EB] | 20 | 1 | 8 (40) | 10 | 0.12 | 0.7 (29) | 4000 (in yr 2100) |

Table 12.1: Comparison of compute and storage requirements of future CERN collider experiments. Luminosity, CPU and storage are quoted as the sum over all planned IPs.

**ing requirements**. Assuming a constant and intense funded R&D programme, they fit extrapolations from the LHC model.

– Longer term facilities have larger computation needs. In particular, FCC-hh would be not be "easy" given a possible advanced (2055+) timescale, but is **considered feasible given adequate funding and personpower investment**. The R&D programme of HL-LHC is a critical first step towards a better understanding of the needs of a future hadron facility. The simple fact that today, some 30-45 years from its expected start, resource needs can be modelled and are not too far off, should be considered a huge success.

– Expectations for experimental needs (and ambitions) tend to be underestimated. Research that goes into making experiments reality opens new opportunities and creates new demands. Taking advantage of new techniques and technologies (today's examples include AI/ML and heterogeneous systems) **will facilitate more ambitious programmes to be implemented as facilities are realized**.

## 12.5 New Technologies

High Energy Physics (HEP) has consistently embraced emerging technologies to advance scientific discovery and demonstrated a strong capacity for technological innovation. Recent examples include software-based triggers on GPUs in LHCb [1071], GPU-accelerated high-level triggers in CMS [1072] and ALICE [1073] and the integration of artificial intelligence in data analysis in all the experiments. Today, quantum computing (QC) offers a fundamental new way to process information and solve complex problems, however, other new technologies can be expected to emerge over the coming decades. Alongside the theoretical efforts in quantum computing to identify problems that can only be solved using quantum algorithms on fault-tolerant quantum computers, there is growing interest in short-term exploration of hybrid classical-quantum systems in which quantum computers serve as hardware accelerators for specific tasks within a larger classical framework. This approach is in accordance with the theme of building future data centres based on heterogeneous architectures. As HEP experiments grow in scale and complexity, the demand for novel computational paradigms grows. At the same time, the integration of possibly disruptive technologies, such as quantum computing, in the HEP computing model requires a robust understanding of both the technological potential and its limitations; through a thorough validation and performance assessment across realistic applications and careful planning. A long term research plan on this topic is therefore required in order to fully evaluate, and then exploit, quantum technologies in HEP. Past experience with adoption of new technologies in HEP has shown (e.g. for HPCs), that it takes at least a decade for a full integration of these in the HEP workflows. It is therefore essential that the new technologies are explored as they emerge.

### 12.5.1 The global quantum technologies landscape

Quantum technologies – including computing, sensing, and communication – are now recognized as strategic sectors by governments and research institutions worldwide. Numerous national and international initiatives are underway, reflecting the broad relevance of quantum technologies. CERN has launched its own Quantum Technology Initiative (QTI [1074]) to coordinate efforts and foster collaboration across its member states. This initiative aims to allow CERN and the HEP community to play a leading role in quantum research for its application to particle physics, and ensure alignment with global developments and maximizing impact.

The HEP community is already exploring and actively contributing to several quantum computing paradigms:

- **Gate-based quantum computers**: Systems that perform computations using quantum gates, enabling the execution of complex algorithms through unitary transformations.
- **Quantum annealers**: Designed for optimization problems, these devices embed problems into Ising or QUBO [1075] models and solve them using static qubit connectivity.
- **Analogue quantum simulators**: These simulators model quantum systems using dynamic qubit arrangements, though they often lack fine-grained control.

In addition, quantum-inspired algorithms – classical methods [1076] influenced by quantum principles – are being developed to harness quantum-like advantages on conventional hardware, particularly in machine learning and optimization tasks.

Given the current limitations of quantum hardware, such as short coherence times, limited qubit counts, and restricted connectivity, hybrid quantum-classical algorithms are essential. These combine quantum processors with classical infrastructure (e.g., cloud computing, high-performance computing). Techniques such as data compression, error mitigation, and pre/post-processing allow researchers to extract results from near-term quantum devices. This hybrid model is already being adopted within the HEP community [1077, 1078].

### 12.5.2 Quantum Algorithms in HEP Applications

Quantum algorithms offer promising solutions to several HEP challenges. A wide range of algorithms has been identified as potentially useful to the community and it covers applications in HEP data processing and theoretical simulations [1079]. Potential improvements range from computational resource efficiency to discovery capabilities. Examples already proven within the community include:

- Quantum associative memory could enhance the execution speed of particle tracking by leveraging exponential memory scaling [1080].
- Quantum reinforcement learning has shown potential in optimizing beam steering in particle accelerators while reducing by several order of magnitude the size of the trainable agents and the number of optimization steps [1081].
- Quantum anomaly detection may improve sensitivity in identifying rare events, contributing to discovery potential [1082].

These applications demonstrate how quantum computers can provide resource efficiency and new analytical capabilities that are difficult to achieve with classical methods alone.

Similarly, theoretical HEP faces significant computational barriers, for which quantum computing could provide a solution, particularly in simulating quantum systems:

- First-principles simulations of nuclear interactions become intractable as the number of quarks increases.
- Dense matter studies and the phase diagram of strong interactions remain largely unexplored due to classical limitations.
- Real-time dynamics, transport properties, and structure functions are not fully accessible with current classical tools. Quantum simulation offers a pathway to overcome these challenges, enabling more accurate and scalable modelling of fundamental physical processes.

### 12.5.3 Recommendations

- Invest in targetted R&D efforts and structured programs in quantum computing, aimed at enhancing the efficiency of solving computationally intensive problems and enabling the analysis of complex systems that remain intractable using conventional methodologies.

## 12.6 Careers

Already in the 2020 ESPP Briefing Book [1083], attention was given to the problems of careers in computing for High Energy Physics, and in general to the need to preserve computing oriented expertise in our domain. The situation, as described in the report, was related to "a flawed academic hiring model in which data analysis contributions are given more weight than instrumentation and/or computing-related research activities". As evident in a large number of inputs, the computing for the next generation of HEP experiments is becoming more complex, more heterogeneous and mode specialized, and a fruitful physics exploitation of the multi billion instruments needs proper R&D, and especially people who realize them. The evolution since ESPP 2020 is not perceived as positive, with the same reported problems and small (if any) improvements. On the negative side, it was noted how our computing experts are increasingly attractive to the private sectors, due a convergence of methods and solutions between the two domains.

The future of the field needs the capability to attract brilliant young physicists with computing skills and aspirations, and to retain a fraction of them for the continuous needs of our multi-decennial initiatives. To date, the importance of computing towards the physics exploitation and the expertise and efforts needed to properly design and operate computing systems are at least on par with activities such as the design and operations of major detector instruments or of critical data analyses, and as such should be valued by the community and rewarded with tenured positions and career paths, in Universities and Research labs.

A proper way to attract young researchers to the field would be to organize extensive and high-level training opportunities, to a dual extent: to form a generation of computing-savvy researchers as needed by the initiatives, and to attract collaborators initially not interested in the specific field, by showing the research opportunities and the highly relevant technical directions.

### 12.6.1 Recommendations

- Create high-profile training opportunities to attract young researchers to the computing and software domains, also in collaboration with academia;
- Strive for a mindset change in the approach to computing, recognizing that the current lack of support via high-profile positions for computing experts in academia and research institutions requires shifting from short-term reliance on their services to a sustainable model that recognizes and empowers them as indispensable drivers of research excellence.

## 12.7 Long Term Data Preservation

The collision data from LHC, expected to end its operations in 2041, will remain unique for many decades and its physics potential will most like be continuously exploited well beyond the end of the collisions. It has been shown that preserved data from previous high energy colliders enhances significantly the physics reach of their respective scientific reach, allows for further collaborations and reinforces the follow-up programs [1084]. These data can be exploited for new ideas, re-analysed or re-interpreted, used in the design of new experiments, combined with complementary data, and potentially used in education and outreach [ID138, ID150]. For instance, converting LEP data to EDM4hep, the data model of the Key4hep software stack [ID269] allows re-analysis in an FCCee compatible framework.

A proper data preservation (DP) process needs to address much more than the digital files. It has to preserve a vivid computing environment, analyses workflows, the software ecosystem, and most importantly the knowledge needed to analyse the data, for which a continued form of collaboration has to be considered as well. Moreover, a comprehensive DP approach should directly support the HEP community's commitment to Open Science and the FAIR[2] data principles. Preserving physics-level access should be planned upfront, ideally during the phases of design and commissioning of experiments, and pursued as a specification during the experiment lifetime. However, the DP upfront planning during the design and the running has not happened at a satisfactory level for past experiments, including the LHC experiments. A DP project for LHC should be implemented and adequately funded while the experiments are in operation, while the necessary computing, detector and analysis expertise can be used to evolve the complex environment towards a stable long term solution. The international collaboration, covering past and future projects, plays a crucial rôle by bringing experiments' DP expertise together and aiming for harmonized and therefore sustainable approaches.

[2]Findable, Accessible, Interoperable, Reusable

### 12.7.1 Recommendations

– Allocate identified resources and define projects at experiment, laboratory and international levels to support data preservation in a efficient and realistic manner for present experiments (in particular LHC) and future projects. Support a structured international cooperation such as DPHEP, with CERN as the host laboratory.

## 12.8 Sustainability

The energy demand of computing centres worldwide has grown significantly in the last 5 years. This growth in demand is expected to continue and even increase also due to the AI revolution[3].

In the field of High Energy Physics, the environmental sustainability of the computing centres (in terms of energy and environmental resources such as $CO_2$) is becoming a central issue, exacerbated by electricity price increase and availability crisis triggered by recent global events. In absolute terms, the energy impact of computing for a large laboratory such as CERN is shadowed by larger consumers (the accelerator complex in primis), and contributes to less than 5% to the total CERN energy budget [1085]; if the CERN WLCG fraction is evaluated to $\sim$ 20% of the overall computing deployment, the computing is responsible for $\sim$ 25% of the total energy budget of the LHC. While not predominant, the contribution is sizable and should be measured and moderated. This effort goes hand in hand with the need for maintaining future computing requirements for CPU and storage within budgets, starting from HL-LHC as described in Section 12.4.

Software is evolving, as described in Section 12.2. Monitoring the operational energy and resource consumption of running software (including AI algorithms) using common metrics is recommended, as it allows for optimization towards energy efficiency e.g. using energy-efficient hardware or by making dedicated choices in computing models.

Importantly, sustainability must be considered across the full lifecycle of computing beyond the operational environmental costs of running software: from hardware procurement and infrastructure planning to daily operations and decommissioning. The transition to more sustainable computing is a multi-dimensional challenge that intersects with funding models, hardware choices, software design, and facility governance.

Energy- and $CO_2$[4]- sustainabilities are related, but along different dimensions, so clear goals must be defined for each aspect. In a scenario in which in a few decades most energy production will be from renewable sources or nuclear, the two aspects could decouple; recent extrapolations [1086, 1087] predict that by 2050 renewable sources plus nuclear are expected to account for the vast majority of electricity and a substantial share of overall energy supply, with fossil fuel use shrinking to about one-fifth.

[3] IEA (2025), Energy and AI, IEA, Paris, Licence: CC BY 4.0.

[4] $CO_2$ is one among many indicators of natural resource depletion, see the European Commission's Green Forum website

Actions towards a concerted plan for HEP to measure and moderate the environmental impact of computing have been discussed in the WLCG Strategy for 2024-2027 [ID9]:

- WLCG to agree on metrics and provide a framework to collect information related to energy efficiency.
- WLCG to facilitate the use of more energy-efficient hardware where possible, depending on the readiness of the experiment software and the common libraries.
- WLCG to develop and promote a sustainability plan to improve energy efficiency and reduce carbon footprint, covering software, computing models, facilities, and hardware technology and lifecycle.

Starting from October 2025, the WLCG has launched a WLCG Environmental Sustainability Forum to discuss and act on the above recommendations.

## 12.9 Conclusion

Computing for High Energy Physics has been extremely successful in supporting the physics operations of the current generation of experiments. Its reach and capabilities far exceed the initial pre-collision design, from all the possible points of view: the capability to deploy physics workflows on a variety of heterogeneous resources, much improved software stacks enabling unprecedented capabilities. Resource needs reduced with respect to a simple back of the envelope extrapolation from the initial models. This was possible due to large investments in technology, ideas and people. The simple fact that today we are able to imagine and model the computing from experiments 50 years in the future is a great achievement, which should not be understated. The future, still, poses important challenges which cannot be met with a simple scaling in operations; at least not in a scenario of constrained economic resources. The past history shows that in order to be functional to the physics exploitation of multi-billion initiatives, computing needs to stay on the top of the technology curve, dedicating effort to continuous R&D processes which in the long term generate large savings OR more physics opportunities (think of the use of GPUs, FPGAs, AI, and their impact on HL-LHC resource needs). **This is possible only in the presence of intense R&D funded activities, without which there is no guarantee computing will be able to support properly the new initiatives within an economically viable budget**. AI retains a particularly important role, and its impact is today probably underestimated; regulating the AI utilization in HEP via common initiatives is highly recommended. For all the R&D under discussion, aimed towards initiatives potentially far away in time (2050 and beyond), even technologies which do not seem to be directly viable in the next decade (like Quantum Computing) can become relevant, and its R&D should be funded. **We are a large community, and blue sky R&Ds should be possible without disrupting critical activities**. Computing in HEP and in LHC in particular (via the WLCG Collaboration) has gained a prominent position in scientific computing, with its large scale distributed system and the ability to execute workflows on a variety of heterogeneous systems. In the near future scientific initiatives from nearby domains are expected to reach a similar level of complexity, and at the same time national, continental and global initiatives are attacking the computing problem from different angles (for example, using supercomputers); **joining forces is imperative to guarantee the sustainability of our field**. Activities, even if properly funded, need person-power (and young and enthusiastic researchers in particular) to be executed. We still see a low commitment level from academia and research institutions in attracting and supporting and offering long term career paths to computing experts; **in the long term, this is more worrying than technological trends, and finding solutions should be a field priority**.

# Acknowledgements

We would like to thank all of those who contributed to the preparation of this Physics Briefing Book, either by providing plots, tables or text, but who do not appear as authors. The organizational effort provided by the CERN Council Secretariat, Vedrana Zorica and Dylan Owen, is gratefully appreciated. The Open Symposium in Venice provided essential input and was excellently prepared by the local organizing team led by Sandra Malvezzi. Support from CERN and the funding agencies for the Open Symposium and Physics Preparatory Group meetings is gratefully acknowledged. The Electroweak Working Group would like to thank the following groups and institutions for generously providing computing resources for our use: the Datacloud INFN team at LNF, the CERN-TH group, the DESY-TH group, and the Kirchhoff Institute at Ruprecht-Karls-Universität Heidelberg.

# Appendices

## A Technical details of the Standard Model Effective Field Theory

Effective theories provide a simplified description of physical systems which admits a separation of scales, focusing on the dynamics of the relevant degrees of freedom at a given scale. This assumptions is applicable to the descriptions of extensions of the SM involving high-energy states, as long as these are too heavy to be produced in experiments. The SM is expected to be completed by a more fundamental theory with new degrees of freedom able to provide answers to the open questions. No clear evidences for new states has been found yet by the direct searches performed at the LHC, hence it is natural to assume that these states lies well above the electroweak scale. In other words, it is natural to assume a separation between the electroweak scale and the scale of new physics. In this perspective, the SM can be viewed as an effective low-energy description of a more general theory (*UV completion*) characterized by new states with masses above a *cut-off* scale $\Lambda$. At energies below such cut-off, the impact of these new degrees of freedom manifests only through contact interactions among SM fields. These can be described in general terms using the formalism of *Effective Field Theories* (EFT).

An EFT is a well-defined quantum field theory with an intrinsic energy limitation ($E < \Lambda$), whose structure is completely determined by two main ingredients: (1) the field content describing the states that can be produced on-shell at energies $E < \Lambda$; (2) the global and local symmetries acting on these states. In an EFT, physical amplitudes are computed in a perturbative expansion in powers of $E/\Lambda$. This expansion defines a *power counting* for the tower of local operators describing the new contact interactions. This power counting can be supplemented by additional rules to classify the relevance of the effective operators based on possible breaking of the global symmetries or other assumptions about the underlying dynamics.

To construct an EFT that reproduces the SM at low energies, the natural starting point is to adopt the field content and local symmetries of the SM. The resulting theory, where the physical Higgs boson is assumed to belong to an $SU(2)$ doublet, is known as the *Standard Model Effective Field Theory* (SMEFT). In principle, however, other possibilities exist. For example, it is not necessary to assume that the massive Higgs boson belongs to a doublet: it can instead be treated as a $SU(2)$ singlet, with the longitudinal components of the $W$ and $Z$ fields arising from Goldstone bosons in a generic non-linear representation of electroweak symmetry breaking. This alternative leads to a more general EFT, referred to as the *Higgs Effective Field Theory* (HEFT). In this Book, however, we will restrict our attention to SMEFT.

In what follows, we briefly describe the notation and conventions used in this study in the use of the SMEFT. For more details we refer to the reviews in [1088, 1089]. We discuss key details about the assumptions used in our comparisons, in particular regarding the flavour structure of the new physics, of relevance for the discussion in Chapters 3 and 5, and the interpretation of the SMEFT results to set bounds on specific UV scenarios, related to the models discussed in Chapter 8.

### Notation and conventions

In the SMEFT, the effective operator of lowest order in the $1/\Lambda$ expansion is a single term of dimension five:

$$\Delta\mathscr{L}^{(5)}_{\text{SMEFT}} = \frac{(C_5)_{ij}}{\Lambda}(\overline{l^{c\,i}_L}\tilde{\phi}^*\tilde{\phi}^\dagger l^j_L) + \text{h.c.}\ . \tag{A.1}$$

This operator violates lepton number and, after electroweak symmetry breaking, generates Majorana neutrino masses, thus being of especial relevance for neutrino physics.

From the point of view of the electroweak and flavour studies, we will assume individual conservation of baryon and lepton number. Under this assumption, the leading terms beyond the SM Lagrangian in the $1/\Lambda$ expansion start at dimensions six. We consider the SMEFT Lagrangian truncated at this level:

$$\Delta\mathscr{L}^{(6)}_{\text{SMEFT}} = \sum_i \frac{C_i}{\Lambda^2}\mathscr{O}_i \,. \tag{A.2}$$

The list of independent dimension-six effective operators contains 59 electroweak structures, resulting in 2499 independent couplings for three generations of fermions. We use the basis in Ref. [264]. To stablish the notation and conventions, we present in Table A.1 the different operators in the basis.

The values of the *Wilson coefficients* $C_i$, or the couplings of the effective operators, encode information about the UV completion of the SM. Their dependence on the masses and couplings of new particles is therefore model-dependent. If the UV completion is known, the Wilson coefficients can be determined by comparing EFT and model predictions for a set of amplitudes, in a process referred to as *matching*, or directly integrating the heavy degrees of freedom via functional methods.

The operators listed in Table A.1 affect physical observables in several ways. Some modify the properties of SM particles after electroweak symmetry breaking, thanks to Higgs fields in the effective operators acquiring vacuum expectation value. Others introduce genuinely new interactions absent in the SM, typically leading to distinct kinematic effects. These effects make differential observables a particularly valuable probe of new physics. In addition, certain operators can break approximate SM symmetries, generating effects forbidden in the SM. Since such processes are already subject to stringent experimental limits, they provide powerful constraints on any scenario that induces them. This is particularly relevant in the flavour sector: the agnostic use of SMEFT naturally gives rise to many flavour-violating processes, which can be tested experimentally. Indeed, the majority of the 2499 couplings appearing in Eq. (A.2) involve fermion fields and are related to the breaking of flavour degeneracy. The strong constraints on these effects suggest not to work with a fully agnostic SMEFT, especially when considering different set of observables at the same time. It is more useful to impose motivated and well-defined hypotheses about the flavour structure of the new physics generating Eq. (A.2). A particularly well-motivated scenario, both theoretically and phenomenologically, is to assume flavour degeneracy among the first two generations and distinguish them from the third one. The importance of such flavour hypotheses will be discussed in the next section.

Aside from the notation for the dimension-six operators $\mathscr{O}_i$, and their corresponding Wilson coefficients $C_i$, introduced in Table A.1, several combinations of operators/coefficients have been introduced in the literature, as they are convenient for particular phenomenological studies. For the purpose of the current document, this is the case for operators involving the top quark, where we follow the recommendations of the conventions of the *LHC Top Working Group* in [53]. The ones relevant for this work are:

$$\begin{aligned}
C^{\pm}_{\phi Q} &= (C^{(1)}_{\phi q})_{33} \pm (C^{(3)}_{\phi q})_{33}, \\
C^{(3)}_{\phi Q} &= (C^{(3)}_{\phi q})_{33}, \\
C_{\phi t} &= (C_{\phi u})_{33}, \\
C_{t\phi} &= (C_{u\phi})_{33}, \\
C_{tZ} &= -\sin\theta_W (C_{uB})_{33} + \cos\theta_W (C_{uW})_{33},
\end{aligned} \tag{A.3}$$

$$
\begin{aligned}
C_{tW} &= (C_{uW})_{33},\\
C_{lQ}^{\pm} &= (C_{lq}^{(1)})_{33} \pm (C_{lq}^{(3)})_{33},\\
C_{lt} &= (C_{lu})_{\ell\ell 33},\\
C_{et} &= (C_{eu})_{\ell\ell 33},\\
C_{eQ} &= (C_{qe})_{33\ell\ell},\\
C_{leqt}^{S} &= (C_{lequ}^{(1)})_{\ell\ell 33},\\
C_{leqt}^{T} &= (C_{lequ}^{(3)})_{\ell\ell 33},
\end{aligned}
$$

where $\ell = e, \mu$. Note that the last two operators are forbidden by the assumed $U(2)^5$ flavour symmetry and are not considered in the fits presented in Chapter 3. They are however used in the presentation of FCNC constraints on top operators in Fig. 3.2 in that same chapter. In that figure, the combination $C_{t\gamma} = \cos\theta_W (C_{uB})_{33} + \sin\theta_W (C_{uW})_{33}$ is also used instead of $C_{tW}$.

Finally, for the interpretation of the EFT results in terms of particular BSM scenarios in Chapter 8 and the comparison of the collider reach in terms of high-energy measurements presented in Chapter 3, we also use a restricted set of SMEFT operators motivated by certain "universal" new physics scenarios. This will be introduced in the last part of this appendix, after discussing the technical details regarding the SMEFT flavour assumptions.

## Flavour assumptions

Stringent constraints from flavour-changing neutral current processes impose very strong bounds on the structure of SMEFT. Without further assumptions—i.e. with generic $\mathcal{O}(1)$ couplings for flavour-violating operators—flavour bounds push the effective scale $\Lambda$ to very high values (see Chapter 5). If this were the only relevant scale, new physics would be unable to address questions related to the electroweak scale. To avoid this, specific hypotheses about the flavour structure of SMEFT, reflecting assumptions about the underlying new physics, must be introduced. In this study, we focus on scenarios in which new physics respects the approximate $U(2)^5 = U(2)_\ell \times U(2)_q \times U(2)_e \times U(2)_u \times U(2)_d$ flavour symmetry acting on the first two families. This symmetry is an exact symmetry of the SM Lagrangian in the limit where we neglect the masses of the light generations. Under this hypothesis, dimension-six operators involving third-generation and light-generation fields are treated separately. We further assume that flavour non-degeneracy among the light families is negligible but for the contribution yielding the light Yukawa couplings. In the CP-conserving limit, the $U(2)^5$ symmetry reduces the number of independent dimension-six operators to 124.

Note that in defining the symmetry we need to specify the direction in flavour space that defines the third generation. For SM leptons and right-handed quarks, a natural and non-ambiguous choice of basis is the one where the Yukawa matrices (or better appropriate products of the type $Y_f^\dagger Y_f$) are diagonal. For the left-handed quarks it is not possible to do so due to the misalignment between the up- and down-quark mass-eigentates basis induced by the CKM matrix. In this work, we present results considering two choices for the definition of the third family in the $U(2)^5$-symmetric limit:

- The third family is aligned with the basis of the up-quark mass eigenstates (UP basis).
- The third family is aligned with the basis of the down-quark mass eigenstates (DOWN basis).

| | Operator | Notation | Operator | Notation |
|---|---|---|---|---|
| $X^3$ | $\varepsilon_{abc} W_\mu^{a\nu} W_\nu^{b\rho} W_\rho^{c\mu}$ | $\mathcal{O}_W$ | $\varepsilon_{abc} \tilde{W}_\mu^{a\nu} W_\nu^{b\rho} W_\rho^{c\mu}$ | $\mathcal{O}_{\tilde{W}}$ |
| | $f_{ABC} G_\mu^{A\nu} G_\nu^{B\rho} G_\rho^{C\mu}$ | $\mathcal{O}_G$ | $f_{ABC} \tilde{G}_\mu^{A\nu} G_\nu^{B\rho} G_\rho^{C\mu}$ | $\mathcal{O}_{\tilde{G}}$ |
| $\phi^6$ | $\left(\phi^\dagger\phi\right)^3$ | $\mathcal{O}_\phi$ | | |
| $\phi^4 D^2$ | $\left(\phi^\dagger\phi\right)\Box\left(\phi^\dagger\phi\right)$ | $\mathcal{O}_{\phi\Box}$ | $\left(\phi^\dagger D_\mu\phi\right)\left(\left(D^\mu\phi\right)^\dagger\phi\right)$ | $\mathcal{O}_{\phi D}$ |
| $\psi^2\phi^2$ | $\left(\phi^\dagger\phi\right)\left(\bar{l}_L\phi e_R\right)$ | $\mathcal{O}_{e\phi}$ | | |
| | $\left(\phi^\dagger\phi\right)\left(\bar{q}_L\phi d_R\right)$ | $\mathcal{O}_{d\phi}$ | $\left(\phi^\dagger\phi\right)\left(\bar{q}_L\tilde{\phi} u_R\right)$ | $\mathcal{O}_{u\phi}$ |
| $X^2\phi^2$ | $\phi^\dagger\phi B_{\mu\nu}B^{\mu\nu}$ | $\mathcal{O}_{\phi B}$ | $\phi^\dagger\phi \tilde{B}_{\mu\nu}B^{\mu\nu}$ | $\mathcal{O}_{\phi\tilde{B}}$ |
| | $\phi^\dagger\phi W^a_{\mu\nu}W^{a\mu\nu}$ | $\mathcal{O}_{\phi W}$ | $\phi^\dagger\phi \tilde{W}^a_{\mu\nu}W^{a\mu\nu}$ | $\mathcal{O}_{\phi\tilde{W}}$ |
| | $\phi^\dagger\sigma_a\phi W^a_{\mu\nu}B^{\mu\nu}$ | $\mathcal{O}_{\phi WB}$ | $\phi^\dagger\sigma_a\phi \tilde{W}^a_{\mu\nu}B^{\mu\nu}$ | $\mathcal{O}_{\phi\tilde{W}B}$ |
| | $\phi^\dagger\phi G^A_{\mu\nu}G^{A\mu\nu}$ | $\mathcal{O}_{\phi G}$ | $\phi^\dagger\phi \tilde{G}^A_{\mu\nu}G^{A\mu\nu}$ | $\mathcal{O}_{\phi\tilde{G}}$ |
| $\psi^2 X\phi$ | $\left(\bar{l}_L\sigma^{\mu\nu}e_R\right)\phi B_{\mu\nu}$ | $\mathcal{O}_{eB}$ | $\left(\bar{l}_L\sigma^{\mu\nu}e_R\right)\sigma^a\phi W^a_{\mu\nu}$ | $\mathcal{O}_{eW}$ |
| | $\left(\bar{q}_L\sigma^{\mu\nu}u_R\right)\tilde{\phi} B_{\mu\nu}$ | $\mathcal{O}_{uB}$ | $\left(\bar{q}_L\sigma^{\mu\nu}u_R\right)\sigma^a\tilde{\phi} W^a_{\mu\nu}$ | $\mathcal{O}_{uW}$ |
| | $\left(\bar{q}_L\sigma^{\mu\nu}d_R\right)\phi B_{\mu\nu}$ | $\mathcal{O}_{dB}$ | $\left(\bar{q}_L\sigma^{\mu\nu}d_R\right)\sigma^a\phi W^a_{\mu\nu}$ | $\mathcal{O}_{dW}$ |
| | $\left(\bar{q}_L\sigma^{\mu\nu}T_A u_R\right)\tilde{\phi} G^A_{\mu\nu}$ | $\mathcal{O}_{uG}$ | $\left(\bar{q}_L\sigma^{\mu\nu}T_A d_R\right)\phi G^A_{\mu\nu}$ | $\mathcal{O}_{dG}$ |
| $\psi^2\phi^2 D$ | $\left(\phi^\dagger i\overleftrightarrow{D}_\mu\phi\right)\left(\bar{l}_L\gamma^\mu l_L\right)$ | $\mathcal{O}^{(1)}_{\phi l}$ | $\left(\phi^\dagger i\overleftrightarrow{D}^a_\mu\phi\right)\left(\bar{l}_L\gamma^\mu\sigma_a l_L\right)$ | $\mathcal{O}^{(3)}_{\phi l}$ |
| | $\left(\phi^\dagger i\overleftrightarrow{D}_\mu\phi\right)\left(\bar{e}_R\gamma^\mu e_R\right)$ | $\mathcal{O}_{\phi e}$ | | |
| | $\left(\phi^\dagger i\overleftrightarrow{D}_\mu\phi\right)\left(\bar{q}_L\gamma^\mu q_L\right)$ | $\mathcal{O}^{(1)}_{\phi q}$ | $\left(\phi^\dagger i\overleftrightarrow{D}^a_\mu\phi\right)\left(\bar{q}_L\gamma^\mu\sigma_a q_L\right)$ | $\mathcal{O}^{(3)}_{\phi q}$ |
| | $\left(\phi^\dagger i\overleftrightarrow{D}_\mu\phi\right)\left(\bar{u}_R\gamma^\mu u_R\right)$ | $\mathcal{O}_{\phi u}$ | $\left(\phi^\dagger i\overleftrightarrow{D}_\mu\phi\right)\left(\bar{d}_R\gamma^\mu d_R\right)$ | $\mathcal{O}_{\phi d}$ |
| | $\left(\tilde{\phi}^\dagger iD_\mu\phi\right)\left(\bar{u}_R\gamma^\mu d_R\right)$ | $\mathcal{O}_{\phi ud}$ | | |
| $(\bar{L}L)(\bar{L}L)$ | $\left(\bar{l}_L\gamma_\mu l_L\right)\left(\bar{l}_L\gamma^\mu l_L\right)$ | $\mathcal{O}_{ll}$ | | |
| | $\left(\bar{q}_L\gamma_\mu q_L\right)\left(\bar{q}_L\gamma^\mu q_L\right)$ | $\mathcal{O}^{(1)}_{qq}$ | $\left(\bar{q}_L\gamma_\mu\sigma_a q_L\right)\left(\bar{q}_L\gamma^\mu\sigma_a q_L\right)$ | $\mathcal{O}^{(3)}_{qq}$ |
| | $\left(\bar{l}_L\gamma_\mu l_L\right)\left(\bar{q}_L\gamma^\mu q_L\right)$ | $\mathcal{O}^{(1)}_{lq}$ | $\left(\bar{l}_L\gamma_\mu\sigma_a l_L\right)\left(\bar{q}_L\gamma^\mu\sigma_a q_L\right)$ | $\mathcal{O}^{(3)}_{lq}$ |
| $(\bar{R}R)(\bar{R}R)$ | $\left(\bar{e}_R\gamma_\mu e_R\right)\left(\bar{e}_R\gamma^\mu e_R\right)$ | $\mathcal{O}_{ee}$ | | |
| | $\left(\bar{u}_R\gamma_\mu u_R\right)\left(\bar{u}_R\gamma^\mu u_R\right)$ | $\mathcal{O}_{uu}$ | $\left(\bar{d}_R\gamma_\mu d_R\right)\left(\bar{d}_R\gamma^\mu d_R\right)$ | $\mathcal{O}_{dd}$ |
| | $\left(\bar{u}_R\gamma_\mu u_R\right)\left(\bar{d}_R\gamma^\mu d_R\right)$ | $\mathcal{O}^{(1)}_{ud}$ | $\left(\bar{u}_R\gamma_\mu T_A u_R\right)\left(\bar{d}_R\gamma^\mu T_A d_R\right)$ | $\mathcal{O}^{(8)}_{ud}$ |
| | $\left(\bar{e}_R\gamma_\mu e_R\right)\left(\bar{u}_R\gamma^\mu u_R\right)$ | $\mathcal{O}_{eu}$ | $\left(\bar{e}_R\gamma_\mu e_R\right)\left(\bar{d}_R\gamma^\mu d_R\right)$ | $\mathcal{O}_{ed}$ |
| $(\bar{L}L)(\bar{R}R)$ | $\left(\bar{l}_L\gamma_\mu l_L\right)\left(\bar{e}_R\gamma^\mu e_R\right)$ | $\mathcal{O}_{le}$ | $\left(\bar{q}_L\gamma_\mu q_L\right)\left(\bar{e}_R\gamma^\mu e_R\right)$ | $\mathcal{O}_{qe}$ |
| | $\left(\bar{l}_L\gamma_\mu l_L\right)\left(\bar{u}_R\gamma^\mu u_R\right)$ | $\mathcal{O}_{lu}$ | $\left(\bar{l}_L\gamma_\mu l_L\right)\left(\bar{d}_R\gamma^\mu d_R\right)$ | $\mathcal{O}_{ld}$ |
| | $\left(\bar{q}_L\gamma_\mu q_L\right)\left(\bar{u}_R\gamma^\mu u_R\right)$ | $\mathcal{O}^{(1)}_{qu}$ | $\left(\bar{q}_L\gamma_\mu T_A q_L\right)\left(\bar{u}_R\gamma^\mu T_A u_R\right)$ | $\mathcal{O}^{(8)}_{qu}$ |
| | $\left(\bar{q}_L\gamma_\mu q_L\right)\left(\bar{d}_R\gamma^\mu d_R\right)$ | $\mathcal{O}^{(1)}_{qd}$ | $\left(\bar{q}_L\gamma_\mu T_A q_L\right)\left(\bar{d}_R\gamma^\mu T_A d_R\right)$ | $\mathcal{O}^{(8)}_{qd}$ |
| $(\bar{L}R)(\bar{R}L)$ | $\left(\bar{l}_L e_R\right)\left(\bar{d}_R q_L\right)$ | $\mathcal{O}_{ledq}$ | | |
| $(\bar{L}R)(\bar{L}R)$ | $\left(\bar{q}_L u_R\right)i\sigma_2\left(\bar{q}_L d_R\right)^{\mathrm{T}}$ | $\mathcal{O}^{(1)}_{quqd}$ | $\left(\bar{q}_L T_A u_R\right)i\sigma_2\left(\bar{q}_L T_A d_R\right)^{\mathrm{T}}$ | $\mathcal{O}^{(8)}_{quqd}$ |
| | $\left(\bar{l}_L e_R\right)i\sigma_2\left(\bar{q}_L u_R\right)^{\mathrm{T}}$ | $\mathcal{O}^{(1)}_{lequ}$ | $\left(\bar{l}_L\sigma_{\mu\nu}e_R\right)i\sigma_2\left(\bar{q}_L\sigma^{\mu\nu}u_R\right)^{\mathrm{T}}$ | $\mathcal{O}^{(3)}_{lequ}$ |

Table A.1: Basis of dimension-six operators. Flavour indices are omitted.

Importantly, despite the naming, these two options are not related by a simple change of basis; they correspond to intrinsically different assumptions about the flavour structure of new physics. More generally, one can introduce an additional parameter to interpolate between them, describing the orientation of the third generation—from the perspective of the new physics—relative to the SM Yukawa couplings. This more general framework will be adopted for specific studies in Chapter 5.

## Mapping Effective Field Theories onto specific models

The power of the EFT approach is to provide a layer of interpretation of experimental data that facilitates the connection to specific models. Indeed, in this formalism, using a bottom-up approach, one only needs to compute all new physics contributions to the measured observables once in terms of the different EFT operators, and obtain a global EFT likelihood that can be used to set bounds on the corresponding Wilson coefficients. More importantly, in a top-down approach one can compute the values of the Wilson coefficients in terms of the parameters of a given model, and use the generic likelihood to set bounds on the masses and coupling of the new states, which we generically refer to as $g_\star$ and $m_\star$. Instead of recomputing the predictions for (many) observables for each model of interest, the calculation of the Wilson coefficients in terms of the new physics model only requires to compute a comparatively smaller number of amplitudes both in the model and in the EFT. The matching of the EFT and model results provides the mapping between the Wilson coefficients $C_i$, $\Lambda$ and $(g_\star, m_\star)$. Furthermore, without going into the actual calculation of the matching, one can impose additional power-counting assumptions for the different Wilson coefficients, following their expected sizes in a given class of models. This allows to reorganize the EFT expansion, keeping only those terms that are relevant for the class of models of interest.

A particular subset of SMEFT interactions that has been widely use in the literature to describe composite Higgs models, but that can also be easily mapped into other types of models is the one in Ref. [403]. Using the assumptions described in [403, 1090], the low-energy signatures of these kind of models can be described by the following effective Lagrangian

$$
\begin{aligned}
\mathscr{L}_{\text{SILH}} =& \frac{c_\phi}{\Lambda^2}\partial_\mu(\phi^\dagger\phi)\partial^\mu(\phi^\dagger\phi) + \frac{c_T}{\Lambda^2}(\phi^\dagger \overleftrightarrow{D}_\mu \phi)(\phi^\dagger \overleftrightarrow{D}^\mu \phi) - \frac{c_6}{\Lambda^2}(\phi^\dagger\phi)^3 + \\
&+ \left(\frac{c_{y_f}}{\Lambda^2} y^f_{ij} \phi^\dagger\phi\, \bar{\psi}_{Li}\phi\,\psi_{Rj} + \text{h.c.}\right) \\
&+ \frac{c_W}{\Lambda^2}\left(\phi^\dagger i\overleftrightarrow{D}^a_\mu \phi\right) D_\nu W^{a\,\mu\nu} + \frac{c_B}{\Lambda^2}\left(\phi^\dagger i\overleftrightarrow{D}_\mu \phi\right)\partial_\nu B^{\mu\nu} \\
&+ \frac{c_{\phi W}}{\Lambda^2} iD_\mu\phi^\dagger \sigma_a D_\nu \phi W^{a\,\mu\nu} + \frac{c_{\phi B}}{\Lambda^2} iD_\mu \phi^\dagger \sigma_a D_\nu \phi B^{\mu\nu} \\
&+ \frac{c_\gamma}{\Lambda^2}\phi^\dagger\phi B^{\mu\nu}B_{\mu\nu} + \frac{c_g}{\Lambda^2}\phi^\dagger\phi G^{A\,\mu\nu}G^A_{\mu\nu} \\
&- \frac{c_{2W}}{\Lambda^2}(D^\mu W^a_{\mu\nu})(D_\rho W^{a\,\rho\nu}) - \frac{c_{2B}}{\Lambda^2}(\partial^\mu B_{\mu\nu})(\partial_\rho B^{\rho\nu}) - \frac{c_{2G}}{\Lambda^2}(D^\mu G^A_{\mu\nu})(D_\rho G^{A\,\rho\nu}) \\
&+ \frac{c_{3W}}{\Lambda^2}\varepsilon_{abc} W^{a\ \nu}_{\mu} W^{b\ \rho}_{\nu} W^{c\ \mu}_{\rho} + \frac{c_{3G}}{\Lambda^2} f_{ABC} G^{A\ \nu}_{\mu} G^{B\ \rho}_{\nu} G^{C\ \mu}_{\rho}.
\end{aligned}
\tag{A.4}
$$

Aside from the contributions from the operators with coefficients $c_{y_f}$, which modify the Higgs Yukawa interactions and are proportional to the SM Yukawa matrices, $y^f_{ij}$, all other operators are flavour-blind at the scale $\Lambda$. The power-counting for the different operators depends on the specific scenario. For the studies presented in Chapter 8 we consider the following scenario for composite Higgs models:

$$\begin{aligned} &\frac{c_{\phi,y_f}}{\Lambda^2}, \frac{c_6}{\lambda_\phi \Lambda^2} \sim \frac{g_\star^2}{m_\star^2}, && \frac{c_T}{\Lambda^2} \sim \frac{y_t^4}{16\pi^2}\frac{1}{m_\star^2}, \\ &\frac{c_B}{g_1\Lambda^2}, \frac{c_W}{g_2\Lambda^2} \sim \frac{1}{m_\star^2}, && \frac{c_{2B}}{g_1^2\Lambda^2}, \frac{c_{2W}}{g_2^2\Lambda^2}, \frac{c_{2G}}{g_3^2\Lambda^2} \sim \frac{1}{g_\star^2}\frac{1}{m_\star^2}, \\ &\frac{c_{\phi B}}{g_1\Lambda^2}, \frac{c_{\phi W}}{g_2\Lambda^2}, \frac{c_\gamma}{g_1^2\Lambda^2}, \frac{c_g}{g_3^2\Lambda^2} \sim \frac{1}{16\pi^2}\frac{1}{m_\star^2}, && \frac{c_{3W}}{g_2^3\Lambda^2}, \frac{c_{3G}}{g_3^3\Lambda^2} \sim \frac{1}{16\pi^2}\frac{1}{g_\star^2}\frac{1}{m_\star^2}, \end{aligned} \tag{A.5}$$

where $\lambda_\phi$ is the quartic coupling in the SM scalar potential and $g_{1,2,3}$ are the hypercharge, weak isosping and QCD gauge coupling constants.

Finally, also some specific models are discussed in the text, in particular in the section describing heavy spin-1 resonances. There we consider the case of a heavy vector singlet, coupling to the SM particles proportionally to the hypercharge current. This $Y'$ model only contributes to the operator with coefficient $c_{2B}$ in Eq. (A.4), generating the following contribution:

$$\frac{c_{2B}}{\Lambda^2} = \frac{1}{2}\frac{g_{Y'}^2}{g_1^2 M_{Y'}^2}, \tag{A.6}$$

where $M_{Y'}$ and $g_{Y'}$ are the mass and coupling of the new resonance.

## B Glossary of acronyms (followed by those from the Accelerator chapter)

| | |
|---|---|
| AD | Antimatter Decelerator, facility at CERN |
| AI | Artificial Intelligence |
| ALP | Axion-like Particle |
| APPEC | Astro Particle Physics European Consortium |
| BDF | Beam Dump Facility, proposed at the CERN SPS |
| BSM | Beyond the SM, i.e. new physics |
| CC | Circular Collider |
| CEPC | Circular Electron Positron Collider, proposed $e^+e^-$ collider (sited in China) |
| CH | Composite Higgs |
| CKM | Cabibbo-Kobayashi-Maskawa, the quark mixing matrix |
| CL | Confidence Level |
| CLIC | Compact Linear Collider, proposed $e^+e^-$ collider (sited at CERN) |
| CMOS | Technique for fabricating integrated circuits in silicon |
| CP | Combination of discrete symmetries: Charge-conjugation (C) and Parity (P) |
| CPV | CP Violation |
| CR | Cosmic Ray |
| DIS | Deep Inelastic Scattering |
| DM | Dark Matter |
| DS | Dark Sector |
| ECFA | European Committee for Future Accelerators |
| EDM | Electric Dipole Moment |
| EFT | Effective Field Theory |
| EIC | Electron-ion Collider |
| ERL | Energy Recovery Linac |
| ESPPU | European Strategy for Particle Physics Update, also sometimes EPPSU or ESU |
| EW | Electroweak |
| EWPO | Electroweak Precision Observables |
| EWSB | Electroweak Symmetry Breaking |
| FCC | Future Circular Collider, proposed 100-km scale collider (sited at CERN) |
| FCC-ee | Version of FCC with $e^+e^-$ collisions |
| FCC-eh | Version of FCC with electron-hadron collisions |
| FCC-hh | Version of FCC with hadron collisions (proton or heavy-ion) |
| FCNC | Flavour Changing Neutral Current |
| FEL | Free Electron Laser, light source |
| FIP | Feebly Interacting Particle |
| GIM | Glashow-Iliopoulos-Maiani, mechanism suppressing some decays |
| GPD | Generalised Parton Distribution (or General Purpose Detector, depending on context) |
| GUT | Grand Unified Theory |
| HE-LHC | High Energy LHC, proposed collider with $\sim$ double LHC energy in same tunnel |
| HEP | High Energy Physics |
| HL-LHC | High Luminosity LHC, upgrade of the LHC to provide higher luminosity |
| HLT | High Level Trigger |
| HNL | Heavy Neutral Lepton |
| HPC | High Performance Computing |
| HTS | High Temperature Superconductor |
| HV | High Voltage |
| ID | Indirect Detection (or Identification, depending on context) |
| ILC | International Linear Collider, proposed $e^+e^-$ collider (sited in Japan) |
| IP | Interaction Point |
| IR | Infrared, i.e. low energy limit |
| KEKB | B factory $e^+e^-$ collider in Japan |
| LC | Linear Collider |
| LCF | Linear Collider Facility (proposed at CERN) |
| LDM | Light Dark Matter |
| LEP | $Z$ factory $e^+e^-$ collider at CERN, used the same tunnel now occupied by LHC |
| LFUV | Lepton Flavour Universality Violation |
| LFV | Lepton Flavour Violation |
| LHC | Large Hadron Collider, hadron collider at CERN |
| LHeC | Proposed electron-hadron collider using hadrons from the LHC plus an ERL |
| LLP | Long-lived Particle |
| LNV | Lepton Number Violation |
| LQCD | Lattice QCD |
| LSP | Lightest Supersymmetric Particle |
| MFV | Minimal Flavour Violation |
| MSSM | Minimal Supersymmetric Model |
| NLL | Next to Leadling Logarithm |
| NLO | Next to Leading Order |
| NP | New Physics, i.e. physics beyond the Standard Model |
| PBC | Physics Beyond Colliders, a study |
| PDF | Parton Distribution Function |
| POT | Protons On Target |
| PP | Particle Physics |
| pQCD | Perturbative QCD |
| QCD | Quantum Chromodynamics, theory of the strong interaction |
| QED | Quantum Electrodynamics, theory of the electromagnetic interaction |
| QFT | Quantum Field Theory |
| QGP | Quark Gluon Plasma |
| RCS | Rapid-cycling Synchrotron |
| RF | Radio Frequency |
| SC | Superconducting |
| SCT | Super Charm Tau (also known as Tau-Charm factory, STC), proposed $e^+e^-$ collider |
| SD | Shutdown |
| SKA | Square Kilometer Array, radio telescope array |
| SLC | Linear Collider previously operating at SLAC |
| SPS | Super Proton Synchrotron, accelerator at CERN |
| SM | Standard Model (of particle physics) |
| SMEFT | Standard Model Effective Field Theory |
| SPPC | Proposed hadron collider to follow the CEPC using the same tunnel |
| SUSY | Supersymmetry, proposed model of NP |
| TDAQ | Trigger and Data Aquisition |
| TDR | Technical Design Report |
| TPC | Time Projection Chamber (or Total Project Cost, depending on context) |
| UHE | Ultra High Energy |
| UT | Unitarity Triangle, relationship between quark mixing matrix elements |
| VBF | Vector Boson Fusion |
| vev | Vacuum Expectation Value |
| WIMP | Weakly Interacting Massive Particle, a candidate for DM |
| WLCG | Worldwide LHC Computing Grid |

# Accelerator acronyms

**ALEGRO** Advanced LinEar collider study GROup. 186
**ALiVE** Advanced Linear collider for Very high Energy. 171, 186
**ALP** Axion-like Particle. 187
**ASSET** Accelerator Test Facility at an SLC beamline. 184
**ATF2** Accelerator Test Facility 2. 170, 189
**AWAKE** Advanced Proton Driven Plasma Wakefield Acceleration Experiment. 186
**BEPC–II** Beijing Electron–Positron Collider II. 173
**BS** Beamstrahlung. 168–171, 174
**$C^3$** Cool Copper Collider. 184–186
**C-band** RF band from 4 GHz to 8 GHz. 184
**c.o.m.** centre-of-mass. 170, 172–179, 186, 189
**CC** Circular Collider. 168–170, 188, 189
**CDR** Conceptual Design Report. 180, 190
**CE** Civil Engineering. 172, 174, 176, 188
**CEPC** Circular Electron Positron Collider. 169, 179
**CERN** European Organization for Nuclear Research. 168, 171–173, 176, 178–180, 187, 189
**CLIC** Compact LInear Collider. 170–173, 175, 184, 185, 188–190
**CTF** CLIC Test Facility. 172
**CTF2** CLIC Test Facility 2. 172
**CTF3** CLIC Test Facility 3. 172, 189
**CW** Continuous Wave. 173, 185
**DAΦNE** Double Annular Φ factory for Nice Experiments. 169, 173, 189
**DC** Direct Current. 182, 188
**DESY** Deutsches Elektronen Synchrotron. 186, 189
**eA** electron-ion. 176, 177
**EN** European Norm. 187
**ep** electron-proton. 176, 177
**ERL** Energy Recovery Linac. 176, 177, 180, 189
**ESPPU** Update of the European Strategy for Particle Physics. 168, 180, 190
**ESS** European Spallation Source. 187
**EU-XFEL** European X-Ray Free-Electron Laser Facility. 175, 182, 183, 189
**EuPRAXIA** European Plasma Research Accelerator with eXcellence in Applications. 172, 185, 186
**FACET** Facility for Advanced Accelerator Experimental Tests. 170, 184
**FAIR** Facility for Antiproton and Ion Research. 187
**FCC** Future Circular Collider. 172, 174, 188

**FCC-ee** Future Circular $e^+e^-$Collider. 169–174, 176, 177, 179, 183–186, 188, 190

**FCC-hh** Future Circular hadron–hadron Collider. 171, 172, 174, 180–182, 190

**FE-FRT** Ferro-Electric Fast Reactive Tuner. 182, 183

**FEL** Free Electron Laser. 186

**FERMI@Elettra** Free Electron laser Radiation for Multidisciplinary Investigations at the Elettra synchrotron light source. 170

**Fermilab** Fermilab. 183, 187

**FFS** Final-Focus System. 170

**FFTB** Final-Focus Test Beam. 170

**FP** Full-Power. 173, 175

**GHG** Greenhouse Gas. 187, 188

**HALHF** Hybrid Asymmetric Linear Higgs Factory. 171, 184–186, 190

**HE** High-Energy. 172, 175, 176, 187, 190

**HEP** High-Energy Physics. 182, 185–188, 190, 191

**HFM** High Field Magnet. 171, 174, 180–182, 187, 188, 190

**HGWFA** High-gradient wakefield accelerator. 186, 187, 190

**HL-LHC** High-Luminosity LHC. 174–177, 180, 185, 190

**HOM** High-Order Mode. 183, 184

**HTS** High Temperature Superconductor. 174, 176, 178, 181, 182, 184, 190

**HW** Hardware. 188

**IBS** Iron Based Superconductor. 182

**ICFA** International Committee for Future Accelerators. 179

**IDT** International Development Team. 179

**IJCLab** Laboratoire de Physique des 2 Infinis Irène Joliot-Curie. 177

**ILC** International Linear Collider. 170, 175, 179, 180, 183, 185, 189, 246

**IMCC** International Muon Collider Collaboration. 177, 178

**IP** Interaction Point. 168–170, 172, 173, 175, 177–180, 189

**IR** Interaction Region. 169, 172, 174, 175, 182

**ISO** International Organization for Standardization. 187

**ISR** Initial State Radiation. 170, 171

**ITN** ILC Technology Network. 180

**J-PARC** Japan Proton Accelerator Research Complex. 187

**JLab** Jefferson Lab. 173

**KEK** High Energy Accelerator Research Organization in Japan. 172

**KEKB** KEK B-factory. 169, 173

**L-band** RF band from 1 GHz to 22 GHz. 190

**LC** Linear Collider. 168–171, 175, 188, 189

**LCA** Life Cycle Assessment. 187, 188

**LCF** Linear Collider Facility. 170, 171, 173, 175, 180, 183, 187–189

**LCLS–II** Linac Coherent Light Source – II. 183, 189

**LDG** Laboratory Directors Group. 180, 190

**LEP** Large Electron Positron (collider). 169, 173, 175, 188, 189

**LEP2** Large Electron Positron (collider) 2. 169, 175, 188, 189

**LEP3** Large Electron Positron (collider) 3. 169, 171, 173, 175–177, 188, 190

**LHC** Large Hadron Collider. 170, 173, 175–178, 190

**LHeC** Large Hadron-electron Collider. 171, 173, 176, 177, 183, 189

**LP** Low-Power. 173, 175

**LSS** Long Straight Section. 176
**LTS** Low Temperature Superconductor. 174, 181, 182
**LURE** Laboratoire d'utilisation du rayonnement électromagnétique. 247
**LWFA** Laser-driven Plasma Wakefield Accelerator. 186
**MAP** Muon Accelerator Programme. 177
**MC** Muon Collider. 171, 177, 178, 180, 182–184, 190
**N/A** Not Available. 173
**NC** Normal Conducting. 170, 171, 178, 180, 184, 188, 189
**P3** PSI Positron Production. 170
**P5** Particle Physics Project Prioritization Panel. 186
**PBC** Physics Beyond Colliders. 187, 190
**pCM** parton-centre-of-momentum. 168
**PEP** Plasma Electro Polishing. 182
**PEP–II** Positron Electron Project–II. 173
**PERLE** Powerful Energy Recovery Linac for Experiments. 177, 189
**PETRA IV** Positron–Electron Tandem Ring Accelerator — IV. 186
**PIP-II** Proton Improvement Plan - II. 183
**PSI** Paul Scherrer Institute. 170, 187, 247
**PWFA** Plasma Wakefield Accelerator. 180, 190
**QC** Quality Control. 183
**R&D** Research and Development. 168, 170, 173–175, 178–183, 186–191
**RCS** Rapid-Cycling Synchrotron. 178, 186
**ReBCO** Rare-earth Barium Copper Oxide. 174, 181, 182
**RF** Radio Frequency. 168, 170–173, 175–178, 180, 183–185, 187–190, 245, 246, 248
**RI** Research Infrastructure. 187, 188, 190
**RLA** Recirculating Linear Accelerator. 178
**RMM1** Racetrack Model Magnet 1. 180
**SBL** Sudden Beam Loss. 169
**SC** Superconducting. 170, 176, 178–180
**SF-QED** Strong Field Quantum Electrodynamics. 186
**SLAC** Stanford Linear Accelerator Center. 172, 175
**SLC** Stanford Linear Collider. 170, 173, 189, 245
**SOLEIL** Source Optimisée de Lumière d'Énergie Intermédiaire du LURE. 185
**SPPC** Super Proton Proton Collider. 179
**SPS** Super Proton Synchrotron. 178, 185
**SR** Synchrotron Radiation. 168, 171, 173–177, 179, 181
**SRF** Superconducting Radio Frequency. 173, 175–177, 179, 182, 183, 187, 189, 190
**SSS** Short Straight Section. 176
**SuperKEKB** Super KEK B-factory. 169, 170, 173, 189
**SwissFEL** Swiss X-ray Free Electron Laser. 184
**TBA** Two-Beam Acceleration. 171, 172, 189
**TESLA** TeV–Energy Superconducting Linear Accelerator. 183, 189
**TRL** Technology Readiness Level. 174, 181, 182, 190
**UHV** Ultra-High Vacuum. 183
**UN** United Nations. 188
**UN-SDG** United Nations Sustainable Development Goals. 187, 188
**US** United States. 177, 180, 186

**US-MDP** US-Magnet Development Programme. 181
**X-band** RF band from 8 GHz to 12 GHz. 171, 172, 184, 185, 189, 190

## C European Strategy Update contributions

There were 281 documents submitted to the European Strategy Update by the deadline of 31 March 2025, of which 15 were withdrawn (mostly due to multiple submission). The remaining 266 submissions are available via the meeting website, and are listed below. The submissions are referenced in the text of this document using the form [ID$n$], and some have addenda which give further details of the community involved, schedule and cost. The contributions can be accessed directly by clicking on the ID number in the list below.

| ID | Title |
|---|---|
| ID2 | *Light Ion Collisions at the LHC* |
| ID4 | *Contribution about Neutrino Physics to the European Strategy for Particle Physics* |
| ID5 | *Prospective report of the French QCD community to the ESPPU 2026 with respect to the program of the LHC Run 5 and beyond and future colliders at CERN* |
| ID6 | *Input for the ESPPU 2026 compiled by the ISOLDE Collaboration Committee* |
| ID7 | *Conclusions of the Town Meeting: Heavy Ion and QGP Physics at CERN* |
| ID9 | *The Worldwide LHC Computing Grid input to the ESPPU 2026* |
| ID10 | *ESPPU 2026 Israeli Input* |
| ID11 | *Fermilab Interest in a Higgs Factory at CERN* |
| ID12 | *Strategy for HPC Integration in WLCG/HEP* |
| ID13 | *Laboratori Nazionali di Frascati of INFN* |
| ID14 | *Input to the ESPPU 2026 from the Hungarian high-energy physics community* |
| ID15 | *French HEP community input to ESPP + comments from CEA and CNRS* |
| ID16 | *Polish national input to the 2026 update of the ESPP* |
| ID17 | *Enabling future detector technology within ePIC at the EIC* |
| ID18 | *CERN openlab: A Flagship Model for Industry-Science Computing R&D* |
| ID19 | *The Forward Physics Facility at the Large Hadron Collider* |
| ID20 | *Future of CERN* |
| ID21 | *Search for the electric dipole moment of the neutron with the n2EDM experiment* |
| ID22 | *Statement by the German Particle Physics Community as Input to the Update of the ESPP* |
| ID23 | *Prospects and Opportunities with an upgraded FASER Neutrino Detector during the HL-LHC era* |
| ID25 | *Countering the biodiversity loss using particle physics research sites* |
| ID26 | *HFLAV input to the 2026 update of the ESPP* |
| ID27 | *Exploring the Dark Universe: A European Strategy for Axions and other WISPs Discovery* |
| ID28 | *The PTOLEMY project* |
| ID29 | *Strategy for the Future of Lattice QCD* |
| ID30 | *European training in instrumentation and particle accelerators* |
| ID31 | *Slovak Particle Physics Community Input to Update of the ESPP* |
| ID32 | *Performance study of the MUSIC detector in* $\sqrt{s} = 10$ *TeV muon collisions* |
| ID33 | *Computer Algebra for Precision Calculations in Particle Physics: the FORM project* |
| ID34 | *Some thoughts on the future of particle physics.* |
| ID35 | *Cusp Spectroscopy, Hyperon-Nucleon Scattering, and Femtoscopy: Pioneering Tools for Next-Generation Hadron Interaction Studies* |
| ID36 | *The JUNO Experiment* |
| ID37 | *Long-Baseline Atom Interferometry* |
| ID38 | *Neutrino Theory in the Precision Era* |
| ID39 | *Clarifications for the Israeli Input for ESG* |

| ID | Title |
|---|---|
| ID40 | *The Linear Collider Facility (LCF) at CERN* |
| ID41 | *Input to the ESPP 2026 update by the Czech particle physics community* |
| ID42 | *Early Career Researcher Input to the ESPP Update – Fifty-five recommendations for the future of our field* |
| ID43 | *REDTOP: Rare Eta Decays TO Explore New Physics* |
| ID44 | *Proposal for a shared transverse LLP detector for FCC-ee and FCC-hh and a forward LLP detector for FCC-hh* |
| ID45 | *Portuguese HEP community input to the update of the ESPP* |
| ID46 | *Search for feebly interacting particles with the Lohengrin experiment at the ELSA accelerator* |
| ID47 | *Input of Ukraine to the 3rd Update of the ESPP* |
| ID48 | *Particle Data Group Input to ESPP Update* |
| ID49 | *High Performance and Cost Effective Superconducting Accelerator Magnet R&D at IHEP* |
| ID50 | *Searching for Light Dark Matter and Dark Sectors with the NA64 experiment at the CERN SPS* |
| ID51 | *Ultra-relativistic Heavy-Ion Collisions: Inputs of the Italian community for the ESPP* |
| ID52 | *The Astro-particle Physics Commission 2* |
| ID53 | *P-ONE: The Pacific Ocean Neutrino Experiment A Next-Generation Deep-Sea Neutrino Detector for Fundamental Physics* |
| ID54 | *Grenoble Axion Haloscopes: From BabyGrAHal to GrAHal for axion dark matter search in the 1-150 micro-eV mass range* |
| ID55 | *Kaon Physics: A Cornerstone for Future Discoveries* |
| ID56 | *PERLE : an ERL facility for future sustainable colliders (LHeC, FCC)* |
| ID57 | *HALHF: a hybrid, asymmetric, linear Higgs factory using plasma- and RF-based acceleration* |
| ID58 | *Spanish astroparticle community input to the European Strategy Group* |
| ID59 | *Croatian national input to ESPP 2026 update* |
| ID60 | *The Critical Role of Particle Physics Education and Outreach* |
| ID61 | *MATHUSLA: An External Long-Lived Particle Detector to Maximize the Discovery Potential of the HL-LHC* |
| ID62 | *INFN National Committee for Particle Physics (CSN1): Input to the ESPP Update* |
| ID63 | *Input from the SND@LHC collaboration to the 2026 Update to the ESPP* |
| ID64 | *Swiss inputs to the 2026 update of the ESPP* |
| ID65 | *Accelerator R&D proposals by INFN* |
| ID66 | *Empowering particle physics beyond discoveries: a unified European communication strategy for the next generation research infrastructures* |
| ID67 | *The Critical Importance of Software for HEP* |
| ID68 | *Input from the ALICE Collaboration* |
| ID69 | *Future Opportunities with Lepton-Hadron Collisions* |
| ID70 | *Frontier sensor R&D for the ALICE 3 apparatus* |
| ID71 | *CERN AD/ELENA Antimatter Program* |
| ID72 | *DESY's role in Europe: A Contribution to the ESPP Update Process* |
| ID73 | *INFN Input on the update of the ESPP: Computing* |

| ID | Title |
|---|---|
| ID74 | *Charged Lepton Flavour Violations searches with muons: present and future* |
| ID75 | *Future perspectives for $\mu \to e\gamma$ searches* |
| ID76 | *Input on the update of the ESPP by the INFN Nuclear and Hadron Physics Community* |
| ID77 | *A Silicon-Tungsten ECAL for Higgs Factory Detectors* |
| ID78 | *The Compact Linear $e^+e^-$ Collider (CLIC)* |
| ID79 | *Ensuring continued operation of INSPIRE as a cornerstone of the HEP information infrastructure* |
| ID80 | *INFN – Gran Sasso National Laboratory – Input for the ESPP* |
| ID81 | *Discovery potential of LHCb Upgrade II* |
| ID82 | *Heavy ion physics with LHCb Upgrade I* |
| ID83 | *Lithuanian Particle Physics Community Input to the 2026 Update of ESPP* |
| ID84 | *The CERN Festival Programme* |
| ID85 | *Higgs Criticality and the Metastability Bound: a target for future colliders* |
| ID87 | *The NEXT Search for Neutrinoless Double Beta Decay in Xe-136* |
| ID88 | *A Survey of the Latin American High Energy Physics community on the future flagship project at CERN for the ESPP Update* |
| ID89 | *Precision cross-sections for advancing cosmic-ray physics* |
| ID90 | *Strategic Imperative: Investing in Accelerator R&D and Technologies for the Future of Particle Physics and Societal Impact — Comments in the framework of the ESPP update by the German Committee of Accelerator Physics (KfB)* |
| ID91 | *United States Early Career Researchers in Collider Physics input to the ESPPU* |
| ID92 | *The LUXE Experiment* |
| ID93 | *Strengthening the Instrumentation Programme* |
| ID94 | *The SiD Detector concept Input to the ESPPU 2026* |
| ID95 | *Expressions of Interest for the Development of Detector Concepts and Sub-detector Systems for the Future Circular Collider FCC* |
| ID97 | *ESPPU INPUT: C3 within the "Linear Collider Vision"* |
| ID98 | *Design Initiative for a 10 TeV pCM Wakefield Collider* |
| ID99 | *Japan's Updated Strategy for High Energy Physics for the ESPP Update 2026* |
| ID100 | *Probing and Knocking with Muons* |
| ID101 | *SBN@CERN: A short-baseline neutrino beam at CERN for high-precision cross-section measurements* |
| ID102 | *The ILD Detector: A Versatile Detector for an Electron-Positron Collider at Energies up to 1 TeV* |
| ID103 | *Nuclear Physics and the ESPP Update 2026* |
| ID104 | *HTS Potential and Needs for Future Accelerator Magnets* |
| ID105 | *Magnet R&D for the Muon Collider* |
| ID106 | *Country input – Serbia* |
| ID107 | *CMS Offline Software and Computing input to the ESPP 2026 update* |
| ID108 | *DRD 6 Calorimetry – Input to the ESPPU* |
| ID109 | *Searching for millicharged particles with the FORMOSA experiment at the CERN LHC* |
| ID110 | *National Input from the Swedish Community to the Update of the ESPP* |
| ID111 | *The Role of Particle Physics in Revolutionising Imaging Technologies: it From Quantum Entanglement to Clinical Applications* |
| ID112 | *Enhancing European Cooperation in the Search for Dark Matter* |

| ID | Title |
| --- | --- |
| ID113 | *Quantum Information meets High-Energy Physics: Input to the update of the ESPP* |
| ID114 | *Synergies between a U.S.-based Electron-Ion Collider and European Research in Particle Physics* |
| ID115 | *PIONEER: a next generation rare pion decay experiment* |
| ID116 | *T2K Experiment: future plans and capabilities* |
| ID117 | *The INFN National Scientific Committee for Theoretical Physics* |
| ID118 | *The DUNE Science Program* |
| ID119 | *The DUNE Phase II Detectors* |
| ID120 | *European Contributions to Fermilab Accelerator Upgrades and Facilities for the DUNE Experiment* |
| ID121 | *DUNE Software and Computing Research and Development* |
| ID122 | *LVK contribution on Gravitational Waves Physics* |
| ID123 | *IUPAP Commission 11 (Particles and Fields) Support Letter on behalf of the High Energy Physics (HEP) Information Infrastructure* |
| ID124 | *JENA White Paper on European Federated Computing* |
| ID125 | *The SNO+ Science Programme: Input to the ESPPU 2026* |
| ID126 | *Input to ESPPU by the German Astroparticle Community* |
| ID127 | *Computing and software for LHCb Upgrade II* |
| ID129 | *The CERN nTOF neutron-time-of-flight facility: present and future* |
| ID130 | *Simplified Models Expose the Interplay of Direct and Indirect Searches* |
| ID131 | *ESPP 2026: the NA60+/DiCE experiment at the SPS* |
| ID132 | *Input to the ESPP – KATRIN collaboration* |
| ID133 | *A High-Precision, Fast, Robust, and Cost-Effective Muon Detector Concept for the FCC-ee* |
| ID134 | *PSI European Strategy Input* |
| ID135 | *Neutrino Scattering: Connections Across Theory and Experiment* |
| ID136 | *Input to the ESPPU 2026 from the Greek Community* |
| ID138 | *Data Preservation in High Energy Physics* |
| ID139 | *Neutrinos@CERN: A Community-Driven Contribution to the ESPPU* |
| ID140 | *A Linear Collider Vision for the Future of Particle Physics* |
| ID141 | *The ECFA Higgs/Electroweak/Top Factory Study* |
| ID142 | *A Straw Tracker for an FCC-ee experiment* |
| ID144 | *Communicating for the future of particle physics* |
| ID145 | *SHiP experiment at the SPS Beam Dump Facility* |
| ID146 | *The International Axion Observatory (IAXO): case, status and plans* |
| ID147 | *Spanish national input to the ESPP* |
| ID148 | *Technology developments for LHCb Upgrade II* |
| ID149 | *MINERvA neutrino interaction experiment input to the ESPP* |
| ID150 | *Reinterpretation and preservation of data and analyses in HEP* |
| ID151 | *ESSnuSB (European Spallation Source neutrino Super Beam)* |
| ID152 | *United States Muon Collider Community White Paper for the ESPPU* |
| ID153 | *The Circular Electron Positron Collider (CEPC)* |
| ID154 | *Midterm Review of the European Accelerator R&D Roadmap* |
| ID155 | *Experimental Study of Rare Kaon Decays at J-PARC with KOTO and KOTO II* |
| ID157 | *Towards instrumentation for future HEP experiments* |

| ID | Title |
|---|---|
| ID158 | *Community input to the ESPP: Searches for Permanent Electric Dipole Moments* |
| ID159 | *Contribution of ALEGRO to the Update of the ESPP* |
| ID160 | *Beam monitoring through beamstrahlung* |
| ID161 | *Call for a new approach to detector developments involving microelectronics* |
| ID162 | *Sustainability Assessment of Future Accelerators* |
| ID163 | *Science4Peace: A Plea for Continued Peaceful International Scientific Cooperation* |
| ID164 | *Science of the LISA mission: A Summary for the ESPP* |
| ID165 | *A Possible Future Use of the LHC Tunnel* |
| ID167 | *AI-RDs: A EuCAIF proposal to structure AI research in Particle Physics* |
| ID168 | *Addressing Scientific Computing for HEP needs beyond Experimental Particle Physics* |
| ID169 | *Pakistan Input to the ESPP 2026* |
| ID170 | *Highlights of the HL-LHC physics projections by ATLAS and CMS* |
| ID171 | *Proposal from the NA61/SHINE Collaboration for the update of the ESPP* |
| ID172 | *AWAKE – Input to the ESPPU on behalf of the AWAKE Collaboration* |
| ID173 | *Input from the Netherlands to ESPP 2026* |
| ID174 | *Phase-One LHeC* |
| ID175 | *XLZD: The low-background observatory for astroparticle physics* |
| ID176 | *National input from the United Kingdom to the 2026 Update to the ESPP* |
| ID177 | *Norwegian Input to the ESPP* |
| ID179 | *The HIBEAM/NNBAR program* |
| ID180 | *SPECTRUM Project: Input to the ESPP Update 2026* |
| ID181 | *HOLMES+ Project* |
| ID182 | *CONUS100: Precision neutrino physics with coherent elastic scattering of reactor neutrinos* |
| ID183 | *Input of the German Nuclear and Hadron Physics Community to the ESPPU 2026 Regarding the Programs at CERN, at FAIR, and Related Activities* |
| ID184 | *Sensitivity study on* $H \to b\bar{b}$, $H \to WW^*$ *and* $HH \to b\bar{b}b\bar{b}$ *cross-sections and trilinear Higgs self-coupling with the MUSIC detector in* $\sqrt{s} = 10$ *TeV muon collisions* |
| ID185 | *EuCAIF Recommendations for Scaling AI Capabilities in Particle Physics* |
| ID186 | *CERN-MEDICIS: A unique facility for the production of innovative biomedical research radionuclides* |
| ID187 | *ATLAS Software and Computing for the Future* |
| ID188 | *LEP3: A High-Luminosity* $e^+e^-$ *Higgs & Electroweak Factory in the LHC Tunnel* |
| ID189 | *The ATLAS Upgrade for the HL-LHC* |
| ID190 | *Particle Physics at the European Spallation Source* |
| ID191 | *Input to the ESPP: Strong-Field Quantum Electrodynamics* |
| ID192 | *National Input to the ESPP 2026 – Slovenia* |
| ID193 | *ESPPU 2026: input from the Romanian experimental particle physics community* |
| ID194 | *Brazilian input to the ESPPU* |
| ID195 | *Estonian national input to the ESPP* |
| ID196 | *Prospects in flavour physics at the FCC* |
| ID197 | *AMoRE Experiment* |
| ID198 | *Einstein Telescope – The future European Gravitational Wave Observatory* |
| ID199 | *CODEX-b: Opening New Windows to the Long-Lived Particle Frontier at the LHC* |
| ID200 | *Neutrinos from Stored Muons (nuSTORM)* |

| ID | Title |
| --- | --- |
| ID201 | *Statement of the Pierre Auger Collaboration as input for the 2026 ESPP* |
| ID202 | *U.S. Higgs Factory Consortium input to ESG on CERN's future collider options* |
| ID203 | *Austrian input to the ESPPU 2026* |
| ID204 | *R&D on quantum sensors for particle physics: the DRD5 collaboration* |
| ID205 | *The Belle II Experiment at SuperKEKB* |
| ID206 | *ESPPU 2026 contribution of Finland* |
| ID207 | *The Muon Collider* |
| ID208 | *Initial INFN Input on the Update of the ESPP* |
| ID209 | *FCC: QCD physics* |
| ID210 | *Proton-Driven Plasma Wakefield Acceleration for Future HEP Colliders* |
| ID211 | *Expression of Interest for the ALLEGRO Full-Detector Concept for FCC-ee* |
| ID213 | *LHCspin: a Polarized Gas Target for LHC* |
| ID214 | *The Large Hadron electron Collider (LHeC) as a bridge project for CERN* |
| ID215 | *A roadmap for astroparticle theory in Europe* |
| ID216 | *Advancing Global Collaboration through the Electron-Ion Collider (EIC)* |
| ID217 | *Prospects in Electroweak, Higgs and Top physics at FCC* |
| ID218 | *COMET – An Experiment to Search for Muon-to-Electron Conversion in Nuclear Field at J-PARC* |
| ID219 | *RF developments for future colliders* |
| ID220 | *Advanced Accelerator and HEP Developments through Networking between the Large Particle Physics Laboratories and CERN* |
| ID221 | *The DArk Messenger Searches at an Accelerator Experiment, A Case of a Table-Top Scale Experiment at a Beam Dump* |
| ID222 | *To build or not to build: FCC* |
| ID223 | *Projections for Key Measurements in Heavy Flavour Physics* |
| ID224 | *Community Support for Physics with high-luminosity proton-nucleus collisions at the LHC* |
| ID225 | *The Project 8 Neutrino Mass Experiment* |
| ID226 | *ICARUS contribution to the European Strategy Process 2026* |
| ID227 | *Prospects for physics at FCC-hh* |
| ID228 | *Quantum Technologies in High Energy Physics: The CERN Quantum Technology Initiative input to the ESPP* |
| ID229 | *Development of Gaseous Detector Technologies: CERN-DRD1 Collaboration Model* |
| ID230 | *US National Input to the ESPPU* |
| ID231 | *Super Tau Charm Facility* |
| ID232 | *The Short-Baseline Near Detector at Fermilab* |
| ID233 | *FCC Integrated Programme Stage 1: The FCC-ee* |
| ID234 | *Enhancing New Physics Searches with a Future Beam Dump Configuration at SBND* |
| ID235 | *Summary Report of the Physics Beyond Colliders Study at CERN* |
| ID236 | *Expanding the neutrino facility at the South Pole: IceCube-Upgrade and IceCube-Gen2* |
| ID237 | *The Elastic Analysis Facility's (EAF) Contribution to the Future of Analysis at Multi-Experiment Institutions and Future Colliders* |
| ID238 | *The Hyper-Kamiokande experiment: input to the update of the ESPP* |

| ID | Title |
|---|---|
| ID239 | *Physics Prospects for a near-term Proton-Proton Collider* |
| ID240 | *Key4hep: A Software Framework for Future Colliders* |
| ID241 | *The FCC integrated programme: a physics manifesto* |
| ID242 | *Prospects in BSM physics at FCC* |
| ID243 | *High Field Magnet Programme – European Strategy Input* |
| ID244 | *The Importance of Test Beams for Particle Physics worldwide* |
| ID245 | *MUonE Contribution to the European Strategy: status of the project* |
| ID246 | *U.S. interest in high-energy nuclear physics at the LHC* |
| ID247 | *FCC Integrated Programme Stage 2: The FCC-hh* |
| ID248 | *Canadian submission to the ESPP* |
| ID249 | *KM3NeT contribution to the ESPP* |
| ID250 | *Rich dark sectors* |
| ID251 | *Physics opportunities with high-brightness, high-intensity muon beams at CERN: a staged approach* |
| ID252 | *Cosmic Microwave Background (CMB) as a window to particle physics* |
| ID253 | *Input to the ESPPU 2026: The COHERENT Experiment* |
| ID254 | *ESPP Contribution – Gravitational Field of Proton Bunches* |
| ID255 | *A Flexible Strategy for the Future of Particle Physics at CERN* |
| ID256 | *Superconductivity Global Alliance (ScGA) European Strategy Input* |
| ID257 | *National input from Belgium to the 2026 Update to the ESPP* |
| ID258 | *The Electron Capture in Ho-163 experiment – ECHo* |
| ID259 | *Equity, Diversity and Inclusion: Strategic Elements of the European Particle Physics Programme* |
| ID260 | *Quantum Sensing for Dark Matter and Gravitational Waves* |
| ID262 | *Expression of Interest Toward MCMOS Time of Flight Tracking Layers for a detector at FCC-ee* |
| ID263 | *Theia: technology and summary of physics program* |
| ID264 | *Coherent Neutrino-Nucleus Scattering Experiment (CONNIE)* |
| ID265 | *Input to ESPPU: Sustainability* |
| ID266 | *Novel 3D-Projection Pixelated Detector for Next-Generation High-Energy Physics Experiments* |
| ID267 | *Constraining the real scalar singlet extension of the SM* |
| ID268 | *Dark Matter Searches with Low-Radioactivity Argon* |
| ID269 | *LEP Data@EDM4HEP: Mitigating Data Loss Risks and Strengthening FCC-ee Potential Studies* |
| ID270 | *LEGEND-1000* |
| ID271 | *A Spin-Based Pathway to Testing the Quantum Nature of Gravity* |
| ID272 | *Neutrino Astronomy in the Peruvian Andes with TAMBO* |
| ID273 | *2026 EPPSU input from the ANUBIS Collaboration* |
| ID274 | *Dark Matter Complementarity: from the Snowmass process to the EPPSU* |
| ID275 | *Status of the International Linear Collider* |
| ID276 | *Input by the AstroParticle Physics European Consortium* |
| ID277 | *Initial input from the Danish CERN Community to the ESPPU* |
| ID279 | *Australian community submission to the 2025 ESPP: Collider recommendation* |
| ID280 | *Bulgarian national input to the ESPP* |
| ID281 | *Future Colliders Comparative Evaluation – Working Group Report* |